\documentclass[manuscript]{aastex}

\begin{document}


\title{Mid-Infrared Galaxy Morphology from the {\it Spitzer Survey of
Stellar Structure in Galaxies} (S$^4$G): The Imprint of the de
Vaucouleurs Revised Hubble-Sandage Classification System at 3.6$\mu$m}


\author{Ronald J. Buta\altaffilmark{1},
Kartik Sheth\altaffilmark{2,3,4},
Michael Regan\altaffilmark{5},
Joannah L. Hinz\altaffilmark{6},
Armando Gil de Paz\altaffilmark{7},
Karin Men{\'e}ndez-Delmestre\altaffilmark{8},
Juan-Carlos Munoz-Mateos\altaffilmark{7},
Mark Seibert\altaffilmark{8},
Eija Laurikainen\altaffilmark{9},
Heikki Salo\altaffilmark{9},
Dimitri A. Gadotti\altaffilmark{10},
E. Athanassoula\altaffilmark{11},
Albert Bosma\altaffilmark{11},
Johan H. Knapen\altaffilmark{12,13},
Luis C. Ho\altaffilmark{8},
Barry F. Madore\altaffilmark{8},
Debra M. Elmegreen\altaffilmark{14}
Karen L. Masters\altaffilmark{15},
Sebastien Comer\'on\altaffilmark{16},
Manuel Aravena\altaffilmark{2}
}
\altaffiltext{1}{Department of Physics and Astronomy, University of Alabama, Box 870324, Tuscaloosa, AL 35487}
\altaffiltext{2}{National Radio Astronomy Observatory / NAASC, 520 Edgemont
 Road, Charlottesville, VA 22903}
\altaffiltext{3}{Spitzer Science Center, 1200 East California Boulevard, Pa
sadena, CA 91125}
\altaffiltext{4}{California Institute of Technology, 1200 East California B
oulevard, Pasadena, CA 91125}
\altaffiltext{5}{Space Telescope Science Institute, 3700 San Martin Drive, Baltimore,MD
212185721}
\altaffiltext{6}{Steward Observatory, University of Arizona, 933 N. Cherry Ave., Tucson,
AZ 85721}
\altaffiltext{7}{Departamento de Astrofísica y CC. de la Atmósfera, Universidad
Complutense de Madrid, Avda. de la Complutense, s/n, E-28040 Madrid, Spain}
\altaffiltext{8}{The Observatories of the Carnegie Institution for Science,
California Institute of Technology, MC 249-17, Pasadena, CA 91125}
\altaffiltext{9}{Division of Astronomy, Department of Physical Sciences, University
of Oulu, Oulu, FIN-90014, Finland}
\altaffiltext{10}{European Southern Observatory, Casilla 19001, Santiago 19, Chile}
\altaffiltext{11}{Laboratoire d'Astrophysique de Marseille, Univ\'ersit\'e de Provence/CNRS UMR 6110, 38 rue Fr\'ed\'eric Joliot Curie, 13388 Marseille C\'ed\'ex 4, France}
\altaffiltext{12}{Instituto de Astrof\'\i sica de Canarias, E-38200 La
Laguna, Tenerife, Spain}
\altaffiltext{13}{Departamento de Astrof\'\i sica, Universidad de La Laguna, E-38205 La Laguna, Tenerife, Spain}
\altaffiltext{14}{Department of Physics \& Astronomy, Vassar College,
Poughkeepsie, NY 12604}
\altaffiltext{15}{Institute for Cosmology and Gravitation, University
of Portsmouth, Dennis Sciama Building, Burnaby Road, Portsmouth, PO1
3FX, UK}
\altaffiltext{16}{Korea Astronomy and Space Science Institute 838, Daedeokdae-ro, Yuseong-
gu, Daejeon Republic of Korea 305-38}



\begin{abstract}
{\it Spitzer Space Telescope} Infrared Array Camera (IRAC) imaging
provides an opportunity to study all known morphological types of
galaxies in the mid-IR at a depth significantly better than ground-based
near-infrared and 
optical images. The goal of this study is to examine the
imprint of the de Vaucouleurs classification volume in the
3.6$\mu$m band, which is the best {\it Spitzer} waveband for galactic stellar
mass morphology owing to its depth and its reddening-free sensitivity
mainly to older stars. For this purpose, we have prepared
classification images for 207 galaxies from the {\it Spitzer} archive, most
of which are formally part of the {\it Spitzer Survey of Stellar Structure
in Galaxies} (S$^4$G), a {\it Spitzer} post-cryogenic (``warm")
mission Exploration Science Legacy Program survey of 2,331 galaxies
closer than 40 Mpc. For the purposes of morphology, the galaxies are
interpreted as if the images are {\it blue light}, the historical
waveband for classical galaxy classification studies. We find that 
3.6$\mu$m classifications are well-correlated with blue-light
classifications, to the point where the essential features of many
galaxies look very similar in the two very different wavelength
regimes. Drastic differences are found only for the most dusty
galaxies. Consistent with a previous study by Eskridge et al. (2002),
the main difference between blue light and mid-IR types is an
$\approx$1 stage interval difference for S0/a to Sbc or Sc galaxies,
which tend to appear ``earlier" in type at 3.6$\mu$m due to the
slightly increased prominence of the bulge, the reduced effects of
extinction, and the reduced (but not completely eliminated)
effect of the extreme population I stellar
component. We present an atlas of all of the 207 galaxies analyzed
here, and bring attention to special features or galaxy types, such as
nuclear rings, pseudobulges, flocculent spiral galaxies, I0 galaxies,
double-stage and double-variety galaxies, and outer rings, that are
particularly distinctive in the mid-IR.

\end{abstract}




\section{Introduction}

The {\it Spitzer Survey of Stellar Structure in Galaxies} (S$^4$G,
Sheth et al. 2010) is a systematic imaging survey with the 
Infrared Array Camera (IRAC, Fazio et al. 2004a) of 2,331 galaxies
in 3.6 and 4.5$\mu$m bands. The goal of the project is to derive
basic photometric parameters for quantitative analysis of these
galaxies for a variety of studies. Independent of these studies,
however, the S$^4$G database is an obvious goldmine for new
investigations of galaxy morphology in the mid-infrared, if only
because the images are deeper than anything achievable from reasonable
ground-based near-IR observations and also because the images are
homogeneous with respect to the point-spread function (PSF).

In this paper, we examine the imprint of the de Vaucouleurs revised
Hubble-Sandage classification system on galaxy morphology at 3.6$\mu$m,
using a subset of {\it Spitzer} archival images of galaxies that meet the
selection criteria of the S$^4$G (or are prominent companions to those
galaxies). These images have been processed through the S$^4$G
Pipeline as described in Sheth et al. (2010), and have significantly
improved quality compared to the post-Basic Calibrated Data (PBCD)
mosaics provided in the SSC archive for photometric analysis. The
goal of our analysis is not merely general Hubble classifications, but
detailed types as described in the {\it de Vaucouleurs Atlas of
Galaxies} (Buta, Corwin \& Odewahn 2007, hereafter the dVA). We want to
know how closely optical galaxy morphology maps into the mid-IR.
Section 2 gives some background to IR morphological studies, while
section 3 describes how we prepared the S$^4$G Pipeline processed
{\it Spitzer} images for detailed morphological study. Section 4 summarizes
our analysis of the {\it Spitzer} archival 3.6$\mu$m images. Our conclusions
are summarized in section 5.

\section{Near- and Mid-IR Galaxy Morphology}

Galaxy morphology in the near- and mid-IR, as in other wavebands, is
important to examine because of the information morphology carries on
processes of galactic evolution. Morphology is strongly correlated with
galactic star formation history, environmental density and
interaction/merger history, and the effectiveness with which internal
perturbations (such as bars) interact with other internal components
(such as the halo and the basic state of the disk). Early evolution was
probably dominated by merger events, and this has found support in
recent studies of the merger rate (e.g., Mihos \& Hernquist 1996). Slower
internal secular evolution has been increasingly invoked to explain the
disk-like properties of ``pseudobulges" (e.g., Kormendy \& Kennicutt
2004, Athanassoula 2005) and to account for the development of other
features, such as rings and spiral arms, in response to perturbations
such as bars and ovals. Together, these processes appear to combine to
produce the wide range of galaxy types that we see nearby.

Galaxy classification is still an important part of modern
extragalactic studies. Catalogues of galaxy Hubble types or
Hubble-related types are often the starting point of observational
investigations, and nearly 80 years of research have not negated or
made obsolete the classification systems (e.g., Hubble 1926, 1936;
Sandage 1961; Sandage \& Bedke 1994; de Vaucouleurs 1959) that form the
basis of catalogued types. The frequency and statistical properties of
structures in galaxies need to be known in order to determine how these
structures fit into the general scheme of galaxy evolution. As noted by
Fukugita et al. (2007), who used Sloan Digital Sky Survey (SDSS) color
images to classify 2253 bright galaxies, visual classification
``remains the best approach for classifying each galaxy into a Hubble
type with high confidence, at least for bright galaxies." This survey
was followed up by a more detailed morphological analysis by Nair \&
Abraham (2010), who recorded rings, bars, Hubble types, and other
features for more than 14,000 SDSS galaxies, as a means of facilitating
automatic classification. In addition, the Galaxy Zoo project (Lintott
et al. 2008) has provided basic morphological information for nearly
a million galaxies, from citizen science participation (see also
Banerj et al. 2010).

Although as a topic of research galaxy morphological studies began
with blue-sensitive photographic plates, in recent years the emphasis
has shifted considerably towards longer wavelengths. Early red and
near-infrared imaging showed that galaxies which, in blue light, show a
patchiness due to dust and complexes of recent star formation, become
smoother at the longer wavelengths owing to the reduced effects of
extinction and the de-emphasis of the younger component. The longer
wavelengths emphasize the older stellar components, including old disk
giants and Population II stars. Although photographic red and
near-infrared imaging were possible in the 1950s-1980s, and were even
the subjects of large sky surveys, no large-scale systematic galaxy
morphological studies were ever based on these surveys. Only the blue
surveys were used for systematic galaxy classification (e.g., Nilson
1973; Lauberts 1982; Corwin et al. 1985; Buta 1995), because the
original Hubble system was based on blue light photographs. Even so, it
was already known that features like inner bars (e.g., Hackwell \&
Schweizer 1983; Scoville et al. 1988; Thronson et al. 1989; Telesco et
al. 1991; Block \& Wainscoat 1991; Rix \& Rieke 1993), triaxial nuclear
bulges (e.g., Zaritsky \& Lo 1986), and regular bars (e.g., Talbot et
al.1979) could be partly obscured or simply less prominent in blue
light images, but become more visible in the red and infrared.
Detailed studies of bar fractions in the near-IR have shown a
comparable (or sometimes slightly larger) fraction to that given by the
early blue light studies. This issue is further discussed in section
4.6.

It was also known that spiral structure which may appear ``flocculent"
in blue light and at 0.8$\mu$m (Elmegreen 1981) could appear more
global (i.e., continuous and large-scale, or ``grand design") at
2.2$\mu$m (Thornley 1996). The differences between optical and near-IR
morphologies for some galaxies appeared to be so great that in early
studies, it was suggested that there is a ``duality" of galactic
structure, in the sense that the Population I and II morphologies are
decoupled (Block \& Wainscoat 1991; Block et al. 1994, 2004).

Several developments brought large-scale digital IR imaging to the
forefront of galaxy morphological studies. The first was extending
near-IR imaging from individual galaxies or parts of galaxies to
statistical-sized samples. This began with near-IR 
surface photometry of 50 spirals by Elmegreen
\& Elmegreen (1984, 1987) and of 86 galaxies by de Jong \& van der Kruit (1994).
The first major near-IR survey designed for large-scale morphological
studies was the Ohio State University Bright Spiral Galaxy Survey 
(OSUBSGS,
Eskridge et al.  2002), which included optical $BVRI$ and near-IR $JHK$
images of 205 bright galaxies of types S0/a to Sm in a statistically
well-defined sample selected to have total blue magnitude
$B_T$$\leq$12.0 and isophotal diameter
$D_{25}$$\leq$6\rlap{.}$^{\prime}$5. This survey allowed a direct
demonstration of how galaxy morphology actually changes from optical to
near-IR wavelengths, not merely for a small, selected sample of
galaxies, but for a large sample covering all spiral subtypes.

The OSUBSGS was later complemented by the {\it Near-Infrared S0 Survey}
(NIRS0S, Laurikainen et al. 2005, 2006, 2010; Buta et al. 2006), a
$K_s$-band imaging survey of about 180 early-type galaxies in the type
range S0$^-$ to Sa, but mostly including S0s, some of which were
misclassified as ellipticals in the Third Reference Catalogue of Bright
Galaxies (RC3, de Vaucouleurs et al. 1991; see below). Although S0
galaxies are dominated by old stars and are smooth even in blue light
images, the $K_s$ band was chosen to complement the OSUBSGS sample of
spirals in order to make a fair comparison between bar strengths and
bulge properties of S0s and spirals. Also, S0 galaxies are not
necessarily dust-free, and near-IR imaging is still necessary to
penetrate what dust they have.

The second development was the {\it Two-Micron All-Sky Survey} (2MASS,
Skrutskie et al. 1997), which provided $JHK_s$ images of a much larger
galaxy sample, although these lack the depth of the OSUBSGS and NIRS0S
images in general. 2MASS provided considerable
information on near-infrared galaxy morphology, which led to the
extensive {\it 2MASS Large Galaxy Atlas} (Jarrett et al. 2003). 2MASS
images were also used by Men{\'e}ndez-Delmestre et al. (2007) to study
quantitative near-IR bar classification and its comparison to 
RC3 classifications.

The third development was the launch of the {\it Spitzer Space
Telescope} (Werner et al. 2004) and subsequent IRAC imaging surveys
such as the {\it Spitzer} Infrared Nearby Galaxies Survey (SINGS,
Kennicutt et al. 2003). SINGS provided both optical and mid-IR imaging
of 75 galaxies of all types at a depth much greater than that of the
OSUBSGS, NIRS0S, or 2MASS. SINGS was followed by the {\it Local Volume
Legacy} project (LVL, Kennicutt et al.  2007), which will provide IRAC
images (as well as images in other passbands) of 258 mostly late-type
galaxies nearer than 11 Mpc.

Among these various sets of data, the NIRS0S is the only IR imaging
survey carried out to a significant extent with 4-m class telescopes.
The full width at half maximum (FWHM) of the stellar PSF for these images
is generally $<$1$^{\prime\prime}$ compared to
1\rlap{.}$^{\prime\prime}$5-2$^{\prime\prime}$ for OSUBSGS images,
3$^{\prime\prime}$ for 2MASS images, and 1\rlap{.}$^{\prime\prime}$75
for SINGS images.

The S$^4$G has been designed to provide a set of very deep mid-IR
images at good spatial resolution of an unprecedentedly large sample
of nearby galaxies: 2,331 galaxies of all Hubble types within a
distance of 40Mpc. The images are being obtained in the IRAC 3.6 and
4.5$\mu$m bands.  Of this sample, 597 galaxies are already in the {\it
Spitzer} archive, and the goal of the S$^4$G is to add the 1,734
remaining sample galaxies as part of the {\it Spitzer} Warm Mission.
The advantages of {\it Spitzer} images lie in their homogeneity and,
most of all, in their considerable depth at wavelengths where
ground-based observations (e.g., at 2.2$\mu$m) would suffer very high
background (sky) emission. Also, the IRAC 3.6 and 4.5$\mu$m bands are
in a regime where extinction is even lower than in the near-IR $JHK$
bands, and where contamination by dust heated by stellar processes is
also still low, so that the bands are largely sampling the backbone of
stellar mass in galaxies (Pahre et al.  2004). However, there
is a still a contribution from young stars, the nature of which
is discussed in section 4.5.

We use a subsample of 167 S$^4$G-pipeline processed {\it Spitzer}
archival 3.6$\mu$m images to classify 207 bright galaxies (including 24
pairs and 16 additional galaxies in small groups) in the formal
framework of the de Vaucouleurs revised Hubble-Sandage classification
system, as revised and updated in the dVA. This subset has no
distinguishing characteristics other than being archival galaxies
selected by S$^4$G team members for initial study and analysis. Such a
sample will have the selections of a variety of programs that were
granted observing time. Figure~\ref{histos} shows the frequency
distributions (solid histograms) of RC3 classifications (stages,
families, and varieties), absolute blue magnitudes $M_B^o$, total
extinction-corrected color indices $(B-V)_T^o$, and
extinction-corrected mean blue light effective surface brightnesses
$(m_e^{\prime})_o$, the latter three parameters also from, or based on,
RC3 data. These plots show that the sample galaxies cover a broad range
of properties. The full range of galaxy types, from E to Im, as well as
giants and dwarfs, is represented. For comparison, Figure~\ref{histos}
also shows the frequency distributions (dashed histograms) of the same
parameters for the full S$^4$G sample. The distrbutions are similar
except that our subset has relatively more E and S0 galaxies ahd high
luminosity galaxies than the full S$^4$G sample.

A partial attempt to classify galaxies using the de Vaucouleurs system
was also made by Eskridge et al. (2002), who applied the system using
OSUBSGS near-IR $H$-band (1.65$\mu$m) images {\it as if they were blue
light images.} That is, the framework of the de Vaucouleurs system was
used without regard to the wavelength used to actually define the
system.  Eskridge et al. showed that although some galaxies can look
very different in the near-IR as compared to the $B$-band, in general
the differences are not so large as to make near-IR types uncorrelated
with optical types. As we will show here also, the imprint of the de
Vaucouleurs $B$-band classification volume holds well in the {\it
Spitzer} 3.6$\mu$m band, and the classifications correlate closely.
The same was also found to hold true in the mid/near-UV, at least for
later type galaxies (e.g., Windhorst et al. 2002). Taylor-Mager et al.
(2007) also found only a mild dependence of
concentration-asymmetry-star formation (CAS) parameters on wavelength,
suggesting that good morphological correlation between different
wavelength regimes is probably valid at all wavelengths dominated by
the emission from stellar photospheres.

\section{Preparation of the Images}

The goals of the S$^4$G project are described by Sheth et al.
(2010). One goal was the processing of the {\it Spitzer} images of all
archival galaxies with the same S$^4$G pipeline as the new Warm Mission
data in the S$^4$G sample. This processing was done to prepare the
images for detailed studies such as photometric decompositions, Fourier
analysis, and color index analysis. Like the SINGS images, the scale of
the final S$^4$G pipeline-processed images is
0\rlap{.}$^{\prime\prime}$75 per pixel, with a FWHM of
$\approx$1\rlap{.}$^{\prime\prime}$8.

The requirements for galaxy morphological classification are no less
stringent than they would be for photometric analysis. The images must
be accurately background-subtracted, and the point-spread function
should be reasonably narrow (FWHM $<$ 2\rlap{.}$^{\prime\prime}$5),
allowing about 100 resolution elements across the major axis.
The background in {\it Spitzer} images is low,
but variations due to zodiacal light are sometimes present. Bright
foreground stars can affect some archival images, but scattered light
from such stars should be less of a problem for the 1,734 Warm Mission
galaxies because special precautions are being taken to prevent bright
stars from falling into scattering zones.

\begin{figure}
\figurenum{1.1}  
\plotone{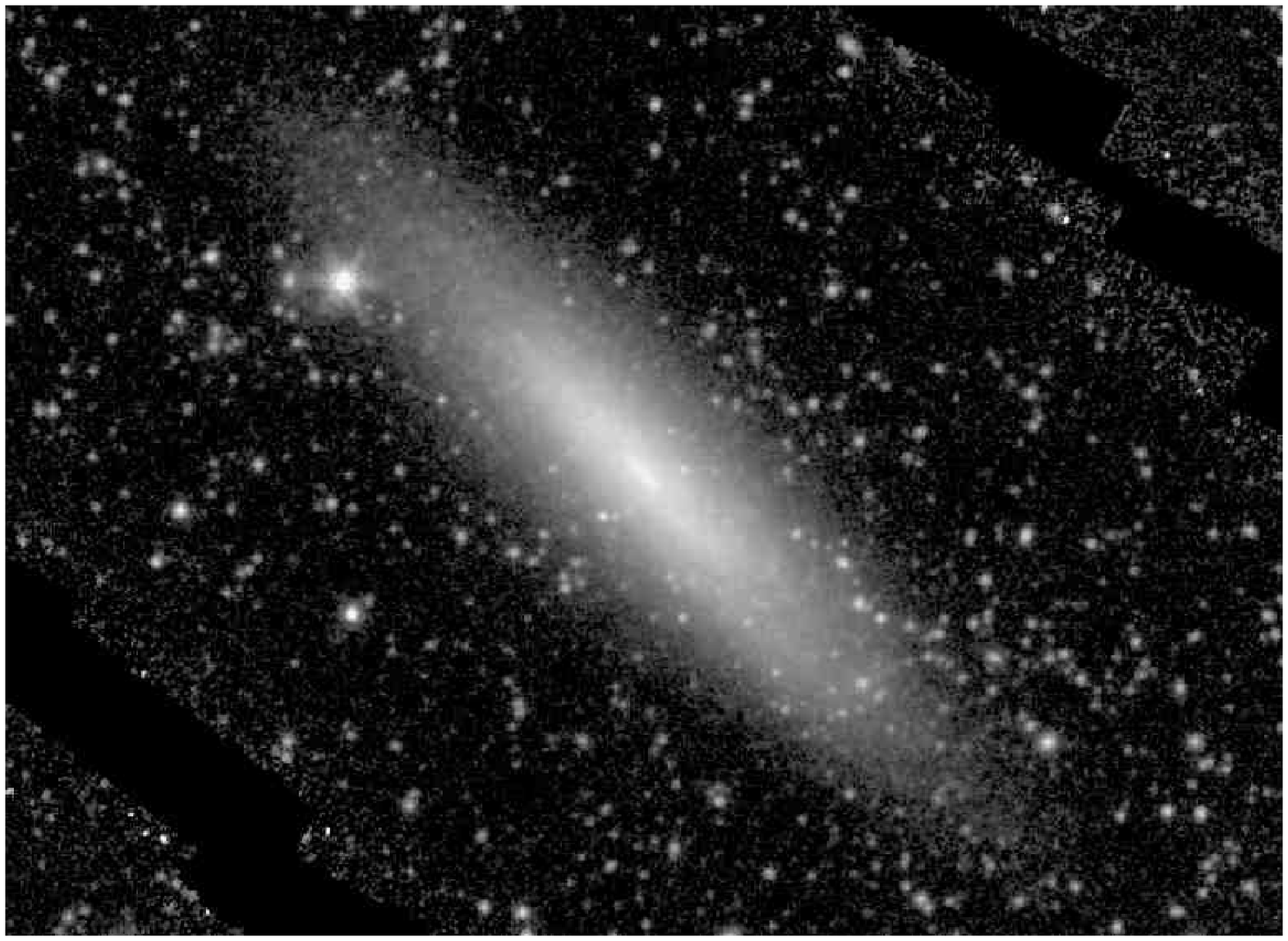}
 \vspace{2.0truecm}
 \caption{
{\bf NGC    24   }              - S$^4$G mid-IR classification:    S(rs)d: sp                                            ; Filter: IRAC 3.6$\mu$m; North:   up, East: left; Field dimensions:   7.9$\times$  5.8 arcmin; Surface brightness range displayed: 16.5$-$28.0 mag arcsec$^{-2}$}                 
\label{NGC0024}     
 \end{figure}

\begin{figure}
\figurenum{1.2}  
\plotone{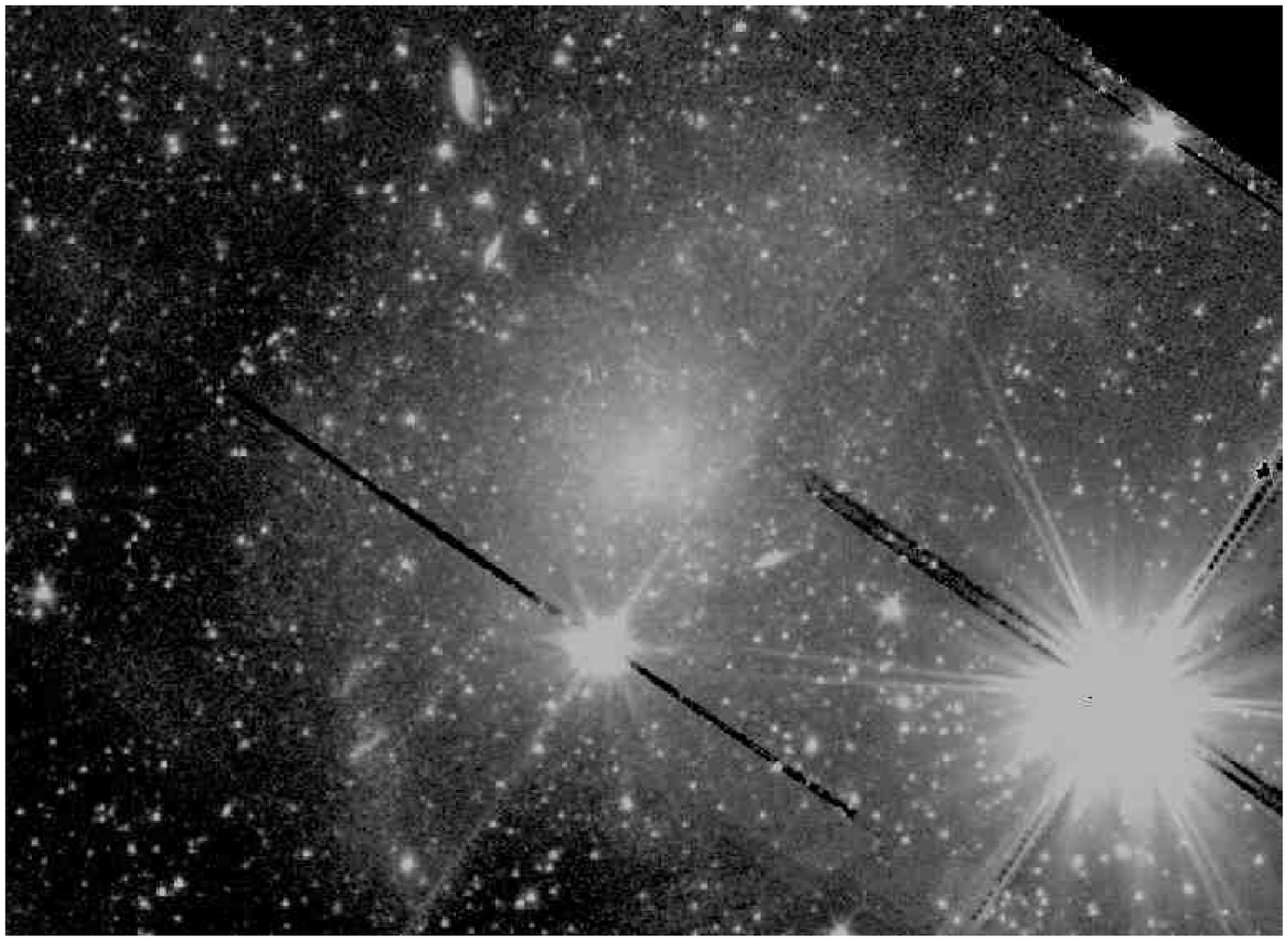}
 \vspace{2.0truecm}
 \caption{
{\bf NGC    45   }              - S$^4$G mid-IR classification:    SA(s)dm                                               ; Filter: IRAC 3.6$\mu$m; North:   up, East: left; Field dimensions:  11.3$\times$  8.2 arcmin; Surface brightness range displayed: 18.7$-$28.0 mag arcsec$^{-2}$}                 
\label{NGC0045}     
 \end{figure}

\begin{figure}
\figurenum{1.3}  
\plotone{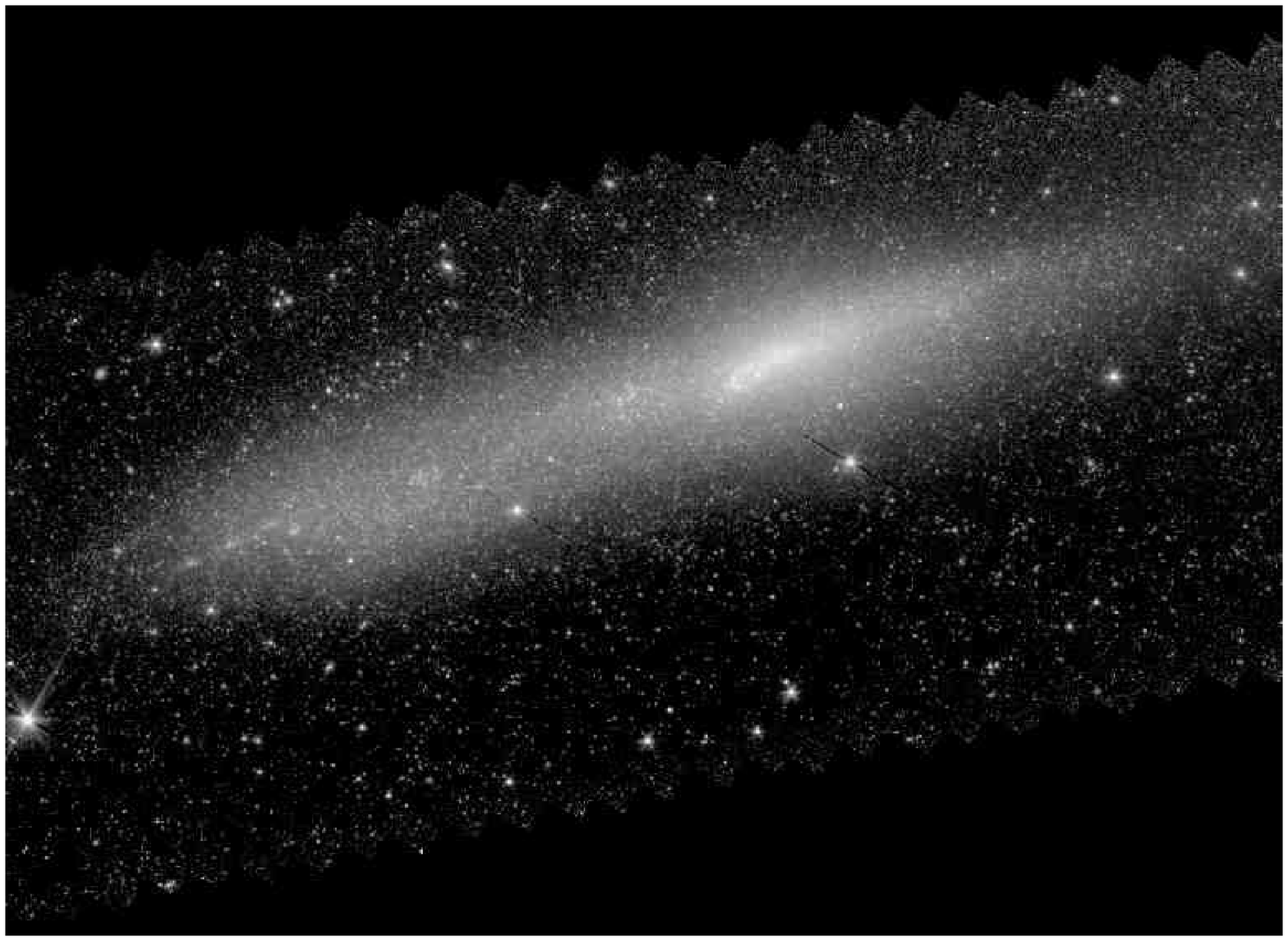}
 \vspace{2.0truecm}
 \caption{
{\bf NGC    55   }              - S$^4$G mid-IR classification:    SB(s)m sp                                             ; Filter: IRAC 3.6$\mu$m; North:   up, East: left; Field dimensions:  31.5$\times$ 23.0 arcmin; Surface brightness range displayed: 16.0$-$28.0 mag arcsec$^{-2}$}                 
\label{NGC0055}     
 \end{figure}

\begin{figure}
\figurenum{1.4}  
\plotone{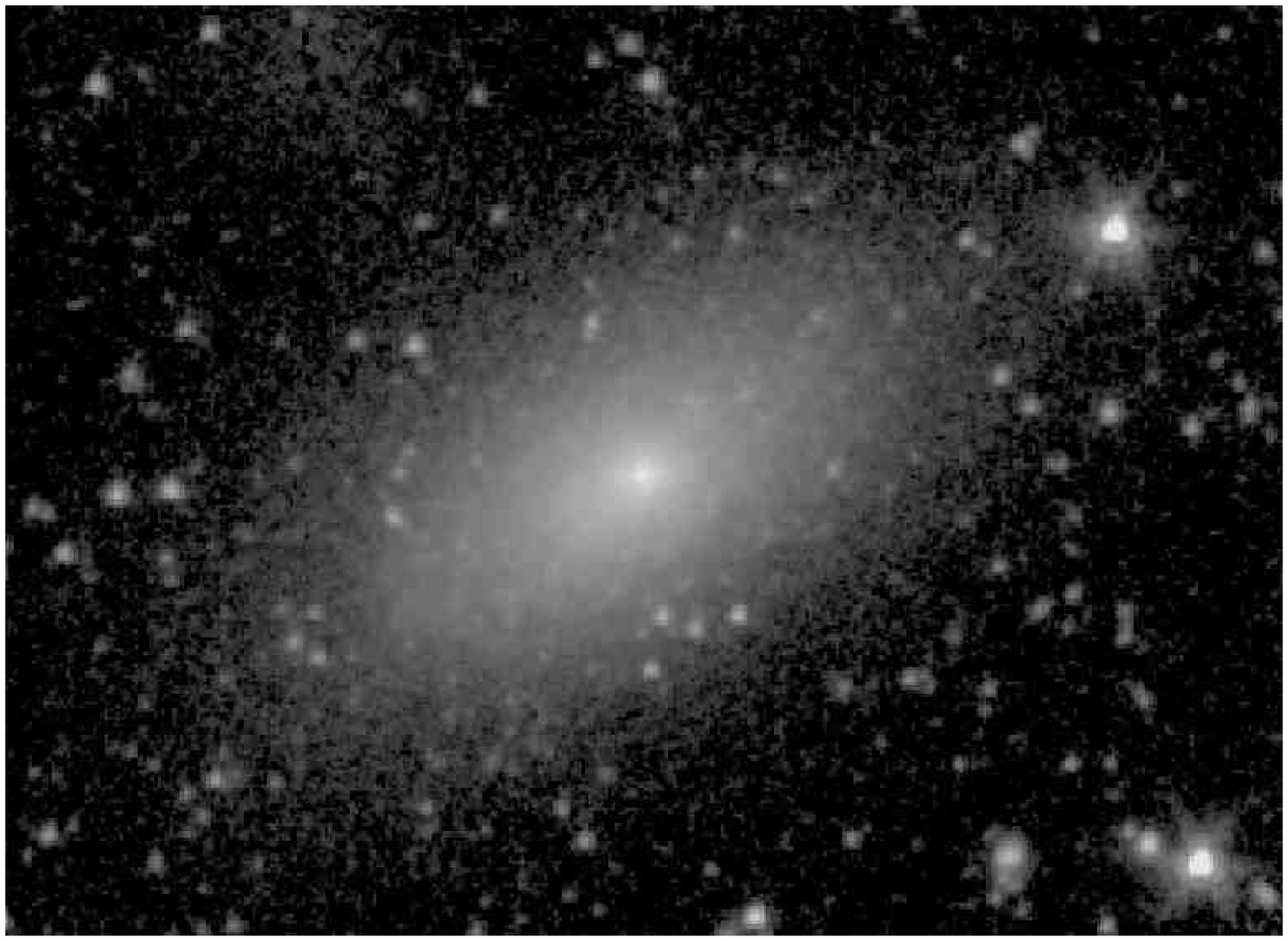}
 \vspace{2.0truecm}
 \caption{
{\bf NGC    59   }              - S$^4$G mid-IR classification:    dE5,N                                                 ; Filter: IRAC 3.6$\mu$m; North:   up, East: left; Field dimensions:   4.2$\times$  3.1 arcmin; Surface brightness range displayed: 16.0$-$28.0 mag arcsec$^{-2}$}                 
\label{NGC0059}     
 \end{figure}

\begin{figure}
\figurenum{1.5}  
\plotone{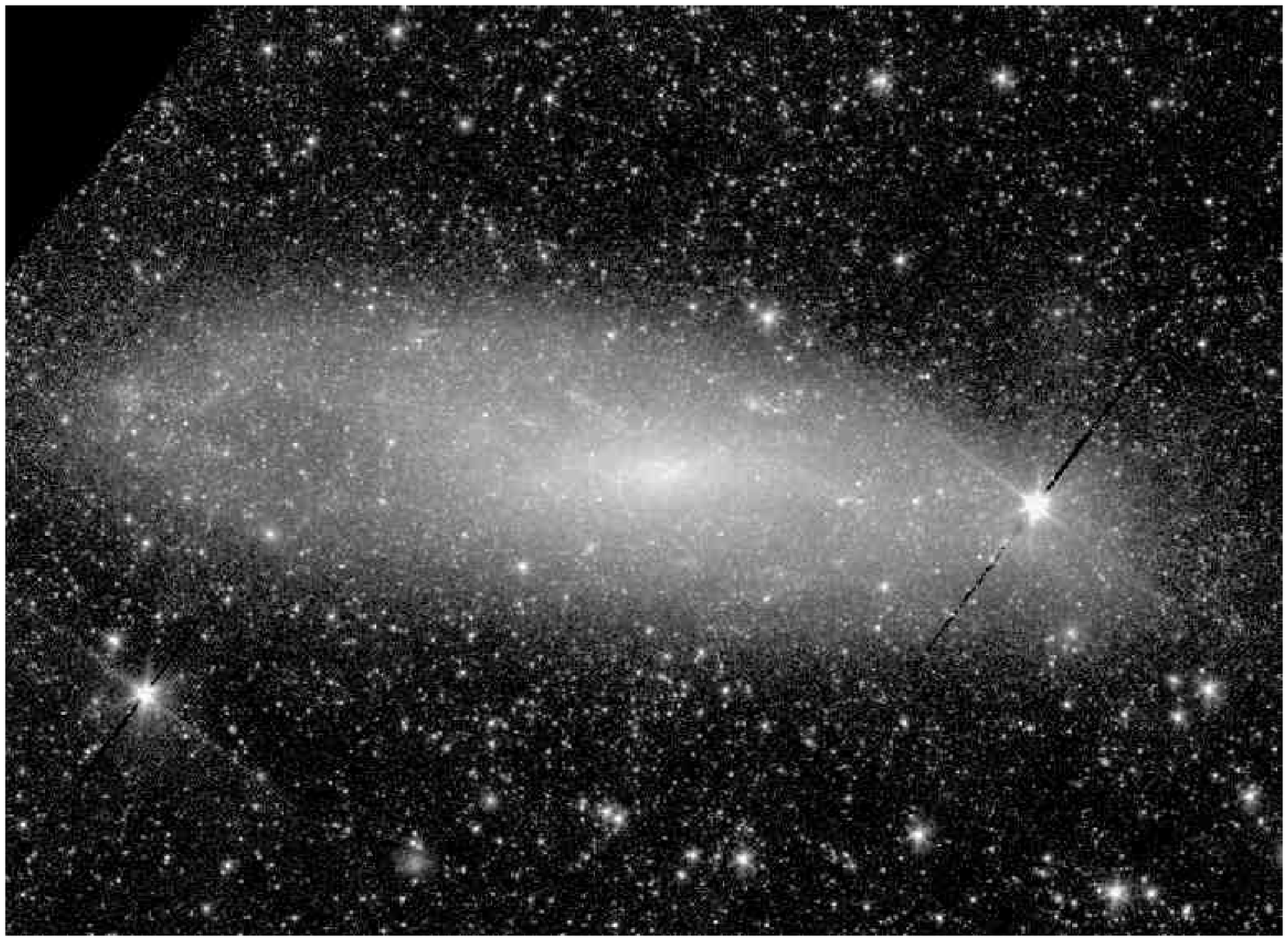}
 \vspace{2.0truecm}
 \caption{
{\bf NGC   247   }              - S$^4$G mid-IR classification:    SAB(s)d                                               ; Filter: IRAC 3.6$\mu$m; North: left, East: down; Field dimensions:  22.6$\times$ 16.4 arcmin; Surface brightness range displayed: 17.5$-$28.0 mag arcsec$^{-2}$}                 
\label{NGC0247}     
 \end{figure}

\begin{figure}
\figurenum{1.6}  
\plotone{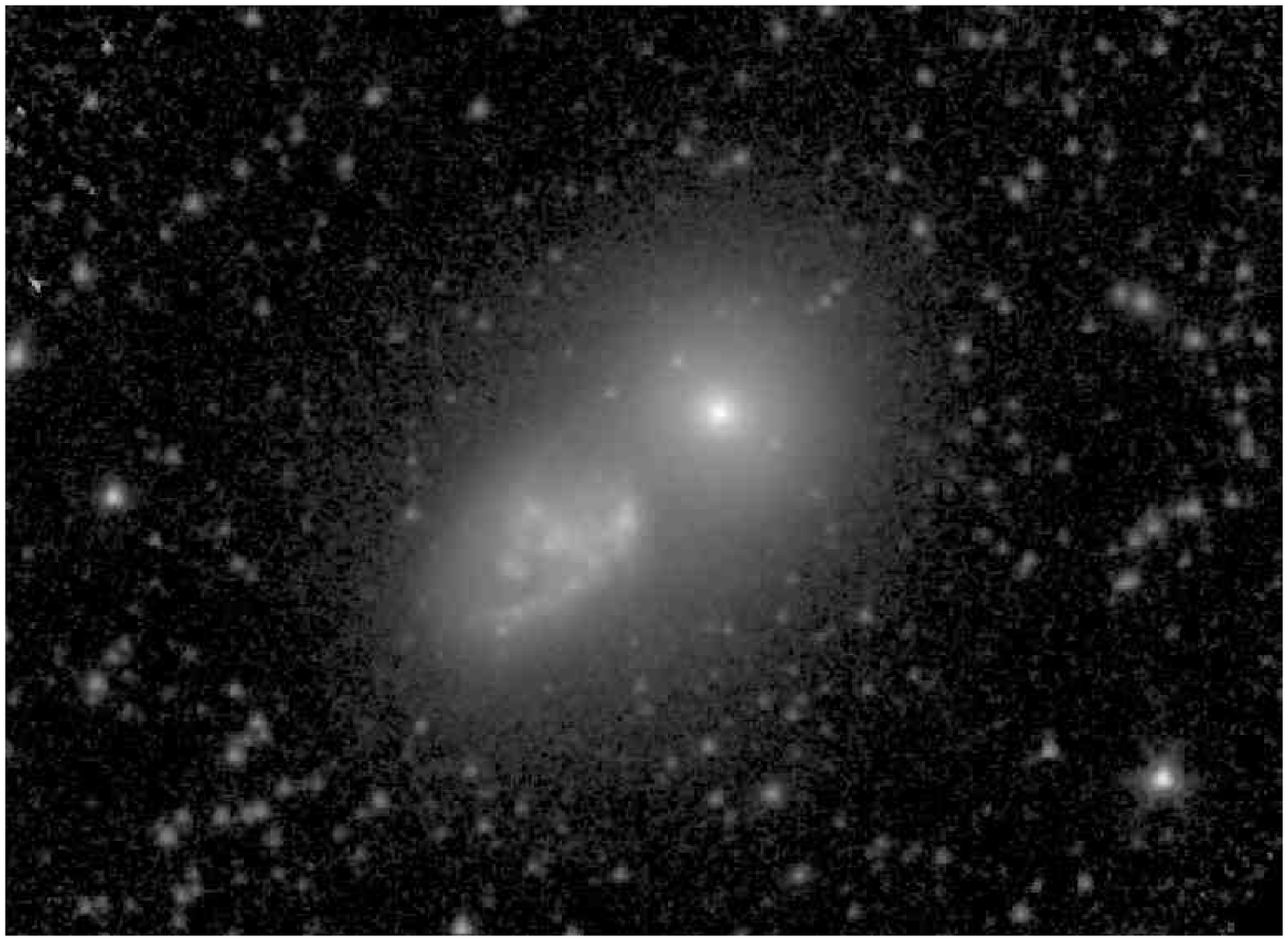}
 \vspace{2.0truecm}
 \caption{
{\bf NGC   274} (upper right) and {\bf NGC   275} (lower left) - S$^4$G mid-IR classifications:    (R)SA(l)0$^-$, S pec, respectively; Filter: IRAC 3.6$\mu$m; North:   up, East: left; Field dimensions:   5.3$\times$  3.8 arcmin; Surface brightness range displayed: 13.5$-$28.0 mag arcsec$^{-2}$}                                                         
\label{NGC0274}     
 \end{figure}

\begin{figure}
\figurenum{1.7}  
\plotone{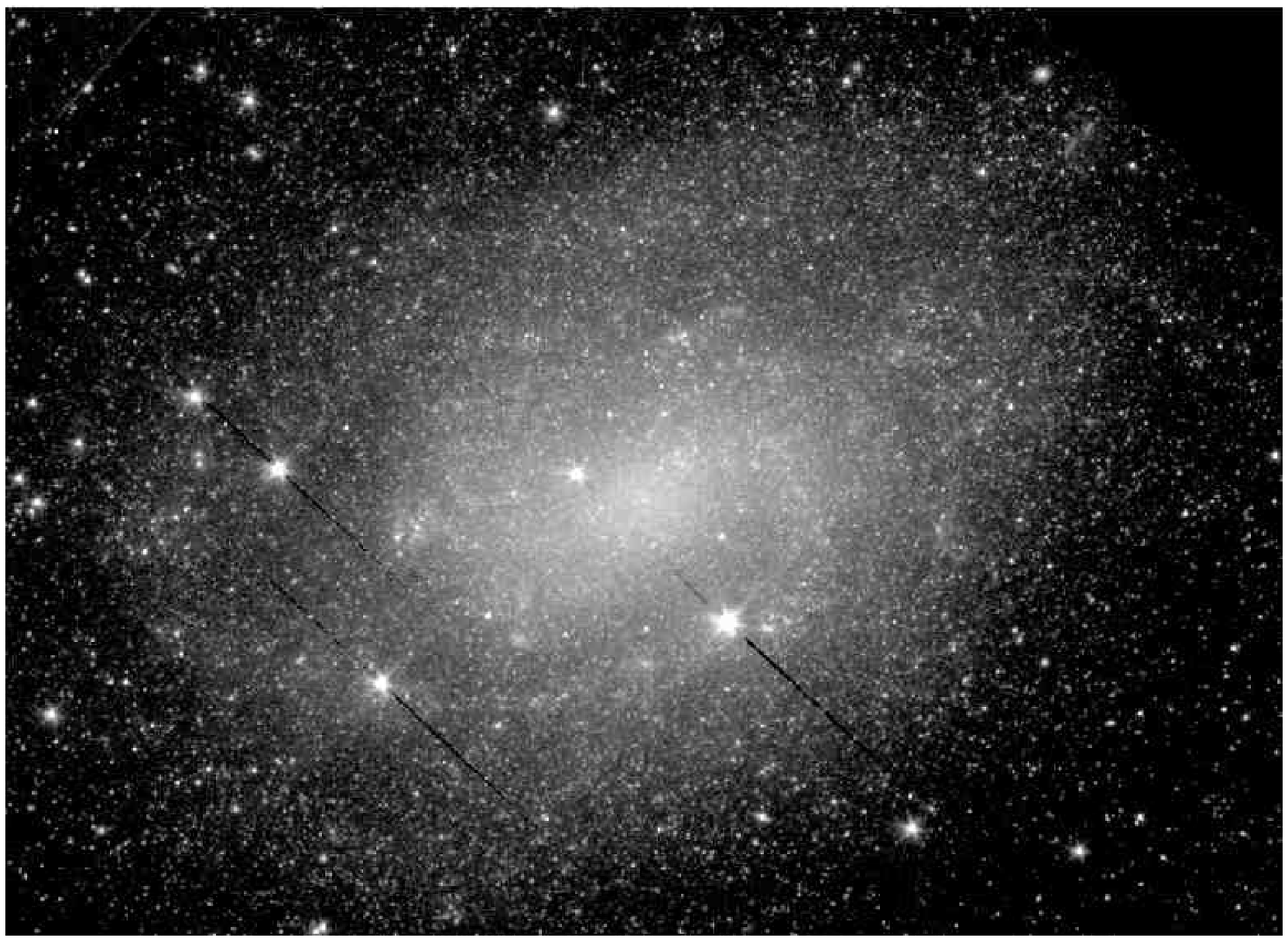}
 \vspace{2.0truecm}
 \caption{
{\bf NGC   300   }              - S$^4$G mid-IR classification:    SA(s)dm                                               ; Filter: IRAC 3.6$\mu$m; North:   up, East: left; Field dimensions:  21.0$\times$ 15.3 arcmin; Surface brightness range displayed: 16.5$-$28.0 mag arcsec$^{-2}$}                 
\label{NGC0300}     
 \end{figure}

\begin{figure}
\figurenum{1.8}  
\plotone{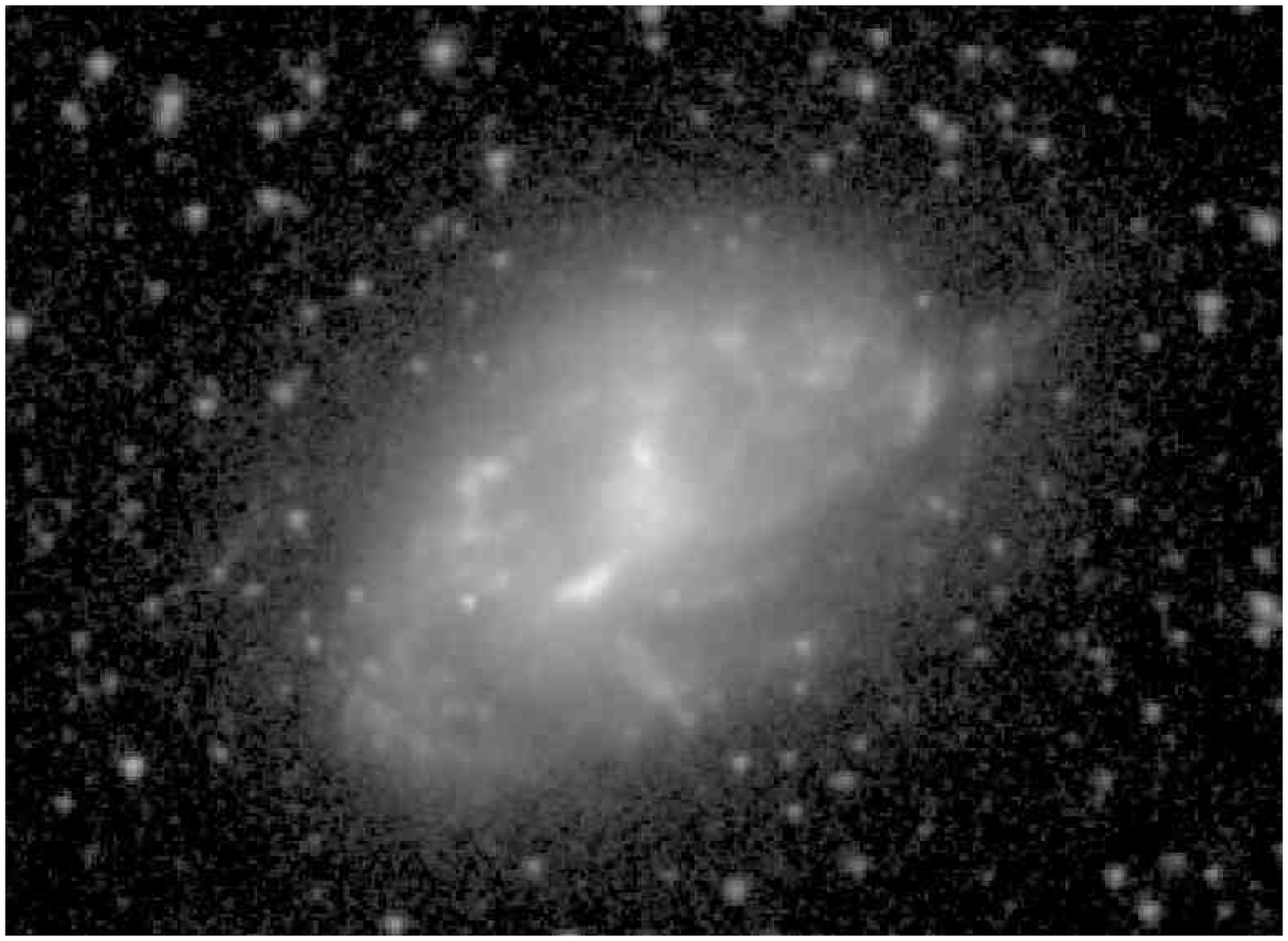}
 \vspace{2.0truecm}
 \caption{
{\bf NGC   337   }              - S$^4$G mid-IR classification:    SAB(s)cd: pec                                         ; Filter: IRAC 3.6$\mu$m; North:   up, East: left; Field dimensions:   4.0$\times$  2.9 arcmin; Surface brightness range displayed: 15.5$-$28.0 mag arcsec$^{-2}$}                 
\label{NGC0337}     
 \end{figure}

\begin{figure}
\figurenum{1.9} 
\plotone{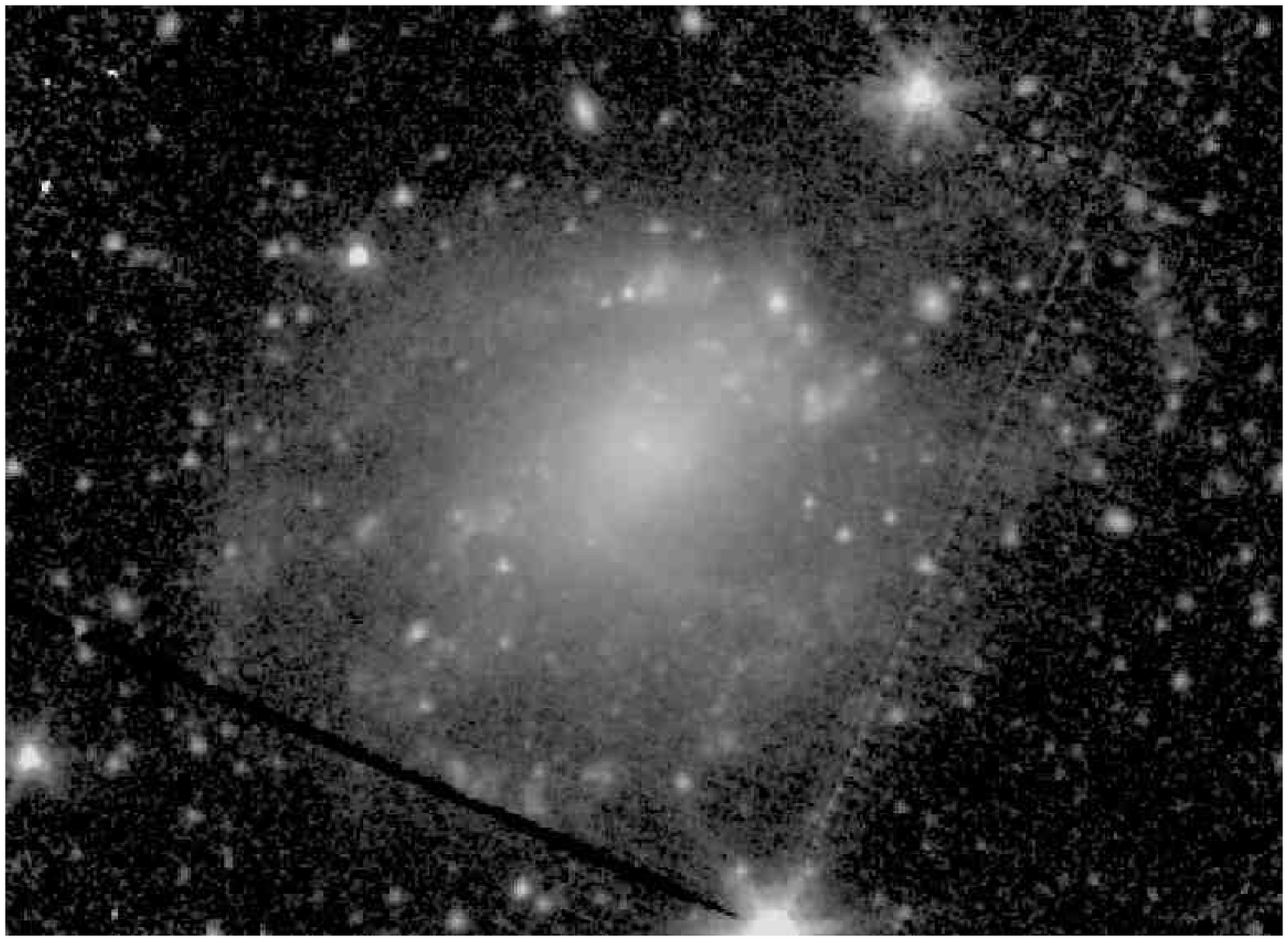}
 \vspace{2.0truecm}
 \caption{
{\bf NGC   428   }              - S$^4$G mid-IR classification:    SAB(s)dm                                              ; Filter: IRAC 3.6$\mu$m; North:   up, East: left; Field dimensions:   5.3$\times$  3.8 arcmin; Surface brightness range displayed: 16.5$-$28.0 mag arcsec$^{-2}$}                 
\label{NGC0428}     
 \end{figure}

\begin{figure}
\figurenum{1.10} 
\plotone{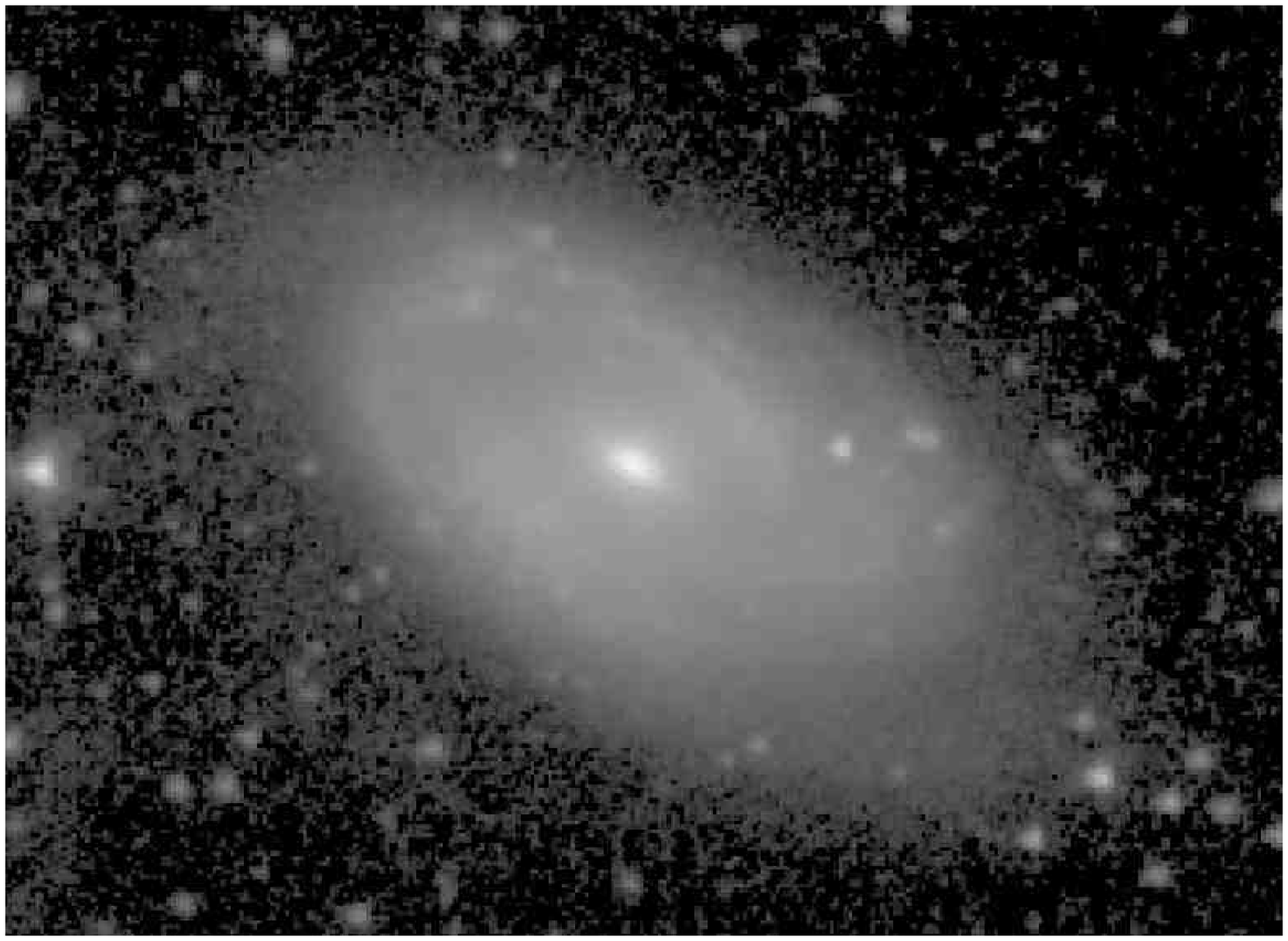}
 \vspace{2.0truecm}
 \caption{
{\bf NGC   470   }              - S$^4$G mid-IR classification:    S$\underline{\rm A}$B(r$\underline{\rm s}$)ab         ; Filter: IRAC 3.6$\mu$m; North: left, East: down; Field dimensions:   3.5$\times$  2.6 arcmin; Surface brightness range displayed: 13.0$-$28.0 mag arcsec$^{-2}$}                 
\label{NGC0470}     
 \end{figure}

\begin{figure}
\figurenum{1.11} 
\plotone{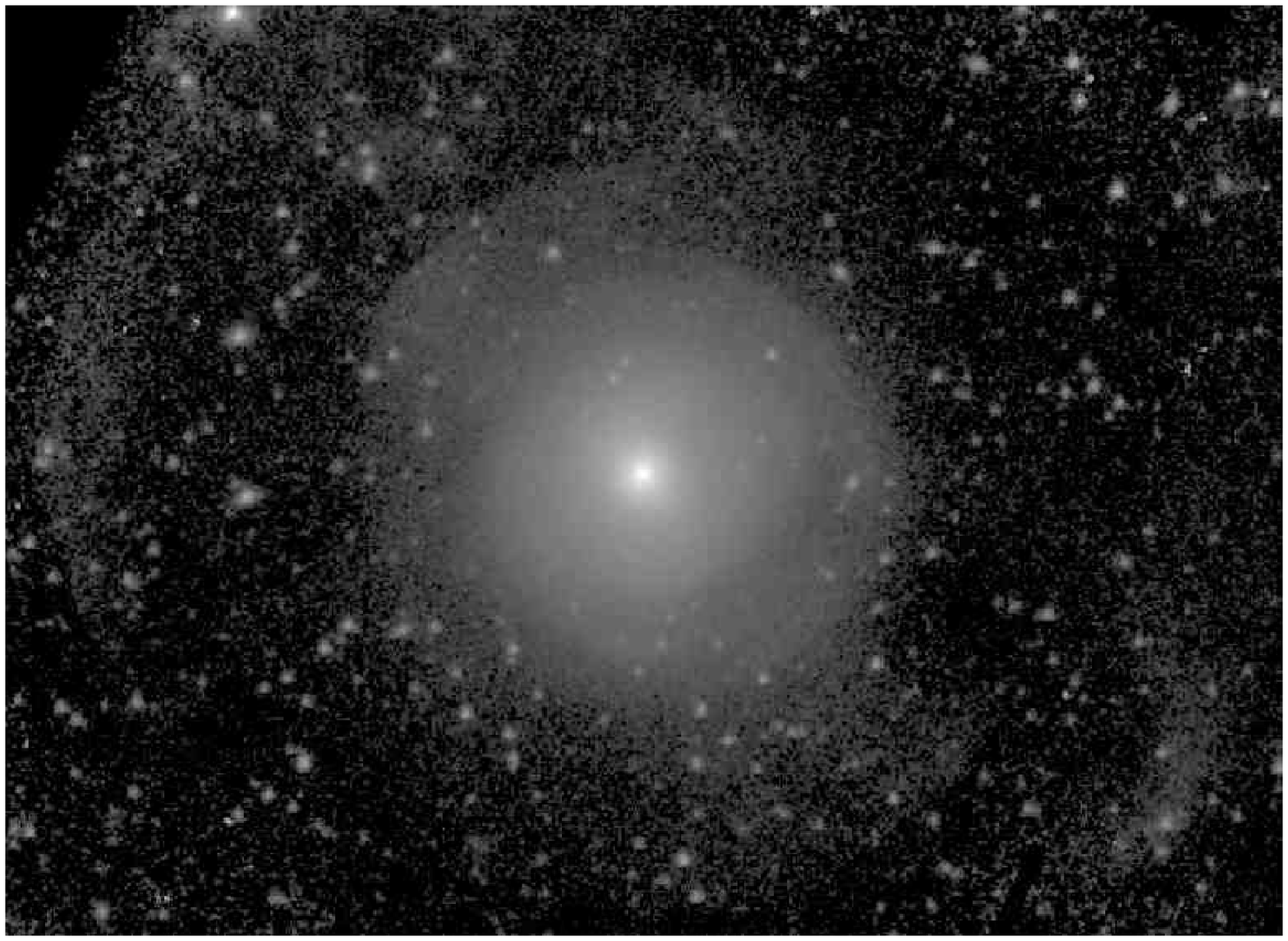}
 \vspace{2.0truecm}
 \caption{
{\bf NGC   474   }              - S$^4$G mid-IR classification:    (R)SAB0/a (shells/ripples) pec                                ; Filter: IRAC 3.6$\mu$m; North:   up, East: left; Field dimensions:   7.0$\times$  5.1 arcmin; Surface brightness range displayed: 13.5$-$28.0 mag arcsec$^{-2}$}                 
\label{NGC0474}     
 \end{figure}

\begin{figure}
\figurenum{1.12} 
\plotone{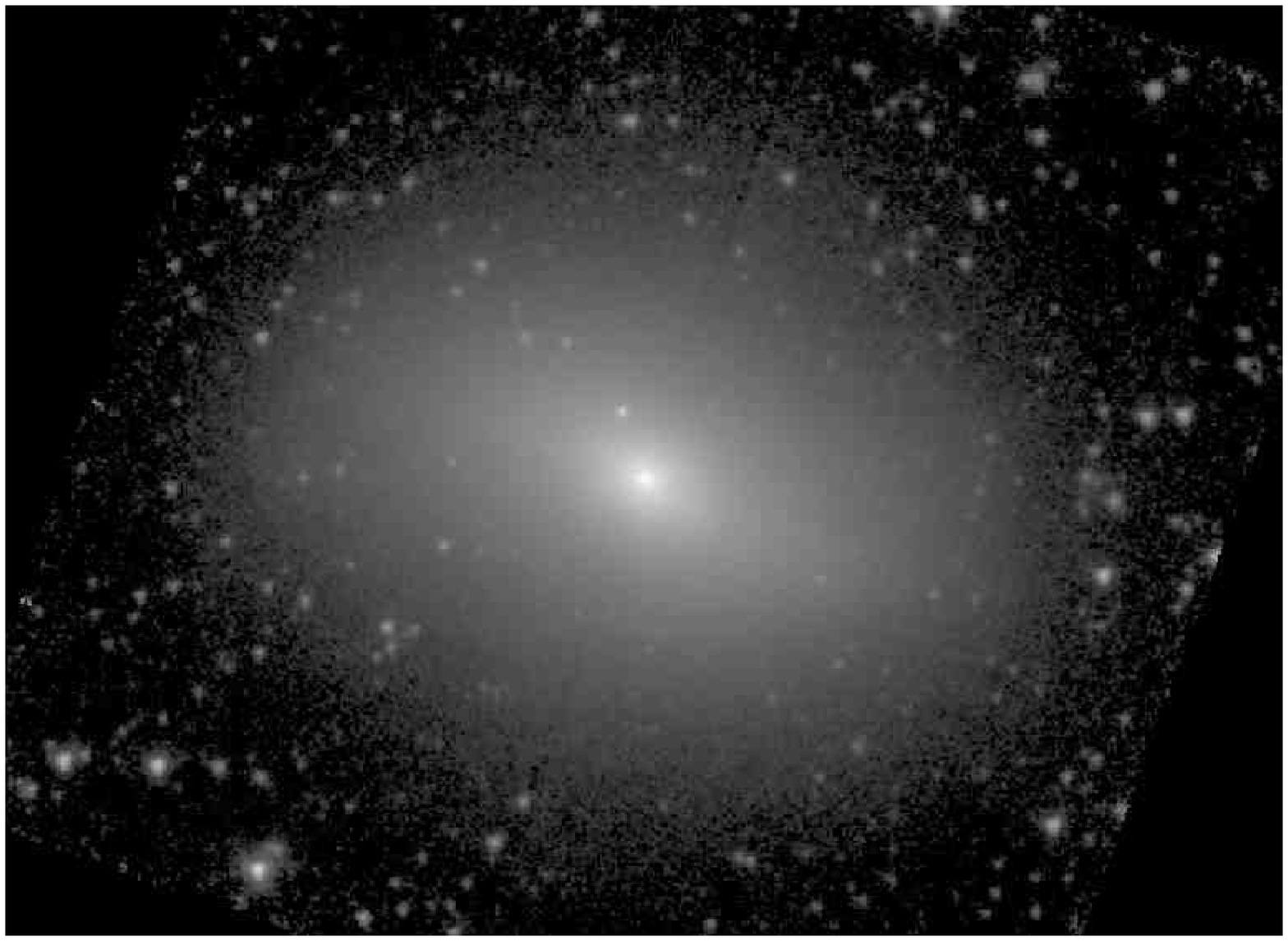}
 \vspace{2.0truecm}
 \caption{
{\bf NGC   584   }              - S$^4$G mid-IR classification:    S$\underline{\rm A}$B0$^-$                            ; Filter: IRAC 3.6$\mu$m; North:   up, East: left; Field dimensions:   6.3$\times$  4.6 arcmin; Surface brightness range displayed: 12.0$-$28.0 mag arcsec$^{-2}$}                 
\label{NGC0584}     
 \end{figure}

\begin{figure}
\figurenum{1.13} 
\plotone{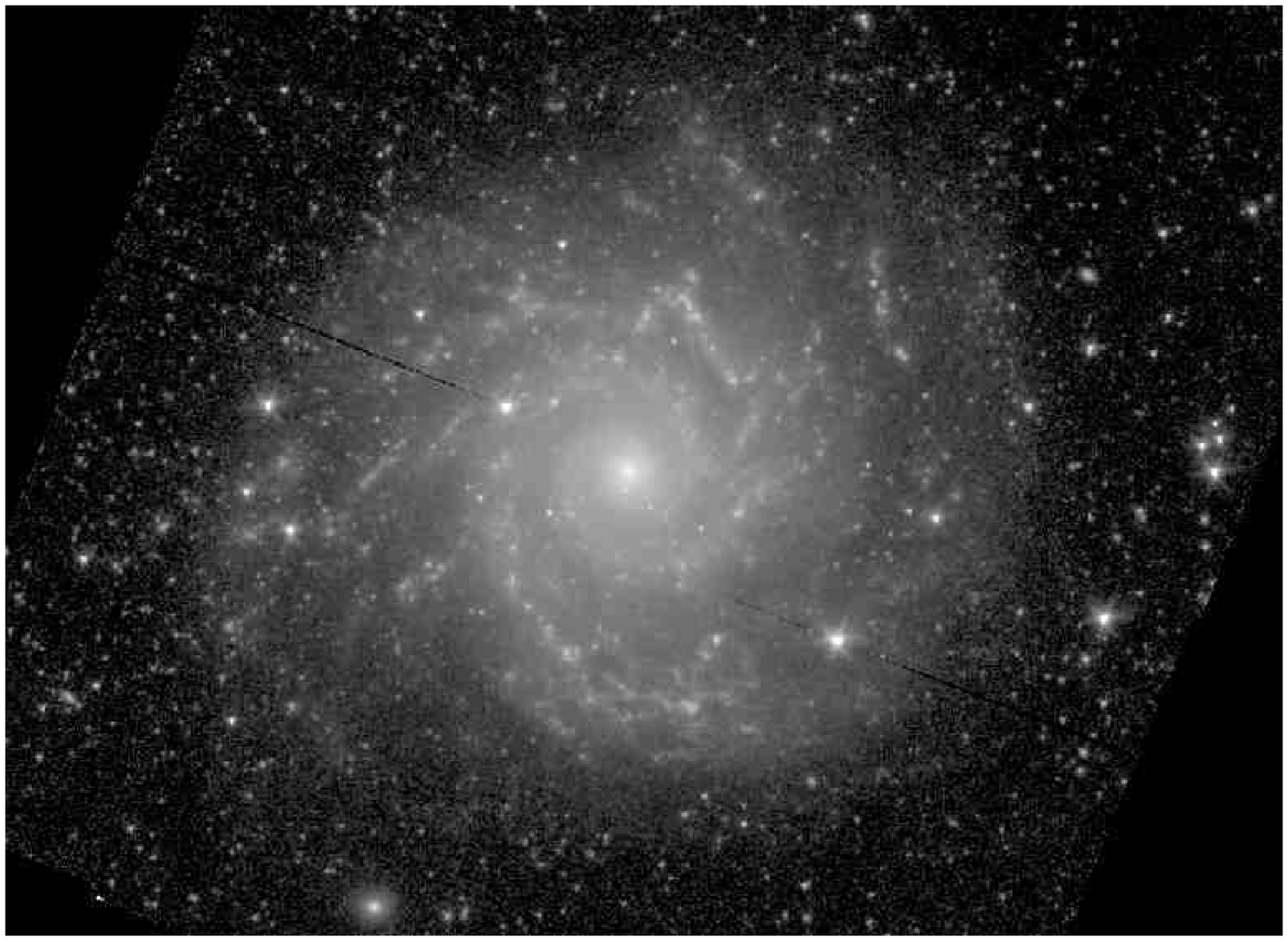}
 \vspace{2.0truecm}
 \caption{
{\bf NGC   628   }              - S$^4$G mid-IR classification:    SA(s)c                                                ; Filter: IRAC 3.6$\mu$m; North:   up, East: left; Field dimensions:  14.4$\times$ 10.5 arcmin; Surface brightness range displayed: 15.0$-$28.0 mag arcsec$^{-2}$}                 
\label{NGC0628}     
 \end{figure}

\begin{figure}
\figurenum{1.14} 
\plotone{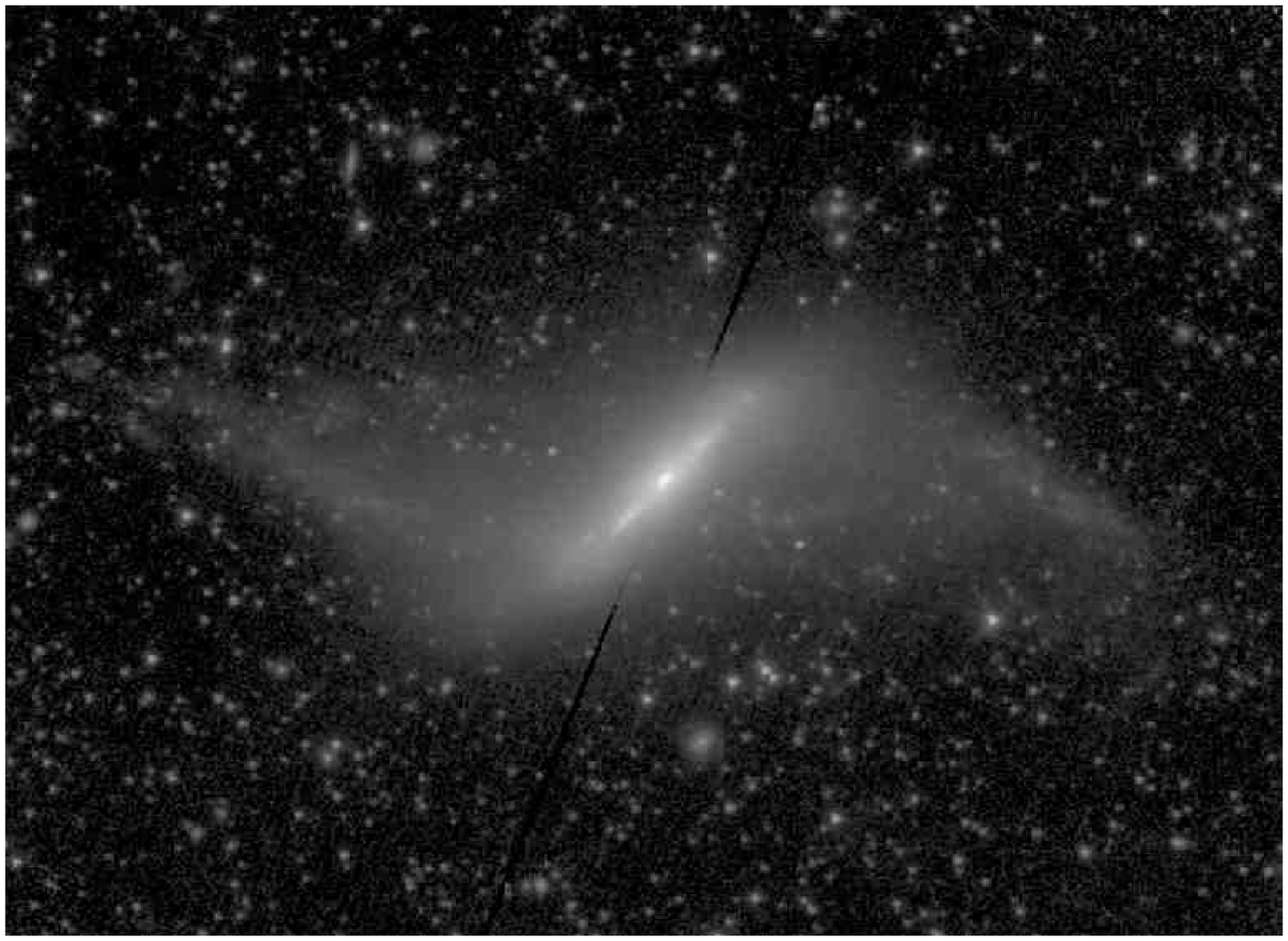}
 \vspace{2.0truecm}
 \caption{
{\bf NGC   660   }              - S$^4$G mid-IR classification:    PRG                                                   ; Filter: IRAC 3.6$\mu$m; North: left, East: down; Field dimensions:  10.5$\times$  7.7 arcmin; Surface brightness range displayed: 11.5$-$28.0 mag arcsec$^{-2}$}                 
\label{NGC0660}     
 \end{figure}

\begin{figure}
\figurenum{1.15} 
\plotone{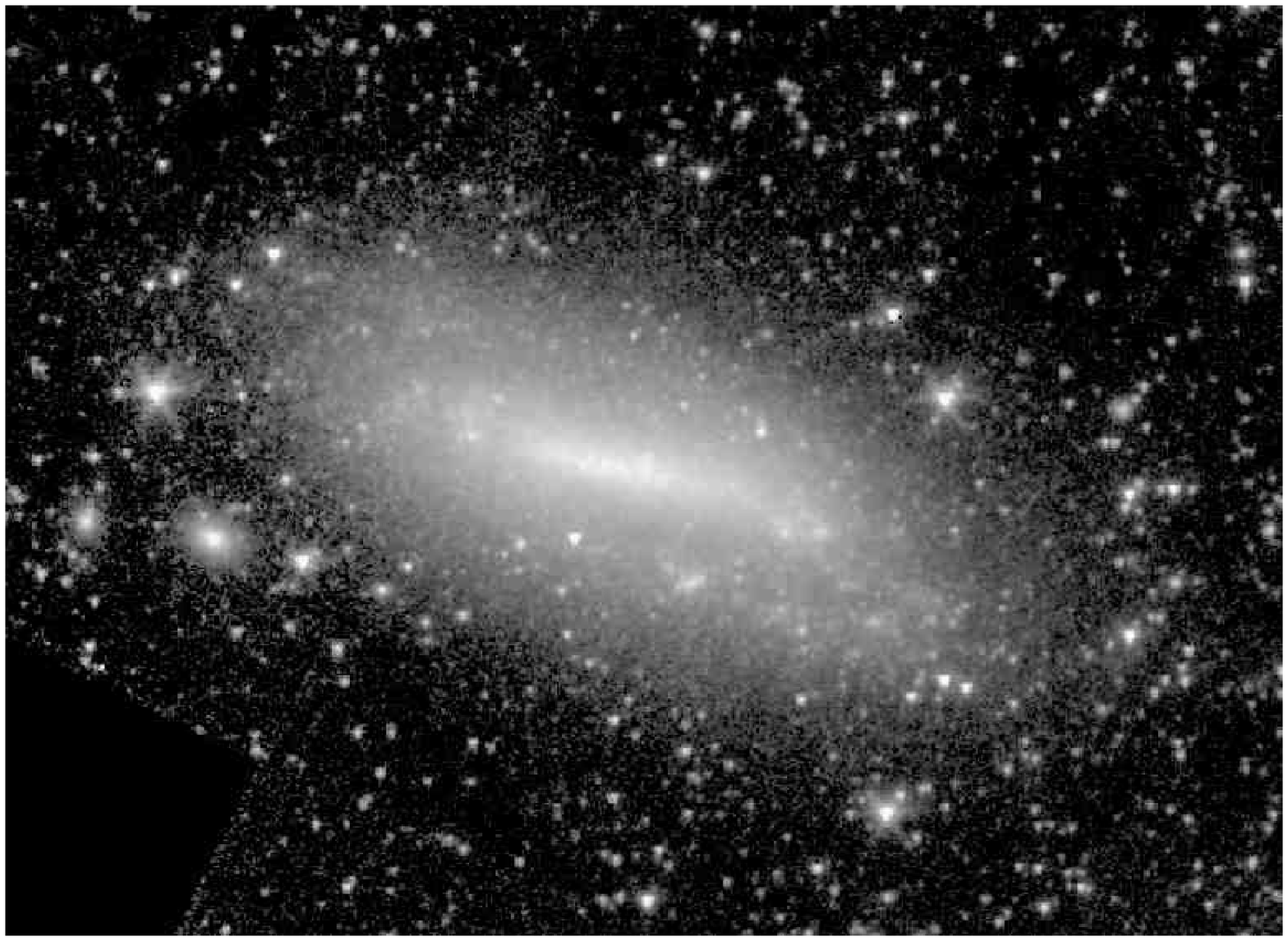}
 \vspace{2.0truecm}
 \caption{
{\bf NGC   672   }              - S$^4$G mid-IR classification:    (R$^{\prime}$)SB(s)d                                            ; Filter: IRAC 3.6$\mu$m; North:   up, East: left; Field dimensions:   8.8$\times$  6.4 arcmin; Surface brightness range displayed: 17.0$-$28.0 mag arcsec$^{-2}$}                 
\label{NGC0672}     
 \end{figure}

\begin{figure}
\figurenum{1.16} 
\plotone{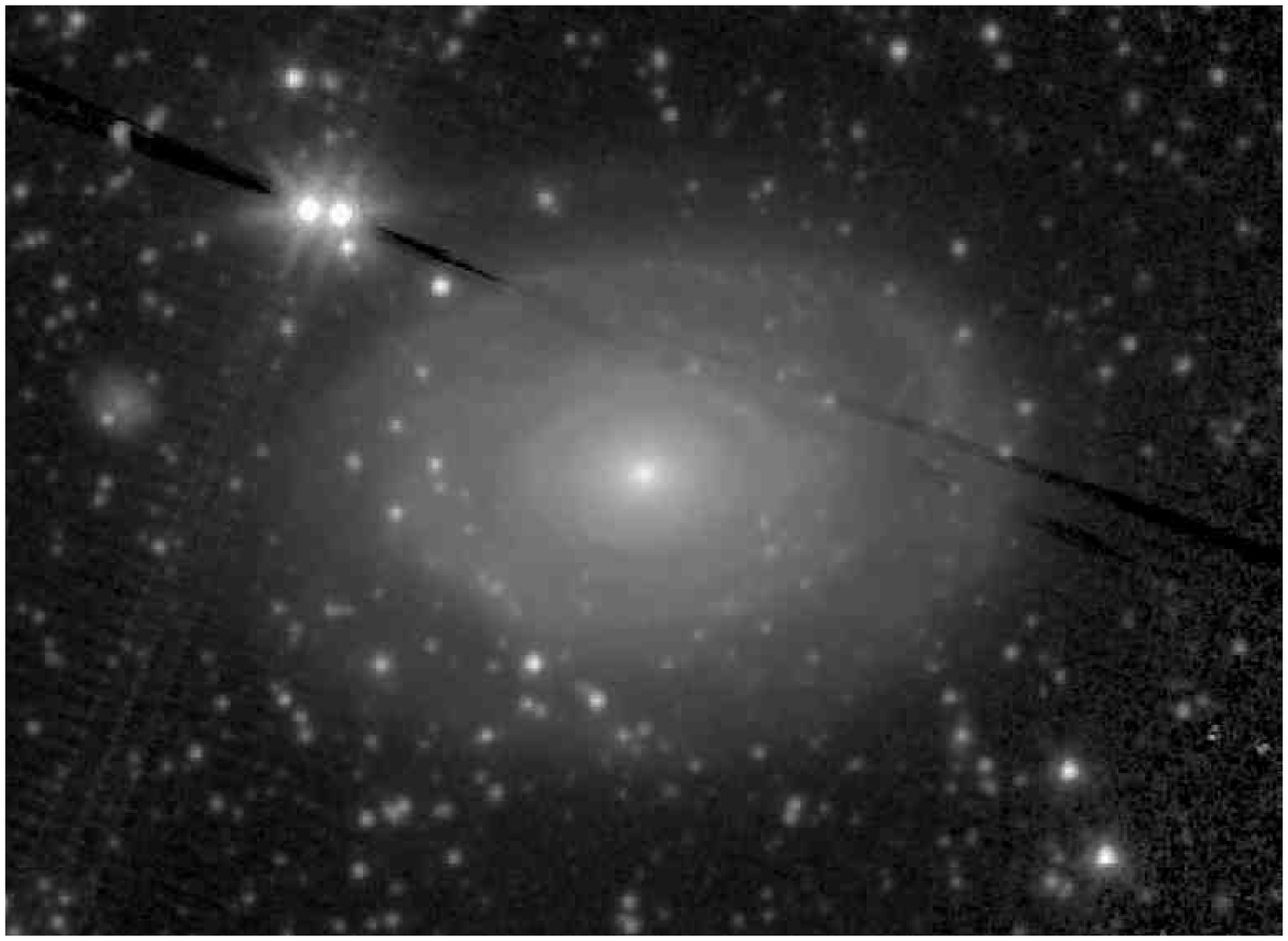}
 \vspace{2.0truecm}
 \caption{
{\bf NGC   691   }              - S$^4$G mid-IR classification:    (R)SA(r)ab                                            ; Filter: IRAC 3.6$\mu$m; North:   up, East: left; Field dimensions:   5.3$\times$  3.8 arcmin; Surface brightness range displayed: 14.5$-$28.0 mag arcsec$^{-2}$}                 
\label{NGC0691}     
 \end{figure}

\begin{figure}
\figurenum{1.17} 
\plotone{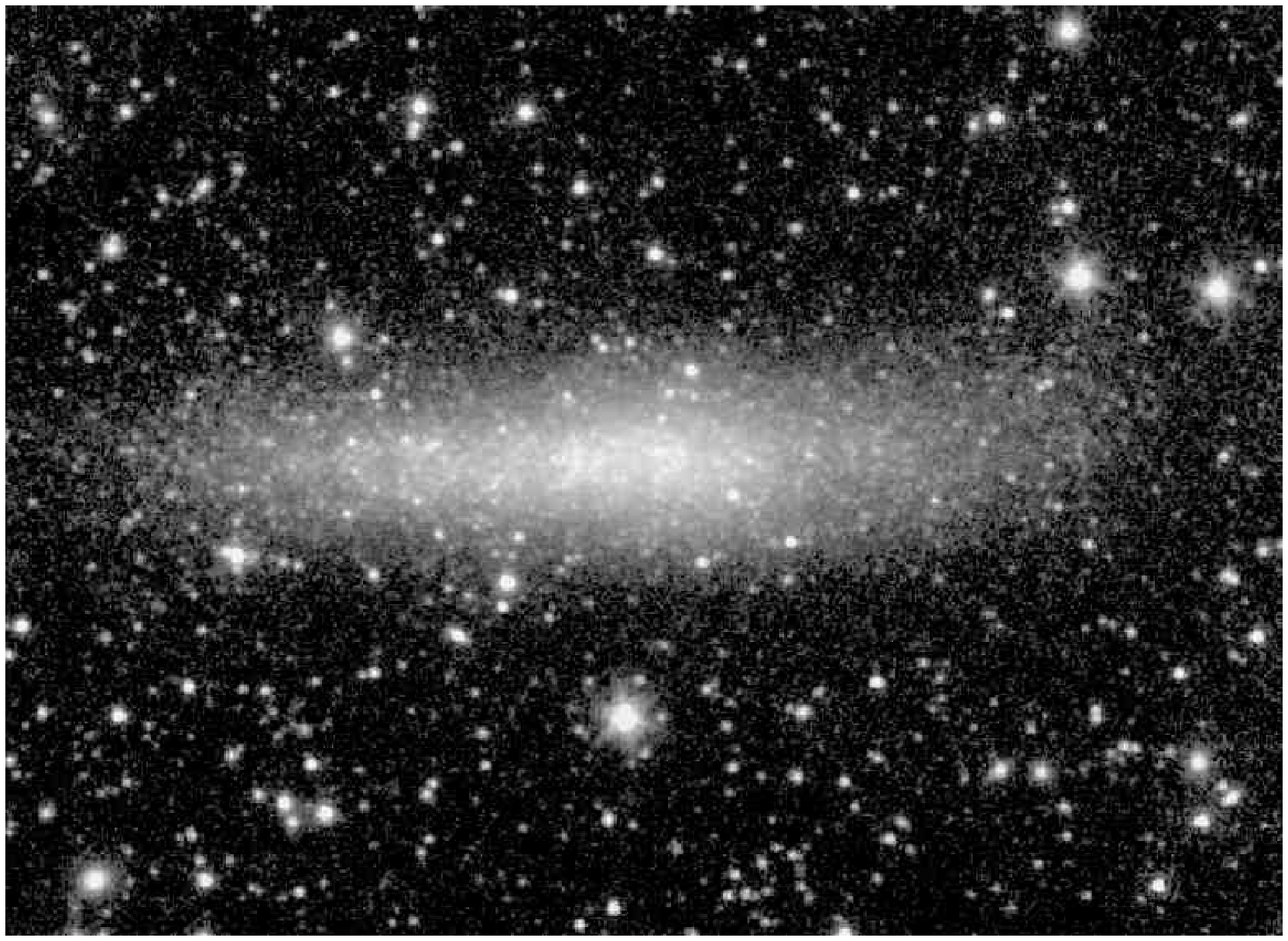}
 \vspace{2.0truecm}
 \caption{
{\bf NGC   784   }              - S$^4$G mid-IR classification:    SBm: sp                                               ; Filter: IRAC 3.6$\mu$m; North: left, East: down; Field dimensions:   7.9$\times$  5.8 arcmin; Surface brightness range displayed: 18.5$-$28.0 mag arcsec$^{-2}$}                 
\label{NGC0784}     
 \end{figure}

\begin{figure}
\figurenum{1.18} 
\plotone{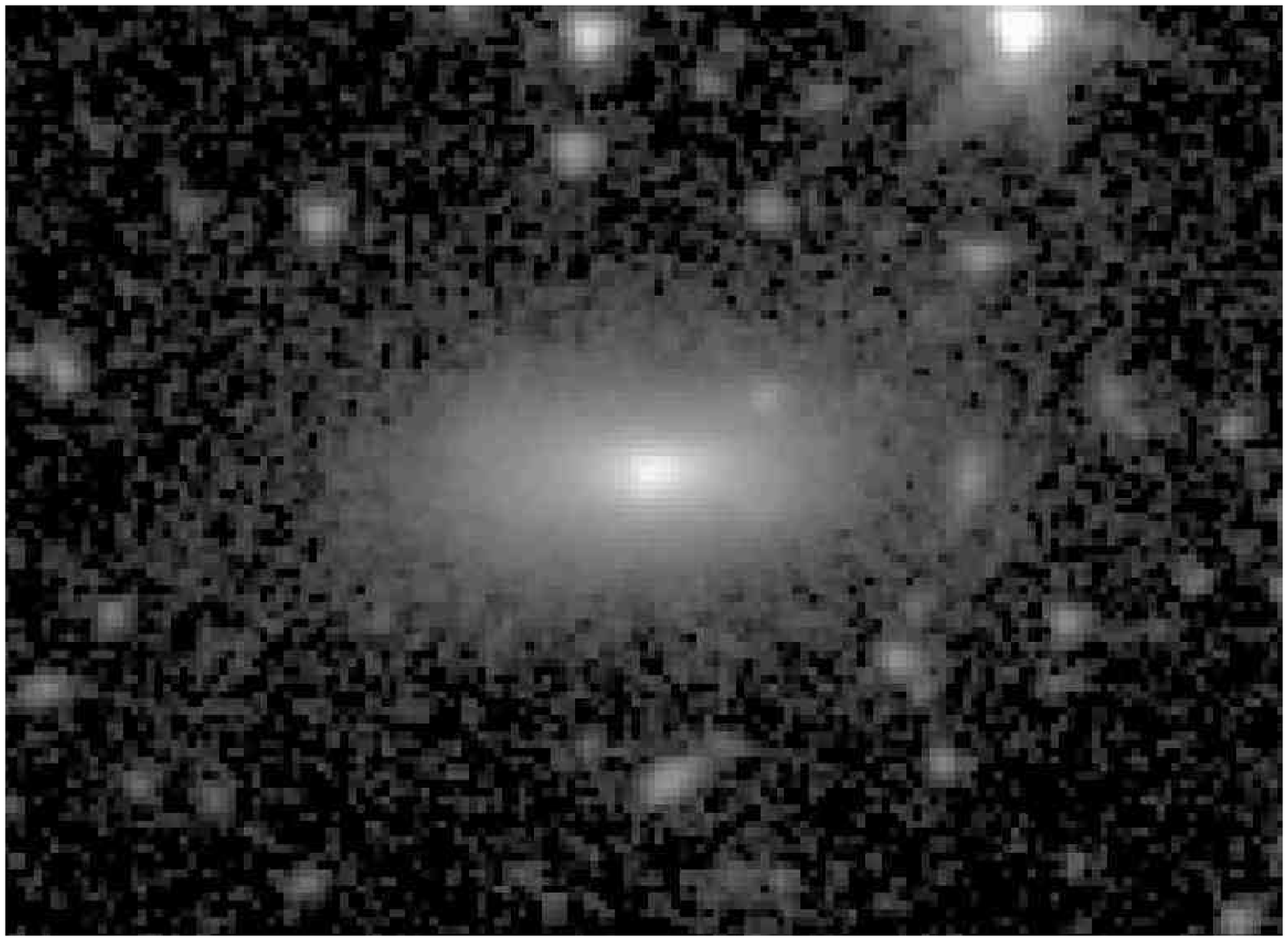}
 \vspace{2.0truecm}
 \caption{
{\bf NGC   814   }              - S$^4$G mid-IR classification:    SAB0$^-$:                                             ; Filter: IRAC 3.6$\mu$m; North: left, East: down; Field dimensions:   2.0$\times$  1.4 arcmin; Surface brightness range displayed: 14.5$-$28.0 mag arcsec$^{-2}$}                 
\label{NGC0814}     
 \end{figure}

\begin{figure}
\figurenum{1.19} 
\plotone{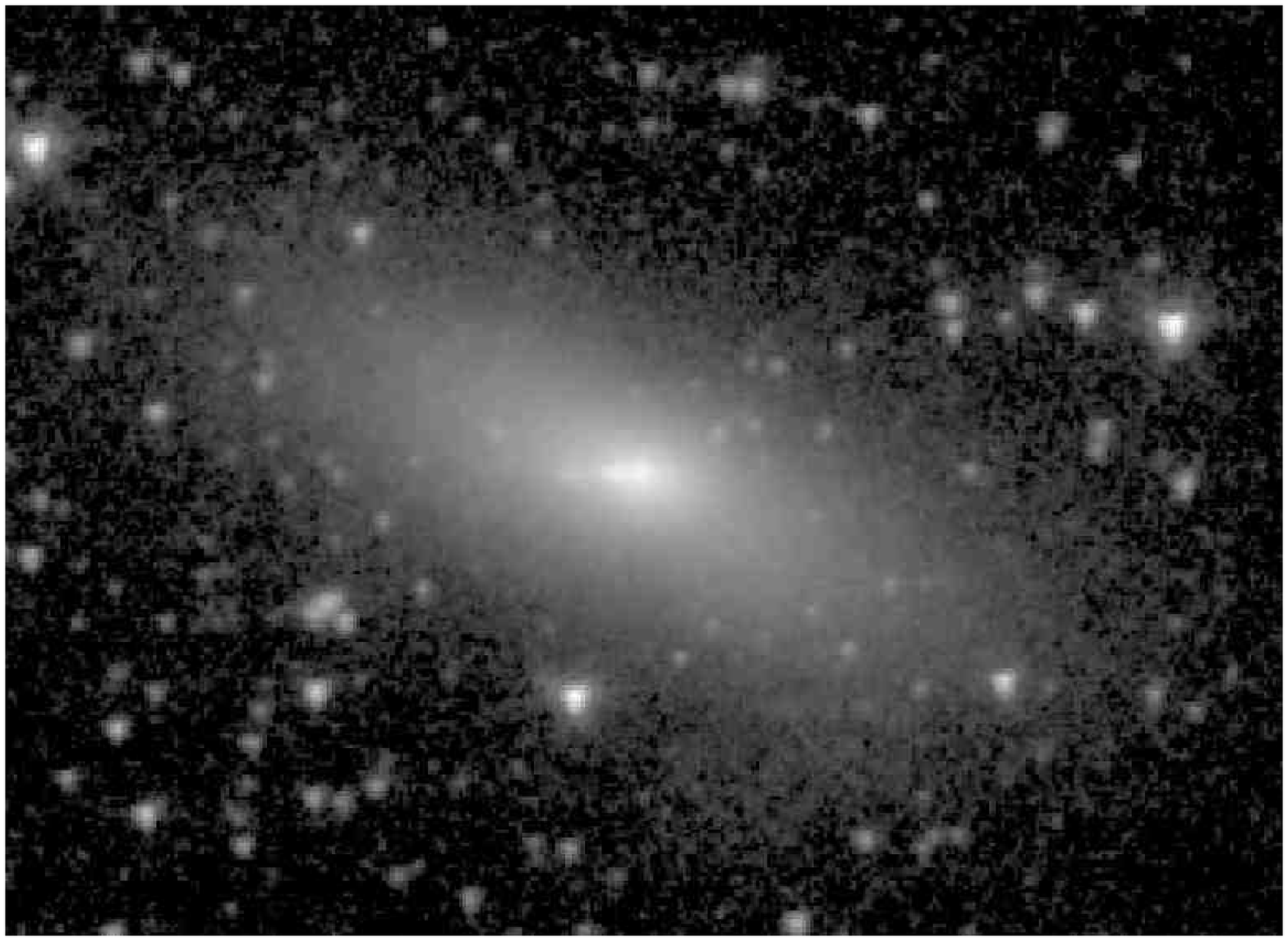}
 \vspace{2.0truecm}
 \caption{
{\bf NGC   855   }              - S$^4$G mid-IR classification:    SA0$^-$                                               ; Filter: IRAC 3.6$\mu$m; North:   up, East: left; Field dimensions:   3.8$\times$  2.7 arcmin; Surface brightness range displayed: 15.0$-$28.0 mag arcsec$^{-2}$}                 
\label{NGC0855}     
 \end{figure}

\begin{figure}
\figurenum{1.20} 
\plotone{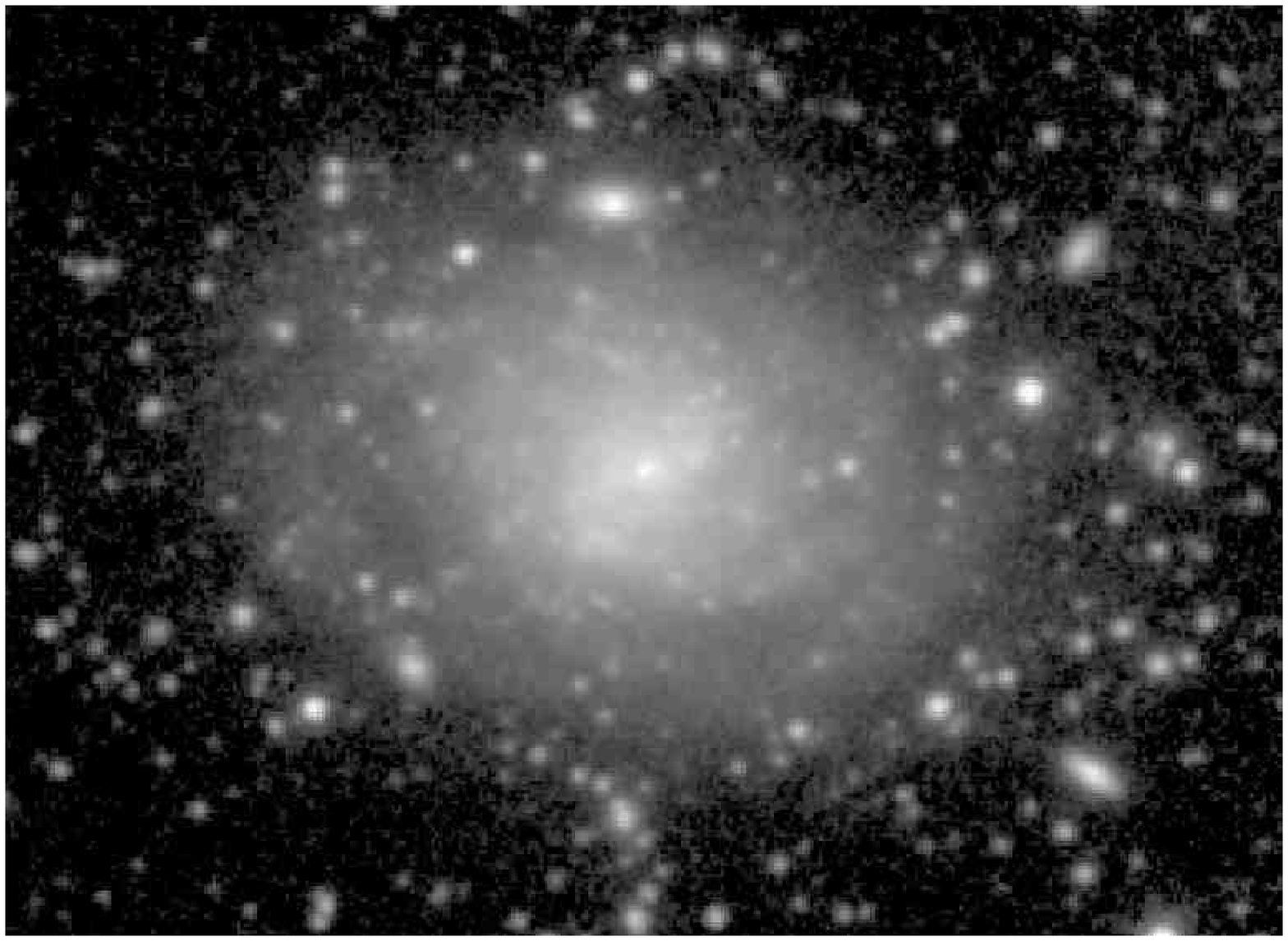}
 \vspace{2.0truecm}
 \caption{
{\bf NGC   941   }              - S$^4$G mid-IR classification:    (R$^{\prime}$:)S$\underline{\rm A}$B(r:)c                       ; Filter: IRAC 3.6$\mu$m; North: left, East: down; Field dimensions:   4.0$\times$  2.9 arcmin; Surface brightness range displayed: 16.5$-$28.0 mag arcsec$^{-2}$}                 
\label{NGC0941}     
 \end{figure}

\begin{figure}
\figurenum{1.21} 
\plotone{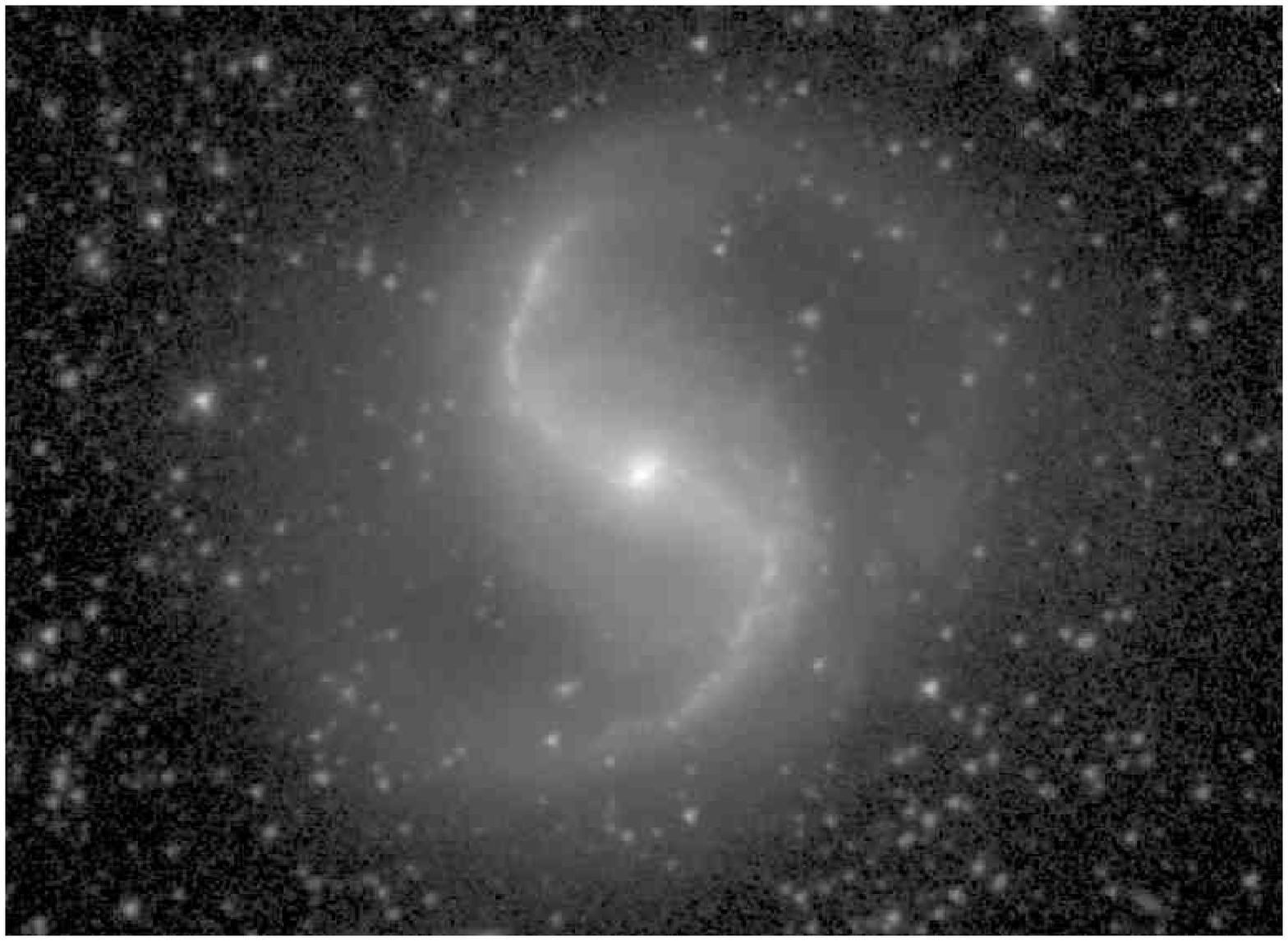}
 \vspace{2.0truecm}
 \caption{
{\bf NGC   986   }              - S$^4$G mid-IR classification:    (R$^{\prime}$)SB(rs,nb)ab                                       ; Filter: IRAC 3.6$\mu$m; North:   up, East: left; Field dimensions:   6.3$\times$  4.6 arcmin; Surface brightness range displayed: 13.0$-$28.0 mag arcsec$^{-2}$}                 
\label{NGC0986}     
 \end{figure}

\begin{figure}
\figurenum{1.22} 
\plotone{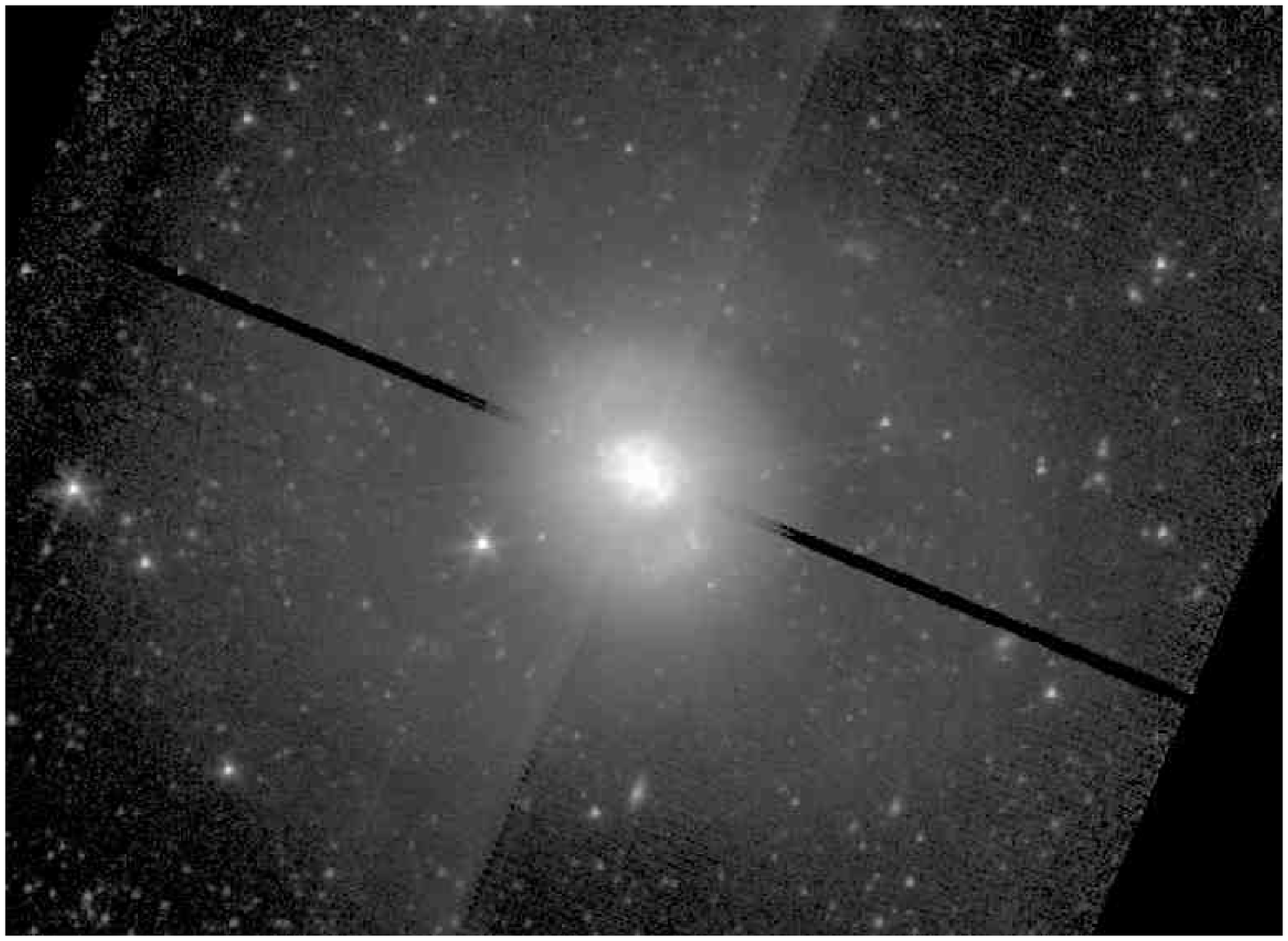}
 \vspace{2.0truecm}
 \caption{
{\bf NGC  1068   }              - S$^4$G mid-IR classification:    (R)SA(s,nr)a                                          ; Filter: IRAC 3.6$\mu$m; North:   up, East: left; Field dimensions:  10.5$\times$  7.7 arcmin; Surface brightness range displayed: 11.0$-$28.0 mag arcsec$^{-2}$}                 
\label{NGC1068}     
 \end{figure}

\begin{figure}
\figurenum{1.23} 
\plotone{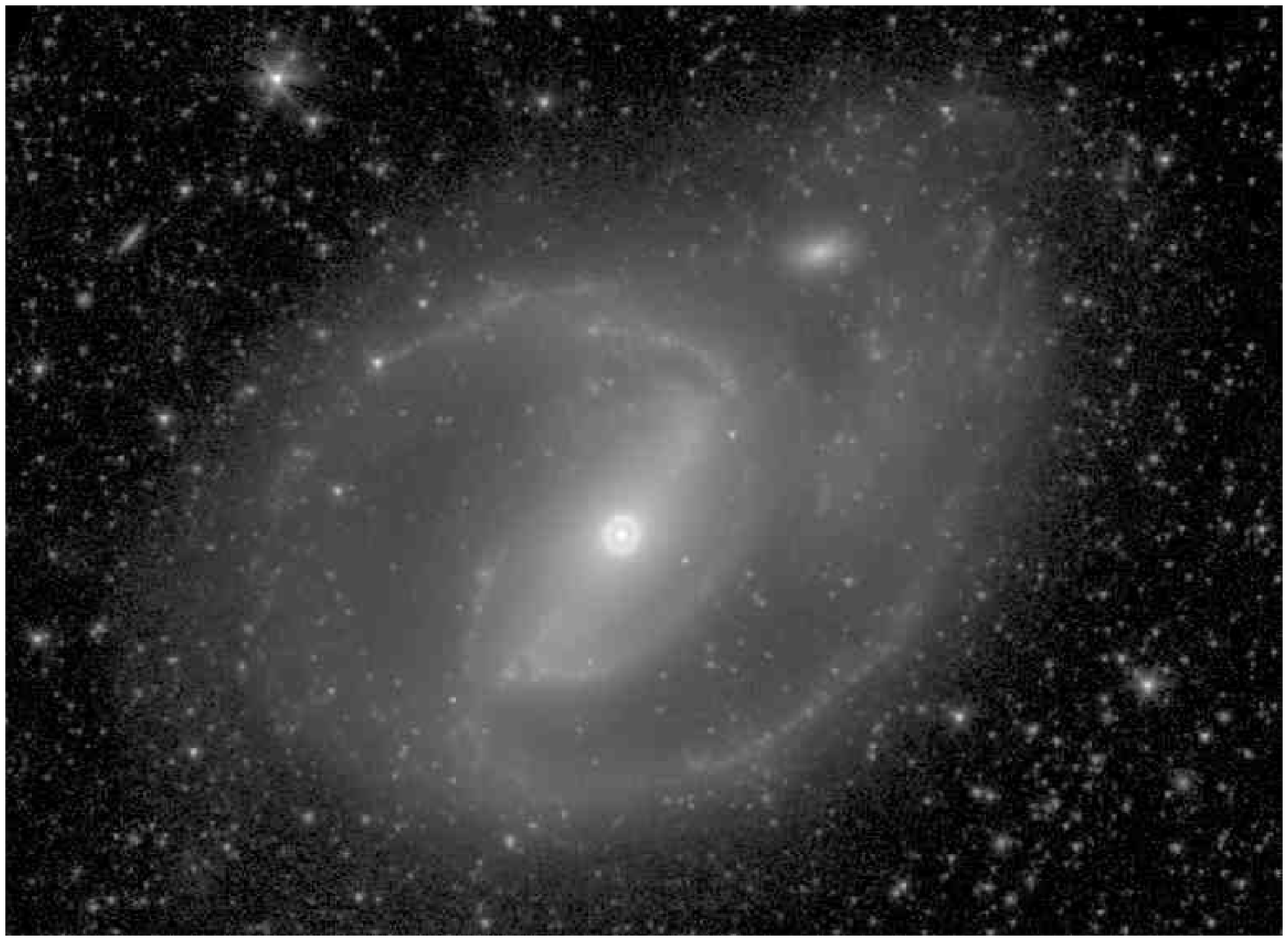}
 \vspace{2.0truecm}
 \caption{
{\bf NGC  1097   }              - S$^4$G mid-IR classification:    (R$^{\prime}$)SB(rs,nr)a$\underline{\rm b}$ pec                 ; Filter: IRAC 3.6$\mu$m; North:   up, East: left; Field dimensions:  12.6$\times$  9.2 arcmin; Surface brightness range displayed: 12.5$-$28.0 mag arcsec$^{-2}$}                 
\label{NGC1097}     
 \end{figure}

\begin{figure}
\figurenum{1.24} 
\plotone{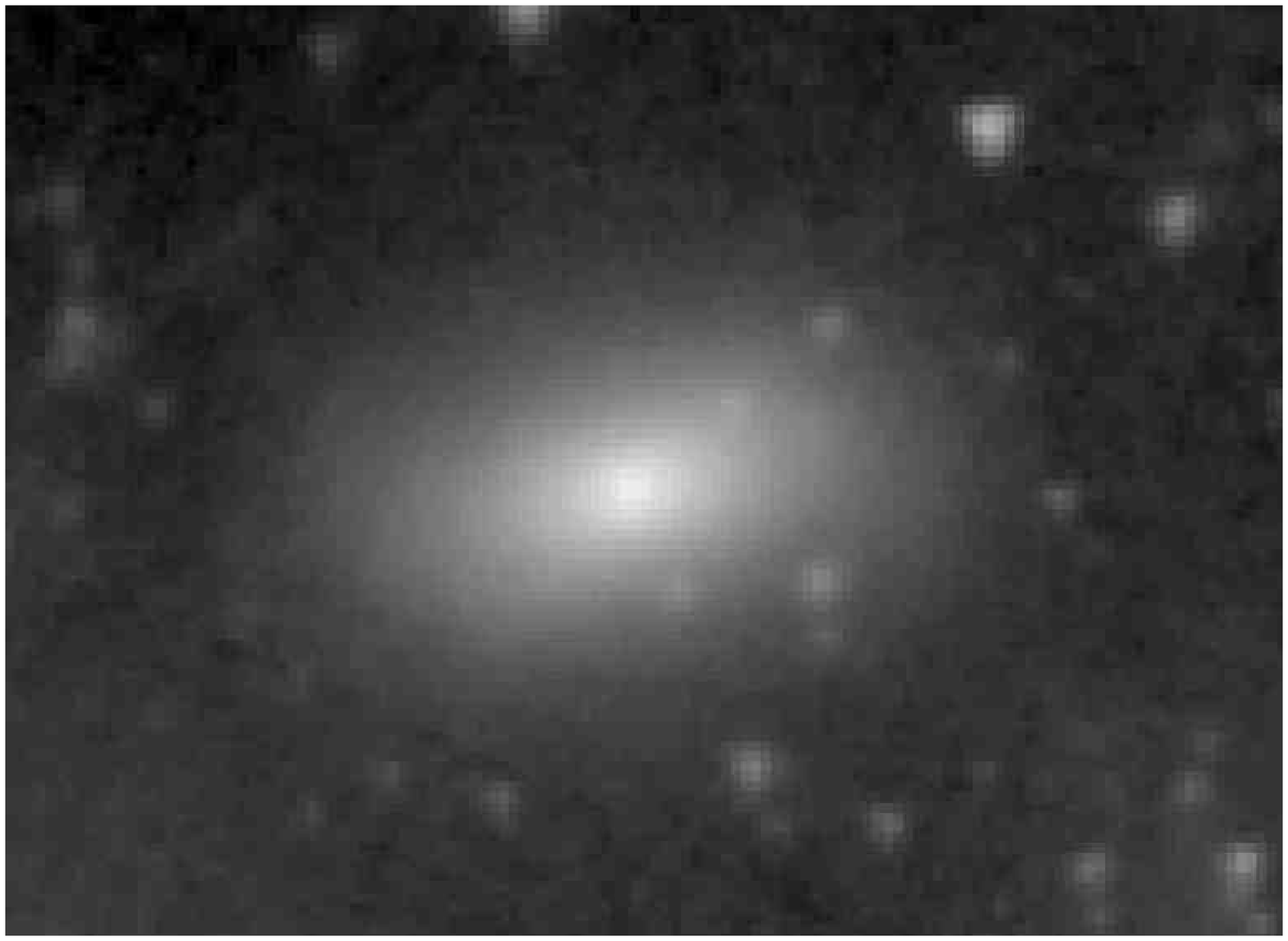}
 \vspace{2.0truecm}
 \caption{
{\bf NGC  1097A  }              - S$^4$G mid-IR classification:    dE5                                                   ; Filter: IRAC 3.6$\mu$m; North:   up, East: left; Field dimensions:   1.6$\times$  1.1 arcmin; Surface brightness range displayed: 15.0$-$28.0 mag arcsec$^{-2}$}                 
\label{NGC1097A}    
 \end{figure}

\begin{figure}
\figurenum{1.25} 
\plotone{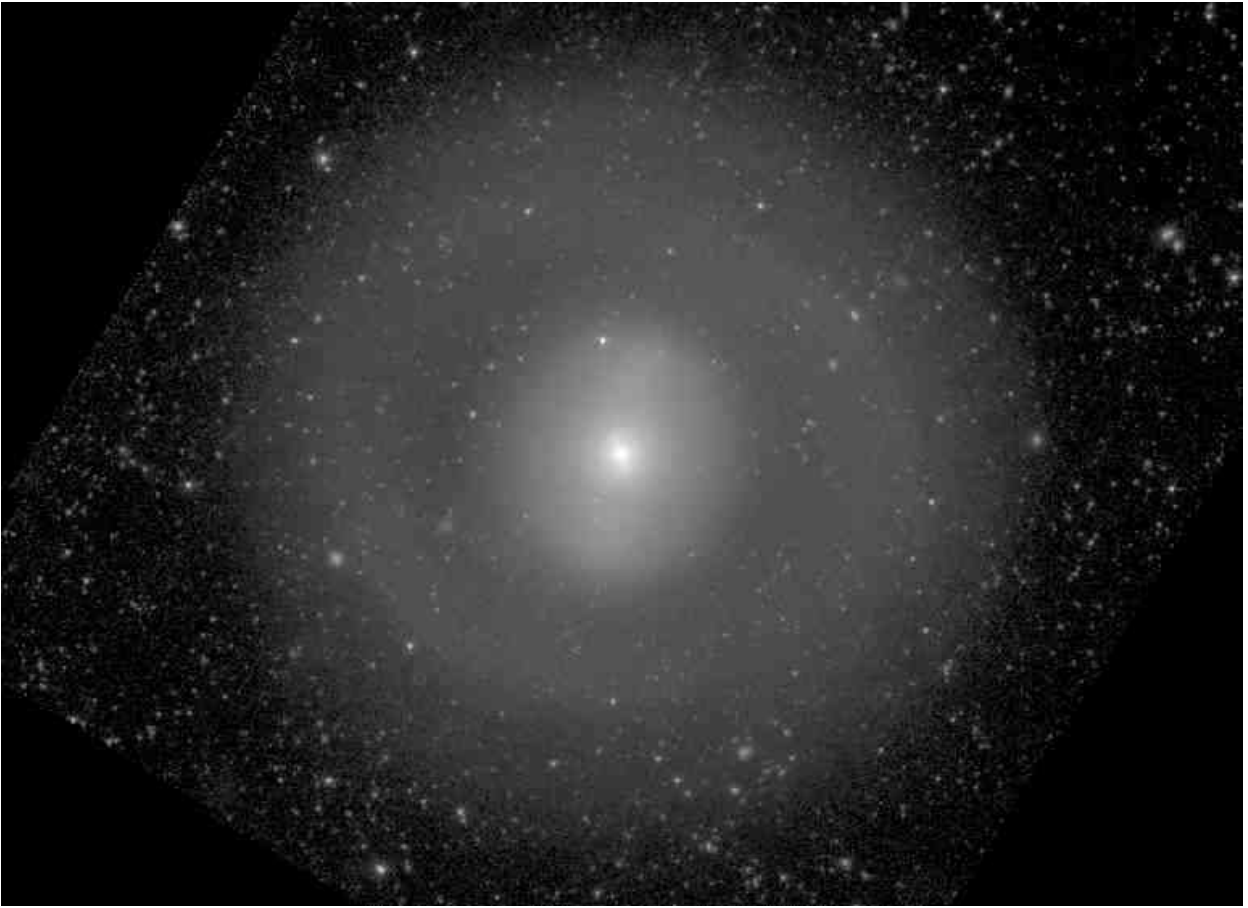}
 \vspace{2.0truecm}
 \caption{
{\bf NGC  1291   }              - S$^4$G mid-IR classification:    (R)SAB(l,nb)0$^+$                                     ; Filter: IRAC 3.6$\mu$m; North:   up, East: left; Field dimensions:  18.5$\times$ 13.5 arcmin; Surface brightness range displayed: 11.5$-$28.0 mag arcsec$^{-2}$}                 
\label{NGC1291}     
 \end{figure}

\begin{figure}
\figurenum{1.26} 
\plotone{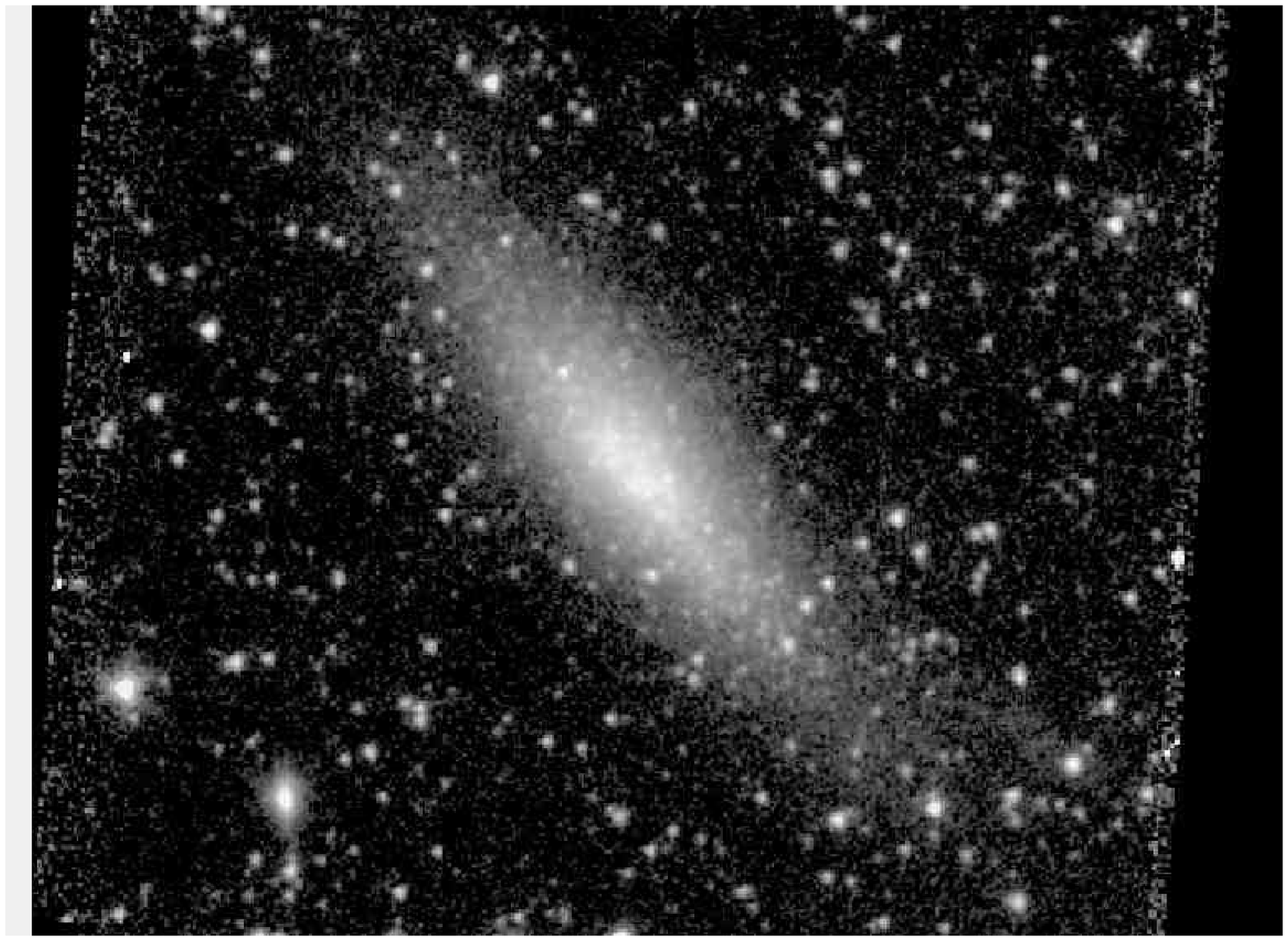}
 \vspace{2.0truecm}
 \caption{
{\bf NGC  1311   }              - S$^4$G mid-IR classification:    SBdm: sp                                              ; Filter: IRAC 3.6$\mu$m; North:   up, East: left; Field dimensions:   6.2$\times$  4.5 arcmin; Surface brightness range displayed: 18.0$-$28.0 mag arcsec$^{-2}$}                 
\label{NGC1311}     
 \end{figure}

\begin{figure}
\figurenum{1.27} 
\plotone{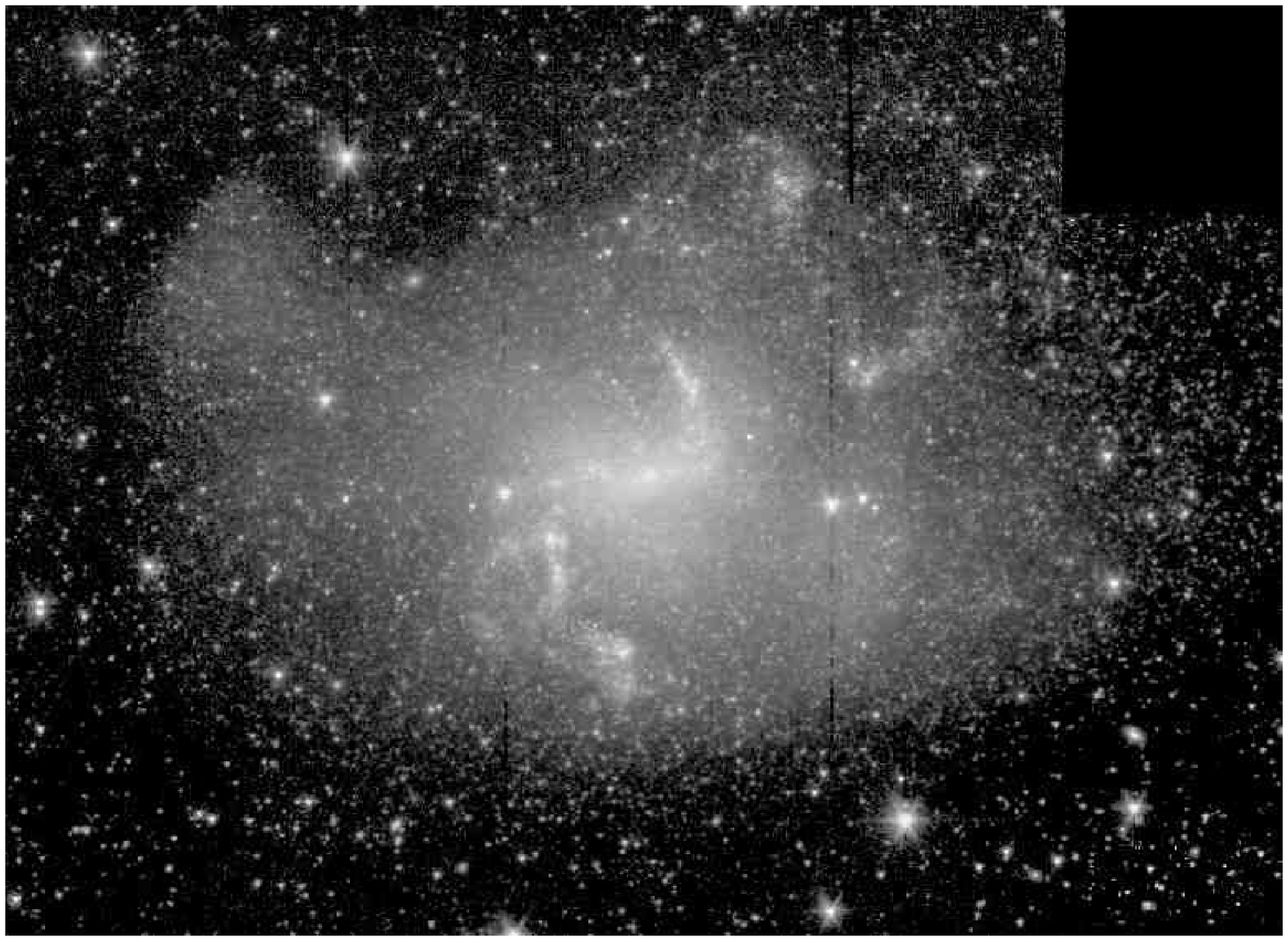}
 \vspace{2.0truecm}
 \caption{
{\bf NGC  1313   }              - S$^4$G mid-IR classification:    SB(s)$\underline{\rm d}$m pec                         ; Filter: IRAC 3.6$\mu$m; North: left, East: down; Field dimensions:  15.8$\times$ 11.5 arcmin; Surface brightness range displayed: 16.5$-$28.0 mag arcsec$^{-2}$}                 
\label{NGC1313}     
 \end{figure}

\begin{figure}
\figurenum{1.28} 
\plotone{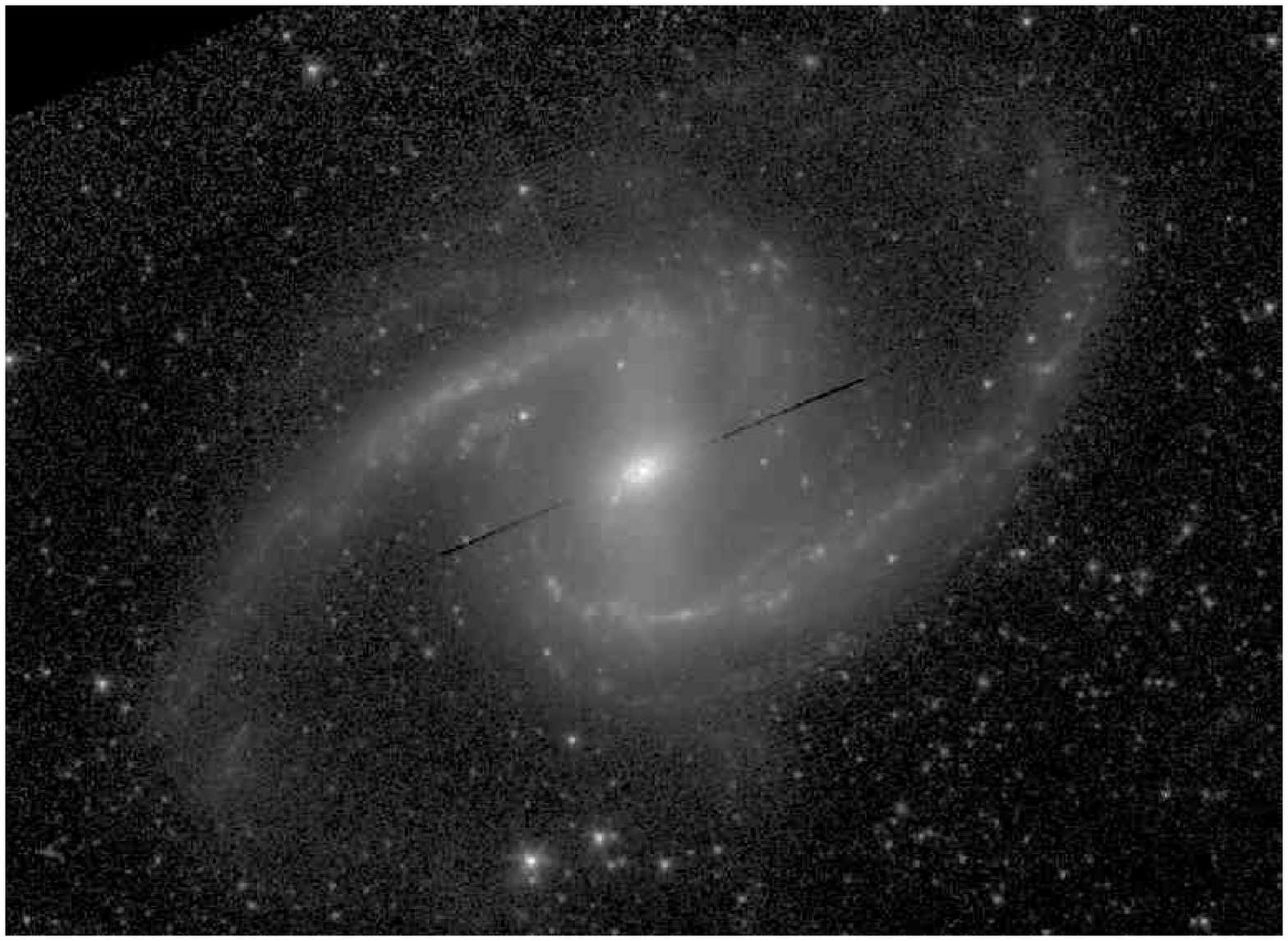}
 \vspace{2.0truecm}
 \caption{
{\bf NGC  1365   }              - S$^4$G mid-IR classification:    SB(r$\underline{\rm s}$,nr)bc                         ; Filter: IRAC 3.6$\mu$m; North: left, East: down; Field dimensions:  12.6$\times$  9.2 arcmin; Surface brightness range displayed: 11.0$-$28.0 mag arcsec$^{-2}$}                 
\label{NGC1365}     
 \end{figure}

\begin{figure}
\figurenum{1.29} 
\plotone{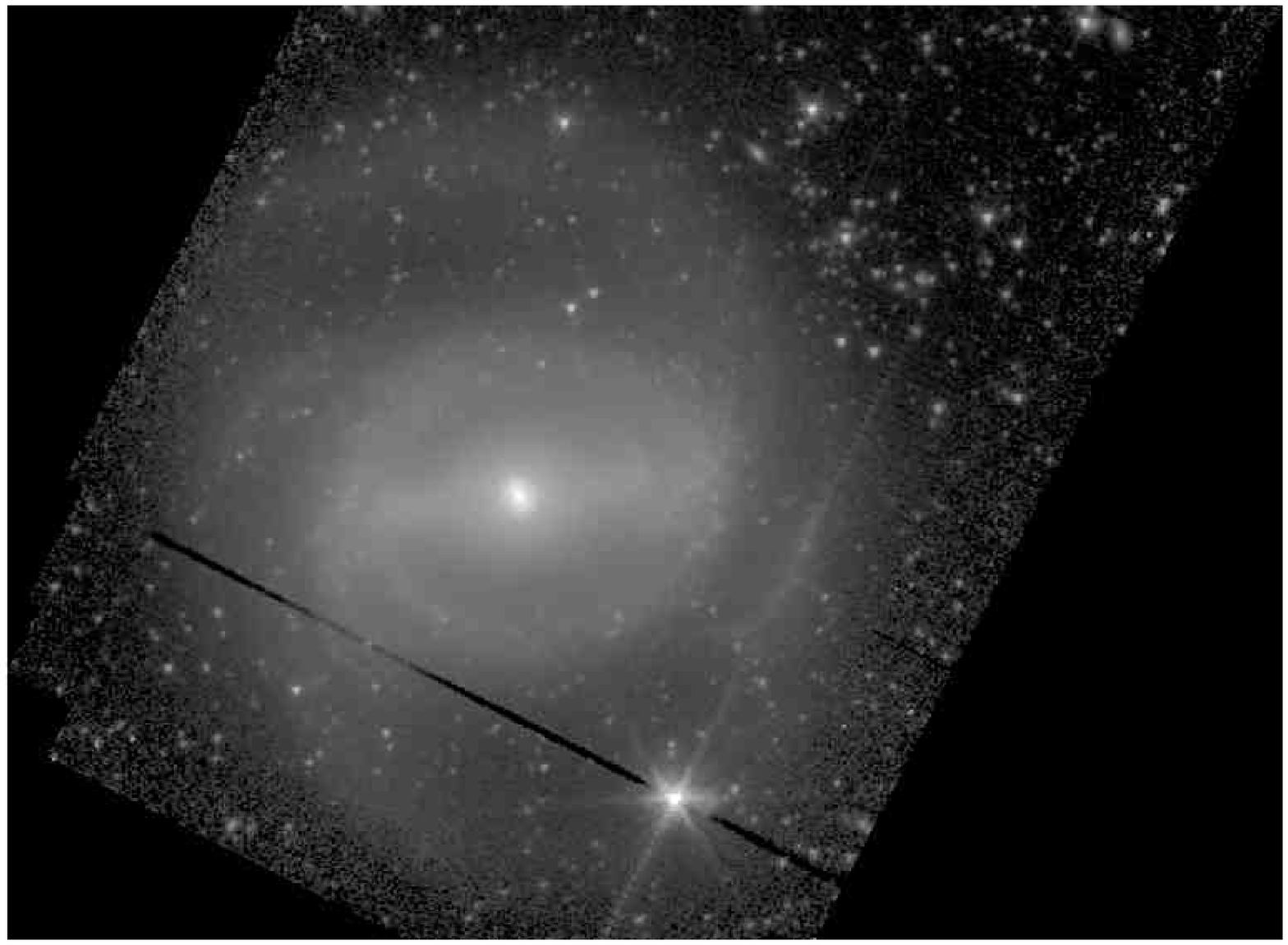}
 \vspace{2.0truecm}
 \caption{
{\bf NGC  1433   }              - S$^4$G mid-IR classification:    (R$_1^{\prime}$)SB(r,nr,nb)a                                  ; Filter: IRAC 3.6$\mu$m; North:   up, East: left; Field dimensions:  10.5$\times$  7.6 arcmin; Surface brightness range displayed: 13.0$-$28.0 mag arcsec$^{-2}$}                 
\label{NGC1433}     
 \end{figure}

\begin{figure}
\figurenum{1.30} 
\plotone{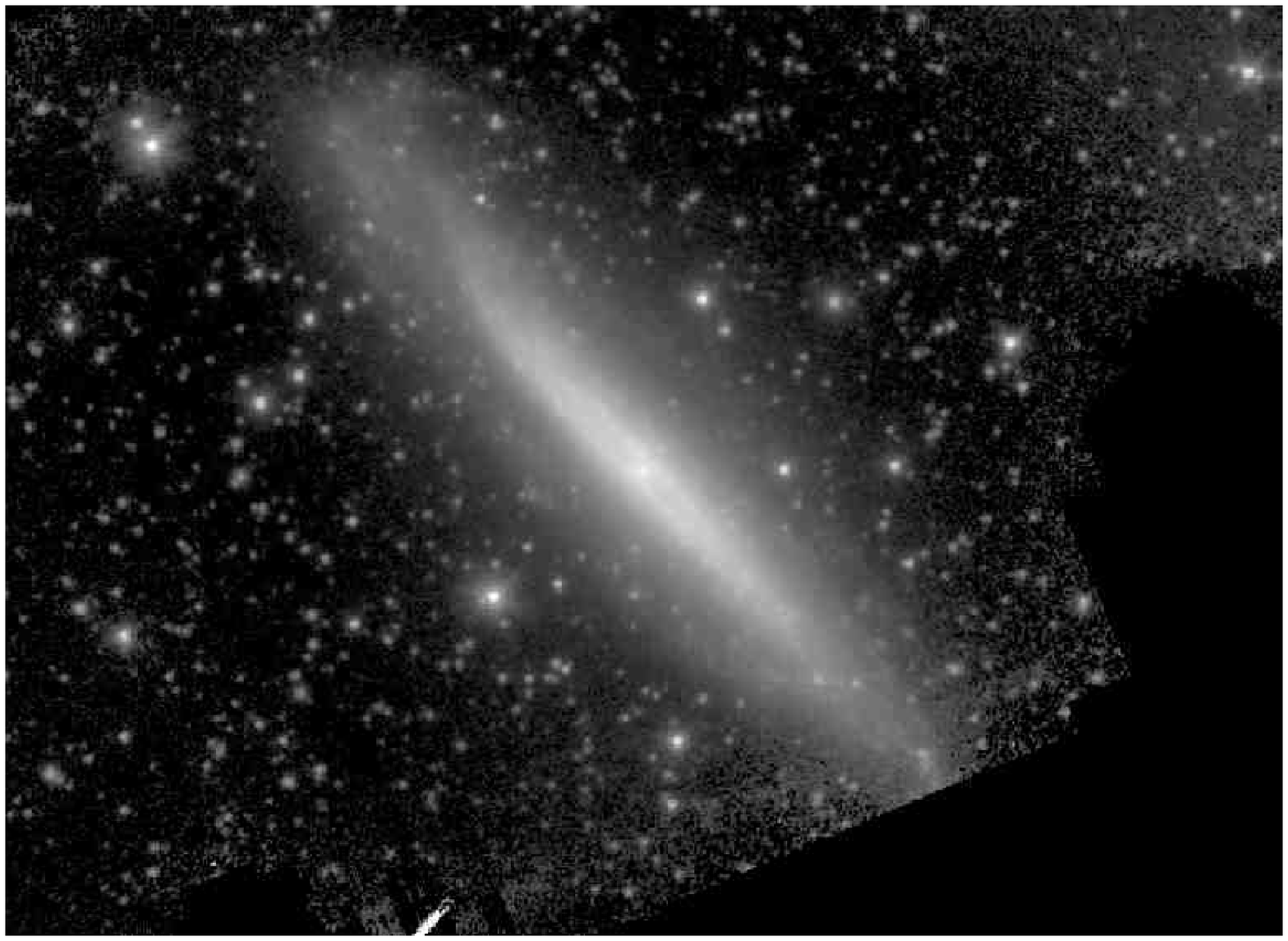}
 \vspace{2.0truecm}
 \caption{
{\bf NGC  1448   }              - S$^4$G mid-IR classification:    SA(rs:)c sp                                           ; Filter: IRAC 3.6$\mu$m; North:   up, East: left; Field dimensions:   9.0$\times$  6.6 arcmin; Surface brightness range displayed: 14.0$-$28.0 mag arcsec$^{-2}$}                 
\label{NGC1448}     
 \end{figure}

\begin{figure}
\figurenum{1.31} 
\plotone{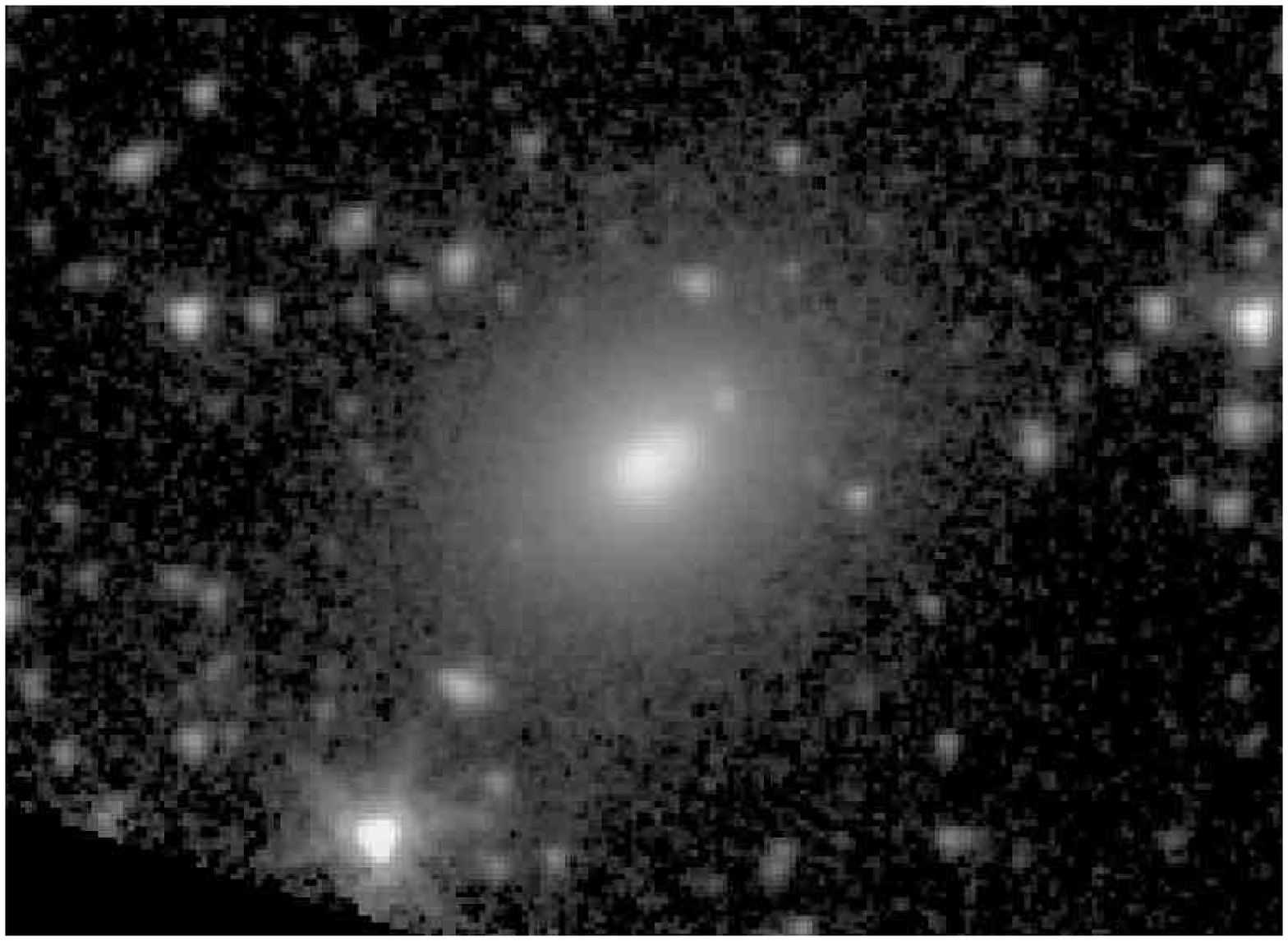}
 \vspace{2.0truecm}
 \caption{
{\bf NGC  1481   }              - S$^4$G mid-IR classification:    SA(l)0$^-$: pec                                       ; Filter: IRAC 3.6$\mu$m; North:   up, East: left; Field dimensions:   2.6$\times$  1.9 arcmin; Surface brightness range displayed: 15.5$-$28.0 mag arcsec$^{-2}$}                 
\label{NGC1481}     
 \end{figure}

\begin{figure}
\figurenum{1.32} 
\plotone{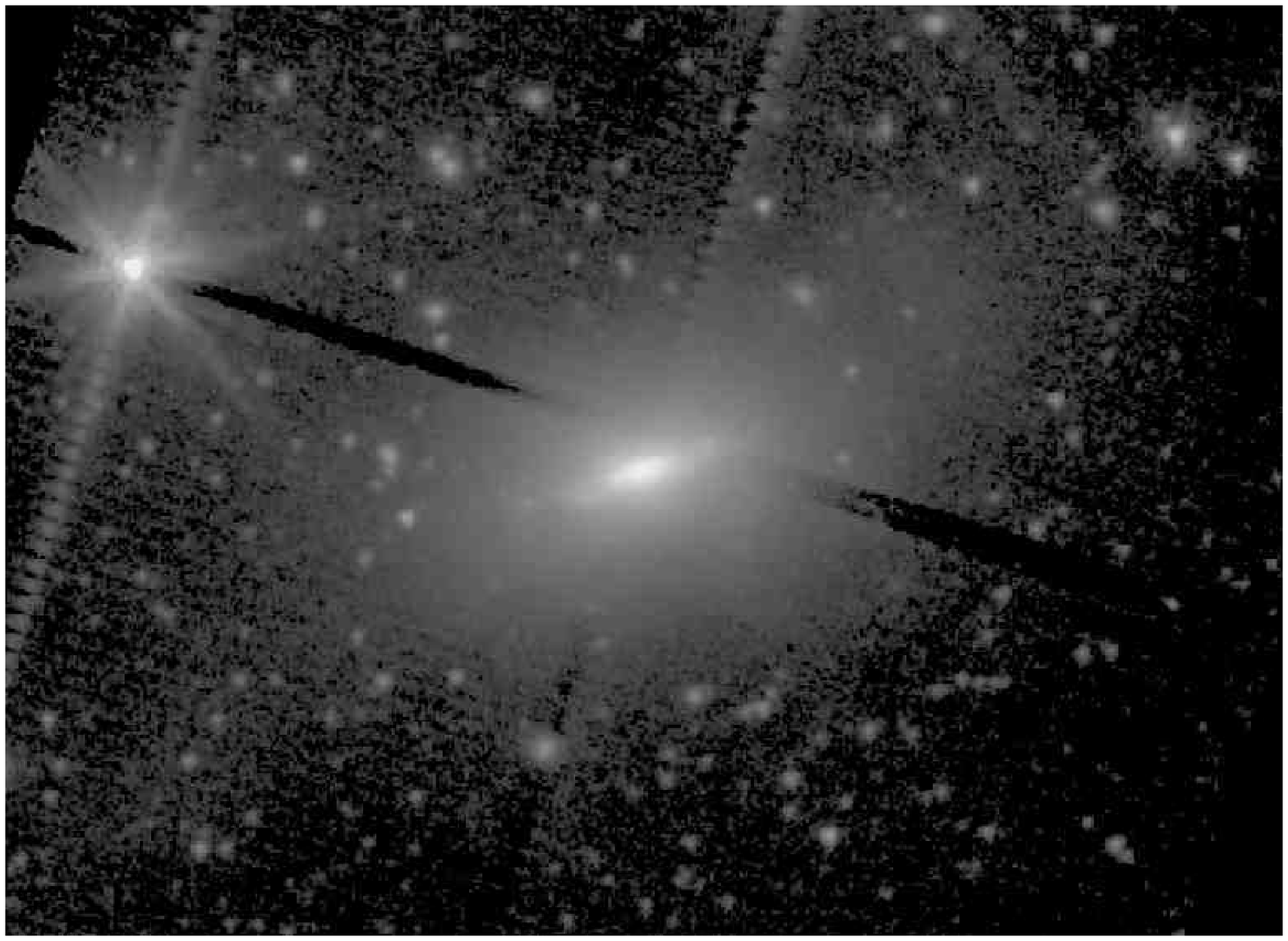}
 \vspace{2.0truecm}
 \caption{
{\bf NGC  1482   }              - S$^4$G mid-IR classification:    Sa: sp                                                ; Filter: IRAC 3.6$\mu$m; North:   up, East: left; Field dimensions:   5.3$\times$  3.8 arcmin; Surface brightness range displayed: 12.0$-$28.0 mag arcsec$^{-2}$}                 
\label{NGC1482}     
 \end{figure}

\begin{figure}
\figurenum{1.33} 
\plotone{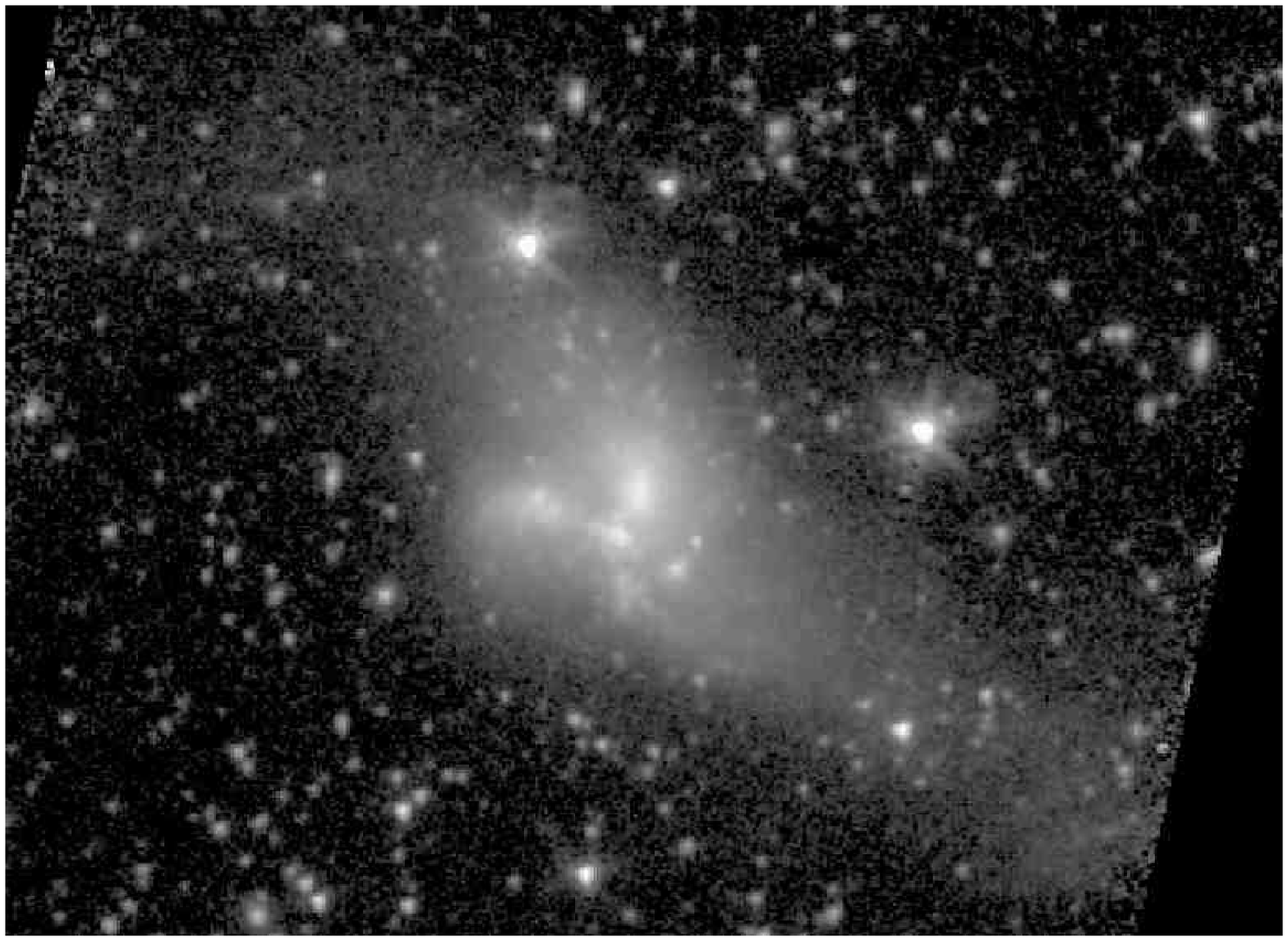}
 \vspace{2.0truecm}
 \caption{
{\bf NGC  1487   }              - S$^4$G mid-IR classification:    Pec                                                   ; Filter: IRAC 3.6$\mu$m; North:   up, East: left; Field dimensions:   5.8$\times$  4.2 arcmin; Surface brightness range displayed: 15.5$-$28.0 mag arcsec$^{-2}$}                 
\label{NGC1487}     
 \end{figure}

\begin{figure}
\figurenum{1.34} 
\plotone{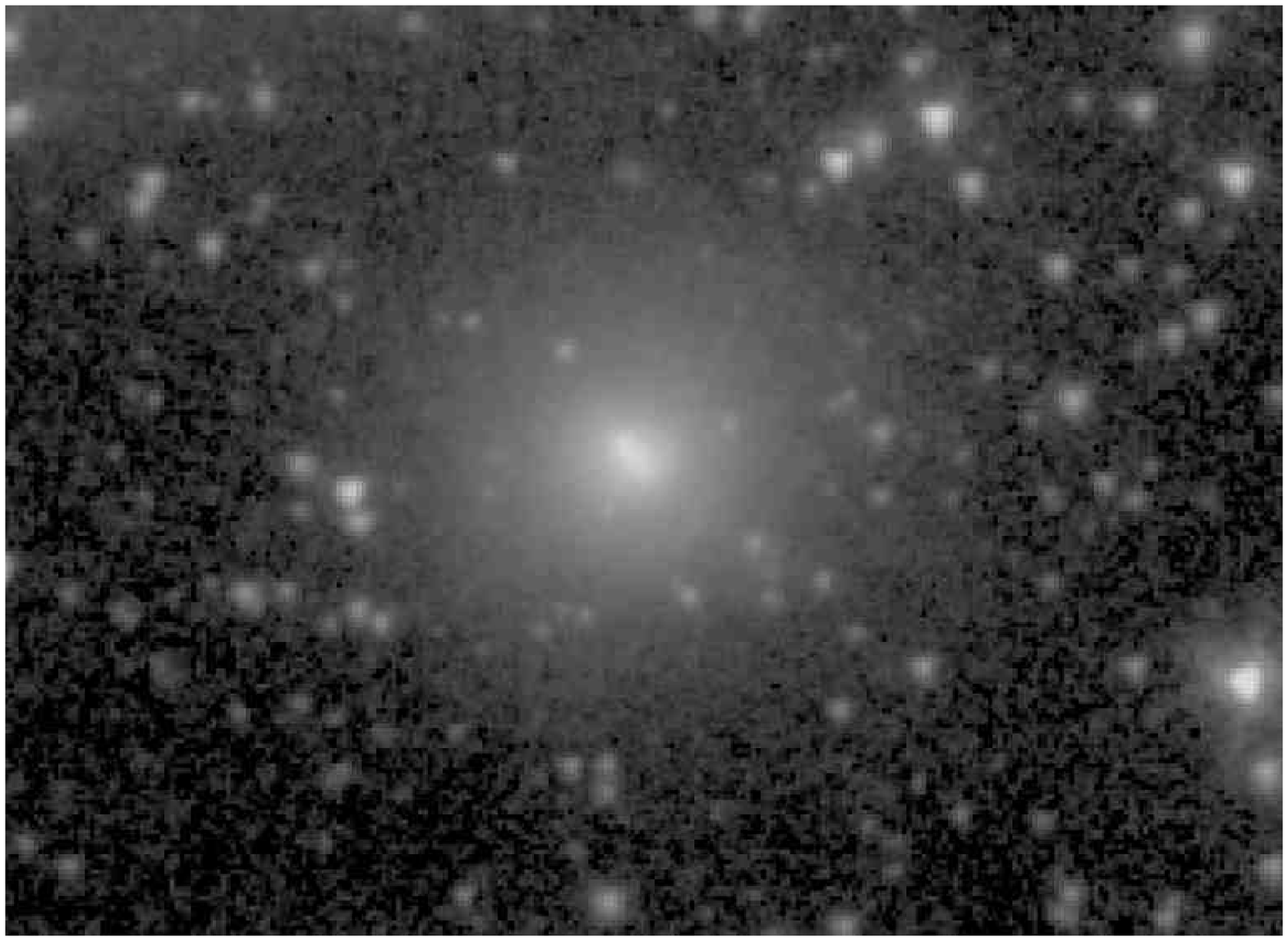}
 \vspace{2.0truecm}
 \caption{
{\bf NGC  1510   }              - S$^4$G mid-IR classification:    SA0$^+$:                                              ; Filter: IRAC 3.6$\mu$m; North:   up, East: left; Field dimensions:   3.2$\times$  2.3 arcmin; Surface brightness range displayed: 15.5$-$28.0 mag arcsec$^{-2}$}                 
\label{NGC1510}     
 \end{figure}

\begin{figure}
\figurenum{1.35} 
\plotone{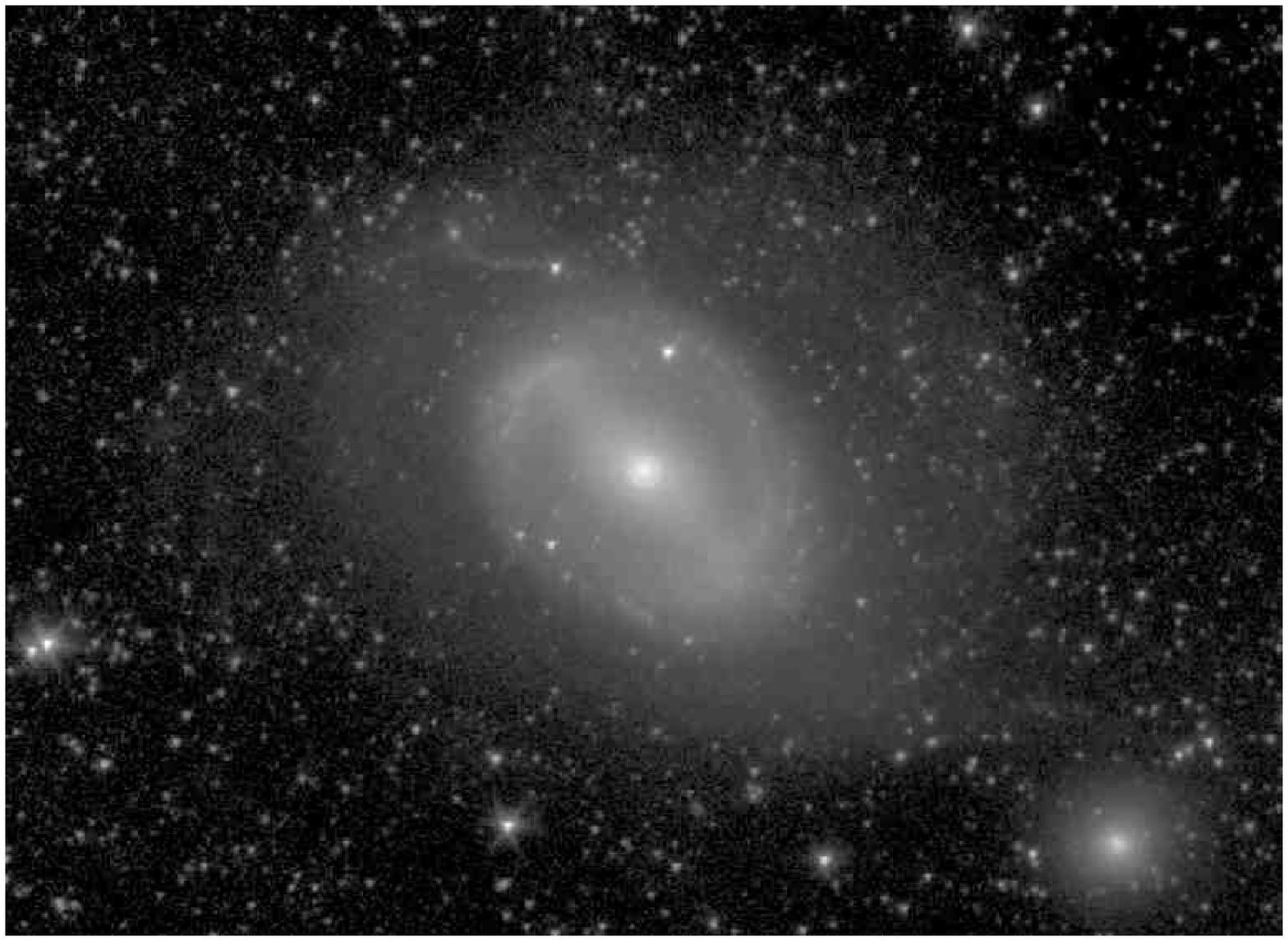}
 \vspace{2.0truecm}
 \caption{
{\bf NGC  1512   }              - S$^4$G mid-IR classification:    (RL)SB(r,nr)a                                        ; Filter: IRAC 3.6$\mu$m; North:   up, East: left; Field dimensions:  10.5$\times$  7.7 arcmin; Surface brightness range displayed: 13.5$-$28.0 mag arcsec$^{-2}$}                  
\label{NGC1512}     
 \end{figure}

\begin{figure}
\figurenum{1.36} 
\plotone{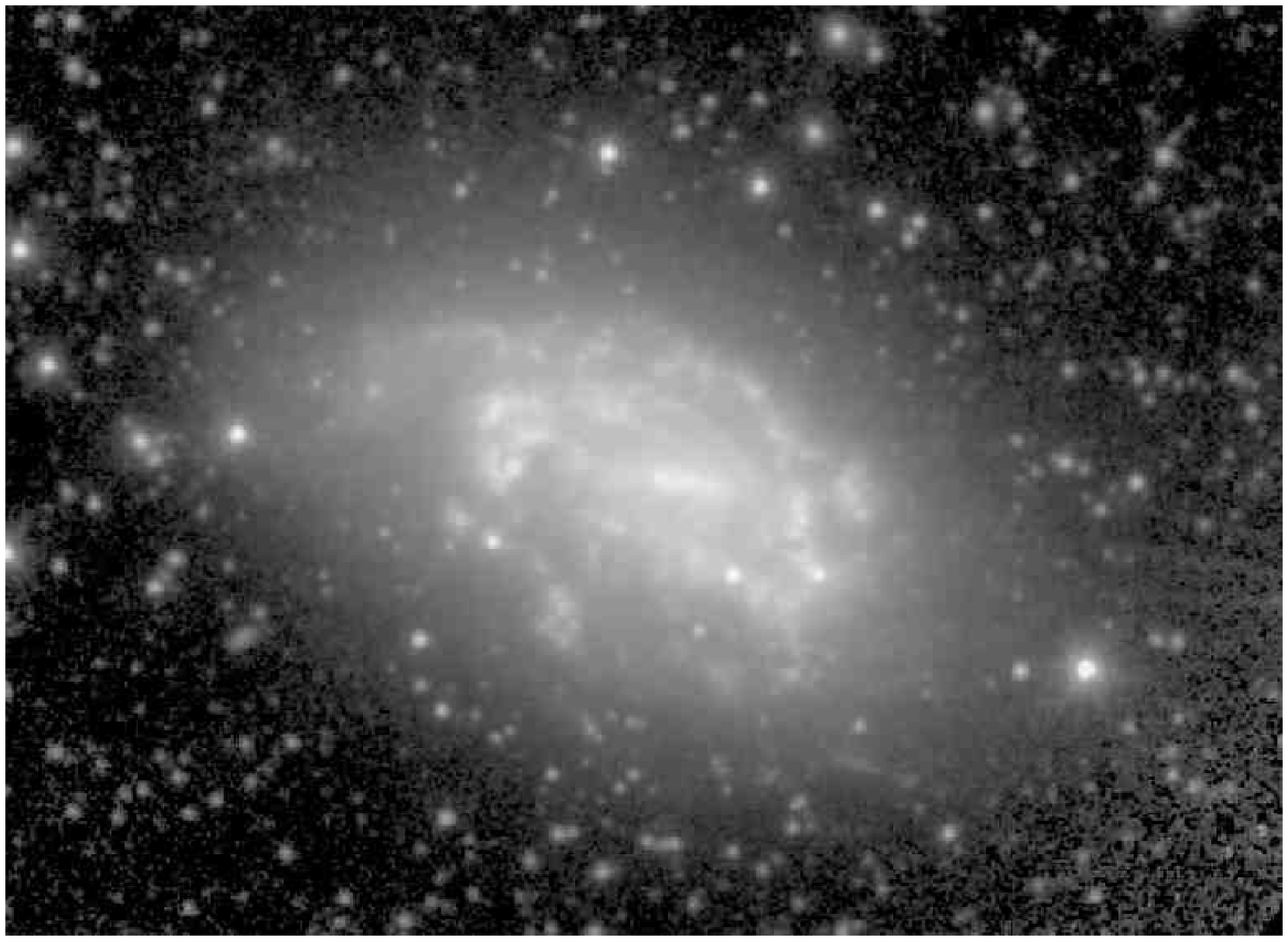}
 \vspace{2.0truecm}
 \caption{
{\bf NGC  1559   }              - S$^4$G mid-IR classification:    SB(s)cd                                               ; Filter: IRAC 3.6$\mu$m; North:   up, East: left; Field dimensions:   5.7$\times$  4.2 arcmin; Surface brightness range displayed: 15.0$-$28.0 mag arcsec$^{-2}$}                 
\label{NGC1559}     
 \end{figure}

\begin{figure}
\figurenum{1.37} 
\plotone{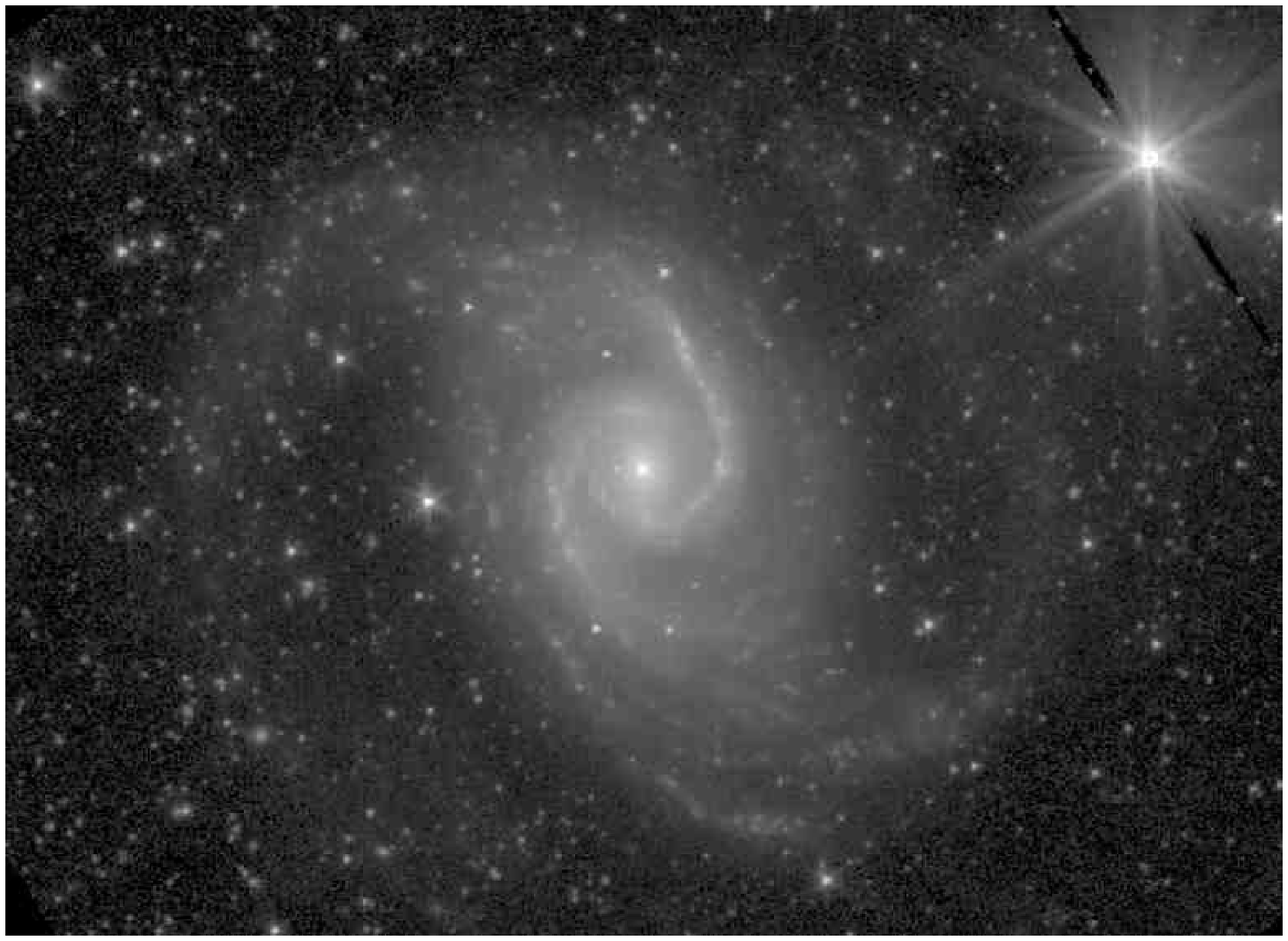}
 \vspace{2.0truecm}
 \caption{
{\bf NGC  1566   }              - S$^4$G mid-IR classification:    (R$_1^{\prime}$)SAB(s)b                                       ; Filter: IRAC 3.6$\mu$m; North:   up, East: left; Field dimensions:  11.5$\times$  8.4 arcmin; Surface brightness range displayed: 12.0$-$28.0 mag arcsec$^{-2}$}                 
\label{NGC1566}     
 \end{figure}

\begin{figure}
\figurenum{1.38} 
\plotone{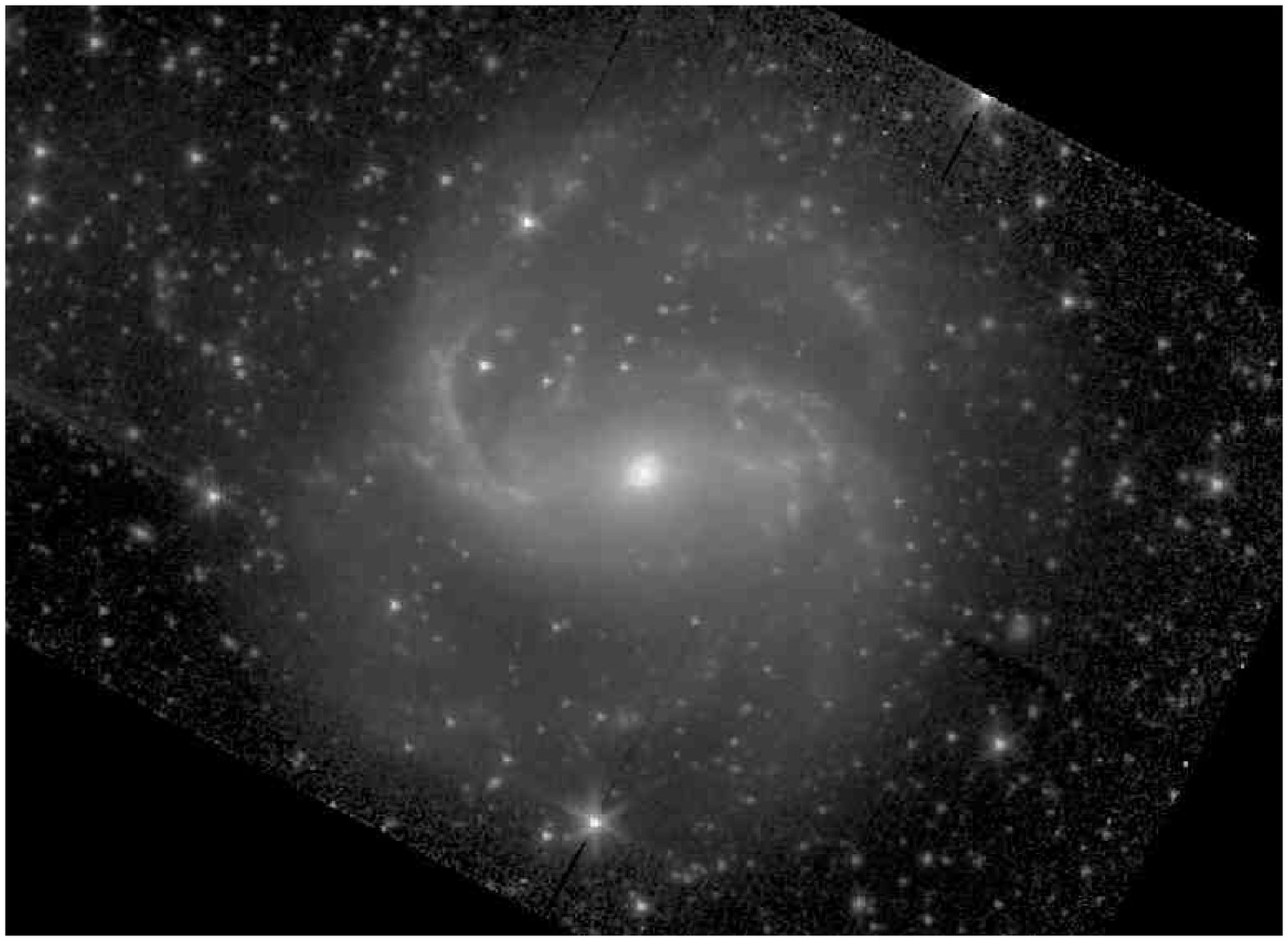}
 \vspace{2.0truecm}
 \caption{
{\bf NGC  1672   }              - S$^4$G mid-IR classification:    (R$^{\prime}$)SA$\underline{\rm B}$(rs,nr)b                     ; Filter: IRAC 3.6$\mu$m; North:   up, East: left; Field dimensions:   9.9$\times$  7.2 arcmin; Surface brightness range displayed: 12.0$-$28.0 mag arcsec$^{-2}$}                 
\label{NGC1672}     
 \end{figure}

\begin{figure}
\figurenum{1.39} 
\plotone{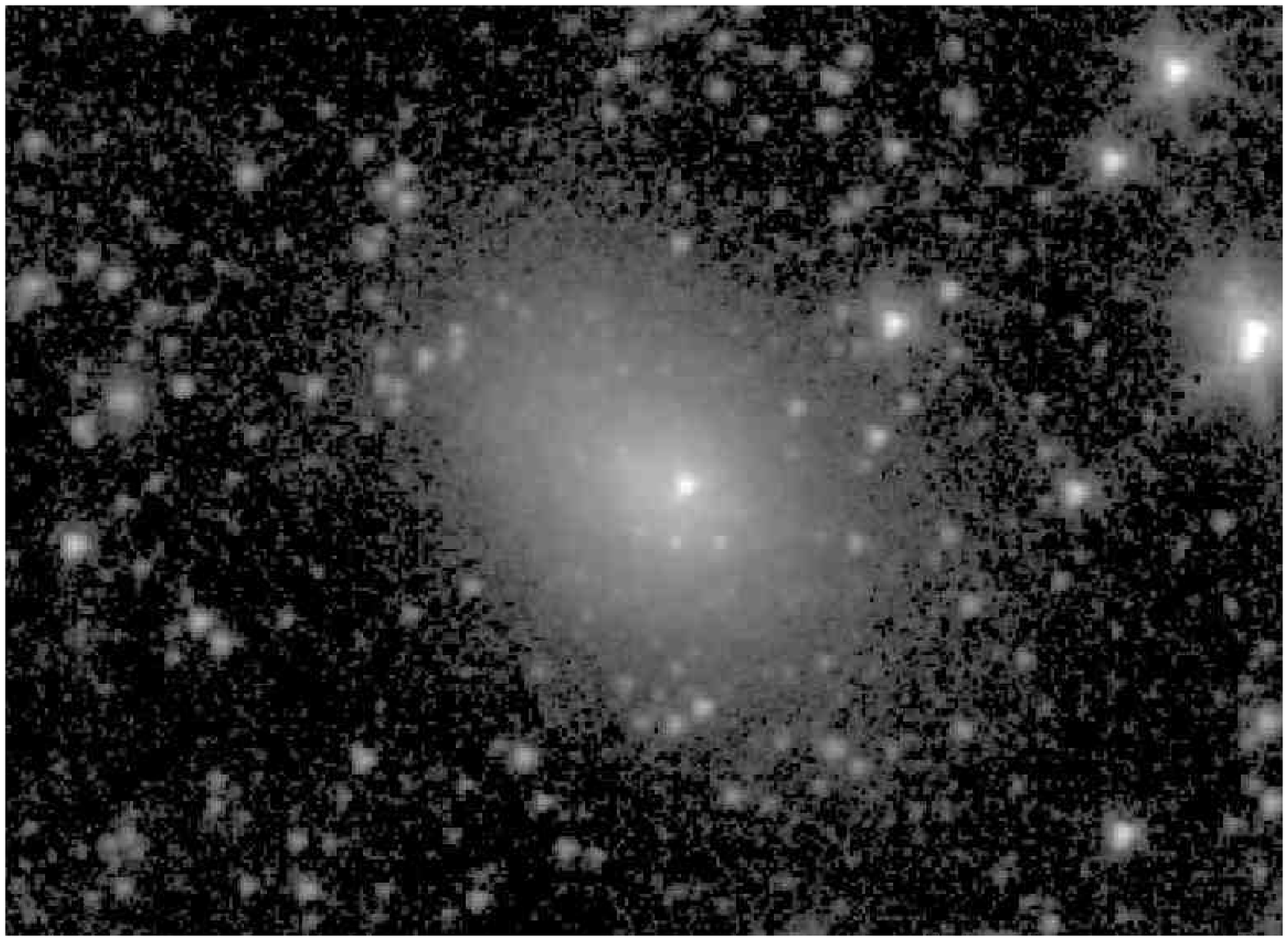}
 \vspace{2.0truecm}
 \caption{
{\bf NGC  1705   }              - S$^4$G mid-IR classification:    dE3,N                                                 ; Filter: IRAC 3.6$\mu$m; North:   up, East: left; Field dimensions:   4.2$\times$  3.0 arcmin; Surface brightness range displayed: 12.0$-$28.0 mag arcsec$^{-2}$}                 
\label{NGC1705}     
 \end{figure}

\begin{figure}
\figurenum{1.40} 
\plotone{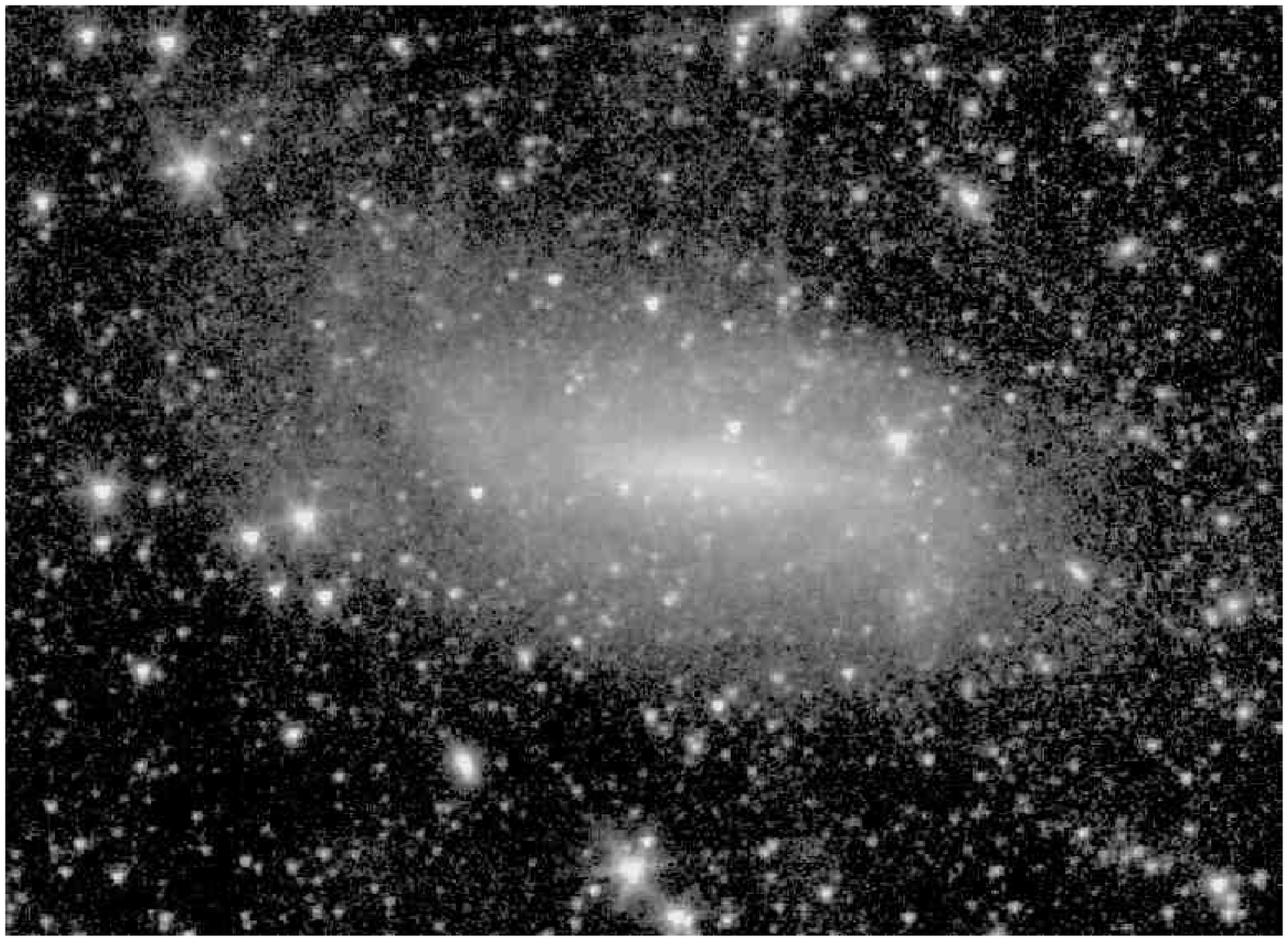}
 \vspace{2.0truecm}
 \caption{
{\bf NGC  1744   }              - S$^4$G mid-IR classification:    SB(s)d                                                ; Filter: IRAC 3.6$\mu$m; North: left, East: down; Field dimensions:   7.9$\times$  5.7 arcmin; Surface brightness range displayed: 17.5$-$28.0 mag arcsec$^{-2}$}                 
\label{NGC1744}     
 \end{figure}

\begin{figure}
\figurenum{1.41} 
\plotone{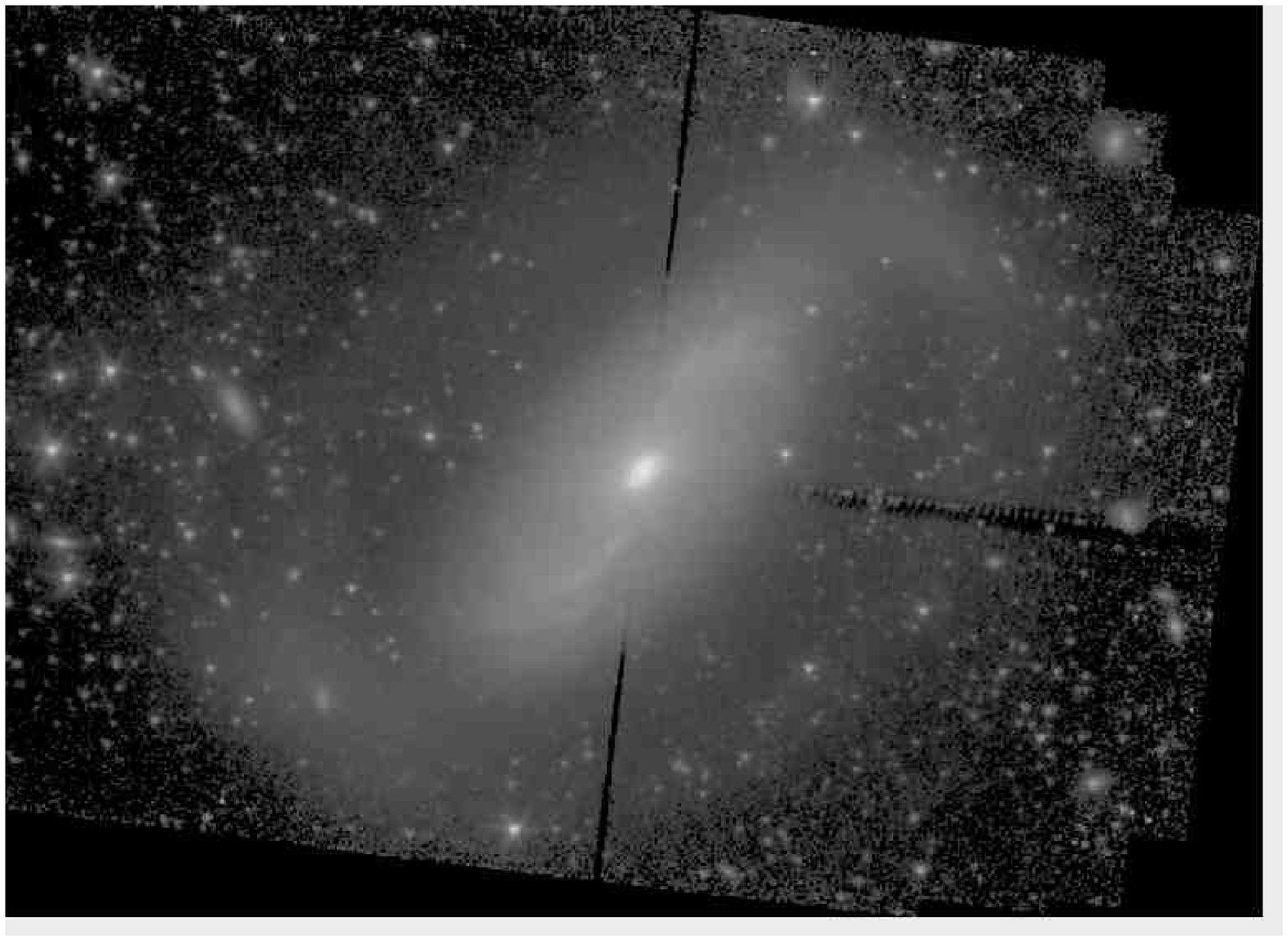}
 \vspace{2.0truecm}
 \caption{
{\bf NGC  1808   }              - S$^4$G mid-IR classification:    (R$_1$)SAB(s,nr)a                                     ; Filter: IRAC 3.6$\mu$m; North:   up, East: left; Field dimensions:  10.4$\times$  7.5 arcmin; Surface brightness range displayed: 11.5$-$28.0 mag arcsec$^{-2}$}                 
\label{NGC1808}     
 \end{figure}

\begin{figure}
\figurenum{1.42} 
\plotone{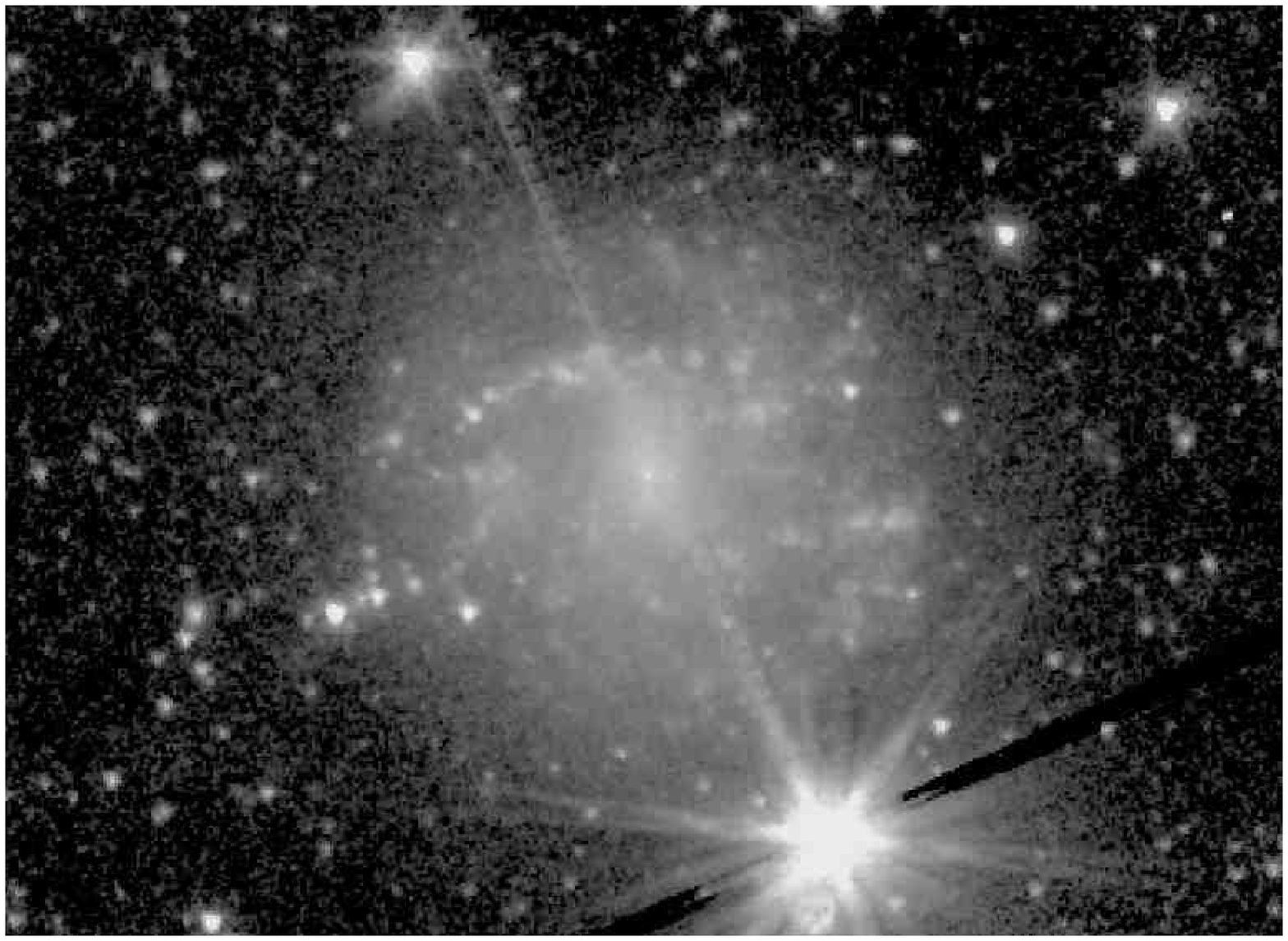}
 \vspace{2.0truecm}
 \caption{
{\bf NGC  2500   }              - S$^4$G mid-IR classification:    SAB(s)d                                               ; Filter: IRAC 3.6$\mu$m; North:   up, East: left; Field dimensions:   5.3$\times$  3.8 arcmin; Surface brightness range displayed: 16.5$-$28.0 mag arcsec$^{-2}$}                 
\label{NGC2500}     
 \end{figure}

\begin{figure}
\figurenum{1.43} 
\plotone{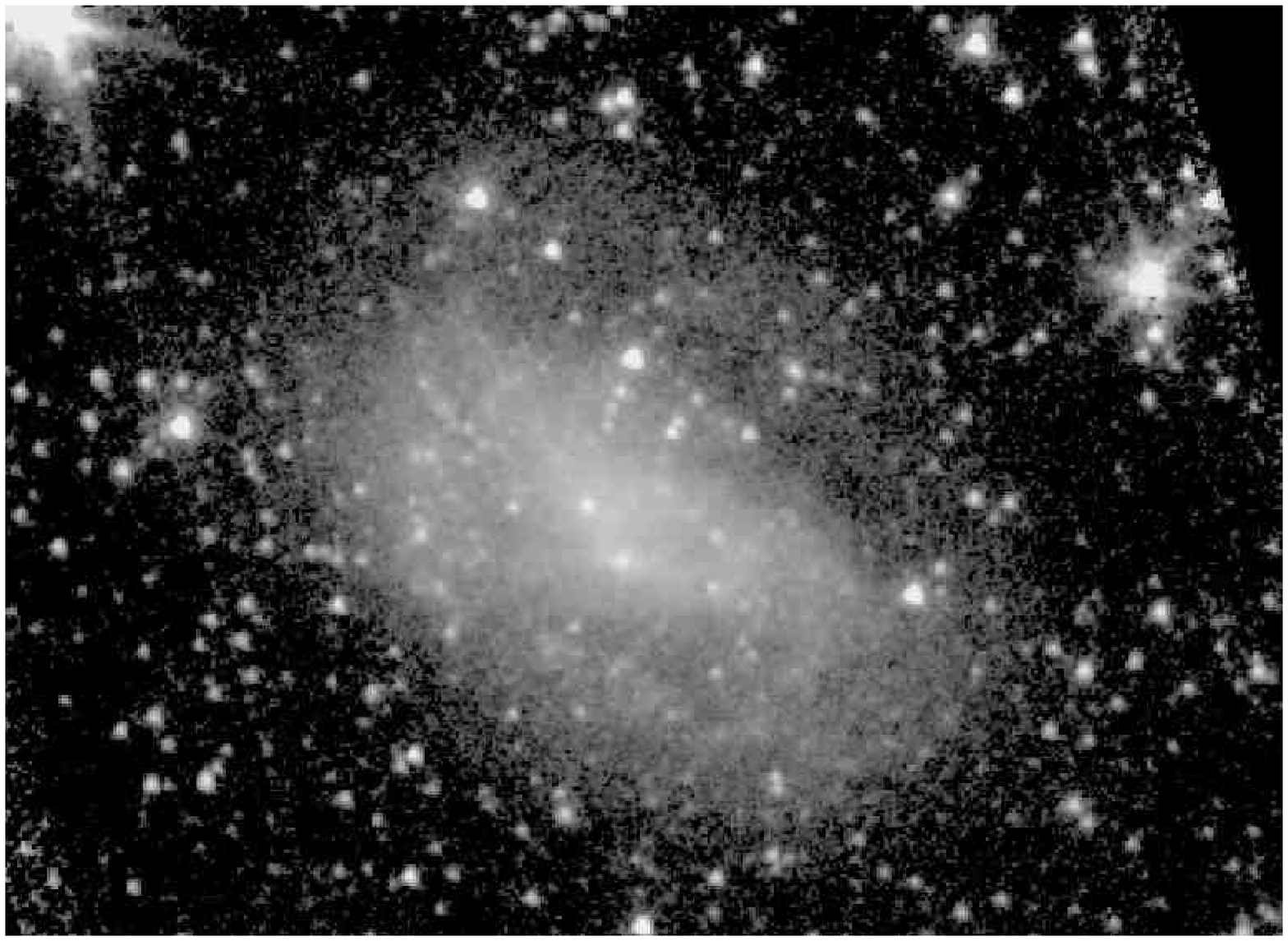}
 \vspace{2.0truecm}
 \caption{
{\bf NGC  2552   }              - S$^4$G mid-IR classification:    (R$^{\prime}$)SAB(s)m                                           ; Filter: IRAC 3.6$\mu$m; North:   up, East: left; Field dimensions:   5.3$\times$  3.8 arcmin; Surface brightness range displayed: 18.0$-$28.0 mag arcsec$^{-2}$}                 
\label{NGC2552}     
 \end{figure}

\begin{figure}
\figurenum{1.44} 
\plotone{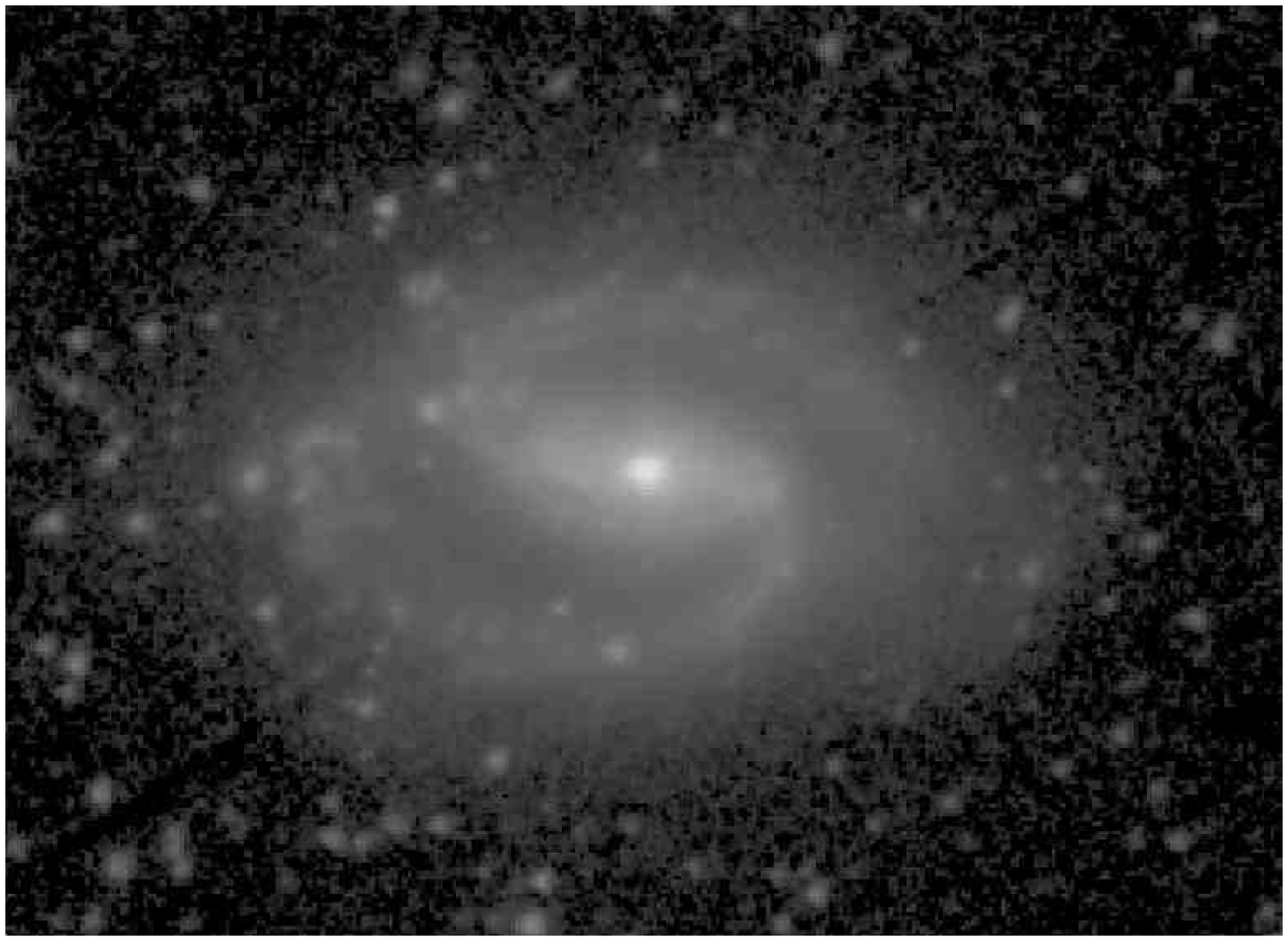}
 \vspace{2.0truecm}
 \caption{
{\bf NGC  2633   }              - S$^4$G mid-IR classification:    SAB(rs)b                                              ; Filter: IRAC 3.6$\mu$m; North: left, East: down; Field dimensions:   3.5$\times$  2.6 arcmin; Surface brightness range displayed: 12.0$-$28.0 mag arcsec$^{-2}$}                 
\label{NGC2633}     
 \end{figure}

\begin{figure}
\figurenum{1.45} 
\plotone{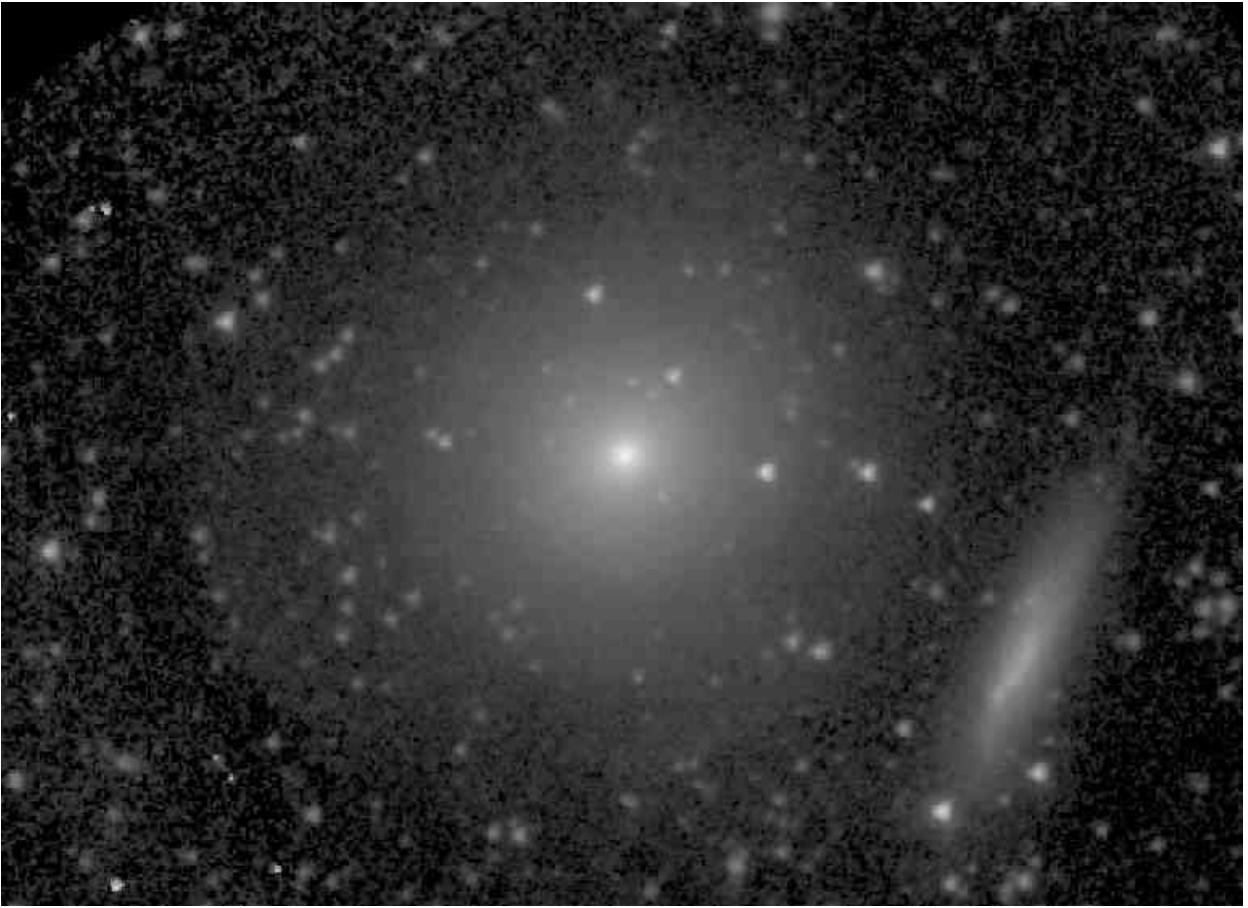}
 \vspace{2.0truecm}
 \caption{
{\bf NGC  2634   }              - S$^4$G mid-IR classification:    SA(nl)0$^-$ (shells/ripples)                                  ; Filter: IRAC 3.6$\mu$m; North: left, East: down; Field dimensions:   5.3$\times$  3.8 arcmin; Surface brightness range displayed: 13.5$-$28.0 mag arcsec$^{-2}$}                 
\label{NGC2634}     
 \end{figure}

\begin{figure}
\figurenum{1.46} 
\plotone{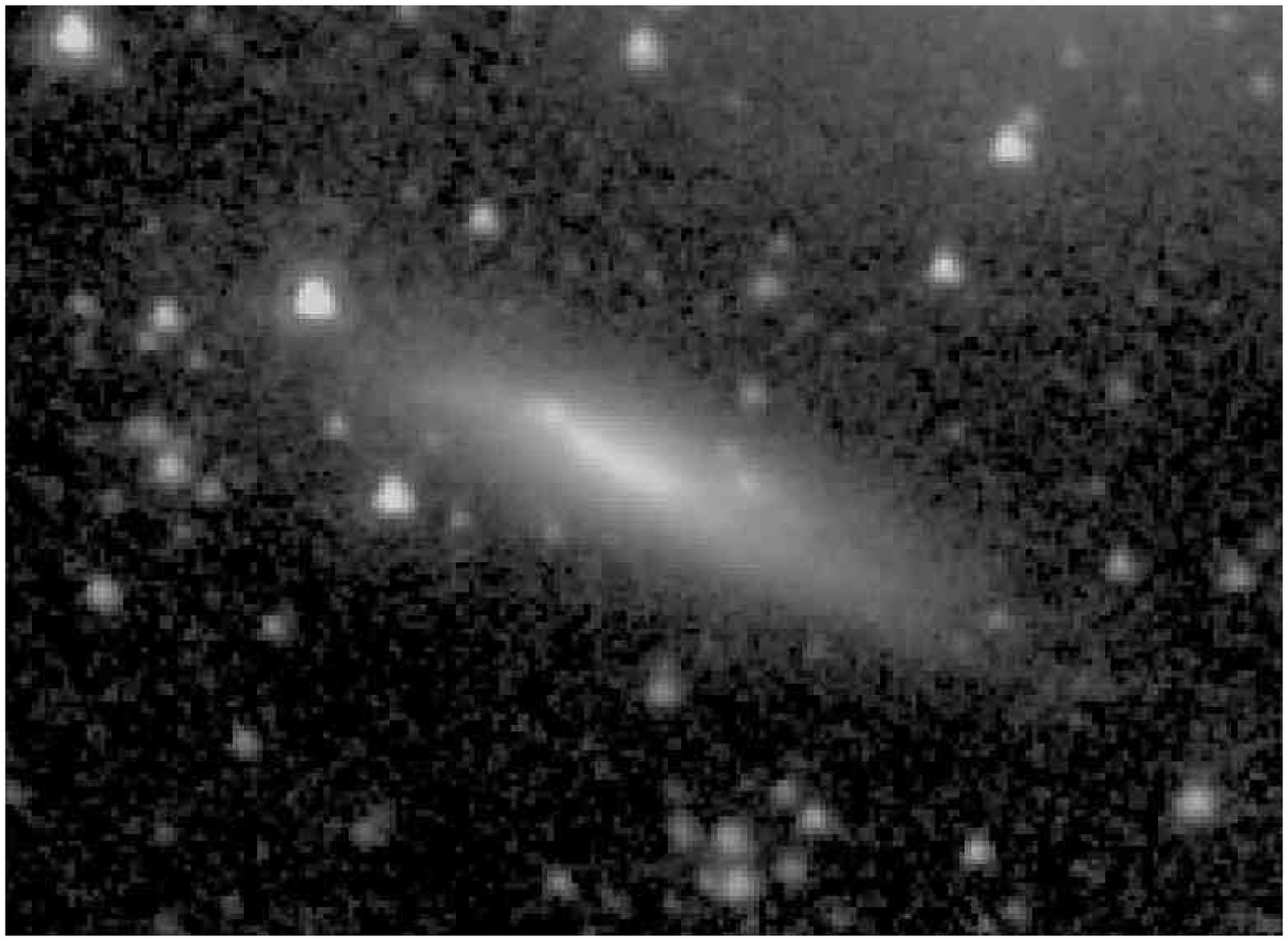}
 \vspace{2.0truecm}
 \caption{
{\bf NGC  2634A  }              - S$^4$G mid-IR classification:    SB(s)m: sp                                            ; Filter: IRAC 3.6$\mu$m; North:   up, East: left; Field dimensions:   2.6$\times$  1.9 arcmin; Surface brightness range displayed: 17.0$-$28.0 mag arcsec$^{-2}$}                 
\label{NGC2634A}    
 \end{figure}

\begin{figure}
\figurenum{1.47} 
\plotone{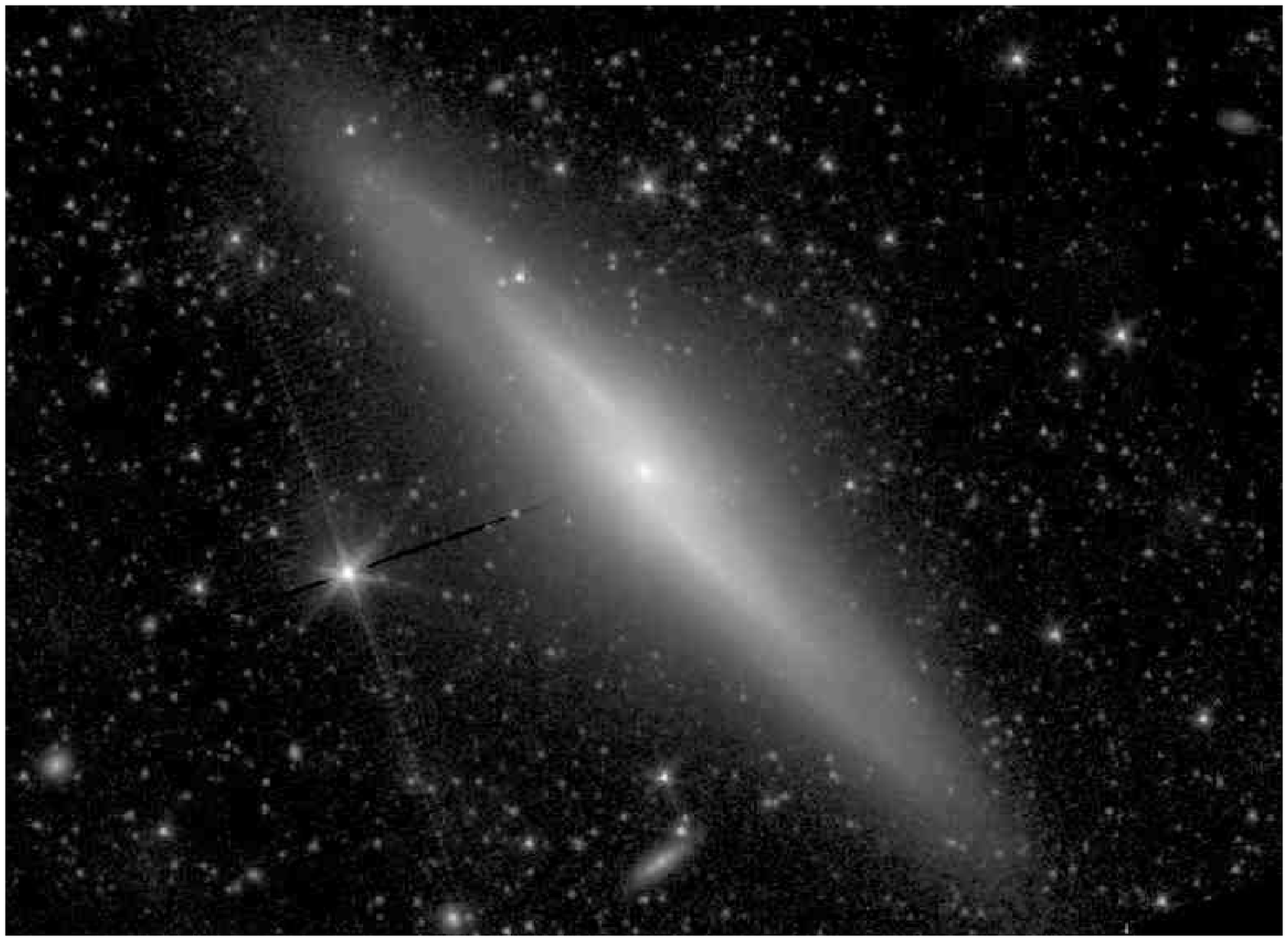}
 \vspace{2.0truecm}
 \caption{
{\bf NGC  2683   }              - S$^4$G mid-IR classification:    (RL)SB$_x$($\underline{\rm r}$s)0/a  sp               ; Filter: IRAC 3.6$\mu$m; North:   up, East: left; Field dimensions:  11.3$\times$  8.2 arcmin; Surface brightness range displayed: 13.5$-$28.0 mag arcsec$^{-2}$}                 
\label{NGC2683}     
 \end{figure}

\begin{figure}
\figurenum{1.48} 
\plotone{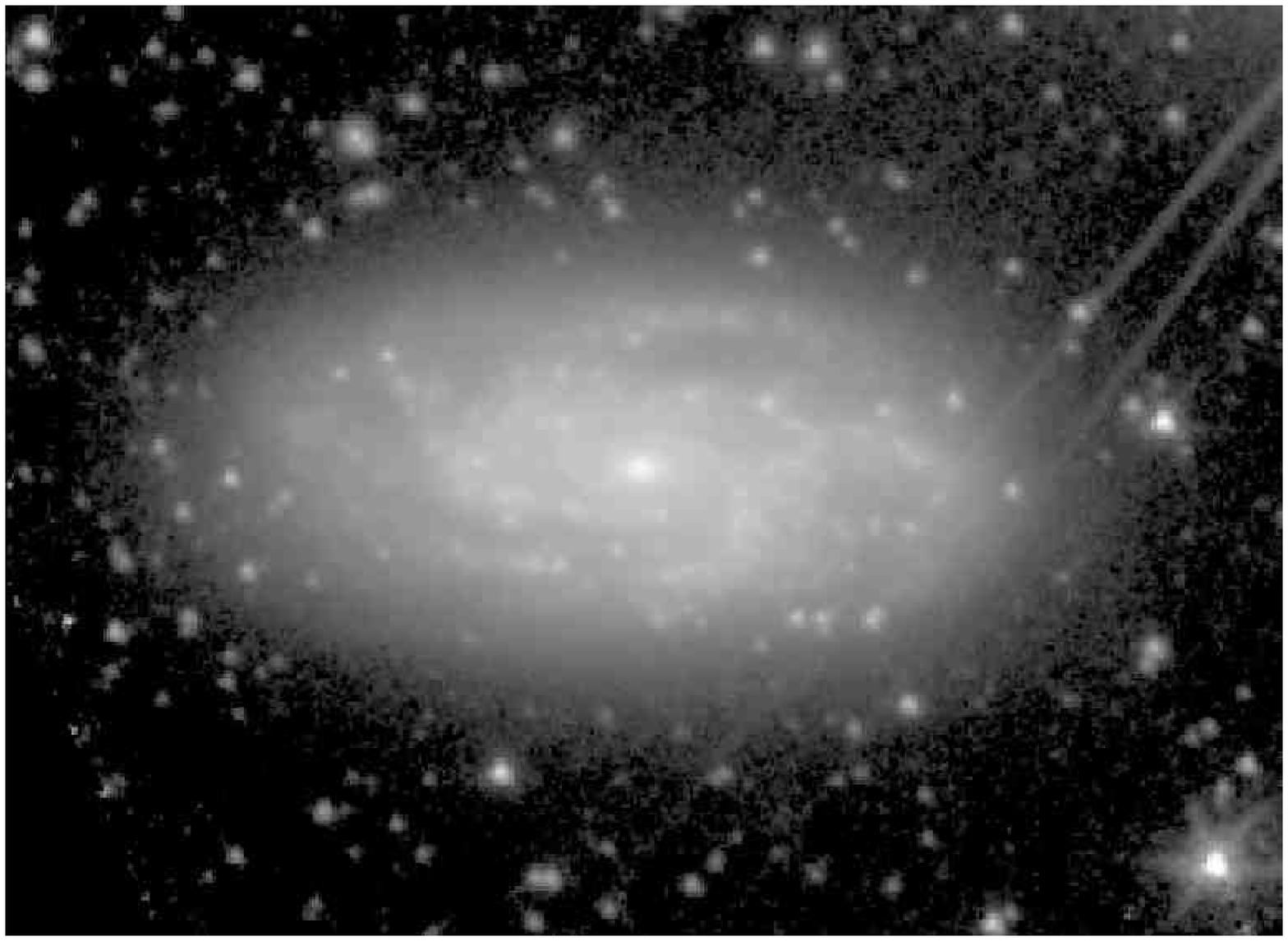}
 \vspace{2.0truecm}
 \caption{
{\bf NGC  2742   }              - S$^4$G mid-IR classification:    SA(s)c                                                ; Filter: IRAC 3.6$\mu$m; North:   up, East: left; Field dimensions:   4.5$\times$  3.3 arcmin; Surface brightness range displayed: 15.0$-$28.0 mag arcsec$^{-2}$}                 
\label{NGC2742}     
 \end{figure}

\begin{figure}
\figurenum{1.49} 
\plotone{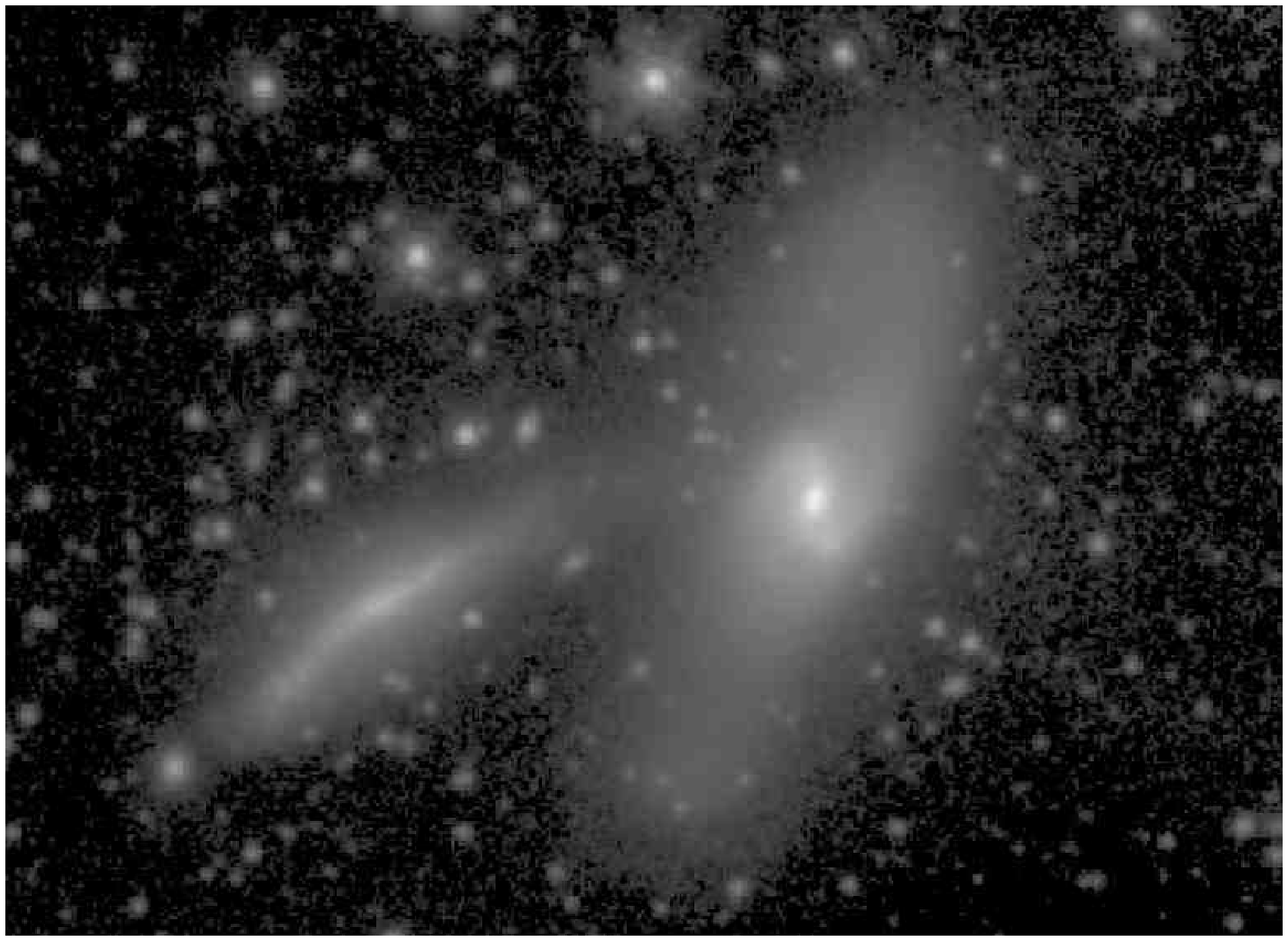}
 \vspace{2.0truecm}
 \caption{
{\bf NGC  2798} (right) and {\bf NGC  2799   } (left) - S$^4$G mid-IR classifications:    S$\underline{\rm A}$B(s)a pec, SB(s)dm? pec sp, respectively; Filter: IRAC 3.6$\mu$m; North:   up, East: left; Field dimensions:   4.5$\times$  3.3 arcmin; Surface brightness range displayed: 12.0$-$28.0 mag arcsec$^{-2}$}                 
\label{NGC2798}     
 \end{figure}

\begin{figure}
\figurenum{1.50} 
\plotone{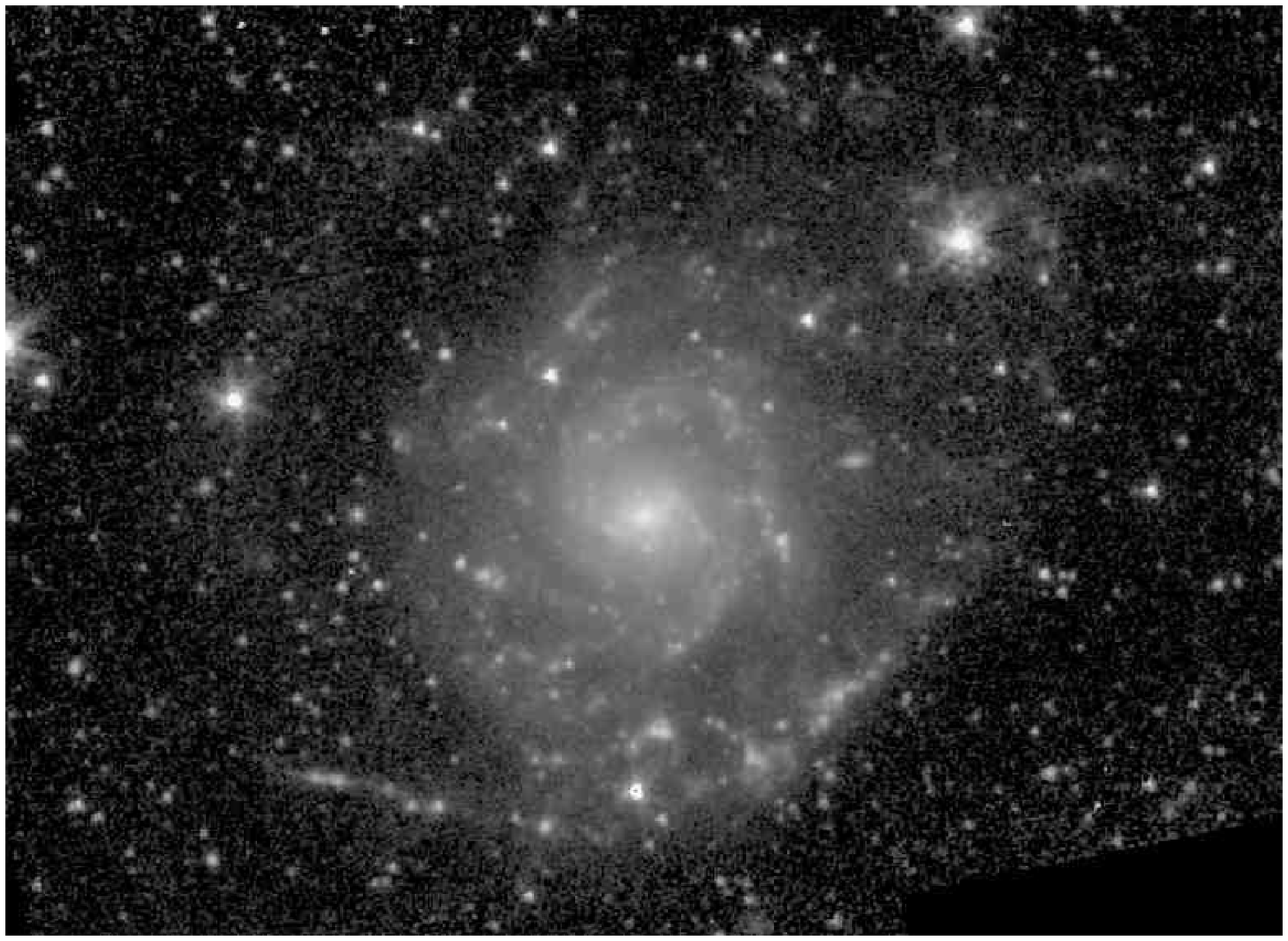}
 \vspace{2.0truecm}
 \caption{
{\bf NGC  2805   }              - S$^4$G mid-IR classification:    (R)S$\underline{\rm A}$B(r$\underline{\rm s}$)c pec   ; Filter: IRAC 3.6$\mu$m; North:   up, East: left; Field dimensions:   7.9$\times$  5.7 arcmin; Surface brightness range displayed: 16.5$-$28.0 mag arcsec$^{-2}$}                 
\label{NGC2805}     
 \end{figure}

\begin{figure}
\figurenum{1.51} 
\plotone{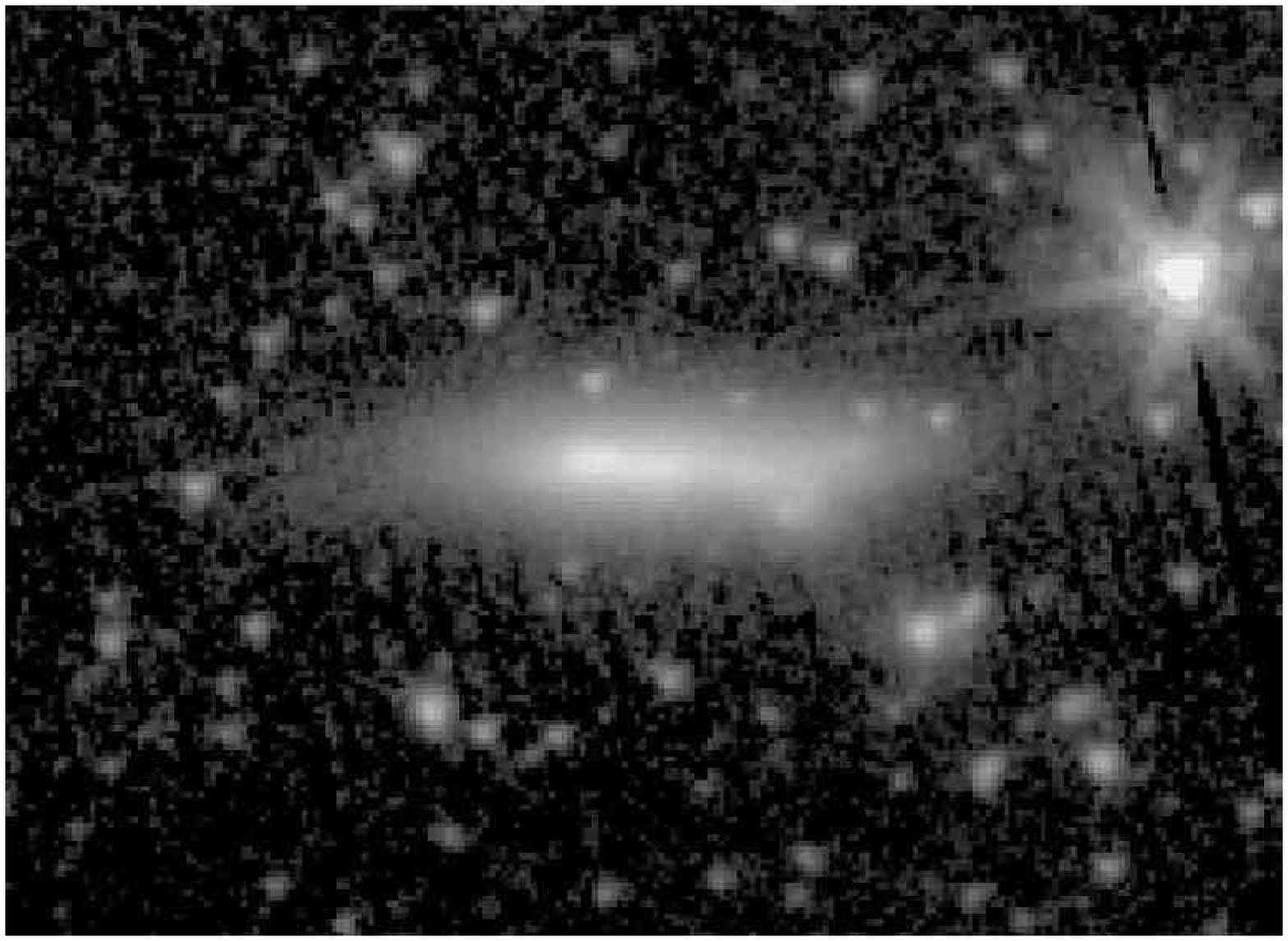}
 \vspace{2.0truecm}
 \caption{
{\bf NGC  2814   }              - S$^4$G mid-IR classification:    S pec sp                                              ; Filter: IRAC 3.6$\mu$m; North: left, East: down; Field dimensions:   2.6$\times$  1.9 arcmin; Surface brightness range displayed: 15.5$-$28.0 mag arcsec$^{-2}$}                 
\label{NGC2814}     
 \end{figure}

In order to classify the galaxies, each processed image was converted
to units of magnitudes per square arcsecond (the same approach as used
in the dVA). Because all the pipeline-processed images are in the same
final physical units, MJy/sr, this conversion was performed using the
same zero point, 17.69, based on the Vega magnitude calibration given
in the {\it IRAC Instrument Handbook}. The images were
then displayed in ds9 using IRAF routine DISPLAY with a faint limit of
28.0 mag arcsec$^{-2}$ and a bright limit depending on the galaxy,
ranging from as bright as 11.0 mag arcsec$^{-2}$ to 18.0 mag
arcsec$^{-2}$ or fainter. The classification images prepared in this
manner are illustrated in Figures~\ref{NGC0024} - ~\ref{CGCG265-55}.

The reason for using this approach is that images in units of mag
arcsec$^{-2}$ reveal the morphology over the whole range of surface
brightnesses, from the center to the outer disk, much better than do
linear intensity images. In early classifications, photographic plates
gave a limited range in the linear relation between photographic
density and the logarithm of the intensity. IRAC data have a much
larger dynamic range, and hence all the details of a galaxy, including
subtle ones, can be displayed at once.  This is useful for accurate
classification, and especially for recognizing the subtle distinctions
between elliptical galaxies and very early S0 galaxies.  We examined
all images using a 24-bit image display, and in
each case a stretch was chosen (by varying contrast and brightness)
that maximized the morphological information visible. This display
setting was then saved in uncompressed jpeg format. Thus, the present
survey is based on a variable dynamic range, as opposed to the surveys
based on homogeneous photographic material, i.e. the dwarfs are
displayed with a shorter dynamic range, and higher contrast, than the
bright spirals.

\section{3.6$\mu$m Morphological Classification}

With a final set of images in hand, the galaxies were classified using
the three dimensions of the de Vaucouleurs revised Hubble-Sandage
system (de Vaucouleurs 1959). Full types include the
{\it stage} (E, E$^+$, S0$^-$, S0$^o$, S0$^+$, Sa, Sab, Sb, Sbc, Sc,
Scd, Sd, Sdm, Sm, and Im), the {\it family} (SA, SAB, SB), the {\it
variety} (r, rs, s), the {\it outer ring or pseudoring} classification
(R or R$^{\prime}$, if present), and indications of a {\it spindle}
shape (sp, meaning edge-on or near edge-on orientation) and the
presence of any {\it peculiarity} (pec, often referring to unusual and
likely interaction-driven asymmetries).  

The classifications were made by RB and independently verified by JK,
and are consistent with the dVA and RC3 classification systems. Any
systematic differences between the actual blue light classification and
the mid-IR classification can then be mostly attributed to the effects
of bandpass. The ``scatter" in type classifications by individual
observers is quantified by Naim et al. (1995).

\subsection{Assigning Stage, Family, and Variety}

In de Vaucouleurs's classification approach, the implication for bars,
inner rings, and stages is a continuum of forms (de Vaucouleurs 1959).
The stage for spirals is based on the appearance of the spiral arms
(degree of openness and resolution) and also on the relative prominence
of the bulge or central concentration. These are the criteria
originally applied by Hubble (1926, 1936). Sa, Sb, and Sc spirals are
mostly as defined by Hubble, with the additional stages Sd, Sm, and Im
as appended by de Vaucouleurs. Intermediate stages (Sab, Sbc, Scd, Sdm)
are almost as common as the main types. In some cases, the three spiral
criteria are inconsistent, or other factors enter in that affect the
type (Sandage 1961; Sandage \& Bedke 1994). In general, the
bulge-to-total luminosity ratio is directly related to Hubble type, but
there is considerable scatter at a given type (see Simien \& de Vaucouleurs 1986;
Laurikainen et al. 2007, Graham \& Worley 2008).

The family classifications SA, SAB, and SB are the purely visual
estimates of bar strength.  The intermediate bar classification SAB is
one of the hallmarks of the de Vaucouleurs system, and is used to
recognize galaxies having characteristics intermediate between barred
and nonbarred galaxies. Underline notation S$\underline {\rm A}$B and
SA$\underline{\rm B}$ (de Vaucouleurs 1963) is used to further
underscore the continuity of this characteristic.

Although a bar that looks relatively weak in blue light can 
appear stronger in the near- or mid-IR, IR imaging does {\it not}
necessarily change the {\it rankings} of bars. A bar which appears
strong in blue light may also appear even stronger in the IR. Studies
of the maximum relative bar torque parameter $Q_b$ (e. g., Buta et
al.  2005), show that what we usually call ``SB" has a wide range of
strengths. $Q_b$ is the maximum relative bar torque per unit mass per
unit square of the circular speed, and has been found to range from 0
for no bar to at least 0.7 for the strongest bars. Buta et al.  (2005;
see also the dVA) defined the $Q_b$ family as follows: SA types have
$Q_b$ $<$ 0.05; S$\underline{\rm A}$B types have 0.05 $\leq$ $Q_b$ $<$
0.10; SAB types have 0.10 $\leq$ $Q_b$ $<$ 0.20; SA$\underline{\rm B}$
types have 0.20 $\leq$ $Q_b$ $<$ 0.25, while SB types have $Q_b$ $\geq$
0.25. The $Q_b$ family is an approximate quantitative representation
of the visual bar strength classes.

Variety is also treated as a continuous classification parameter,
ranging from closed inner rings (r) to open spirals (s).  The
intermediate variety (rs) is also well-defined. As for family,
underline notation $\underline{\rm r}$s and r$\underline{\rm s}$ is
used to underscore further continuity.

The classification of S0 galaxies depends on recognizing the presence
of a disk and a bulge at minimum, and usually a lens as well, and no
spiral arms.  A lens is a galaxy component having a shallow brightness
distribution interior to a sharp edge.  Even if a lens isn't obvious, a
galaxy could still be an S0.  (Lenses are also not exclusive to S0s.)
Other structures, such as bars and rings, can enter in the
classification of S0s. The stage sequence S0$^-$, S0$^o$, S0$^+$ is a
sequence of increasing detail. Exceptionally early nonbarred S0s can
look very much like ellipticals, and in fact some galaxies classified
as ellipticals in RC3 are classified as S0 galaxies in the {\it Revised
Shapley-Ames Catalogue} (RSA, Sandage \& Tammann 1981). The transition
type S0/a is the formal beginning of the spiral stage sequence.

In mag arcsec$^{-2}$ units, luminous elliptical galaxies usually have
very smooth light distributions with no trace of a lens or any other
structures. In principle, we should be able to classify ellipticals
more consistently with digital images than with photographic plates,
but the distinction from early S0s can still be very subtle in some
cases as we have noted. de Souza et al. (2004) found that about a third
of elliptical galaxies can be misclassified as S0s, and that it is
equally easy to misclassify an elliptical as an S0 as the other way
around. Type E$^+$ was originally intended by de Vaucouleurs (1959) to
describe ``late" ellipticals, or ``the first stage of the transition to
the S0 class."

Outer ring and pseudoring classifications are made in the same manner
as in blue light: the more closed outer rings are classified as (R)
preceding the main type symbols, while large pseudorings made of
outer spiral arms whose variable pitch angle causes them to close
together are classified as (R$^{\prime}$). In general, these
classifications are not very sensitive to the difference between
blue and near-IR bands, although one or the other may facilitate 
detecting the structures better.

A {\it spindle} is a highly inclined disk galaxy. For blue-light
images, usually an ``sp" after the classification would almost
automatically imply considerable uncertainty in the interpretation,
because stage, family, and variety are not easily distinguished when
the inclination is high. Even in the near-IR, classifying spindles is
still difficult, but nevertheless can be better than in blue light
because planar absorption lanes are far less significant. One important
development in the classification of edge-on galaxies has been the
ability to recognize edge-on bars through boxy/peanut and ``X"-shapes.
Boxy/peanut bulges in edge-on galaxies were proven to be bars seen
edge-on from kinematic considerations (cf. Kuijken \& Merrifield 1995;
see also Bureau \& Freeman 1999). These shapes can be more clearly
evident in {\it Spitzer} images than in blue-light images.  An example
here is NGC 2683 (Figure~\ref{NGC2683}).

A few classification details are used here that were not originally
part of the de Vaucouleurs system, but were used or discussed in the
dVA. For example, the notation ``E (shells/ripples)" or ``S0
(shells/ripples)" is used to denote an elliptical or S0 galaxy that
shows faint arc-like or curved enhancements (Malin \& Carter 1980,
1983). The term ``shells" implies a particular three-dimensional
geometry that Schweizer \& Seitzer (1988) argued imposes a prejudice on
the interpretation of the structures. They proposed instead the
alternate term ``ripples," which implies less of a restrictive
geometry.  The {\it Spitzer} images are deep enough to reveal even the
inner shells in ellipticals well, if they are there, and at least 4
shell galaxies were identified in the relatively small subsample of the
S$^4$G studied here. We use the notation E(d) for disky ellipticals and
E(b) for boxy ellipticals, after Kormendy \& Bender (1996). Although
quantifiable in terms of Fourier analysis, our classification is by eye
and thus selects the most obvious cases. These distinctions, which
relate to velocity anisotropy, can be seen mainly in
ellipticals harboring edge-on disks.

Following Kormendy (1979), we also recognize both inner lenses (l) and
outer lenses (L). We also recognize the ``outer Lindblad resonance
(OLR)" subcategories of outer rings and pseudorings, R$_1$,
R$_1^{\prime}$, R$_2^{\prime}$, and R$_1$R$_2^{\prime}$, following Buta
\& Crocker (1991) and Buta (1995).  An R$_1^{\prime}$ outer pseudoring
is defined by a 180$^o$ winding of two spiral arms relative to the bar
ends, while an R$_2^{\prime}$ outer pseudoring is defined by a 270$^o$
winding. R$_1$ outer rings are more detached versions of R$_1^{\prime}$
outer pseudorings, while the double outer ring/pseudoring morphology
R$_1$R$_2^{\prime}$ is a distinctive combination of features including
an R$_1$ component surrounded by an R$_2^{\prime}$ pseudoring.

The prominence of nuclear rings and bars in 3.6$\mu$m images further
necessitates additional classification symbols beyond those just
described. The nuclear rings in our sample are so distinctive that we
have used notation suggested by Buta \& Combes (1996) to recognize
them. For example, NGC 3351 is classified as (R$^{\prime}$)SB(r,nr)a,
where ``nr" is the symbol for nuclear ring (``nl" if a lens instead).
Comer\'on et al. (2010) have compiled an atlas of known nuclear rings
and made a statistical study of their sizes and other characteristics.
In addition, nuclear bars are seen in several of the sample galaxies,
and are distinctive enough features that we have also recognized them
with the type classification ``nb." Thus, the classification of NGC
5850 is (R$^{\prime}$)SB(r,nr,nb)ab. We caution, however, that we
cannot provide these classifications for all S$^4$G galaxies where
nuclear rings and bars may be present because the resolution (in
parsecs) is critical for detecting them, and as a result we can never
make a claim that the classification is complete.

\subsection{Comparison of Classifications}

Our 3.6$\mu$m classifications are listed in Table 1, together with the
classifications for the same galaxies given in the RC3, the RSA, the
dVA, and also in Eskridge et al. (2002), which has 22 galaxies in
common with our sample. Following de Vaucouleurs (1963), we use ":"
or "?" to indicate different levels of uncertainty, with "?" implying
greater uncertainty. In order to compare the 3.6$\mu$m stages,
families, and varieties with those listed in the RC3, the RSA, and the
dVA, we use convenient numerical indices. For the stage, the 15 types
from E to Im are assigned numerical $T$ values as follows: $-$5 (E),
$-$4 (E$^+$), $-$3 (S0$^-$), $-$2 (S0$^o$), $-$1 (S0$^+$), 0 (S0/a), 1
(Sa), 2 (Sab), 3 (Sb), 4 (Sbc), 5 (Sc), 6 (Scd), 7 (Sd), 8 (Sdm), 9
(Sm), and 10 (Im) (de Vaucouleurs \& de Vaucouleurs 1964). For family
and variety, we use the numerical indices adopted by de Vaucouleurs \&
Buta (1980):  $F$=$-1$ for SA, 0 for SAB, and +1 for SB familes, and
$V$= $-1$ for (r), 0 for (rs), and +1 for (s) varieties. Underline
notations in all cases are assigned half steps. The comparisons are
plotted as histograms of the numerical index difference, $\Delta
(T,F,V) = T,F,V(S^4G)-T,F,V(other)$ in
Figures~\ref{rc3comp}~-~\ref{dvacomp}.  All of the top frames in these
figures are $\Delta (T)$ comparisons. If $\Delta (T)$ $<$ 0, the S$^4$G
classification is earlier than the other source's classification, while
if $\Delta (T)$ $>$ 0, the S$^4$G classification is later. The $\Delta
(F)$ comparisons are in the two lower left frames.  If $\Delta (F)$ $<$
0, the S$^4$G bar classification is more nonbarred than the other
source's classification, while if $\Delta (F)$ $>$ 0, the S$^4$G bar
classification is more barred.  The $\Delta (V)$ comparisons are in the
two lower right frames.  If $\Delta (V)$ $<$ 0, the S$^4$G variety
classification is more ringed than the other source's classification,
while if $\Delta (V)$ $>$ 0, the S$^4$G variety classification is more
spiral-shaped.

The RC3 stage comparisons in Figure~\ref{rc3comp} (top left frame and
lower three frames) show
first that the bin with the largest number of galaxies has $\Delta (T)$
= 0, meaning the 3.6$\mu$m stage and the RC3 stage are the same.  The
largest difference is found for RC3 stages S0/a-Sc, where many galaxies
are classified 1 stage interval {\it earlier} than in RC3.  This is
very similar to what was found by Eskridge et al. (2002) using
ground-based 1.65$\mu$m $H$-band images for classifications, where
galaxies with RC3 types Sab to Sc were classified about 1 stage interval
earlier at 1.65$\mu$m than in RC3. The same systematic effect was found
for types Sa to Scd when these authors compared their 1.65$\mu$m types
with their own $B$-band types estimated from OSUBSGS images.

In the family and variety comparisons (upper right panels of
Figure~\ref{rc3comp}), the most populated bins again have $\Delta$=0.
In fact, in these the concentration in the $\Delta$=0 bin is much
larger than that in the stage comparisons in the upper panels. The plots show
considerable consistency between RC3 families and varieties and our
Table 1 judgments. Surprisingly, in this comparison we do not see a
tendency for $\Delta (F)$ to be greater than zero (the ``stronger bar"
effect), which is what we would see if many 3.6$\mu$m bar
classifications advanced to higher bar strengths compared to RC3
classifications. That is, a big shift of SA to SAB and SAB to SB is not
seen. In contrast, Eskridge et al. (2000) report a factor of 2 more SBs
than in RC3. The problem with this kind of comparison is that blue
light SB-type bars, as seen in the IR, do not have a new classification
bin to be placed in if they look stronger in the IR, while blue light
SAB-type bars do have a new bin - type SB. This appears to increase the
number of strong bars when in fact there is little change in bar
rankings. In Table 1 there are 16 galaxies in common with Eskridge et
al., and of those, 12/16 are classified as SB by Eskridge et al., 8/16
are SB in Table 1, and 7/16 are SB in RC3. The difference between
Eskridge et al. on one hand, and the Table 1 classifications on the
other, is likely not due to small number statistics in the Eskridge et
al. sample, but to a difference in what is called an SB.  Our result is
more consistent with Men{\'e}ndez-Delmestre et al.  (2007) and Sheth et
al.  (2008), who found the same bar frequency from the $B$-band to the
$K$-band, based on a semi-automated method of bar detection.

The comparisons shown in Figure~\ref{rsacomp} for RSA galaxies are
different because the RSA represents a different classification
system. There are very few RSA galaxies classified as later than Sc,
and the classification of S0s is somewhat different.  Nevertheless, we
assigned the same numerical indices to types S0/a to Im as for RC3, and
we assigned values of $-$3, $-$2, and $-$1 for types S0$_1$, S0$_2$,
and S0$_3$, the same as for types S0$^-$, S0$^o$, and S0$^+$,
respectively. Ellipticals are assigned $T$=$-$5, just as in RC3. 
Classifications like ``E7/S0$^-$ were assigned $T$=$-$4. The
comparison shows that for the full range of RSA types, there is little
systematic difference between RSA and 3.6$\mu$m stages. However, for
RSA stages Sab-Sbc, the ``earlier effect" is definitely seen, while for
RSA stages Sc-Im, the 3.6$\mu$m types are actually later. Only for
types E-Sa is little or no systematic difference seen. The family
comparison shows that bar classifications are definitely stronger
on average in the 3.6$\mu$m types as compared to the RSA, and
inner rings are classified more often in the 3.6$\mu$m sample than in
the RSA. Most of these differences are due to differences between the
RSA and de Vaucouleurs classification systems, and not to wavelength
effects. The use of the SAB symbolism allows RC3 types to be more
discriminatory on bar classifications than RSA types; a ``stronger bar"
effect is seen even in a comparison of RC3 and RSA blue light
classifications, due mainly to RC3 SAB galaxies classified mostly as S
in the RSA. In the case of variety, any inner ring that is made of
tightly-wrapped spiral arms is classified as (s) variety in the RSA,
and usually as (r) in RC3 or (r) or ($\underline{\rm r}$s) in Table 1.

The dVA comparisons in Figure~\ref{dvacomp} are similar to the RC3
comparisons, but show slightly more prominent ``earlier type" and
``stronger bar" effects. The comparisons of dVA and 3.6$\mu$m types
show less scatter not because of greater precision, but because the
same observer classified both data sets. Thus, Figure~\ref{dvacomp} is
more of an internal, rather than an external, comparison. 

\subsection{Noteworthy Examples}

The results from the previous section are illustrated in
Figures~\ref{montage1}-~\ref{montage5}. These compare S$^4$G 3.6$\mu$m
images in mag arcsec$^{-2}$ units with $B$-band images in the same
units. Most of the $B$-band images are from the dVA.

The four galaxies shown in Figure~\ref{montage1} cover a wide range of
types. The images of the Sdm/Sm galaxy NGC 428 are barely different.
This is true also for the Sc galaxy NGC 628, although its inner arms
are smoother in the 3.6$\mu$m image than in the $B$-band image.  The
image of NGC 1097 definitely looks a little earlier at 3.6$\mu$m, but
still the differences are relatively small. NGC 584 is an early-type
galaxy shown in the figure that highlights the greater depth of the
3.6$\mu$m image compared to a typical optical image. Although classified
as type E in RC3, NGC 584 is definitely an early S0 galaxy.

The four ringed galaxies in Figure~\ref{montage2} all show the
``earlier effect": S0/a galaxy NGC 1291 becomes type S0$^+$; SBab
galaxies NGC 1433 and NGC 1512 become type SBa; and SBb
galaxy NGC 3351 becomes type SBa. Even with these type changes, the
overall morphology of all four galaxies looks nearly the same in the
two filters.

Figure~\ref{montage3} and the top panels of Figure~\ref{montage4} show
three classic ``flocculent" spirals, NGC 2841, 5055, and 7793
(Elmegreen 1981). In blue light all three have rather piece-wise
continuous spiral structures, but not the global patterns
characteristic of grand design spirals. In NGC 2841 and 5055, dust is a
major factor in the appearance of the spiral structure. 

NGC 2841 changes from type Sb in the $B$-band to type Sa at 3.6$\mu$m,
mostly because of the penetration of this dust. Even so, the overall
appearance of the more coherent $B$-band spiral features in NGC 2841 is
about the same at 3.6$\mu$m. The 3.6$\mu$m image also reveals a weak
bar in NGC 2841, $\approx$30$^{\prime\prime}$ in radius and with a
position angle of $\approx$160$^{\circ}$ (compared to the galaxy major
axis position angle of 147$^{\circ}$ listed in RC3). This is not the
same bar-like feature described by Varela et al. (1996), who identified
a nuclear bar-like structure about 10$^{\prime\prime}$ in radius. The
inner ring in NGC 2841 recognized in Table 1 is a large feature,
3\rlap{.}$^{\prime}$1 in angular diameter, but the structure of the
whole galaxy is very much a series of ring-like features.

In NGC 5055, the flocculent spiral pattern looks more global at 3.6$\mu$m.
However, the appearance of the main arms still favors an Sbc classification,
the same as the RC3 $B$-band type. The main difference is that the
feature classified as an inner pseudoring in RC3 [the type is SA(rs)bc]
is a smooth inner spiral at 3.6$\mu$m. The ring/lens (rl) feature recognized
in the Table 1 classification lies {\it inside} the RC3 pseudoring (as seen
also in the 2.2$\mu$m image shown in Thornley 1996). In the inner parts of
NGC 5055, and throughout the disk of NGC 2841, star-forming regions are
not prominent, but throughout the outer disk of NGC 5055, the 3.6$\mu$m
image reveals many large star-forming complexes.

The images of NGC 7793 in Figure~\ref{montage4} show a slightly more
coherent spiral pattern in the 3.6$\mu$m as compared to $B$, but which
is still largely flocculent. The most noticeable difference is in the size of
the bulge, which seems more prominent at 3.6$\mu$m. However, in this
case the IRAC image is not as deep as the $B$-band image. Because of
the more prominent bulge and more coherent pattern at 3.6$\mu$m, the
classification given in Table 1 is Sc compared to the $B$-band type of
Sd.

The second row of Figure~\ref{montage4} shows NGC 4527. This highly
inclined, dusty spindle shows a well-defined nuclear ring and a weak
bar and partial inner ring. Similar to NGC 5055, NGC 4527 does not look
like an earlier type at 3.6$\mu$m.  The third row shows NGC 4579, an
example where the $B$-band bar (type SA$\underline{\rm B}$ in the dVA)
becomes type SB at 3.6$\mu$m. The 3.6$\mu$m image also reveals a faint
outer ring, not recognized in the RC3, RSA, or the dVA, tightly
surrounding the inner spiral. The ring has dimensions of
4\rlap{.}$^{\prime}$4 $\times$ 3\rlap{.}$^{\prime}$2, and may have been
missed in the earlier blue light classifications owing to the uncertain
effects of extinction, not because of faintness. The fourth row shows
NGC 4736, a galaxy with a very wide range of surface brightness from
the center to the outer disk. The change in type from $B$ to 3.6$\mu$m
is only Sab to Sa. The most striking aspect of the 3.6$\mu$m morphology
of NGC 4736 is how the broad, intermediate oval zone (the basis for the
Table 1 classification S$\underline{\rm A}$B) becomes a large ring/lens
feature. The prominent $B$-band inner pseudoring is still a pseudoring
at 3.6$\mu$m, but following Knapen (2005) and Comer\'on et al. (2010),
this feature is interpreted as a nuclear pseudoring (nr$^{\prime}$) in
Table 1. The nuclear lens (nl) and nuclear bar (nb) in the Table 1
classification are inside the nr$^{\prime}$.

Figure~\ref{montage5} shows images of four very late-type galaxies.
The top row shows NGC 1705, typed as SA0$^-$: pec in the RC3 and
Amorphous in the RSA. Meurer et al. (1989) interpreted the galaxy as a
nucleated blue compact dwarf (BCD), where the nucleus in this case is a
super star cluster having absolute blue magnitude $-$15 (Melnick et al.
1985). The peculiar morphology with strange filaments is evident
in the $B$-band image, and accounts partly for the dVA type of
I0/BCD, but at 3.6$\mu$m, NGC 1705 looks more like a
dwarf elliptical with a slightly miscentered nucleus.  This accounts
for the classification ``dE3, N" in Table 1, meaning nucleated
dwarf elliptical (Bingelli et al. 1985).  Table 1 also includes a
few galaxies classified as ``dE (Im)"; NGC 3738, shown in the second row
of Figure~\ref{montage5}, is an example. This refers to
an object where the structure in blue light consists of an irregular,
bright inner star-forming zone, surrounded by smoother elliptical
isophotes. At 3.6$\mu$m, the inner star-forming zone may be subdued,
and the smooth elliptical background takes prominence.

The third and fourth rows of Figure~\ref{montage5} show NGC 3906 and
NGC 4618, two examples of very late type barred spirals where the bar
is miscentered within very regular outer isophotes.  These offcentered
barred galaxies have been extensively discussed by de Vaucouleurs \&
Freeman (1972; see also Freeman 1975). The asymmetry of these galaxies
is characteristic of these very late types and it is possible to still
recognize them in the edge-on view. For example, we interpret IC 2233
(Figure~\ref{IC2233}) as an edge-on SBdm galaxy, because it has a
bright elongated, somewhat offset inner zone that is likely to be an
off-center bar as in NGC 3906. By the same token, NGC 55
(Figure~\ref{NGC0055}) is interpreted to be a nearly edge-on view of
NGC 4618. In this case, the bar is seen on the northwest side of the
major axis, while the considerable amount of light to the northeast
would be the single main spiral arm. In many other of the very
late-types in Table 1, we suspect that inner boxy zones, or bright
miscentered zones of limited extent, are edge-on views of bars.

\subsection{Special Cases}

We note the following special cases in our small subsample of
207 mostly S$^4$G galaxies.

\noindent
{\it I0 Galaxies} - Two of the galaxies in our sample are
classified as types I0 in RC3. The top panels of Figure~\ref{montage6}
show NGC 5195, the familiar companion of M51. Hidden behind the dust in
the $B$-band image is a very regular early-type galaxy with a
relatively weak bar oriented north-south and a broad and diffuse inner
ring. The lower panels in Figure~\ref{montage6} show NGC 2968. We do
not have a $B$-band digital image of NGC 2968, and the figure shows the
photograph of the galaxy from the Carnegie Atlas of Galaxies (Sandage
\& Bedke 1994), as downloaded from the NASA/IPAC Extragalactic Database
(NED) website. Although the Carnegie Atlas photograph is not very deep, it
is sufficient to show a complex, dusty early-type system typical of
I0 galaxies. The 3.6$\mu$m image of NGC 2968 reveals a beautifully
symmetric late barred S0. These are the only galaxies in the sample
of 207 where the 3.6$\mu$m morphology looks drastically different from
the $B$-band. For these galaxies, the I0 class is not needed at 3.6
$\mu$m.

\noindent
{\it Pseudobulges} - The top two panels of Figure 12 show two galaxies
having prominent pseudobulges (Kormendy \& Kennicutt 2004) or disky
bulges (Athanassoula 2005). Both NGC 470 and 4536 have bulge isophotes
that align with the major axis, indicating significant bulge
flattening. This argues that they are very disk-like, which is one of
the criteria for recognizing such bulges.

Another criterion for pseudobulges is the presence of a nuclear bar.
These are identified using the notation ``nb" in the classification,
and include NGC 1291, 1433, 4725, 5728, and 5850.  Each of these
galaxies has a clear primary bar, and four have an inner ring.

A third criterion for a pseudobulge is the presence of a boxy or
boxy/peanut inner structure. In our sample, these are evident in NGC
2683 (Figure~\ref{NGC2683}), 4527 (Figure~\ref{NGC4527}), and 5353
(Figure~\ref{NGC5353}), among others. In the case of the spindle galaxy
NGC 2683, the inner boxy/X zone is so distinctive that in Table 1 we
use the notation ``SB$_{\rm X}$" to recognize it is as a barred galaxy
without actually seeing the extent of the bar.

\noindent
{\it Double-Stage Galaxies} - Some galaxies in the S$^4$G sample are
large-scale S0 or S0/a galaxies with smaller-scale inner spirals.  Four
are shown in Figure~\ref{montage7} (lower two rows). The inner spirals
in these galaxies are only a small fraction of the size of the system.
In effect, these are ``double-stage" galaxies, because the large-scale
structure is that of an early-type galaxy and the inner structure is
that of a later-type galaxy. The final type we have adopted is usually
a compromise unless one characteristic dominates (see notes to
Table 1). The double-stage
character in NGC 5713, where an inner asymmetric star-forming component
lies within a smooth, asymmetric outer ring, could be linked to an
interaction (Vergani et al. 2007). Double-stage spirals were described
by Vorontsov-Velyaminov (1987) as cases where an inner spiral pattern
gives a different type from an outer spiral pattern.

\noindent
{\it Double-Variety Galaxies} - We have already described NGC 5055 as
having an inner pseudoring and an inner ring/lens in the 3.6$\mu$m
image, which makes the galaxy an unusual case of a double variety
system. This characteristic was not readily evident in blue light.
Both features, especially the inner pseudoring, are visible in the
3.6$\mu$m radial luminosity profile of NGC 5055 shown by de Blok et
al. (2008). While the two variety patterns are completely separated in
this case, they can also overlap. For example, NGC 986
(Figure~\ref{NGC0986}) has a very strong s-shaped spiral superposed on
a diffuse inner ring. The two varieties in this case do not seem
related, i.e., there is an inner ring but the main arms are not related
to the ring.

\noindent
{\it Inner Disks} - Among the spindles in our sample, NGC 24
(Figure~\ref{NGC0024}) is unique in showing a very bright, well-centered
inner component that aligns almost exactly with the major axis,
suggesting that the component is an inner disk and not an inner bar.
The galaxy is not exactly edge-on, and a ring is recognizable in the
spiral morphology. The inner disk component greatly underfills this
ring, another characteristic that suggests the feature is not a bar.
This shows the advantage of the mid-IR for morphological studies of
edge-on disks (see also Fazio et al. 2004b).

\noindent
{\it Outer Rings} - One of the stated goals of the S$^4$G project was
to take advantage of the considerable depth of Warm Mission images to
detect previously unrecognized outer rings, especially rings so faint
that optical images failed to reveal them. In blue light, such rings
tend to have surface brightnesses $\mu_B$ $\geq$ 25.0 mag
arcsec$^{-2}$, and some could be lost due to extinction or exceptional
intrinsic faintness. Although we expect that some faint outer rings
will be found when the entire survey is completed, none were identified
in the subsample of S$^4$G that we have examined here. The main new
outer ring we detected is in NGC 4579 (Figure~\ref{montage4}), but, as
we have already noted, it was missed in the RC3, RSA, and dVA only
because of the uncertain effects of internal extinction, not because of
exceptional faintness.

The main reason few outer rings were detected in our sample here is
probably because the sample has a large fraction of late-type galaxies,
where outer rings are infrequently seen (Buta \& Combes 1996).
The images of NGC 1291 in Figure~\ref{montage2} show that
even an exceptionally bright outer ring in blue light can have
considerably reduced contrast at 3.6$\mu$m. The outer ring in NGC 1291
is where most of the recent star formation is taking place, and is
prominent in blue light as a result, but at 3.6$\mu$m, the feature
barely stands out as a broadly oval enhancement in a rounder diffuse
background.

\subsection{The Nature of Resolved 3.6$\mu$m Objects}

The strong similarity between $B$-band and 3.6 micron images in many of the
galaxies we show here is one of the most important findings from this
study. It is perhaps surprising that at 3.6$\mu$m, the ``degree of
resolution" effect can still play a role in the classification of
spirals, when the common assumption is that IR light traces mass and
de-emphasizes the massive OB star complexes that line the arms in blue
light. Nevertheless, young stars can impact morphology in the near- and
mid-IR. Rix \& Rieke (1993) and Rhoads (1998) showed using the CO index that young red
supergiant stars no more than 10$^7$ yr old may contribute substantial
local flux at 2$\mu$m even if the global 2$\mu$m flux is dominated by
old stars.  

Detailed comparison of resolved objects in $B$-band and 3.6$\mu$m
images shows that, in many cases, the same complexes are being seen. An
especially good illustration of this is provided by NGC 1559, whose
type [SB(s)cd] is essentially the same in the two filters
(Figure~\ref{ngc1559}). Although the considerably reduced sky
background in 3.6$\mu$m images allows a flurry of faint foreground
stars to appear, the resolved objects we see in NGC 1559 follow the
spiral arms closely and are not randomly scattered. A color
index map (not shown) reveals that most of the resolved sources are
redder in the [3.6]$-$[4.5] color index than the galaxy background
light.

Figure~\ref{ngc1559objects} shows a color-magnitude plot of 159 of
these objects as compared to a similar-sized sample of surrounding
field objects (foreground stars and background galaxies). A
redshift-independent distance modulus of 30.95 (NED) has been used for
the absolute magnitude scale. Photometry was performed with IRAF
routine PHOT using a measuring aperture of 2 pix and background
estimates taken from 3-5 pix for each object. A bright foreground star
off the galaxy was used to determine an aperture correction of 0.88
mag, which has been subtracted from the 2 pix radius 3.6$\mu$m
magnitude [3.6]. These graphs can be compared to Figures 7 and 8 of
Mould et al. (2008), who analyzed IRAC data for the resolved stellar
population in M31 and a comparison SWIRE control field. Mould et al.
detected luminous red supergiants in M31, some showing evidence of mass
loss. As also shown by Mould et al., field objects will overlap the
distribution of galaxy objects in such plots. The asymptotic giant
branch reaches to only $M$[3.6]$\approx$$-$10 (see also Jackson et al.
2007), while the brightest evolved supergiants in M31 reach [3.6]=9.5,
or $M$[3.6]$\approx$$-$15.  Figure~\ref{ngc1559objects} shows that most
of the resolved objects in NGC 1559 are more luminous than this, which
is consistent with their slightly fuzzy appearance; these are complexes
rather than merely individual stars, although some of the faintest
objects could be individual stars. The individual AGB population is
largely out of reach at the distance of NGC 1559, and we conclude that
the resolved objects in the arms are likely groups of massive young
stars. However, it is not clear that we can think of the resolved
objects entirely in terms of groups of young red stars.

Mentuch et al. (2009) modeled the spectral energy distributions of high
redshift galaxies and concluded that a previously discovered excess of
infrared flux in the 2-5$\mu$m region can be best interpreted in terms
of modified blackbody emission from dust at 850K together with the
well-known polycyclic aromatic hydrocarbon (PAH) emission at
3.3$\mu$m.  The excess emission is believed to come from the puffed up
inner edges of circumstellar disks in massive star-forming regions.
Given what we have described above, this mechanism could account for
some of the light of the resolved sources we
see in S$^4$G images. Further studies should be able to clarify this
issue. 

Note that use of 4.5$\mu$m images will not necessarily avoid the
effects of the resolved objects on morphological interpretations.
Comparison between the 3.6 and 4.5$\mu$m images of virtually all the
galaxies in our sample shows little difference in apparent morphology.
The resolved objects are just as conspicuous at 4.5$\mu$m as they are
at 3.6$\mu$m. Given this, the question arises: which IR domain, near or
mid, might be better for revealing the true ``stellar backbone," or
what is called the ``star-dominated Population II disk" by Block et
al.  (2004). Figure~\ref{m51} shows the IRAC 3.6$\mu$m image of M51
(right frame) as compared to a 2.2$\mu$m $K_s$-band image obtained in
May 2009 by RB using the FLAMINGOS IR imaging camera attached to the
Kitt Peak 2.1m telescope. The on-source exposure time of the latter
image was 30min. The comparison shows that the spiral arms of M51 are
smoother and less affected by star-forming regions in the $K_s$ band
than in the 3.6 $\mu$m {\it Spitzer} band, perhaps arguing in favor of
the $K_s$ band for stellar mass studies. However, this comparison also
displays the considerably greater depth of the 3.6$\mu$m image compared
with a typical, $K_s$-band image. Although the 3.6$\mu$m image is
affected by star formation, the light is still dominated by the
``stellar backbone" of old stars and the best approach to getting at
this background would be to find ways to correct 3.6$\mu$m images for
the younger ``contaminants." This will be extensively discussed in
a future S$^4$G paper (S. Meidt et al., in preparation).

\subsection{Bar Fraction}

We conclude our analysis by examining the bar fraction in our subsample
of S$^4$G galaxies. The bar fraction has cosmological signficance
(Sheth et al. 2008), and has been the topic of many recent optical and
near-IR studies (e.g., Knapen et al. 2000; Eskridge et al. 2000;
Laurikainen et al. 2004; Men{\'e}ndez-Delmestre et al. 2007; Marinova \&
Jogee 2007; Barazza et al. 2008; Sheth et al. 2008; Aguerri et al.
2009; Marinova et al.  2009; M\`endez-Abreu et al. 2010; Masters et
al.  2010). In general, these studies are in fairly good agreement on
this parameter, in spite of different methodologies.  Some find a
slightly higher bar fraction in the near-IR than in the optical (e. g.,
Knapen et al 2000, Eskridge et al 2000), while others do not  (e. g.,
Men{\'e}ndez-Delmestre et al. 2007; Sheth et al. 2008). Eskridge et al.
(2000) found that the fraction of {\em strongly} barred galaxies (i.e.,
classified as SB) rises by almost a factor of 2 from the optical to the
near-IR. The rise is mainly due to reclassification of SAB-type bars to
SBs because bars are often more prominent at near-IR than at blue
wavelengths, but the overall SAB+SB bar fraction is similar in the two
wavelength domains as shown by Whyte et al. (2002),
Men{\'e}ndez-Delmestre et al. (2007), and Sheth et al. (2008), among
others.  Men{\'e}ndez-Delmestre et al.
(2007) and Sheth et al. (2008) also found that almost as many galaxies in
the near-IR get classified from barred to nonbarred as are classified
from nonbarred to barred. However, these constituted only a small
percentage of the sample. In fact, 127/139 galaxies did not change bar
type when these authors compared $G$ to IR images, 7 went from SA to SAB,
4 from SAB to SA, and 1 from SB to SA. Thus, over 90\% looked the
same. The ones that did not were faint, small, or close to the
inclination cut-off, and as a consequence were hard to classify.

Table 2 compiles counts of the galaxies in Table 1 classified
as types SA, S$\underline{\rm A}$B, SAB, SA$\underline{\rm B}$, and SB,
for the full sample of 207 galaxies (Col. 2, including all types), and
for several subsamples. Column 3 lists the counts for galaxies
restricted to the Table 1 type range S0/a-Sm. Columns 4 and 5 are the
same as columns 2 and 3, but after rejection of any galaxy classified
as a spindle (sp) in Table 1. This is reasonable since morphology-based
bar classification is difficult in highly inclined galaxies. We
consider any galaxy classified as SAB, SA$\underline{\rm B}$, or SB to
be a barred galaxy, and define the bar fraction as the number of those
galaxies relative to the number in the given subset in Table 2. Over
all types and even after removal of spindles, the bar fraction is about
50\%. However, when restricted to spirals, the fraction is 64-66\%,
similar to previous studies (e.g., Laurikainen et al. 2004). In the
OSUBSGS sample, Eskridge et al. (2000) found a bar fraction of 73\%
based on $H$-band images. The SB bar fraction in our (non-spindle)
spiral sample, 36\%$\pm$4\%, is higher than the same fraction,
29.4\%$\pm$0.5\% estimated by Masters et al. (2010) from the optical
Galaxy Zoo project, which could partly be the ``stronger bar effect."
The bar fraction in our subsample is consistent with that estimated
from RC3 classifications (see Table 1 of Eskridge et al. 2000).

\section{Conclusions}

We have used a subsample of S$^4$G pipeline-processed archival {\it
Spitzer Space Telescope} images to examine the imprint of the de
Vaucouleurs revised Hubble-Sandage classification system in the IRAC
3.6$\mu$m band.  Although our sample includes less than 10\% of what
will be the final S$^4$G sample, it is sufficient for us to achieve some
interesting results.  We find that 3.6$\mu$m classifications are
well-correlated with blue light classifications in RC3. The most
significant difference occurs for S0/a to Sc galaxies, where 3.6$\mu$m
types average about 1 stage interval earlier than $B$-band types. This
is consistent with the previous findings of Eskridge et al. (2002),
which were based on OSUBSGS $H$-band 1.65$\mu$m images. The great
advantage of the IRAC 3.6$\mu$m images is their considerably better
depth than any previous ground-based IR images, allowing a more
complete picture of global IR galaxy morphology, rather just the inner
regions which were all that were revealed in earlier studies.

The use of the same classification methods for 3.6$\mu$m images as for
blue light images is a strength of our analysis. Had we used different
methods we likely would have found spurious differences. This is
actually illustrated by our comparison of our 3.6$\mu$m
classifications, evaluated as in the RC3/dVA, and the classifications
for the same galaxies given in the RSA. This comparison revealed large
systematic differences that are only partly attributable to the
wavelength difference between blue light photographic plates
and S$^4$G images. Instead, most of the differences are due to
methodological differences between RSA and RC3/dVA classifications.
Another issue is that morphological classification is based on
important properties, not minor details. In classifying a galaxy, we
look at whether a particular feature exists, is predominant, secondary
or non-existent, and to the extent that we see these features, the
structure of the old and the young populations do not differ that
wildly.

Deep 3.6$\mu$m galaxy morphology provides an effective, dust-penetrated
alternative to historical blue light morphology, and in principle could
one day fully replace blue light as the standard passband for galaxy
classification. However, the band is not completely free of the effects
of young star forming regions, and thus may not show the
``stellar backbone" of galactic disks as effectively as in
other bands, such as the $K$-band (e.g., Block et al. 2004). For the
future, our goal is to classify the entire S$^4$G sample in the same
manner as described here, to facilitate statistical studies of various
morphological features, and to provide a morphological backdrop to the
quantitative analyses that will come out of the survey.

We thank the referee for many helpful comments that improved the
presentation of this paper. We also thank D. L. Block for additional comments
and for pointing us to a useful reference, as well as the other members of the
S$^4$G Team for helping to make the project possible. RB acknowledges
the support of NSF Grant AST 05-07140. EA and AB thank the Centre
National d'Etudes Spatiales for financial support.  K.L.M.
acknowledges funding from the Peter and Patricia Gruber Foundation as
the 2008 IAU Fellow, from the University of Portsmouth and from SEPnet
(www.sepnet.ac.uk). HS and EL acknowledge the Academy of Finland for
support. JHK acknowledges support by the Instituto de Astrofisica de
Canarias (312407). KMD is supported by an NSF Astronomy and
Astrophysics Postdoctoral Fellowship under award AST-0802399.  
A.G.dP and J.C.M.M are partially financed by the Spanish Programa
Nacional de Astronom\'ia y Astrof\'isica under grants AyA2006-02358
and AyA2009-10368, and by the Spanish MEC under the Consolider-Ingenio
2010 Program grant CSD2006-00070: first Science with the GTC. A.G.dP
is also financed by the Spanish Ram\'on y Cajal program. J.C.M.M.
acknowledges the receipt of a Formaci\'on del Profesorado
Universitario fellowship.
SC is supported by a KASI Postdoctoral Fellowship.
This work is based [in part] on archival data obtained with the
Spitzer Space Telescope, which is operated by the Jet Propulsion
Laboratory, California Institute of Technology under a contract with
NASA. Support for this work was provided by an award issued by
JPL/Caltech. Funding
for the OSUBSGS was provided by grants from the NSF (grants AST 92-17716
and AST 96-17006), with additional funding from the Ohio State
University. Funding for the creation and distribution of the SDSS
Archive has been provided by the Alfred P. Sloan Foundation, the
Participating Institutions, NASA, NSF, the U.S.  Department of Energy,
the Japanese Monbukagakusho, and Max Planck Society. NED is operated by
the Jet Propulsion Laboratory, California Institute of Technology,
under contract with NASA.

\clearpage
REFERENCES

\noindent
Athanassoula, E. 2005, \mnras, 358, 1477

\noindent
Aguerri, J. A. L., M\'endez-Abreu, J., \& Corsini, E. M. 2009, \aap, 495,
491

\noindent
Banerj, M. et al. 2010, astro-ph 0908.2033

\noindent
Barazza, F. D., Jogee, S., \& Marinova, I. 2008, \apj, 675, 1194

\noindent
Bingelli, B., Sandage, A., \& Tammann, G. A. 1985, \aj, 90, 1681

\noindent
Block, D. L., Bertin, G., Stockton, A., Grosbol, P., Moorwood, A. F. M.,
Peletier, R. F. 1994, \aap, 288, 365

\noindent
Block, D., Freeman, K. C., Puerari, I., Combes, F., Buta, R., Jarrett,
T., \& Worthey, G. 2004, in Penetrating Bars Through Masks of Cosmic
Dust, eds. D. L. Block, I. Puerari, K. C. Freeman, R. Groess, \&
E. Block, Springer, Kluwer, p. 15

\noindent
Block, D. L. \& Wainscoat, R. J. 1991, Nature, 353, 48

\noindent
Bureau, M. \& Freeman, K. C. 1999, \aj, 118, 126

\noindent
Buta, R. 1995, \apjs, 96, 39

\noindent
Buta, R. \& Combes, F. 1996, Fund. Cosmic Phys. 17, 95

\noindent
Buta, R. \& Crocker, D. A. 1991, \aj, 102, 1715

\noindent
Buta, R. J., Corwin, H. G., \& Odewahn, S. C. 2007, The de Vaucouleurs
Atlas of Galaxies, Cambridge: Cambridge U. Press (dVA)

\noindent
Buta, R., Vasylyev, S., Salo, H., \& Laurikainen, E. 2005, \aj, 130, 506

\noindent
Buta, R., Laurikainen, E., Salo, H., Block, D. L., \& Knapen, J. H.
2006, \aj, 132, 1859

\noindent
Comer\'on, S., Knapen, J. H., Beckman, J. E., Laurikainen, E., Salo, H.,
Martinez-Valpuesta, I., \& Buta, R. J. 2010, \mnras, 402, 2462

\noindent
Corwin, H., de Vaucouleurs, A., \& de Vaucouleurs, G. 1985, Southern
Galaxy Catalogue, Univ. Texas Monographs, No. 4.

\noindent
Davidge, T. J. 2003, \aj, 125, 30

\noindent
de Blok, W. J. G., Walter, F., Brinks, E., Trachternach, C., Oh, S.-H., \&
Kennicutt, R. C. 2008, \aj, 136, 2648

\noindent
de Jong, R. S. \& van der Kruit, P. C. 1994, \aaps, 106, 451

\noindent
de Souza, R. E., Gadotti, D. A., \& dos Anjos, S. 2004, \apjs, 153, 411

\noindent
de Vaucouleurs, G. 1959, Handbuch der Physik, 53, 275

\noindent
de Vaucouleurs, G. 1963, \apjs, 8, 31

\noindent
de Vaucouleurs, G. \& Buta, R. 1980, \apjs, 44, 451

\noindent
de Vaucouleurs, G. \& de Vaucouleurs 1964, Reference Catalog of Bright
Galaxies, University of Texas Monographs in Astronomy, No. 1 (RC1)

\noindent
de Vaucouleurs, G. \& Freeman, K. C. 1972, Vista in Astr. 14, 163

\noindent
de Vaucouleurs, G., de Vaucouleurs, A., Corwin, H. G., Buta, R. J.,
Paturel, G., \& Fouque, P. 1991, Third Reference Catalog of Bright Galaxies (New
York: Springer) (RC3)

\noindent
Elmegreen, D. 1981, \apjs, 47, 229

\noindent
Elmegreen, D. M. \& Elmegreen, B. G. 1984, \apjs, 54, 127

\noindent
Elmegreen, D. M. \& Elmegreen, B. G. 1987, \apj, 314, 3

\noindent
Eskridge, P. et al. 2000, \aj, 119, 536

\noindent
Eskridge, P. et al. 2002, \apjs, 143, 73

\noindent
Fazio, G. G. et al. 2004, \apjs, 154, 10

\noindent
Fazio, G. G., Pahre, M. A., Willner, S. P., \& Ashby, M. L. N.  2004b,
in Penetrating Bars Through Masks of Cosmic Dust, eds. D. L. Block, I.
Puerari, K. C. Freeman, R. Groess, \& E. Block, Springer, Kluwer, p. 389

\noindent
Freeman, K.C.1975, in Galaxies and the Universe, A. Sandage,
M. Sandage, \& J. Kristian, eds., cHIcago, University of Chicago Press, 
p. 409

\noindent
Fukugita, M. et al. 2007, \aj, 134, 570

\noindent
Graham, A. W. \& Worley, C. C. 2008, \mnras, 388, 1708

\noindent
Hackwell, J. \& Schweizer, F. 1983, \apj, 265, 643

\noindent
Holmberg, E. 1950, Medd. Lunds. Astron. Obs., Ser. II, No. 128

\noindent
Hubble, E. 1926, \apj, 64, 321

\noindent
Hubble, E. 1936, The Realm of the Nebulae, Yale, Yale University Press

\noindent
Jackson, D. C., Skillman, E. D., Gehrz, R. D., Polomski, E., \&
Woodward, C. E. 2007, \apj, 656, 818

\noindent
Jarrett, T. H. et al. 2003, \aj, 125, 525

\noindent
Kennicutt, R. C. et al. 2003, \pasp, 115, 928

\noindent
Kennicutt, R. C. et al. 2007, BAAS, 211, 9502

\noindent
Knapen, J. H. 2005, \aap, 429, 141

\noindent
Knapen, J., Shlosman, I., \& Peletier, R. 2007, \apj, 529, 93

\noindent
Kormendy, J. 1979, \apj, 227, 714

\noindent
Kormendy, J. \& Bender, R. 1996, \apj, 464, L119

\noindent
Kormendy, J. \& Kennicutt, R. 2004, ARAA, 42, 603 (KK04)

\noindent
Lauberts, A. 1982, The ESO-Uppsala Survey of the ESO (B) Atlas,
ESO

\noindent
Laurikainen, E., Salo, H., \& Buta, R. 2004, \apj, 607, 103

\noindent
Laurikainen, E., Salo, H., \& Buta, R. 2005, \mnras, 362, 1319

\noindent
Laurikainen, E., Salo, H., Buta, R., Knapen, J., Speltincx, T.,
\& Block, D. L.  2006, \mnras, 132, 2634

\noindent
Laurikainen, E., Salo, H., Buta, R., Knapen, J. H., \& Comer\'on, S. 2010, \mnras, in press (astro-ph 1002.4370)

\noindent
Lintott, C. et al. 2008, \mnras, 389, 1179

\noindent
Malin, D. \& Carter, D. 1980, Nature, 285, 643

\noindent
Malin, D. \& Carter, D. 1983, \apj, 274, 534

\noindent
Marinova, I. \& Jogee, S. 2007, \apj, 659, 1176

\noindent
Marinova, I. et al. 2009, \apj, 698, 1639

\noindent
Masters, K. et al. 2010, astro-ph 1003.0449

\noindent
M\'endez-Abreu, J. S\'anchez-Janssen, \& Aguerri, J. A. L. 2010, \apj,
711, L61

\noindent
Mentuch, E. et al. 2009, \apj, 706, 1020

\noindent
Men{\'e}ndez-Delmestre, K., Sheth, K., Schinnerer, E., Jarrett, T., \& Scoville, N. Z. 2007,
\apj, 657, 790

\noindent
Meurer, G., Freeman, K. C., \& Dopita, M. A. 1989, Ap\&SS, 156, 141

\noindent
Melnick, J., Moles, M., \& Terlevich, R. 1985, \aj, 149, L24

\noindent
Mihos, J. C. \& Hernquist, L. 1996, \apj, 464, 641

\noindent
Naim, A. et al. 1995, \mnras, 274, 1107

\noindent
Nair, P. B. \& Abraham, R. G. 2010, \apjs, 186, 427

\noindent
Pahre, M., Ashby, M. L. N., Fazio, G. G., \& Willner, S. P. 2004,
\apjs, 154, 235

%
\noindent
Rhoads, J. E. 1998, \aj, 115, 472

\noindent
Rix, H.-W. \& Rieke, M. J. 1993, \apj, 418, 123

%
\noindent
Sandage, A. 1961, The Hubble Atlas of Galaxies, Carnegie Inst. of
Wash. Publ. No. 618

\noindent 
Sandage, A. and Bedke, J. 1994, {\it The Carnegie Atlas of Galaxies},
Carnegie Institute of Washington Publ. No. 638 

\noindent
Schweizer, F. \& Seitzer, P. 1988, \apj, 328, 88

\noindent
Scoville, N. Z., Matthews, K., Carico, D. P., \& Sanders, D. B. 1988, \apj, 327, L61

\noindent
Sheth, K. et al. 2008, \apj, 675, 1141

\noindent
Sheth, K. et al. 2010, \pasp, submitted

\noindent
Simien, F. \& de Vaucouleurs, G. 1986, \apj, 302, 564

\noindent
Skrutskie, M. et al. 1997, ASSL, 210, 25

\noindent
Talbot, R. J., Jensen, E. B., \& Dufour, R. J. 1979, \apj, 229, 91

\noindent
Taylor, V. A., Jansen, R. A., Windhorst, R. A., Odewahn, S. C., \& Hibbard,
J. E. 2005, \apj, 630, 784

\noindent
Taylor-Mager, V. A., Conselice, C. J., Windhorst, R. A., \& Jansen,
R. A. \apj, 659, 162

\noindent
Telesco, C. M., Joy, M., Dietz, K., Decher, R., \& Campins, H. 1991, \apj, 
369, 135

\noindent
Thornley, M. 1996, \apj, 469, L45

\noindent
Thronson, H. A., Hereld, M., Majewski, S., Greenhouse, M., Johnson, P.,
Spillar, E., Woodward, C. E., Harper, D. A., \& Rauscher, B. J. 1989,
\apj, 343, 158

\noindent
Varela, A. M., Munoz-Tunon, C., \& Simmoneau, E. 1996, \aap, 306, 381

\noindent
Vergani, D., Pizzella, A., Corsini, E. M., van Driel, W., Buson, L. M.,
Dettmar, R.-J., \& Bertola, F. 2007, \aap, 463, 883

\noindent
Vorontsov-Velyaminov, B. A. 1987, Extragalactic Astronomy, New york, Harwood
Academic Publishers

\noindent
Werner, M. W. et al. 2004, \apjs, 154, 1

\noindent
Whitmore, B. C., Lucas, R. A., McElroy, D. B., Steiman-Cameron, T. Y.,
Sackett, P. D., \& Olling, R. P. 1990, \aj, 100, 1489

\noindent
Whyte, L. F., Abraham, R. G., Merrifield, M. R., Eskridge, P. B., Frogel, J. A.,
Pogge, R. W. 2002, \mnras, 336, 1281

\noindent
Windhorst, R. et al. 2002, \apjs, 143, 113

\noindent
Zaritsky, D. \& Lo, K. Y. 1986, \apj, 303, 66

\clearpage
\makeatletter
\def\jnl@aj{AJ}
\ifx\revtex@jnl\jnl@aj\let\tablebreak=\nl\fi
\makeatother

\begin{deluxetable}{lrlllll}
\tablenum{1}
\tablewidth{45pc}
\tablecaption{3.6$\mu$m Morphological Classifications\tablenotemark{a}}
\tablehead{
\colhead{Galaxy} &
\colhead{Fig 1.} &
\colhead{S$^4$G} &
\colhead{RC3} &
\colhead{RSA} &
\colhead{dVA} &
\colhead{OSU}
\\
\colhead{} &
\colhead{} &
\colhead{mid-IR} &
\colhead{blue} &
\colhead{blue} &
\colhead{blue} &
\colhead{near-IR} 
\\
\colhead{} &
\colhead{} &
\colhead{type} &
\colhead{type} &
\colhead{type} &
\colhead{type} &
\colhead{type} 
\\
\colhead{1} &
\colhead{2} &
\colhead{3} &
\colhead{4} &
\colhead{5} &
\colhead{6} &
\colhead{7} 
}
\startdata
NGC    24    &   1    &    S(rs)d: sp                                             & SA(s)c                   & Sc(s)                & .....                                                                       & .....                                                                       \\            
NGC    45    &   2    &    SA(s)dm                                                & SA(s)dm                  & Scd(s)               & SA(s)d                                                                      & .....                                                                       \\            
NGC    55    &   3    &    SB(s)m sp                                              & SB(s)m: sp               & Sc                   & .....                                                                       & .....                                                                       \\            
NGC    59    &   4    &    dE5,N                                                  & SA(rs)0$^-$:             & .....                & .....                                                                       & .....                                                                       \\            
NGC   247    &   5    &    SAB(s)d                                                & SAB(s)d                  & Sc(s)                & S$\underline{\rm A}$B(s)d                                                   & .....                                                                       \\            
NGC   274    &   6    &    (R)SA(l)0$^-$                                          & SAB(r)0$^-$ pec          & S0$_1$(0)            & .....                                                                       & .....                                                                       \\
\enddata
\tablenotetext{a}{Col. 1: galaxy name; col. 2: Figure 1 plate
number; col. 3: mid-IR classification
using the precepts and notation of the dVA; col. 4: decoded RC3
blue light classification;
index; col 5: RSA blue light classification; col. 6: dVA classification
(mostly blue, some visual); col. 7: Eskridge et al. (2002) OSUBGS $H$-band 
classification}
\tablenotetext{b}{PRG=polar ring galaxy (see Whitmore et al. 1990)}
\tablenotetext{c}{double-stage: SB(s)cd/S0$^-$}
\tablenotetext{d}{NGC 3804 in RC3}
\tablenotetext{e}{double-stage: SB(s)m/S0/a}
\tablenotetext{f}{NGC 4517 in RC3 and RSA}
\tablenotetext{g}{double-stage: SB(rs)m/S0/a}
\tablenotetext{h}{double-stage: SA(rs)ab/S0/a}
\tablenotetext{i}{double-stage: SA(s)b/S0$^o$}
\tablenotetext{j}{disky inner regions only; boxy at larger radii}
\tablenotetext{k}{apparent bar could also be inner disk}
\end{deluxetable}

\makeatletter
\def\jnl@aj{AJ}
\ifx\revtex@jnl\jnl@aj\let\tablebreak=\nl\fi
\makeatother

\begin{deluxetable}{lrrrr}
\tablenum{2}
\tablewidth{35pc}
\tablecaption{Bar Fraction Analysis\tablenotemark{a}}
\tablehead{
\colhead{} &
\colhead{All Types} &
\colhead{0$\leq T \leq$9} &
\colhead{All Types} &
\colhead{0$\leq T \leq$9} 
\\
\colhead{} &
\colhead{} &
\colhead{} &
\colhead{no spindles} &
\colhead{no spindles} 
\\
\colhead{1} &
\colhead{2} &
\colhead{3} &
\colhead{4} &
\colhead{5} 
}
\startdata
SA                        &     17$\pm$     3 &     17$\pm$     3 &
19$\pm$
    3 &     20$\pm$     4\\
S$\underline{\rm A}$B     &      8$\pm$     2 &     11$\pm$     3 &
10$\pm$
    2 &     13$\pm$     3\\
SAB                       &     18$\pm$     3 &     22$\pm$     4 &
21$\pm$
    3 &     26$\pm$     4\\
SA$\underline{\rm B}$     &      2$\pm$     1 &      3$\pm$     1 &
2$\pm$
    1 &      4$\pm$     2\\
SB                        &     31$\pm$     3 &     40$\pm$     4 &
28$\pm$
    3 &     36$\pm$     4\\
no family                 &     24$\pm$     3 &      7$\pm$     2 &
19$\pm$
    3 &      1$\pm$     1\\
bar fraction\tablenotemark{b}              &     51$\pm$     3 &     64$\pm$     4 &
52$\pm$
    4 &     66$\pm$     4\\
Number                    &    207 &    138 &    173 &    114\\
\enddata
\tablenotetext{a}{Col. 1: category or parameter; cols. 2-5: 
percentage of that category out of the number listed on line 8.}
\tablenotetext{b}{The sum of SAB, SA$\underline{\rm B}$, and SB classifications.}
\end{deluxetable}

\clearpage
 
\clearpage
 
\clearpage
 
\clearpage
 
\clearpage
 
\clearpage
 
\clearpage
 
\clearpage
 
\clearpage
 
\clearpage
 
\clearpage
 
\clearpage
 
\clearpage
 
\clearpage
 
\clearpage
 
\clearpage
 
\clearpage
 
\clearpage
 
\clearpage
 
\clearpage
 
\clearpage
 
\clearpage
 
\clearpage
 
\clearpage
 
\clearpage
 
\clearpage
 
\clearpage
 
\clearpage
 
\clearpage
 
\clearpage
 
\clearpage
 
\clearpage
 
\clearpage
 
\clearpage
 
\clearpage
 
\clearpage
 
\clearpage
 
\clearpage
 
\clearpage
 
\clearpage
 
\clearpage
 
\clearpage
 
\clearpage
 
\clearpage
 
\clearpage
 
\clearpage
 
\clearpage
 
\clearpage
 
\clearpage
 
\clearpage
 
\clearpage
 
\clearpage
\begin{figure}
\figurenum{1.52} 
\plotone{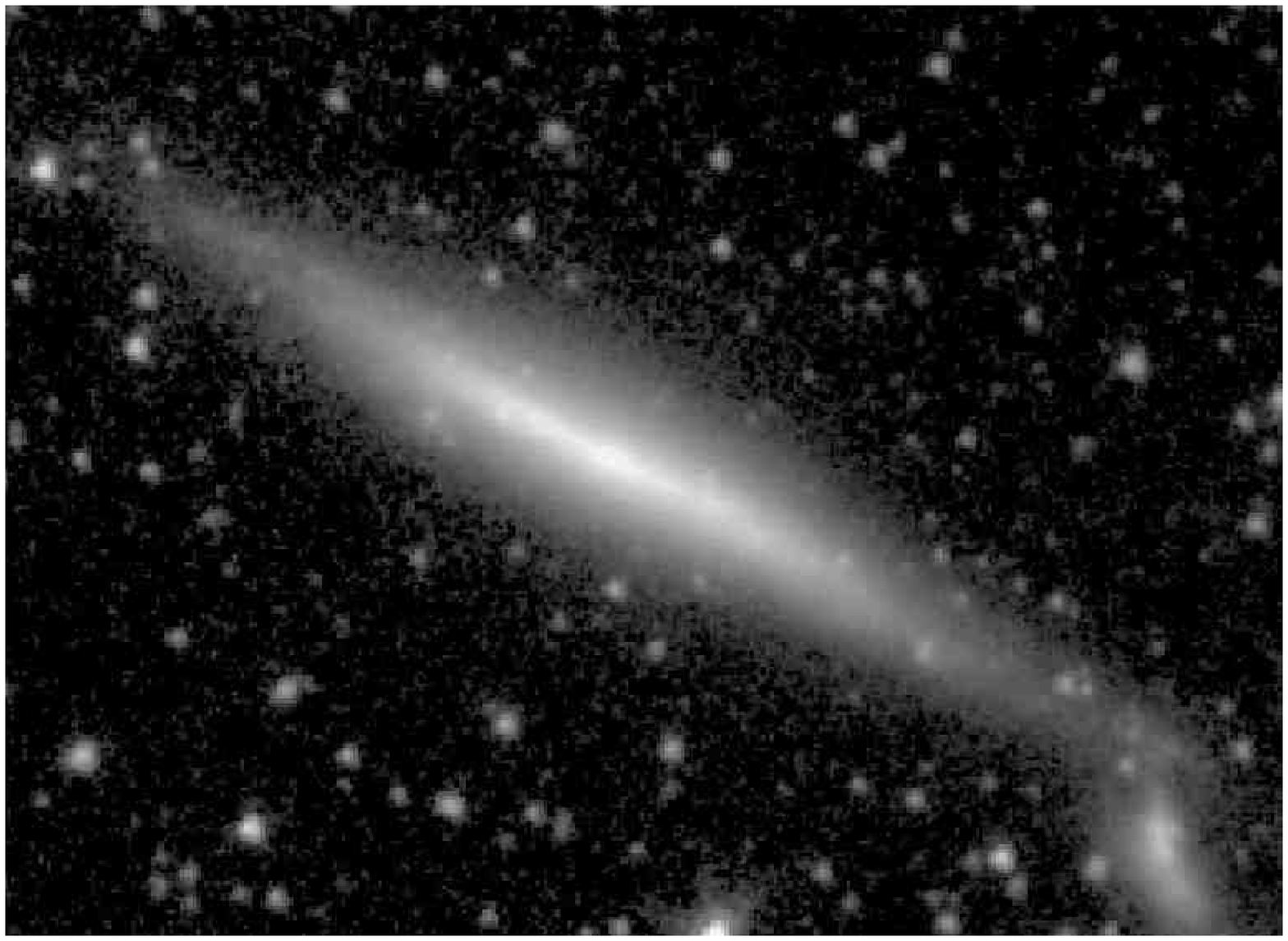}
 \vspace{2.0truecm}
 \caption{
{\bf NGC  2820   }              - S$^4$G mid-IR classification:    Sc sp                                                 ; Filter: IRAC 3.6$\mu$m; North:   up, East: left; Field dimensions:   3.9$\times$  2.9 arcmin; Surface brightness range displayed: 15.0$-$28.0 mag arcsec$^{-2}$}                 
\label{NGC2820}     
 \end{figure}
 
\clearpage
\begin{figure}
\figurenum{1.53} 
\plotone{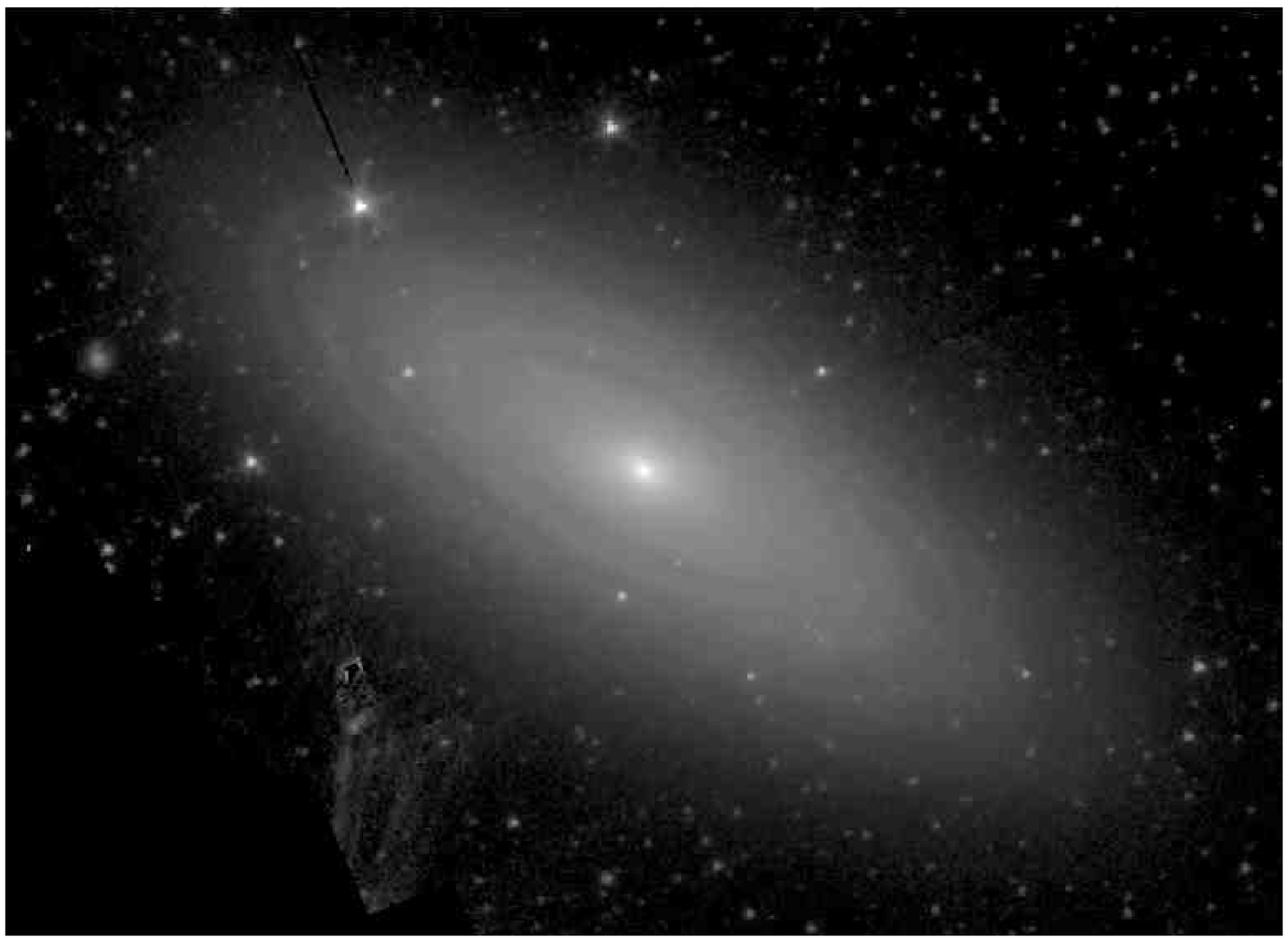}
 \vspace{2.0truecm}
 \caption{
{\bf NGC  2841   }              - S$^4$G mid-IR classification:    S$\underline{\rm A}$B(r)a                             ; Filter: IRAC 3.6$\mu$m; North: left, East: down; Field dimensions:   9.6$\times$  7.0 arcmin; Surface brightness range displayed: 12.0$-$28.0 mag arcsec$^{-2}$}                 
\label{NGC2841}     
 \end{figure}
 
\clearpage
\begin{figure}
\figurenum{1.54} 
\plotone{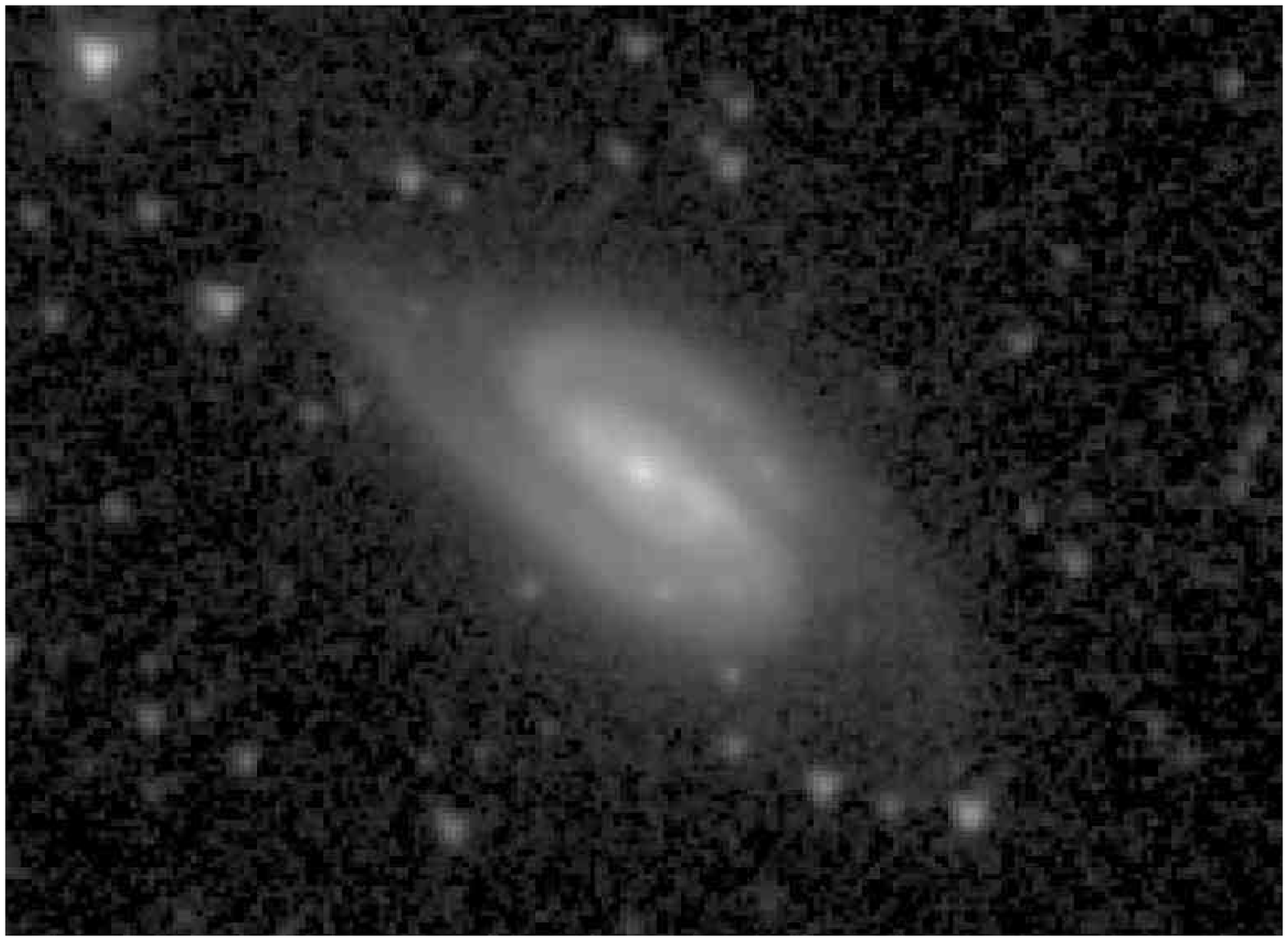}
 \vspace{2.0truecm}
 \caption{
{\bf NGC  2854   }              - S$^4$G mid-IR classification:    SAB(r)a                                               ; Filter: IRAC 3.6$\mu$m; North:   up, East: left; Field dimensions:   2.6$\times$  1.9 arcmin; Surface brightness range displayed: 14.0$-$28.0 mag arcsec$^{-2}$}                 
\label{NGC2854}     
 \end{figure}
 
\clearpage
\begin{figure}
\figurenum{1.55} 
\plotone{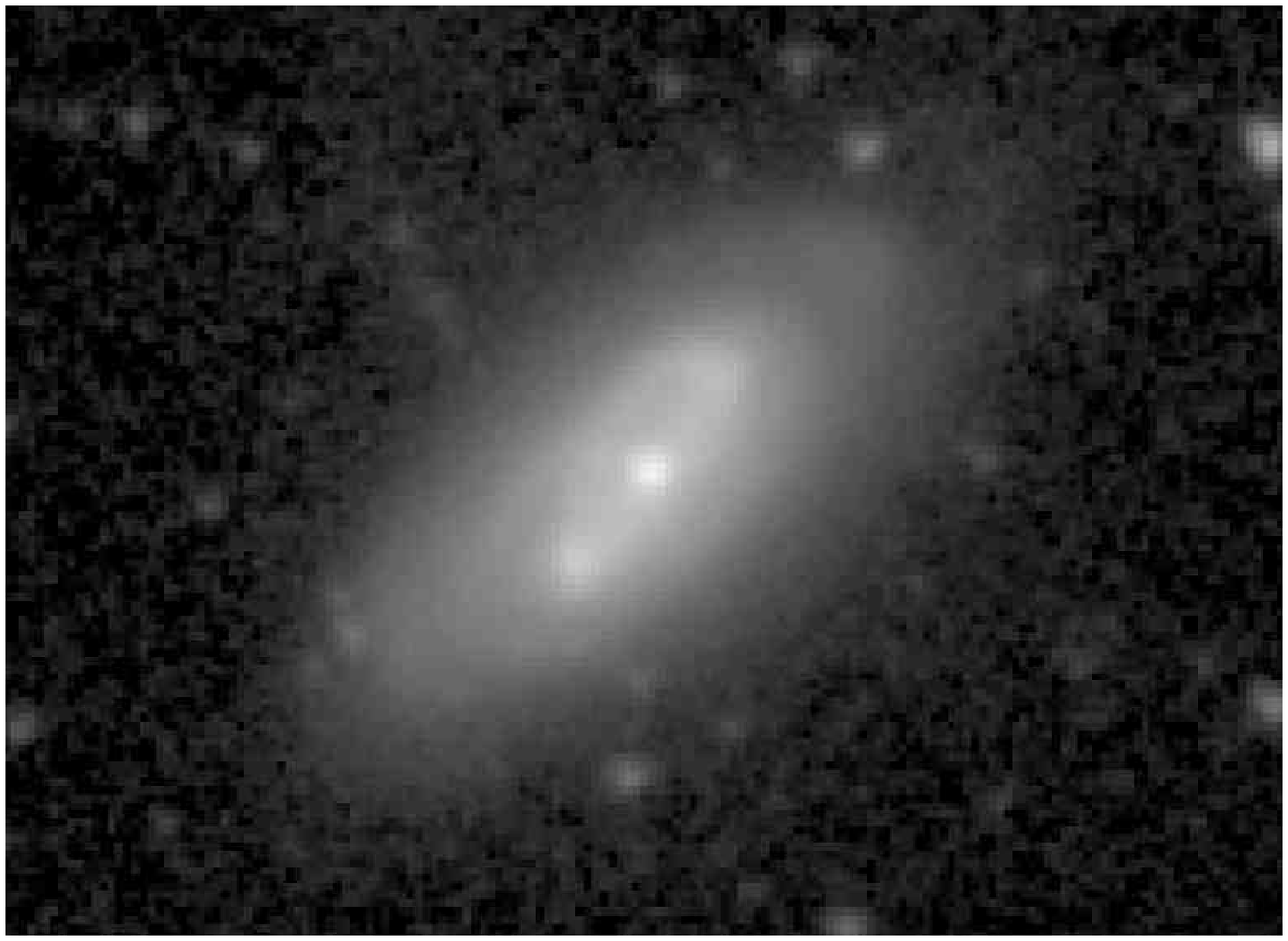}
 \vspace{2.0truecm}
 \caption{
{\bf NGC  2856   }              - S$^4$G mid-IR classification:    SAB(r)a:                                              ; Filter: IRAC 3.6$\mu$m; North:   up, East: left; Field dimensions:   2.0$\times$  1.4 arcmin; Surface brightness range displayed: 13.0$-$28.0 mag arcsec$^{-2}$}                 
\label{NGC2856}     
 \end{figure}
 
\clearpage
\begin{figure}
\figurenum{1.56} 
\plotone{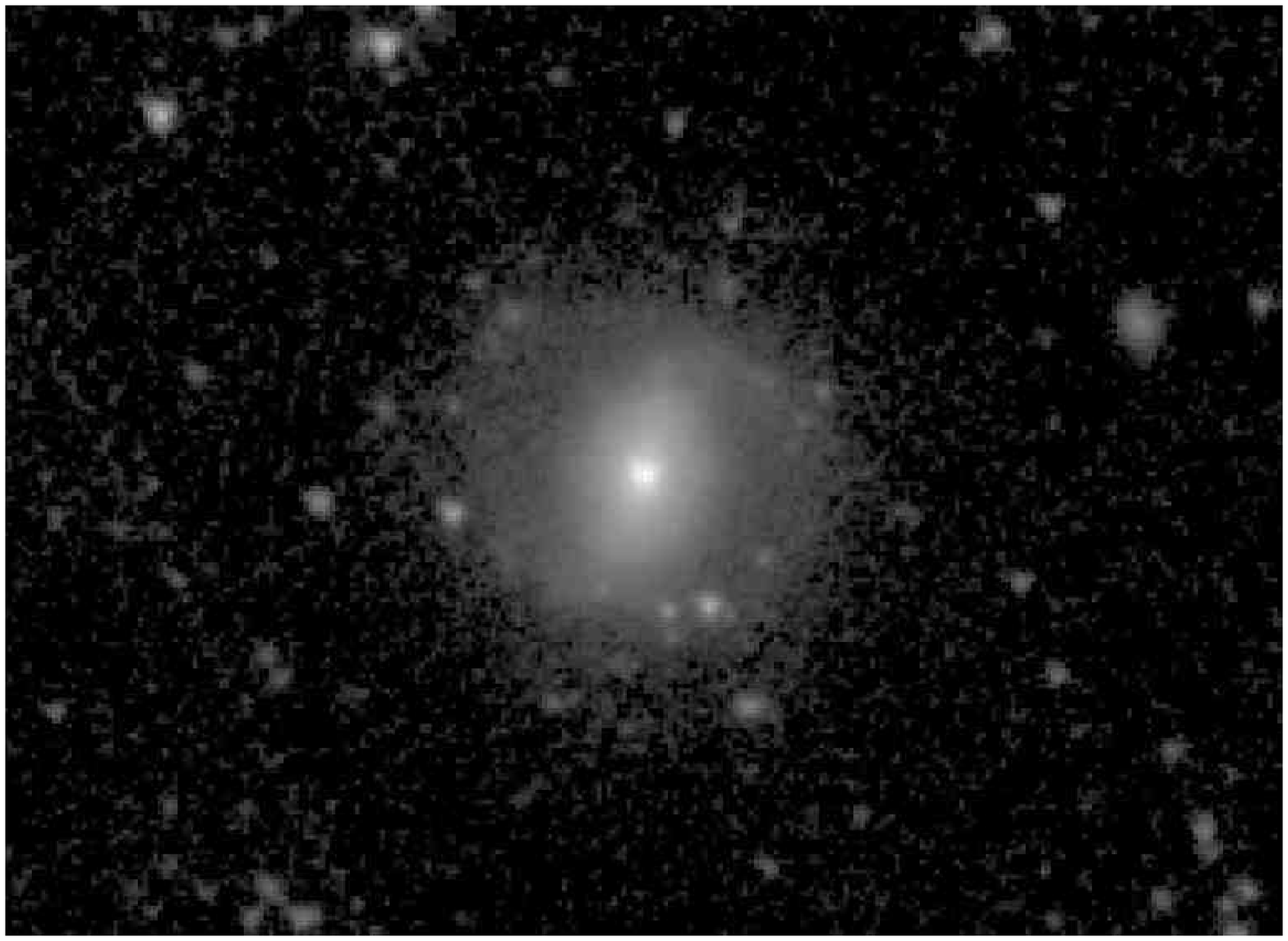}
 \vspace{2.0truecm}
 \caption{
{\bf NGC  2893   }              - S$^4$G mid-IR classification:    (RL)SAB(l)0$^+$                                       ; Filter: IRAC 3.6$\mu$m; North:   up, East: left; Field dimensions:   3.2$\times$  2.3 arcmin; Surface brightness range displayed: 13.0$-$28.0 mag arcsec$^{-2}$}                 
\label{NGC2893}     
 \end{figure}
 
\clearpage
\begin{figure}
\figurenum{1.57} 
\plotone{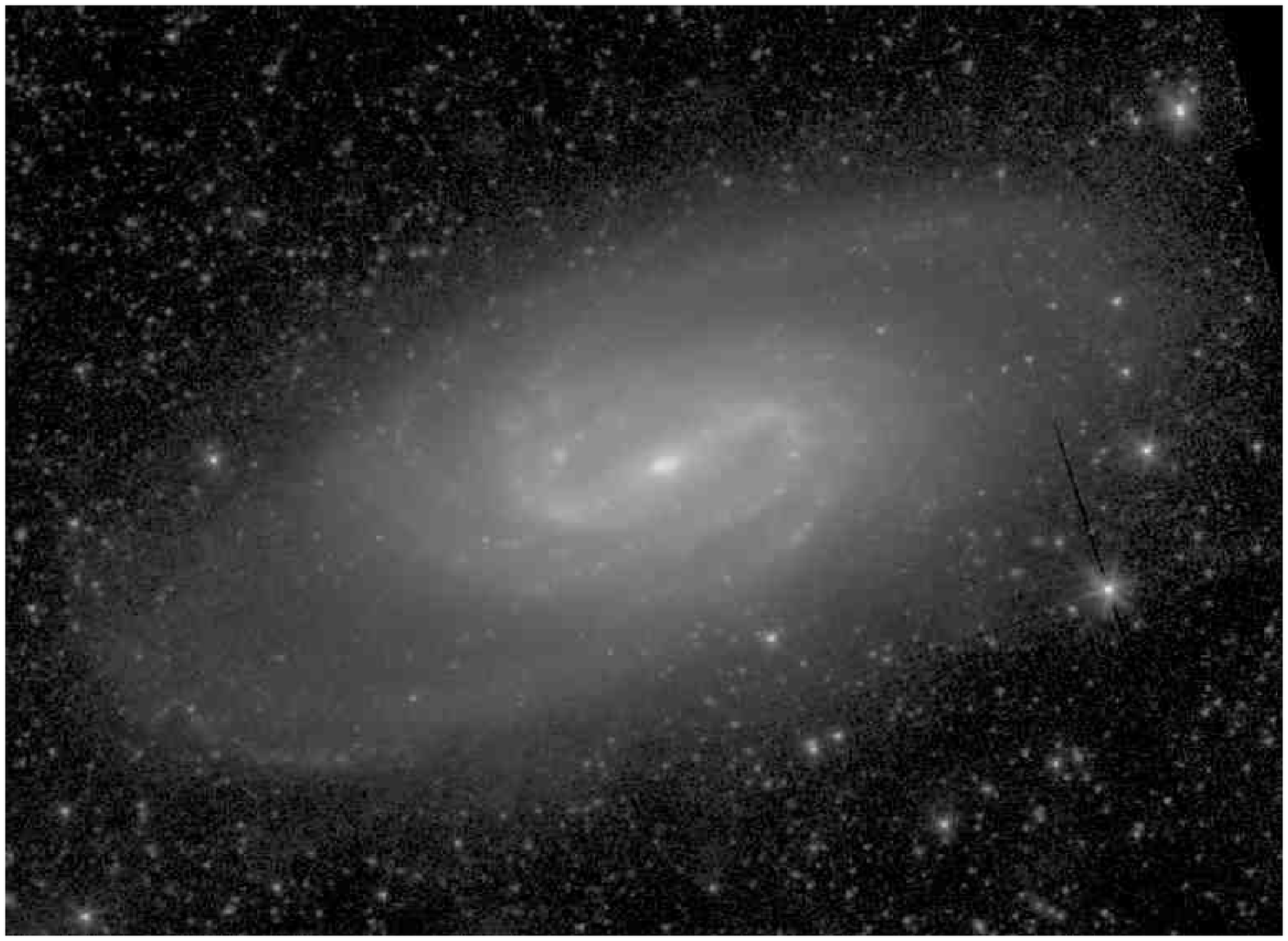}
 \vspace{2.0truecm}
 \caption{
{\bf NGC  2903   }              - S$^4$G mid-IR classification:    (R$^{\prime}$)SB(rs,nr)b                                        ; Filter: IRAC 3.6$\mu$m; North: left, East: down; Field dimensions:  12.6$\times$  9.2 arcmin; Surface brightness range displayed: 12.8$-$28.0 mag arcsec$^{-2}$}                 
\label{NGC2903}     
 \end{figure}
 
\clearpage
\begin{figure}
\figurenum{1.58} 
\plotone{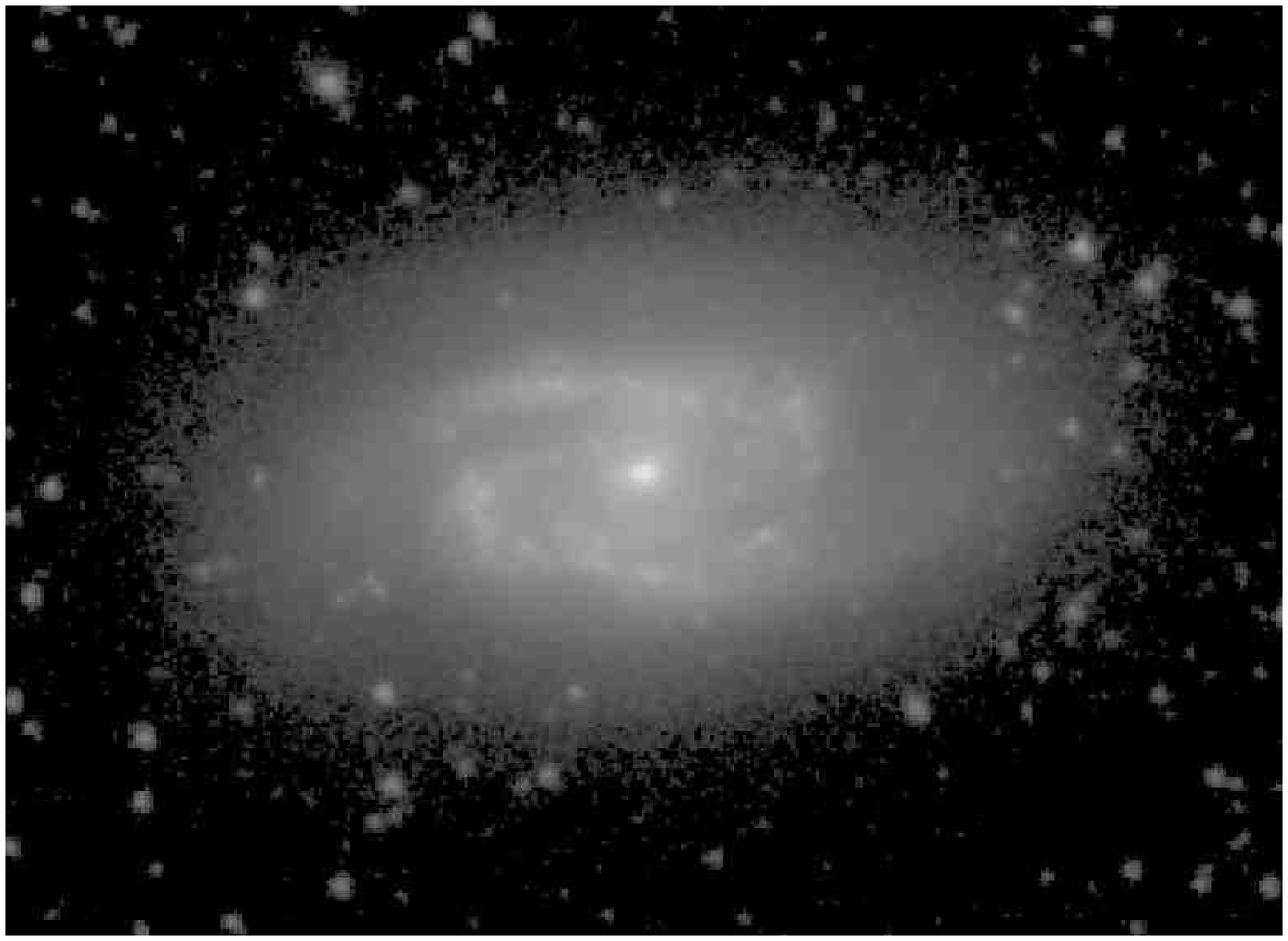}
 \vspace{2.0truecm}
 \caption{
{\bf NGC  2964   }              - S$^4$G mid-IR classification:    S$\underline{\rm A}$B(r$\underline{\rm s}$)b          ; Filter: IRAC 3.6$\mu$m; North:   up, East: left; Field dimensions:   4.0$\times$  2.9 arcmin; Surface brightness range displayed: 13.0$-$28.0 mag arcsec$^{-2}$}                 
\label{NGC2964}     
 \end{figure}
 
\clearpage
\begin{figure}
\figurenum{1.59} 
\plotone{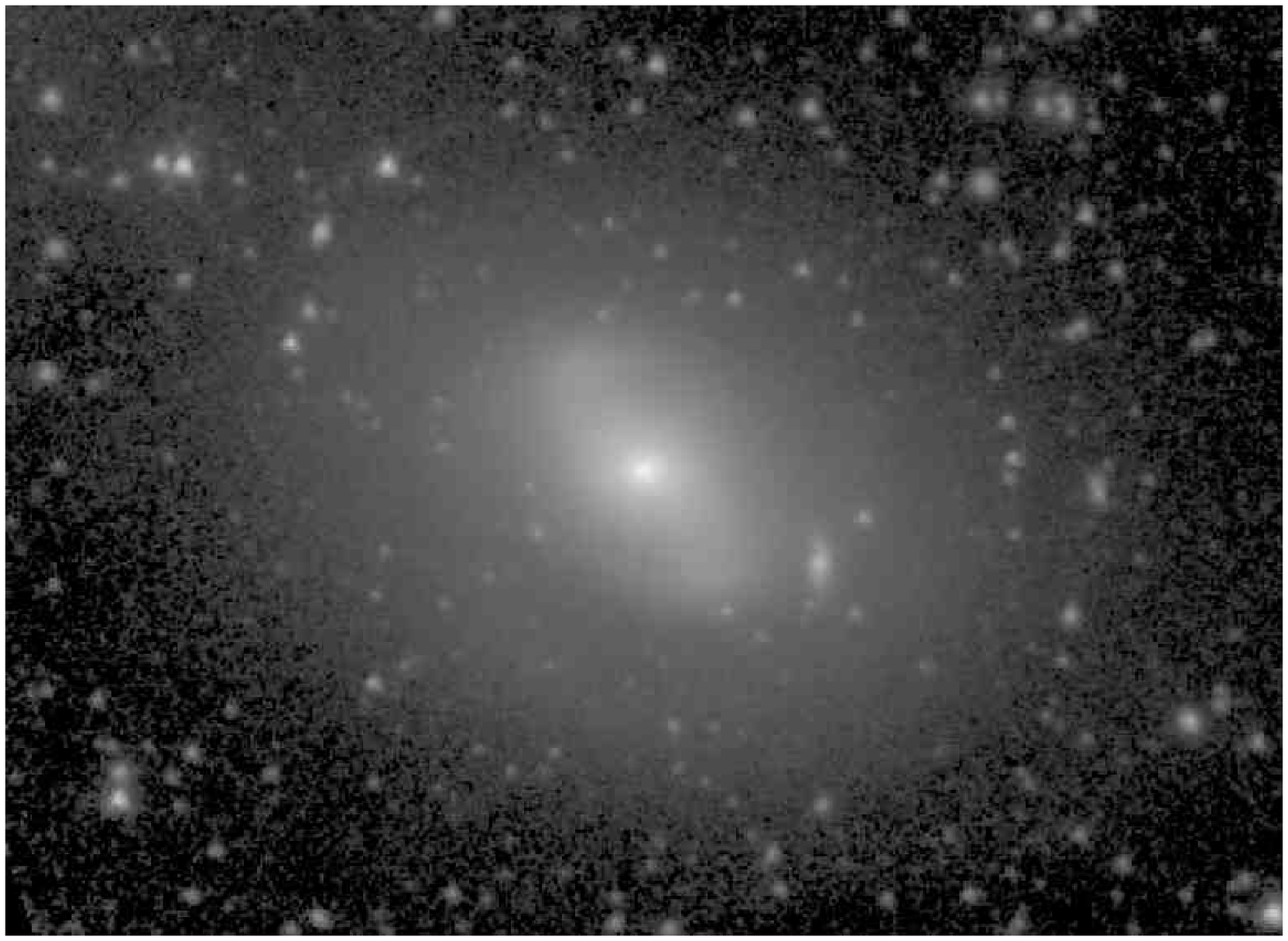}
 \vspace{2.0truecm}
 \caption{
{\bf NGC  2968   }              - S$^4$G mid-IR classification:    SB(r$\underline{\rm s}$)0$^+$                         ; Filter: IRAC 3.6$\mu$m; North:   up, East: left; Field dimensions:   5.3$\times$  3.8 arcmin; Surface brightness range displayed: 13.0$-$28.0 mag arcsec$^{-2}$}                 
\label{NGC2968}     
 \end{figure}
 
\clearpage
\begin{figure}
\figurenum{1.60} 
\plotone{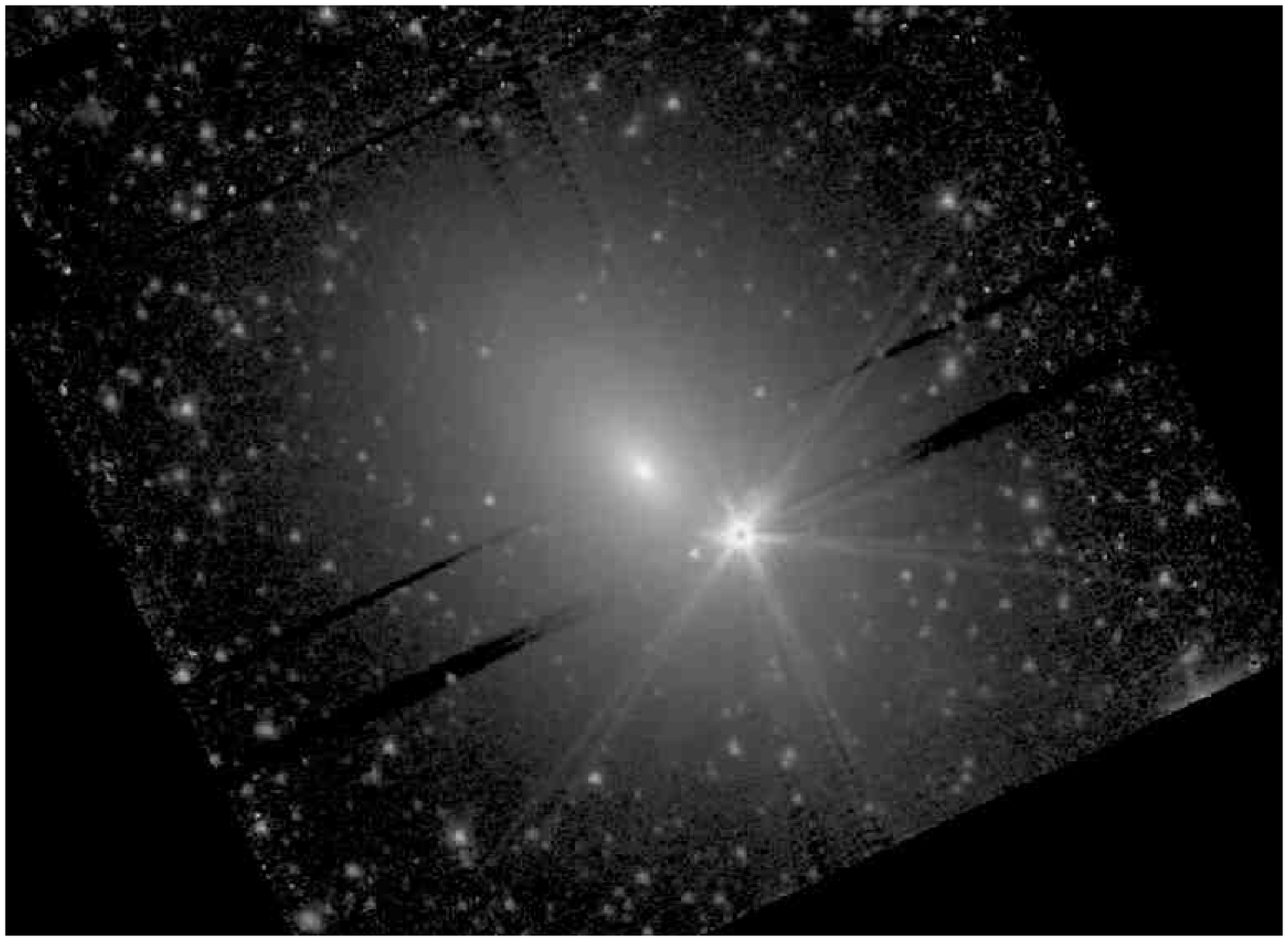}
 \vspace{2.0truecm}
 \caption{
{\bf NGC  2974   }              - S$^4$G mid-IR classification:    SA(r)0/a                                              ; Filter: IRAC 3.6$\mu$m; North:   up, East: left; Field dimensions:   7.9$\times$  5.8 arcmin; Surface brightness range displayed: 13.0$-$28.0 mag arcsec$^{-2}$}                 
\label{NGC2974}     
 \end{figure}
 
\clearpage
\begin{figure}
\figurenum{1.61} 
\plotone{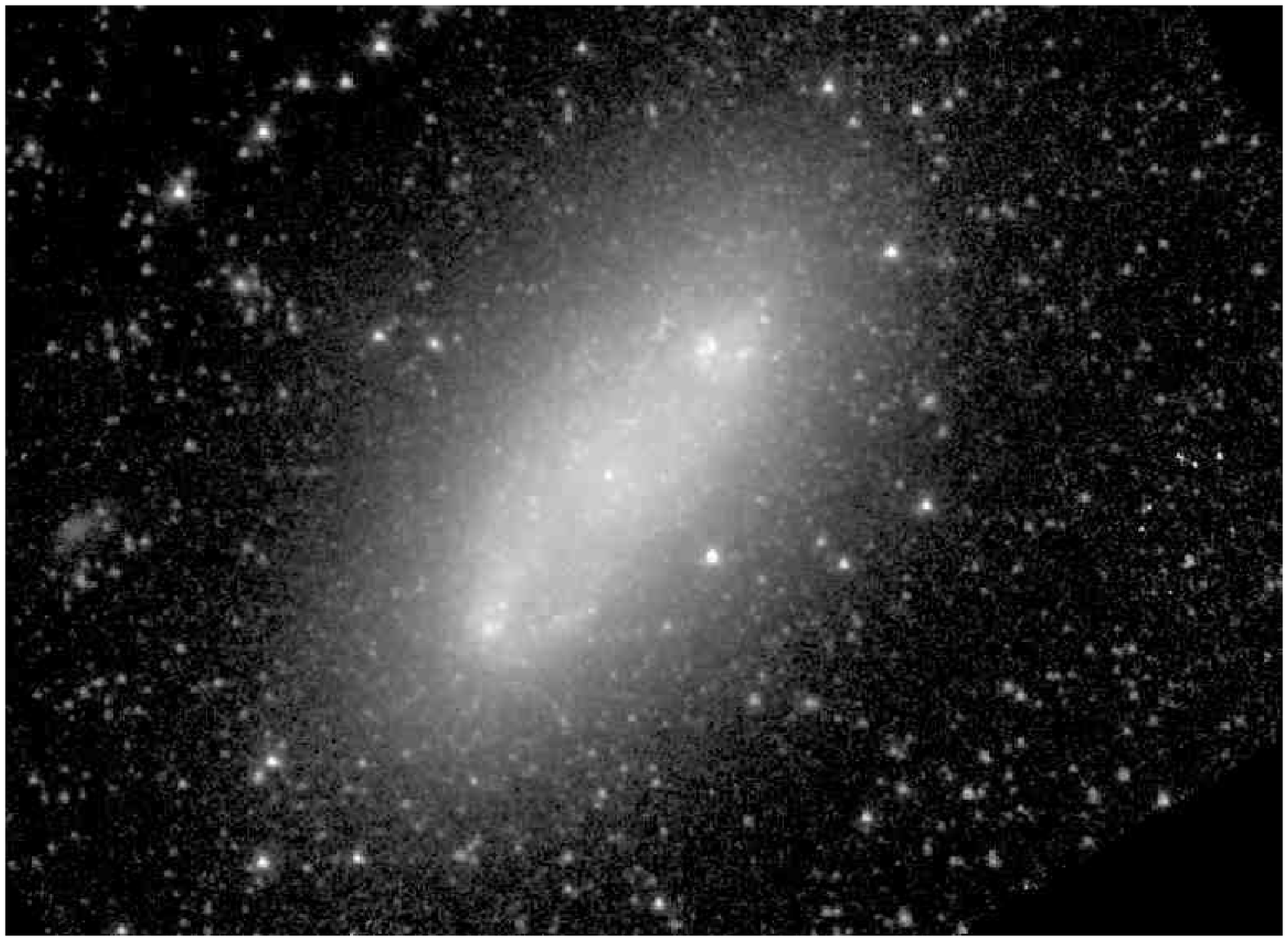}
 \vspace{2.0truecm}
 \caption{
{\bf NGC  2976   }              - S$^4$G mid-IR classification:    SAB(s:)d                                              ; Filter: IRAC 3.6$\mu$m; North:   up, East: left; Field dimensions:   9.0$\times$  6.6 arcmin; Surface brightness range displayed: 15.0$-$28.0 mag arcsec$^{-2}$}                 
\label{NGC2976}     
 \end{figure}
 
\clearpage
\begin{figure}
\figurenum{1.62} 
\plotone{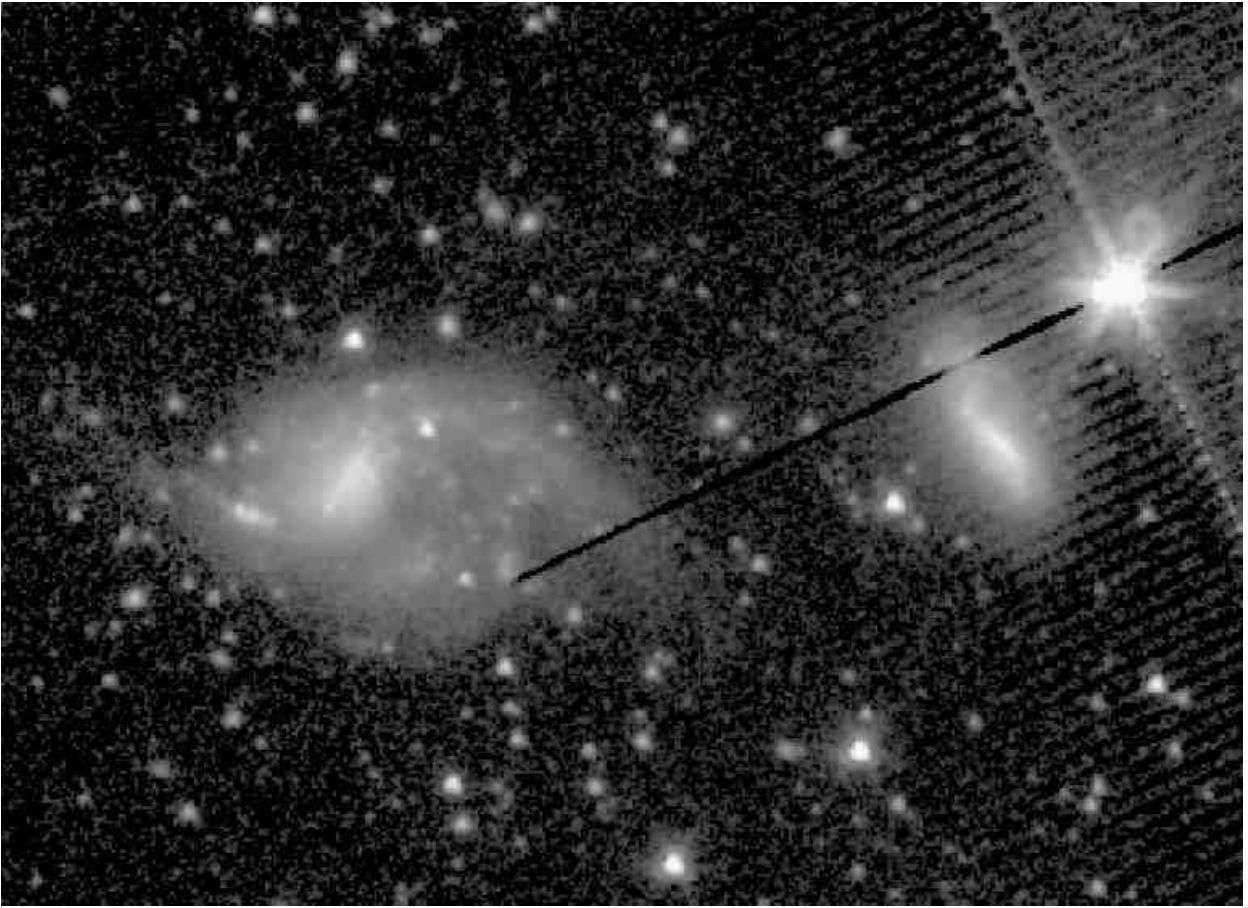}
 \vspace{2.0truecm}
 \caption{
{\bf NGC  3018} (right) and {\bf NGC  3023} (left) - S$^4$G mid-IR classifications:    SB(s)d, SB(s)dm, respectively; Filter: IRAC 3.6$\mu$m; North:   up, East: left; Field dimensions:   5.3$\times$  3.8 arcmin; Surface brightness range displayed: 16.0$-$28.0 mag arcsec$^{-2}$}                 
\label{NGC3018}     
 \end{figure}
 
\clearpage
\begin{figure}
\figurenum{1.63} 
\plotone{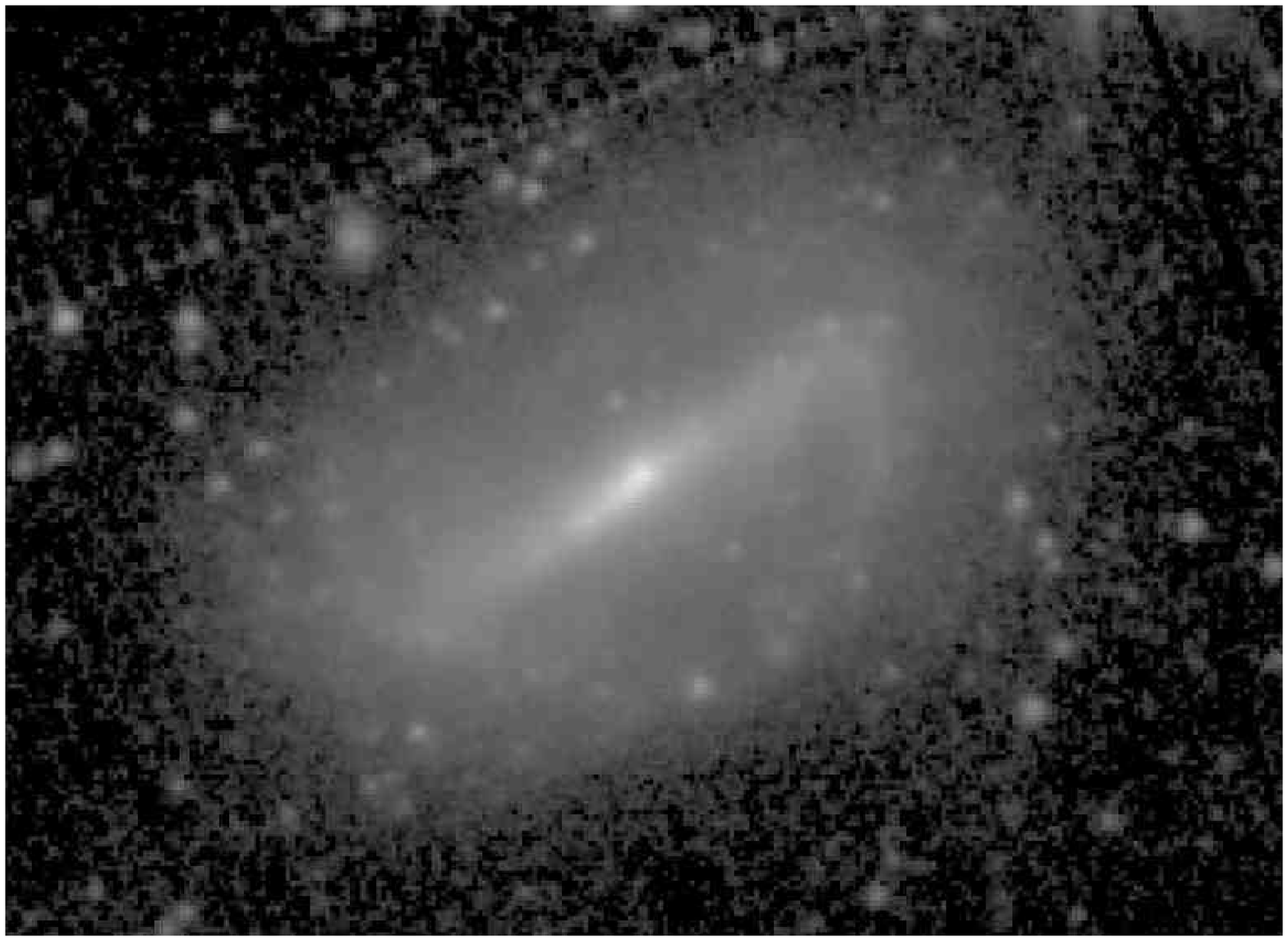}
 \vspace{2.0truecm}
 \caption{
{\bf NGC  3049   }              - S$^4$G mid-IR classification:    SB(s)ab:                                              ; Filter: IRAC 3.6$\mu$m; North: left, East: down; Field dimensions:   3.2$\times$  2.3 arcmin; Surface brightness range displayed: 14.5$-$28.0 mag arcsec$^{-2}$}                 
\label{NGC3049}     
 \end{figure}
 
\clearpage
\begin{figure}
\figurenum{1.64} 
\plotone{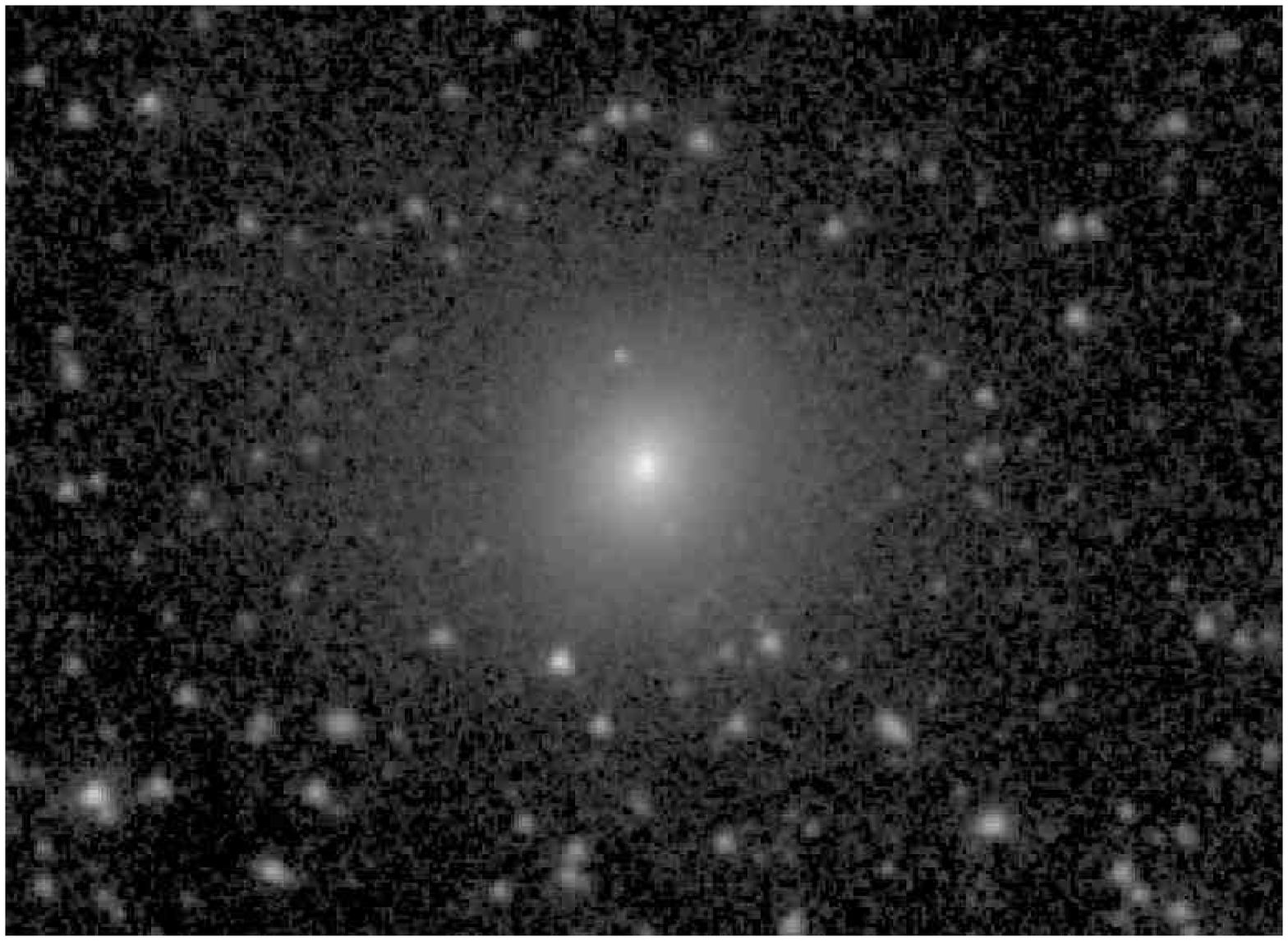}
 \vspace{2.0truecm}
 \caption{
{\bf NGC  3073   }              - S$^4$G mid-IR classification:    S0$^-$                                                ; Filter: IRAC 3.6$\mu$m; North:   up, East: left; Field dimensions:   4.0$\times$  2.9 arcmin; Surface brightness range displayed: 15.0$-$28.0 mag arcsec$^{-2}$}                 
\label{NGC3073}     
 \end{figure}
 
\clearpage
\begin{figure}
\figurenum{1.65} 
\plotone{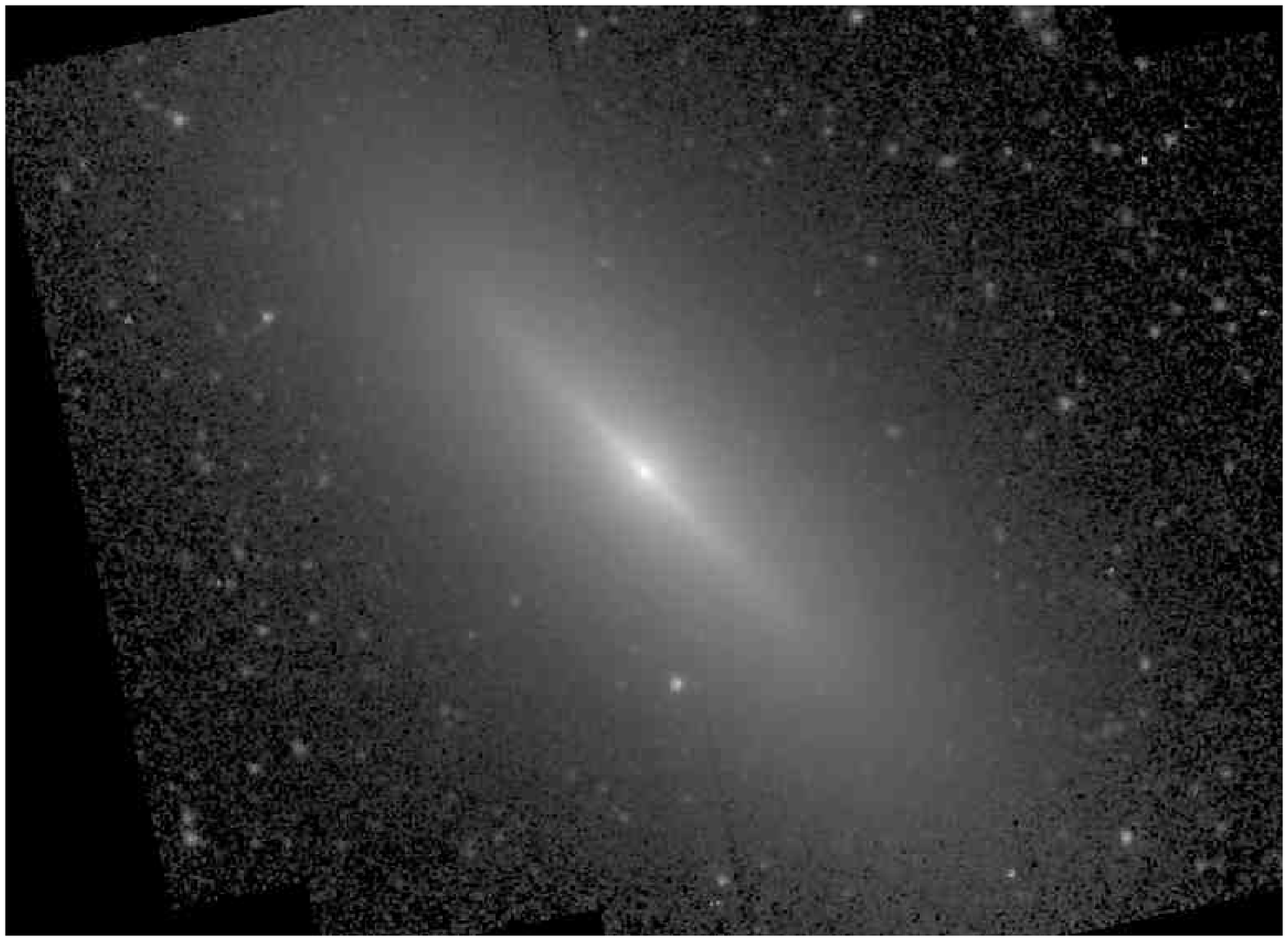}
 \vspace{2.0truecm}
 \caption{
{\bf NGC  3115   }              - S$^4$G mid-IR classification:    S0$^-$/E7 sp                                         ; Filter: IRAC 3.6$\mu$m; North:   up, East: left; Field dimensions:   7.9$\times$  5.8 arcmin; Surface brightness range displayed: 11.0$-$28.0 mag arcsec$^{-2}$}                 
\label{NGC3115}     
 \end{figure}
 
\clearpage
\begin{figure}
\figurenum{1.66} 
\plotone{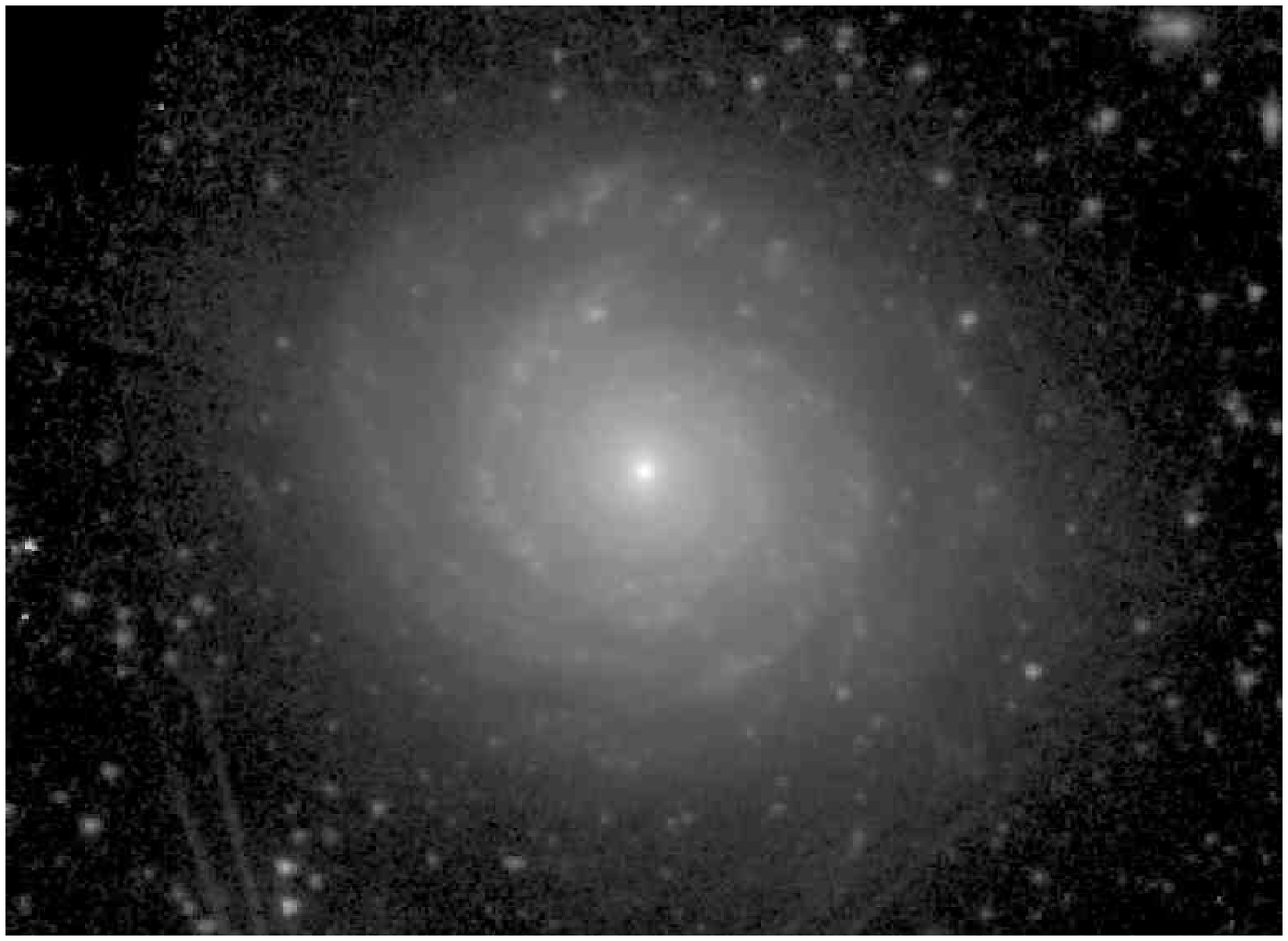}
 \vspace{2.0truecm}
 \caption{
{\bf NGC  3147   }              - S$^4$G mid-IR classification:    S$\underline{\rm A}$B(r$\underline{\rm s}$)b          ; Filter: IRAC 3.6$\mu$m; North: left, East: down; Field dimensions:   5.6$\times$  4.1 arcmin; Surface brightness range displayed: 12.0$-$28.0 mag arcsec$^{-2}$}                 
\label{NGC3147}     
 \end{figure}
 
\clearpage
\begin{figure}
\figurenum{1.67} 
\plotone{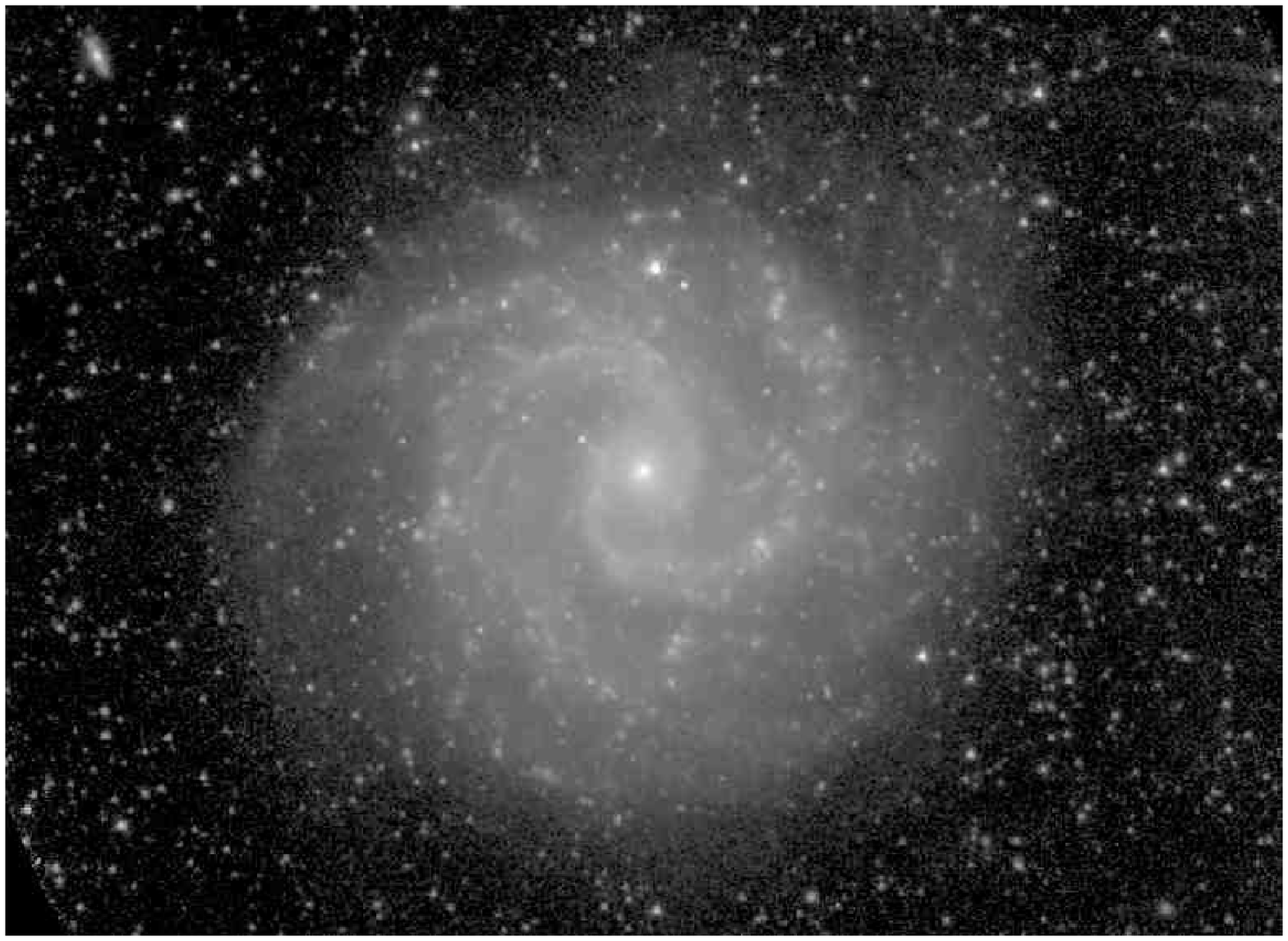}
 \vspace{2.0truecm}
 \caption{
{\bf NGC  3184   }              - S$^4$G mid-IR classification:    SA(r$\underline{\rm s}$)bc                            ; Filter: IRAC 3.6$\mu$m; North:   up, East: left; Field dimensions:  11.3$\times$  8.2 arcmin; Surface brightness range displayed: 14.5$-$28.0 mag arcsec$^{-2}$}                 
\label{NGC3184}     
 \end{figure}
 
\clearpage
\begin{figure}
\figurenum{1.68} 
\plotone{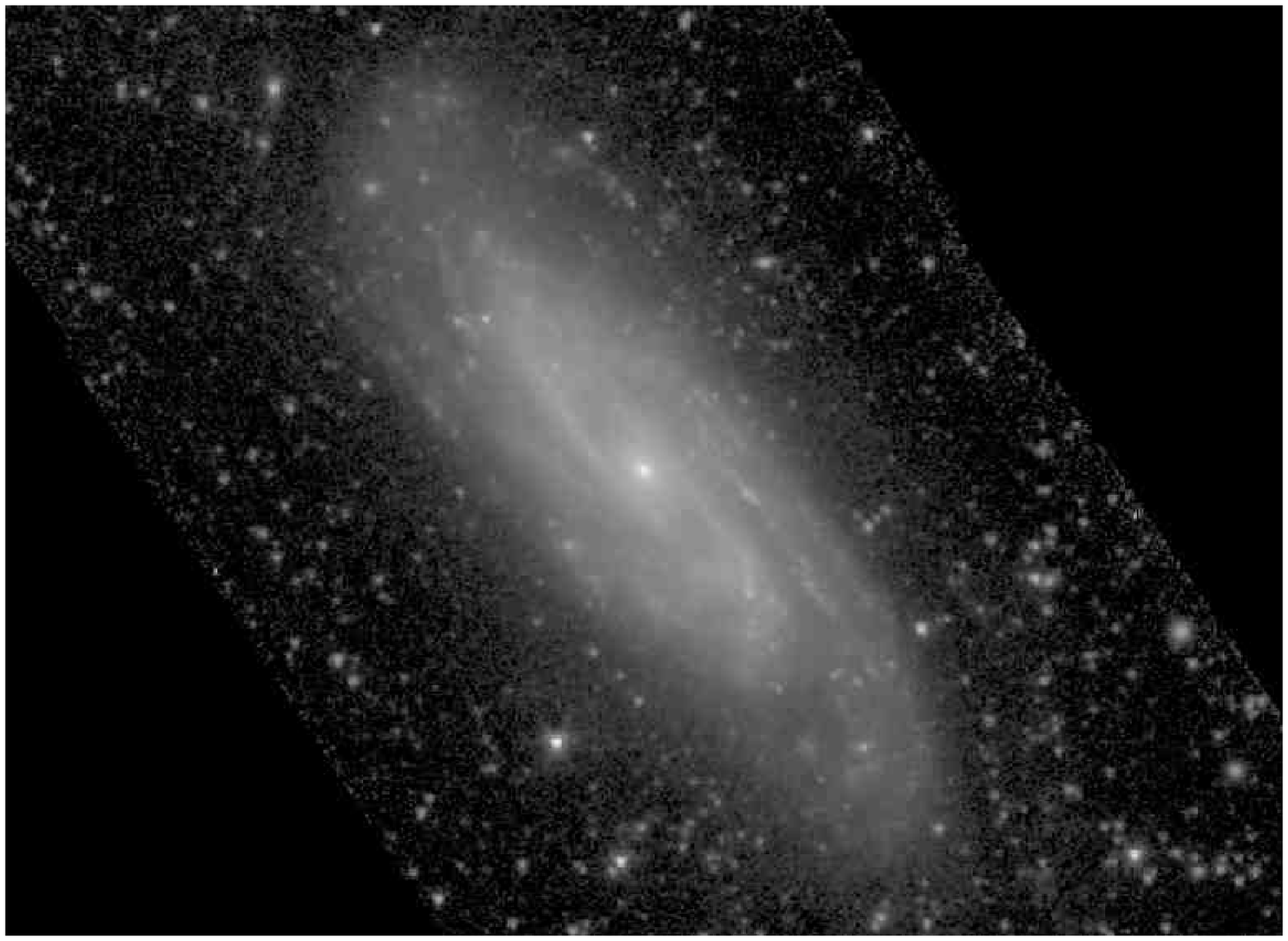}
 \vspace{2.0truecm}
 \caption{
{\bf NGC  3198   }              - S$^4$G mid-IR classification:    SAB(r$\underline{\rm s}$)bc                           ; Filter: IRAC 3.6$\mu$m; North:   up, East: left; Field dimensions:   8.8$\times$  6.4 arcmin; Surface brightness range displayed: 14.0$-$28.0 mag arcsec$^{-2}$}                 
\label{NGC3198}     
 \end{figure}
 
\clearpage
\begin{figure}
\figurenum{1.69} 
\plotone{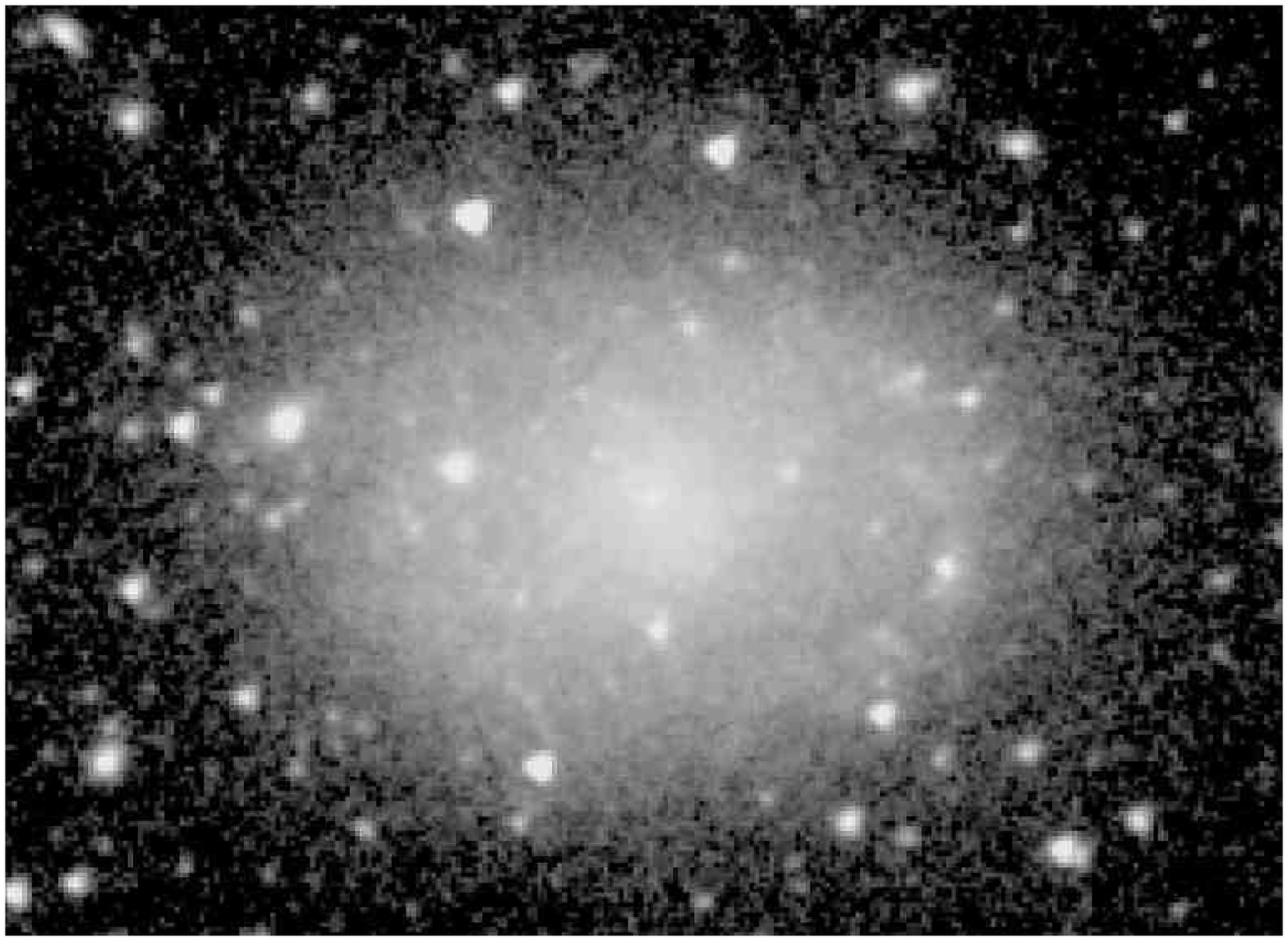}
 \vspace{2.0truecm}
 \caption{
{\bf NGC  3299   }              - S$^4$G mid-IR classification:    S$\underline{\rm A}$Bd:                              ; Filter: IRAC 3.6$\mu$m; North: left, East: down; Field dimensions:   3.2$\times$  2.3 arcmin; Surface brightness range displayed: 18.5$-$28.0 mag arcsec$^{-2}$}                 
\label{NGC3299}     
 \end{figure}
 
\clearpage
\begin{figure}
\figurenum{1.70} 
\plotone{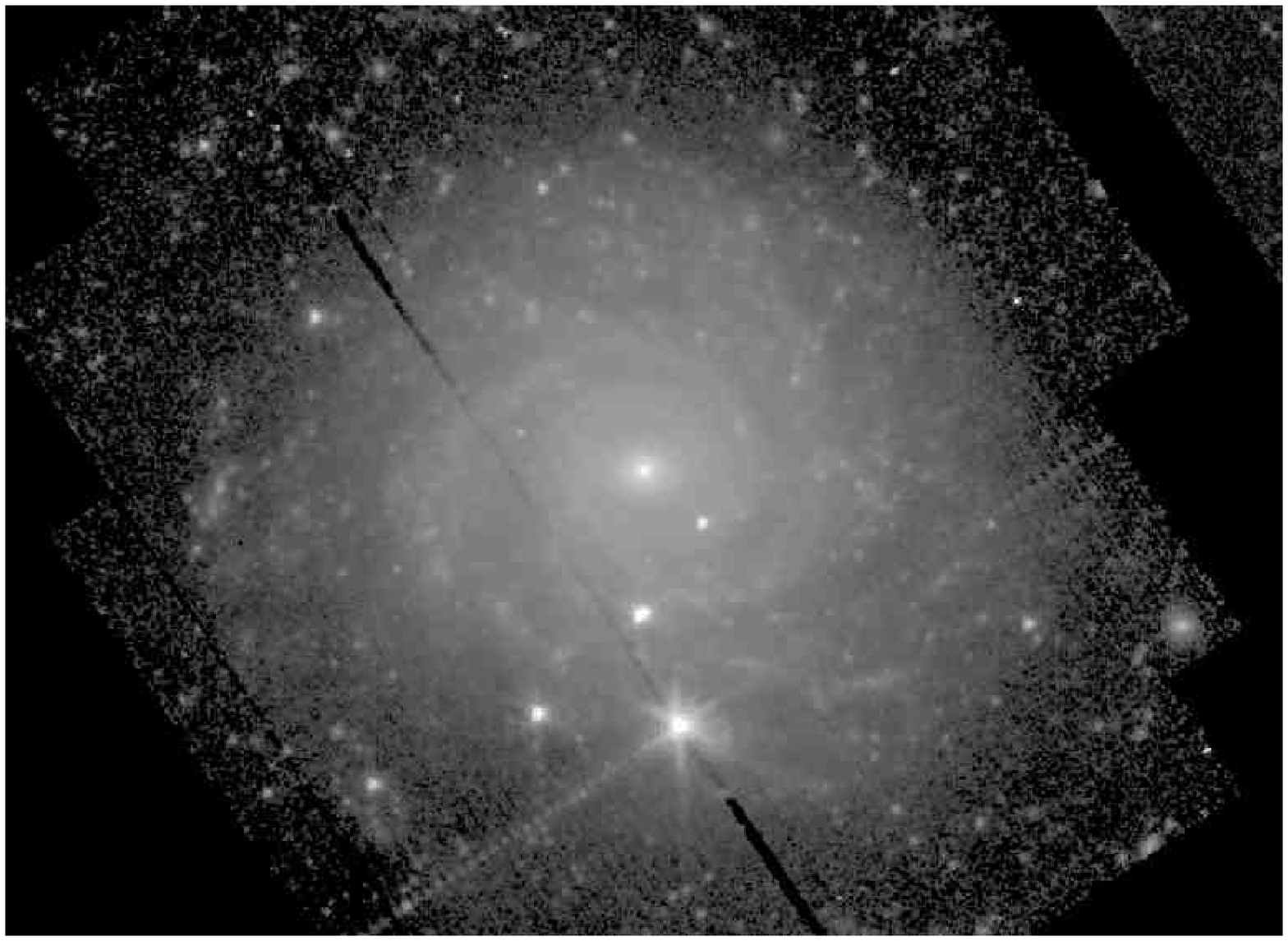}
 \vspace{2.0truecm}
 \caption{
{\bf NGC  3344   }              - S$^4$G mid-IR classification:    SAB(r)bc                                              ; Filter: IRAC 3.6$\mu$m; North: left, East: down; Field dimensions:   7.9$\times$  5.8 arcmin; Surface brightness range displayed: 13.0$-$28.0 mag arcsec$^{-2}$}                 
\label{NGC3344}     
 \end{figure}
 
\clearpage
\begin{figure}
\figurenum{1.71} 
\plotone{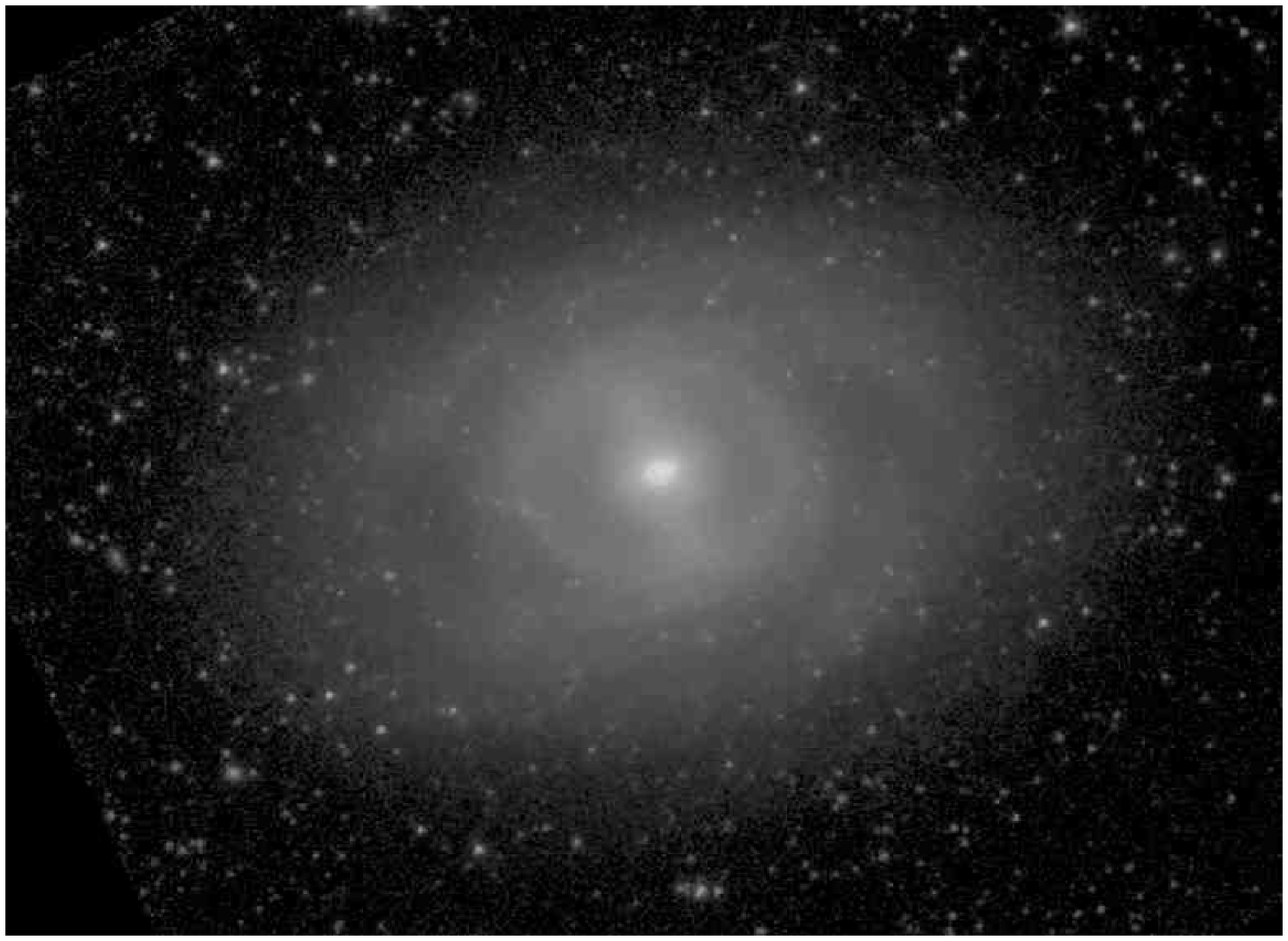}
 \vspace{2.0truecm}
 \caption{
{\bf NGC  3351   }              - S$^4$G mid-IR classification:    (R$^{\prime}$)SB(r,nr)a                                         ; Filter: IRAC 3.6$\mu$m; North: left, East: down; Field dimensions:  10.5$\times$  7.6 arcmin; Surface brightness range displayed: 13.5$-$28.0 mag arcsec$^{-2}$}                 
\label{NGC3351}     
 \end{figure}
 
\clearpage
\begin{figure}
\figurenum{1.72} 
\plotone{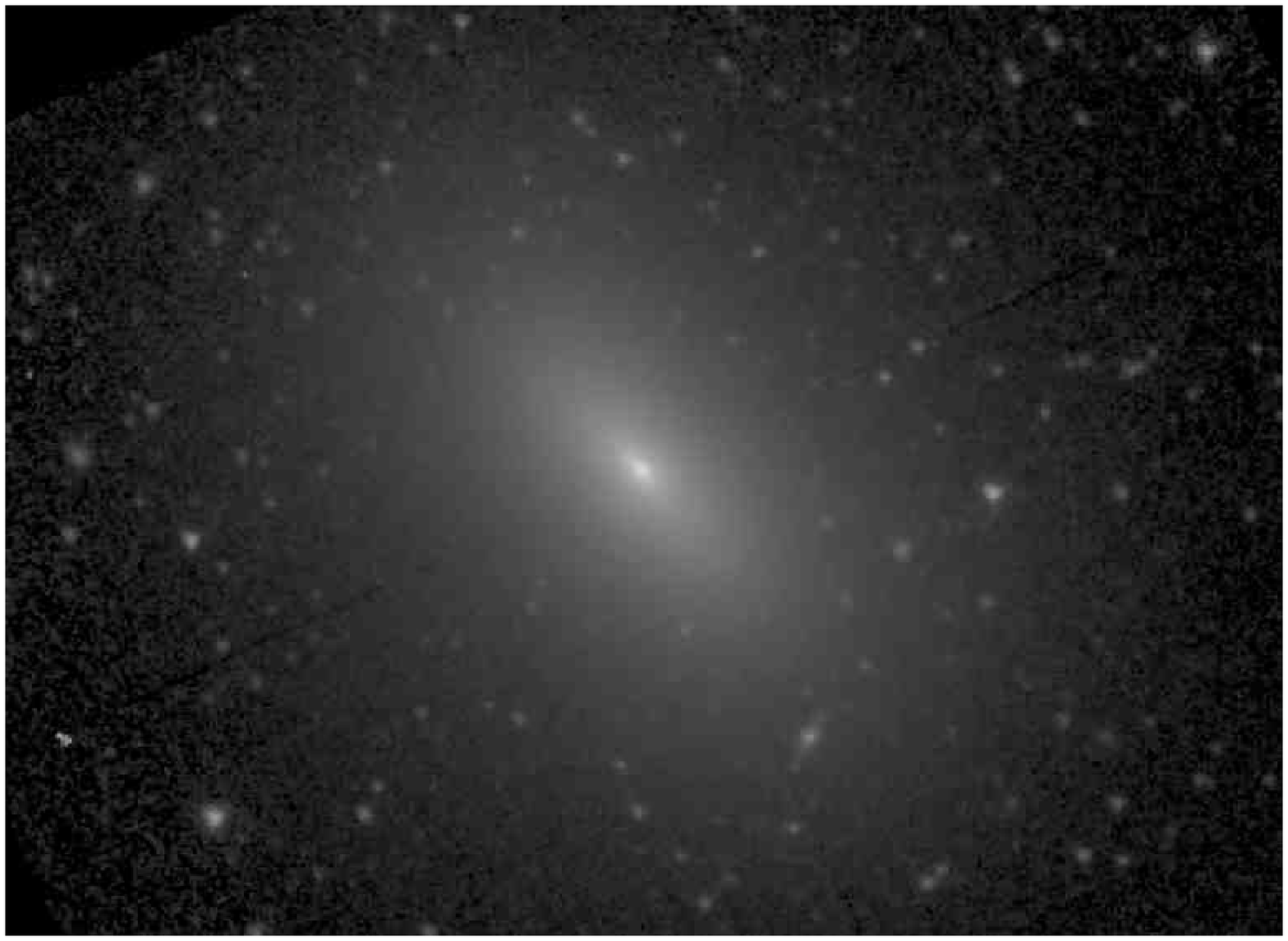}
 \vspace{2.0truecm}
 \caption{
{\bf NGC  3377   }              - S$^4$G mid-IR classification:    E(d)5                                                   ; Filter: IRAC 3.6$\mu$m; North:   up, East: left; Field dimensions:   5.6$\times$  4.1 arcmin; Surface brightness range displayed: 12.0$-$28.0 mag arcsec$^{-2}$}                 
\label{NGC3377}     
 \end{figure}
 
\clearpage
\begin{figure}
\figurenum{1.73} 
\plotone{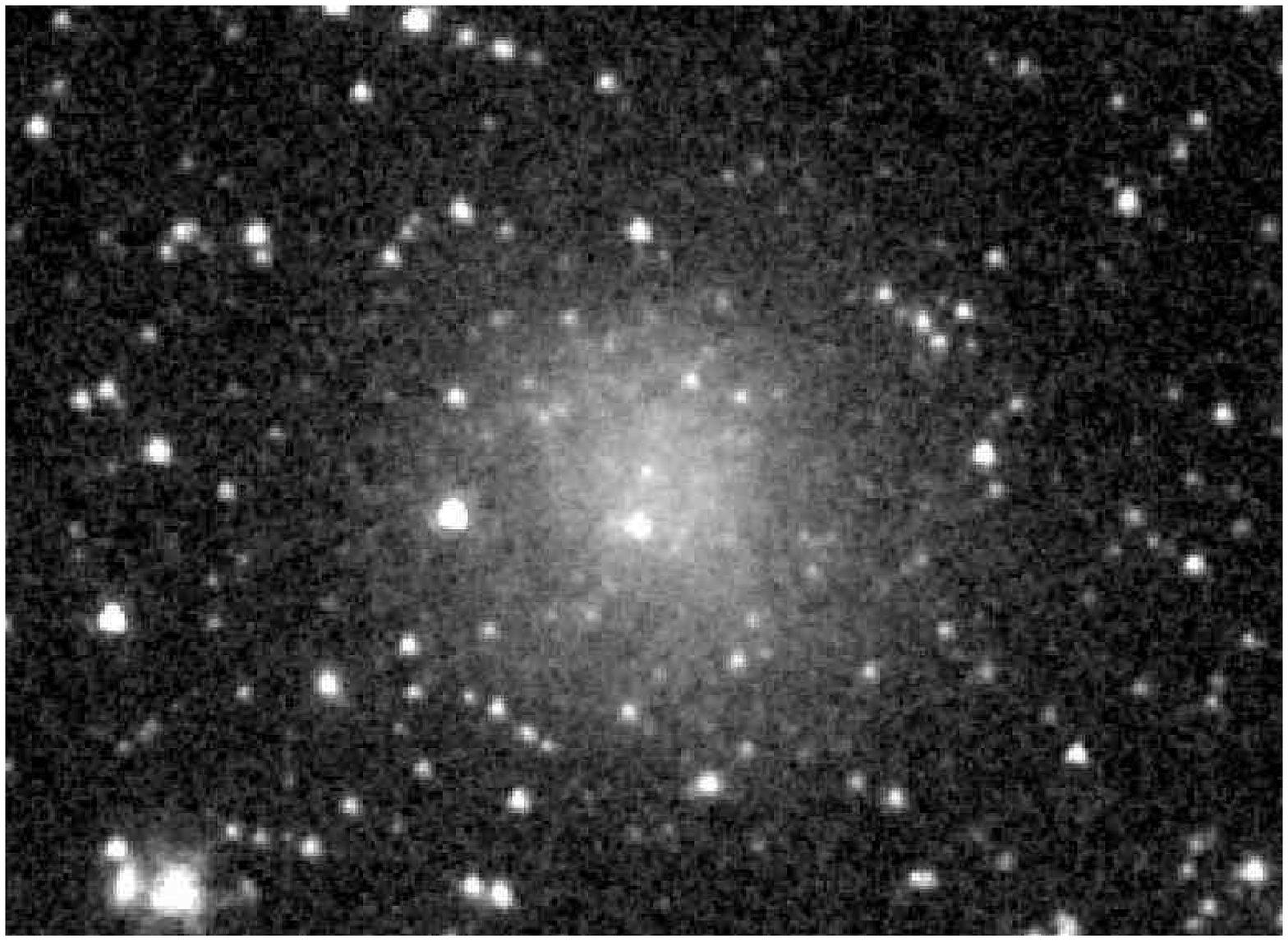}
 \vspace{2.0truecm}
 \caption{
{\bf NGC  3377A  }              - S$^4$G mid-IR classification:    Im                                                    ; Filter: IRAC 3.6$\mu$m; North:   up, East: left; Field dimensions:   4.0$\times$  2.9 arcmin; Surface brightness range displayed: 18.5$-$28.0 mag arcsec$^{-2}$}                 
\label{NGC3377A}    
 \end{figure}
 
\clearpage
\begin{figure}
\figurenum{1.74} 
\plotone{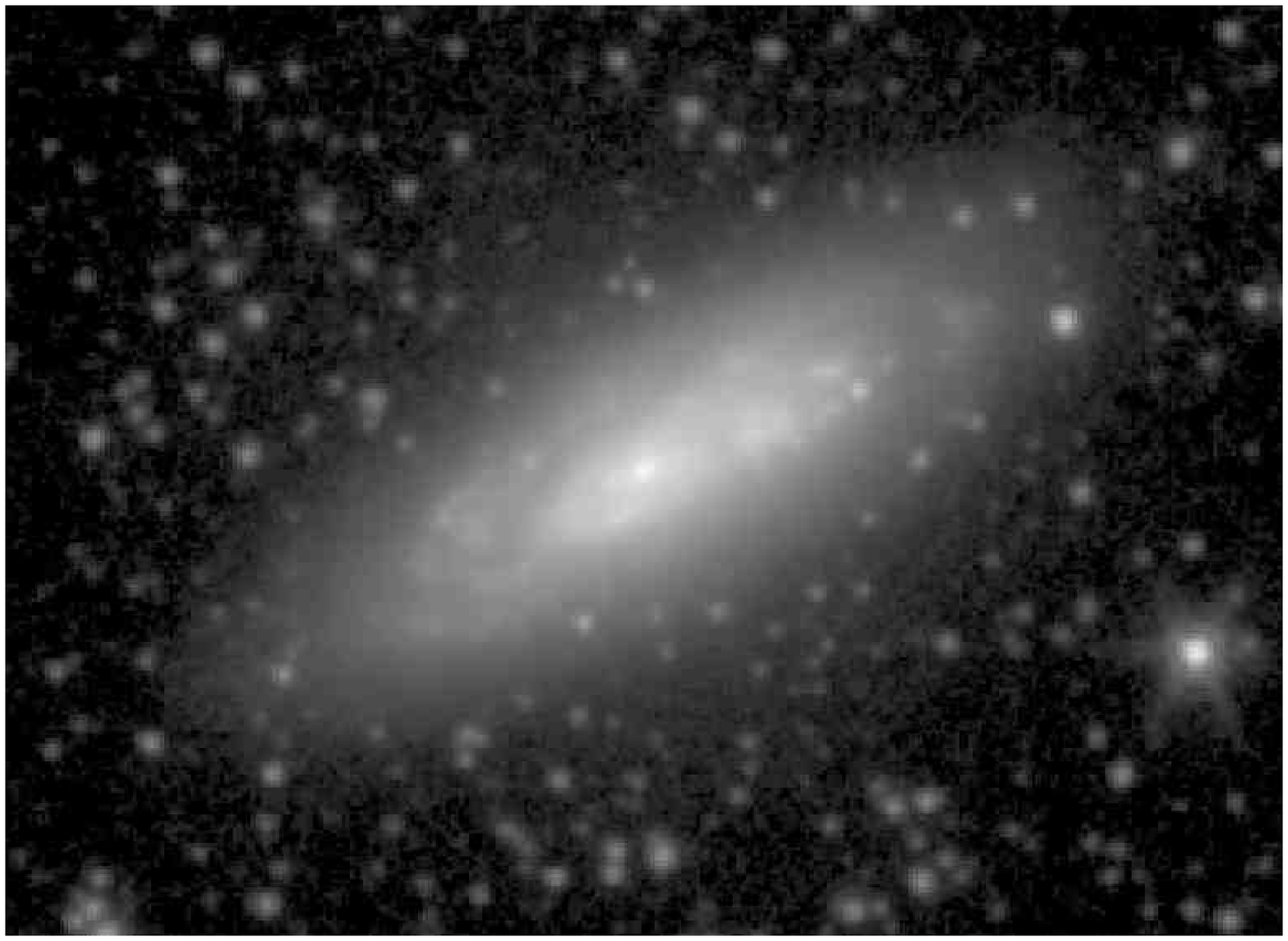}
 \vspace{2.0truecm}
 \caption{
{\bf NGC  3437   }              - S$^4$G mid-IR classification:    SA(rs)c                                               ; Filter: IRAC 3.6$\mu$m; North:   up, East: left; Field dimensions:   3.5$\times$  2.6 arcmin; Surface brightness range displayed: 13.5$-$28.0 mag arcsec$^{-2}$}                 
\label{NGC3437}     
 \end{figure}
 
\clearpage
\begin{figure}
\figurenum{1.75} 
\plotone{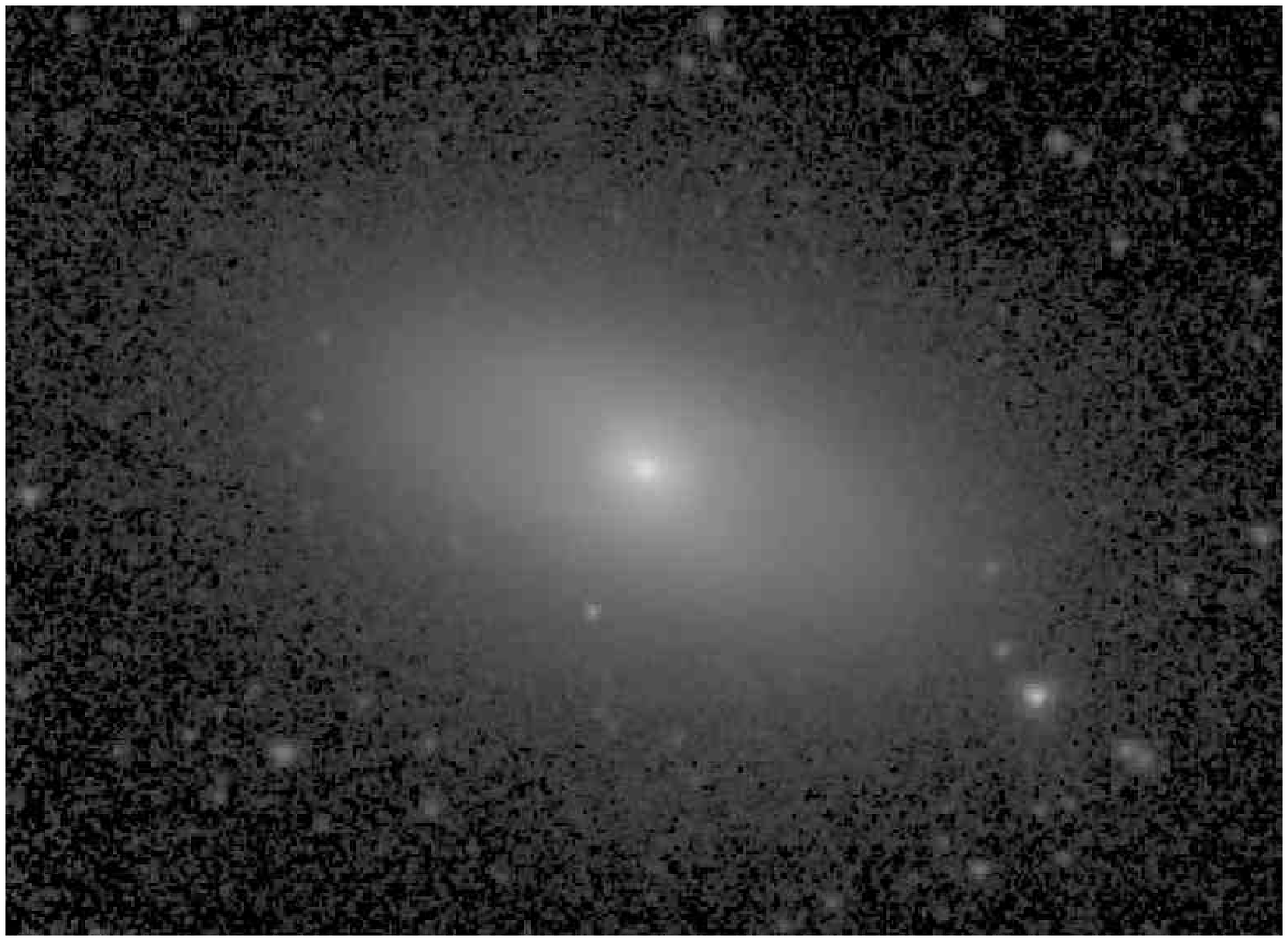}
 \vspace{2.0truecm}
 \caption{
{\bf NGC  3489   }              - S$^4$G mid-IR classification:    (R)SB(r:)0$^o$                                        ; Filter: IRAC 3.6$\mu$m; North:   up, East: left; Field dimensions:   4.5$\times$  3.3 arcmin; Surface brightness range displayed: 11.5$-$28.0 mag arcsec$^{-2}$}                 
\label{NGC3489}     
 \end{figure}
 
\clearpage
\begin{figure}
\figurenum{1.76} 
\plotone{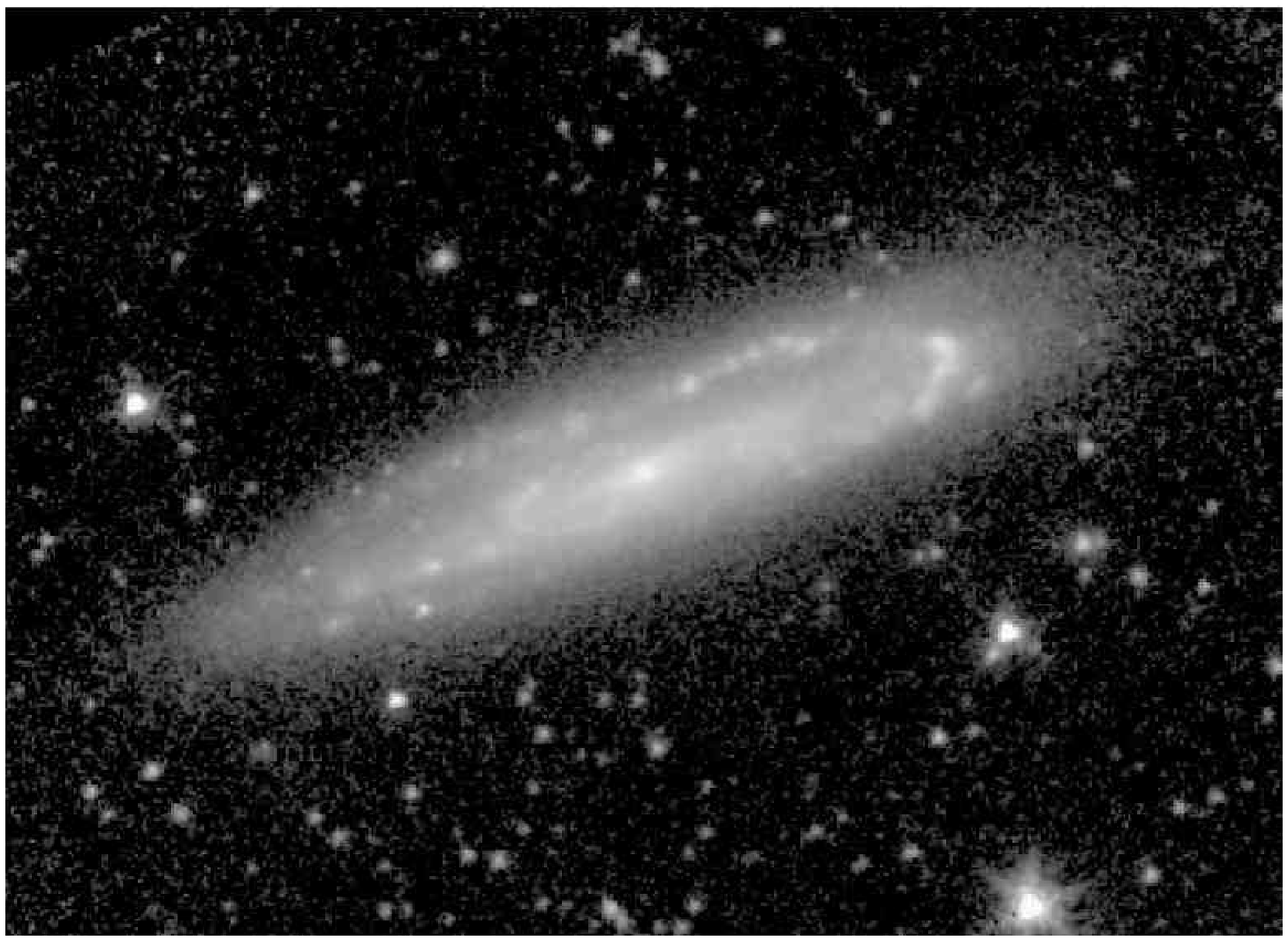}
 \vspace{2.0truecm}
 \caption{
{\bf NGC  3495   }              - S$^4$G mid-IR classification:    (R$^{\prime}$)SB(rs)c:                                          ; Filter: IRAC 3.6$\mu$m; North: left, East: down; Field dimensions:   5.3$\times$  3.8 arcmin; Surface brightness range displayed: 16.0$-$28.0 mag arcsec$^{-2}$}                 
\label{NGC3495}     
 \end{figure}
 
\clearpage
\begin{figure}
\figurenum{1.77} 
\plotone{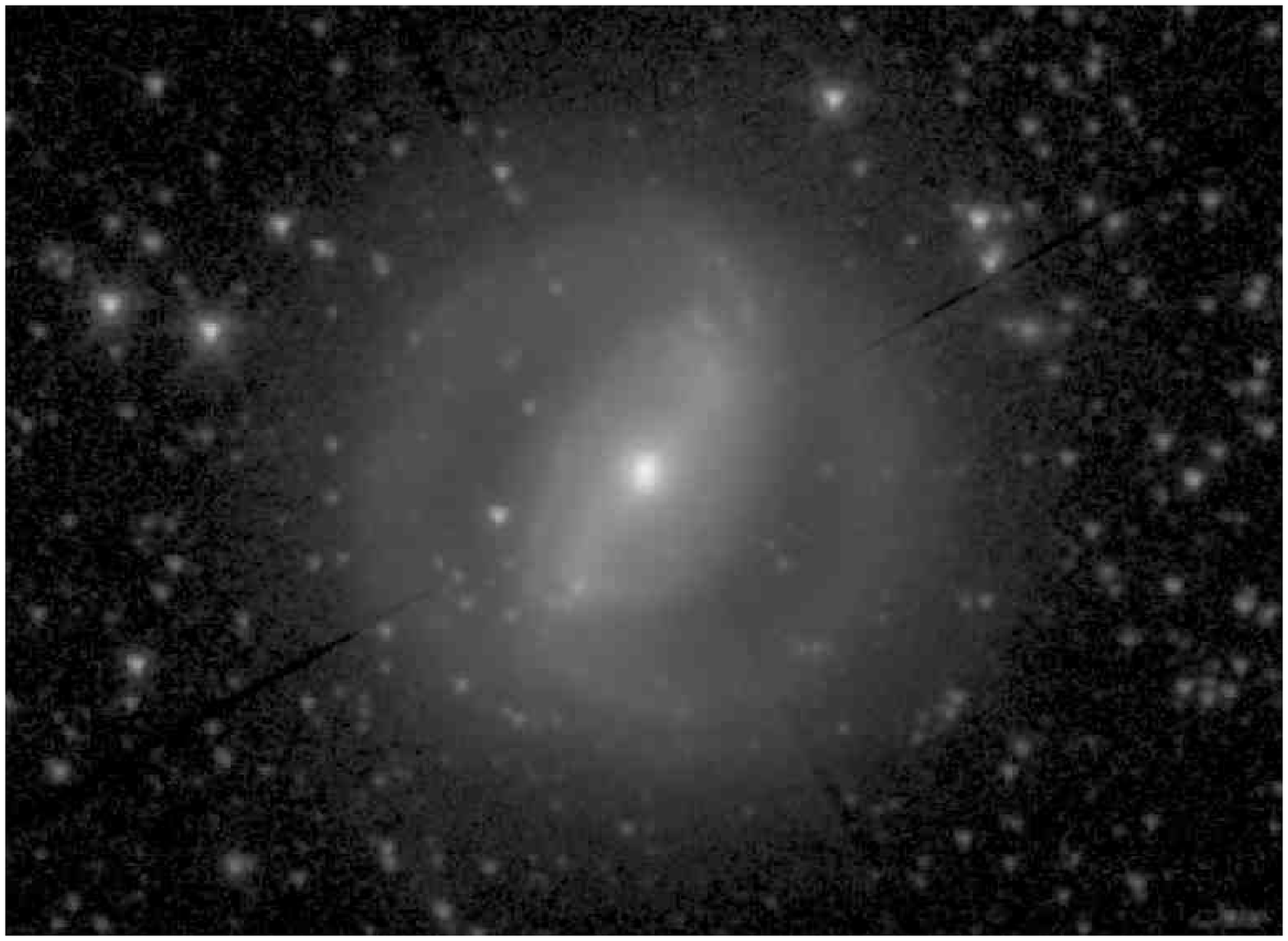}
 \vspace{2.0truecm}
 \caption{
{\bf NGC  3504   }              - S$^4$G mid-IR classification:    (R$_1^{\prime}$)SA$\underline{\rm B}$($\underline{\rm r}$s,nl)a                  ; Filter: IRAC 3.6$\mu$m; North:   up, East: left; Field dimensions:   5.3$\times$  3.8 arcmin; Surface brightness range displayed: 12.0$-$28.0 mag arcsec$^{-2}$}
\label{NGC3504}     
 \end{figure}
 
\clearpage
\begin{figure}
\figurenum{1.78} 
\plotone{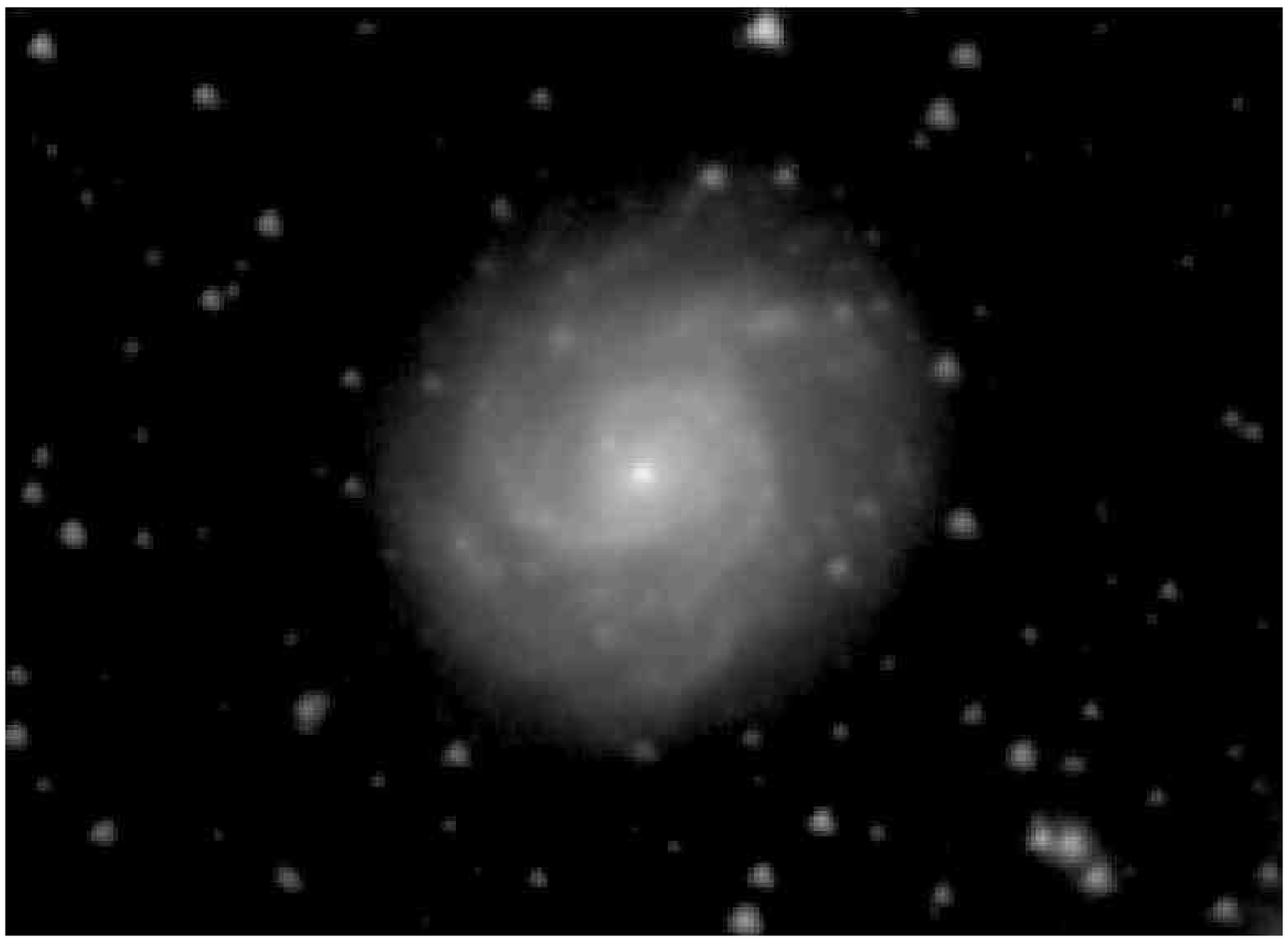}
 \vspace{2.0truecm}
 \caption{
{\bf NGC  3512   }              - S$^4$G mid-IR classification:    SA(s)bc                                               ; Filter: IRAC 3.6$\mu$m; North:   up, East: left; Field dimensions:   2.9$\times$  2.1 arcmin; Surface brightness range displayed: 14.5$-$28.0 mag arcsec$^{-2}$}                 
\label{NGC3512}     
 \end{figure}
 
\clearpage
\begin{figure}
\figurenum{1.79} 
\plotone{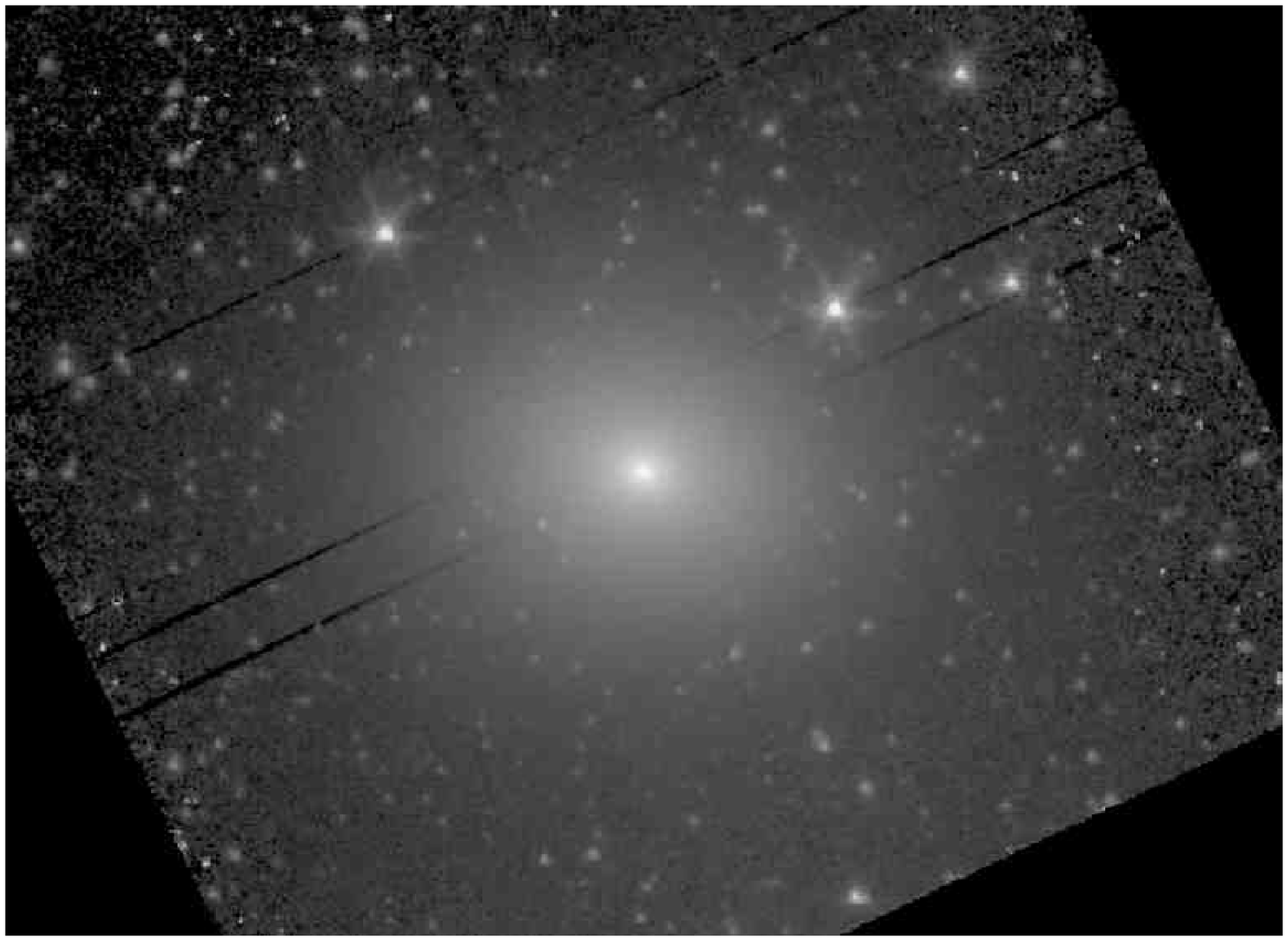}
 \vspace{2.0truecm}
 \caption{
{\bf NGC  3608   }              - S$^4$G mid-IR classification:    E2                                                    ; Filter: IRAC 3.6$\mu$m; North:   up, East: left; Field dimensions:   7.0$\times$  5.1 arcmin; Surface brightness range displayed: 13.0$-$26.0 mag arcsec$^{-2}$}                 
\label{NGC3608}     
 \end{figure}
 
\clearpage
\begin{figure}
\figurenum{1.80} 
\plotone{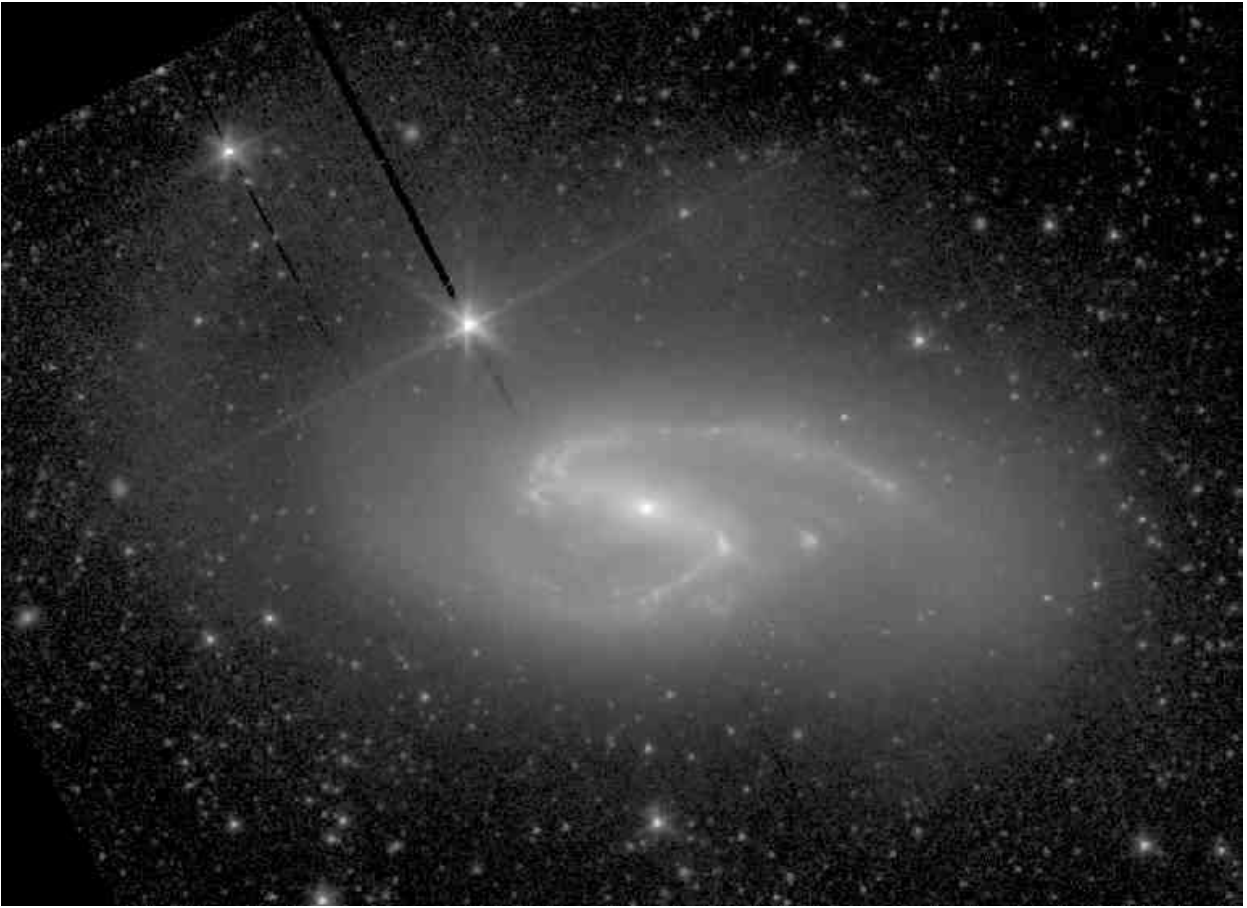}
 \vspace{2.0truecm}
 \caption{
{\bf NGC  3627   }              - S$^4$G mid-IR classification:    SB(s)b pec                                            ; Filter: IRAC 3.6$\mu$m; North: left, East: down; Field dimensions:  12.6$\times$  9.2 arcmin; Surface brightness range displayed: 12.0$-$28.0 mag arcsec$^{-2}$}                 
\label{NGC3627}     
 \end{figure}
 
\clearpage
\begin{figure}
\figurenum{1.81} 
\plotone{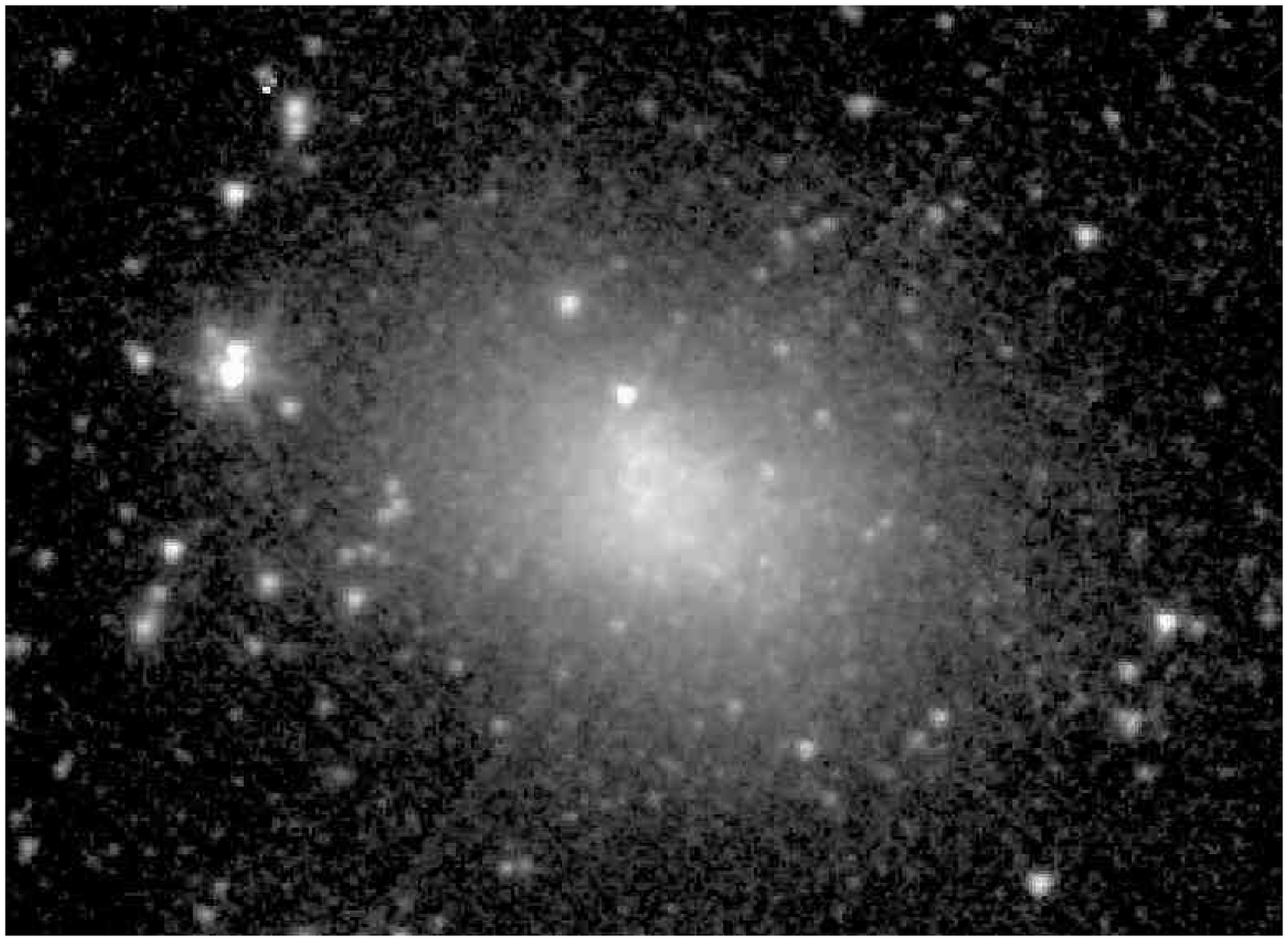}
 \vspace{2.0truecm}
 \caption{
{\bf NGC  3738   }              - S$^4$G mid-IR classification:    dE (Im)                                               ; Filter: IRAC 3.6$\mu$m; North: left, East: down; Field dimensions:   4.5$\times$  3.3 arcmin; Surface brightness range displayed: 16.5$-$26.0 mag arcsec$^{-2}$}                 
\label{NGC3738}     
 \end{figure}
 
\clearpage
\begin{figure}
\figurenum{1.82} 
\plotone{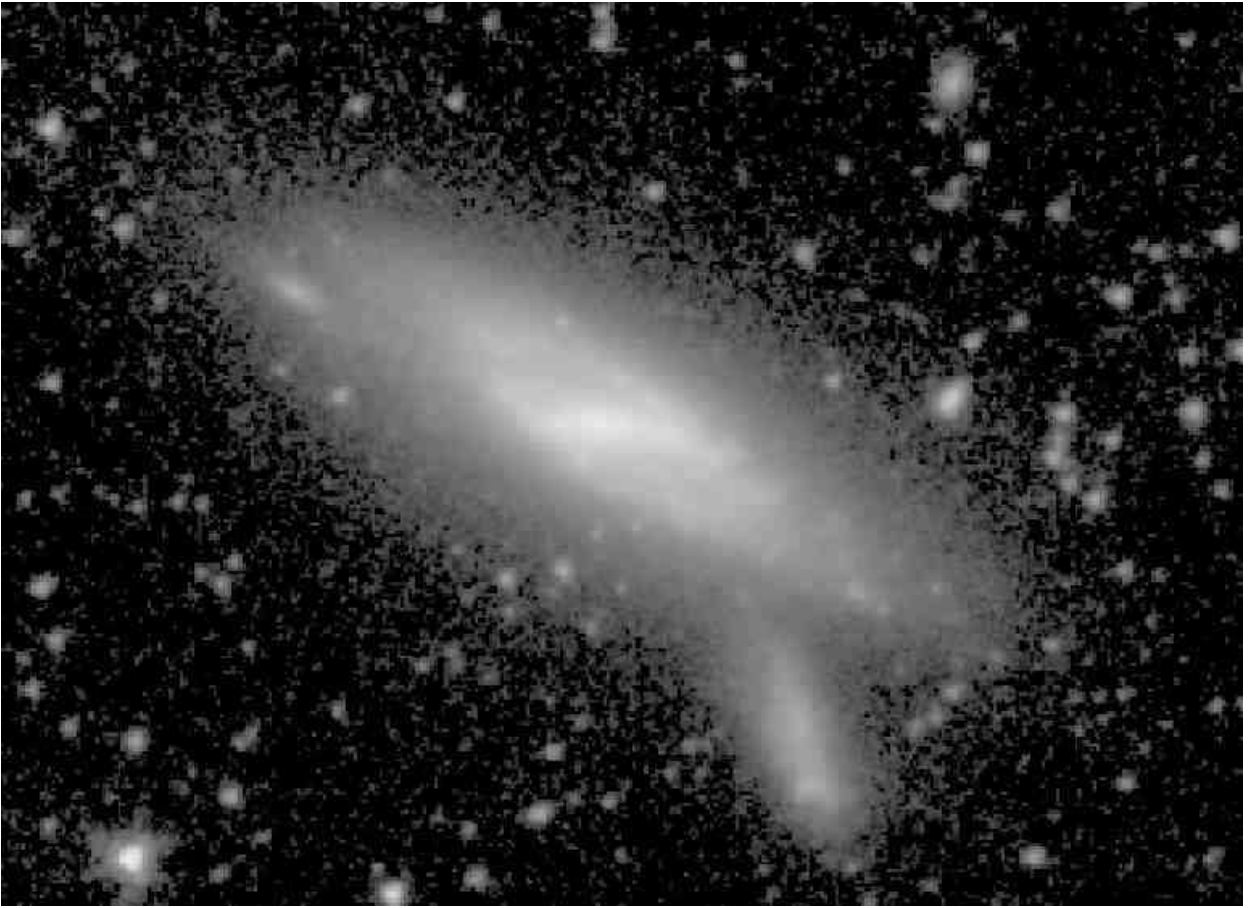}
 \vspace{2.0truecm}
 \caption{
{\bf NGC  3769} (center) and {\bf MCG8-21-76  } (lower right) - S$^4$G mid-IR classifications:    (R$^{\prime}$)SB(s)cd, IBm:, respectively; Filter: IRAC 3.6$\mu$m; North: left, East: down; Field dimensions:   4.2$\times$  3.1 arcmin; Surface brightness range displayed: 15.5$-$26.0 mag arcsec$^{-2}$}                 
\label{NGC3769}     
 \end{figure}
 
\clearpage
\begin{figure}
\figurenum{1.83} 
\plotone{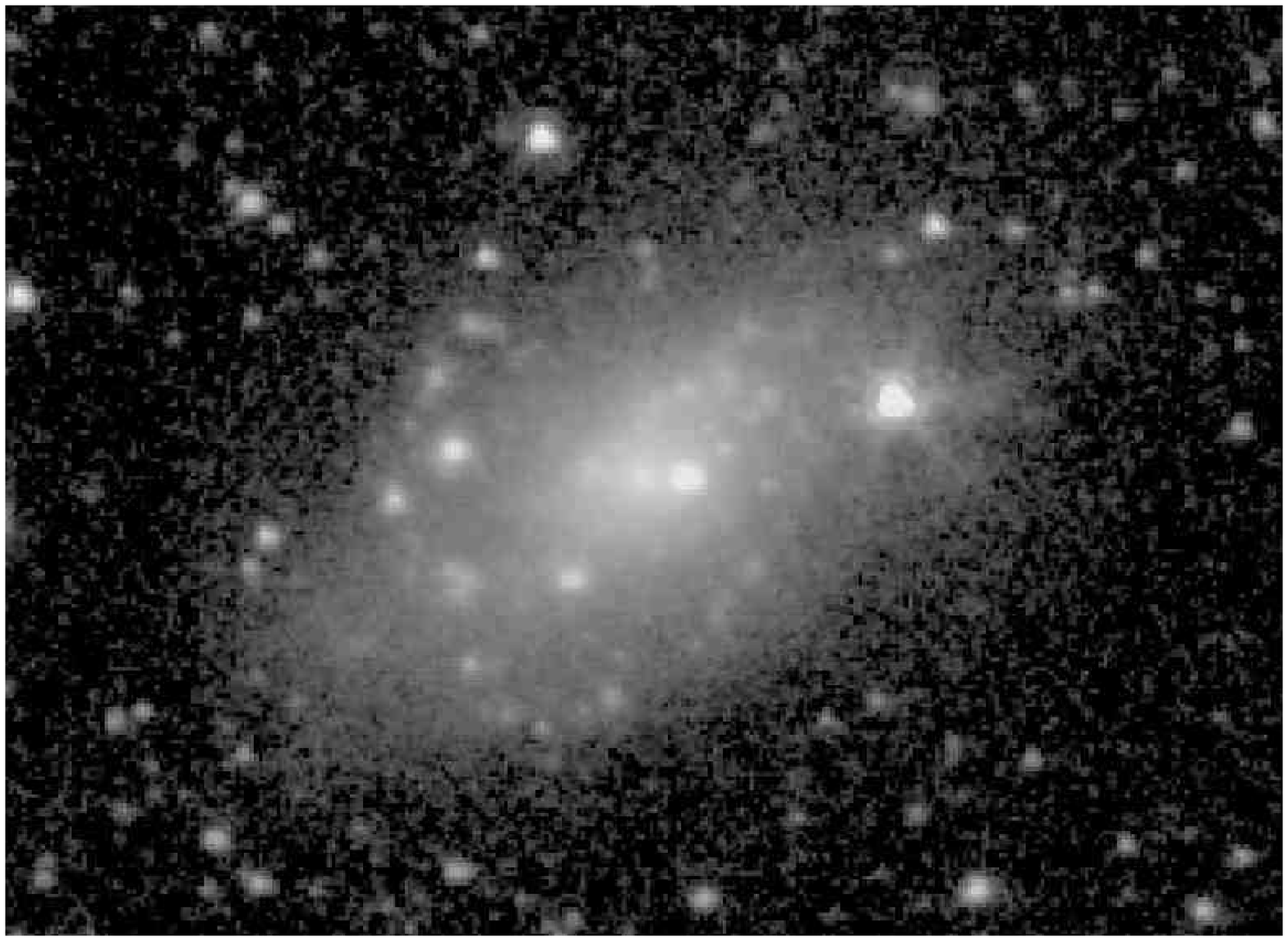}
 \vspace{2.0truecm}
 \caption{
{\bf NGC  3794   }              - S$^4$G mid-IR classification:    S$\underline{\rm A}$B(s)dm                            ; Filter: IRAC 3.6$\mu$m; North:   up, East: left; Field dimensions:   3.5$\times$  2.6 arcmin; Surface brightness range displayed: 17.0$-$26.0 mag arcsec$^{-2}$}                 
\label{NGC3794}     
 \end{figure}
 
\clearpage
\begin{figure}
\figurenum{1.84} 
\plotone{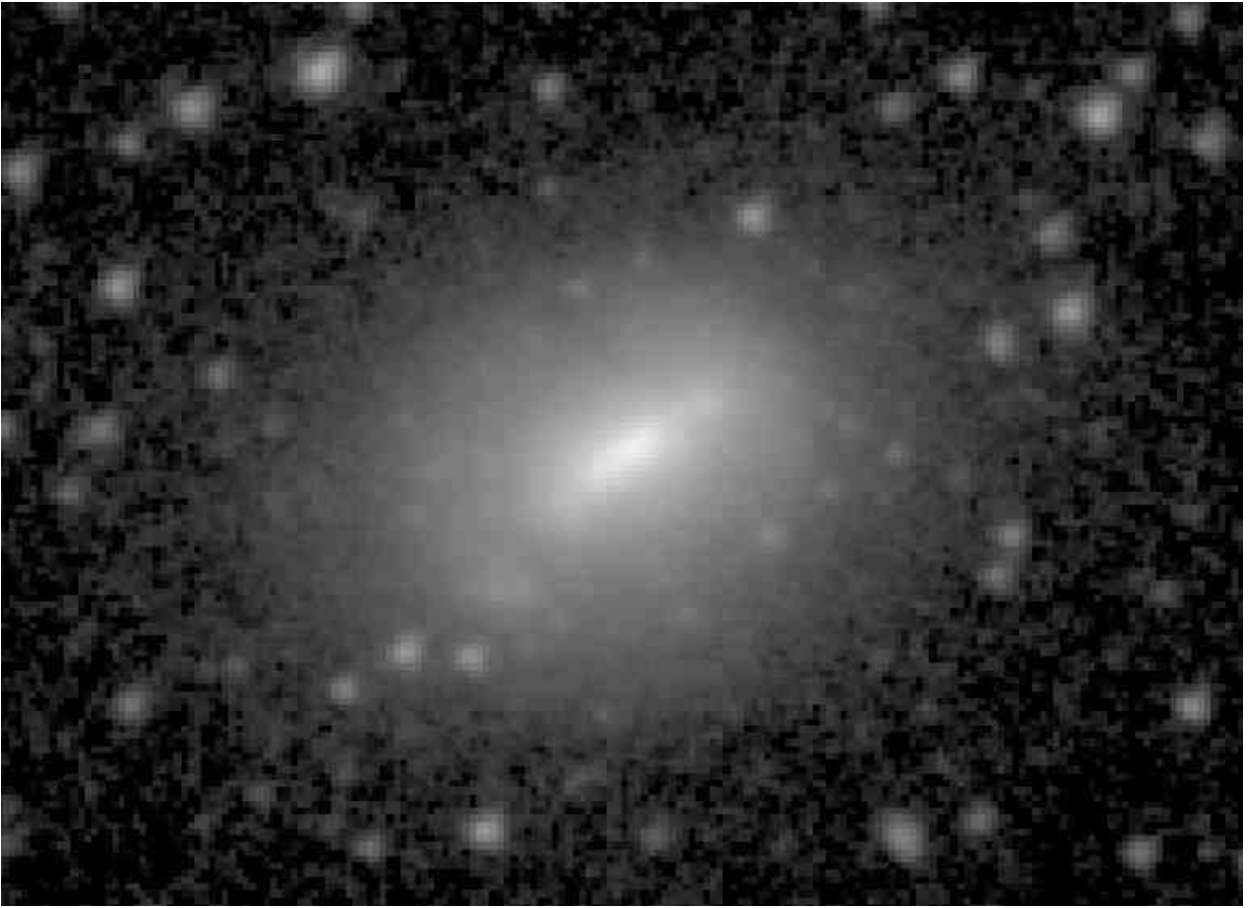}
 \vspace{2.0truecm}
 \caption{
{\bf NGC  3870   }              - S$^4$G mid-IR classification:    SB(rs)0$^o$?                                          ; Filter: IRAC 3.6$\mu$m; North: left, East: down; Field dimensions:   4.5$\times$  3.3 arcmin; Surface brightness range displayed: 15.0$-$28.0 mag arcsec$^{-2}$}                 
\label{NGC3870}     
 \end{figure}
 
\clearpage
\begin{figure}
\figurenum{1.85} 
\plotone{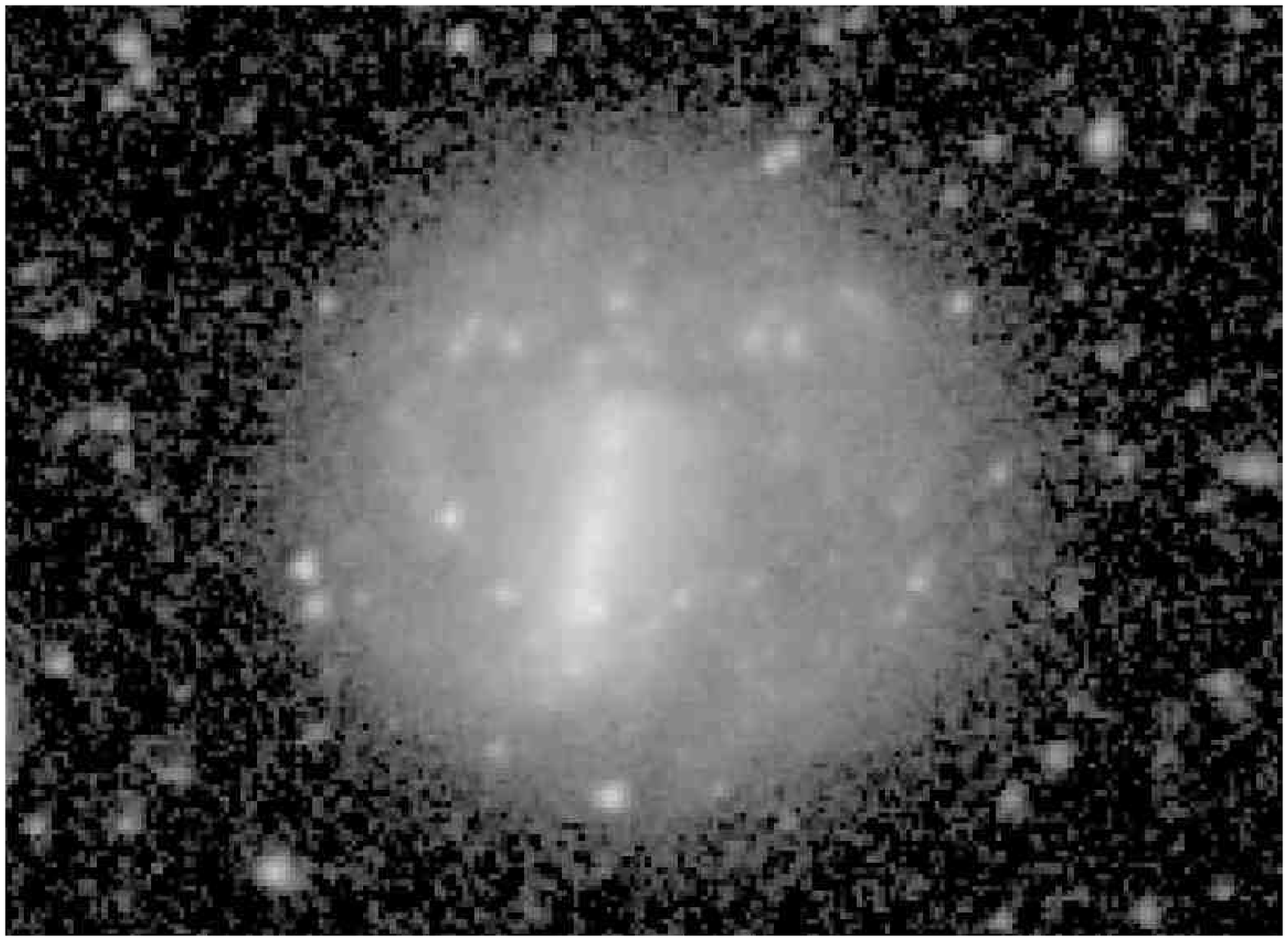}
 \vspace{2.0truecm}
 \caption{
{\bf NGC  3906   }              - S$^4$G mid-IR classification:    SB(l)dm                                               ; Filter: IRAC 3.6$\mu$m; North: left, East: down; Field dimensions:   2.9$\times$  2.1 arcmin; Surface brightness range displayed: 17.0$-$28.0 mag arcsec$^{-2}$}                 
\label{NGC3906}     
 \end{figure}
 
\clearpage
\begin{figure}
\figurenum{1.86} 
\plotone{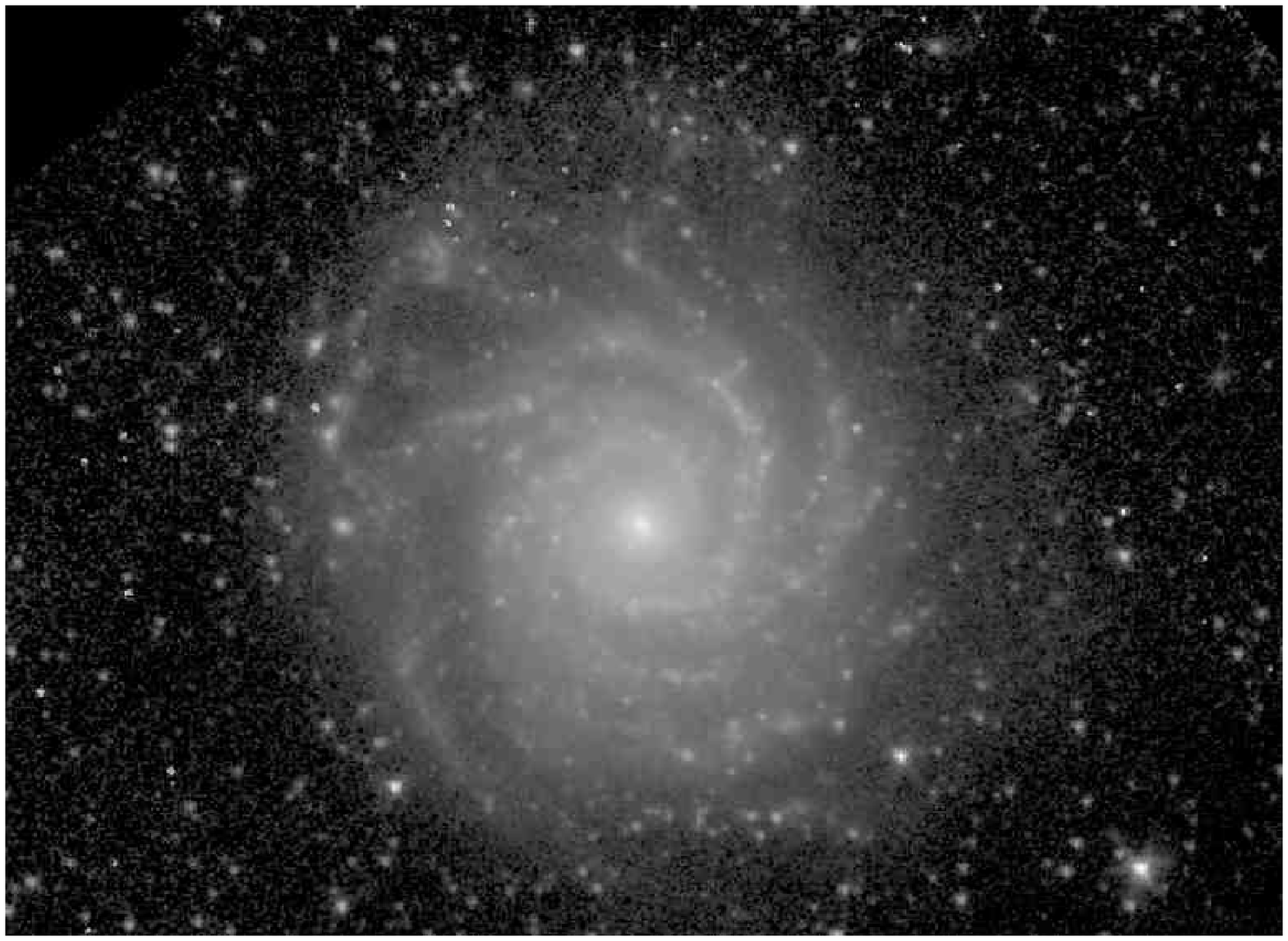}
 \vspace{2.0truecm}
 \caption{
{\bf NGC  3938   }              - S$^4$G mid-IR classification:    SA(s)c                                                ; Filter: IRAC 3.6$\mu$m; North:   up, East: left; Field dimensions:   7.9$\times$  5.8 arcmin; Surface brightness range displayed: 14.5$-$28.0 mag arcsec$^{-2}$}                 
\label{NGC3938}     
 \end{figure}
 
\clearpage
\begin{figure}
\figurenum{1.87} 
\plotone{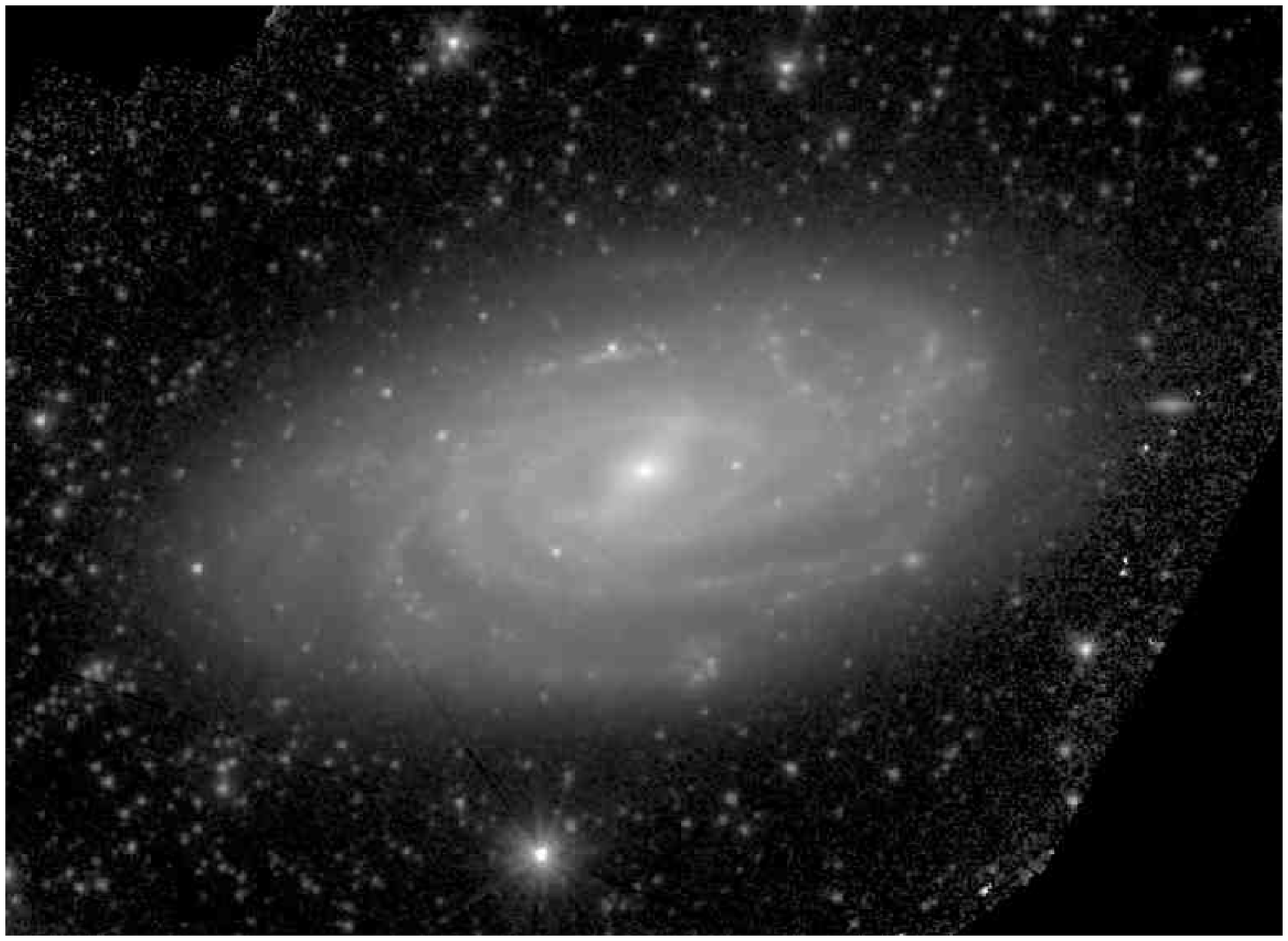}
 \vspace{2.0truecm}
 \caption{
{\bf NGC  3953   }              - S$^4$G mid-IR classification:    SB(r)$\underline{\rm b}$c                             ; Filter: IRAC 3.6$\mu$m; North: left, East: down; Field dimensions:   9.0$\times$  6.6 arcmin; Surface brightness range displayed: 13.0$-$28.0 mag arcsec$^{-2}$}                 
\label{NGC3953}     
 \end{figure}
 
\clearpage
\begin{figure}
\figurenum{1.88} 
\plotone{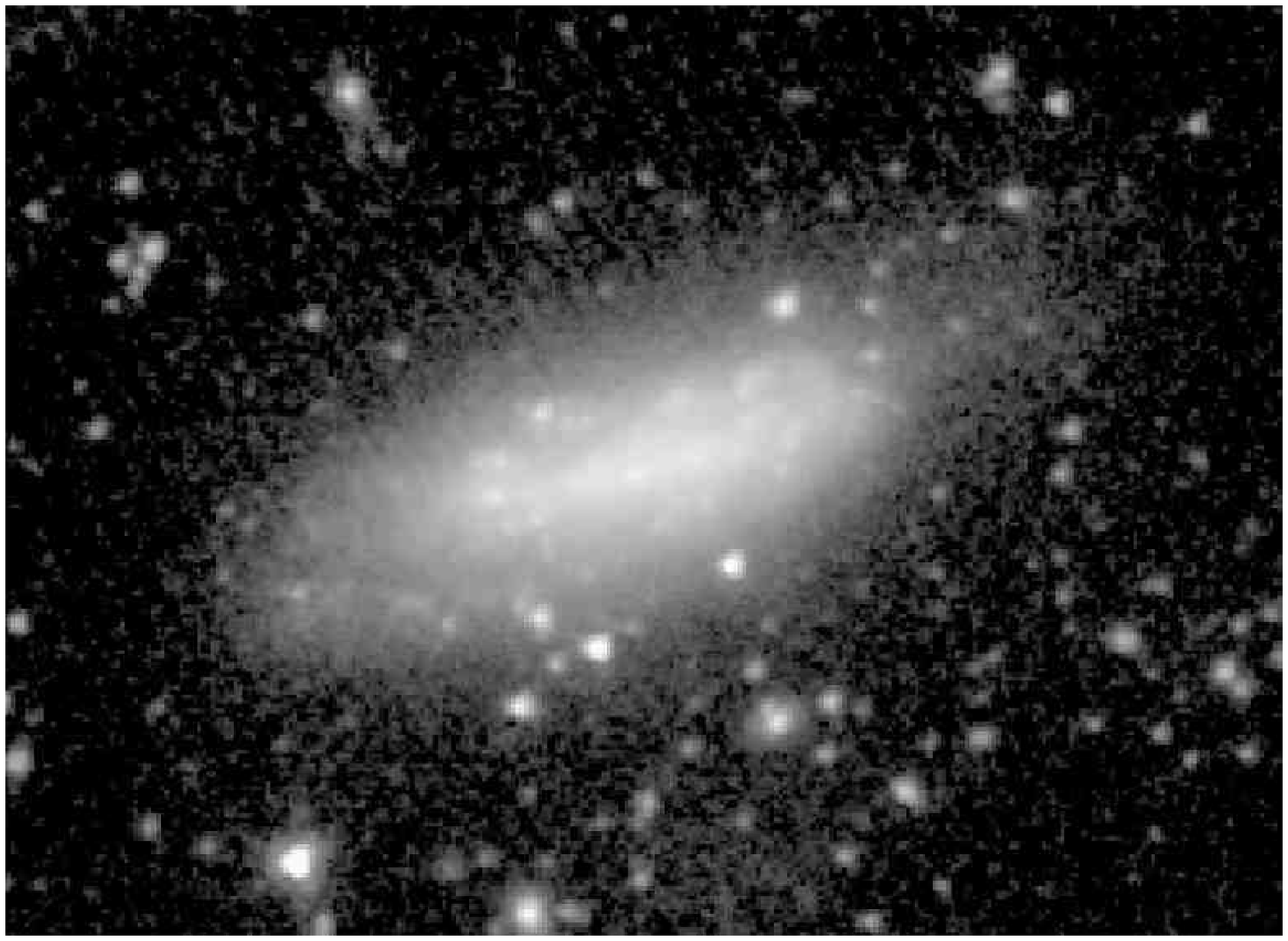}
 \vspace{2.0truecm}
 \caption{
{\bf NGC  4020   }              - S$^4$G mid-IR classification:    SAB(s)d                                               ; Filter: IRAC 3.6$\mu$m; North: left, East: down; Field dimensions:   3.5$\times$  2.6 arcmin; Surface brightness range displayed: 17.0$-$28.0 mag arcsec$^{-2}$}                 
\label{NGC4020}     
 \end{figure}
 
\clearpage
\begin{figure}
\figurenum{1.89} 
\plotone{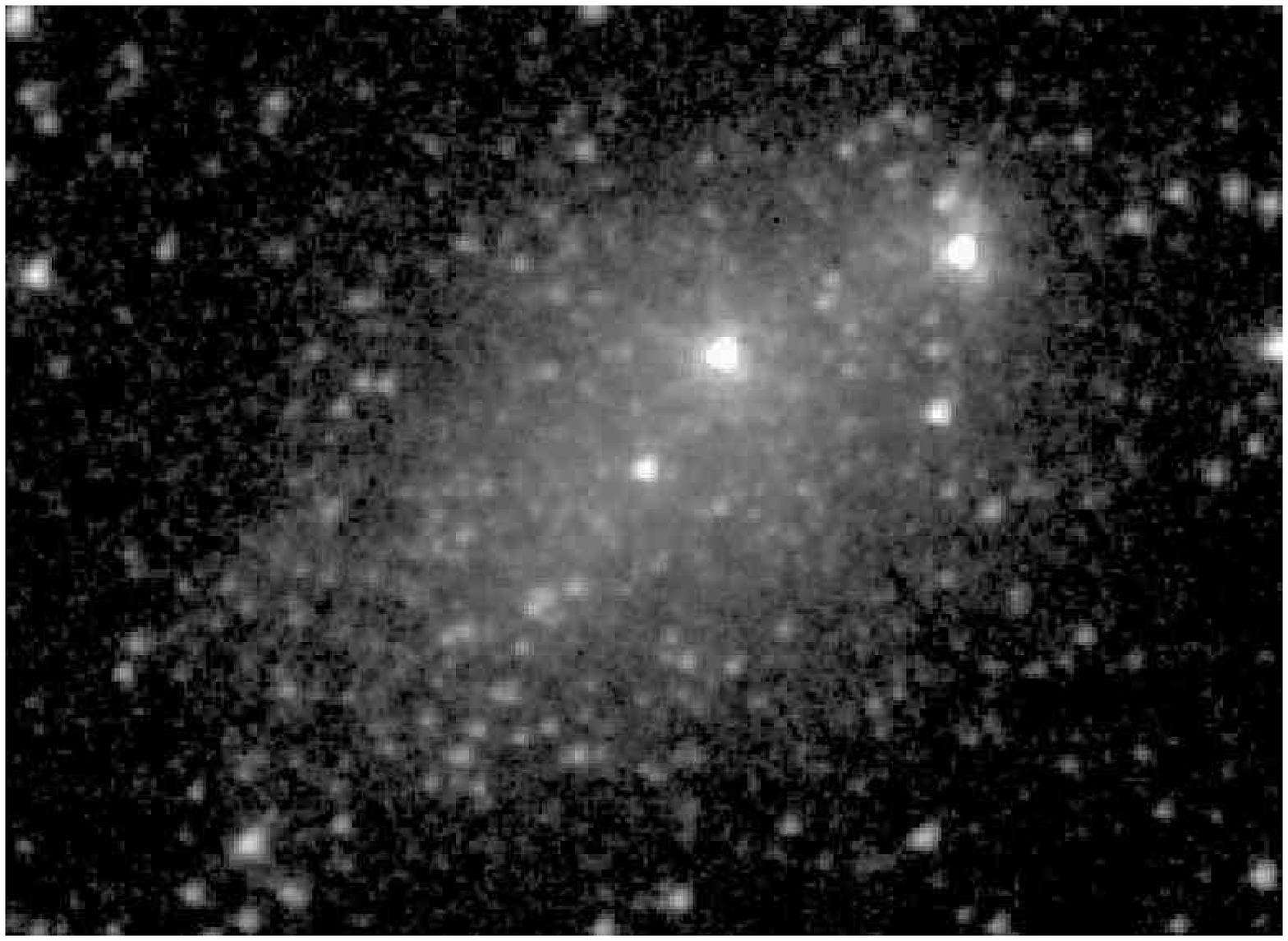}
 \vspace{2.0truecm}
 \caption{
{\bf NGC  4068   }              - S$^4$G mid-IR classification:    Im                                                    ; Filter: IRAC 3.6$\mu$m; North:   up, East: left; Field dimensions:   2.7$\times$  2.0 arcmin; Surface brightness range displayed: 16.5$-$28.0 mag arcsec$^{-2}$}                 
\label{NGC4068}     
 \end{figure}
 
\clearpage
\begin{figure}
\figurenum{1.90} 
\plotone{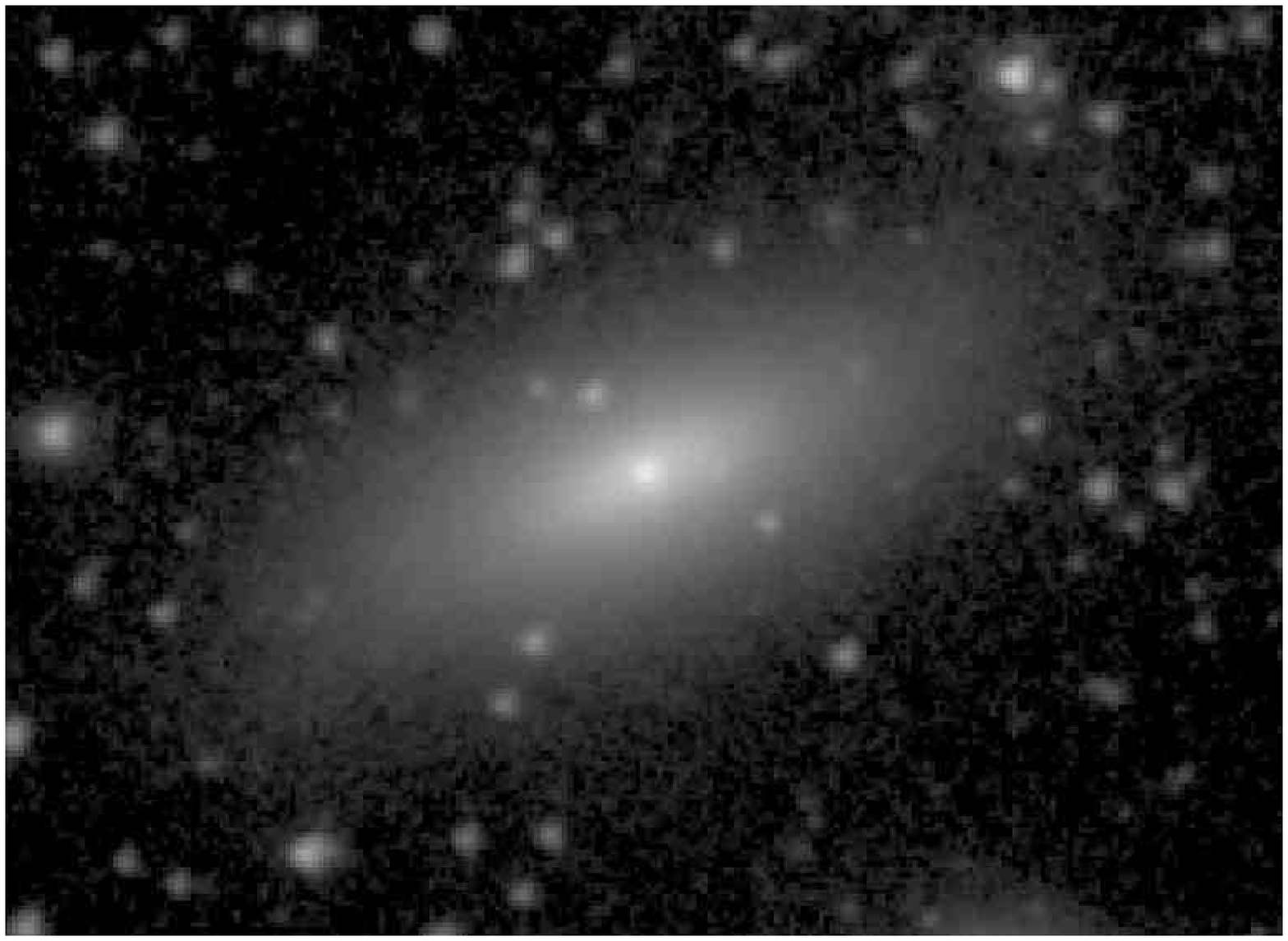}
 \vspace{2.0truecm}
 \caption{
{\bf NGC  4117   }              - S$^4$G mid-IR classification:    S0$^-$ sp                                             ; Filter: IRAC 3.6$\mu$m; North: left, East: down; Field dimensions:   2.9$\times$  2.1 arcmin; Surface brightness range displayed: 14.0$-$28.0 mag arcsec$^{-2}$}                 
\label{NGC4117}     
 \end{figure}
 
\clearpage
\begin{figure}
\figurenum{1.91} 
\plotone{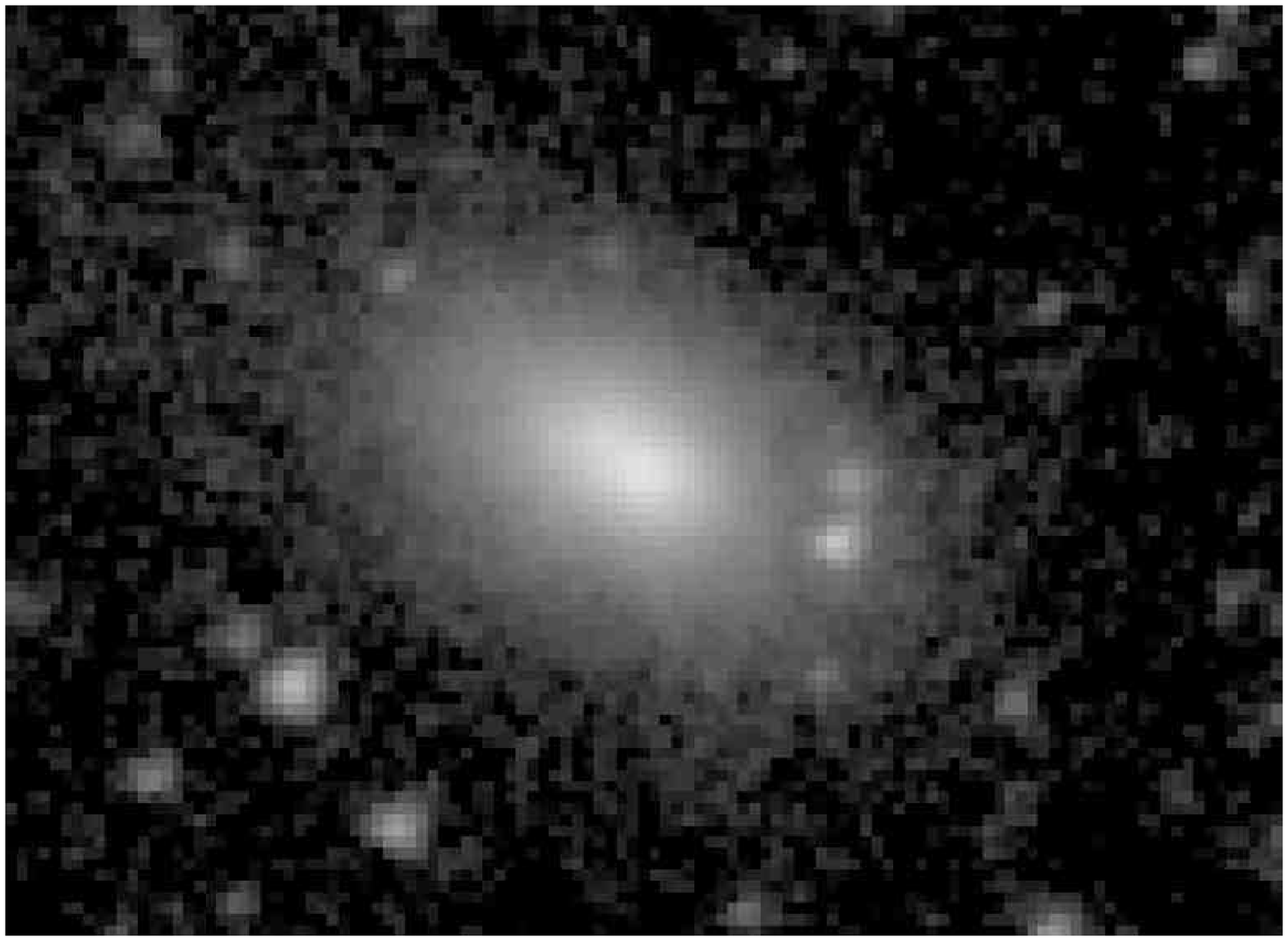}
 \vspace{2.0truecm}
 \caption{
{\bf NGC  4118   }              - S$^4$G mid-IR classification:    dE3-4:                                                ; Filter: IRAC 3.6$\mu$m; North: left, East: down; Field dimensions:   1.4$\times$  1.0 arcmin; Surface brightness range displayed: 17.0$-$28.0 mag arcsec$^{-2}$}                 
\label{NGC4118}     
 \end{figure}
 
\clearpage
\begin{figure}
\figurenum{1.92} 
\plotone{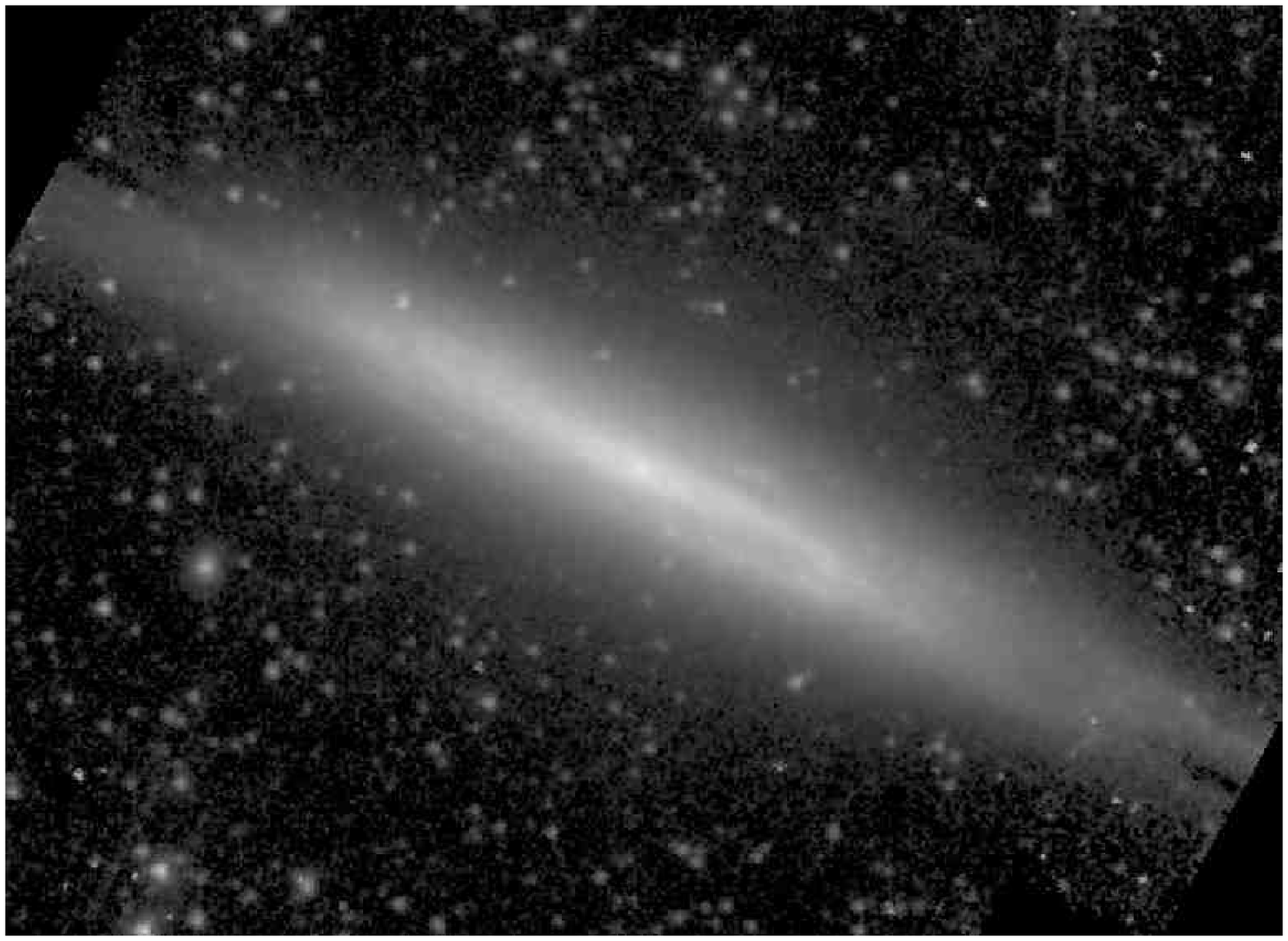}
 \vspace{2.0truecm}
 \caption{
{\bf NGC  4157   }              - S$^4$G mid-IR classification:    SAB(s)c:                                              ; Filter: IRAC 3.6$\mu$m; North:   up, East: left; Field dimensions:   6.3$\times$  4.6 arcmin; Surface brightness range displayed: 14.0$-$28.0 mag arcsec$^{-2}$}                 
\label{NGC4157}     
 \end{figure}
 
\clearpage
\begin{figure}
\figurenum{1.93} 
\plotone{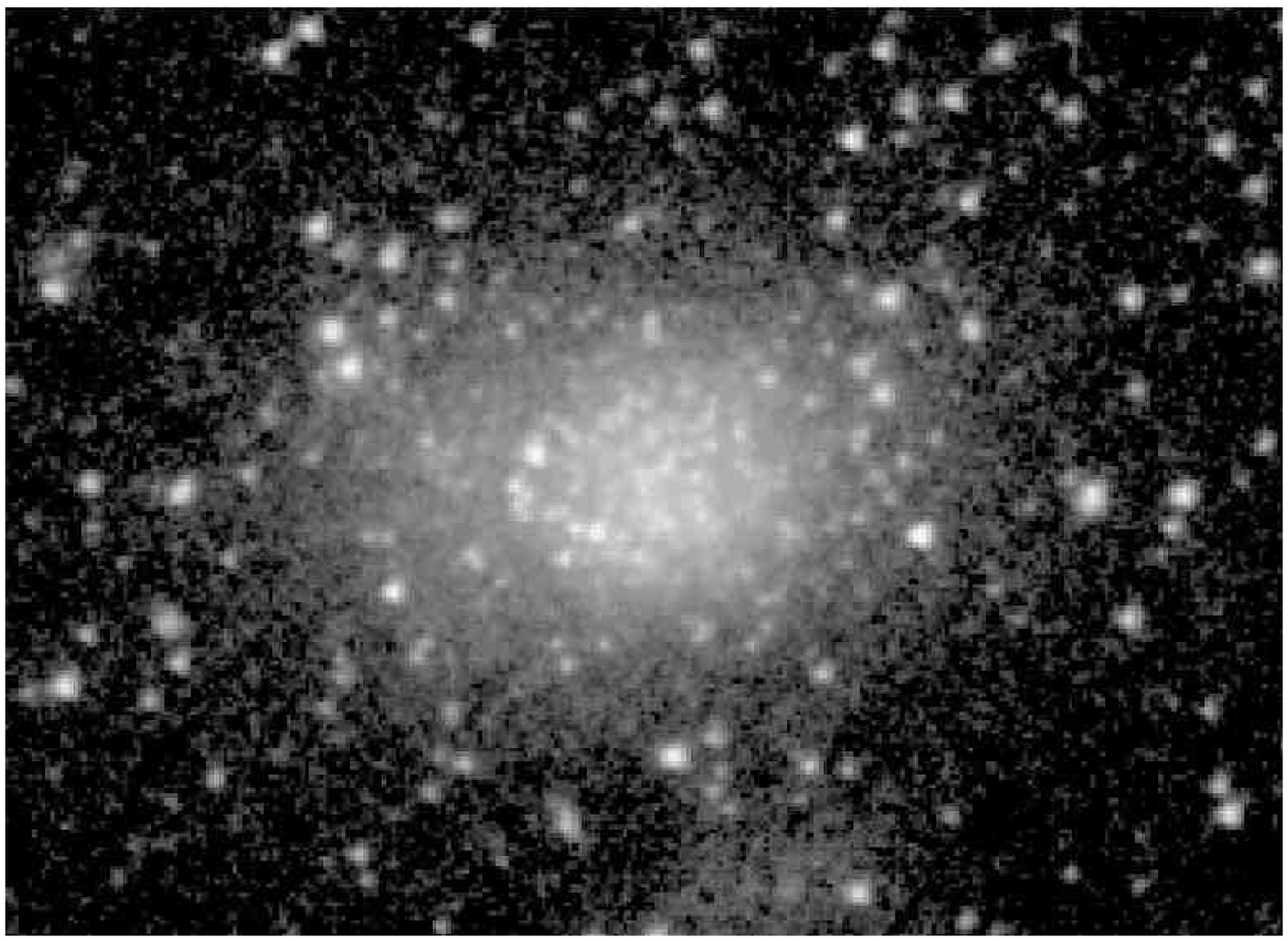}
 \vspace{2.0truecm}
 \caption{
{\bf NGC  4163   }              - S$^4$G mid-IR classification:    dE (Im)                                               ; Filter: IRAC 3.6$\mu$m; North: left, East: down; Field dimensions:   3.5$\times$  2.6 arcmin; Surface brightness range displayed: 18.0$-$28.0 mag arcsec$^{-2}$}                 
\label{NGC4163}     
 \end{figure}
 
\clearpage
\begin{figure}
\figurenum{1.94} 
\plotone{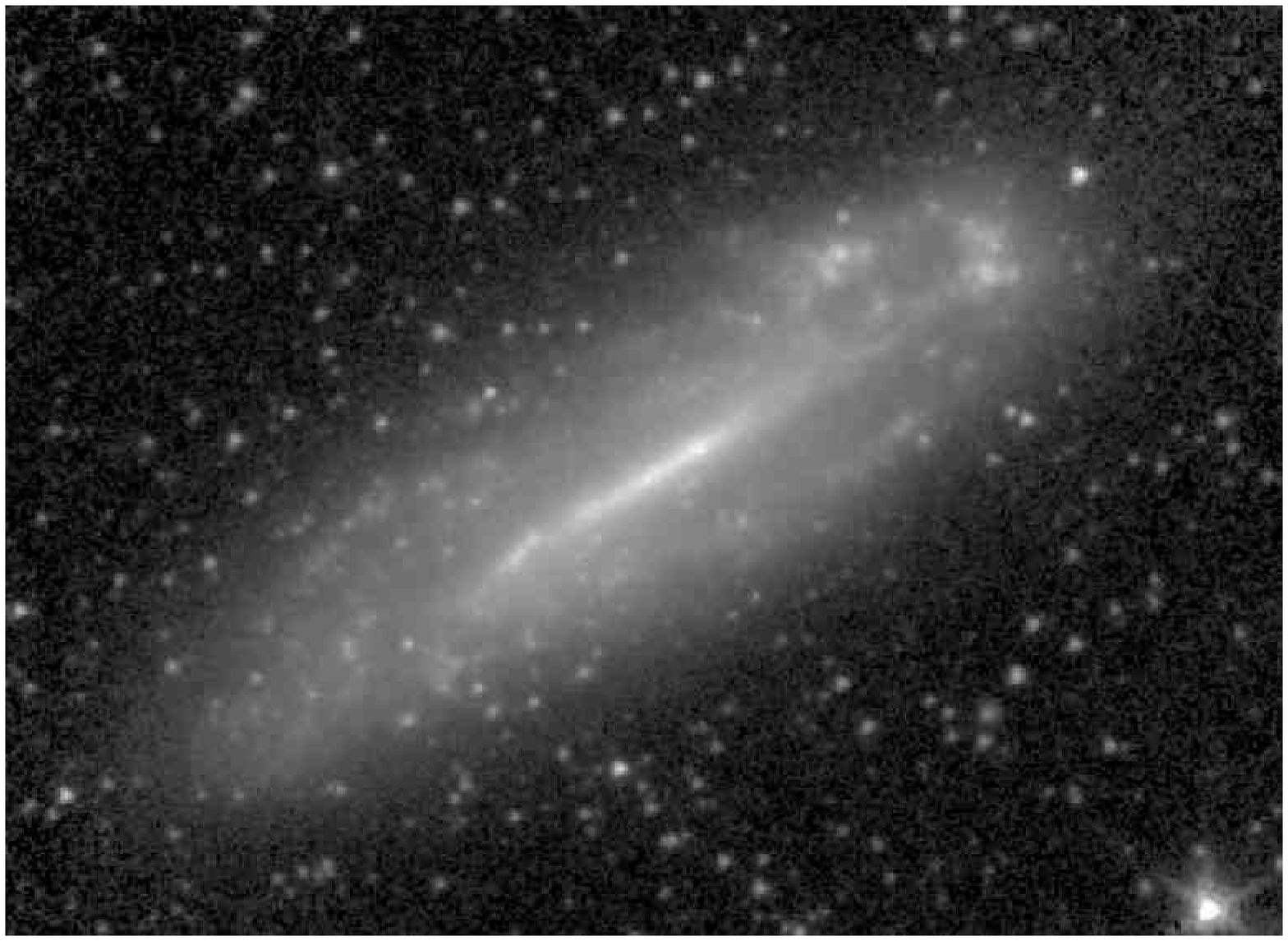}
 \vspace{2.0truecm}
 \caption{
{\bf NGC  4178   }              - S$^4$G mid-IR classification:    (R$_2^{\prime}$)SB(s)d                                        ; Filter: IRAC 3.6$\mu$m; North: left, East: down; Field dimensions:   5.8$\times$  4.2 arcmin; Surface brightness range displayed: 15.5$-$28.0 mag arcsec$^{-2}$}                 
\label{NGC4178}     
 \end{figure}
 
\clearpage
\begin{figure}
\figurenum{1.95} 
\plotone{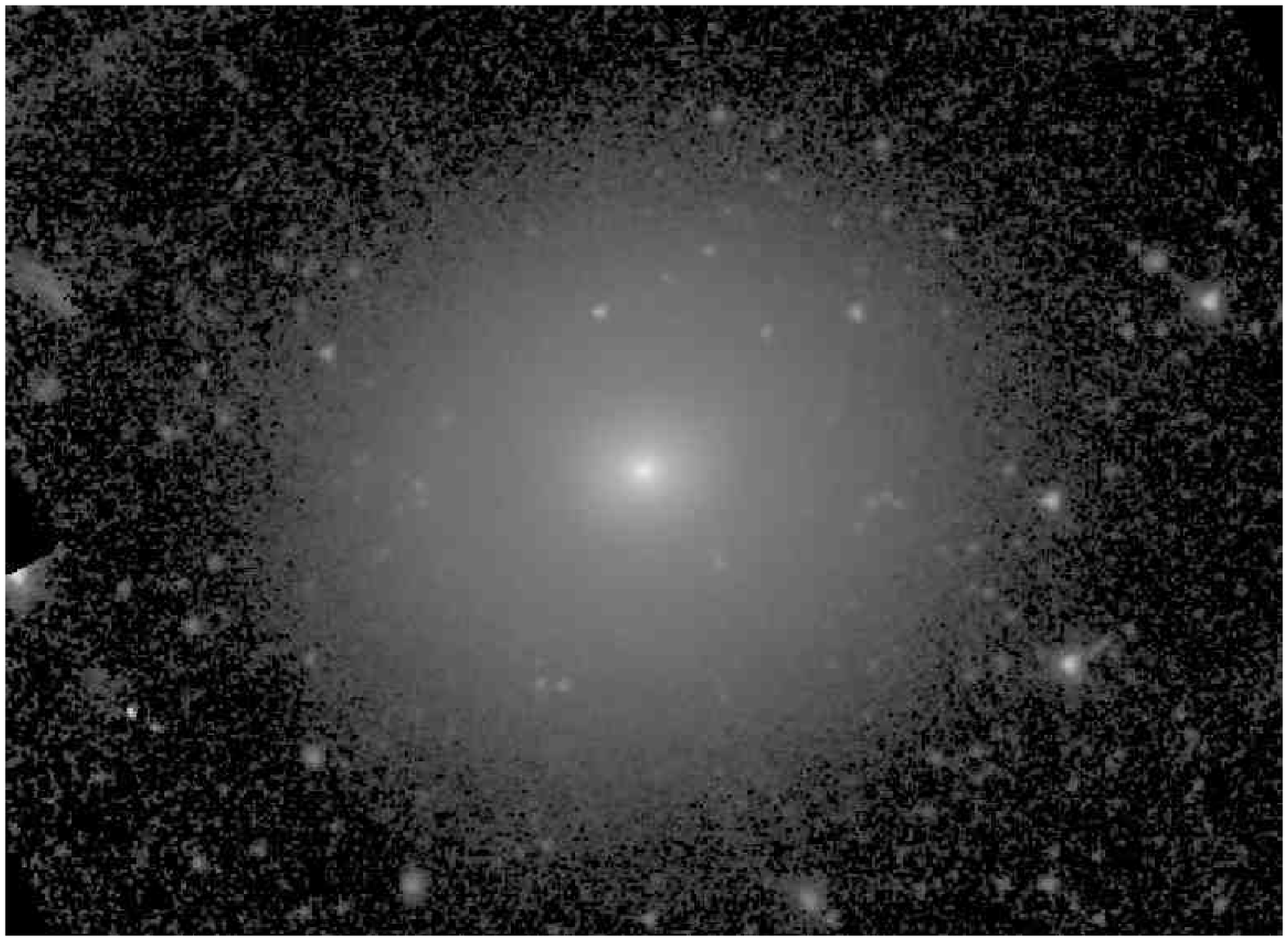}
 \vspace{2.0truecm}
 \caption{
{\bf NGC  4203   }              - S$^4$G mid-IR classification:    S$\underline{\rm A}$B0$^-$                            ; Filter: IRAC 3.6$\mu$m; North: left, East: down; Field dimensions:   5.7$\times$  4.2 arcmin; Surface brightness range displayed: 12.5$-$28.0 mag arcsec$^{-2}$}                 
\label{NGC4203}     
 \end{figure}
 
\clearpage
\begin{figure}
\figurenum{1.96} 
\plotone{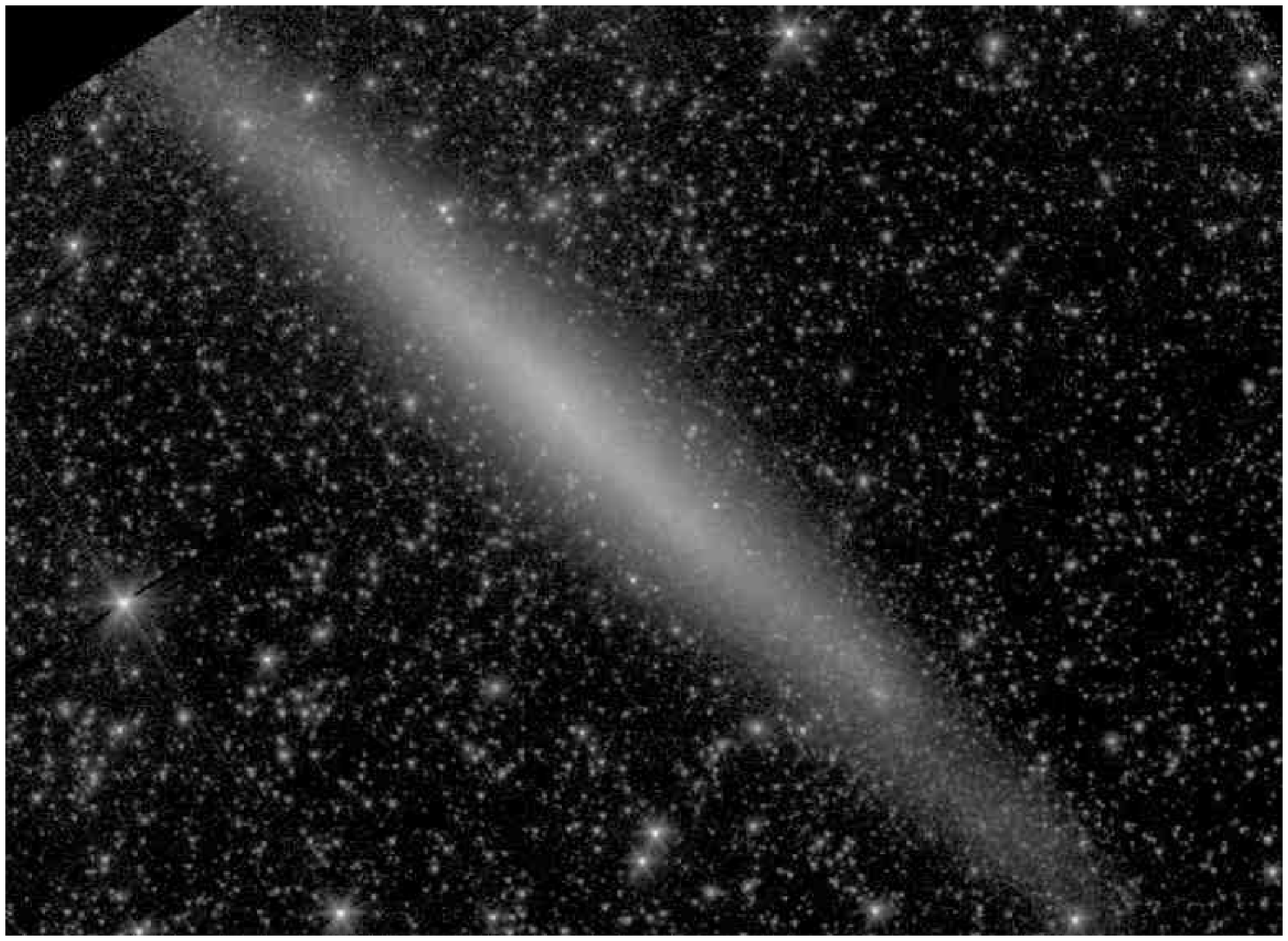}
 \vspace{2.0truecm}
 \caption{
{\bf NGC  4244   }              - S$^4$G mid-IR classification:    Sd sp                                                 ; Filter: IRAC 3.6$\mu$m; North:   up, East: left; Field dimensions:  15.8$\times$ 11.5 arcmin; Surface brightness range displayed: 15.0$-$28.0 mag arcsec$^{-2}$}                 
\label{NGC4244}     
 \end{figure}
 
\clearpage
\begin{figure}
\figurenum{1.97}
\plotone{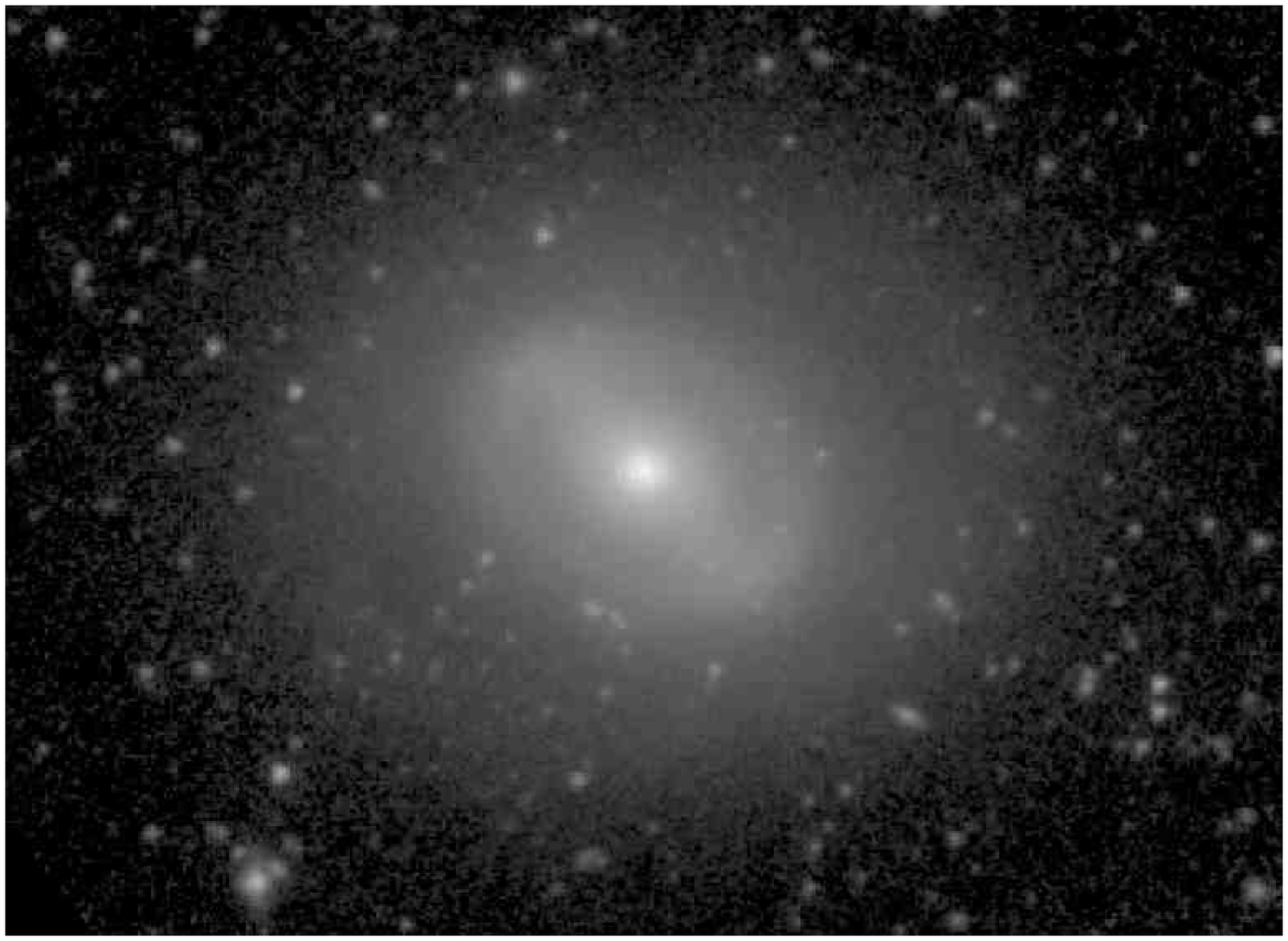}
 \vspace{2.0truecm}
 \caption{
{\bf NGC  4245   }              - S$^4$G mid-IR classification:    (RL)SB(r,nr)0$^+$                                     ; Filter: IRAC 3.6$\mu$m; North: left, East: down; Field dimensions:   4.8$\times$  3.5 arcmin; Surface brightness range displayed: 13.5$-$28.0 mag arcsec$^{-2}$}                 
\label{NGC4245}     
 \end{figure}
 
\clearpage
\begin{figure}
\figurenum{1.98}
\plotone{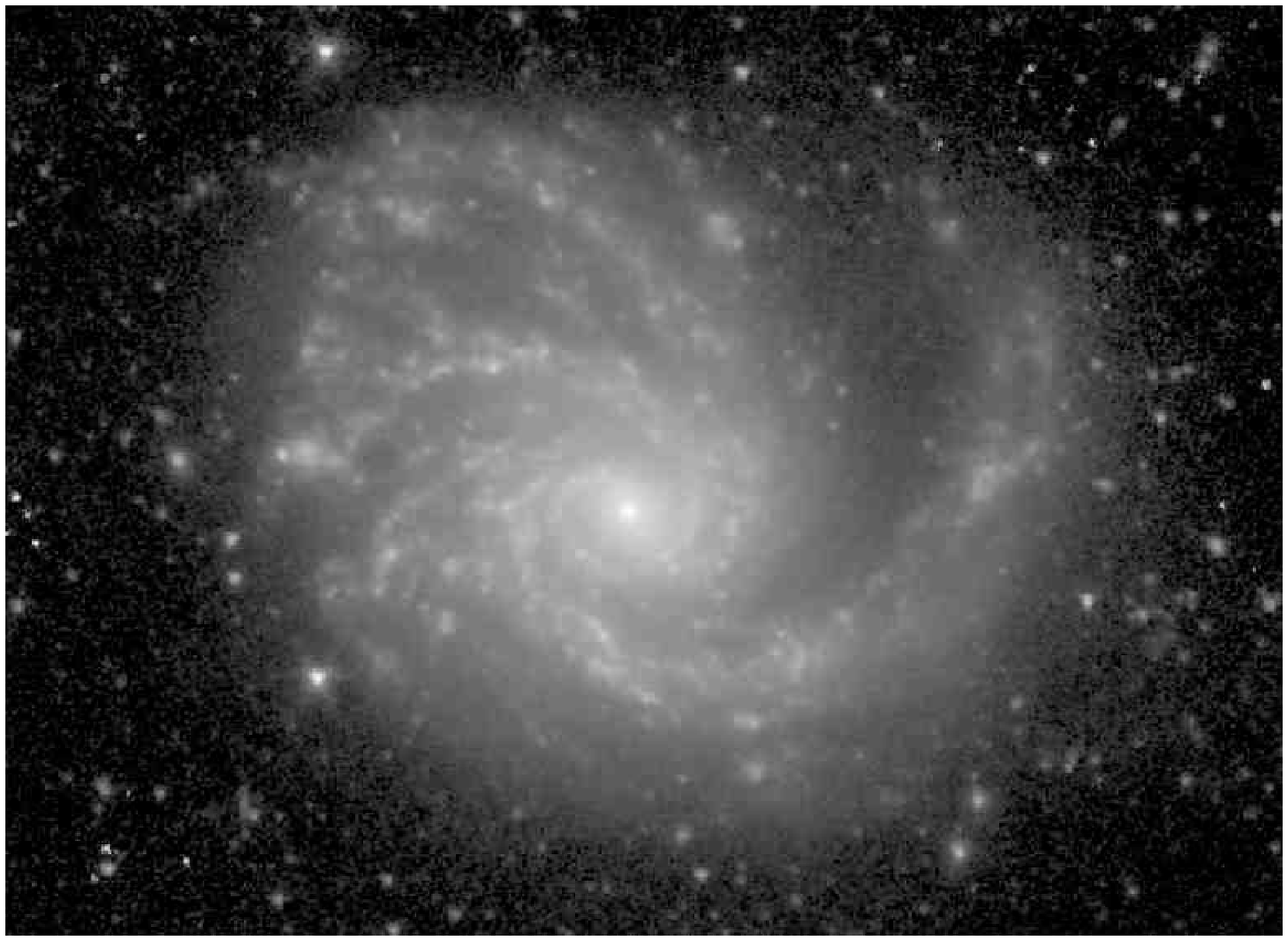}
 \vspace{2.0truecm}
 \caption{
{\bf NGC  4254   }              - S$^4$G mid-IR classification:    SA(s)c pec                                            ; Filter: IRAC 3.6$\mu$m; North:   up, East: left; Field dimensions:   7.0$\times$  5.1 arcmin; Surface brightness range displayed: 14.0$-$28.0 mag arcsec$^{-2}$}                 
\label{NGC4254}     
 \end{figure}
 
\clearpage
\begin{figure}
\figurenum{1.99}
\plotone{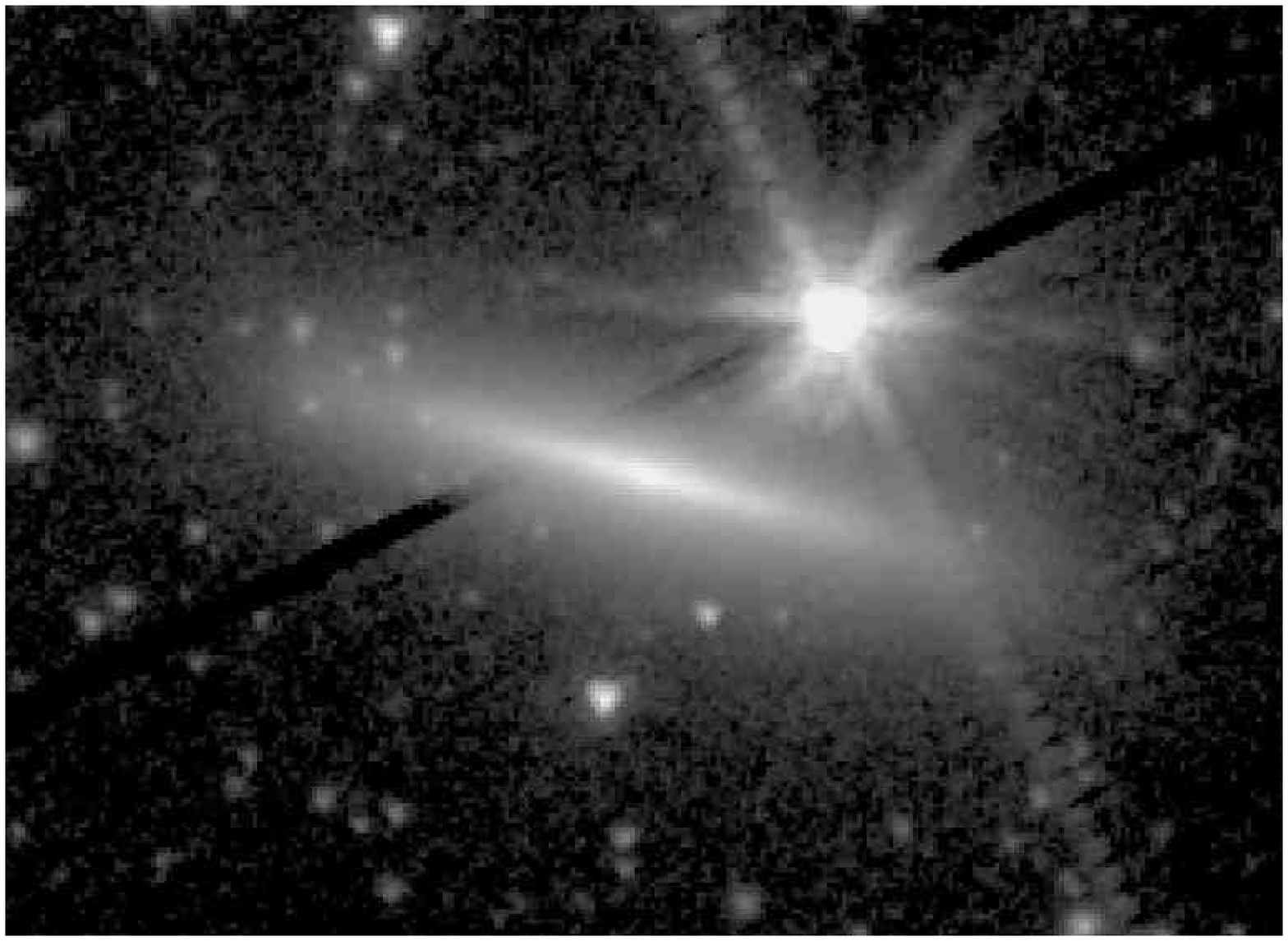}
 \vspace{2.0truecm}
 \caption{
{\bf NGC  4266   }              - S$^4$G mid-IR classification:    S0$^o$ sp                                             ; Filter: IRAC 3.6$\mu$m; North:   up, East: left; Field dimensions:   3.2$\times$  2.3 arcmin; Surface brightness range displayed: 15.0$-$28.0 mag arcsec$^{-2}$}                 
\label{NGC4266}     
 \end{figure}
 
\clearpage
\begin{figure}
\figurenum{1.100}
\plotone{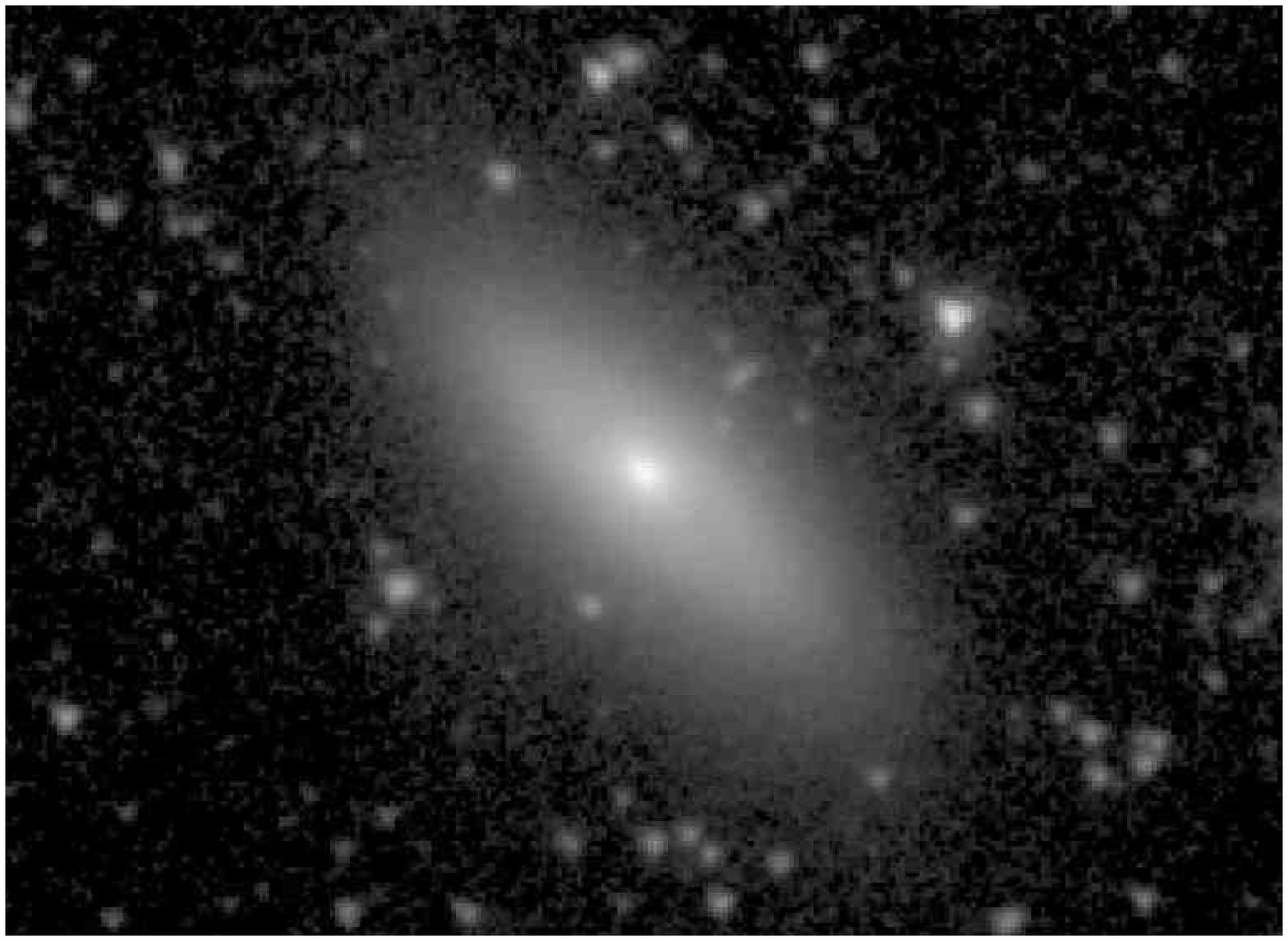}
 \vspace{2.0truecm}
 \caption{
{\bf NGC  4268   }              - S$^4$G mid-IR classification:    SA:($\underline{\rm r}$s)0$^+$ sp                     ; Filter: IRAC 3.6$\mu$m; North:   up, East: left; Field dimensions:   3.2$\times$  2.3 arcmin; Surface brightness range displayed: 13.5$-$28.0 mag arcsec$^{-2}$}                 
\label{NGC4268}     
 \end{figure}
 
\clearpage
\begin{figure}
\figurenum{1.101}
\plotone{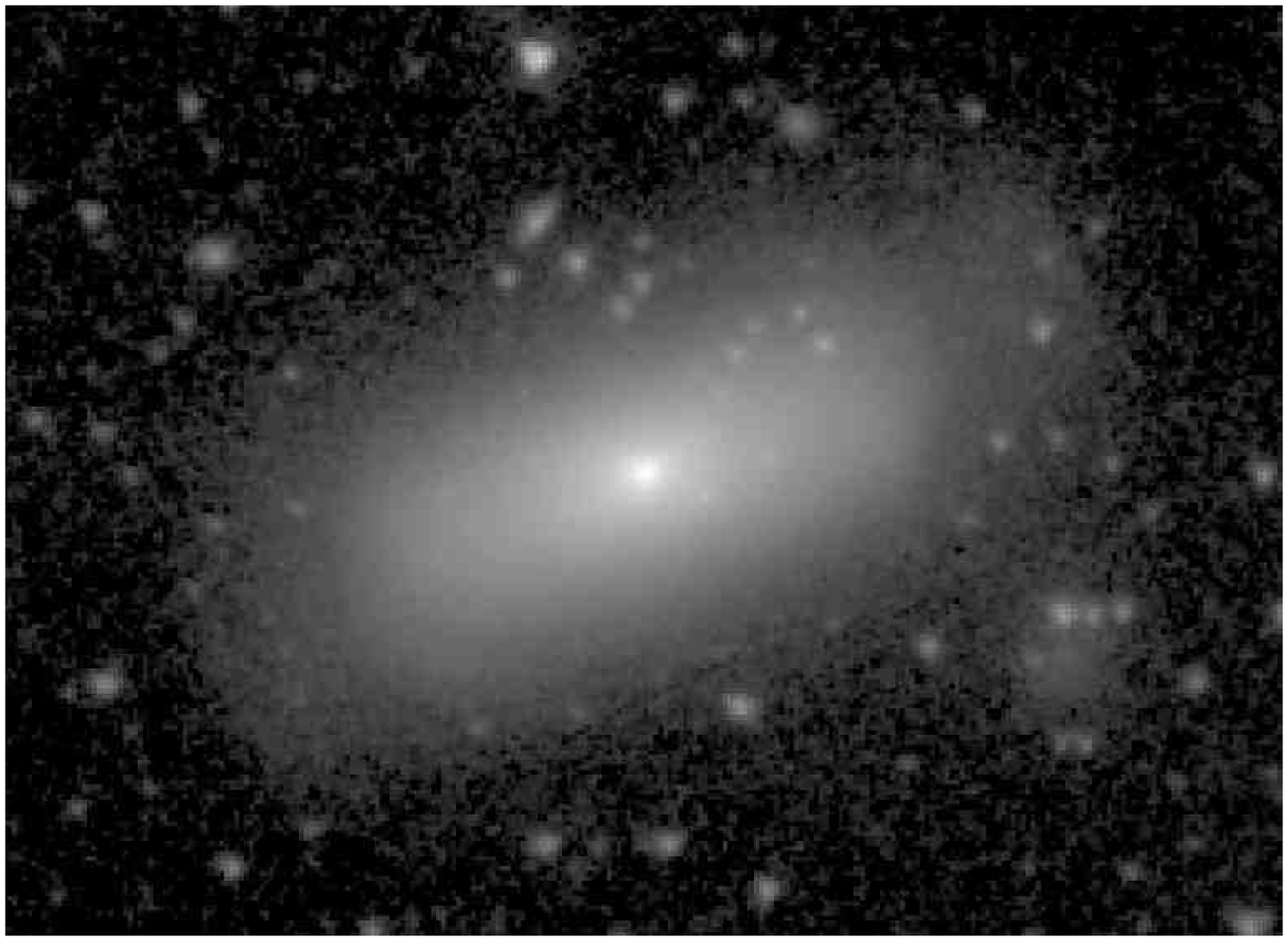}
 \vspace{2.0truecm}
 \caption{
{\bf NGC  4270   }              - S$^4$G mid-IR classification:    S0$^+$ sp?                                            ; Filter: IRAC 3.6$\mu$m; North:   up, East: left; Field dimensions:   5.3$\times$  3.8 arcmin; Surface brightness range displayed: 13.5$-$28.0 mag arcsec$^{-2}$}                 
\label{NGC4270}     
 \end{figure}
 
\clearpage
\begin{figure}
\figurenum{1.102}
\plotone{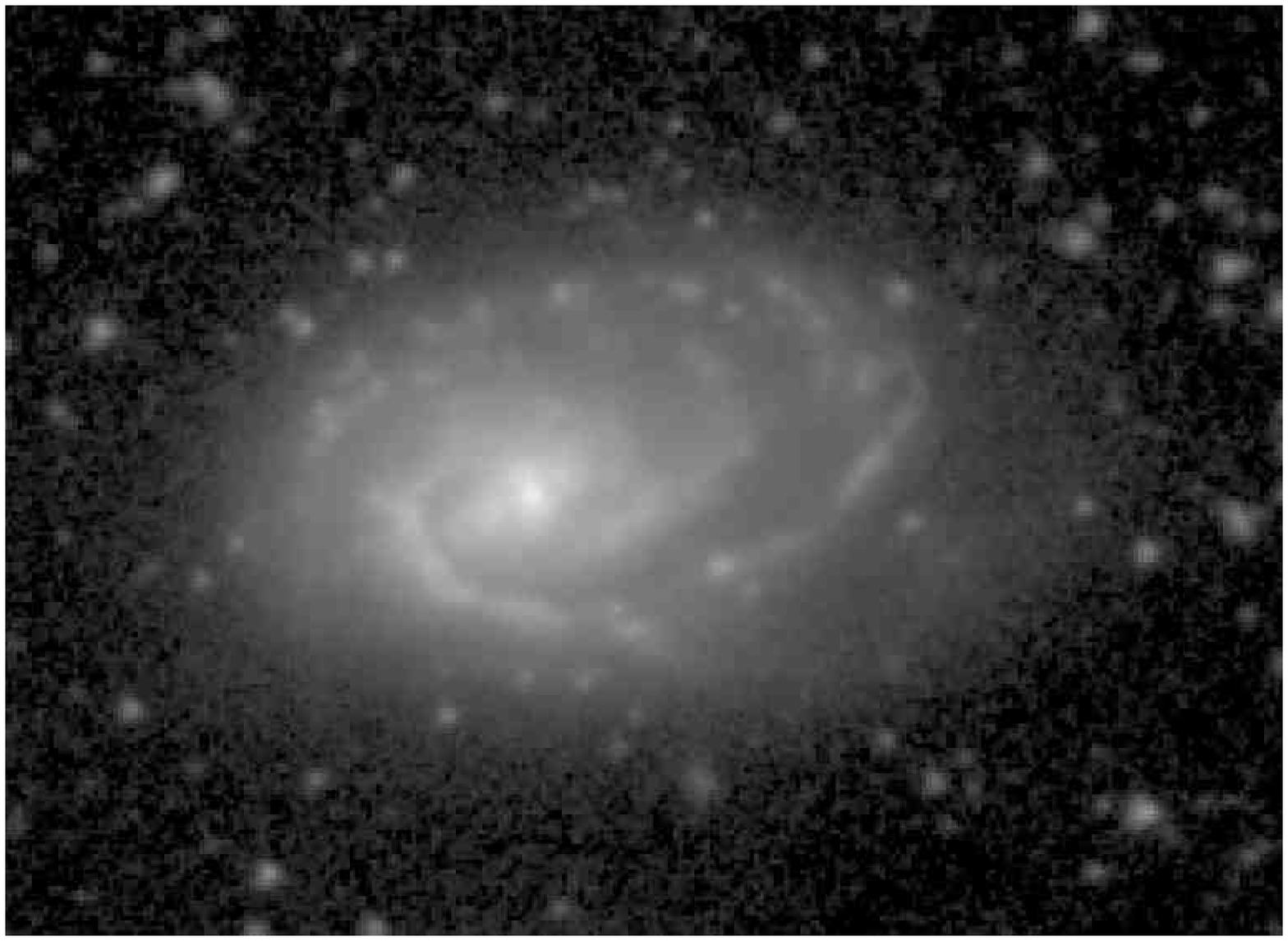}
 \vspace{2.0truecm}
 \caption{
{\bf NGC  4273   }              - S$^4$G mid-IR classification:    SA$\underline{\rm B}$(s)c pec                         ; Filter: IRAC 3.6$\mu$m; North: left, East: down; Field dimensions:   3.2$\times$  2.3 arcmin; Surface brightness range displayed: 13.5$-$28.0 mag arcsec$^{-2}$}                 
\label{NGC4273}     
 \end{figure}
 
\clearpage
\begin{figure}
\figurenum{1.103}
\plotone{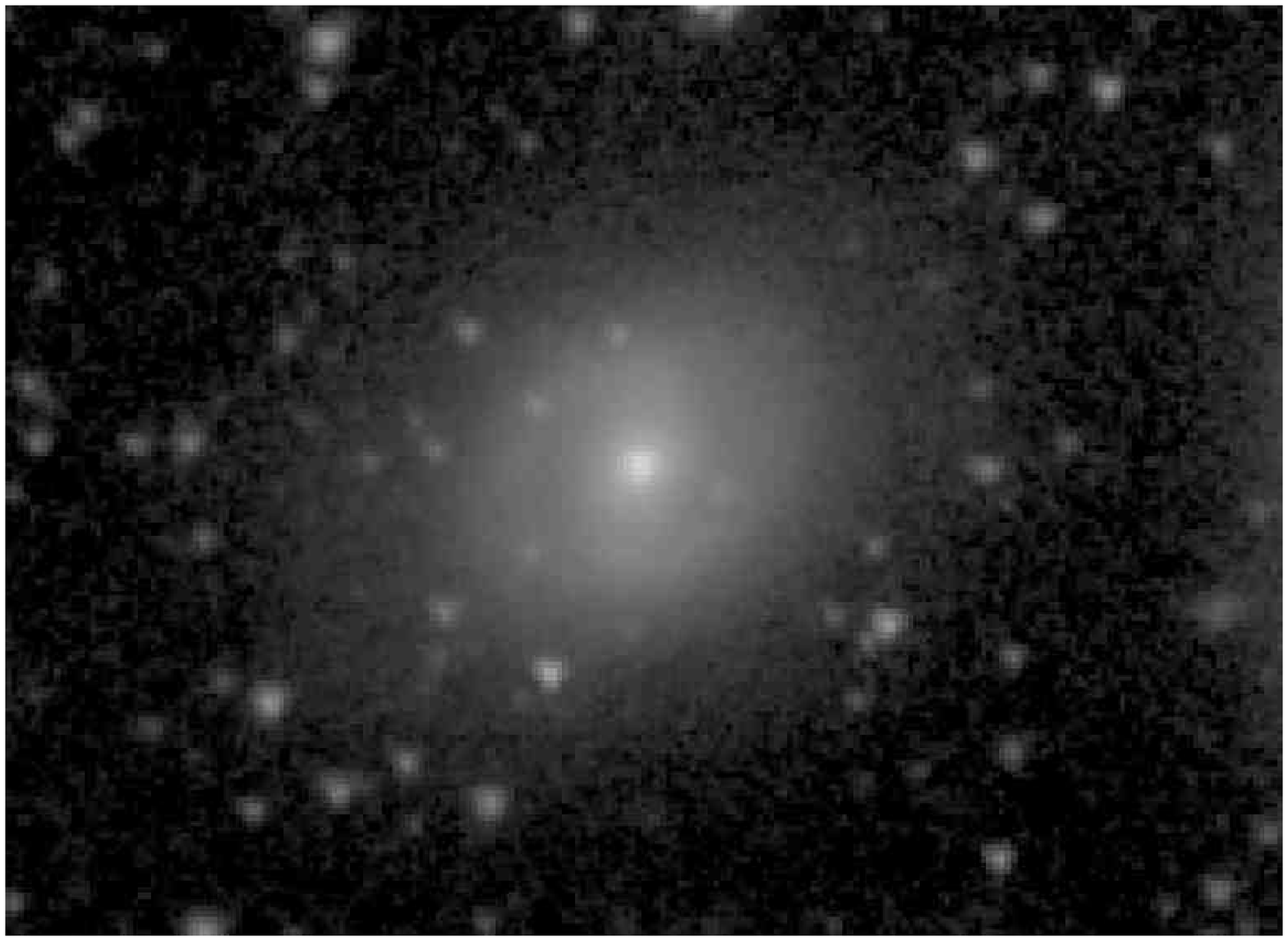}
 \vspace{2.0truecm}
 \caption{
{\bf NGC  4277   }              - S$^4$G mid-IR classification:    SAB(rs)0$^+$                                          ; Filter: IRAC 3.6$\mu$m; North:   up, East: left; Field dimensions:   2.6$\times$  1.9 arcmin; Surface brightness range displayed: 13.5$-$28.0 mag arcsec$^{-2}$}                 
\label{NGC4277}     
 \end{figure}
 
\clearpage
\begin{figure}
\figurenum{1.104}
\plotone{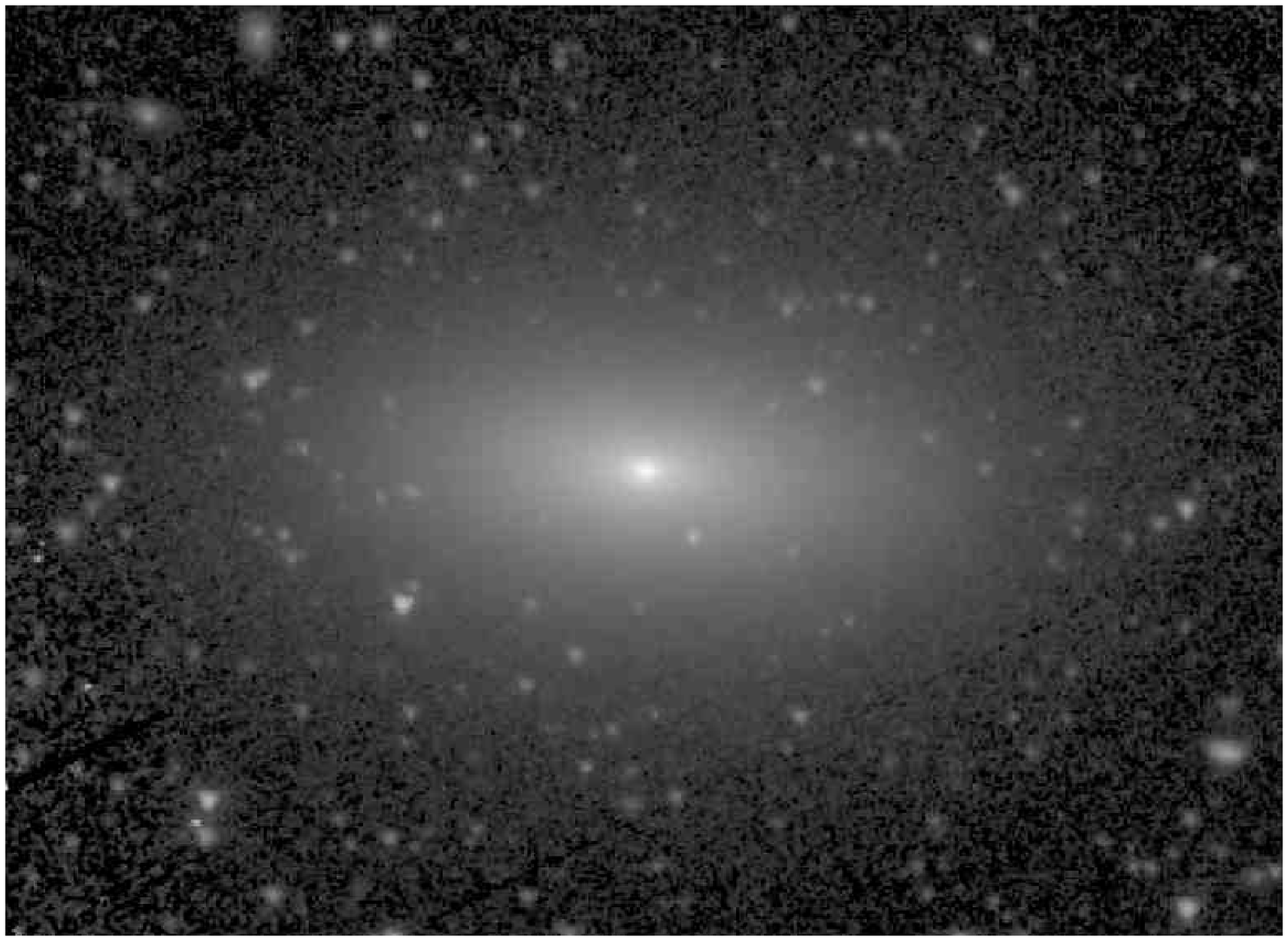}
 \vspace{2.0truecm}
 \caption{
{\bf NGC  4281   }              - S$^4$G mid-IR classification:    E(d)5                                                   ; Filter: IRAC 3.6$\mu$m; North:   up, East: left; Field dimensions:   5.3$\times$  3.8 arcmin; Surface brightness range displayed: 12.5$-$28.0 mag arcsec$^{-2}$}                 
\label{NGC4281}     
 \end{figure}
 
\clearpage
\begin{figure}
\figurenum{1.105}
\plotone{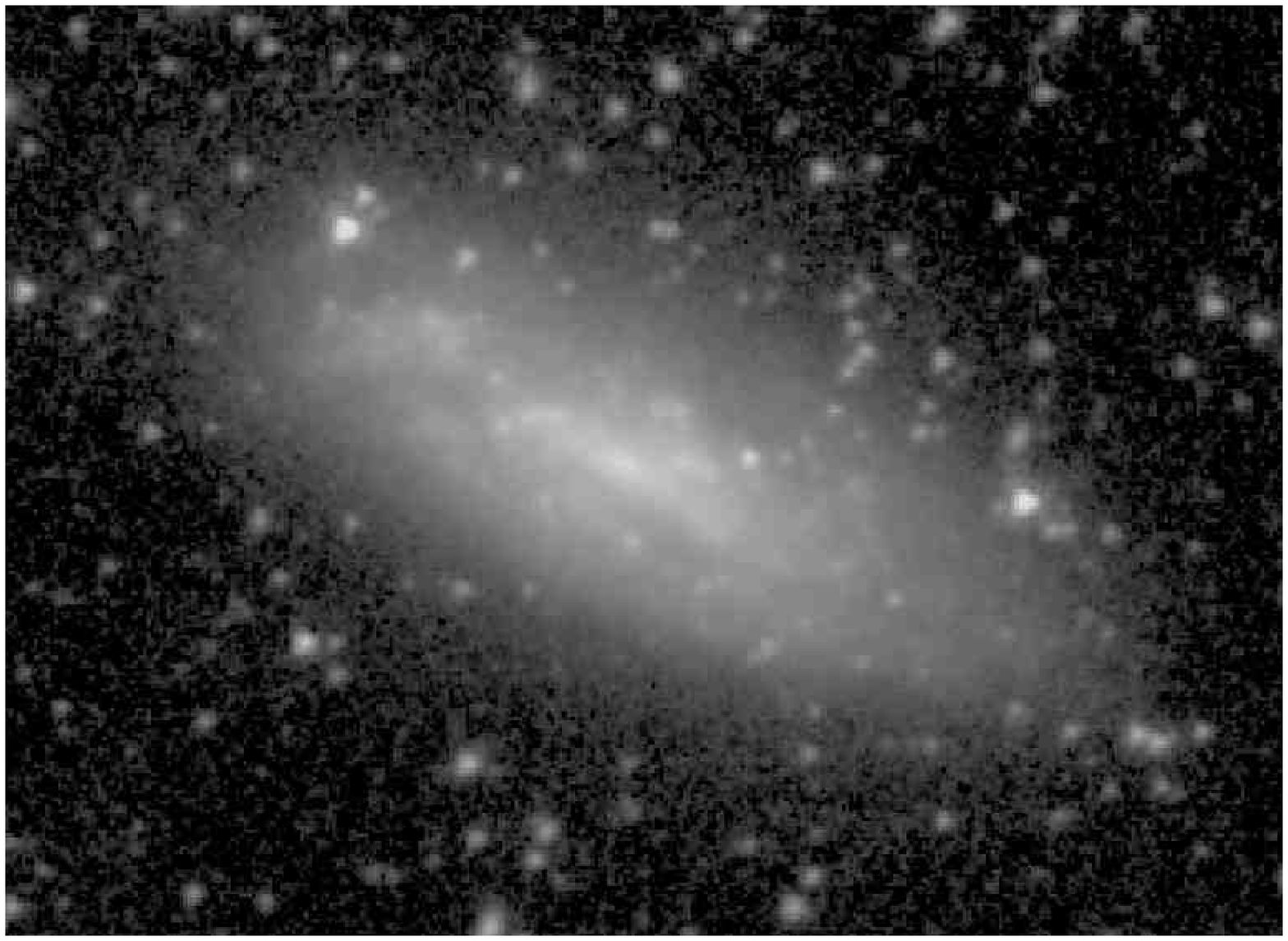}
 \vspace{2.0truecm}
 \caption{
{\bf NGC  4294   }              - S$^4$G mid-IR classification:    SB(s)d                                                ; Filter: IRAC 3.6$\mu$m; North: left, East: down; Field dimensions:   4.0$\times$  2.9 arcmin; Surface brightness range displayed: 16.0$-$28.0 mag arcsec$^{-2}$}                 
\label{NGC4294}     
 \end{figure}
 
\clearpage
\begin{figure}
\figurenum{1.106}
\plotone{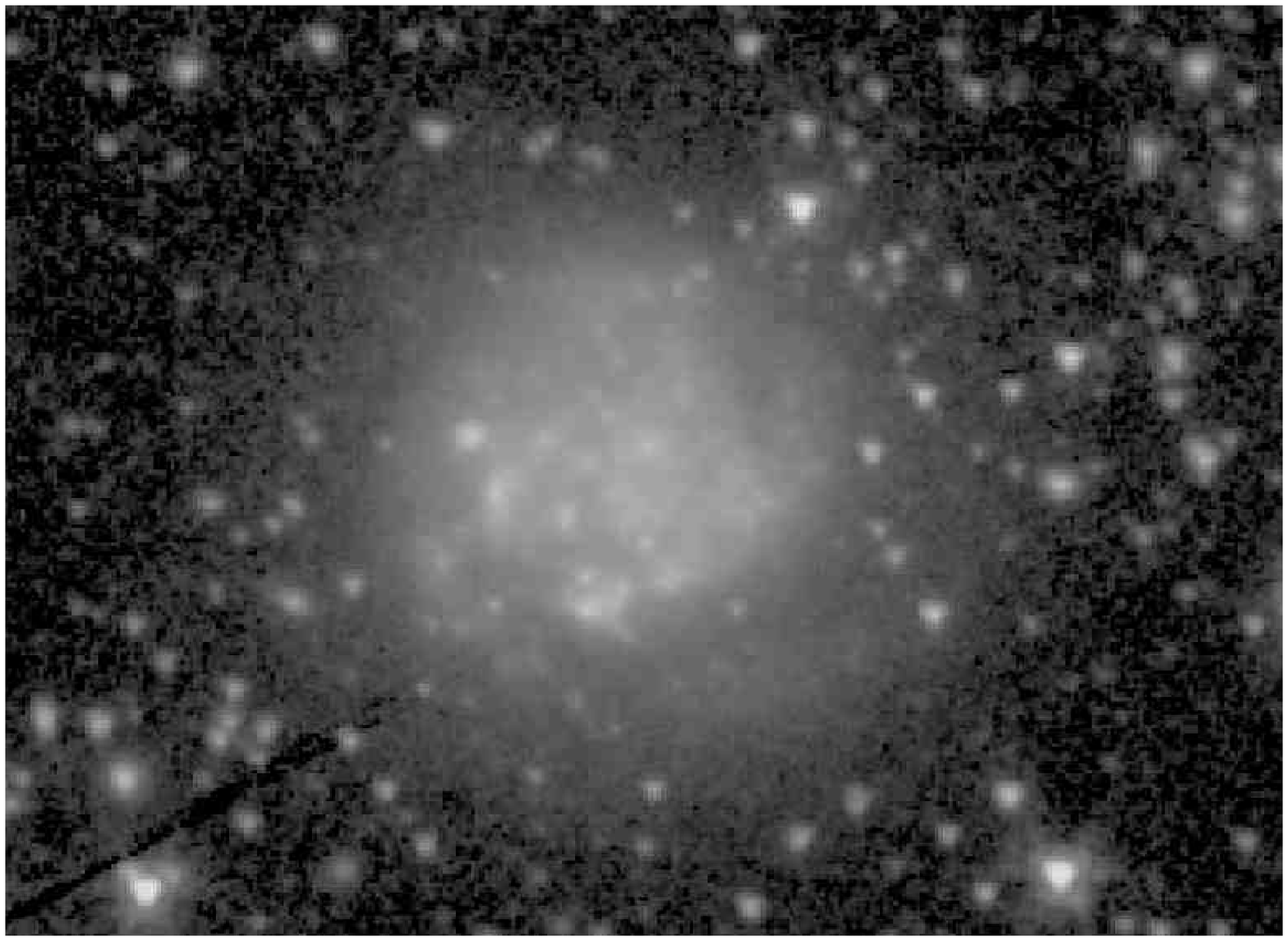}
 \vspace{2.0truecm}
 \caption{
{\bf NGC  4299   }              - S$^4$G mid-IR classification:    SA(s)dm                                               ; Filter: IRAC 3.6$\mu$m; North:   up, East: left; Field dimensions:   3.5$\times$  2.6 arcmin; Surface brightness range displayed: 17.0$-$28.0 mag arcsec$^{-2}$}                 
\label{NGC4299}     
 \end{figure}
 
\clearpage
\begin{figure}
\figurenum{1.107}
\plotone{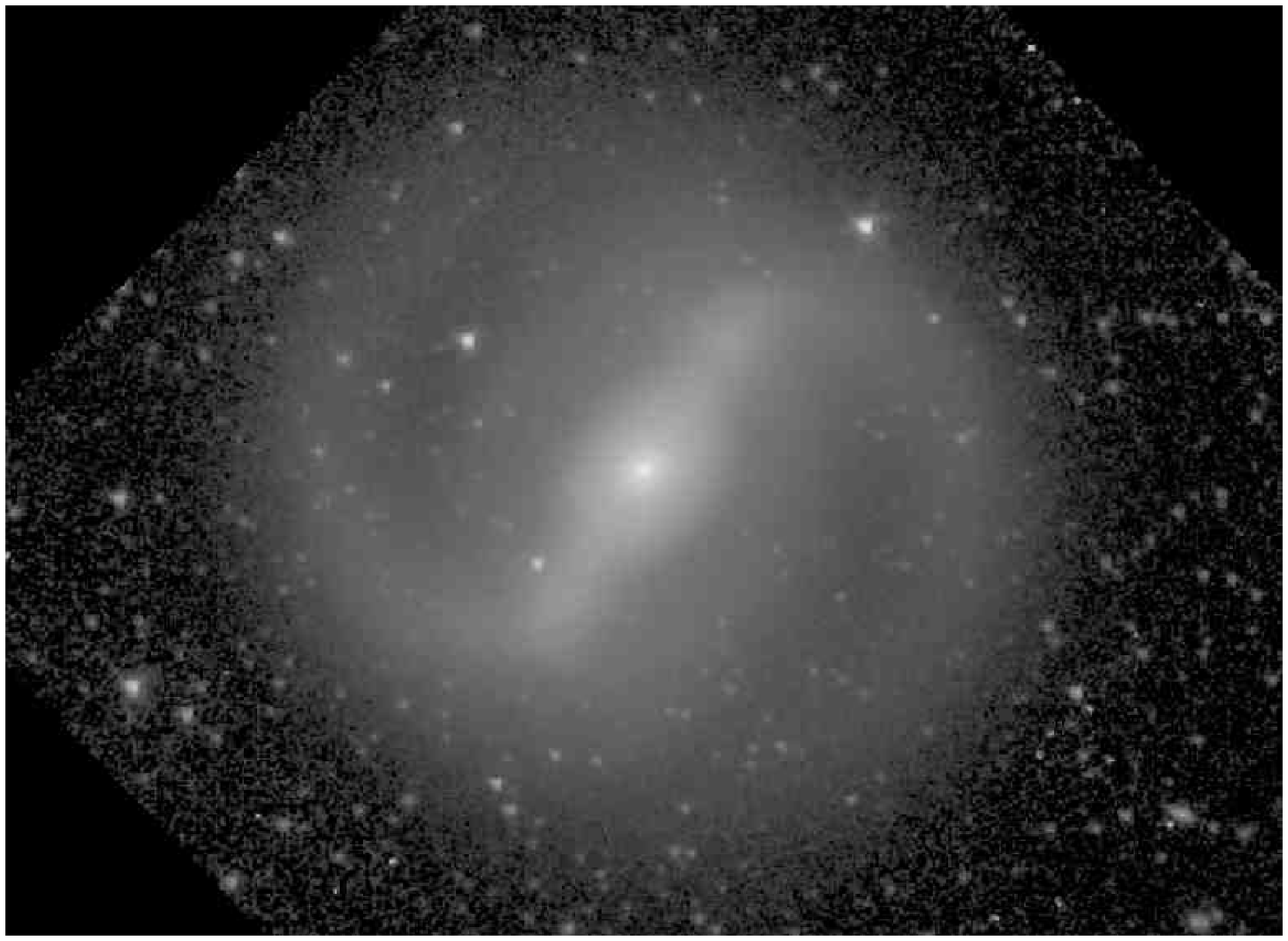}
 \vspace{2.0truecm}
 \caption{
{\bf NGC  4314   }              - S$^4$G mid-IR classification:    (R$_1^{\prime}$)SB(rl,nr)a                                    ; Filter: IRAC 3.6$\mu$m; North:   up, East: left; Field dimensions:   7.0$\times$  5.1 arcmin; Surface brightness range displayed: 13.0$-$28.0 mag arcsec$^{-2}$}                 
\label{NGC4314}     
 \end{figure}
 
\clearpage
\begin{figure}
\figurenum{1.108}
\plotone{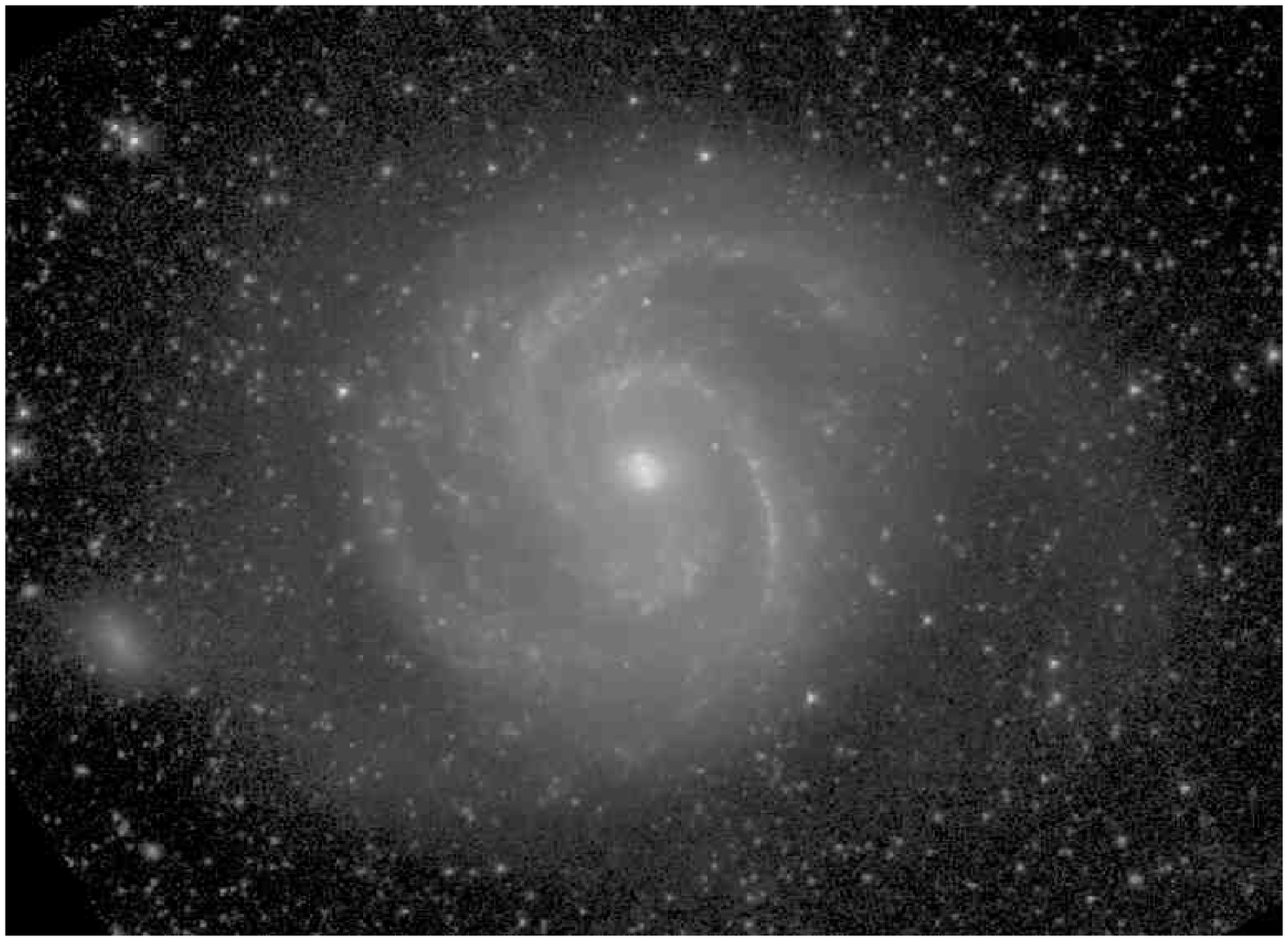}
 \vspace{2.0truecm}
 \caption{
{\bf NGC  4321   }              - S$^4$G mid-IR classification:    SAB(r$\underline{\rm s}$,nr)bc                        ; Filter: IRAC 3.6$\mu$m; North: left, East: down; Field dimensions:  12.1$\times$  8.9 arcmin; Surface brightness range displayed: 13.0$-$28.0 mag arcsec$^{-2}$}                 
\label{NGC4321}     
 \end{figure}
 
\clearpage
\begin{figure}
\figurenum{1.109}
\plotone{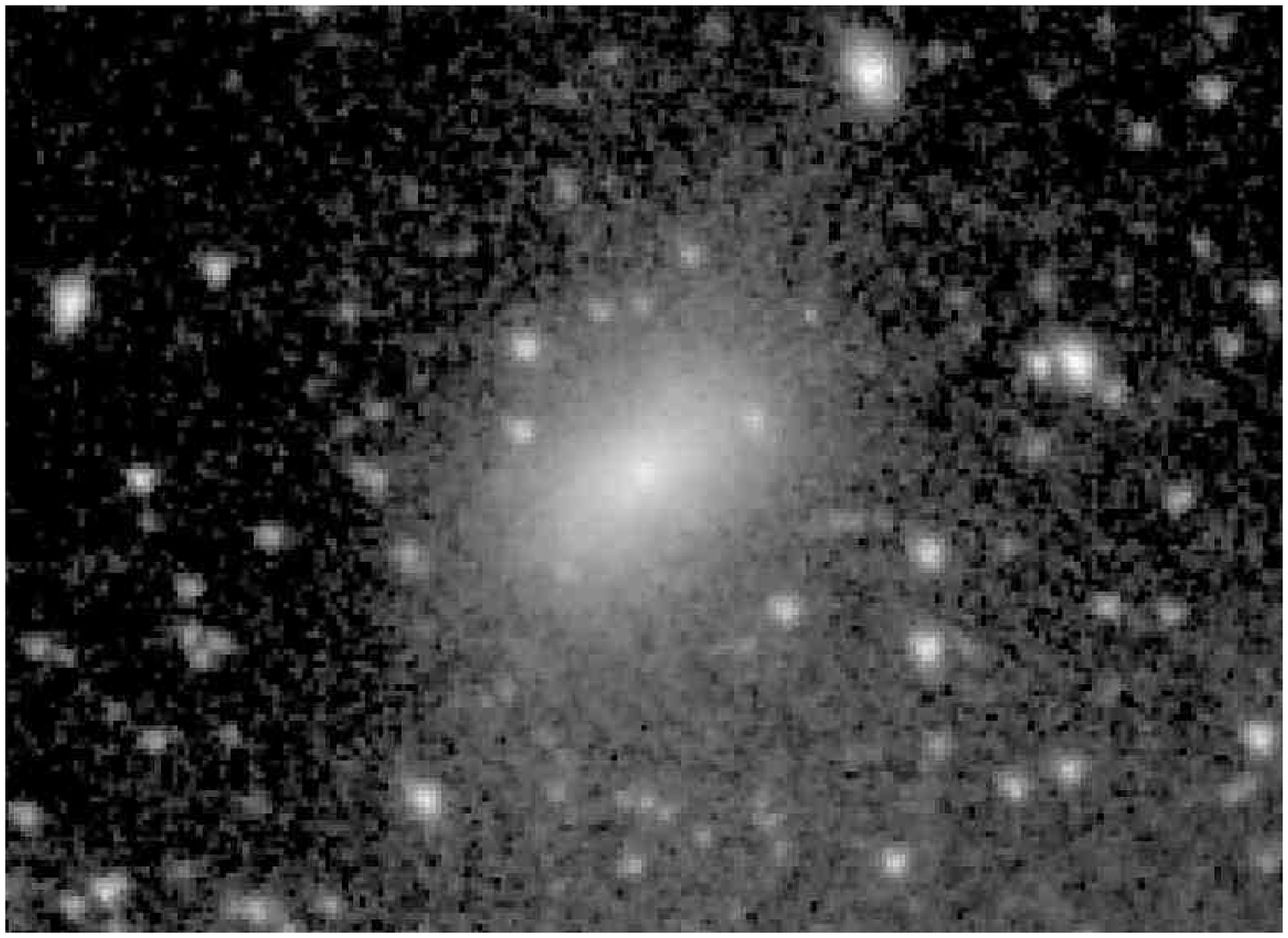}
 \vspace{2.0truecm}
 \caption{
{\bf NGC  4323   }              - S$^4$G mid-IR classification:    dE3,N                                                 ; Filter: IRAC 3.6$\mu$m; North:   up, East: left; Field dimensions:   2.7$\times$  2.0 arcmin; Surface brightness range displayed: 17.5$-$28.0 mag arcsec$^{-2}$}                 
\label{NGC4323}     
 \end{figure}
 
\clearpage
\begin{figure}
\figurenum{1.110}
\plotone{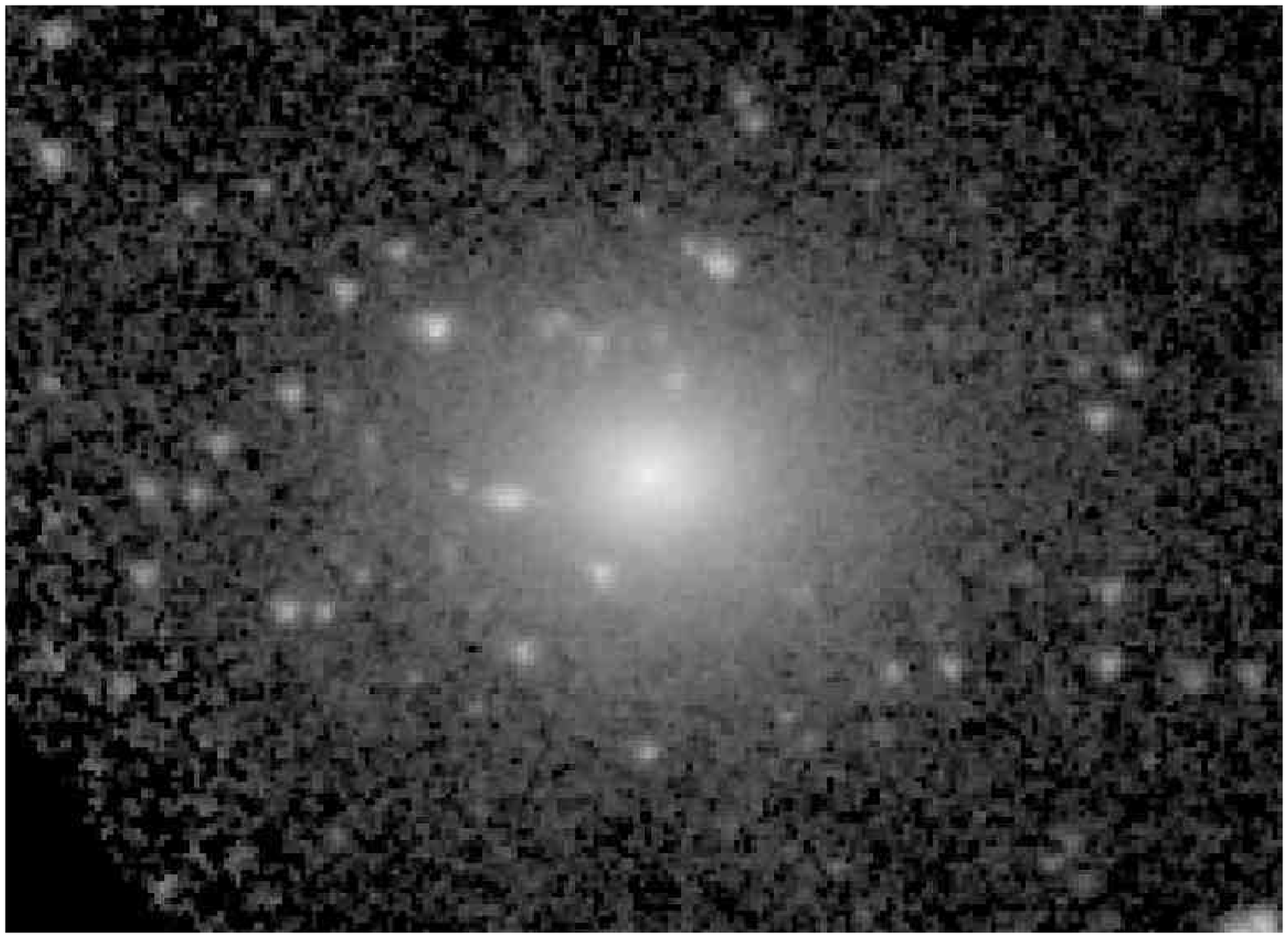}
 \vspace{2.0truecm}
 \caption{
{\bf NGC  4328   }              - S$^4$G mid-IR classification:    dSA(l)0$^-$                                           ; Filter: IRAC 3.6$\mu$m; North:   up, East: left; Field dimensions:   2.7$\times$  2.0 arcmin; Surface brightness range displayed: 17.0$-$28.0 mag arcsec$^{-2}$}                 
\label{NGC4328}     
 \end{figure}
 
\clearpage
\begin{figure}
\figurenum{1.111}
\plotone{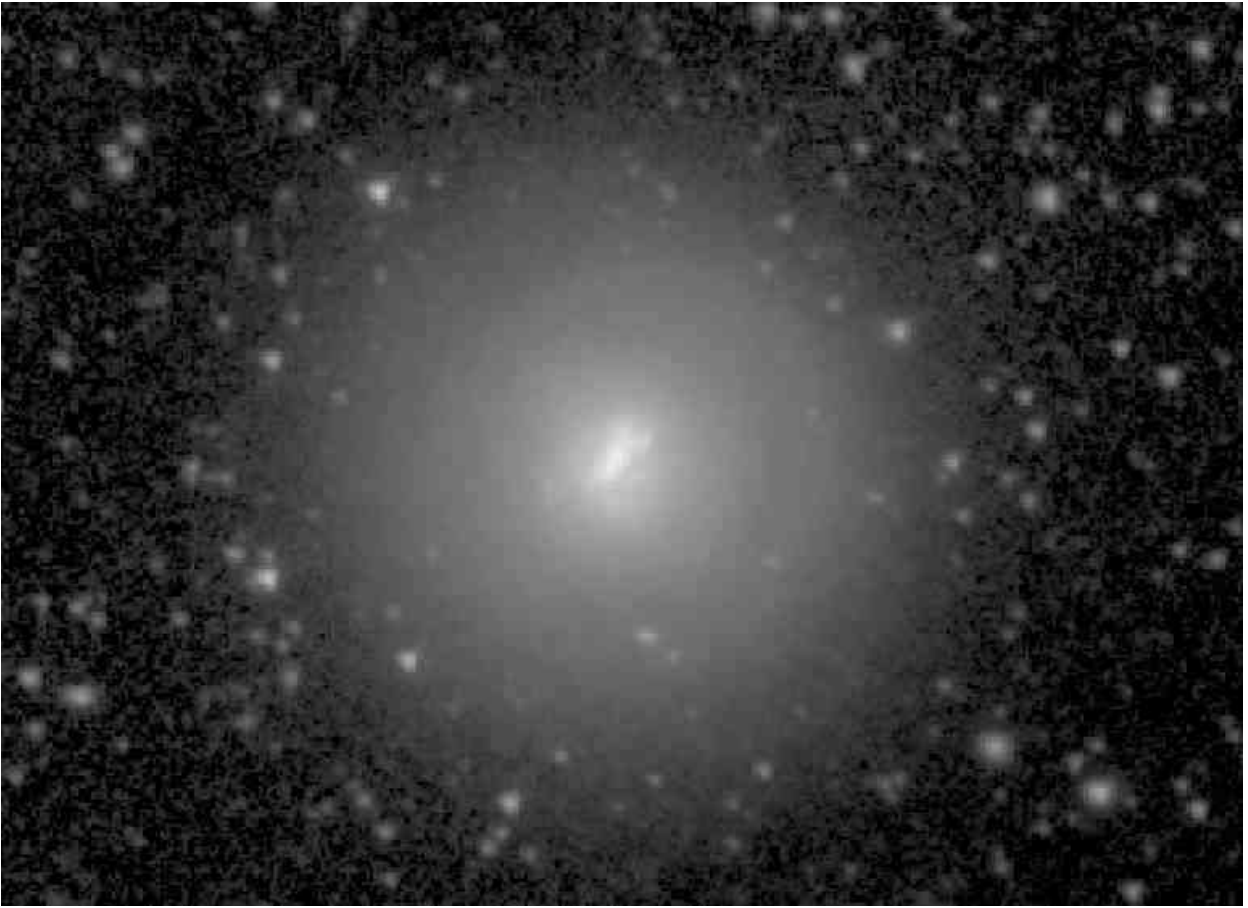}
 \vspace{2.0truecm}
 \caption{
{\bf NGC  4369   }              - S$^4$G mid-IR classification:    (R)SB(rs)0/a: pec                                     ; Filter: IRAC 3.6$\mu$m; North:   up, East: left; Field dimensions:   4.5$\times$  3.3 arcmin; Surface brightness range displayed: 14.0$-$28.0 mag arcsec$^{-2}$}                 
\label{NGC4369}     
 \end{figure}
 
\clearpage
\begin{figure}
\figurenum{1.112}
\plotone{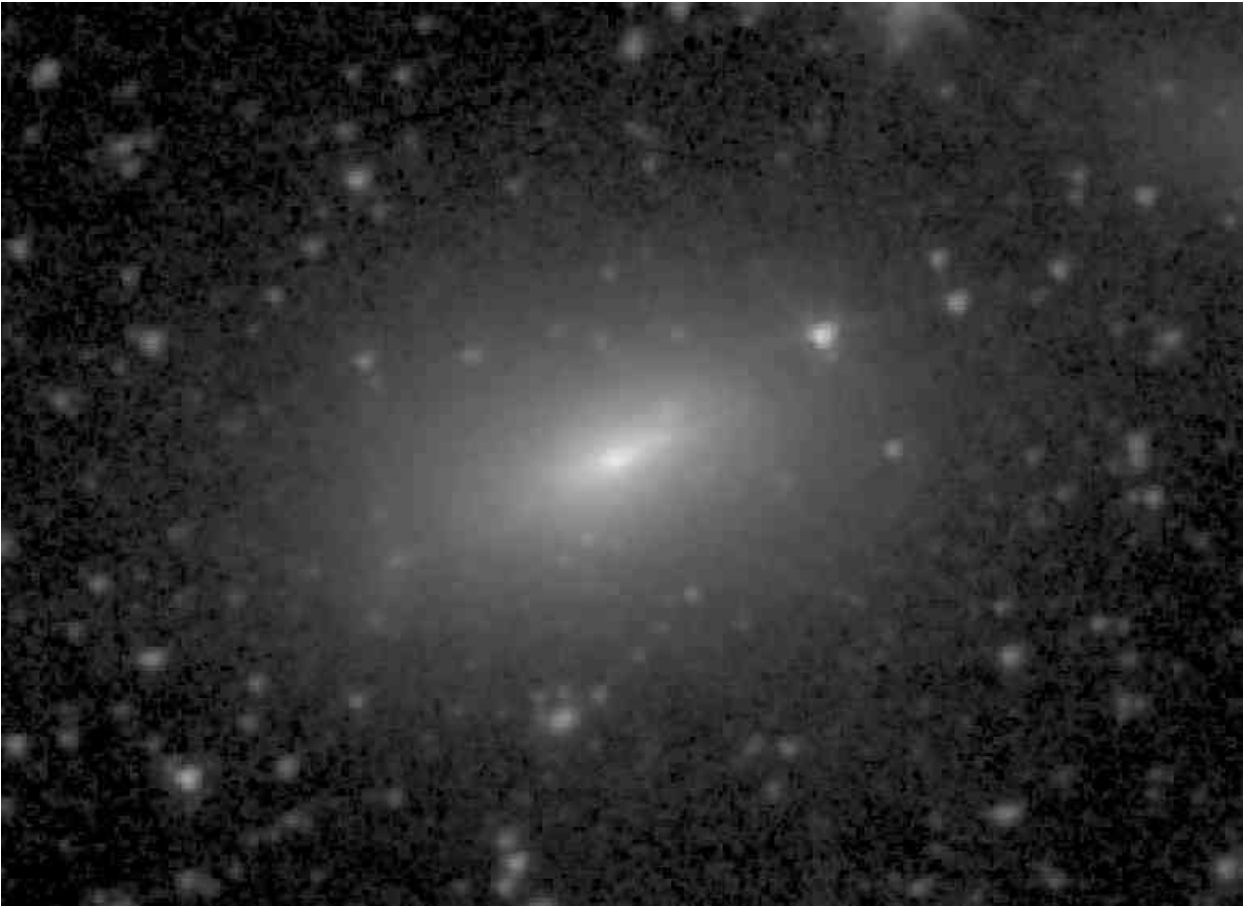}
 \vspace{2.0truecm}
 \caption{
{\bf NGC  4383   }              - S$^4$G mid-IR classification:    E(twist?)/SAB0$^-$:                                   ; Filter: IRAC 3.6$\mu$m; North: left, East: down; Field dimensions:   4.0$\times$  2.9 arcmin; Surface brightness range displayed: 13.0$-$28.0 mag arcsec$^{-2}$}                 
\label{NGC4383}     
 \end{figure}
 
\clearpage
\begin{figure}
\figurenum{1.113}
\plotone{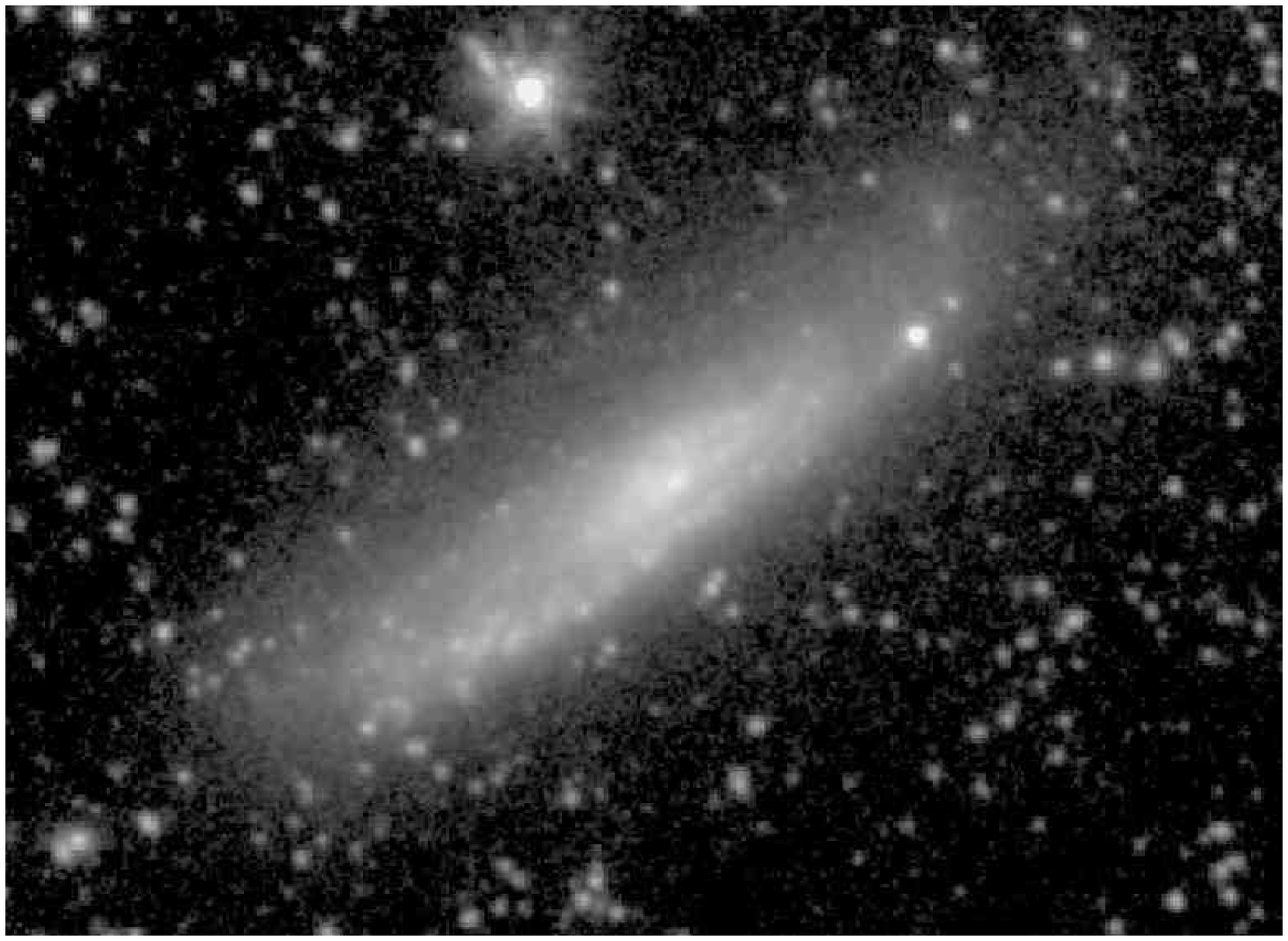}
 \vspace{2.0truecm}
 \caption{
{\bf NGC  4396   }              - S$^4$G mid-IR classification:    Scd: pec sp                                           ; Filter: IRAC 3.6$\mu$m; North:   up, East: left; Field dimensions:   4.5$\times$  3.3 arcmin; Surface brightness range displayed: 16.5$-$28.0 mag arcsec$^{-2}$}                 
\label{NGC4396}     
 \end{figure}
 
\clearpage
\begin{figure}
\figurenum{1.114}
\plotone{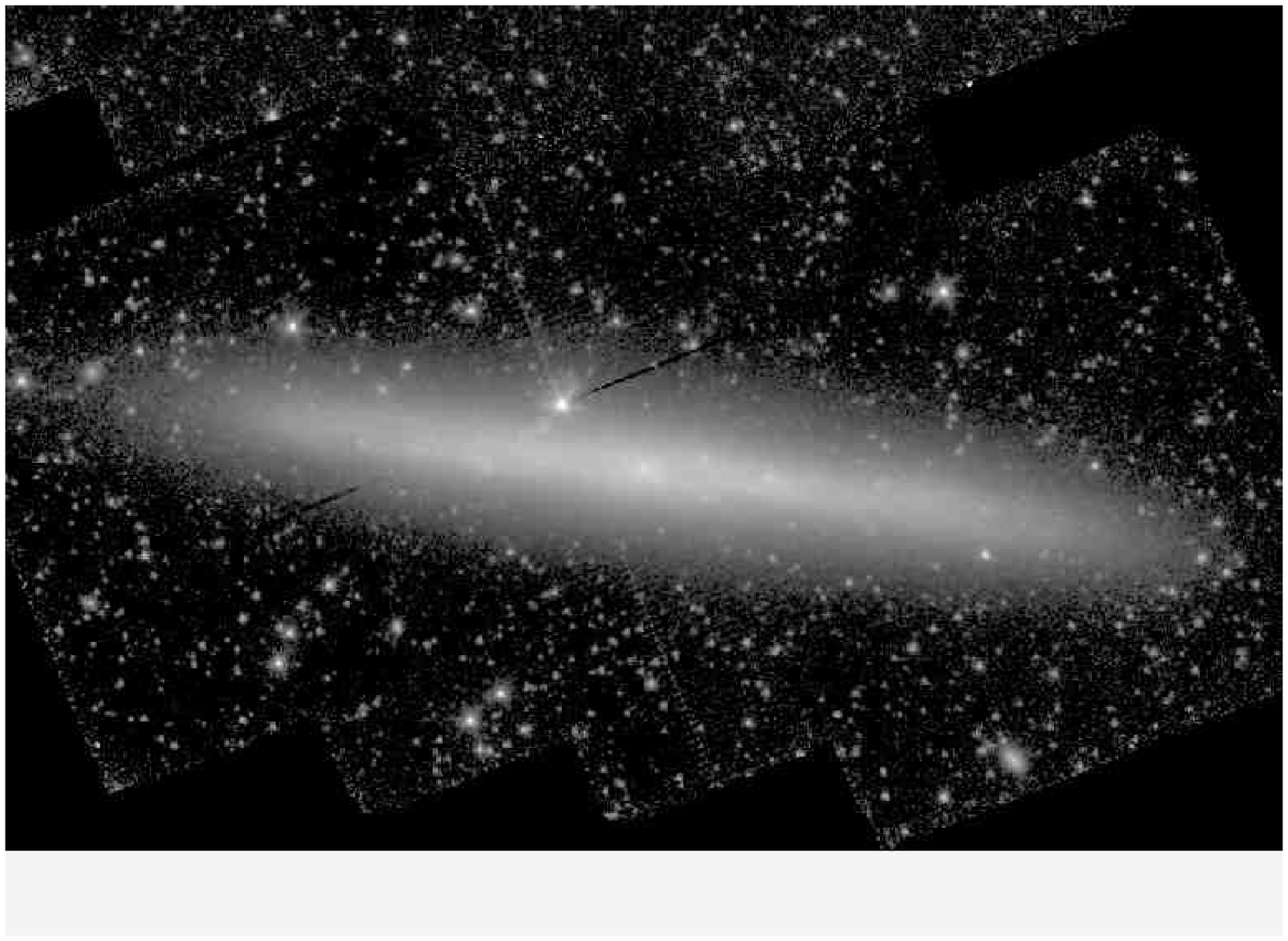}
 \vspace{2.0truecm}
 \caption{
{\bf NGC  4437   }              - S$^4$G mid-IR classification:    Sc sp                                                 ; Filter: IRAC 3.6$\mu$m; North:   up, East: left; Field dimensions:  12.6$\times$  9.2 arcmin; Surface brightness range displayed: 14.5$-$28.0 mag arcsec$^{-2}$}                 
\label{NGC4437}     
 \end{figure}
 
\clearpage
\begin{figure}
\figurenum{1.115}
\plotone{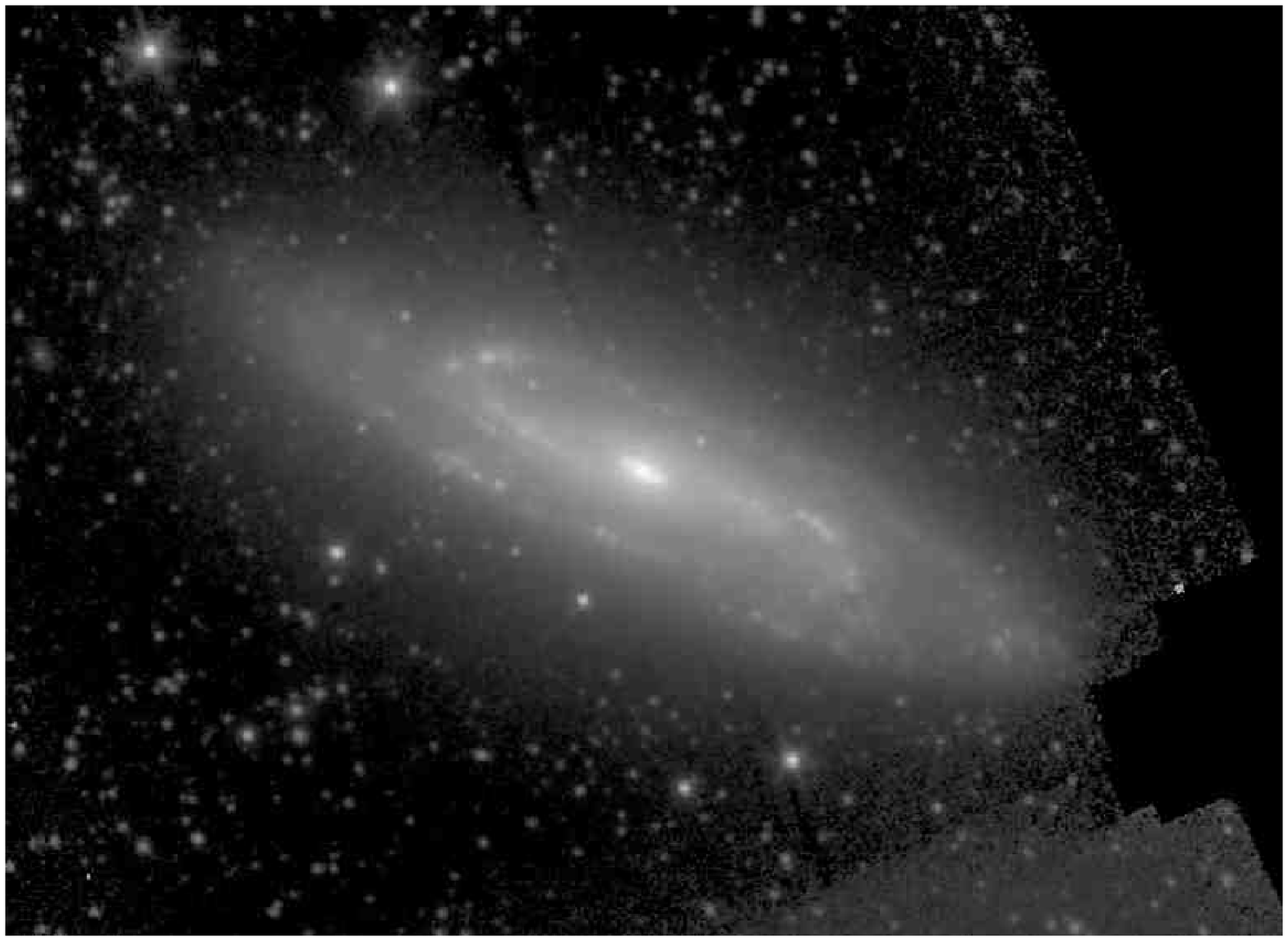}
 \vspace{2.0truecm}
 \caption{
{\bf NGC  4527   }              - S$^4$G mid-IR classification:    (R$_2^{\prime}$)SAB(rs,nr)b$\underline{\rm c}$                ; Filter: IRAC 3.6$\mu$m; North:   up, East: left; Field dimensions:   7.9$\times$  5.7 arcmin; Surface brightness range displayed: 12.0$-$28.0 mag arcsec$^{-2}$}                 
\label{NGC4527}     
 \end{figure}
 
\clearpage
\begin{figure}
\figurenum{1.116}
\plotone{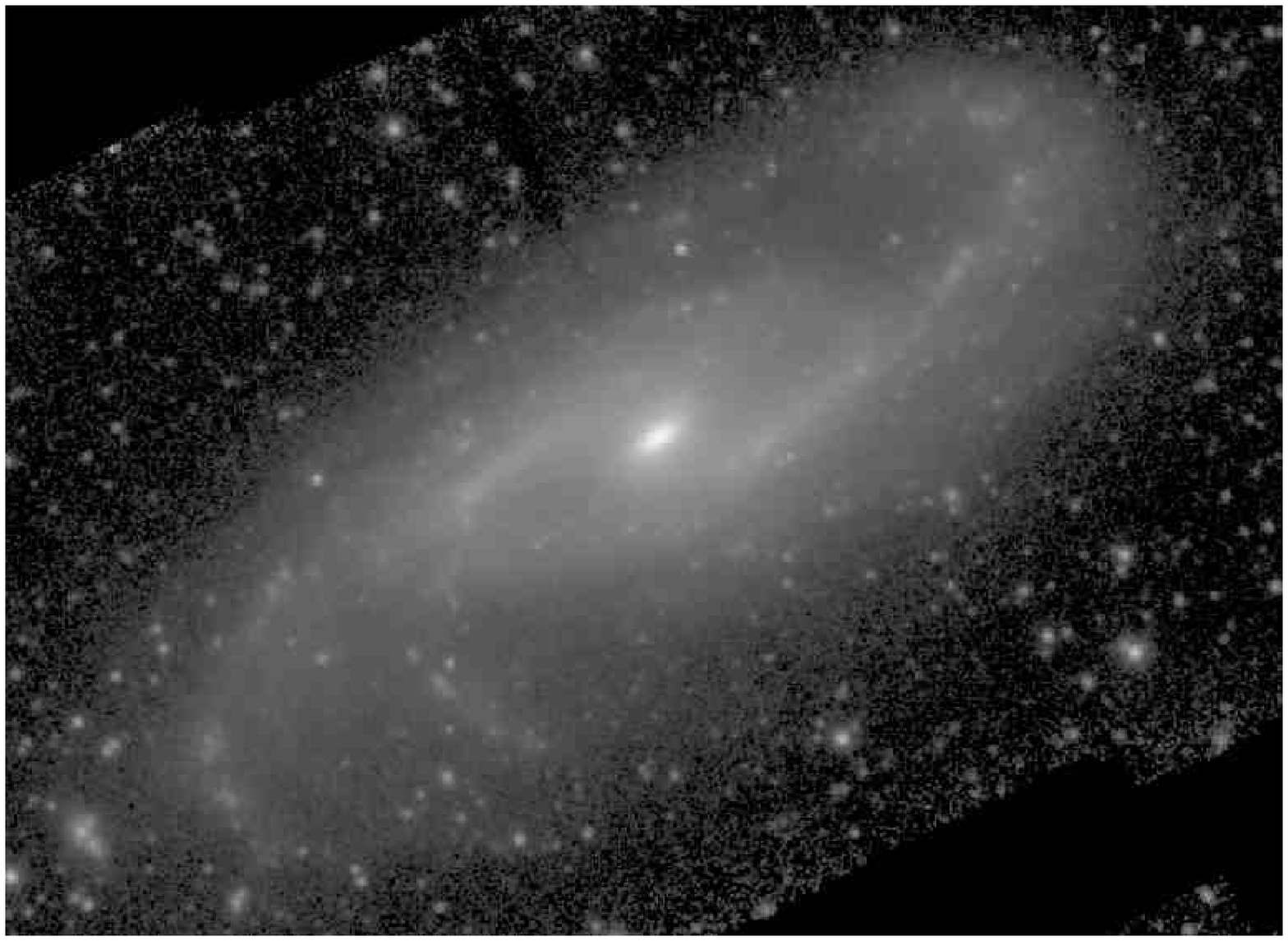}
 \vspace{2.0truecm}
 \caption{
{\bf NGC  4536   }              - S$^4$G mid-IR classification:    SAB(rs)$\underline{\rm b}$c                           ; Filter: IRAC 3.6$\mu$m; North:   up, East: left; Field dimensions:   7.9$\times$  5.8 arcmin; Surface brightness range displayed: 12.5$-$28.0 mag arcsec$^{-2}$}                 
\label{NGC4536}     
 \end{figure}
 
\clearpage
\begin{figure}
\figurenum{1.117}
\plotone{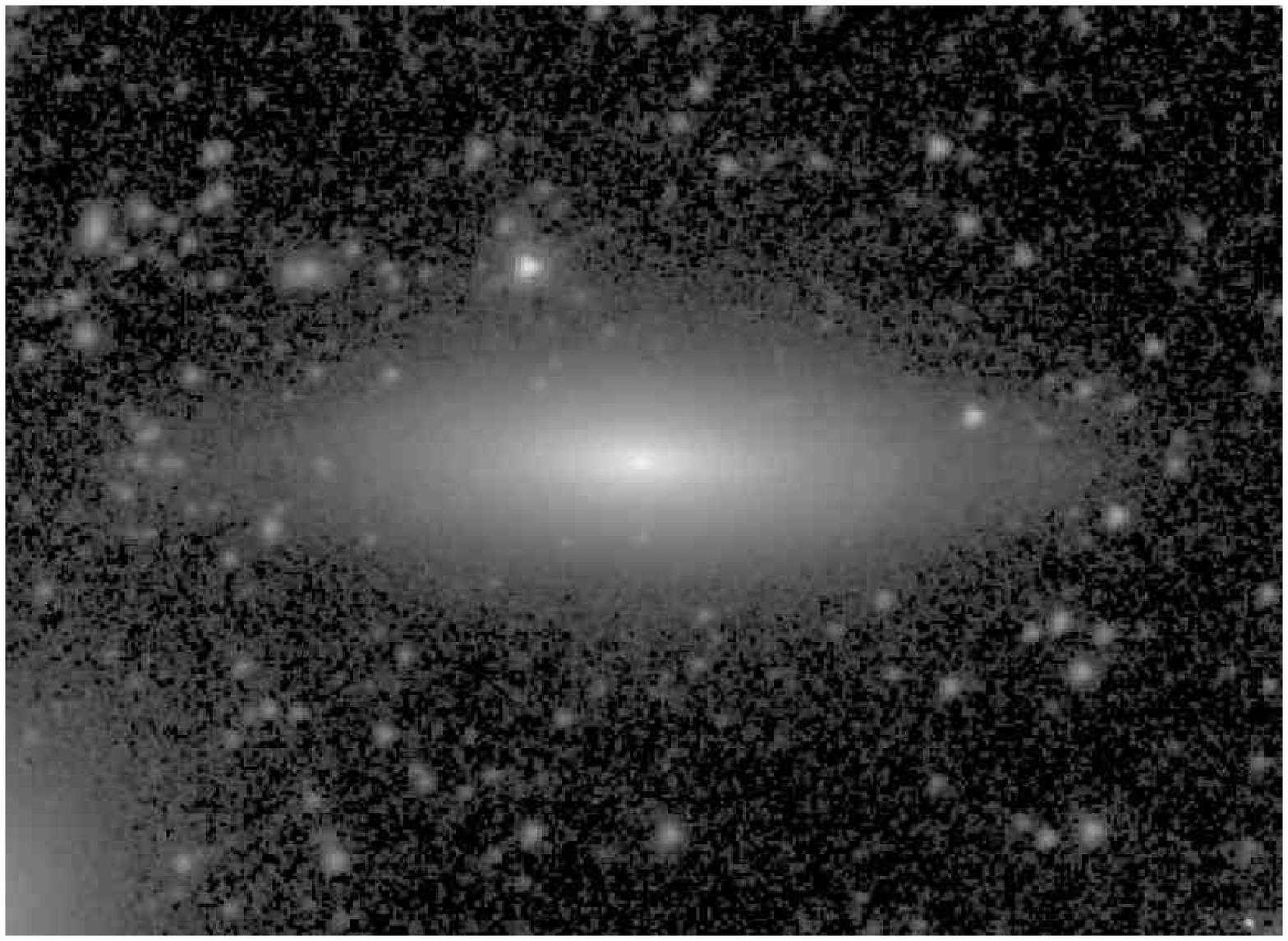}
 \vspace{2.0truecm}
 \caption{
{\bf NGC  4550   }              - S$^4$G mid-IR classification:    S0$^-$ sp                                             ; Filter: IRAC 3.6$\mu$m; North: left, East: down; Field dimensions:   4.5$\times$  3.3 arcmin; Surface brightness range displayed: 13.5$-$28.0 mag arcsec$^{-2}$}                 
\label{NGC4550}     
 \end{figure}
 
\clearpage
\begin{figure}
\figurenum{1.118}
\plotone{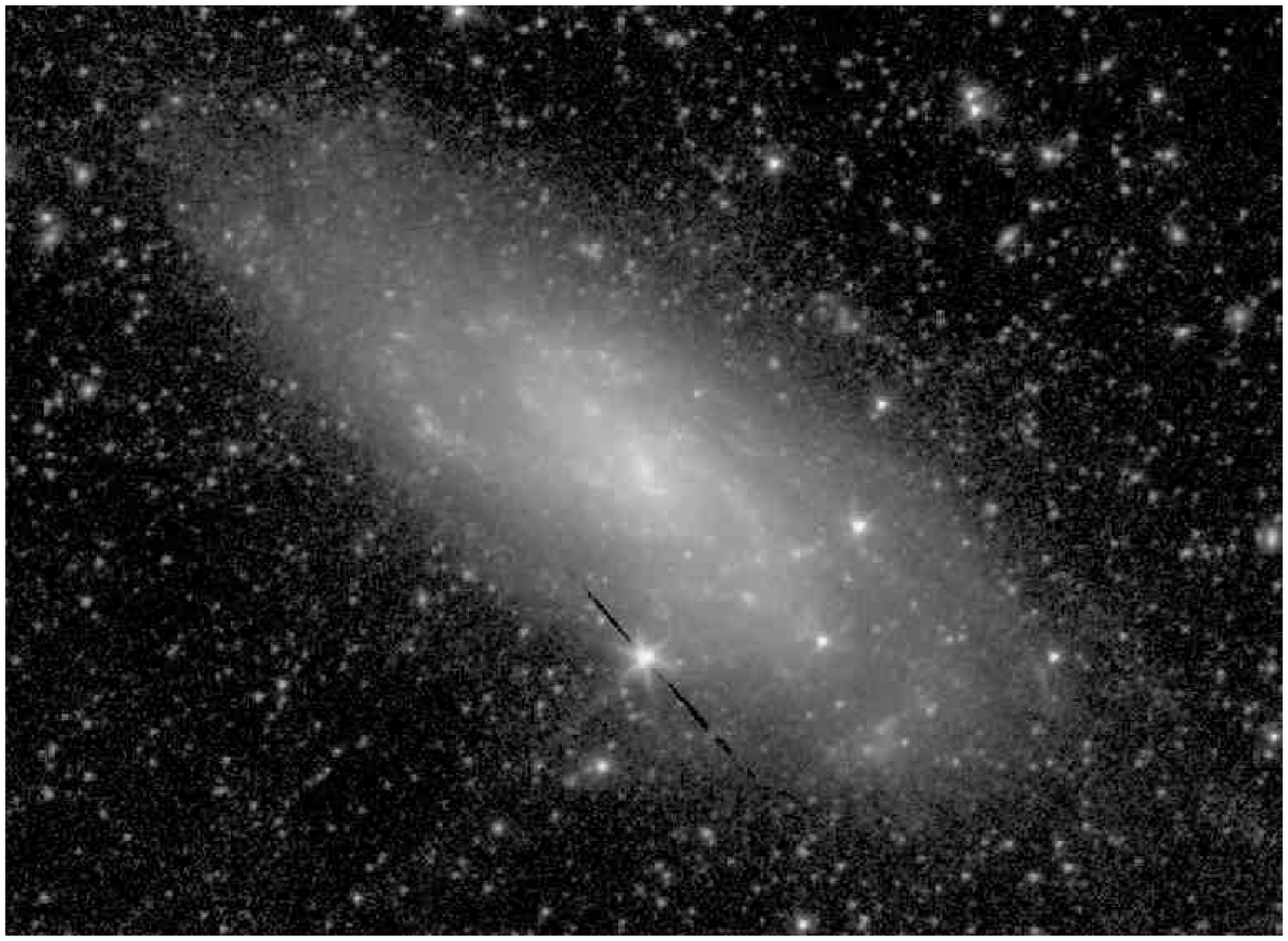}
 \vspace{2.0truecm}
 \caption{
{\bf NGC  4559   }              - S$^4$G mid-IR classification:    SB(s)cd                                               ; Filter: IRAC 3.6$\mu$m; North: left, East: down; Field dimensions:  10.5$\times$  7.7 arcmin; Surface brightness range displayed: 15.5$-$28.0 mag arcsec$^{-2}$}                 
\label{NGC4559}     
 \end{figure}
 
\clearpage
\begin{figure}
\figurenum{1.119}
\plotone{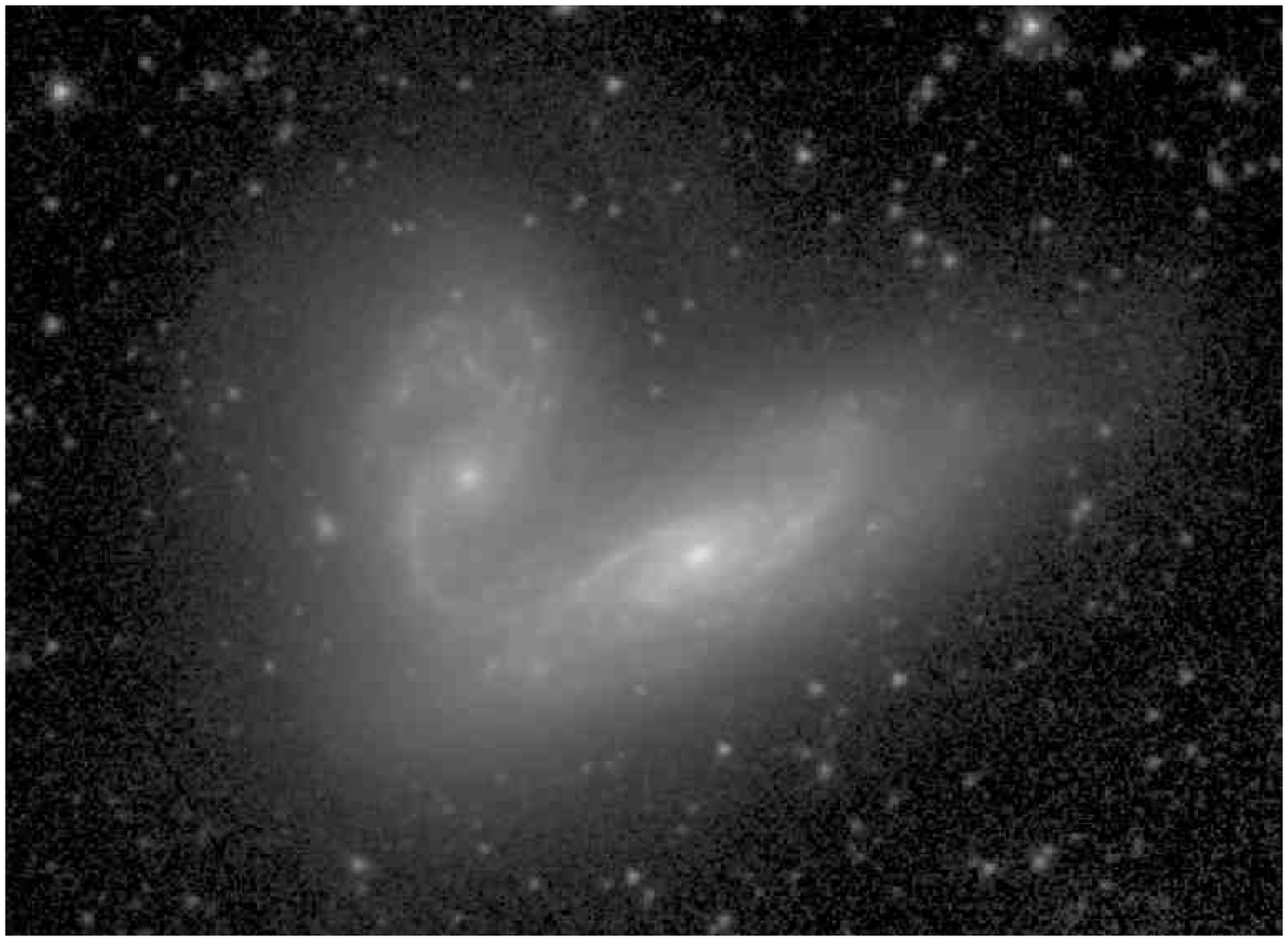}
 \vspace{2.0truecm}
 \caption{
{\bf NGC  4567} (left) and {\bf NGC  4568} (right)      - S$^4$G mid-IR classifications:    SA(rs)bc, SA(r$\underline{\rm s}$)bc, respectively; Filter: IRAC 3.6$\mu$m; North: left, East: down; Field dimensions:   6.3$\times$  4.6 arcmin; Surface brightness range displayed: 13.0$-$28.0 mag arcsec$^{-2}$}                 
\label{NGC4567}     
 \end{figure}
 
\clearpage
\begin{figure}
\figurenum{1.120}
\plotone{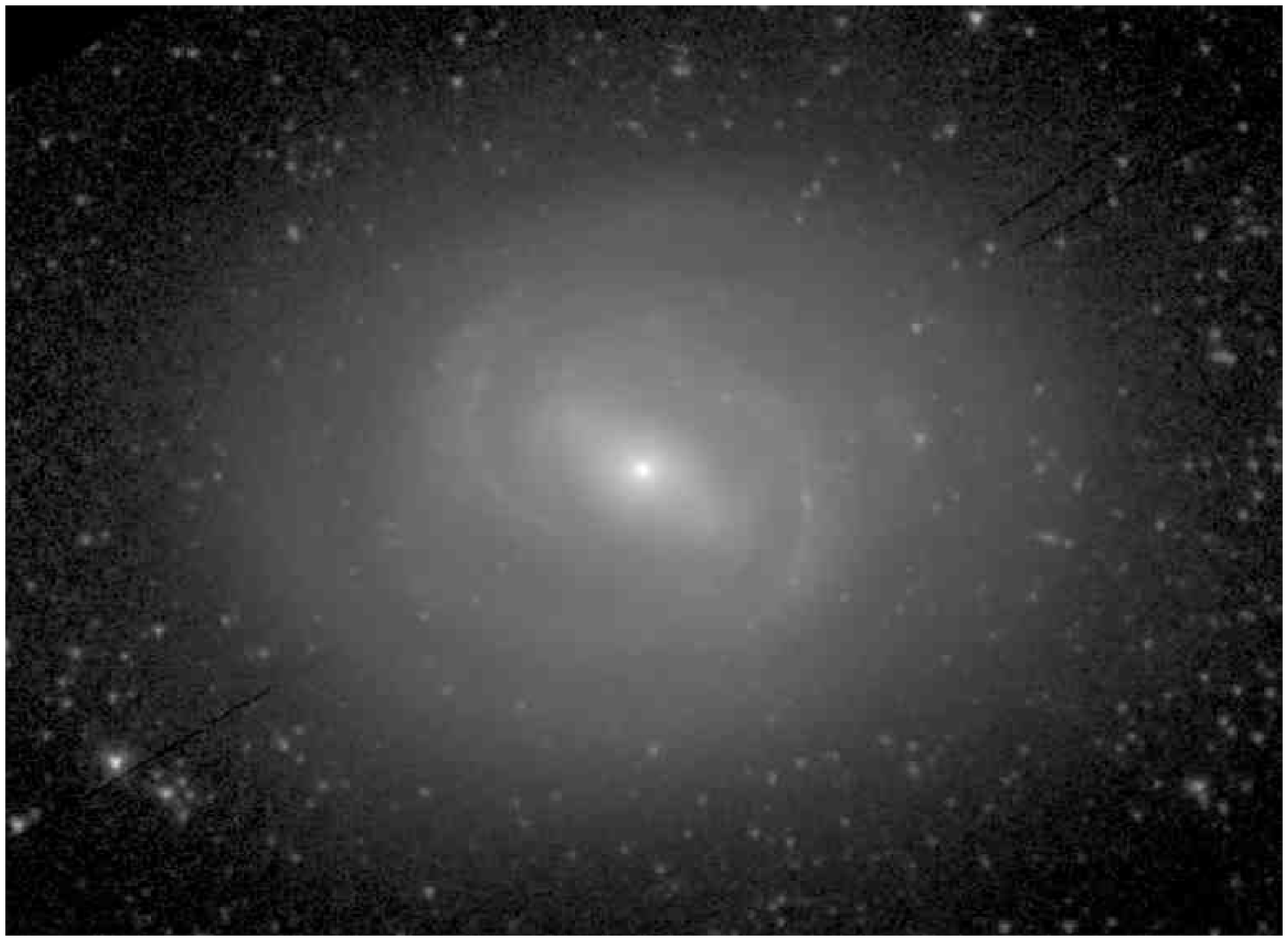}
 \vspace{2.0truecm}
 \caption{
{\bf NGC  4579   }              - S$^4$G mid-IR classification:    (R)SB(r$\underline{\rm s}$)a                          ; Filter: IRAC 3.6$\mu$m; North:   up, East: left; Field dimensions:   7.9$\times$  5.7 arcmin; Surface brightness range displayed: 12.0$-$28.0 mag arcsec$^{-2}$}                 
\label{NGC4579}     
 \end{figure}
 
\clearpage
\begin{figure}
\figurenum{1.121}
\plotone{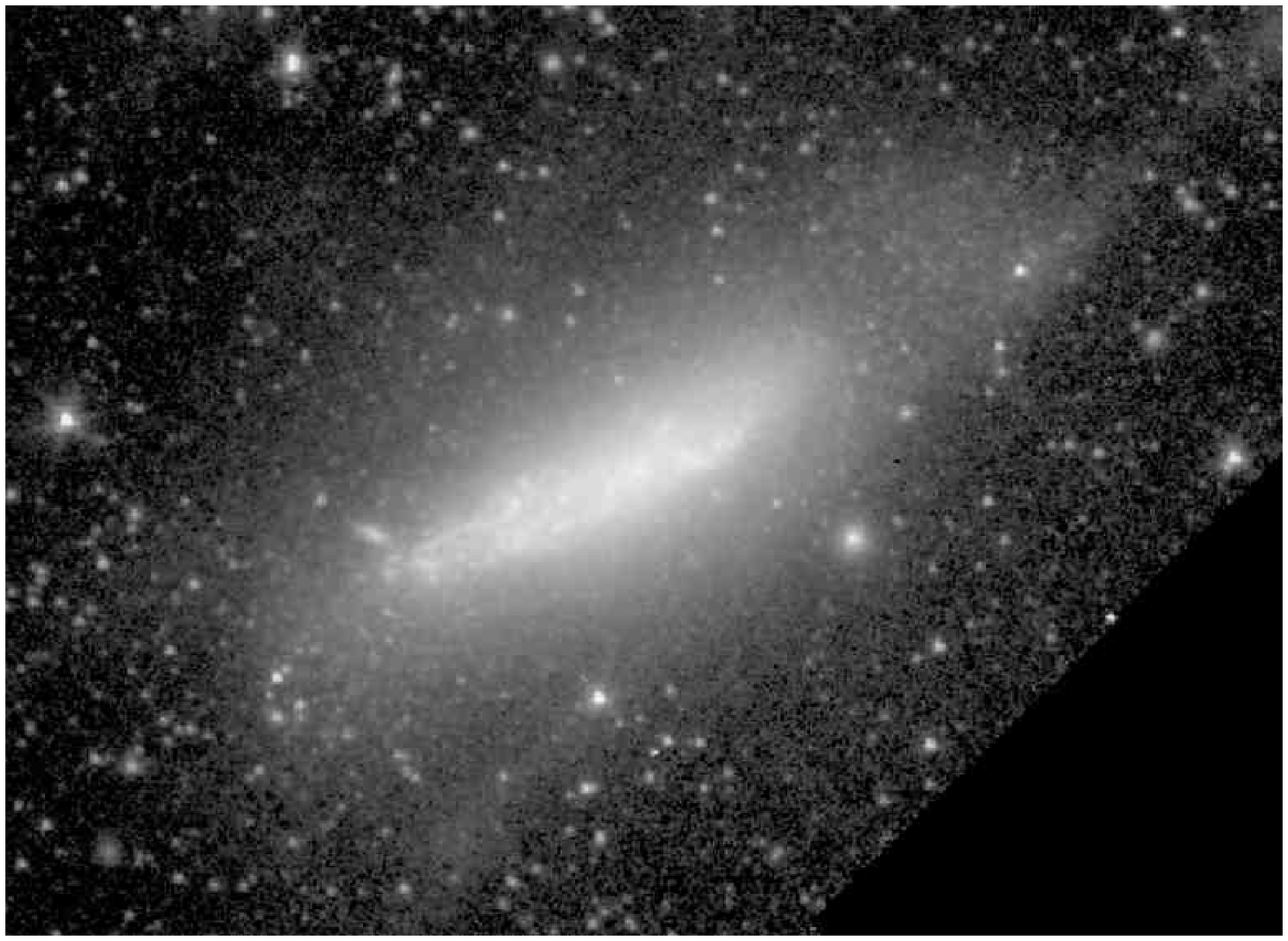}
 \vspace{2.0truecm}
 \caption{
{\bf NGC  4605   }              - S$^4$G mid-IR classification:    IB(s)m sp                                             ; Filter: IRAC 3.6$\mu$m; North:   up, East: left; Field dimensions:   7.9$\times$  5.7 arcmin; Surface brightness range displayed: 16.0$-$28.0 mag arcsec$^{-2}$}                 
\label{NGC4605}     
 \end{figure}
 
\clearpage
\begin{figure}
\figurenum{1.122}
\plotone{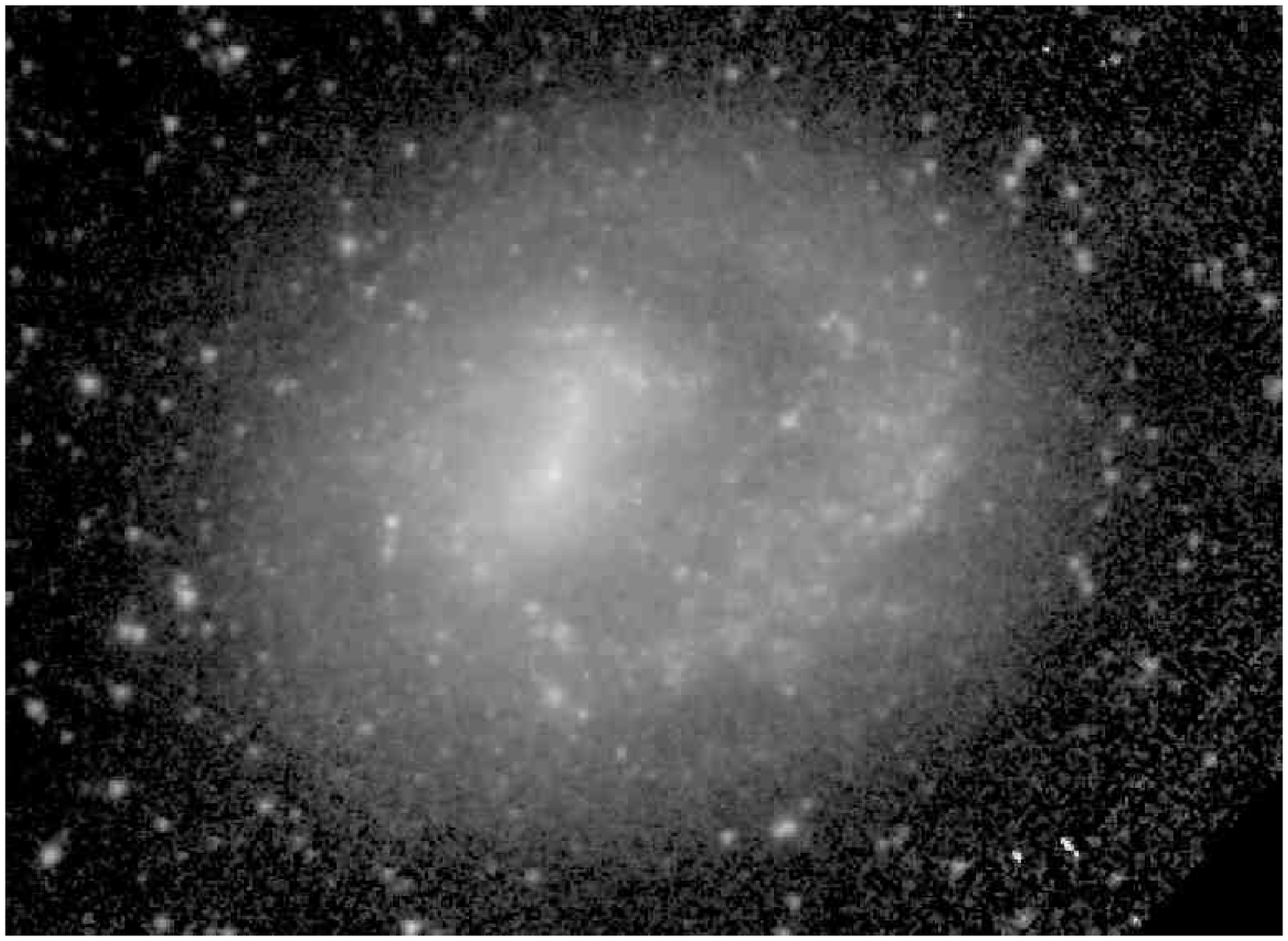}
 \vspace{2.0truecm}
 \caption{
{\bf NGC  4618   }              - S$^4$G mid-IR classification:    (R$^{\prime}$)SB(r$\underline{\rm s}$)m                         ; Filter: IRAC 3.6$\mu$m; North: left, East: down; Field dimensions:   5.7$\times$  4.1 arcmin; Surface brightness range displayed: 16.0$-$28.0 mag arcsec$^{-2}$}                 
\label{NGC4618}     
 \end{figure}
 
\clearpage
\begin{figure}
\figurenum{1.123}
\plotone{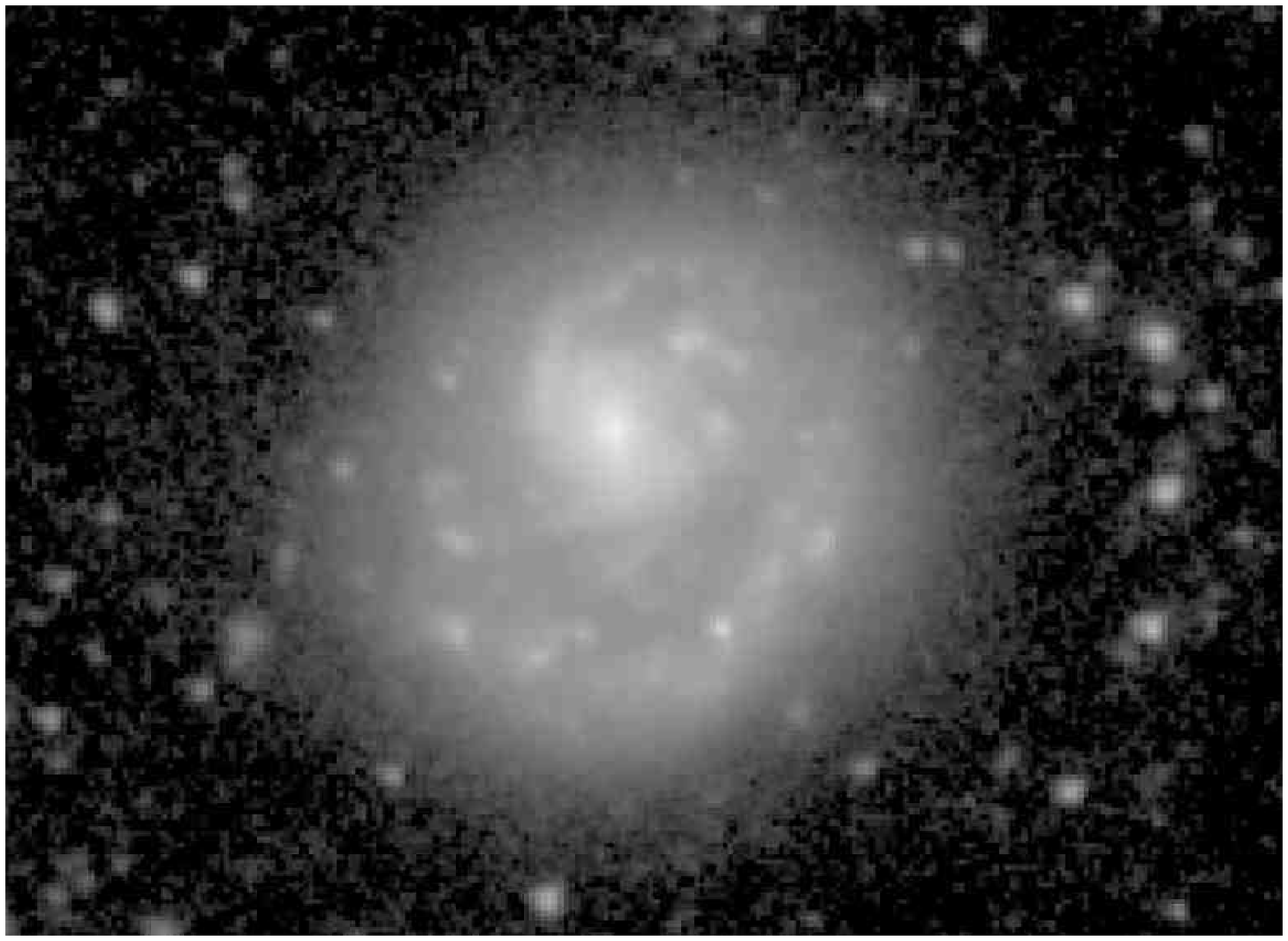}
 \vspace{2.0truecm}
 \caption{
{\bf NGC  4625   }              - S$^4$G mid-IR classification:    (R$^{\prime}$)SAB(rs)m                                          ; Filter: IRAC 3.6$\mu$m; North:   up, East: left; Field dimensions:   2.8$\times$  2.1 arcmin; Surface brightness range displayed: 16.0$-$28.0 mag arcsec$^{-2}$}                 
\label{NGC4625}     
 \end{figure}
 
\clearpage
\begin{figure}
\figurenum{1.124}
\plotone{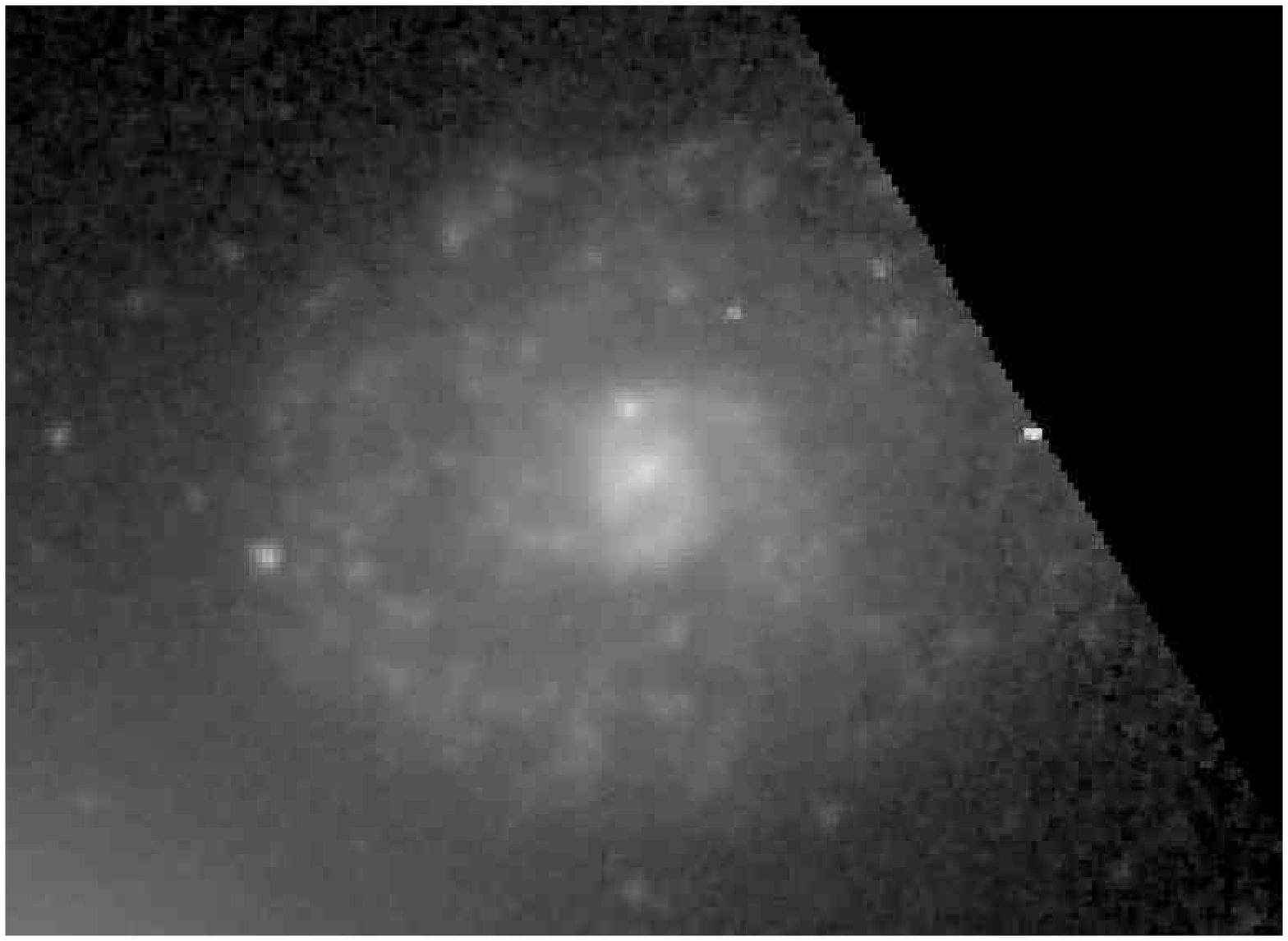}
 \vspace{2.0truecm}
 \caption{
{\bf NGC  4647   }              - S$^4$G mid-IR classification:    SAB(rs)cd                                             ; Filter: IRAC 3.6$\mu$m; North:   up, East: left; Field dimensions:   2.6$\times$  1.9 arcmin; Surface brightness range displayed: 15.0$-$28.0 mag arcsec$^{-2}$}                 
\label{NGC4647}     
 \end{figure}
 
\clearpage
\begin{figure}
\figurenum{1.125}
\plotone{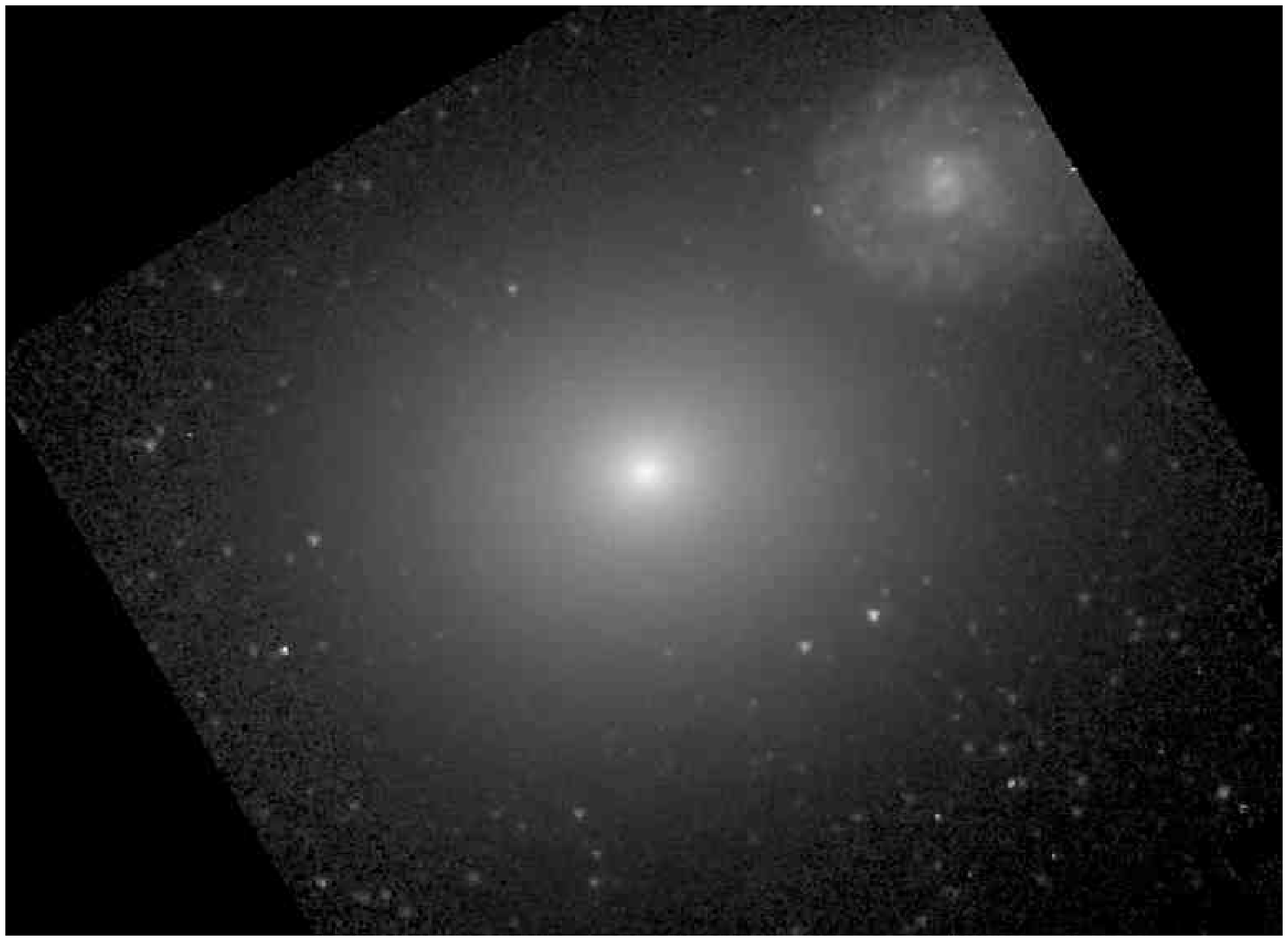}
 \vspace{2.0truecm}
 \caption{
{\bf NGC  4649   }              - S$^4$G mid-IR classification:    SA0$^-$                                        ; Filter: IRAC 3.6$\mu$m; North:   up, East: left; Field dimensions:   7.9$\times$  5.7 arcmin; Surface brightness range displayed: 12.5$-$28.0 mag arcsec$^{-2}$}                 
\label{NGC4649}     
 \end{figure}
 
\clearpage
\begin{figure}
\figurenum{1.126}
\plotone{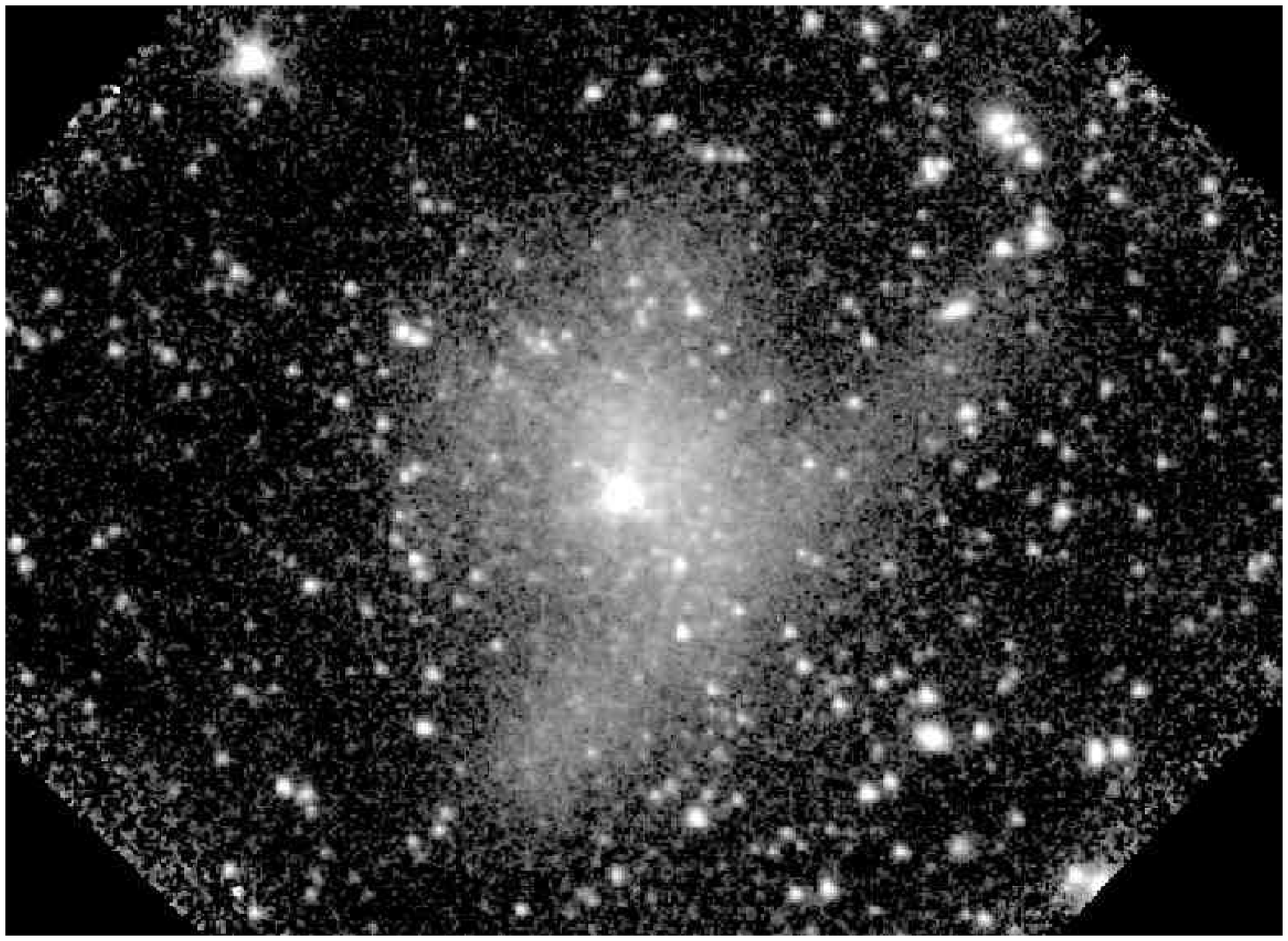}
 \vspace{2.0truecm}
 \caption{
{\bf NGC  4707   }              - S$^4$G mid-IR classification:    Im                                                    ; Filter: IRAC 3.6$\mu$m; North:   up, East: left; Field dimensions:   5.6$\times$  4.1 arcmin; Surface brightness range displayed: 18.5$-$28.0 mag arcsec$^{-2}$}                 
\label{NGC4707}     
 \end{figure}
 
\clearpage
\begin{figure}
\figurenum{1.127}
\plotone{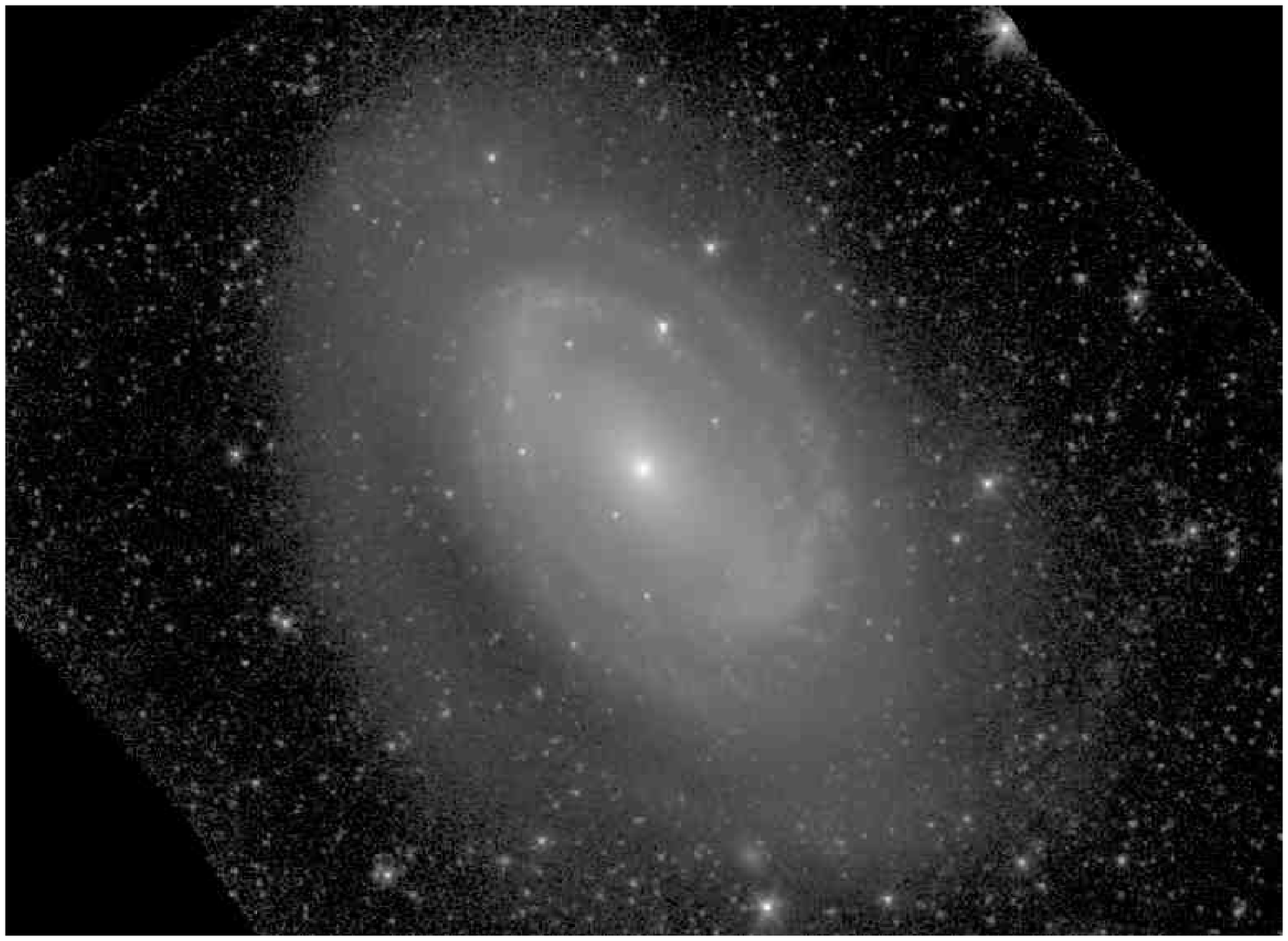}
 \vspace{2.0truecm}
 \caption{
{\bf NGC  4725   }              - S$^4$G mid-IR classification:    SAB(r,nb)a                                            ; Filter: IRAC 3.6$\mu$m; North:   up, East: left; Field dimensions:  14.3$\times$ 10.4 arcmin; Surface brightness range displayed: 12.5$-$28.0 mag arcsec$^{-2}$}                 
\label{NGC4725}     
 \end{figure}
 
\clearpage
\begin{figure}
\figurenum{1.128}
\plotone{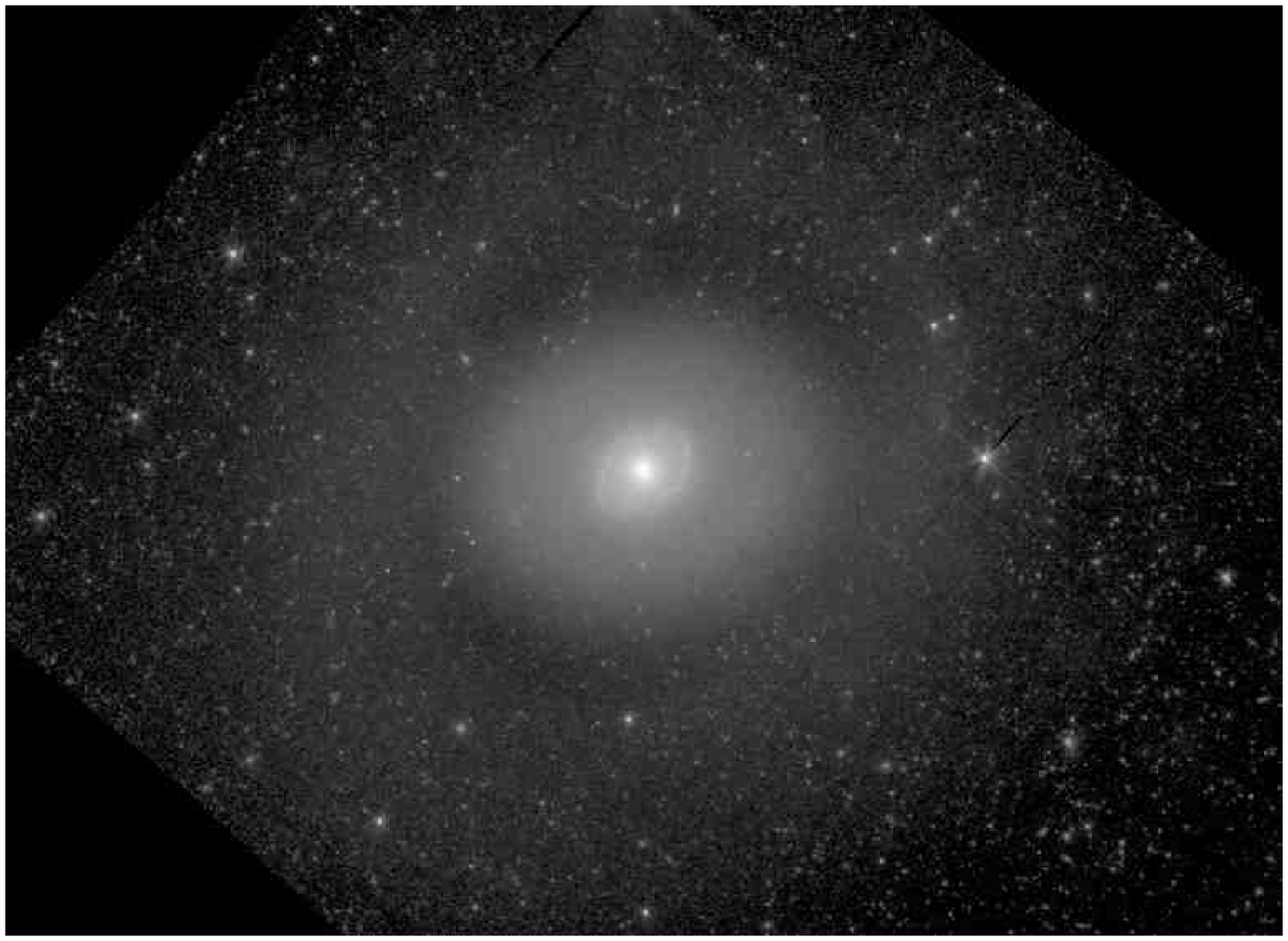}
 \vspace{2.0truecm}
 \caption{
{\bf NGC  4736   }              - S$^4$G mid-IR classification:    (R)S$\underline{\rm A}$B(rl,nr',nl,nb)a               ; Filter: IRAC 3.6$\mu$m; North:   up, East: left; Field dimensions:  21.0$\times$ 15.3 arcmin; Surface brightness range displayed: 12.0$-$28.0 mag arcsec$^{-2}$}                 
\label{NGC4736}     
 \end{figure}
 
\clearpage
\begin{figure}
\figurenum{1.129}
\plotone{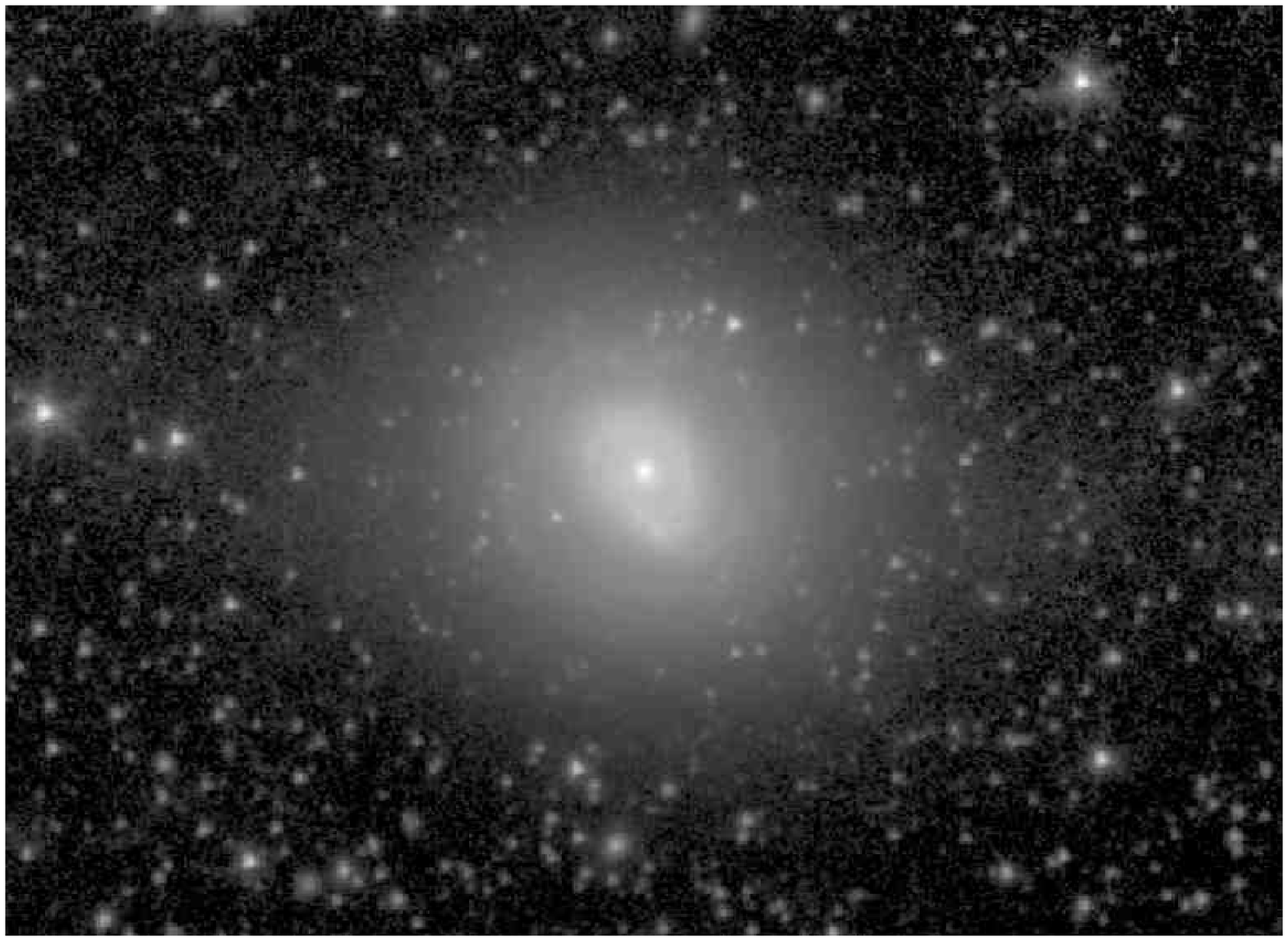}
 \vspace{2.0truecm}
 \caption{
{\bf NGC  4750   }              - S$^4$G mid-IR classification:    (R$^{\prime}$)SA($\underline{\rm r}$s)a                         ; Filter: IRAC 3.6$\mu$m; North: left, East: down; Field dimensions:   6.7$\times$  4.9 arcmin; Surface brightness range displayed: 13.0$-$28.0 mag arcsec$^{-2}$}                 
\label{NGC4750}     
 \end{figure}
 
\clearpage
\begin{figure}
\figurenum{1.130}
\plotone{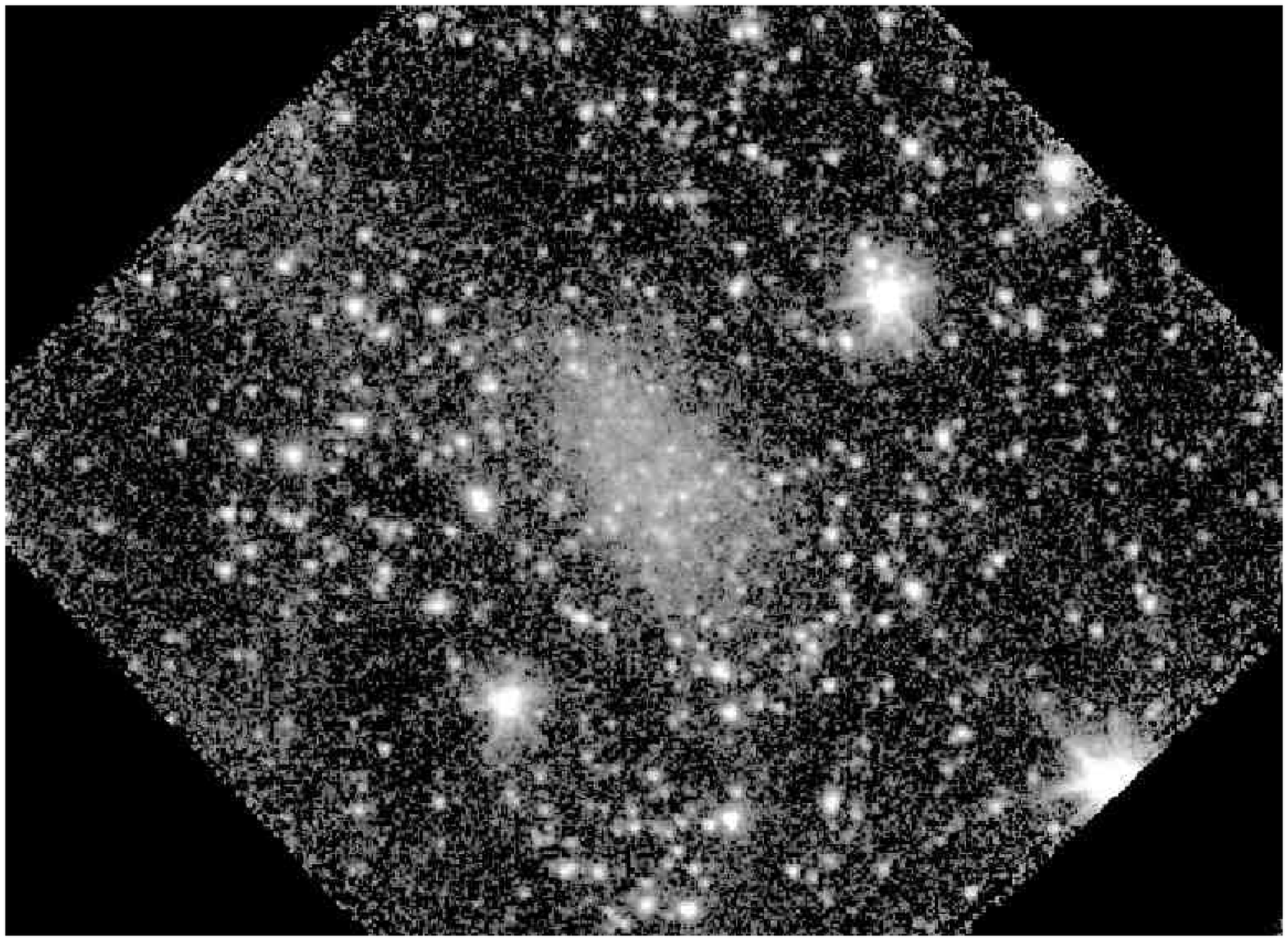}
 \vspace{2.0truecm}
 \caption{
{\bf NGC  4789A  }              - S$^4$G mid-IR classification:    Im                                                    ; Filter: IRAC 3.6$\mu$m; North:   up, East: left; Field dimensions:   6.7$\times$  4.9 arcmin; Surface brightness range displayed: 18.5$-$28.0 mag arcsec$^{-2}$}                 
\label{NGC4789A}    
 \end{figure}
 
\clearpage
\begin{figure}
\figurenum{1.131}
\plotone{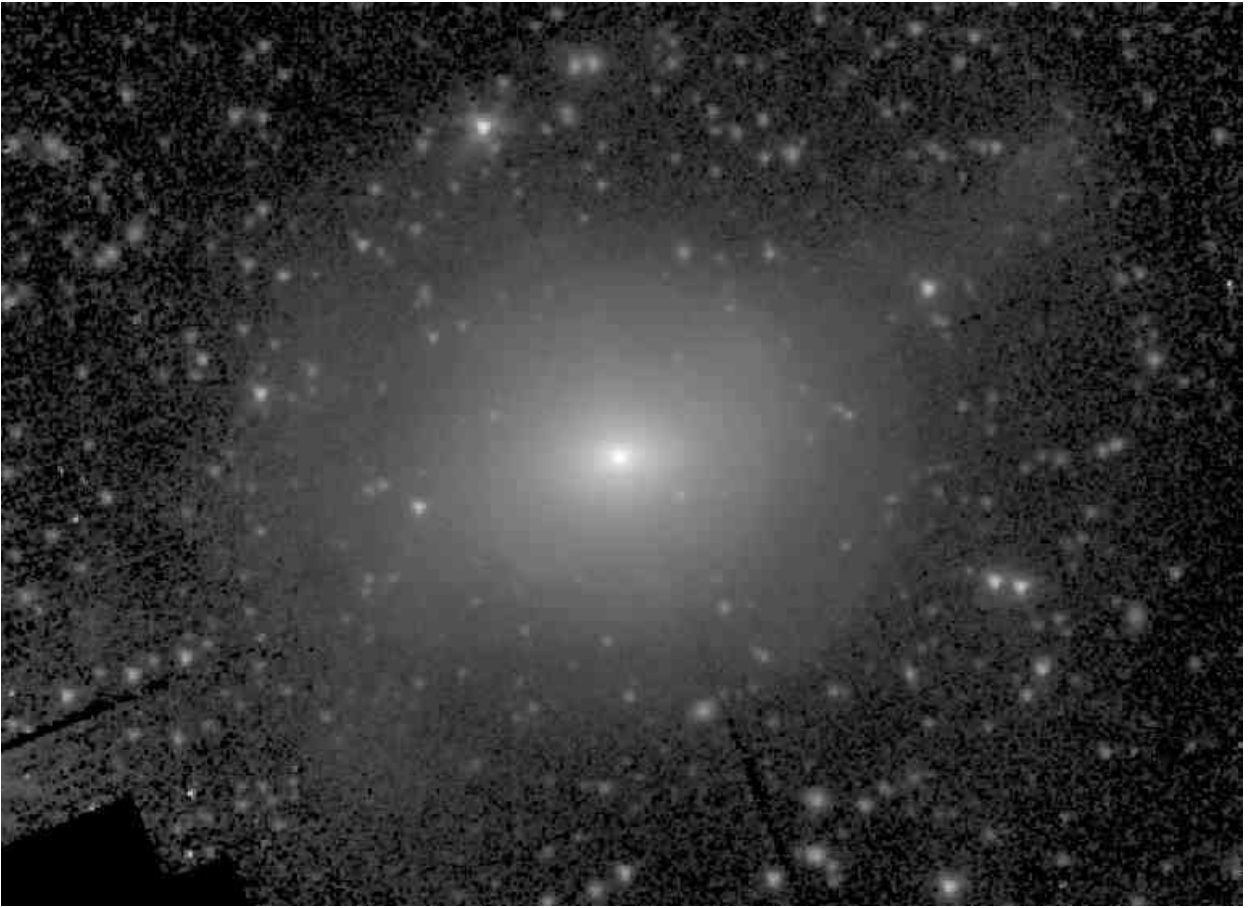}
 \vspace{2.0truecm}
 \caption{
{\bf NGC  5018   }              - S$^4$G mid-IR classification:    SAB0$^-$ (shells/ripples)  pec                                ; Filter: IRAC 3.6$\mu$m; North:   up, East: left; Field dimensions:   6.7$\times$  4.9 arcmin; Surface brightness range displayed: 12.0$-$28.0 mag arcsec$^{-2}$}                 
\label{NGC5018}     
 \end{figure}
 
\clearpage
\begin{figure}
\figurenum{1.132}
\plotone{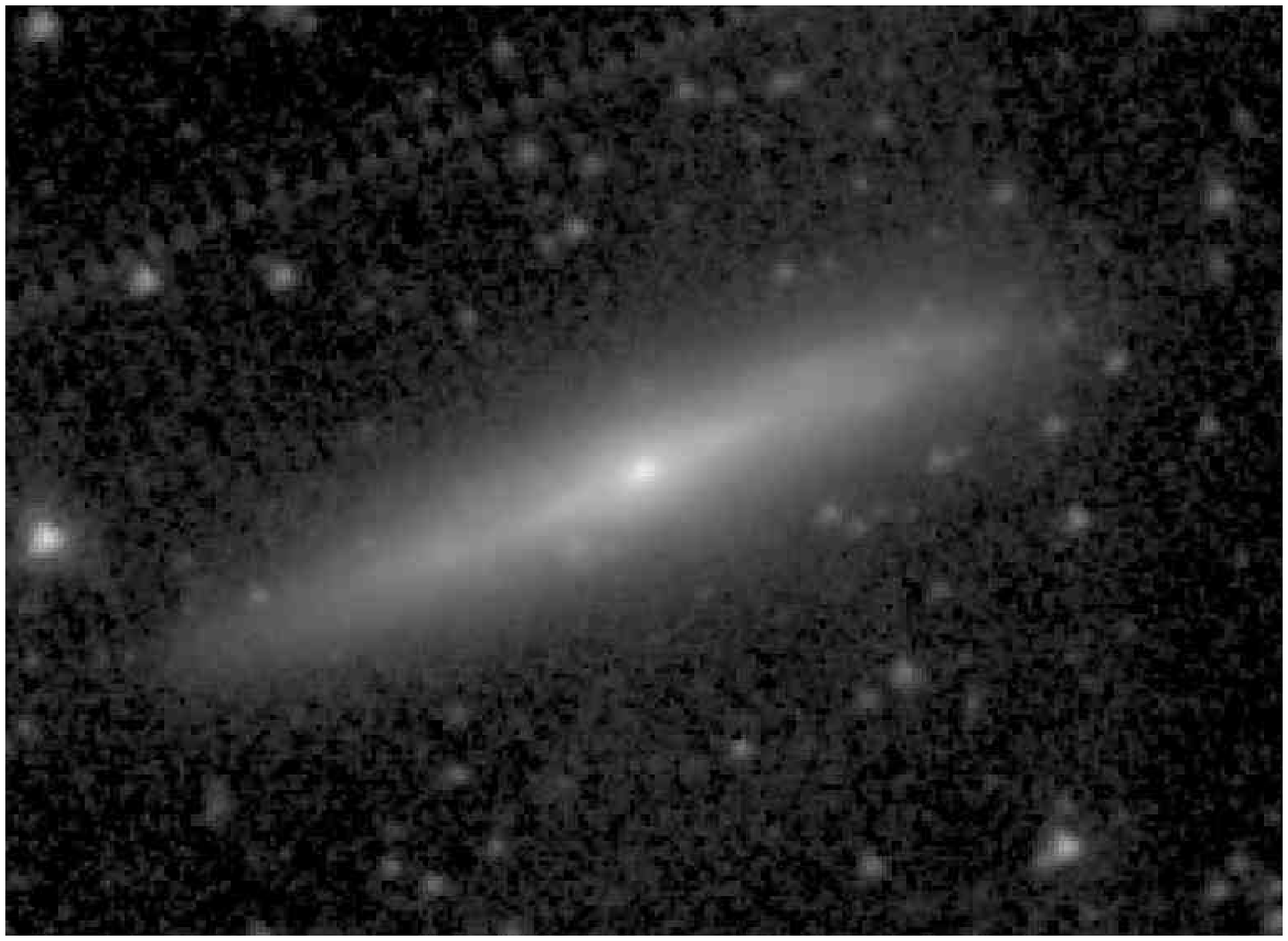}
 \vspace{2.0truecm}
 \caption{
{\bf NGC  5022   }              - S$^4$G mid-IR classification:    Sab sp                                                ; Filter: IRAC 3.6$\mu$m; North: left, East: down; Field dimensions:   3.2$\times$  2.3 arcmin; Surface brightness range displayed: 13.5$-$28.0 mag arcsec$^{-2}$}                 
\label{NGC5022}     
 \end{figure}
 
\clearpage
\begin{figure}
\figurenum{1.133}
\plotone{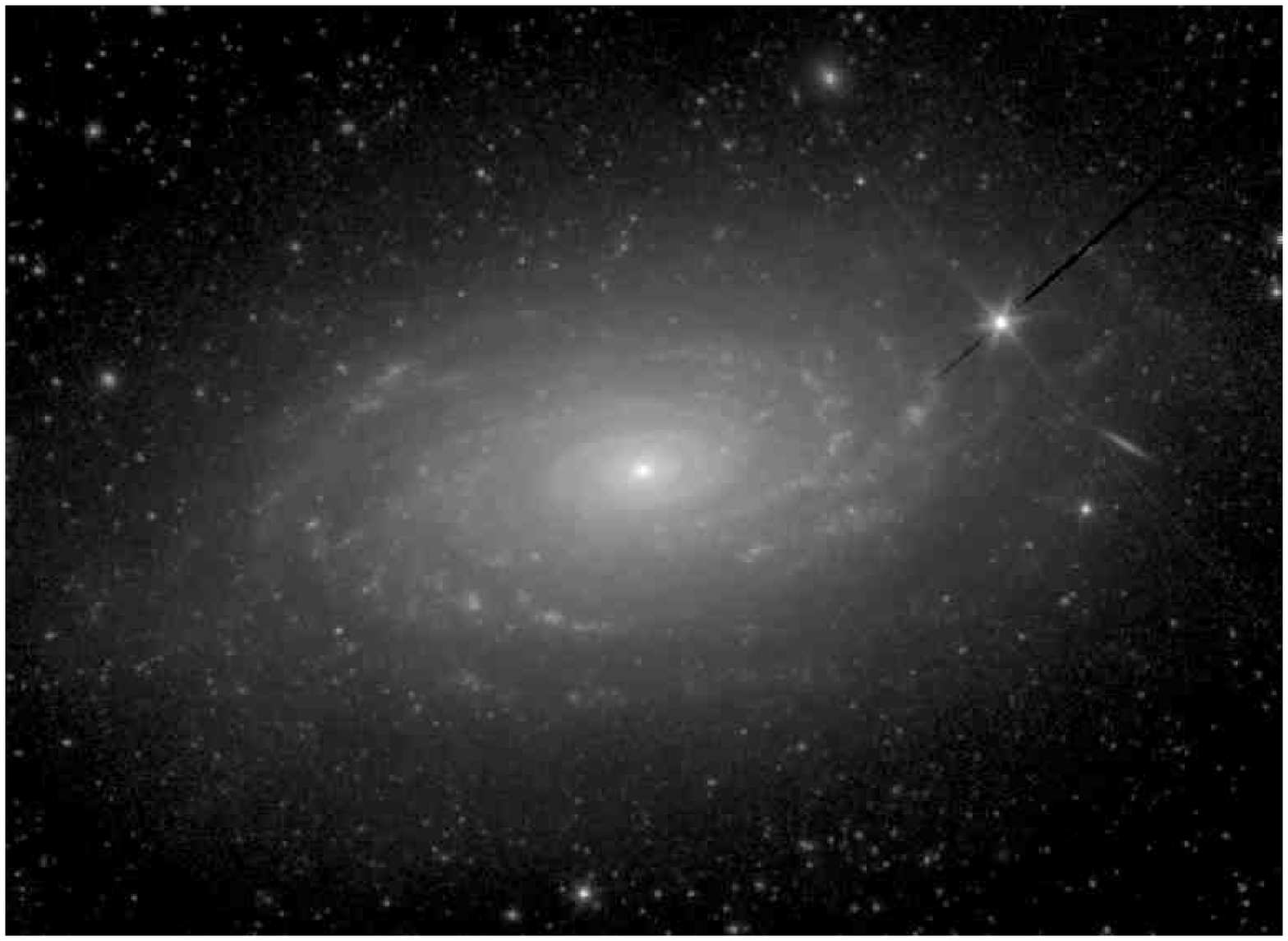}
 \vspace{2.0truecm}
 \caption{
{\bf NGC  5055   }              - S$^4$G mid-IR classification:    SA(rs,rl)bc                                           ; Filter: IRAC 3.6$\mu$m; North:   up, East: left; Field dimensions:  12.6$\times$  9.2 arcmin; Surface brightness range displayed: 12.0$-$28.0 mag arcsec$^{-2}$}                 
\label{NGC5055}     
 \end{figure}
 
\clearpage
\begin{figure}
\figurenum{1.134}
\plotone{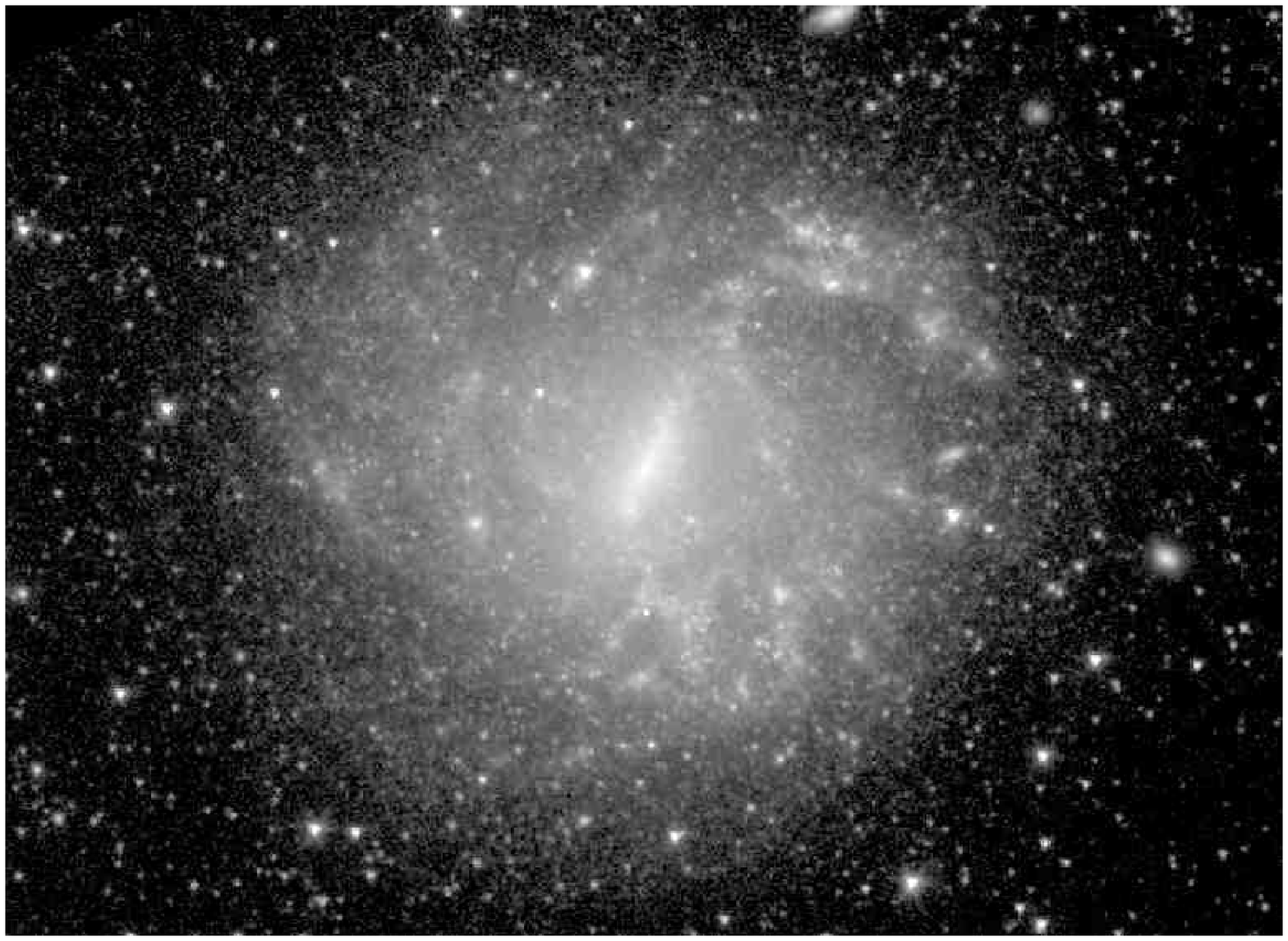}
 \vspace{2.0truecm}
 \caption{
{\bf NGC  5068   }              - S$^4$G mid-IR classification:    SB(r$\underline{\rm s}$)d                             ; Filter: IRAC 3.6$\mu$m; North:   up, East: left; Field dimensions:  10.5$\times$  7.7 arcmin; Surface brightness range displayed: 16.0$-$28.0 mag arcsec$^{-2}$}                 
\label{NGC5068}     
 \end{figure}
 
\clearpage
\begin{figure}
\figurenum{1.135}
\plotone{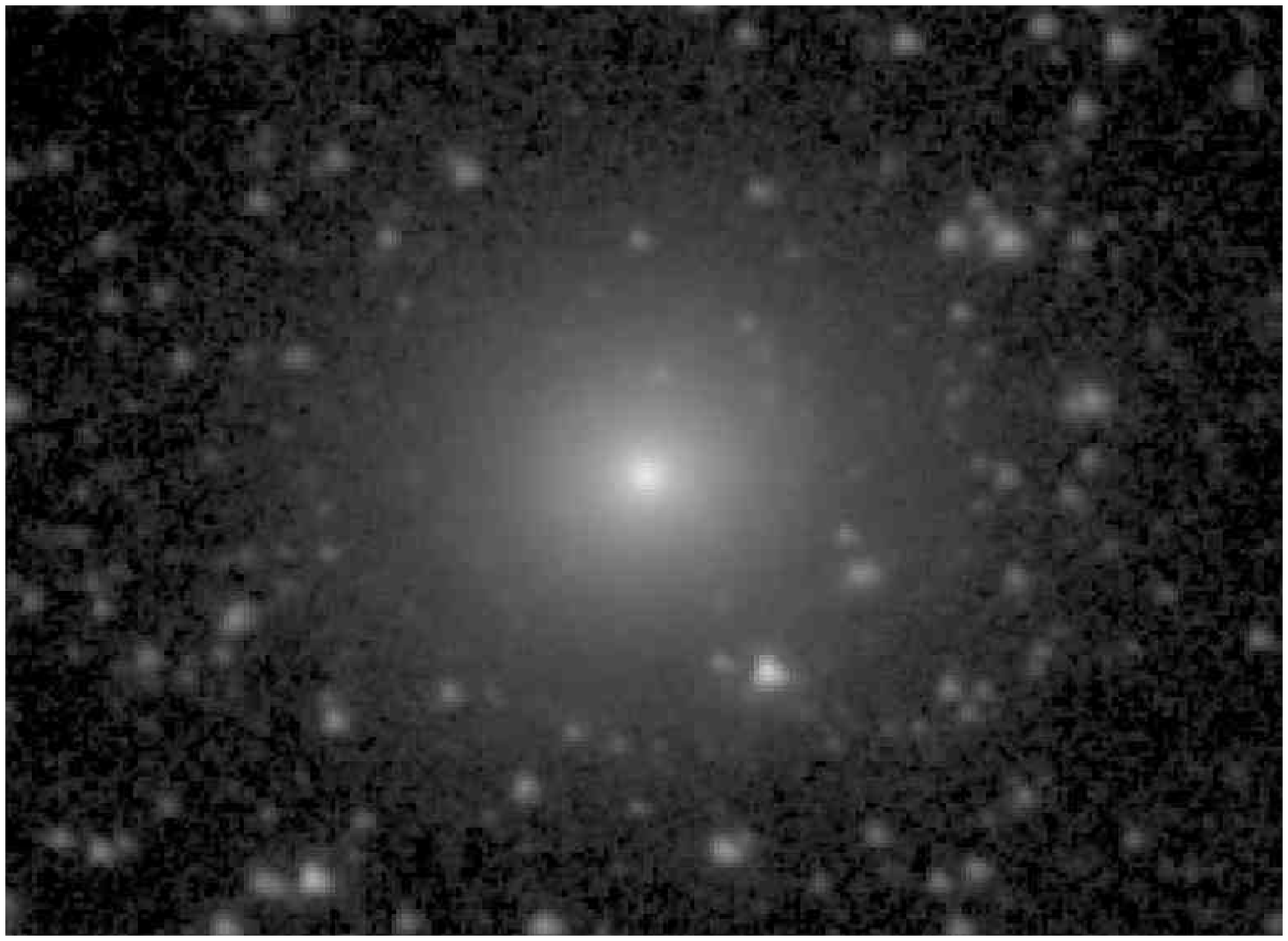}
 \vspace{2.0truecm}
 \caption{
{\bf NGC  5173   }              - S$^4$G mid-IR classification:    E$^+$1                                                    ; Filter: IRAC 3.6$\mu$m; North:   up, East: left; Field dimensions:   4.4$\times$  3.2 arcmin; Surface brightness range displayed: 13.5$-$28.0 mag arcsec$^{-2}$}                 
\label{NGC5173}     
 \end{figure}
 
\clearpage
\begin{figure}
\figurenum{1.136}
\plotone{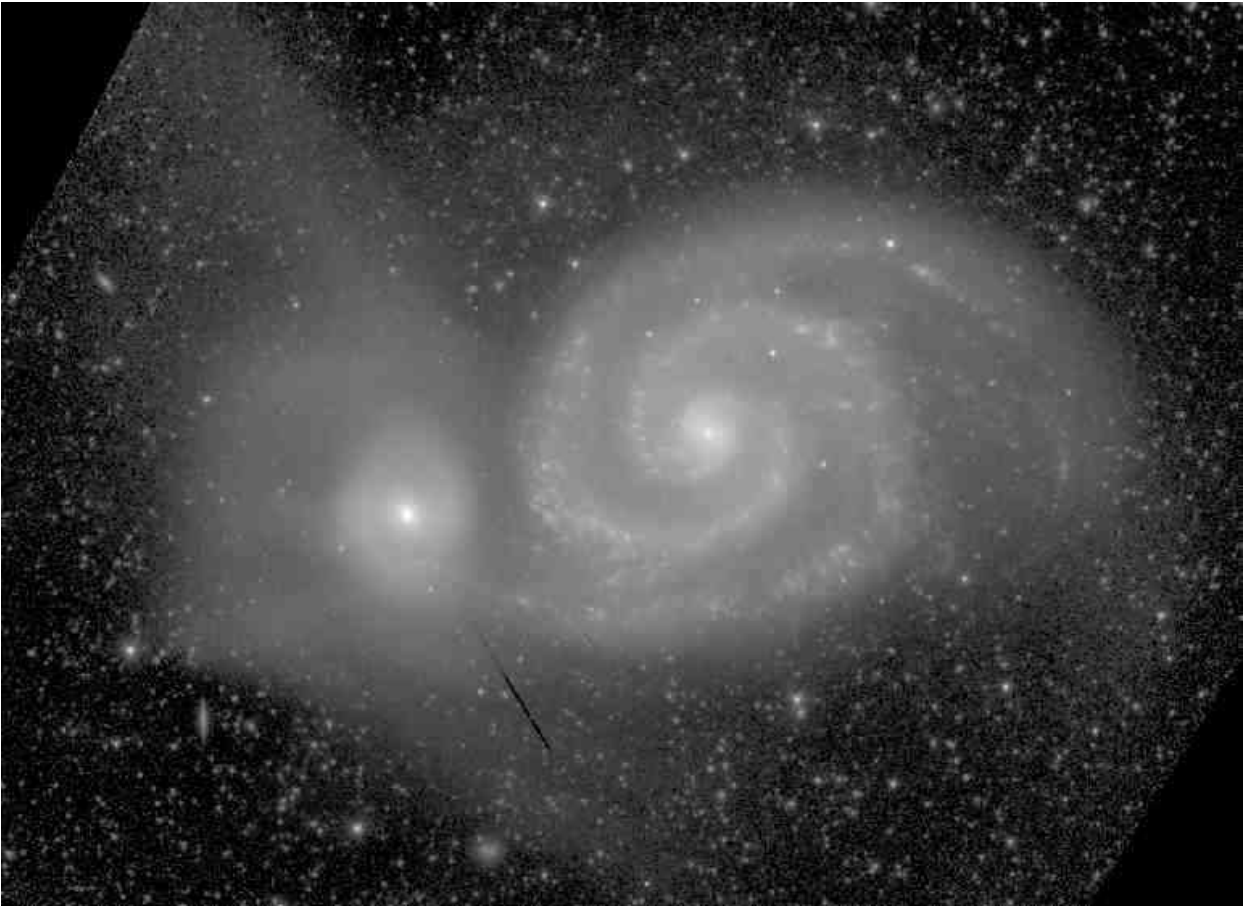}
 \vspace{2.0truecm}
 \caption{
{\bf NGC  5194} (right) and {\bf NGC  5195} (left) - S$^4$G mid-IR classification:    S$\underline{\rm A}$B(rs,nr)bc, SAB(r)0/a; Filter: IRAC 3.6$\mu$m; North: left, East: down; Field dimensions:  17.5$\times$ 12.8 arcmin; Surface brightness range displayed: 12.5$-$28.0 mag arcsec$^{-2}$}                 
\label{NGC5194}     
 \end{figure}
 
\clearpage
\begin{figure}
\figurenum{1.137}
\plotone{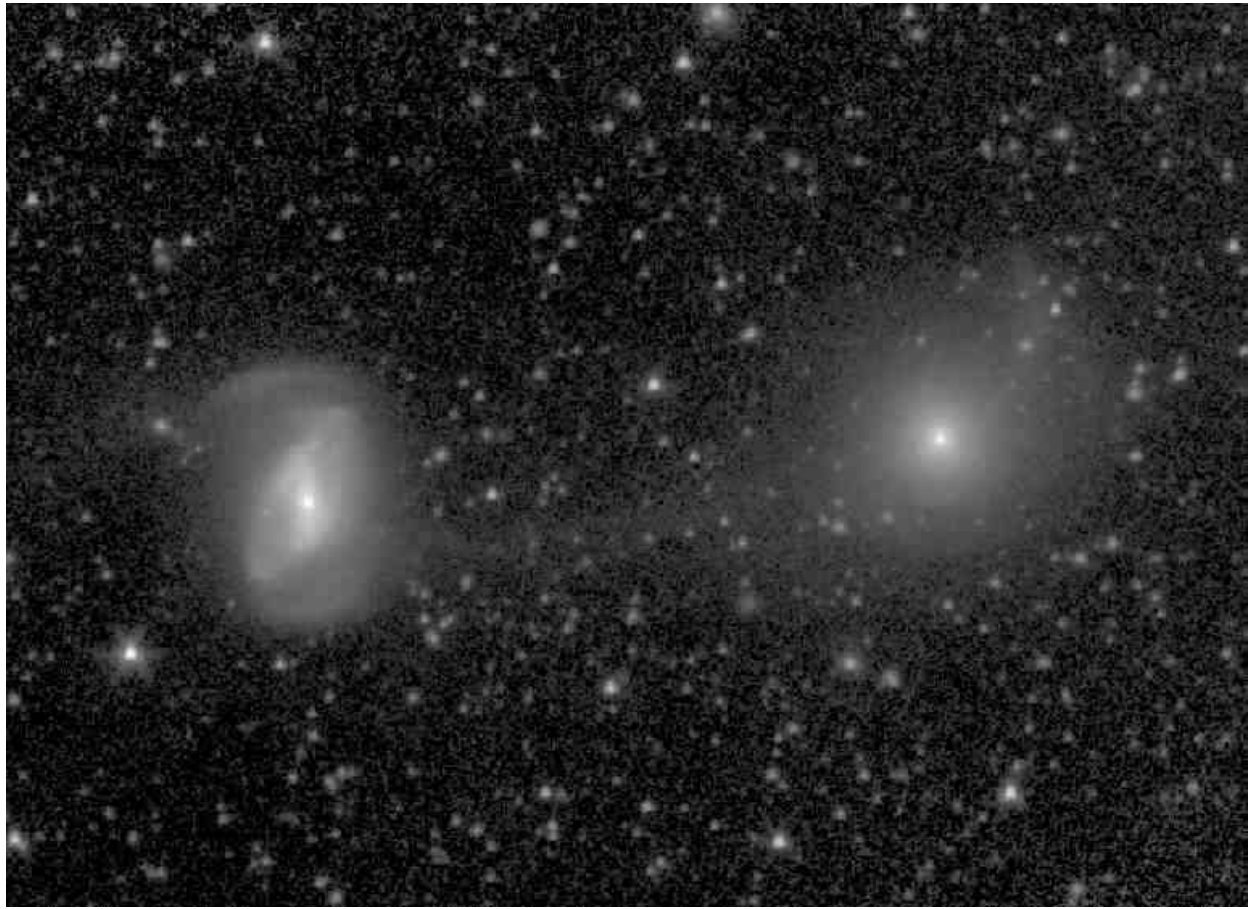}
 \vspace{2.0truecm}
 \caption{
{\bf NGC  5216} (right) and {\bf NGC  5218   } (left) - S$^4$G mid-IR classifications:    E0(shells?) pec, SB(rs)a pec, respectively; Filter: IRAC 3.6$\mu$m; North: left, East: down; Field dimensions:   7.9$\times$  5.7 arcmin; Surface brightness range displayed: 14.0$-$28.0 mag arcsec$^{-2}$}                 
\label{NGC5216}     
 \end{figure}
 
\clearpage
\begin{figure}
\figurenum{1.138}
\plotone{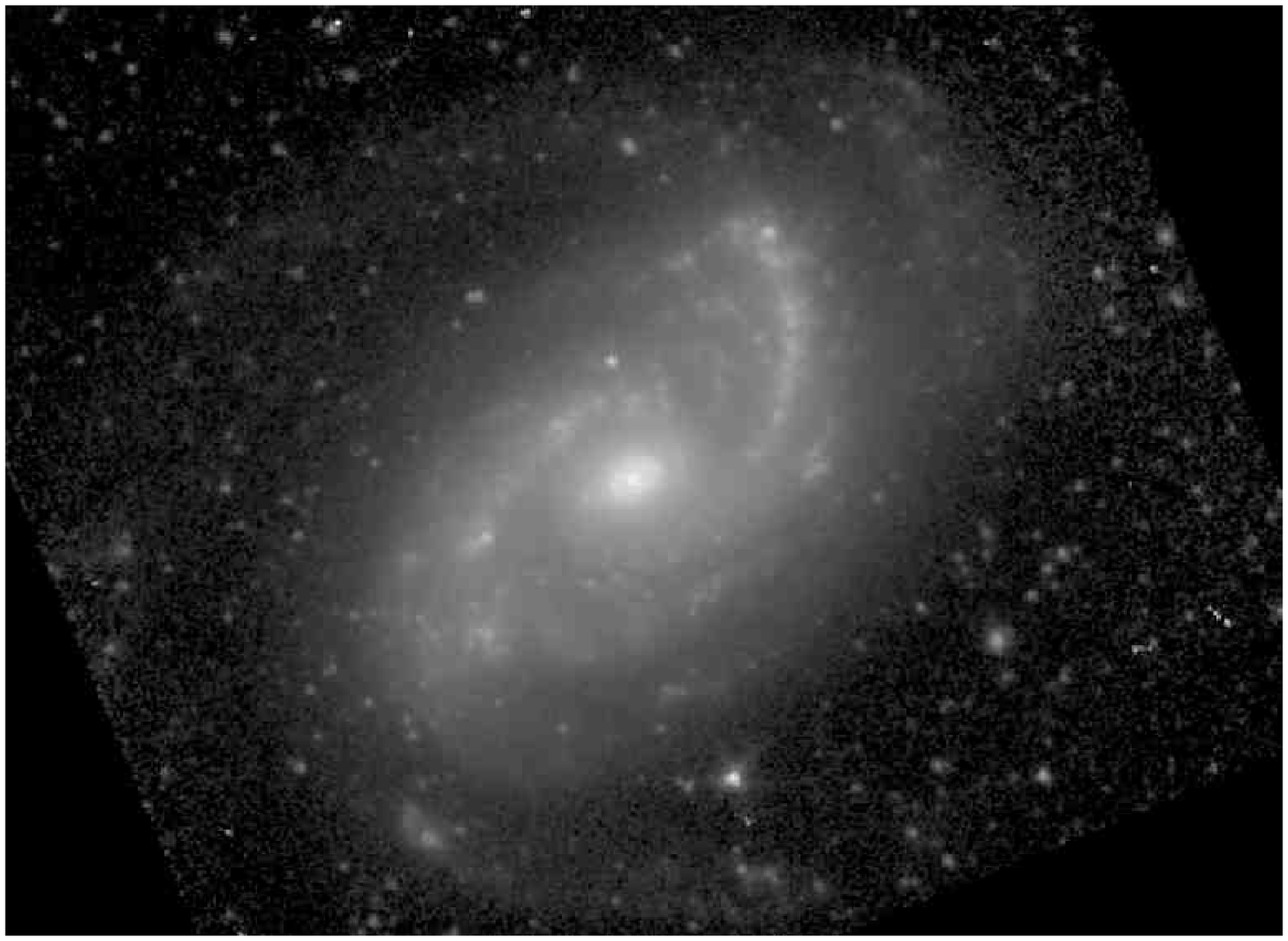}
 \vspace{2.0truecm}
 \caption{
{\bf NGC  5248   }              - S$^4$G mid-IR classification:    SAB(s,nr)b$\underline{\rm c}$                         ; Filter: IRAC 3.6$\mu$m; North:   up, East: left; Field dimensions:   7.0$\times$  5.1 arcmin; Surface brightness range displayed: 13.5$-$28.0 mag arcsec$^{-2}$}                 
\label{NGC5248}     
 \end{figure}
 
\clearpage
\begin{figure}
\figurenum{1.139}
\plotone{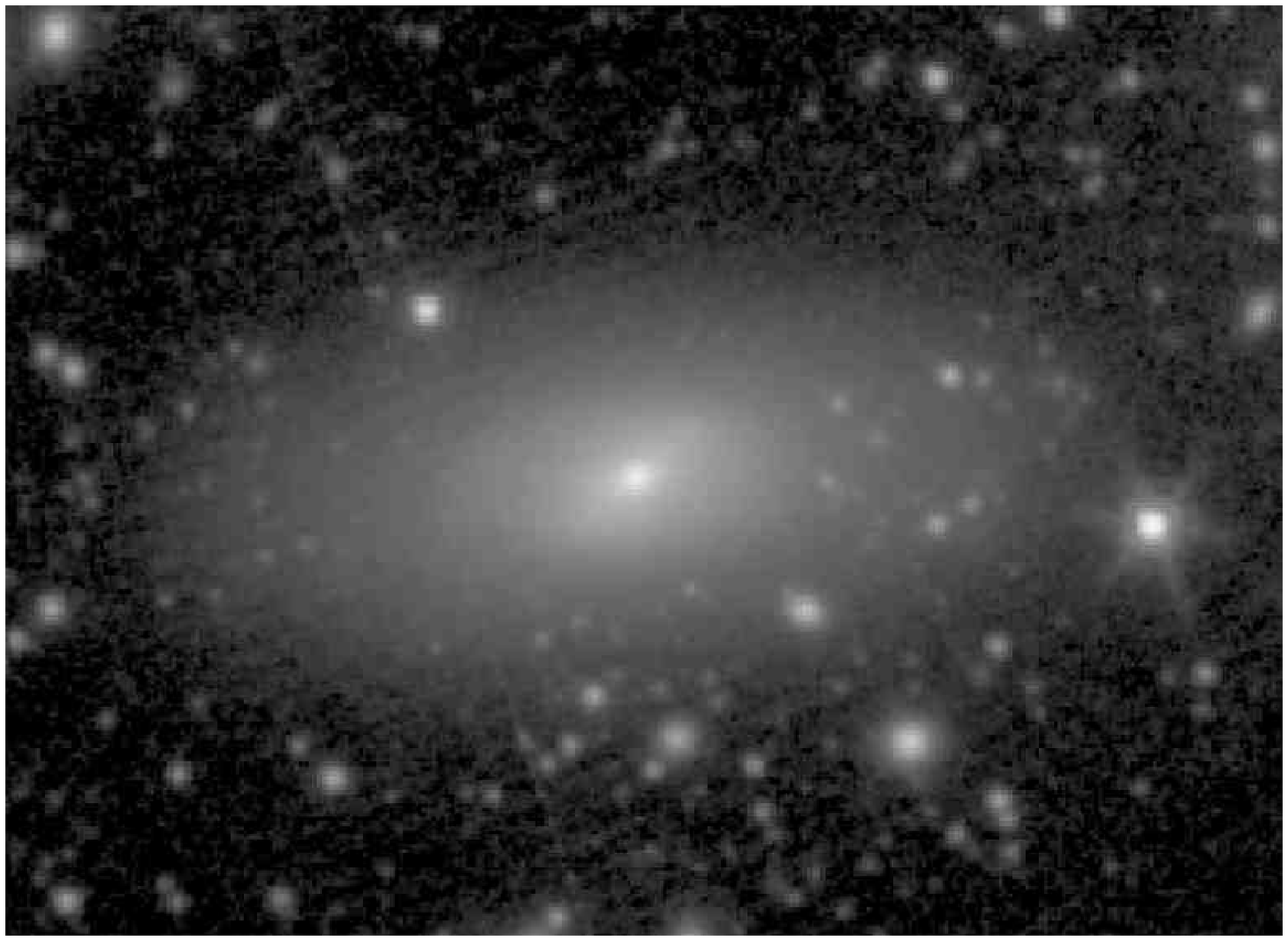}
 \vspace{2.0truecm}
 \caption{
{\bf NGC  5338   }              - S$^4$G mid-IR classification:    SB($\underline{\rm r}$s)0$^o$                         ; Filter: IRAC 3.6$\mu$m; North:   up, East: left; Field dimensions:   3.5$\times$  2.6 arcmin; Surface brightness range displayed: 15.0$-$28.0 mag arcsec$^{-2}$}                 
\label{NGC5338}     
 \end{figure}
 
\clearpage
\begin{figure}
\figurenum{1.140}
\plotone{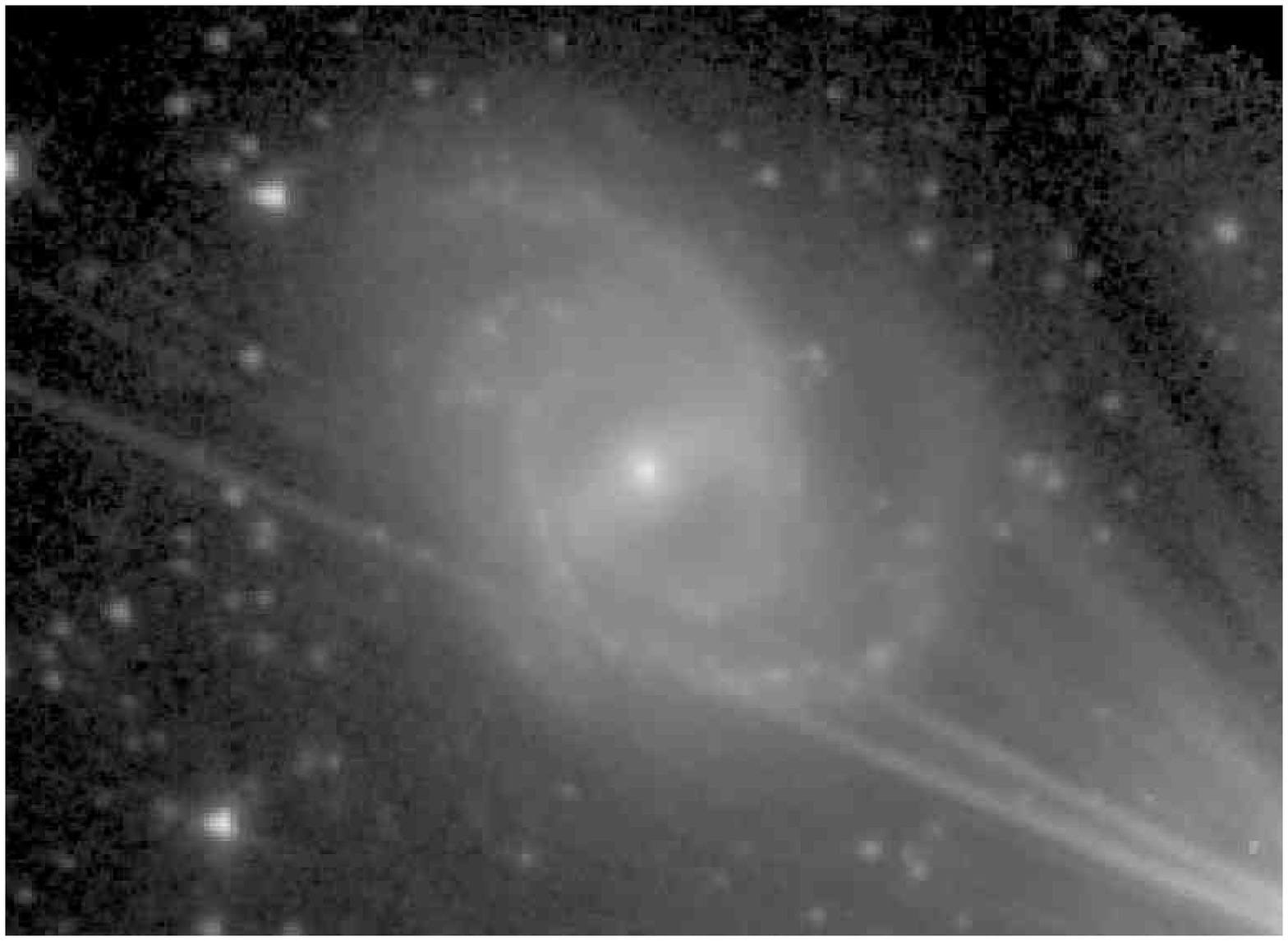}
 \vspace{2.0truecm}
 \caption{
{\bf NGC  5350   }              - S$^4$G mid-IR classification:    SB($\underline{\rm r}$s)a$\underline{\rm b}$          ; Filter: IRAC 3.6$\mu$m; North:   up, East: left; Field dimensions:   3.5$\times$  2.6 arcmin; Surface brightness range displayed: 14.0$-$28.0 mag arcsec$^{-2}$}                 
\label{NGC5350}     
 \end{figure}
 
\clearpage
\begin{figure}
\figurenum{1.141}
\plotone{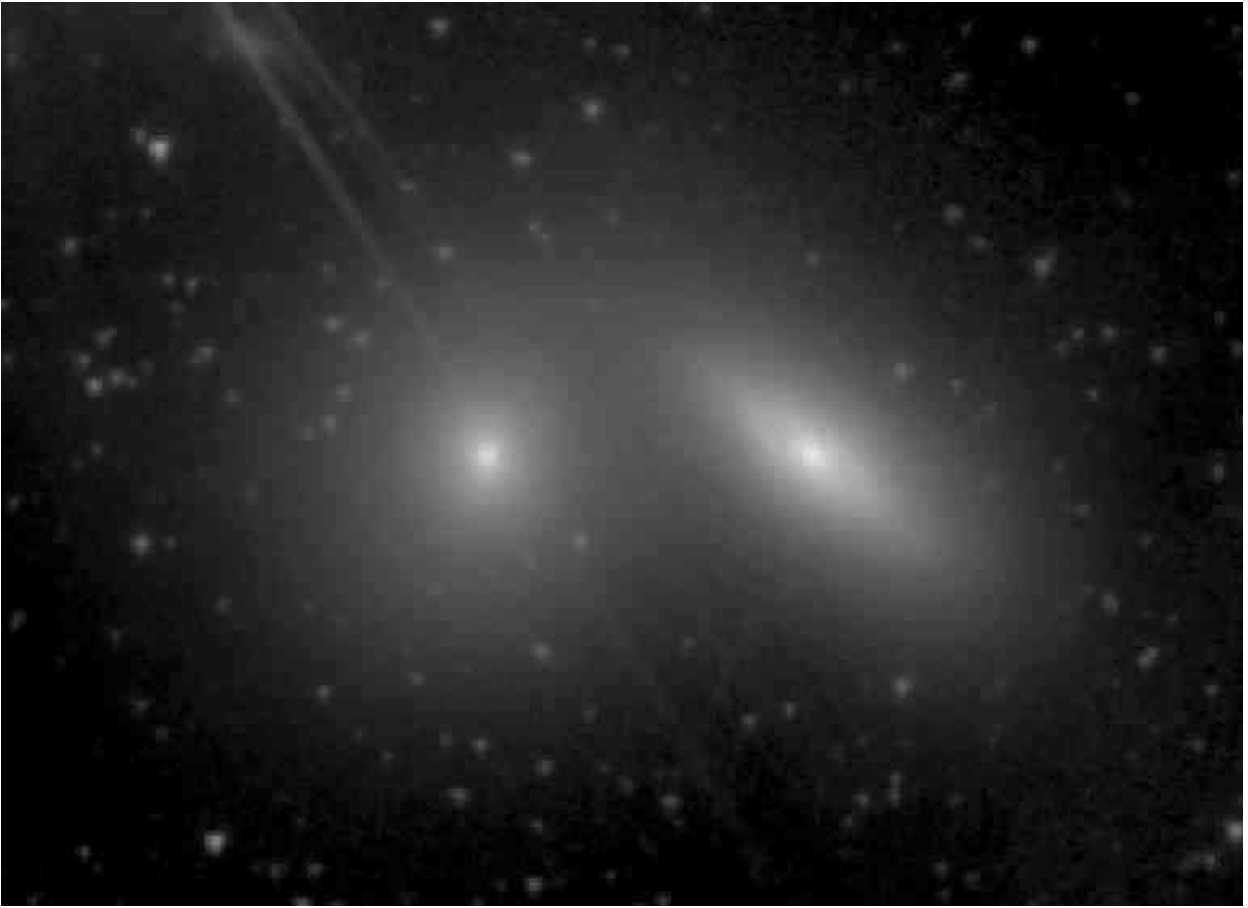}
 \vspace{2.0truecm}
 \caption{
{\bf NGC  5353} (right) and {\bf NGC  5354} (left) - S$^4$G mid-IR classifications:    S0$^+$ sp, SA0$^-$, respectively; Filter: IRAC 3.6$\mu$m; North: left, East: down; Field dimensions:   4.5$\times$  3.3 arcmin; Surface brightness range displayed: 12.5$-$28.0 mag arcsec$^{-2}$}                 
\label{NGC5353}     
 \end{figure}
 
\clearpage
\begin{figure}
\figurenum{1.142}
\plotone{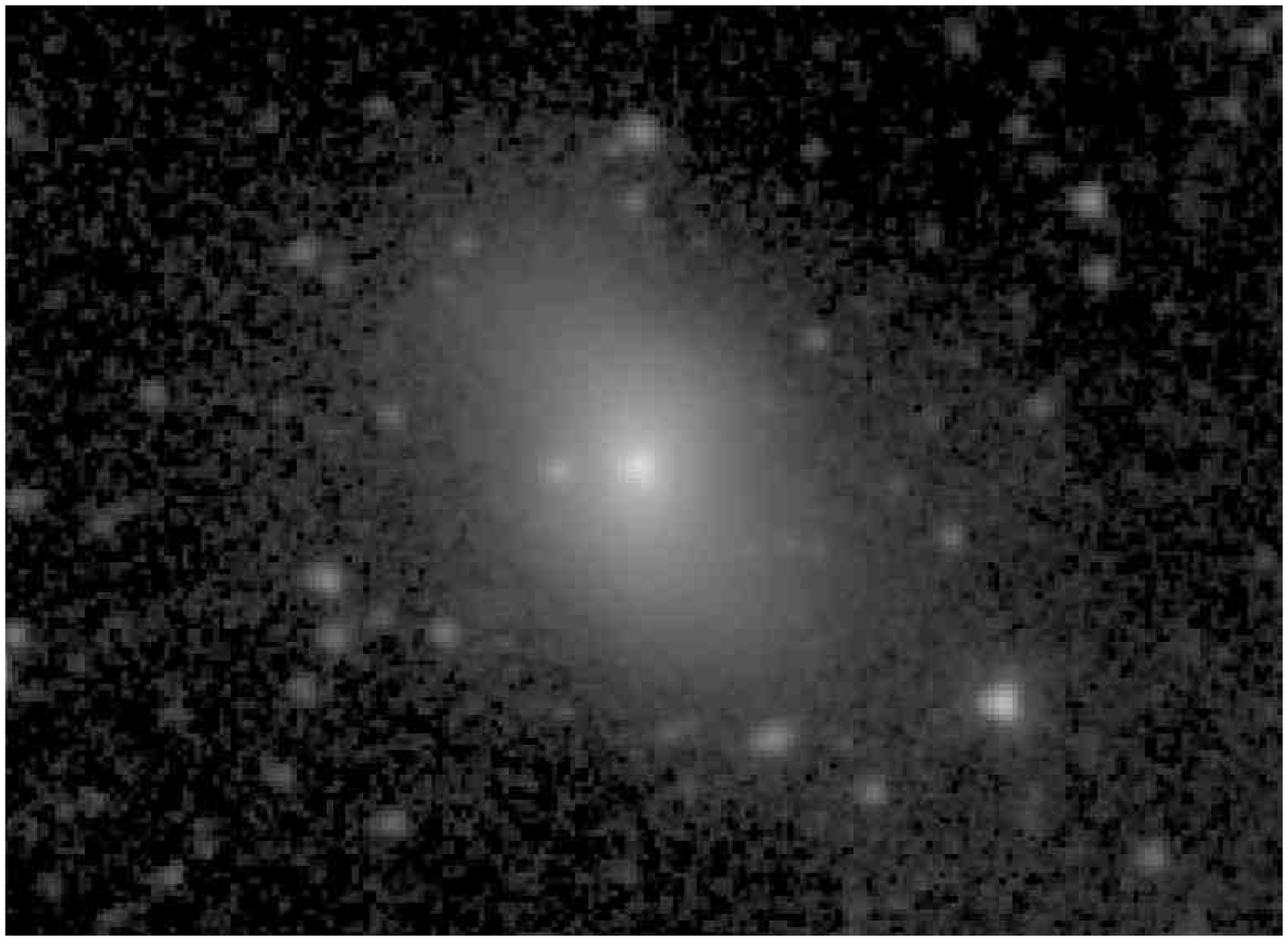}
 \vspace{2.0truecm}
 \caption{
{\bf NGC  5355   }              - S$^4$G mid-IR classification:    SAB(s)0$^o$                                           ; Filter: IRAC 3.6$\mu$m; North:   up, East: left; Field dimensions:   2.6$\times$  1.9 arcmin; Surface brightness range displayed: 14.0$-$28.0 mag arcsec$^{-2}$}                 
\label{NGC5355}     
 \end{figure}
 
\clearpage
\begin{figure}
\figurenum{1.143}
\plotone{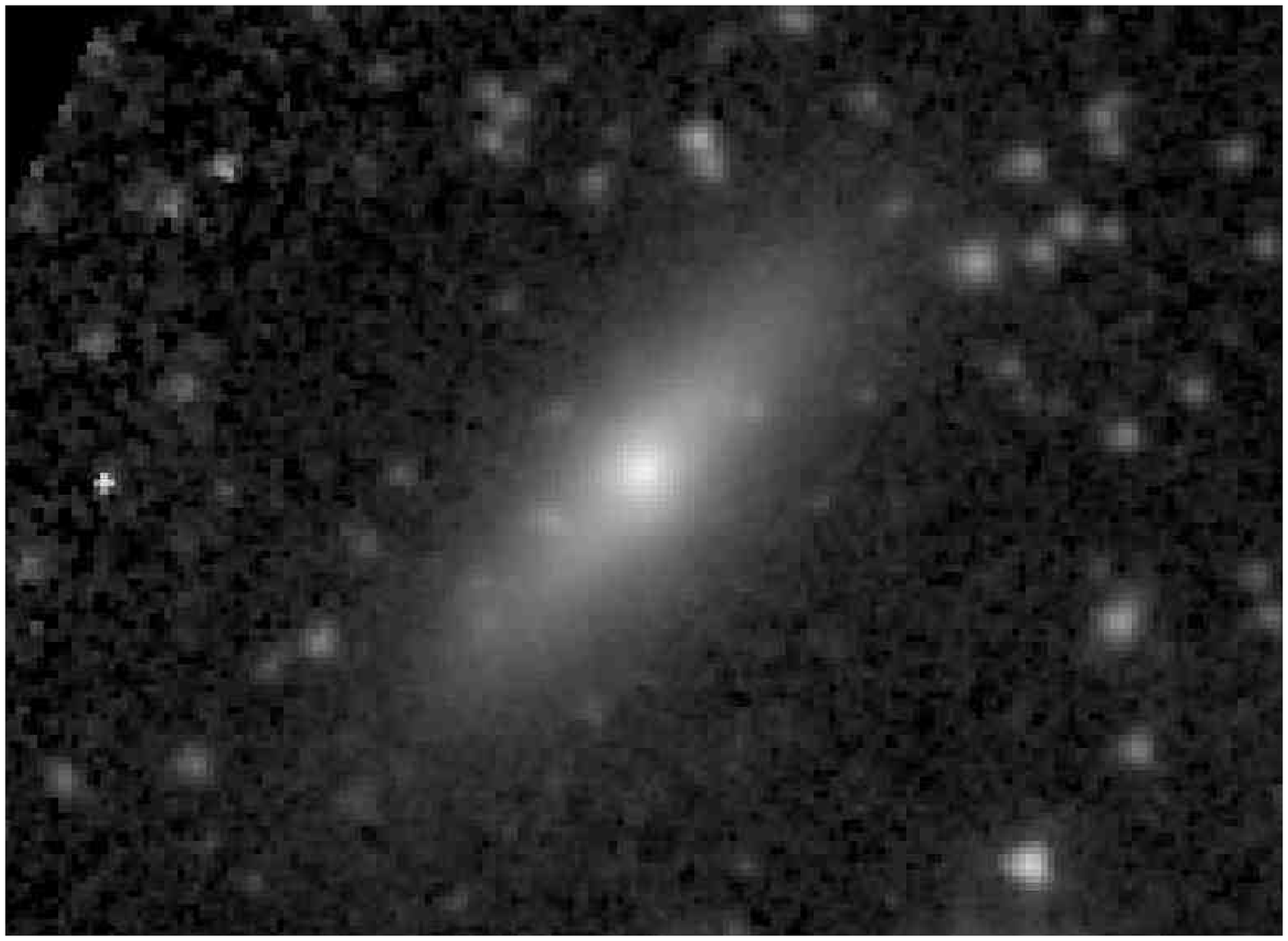}
 \vspace{2.0truecm}
 \caption{
{\bf NGC  5358   }              - S$^4$G mid-IR classification:    S0$^o$ sp                                             ; Filter: IRAC 3.6$\mu$m; North:   up, East: left; Field dimensions:   2.3$\times$  1.6 arcmin; Surface brightness range displayed: 15.0$-$28.0 mag arcsec$^{-2}$}                 
\label{NGC5358}     
 \end{figure}
 
\clearpage
\begin{figure}
\figurenum{1.144}
\plotone{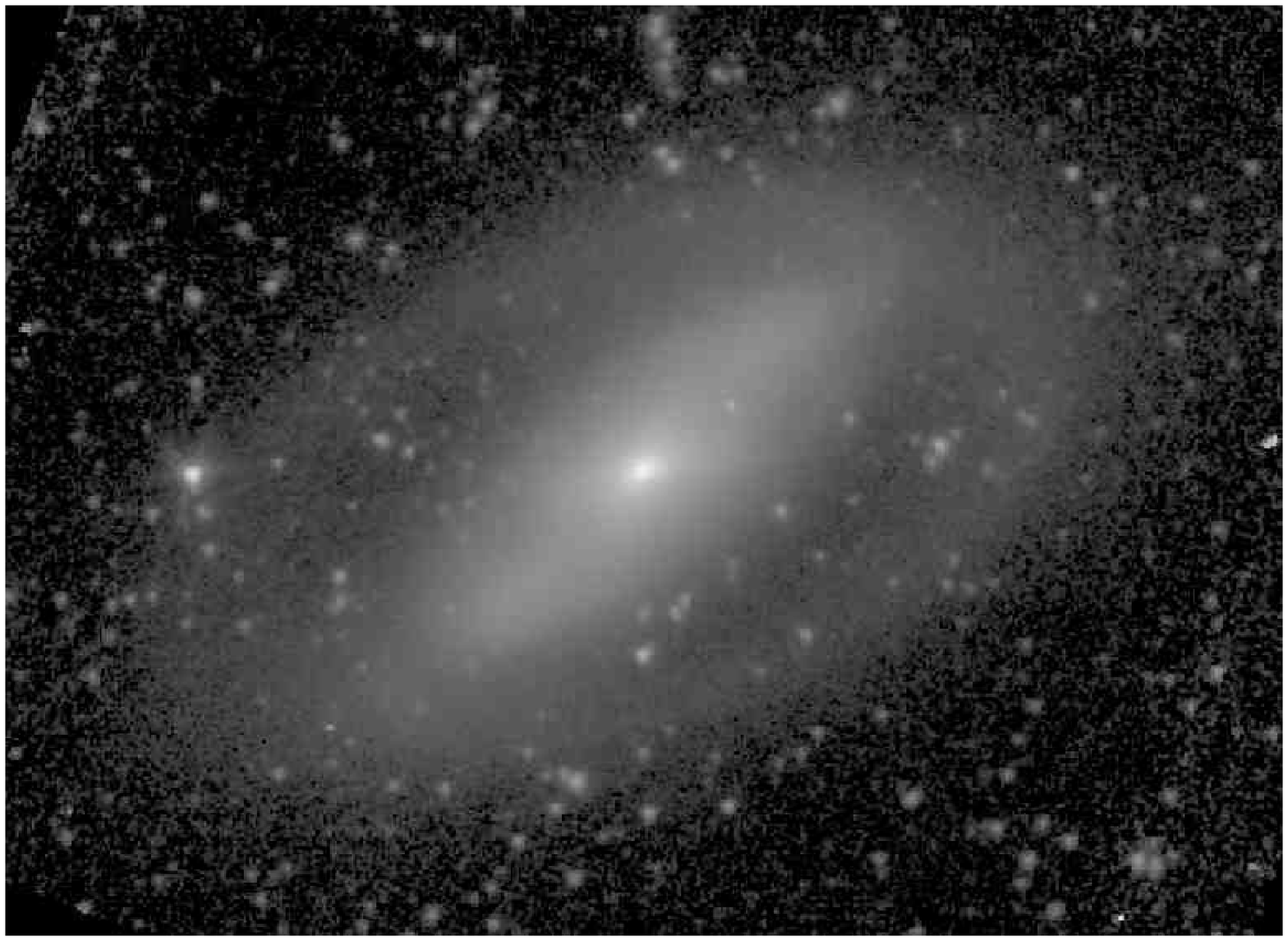}
 \vspace{2.0truecm}
 \caption{
{\bf NGC  5377   }              - S$^4$G mid-IR classification:    (R$_1$)SA$\underline{\rm B}$(s)0/a                    ; Filter: IRAC 3.6$\mu$m; North: left, East: down; Field dimensions:   5.8$\times$  4.2 arcmin; Surface brightness range displayed: 13.0$-$28.0 mag arcsec$^{-2}$}                 
\label{NGC5377}     
 \end{figure}
 
\clearpage
\begin{figure}
\figurenum{1.145}
\plotone{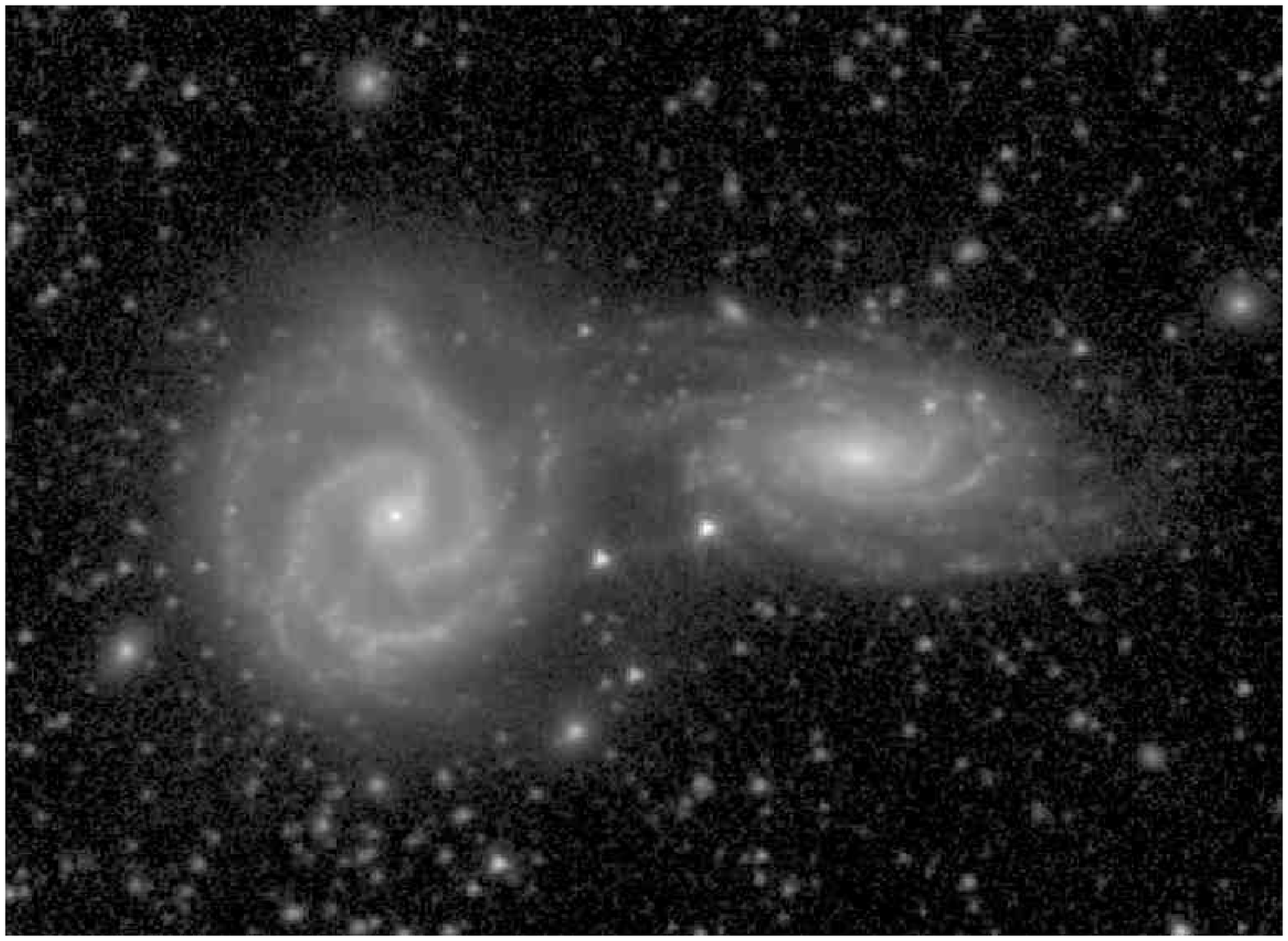}
 \vspace{2.0truecm}
 \caption{
{\bf NGC  5426} (right) and {\bf NGC  5427   } (left) - S$^4$G mid-IR classifications:    S$\underline{\rm A}$B(rs)c, SA(r)bc, respectively; Filter: IRAC 3.6$\mu$m; North: left, East: down; Field dimensions:   6.3$\times$  4.6 arcmin; Surface brightness range displayed: 13.5$-$28.0 mag arcsec$^{-2}$}                 
\label{NGC5426}     
 \end{figure}
 
\clearpage
\begin{figure}
\figurenum{1.146}
\plotone{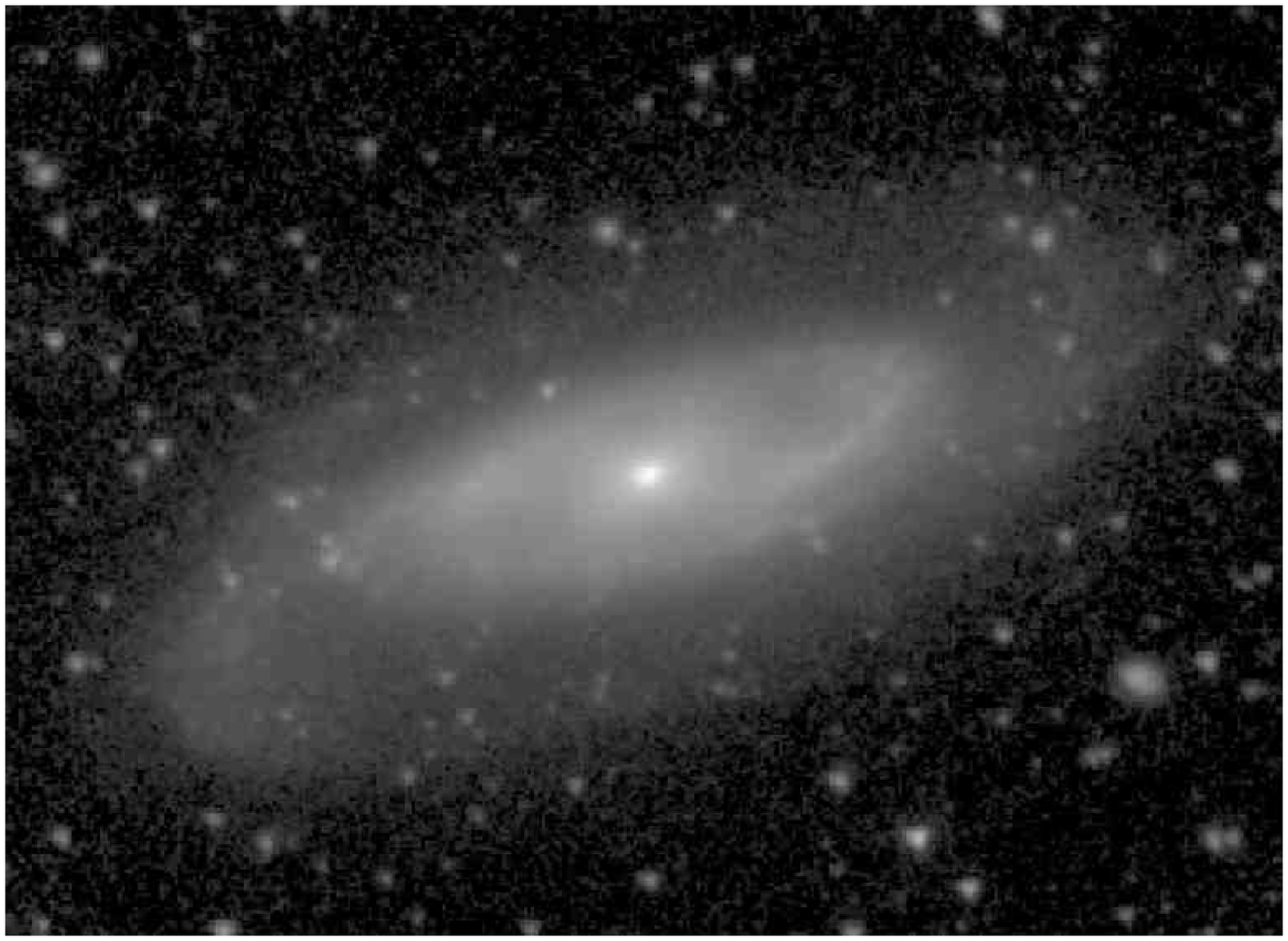}
 \vspace{2.0truecm}
 \caption{
{\bf NGC  5448   }              - S$^4$G mid-IR classification:    (R$_1$)SAB($\underline{\rm r}$s)a                     ; Filter: IRAC 3.6$\mu$m; North:   up, East: left; Field dimensions:   4.5$\times$  3.3 arcmin; Surface brightness range displayed: 13.0$-$28.0 mag arcsec$^{-2}$}                 
\label{NGC5448}     
 \end{figure}
 
\clearpage
\begin{figure}
\figurenum{1.147}
\plotone{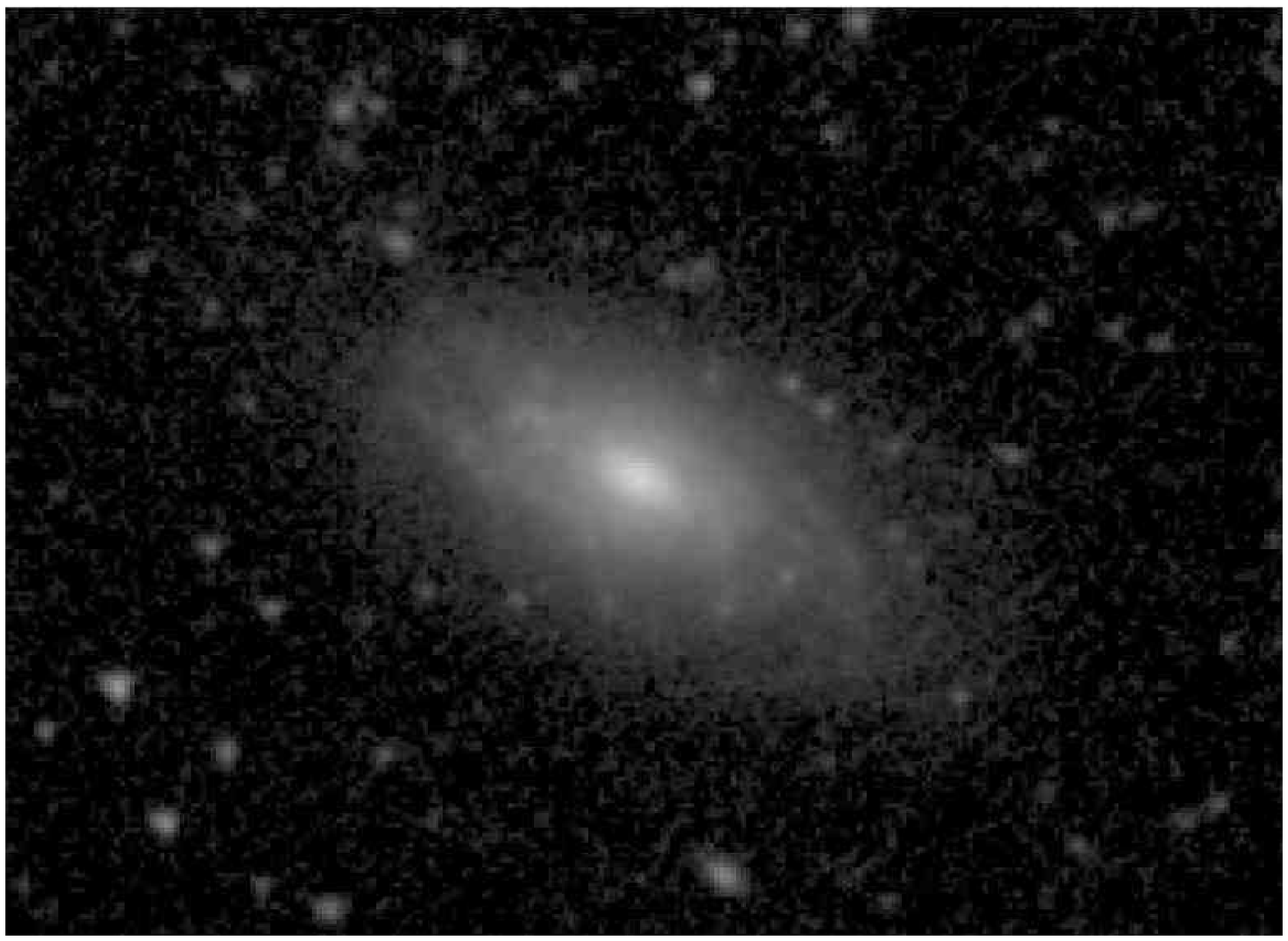}
 \vspace{2.0truecm}
 \caption{
{\bf NGC  5520   }              - S$^4$G mid-IR classification:    SA(l)bc                                               ; Filter: IRAC 3.6$\mu$m; North:   up, East: left; Field dimensions:   2.3$\times$  1.6 arcmin; Surface brightness range displayed: 14.0$-$28.0 mag arcsec$^{-2}$}                 
\label{NGC5520}     
 \end{figure}
 
\clearpage
\begin{figure}
\figurenum{1.148}
\plotone{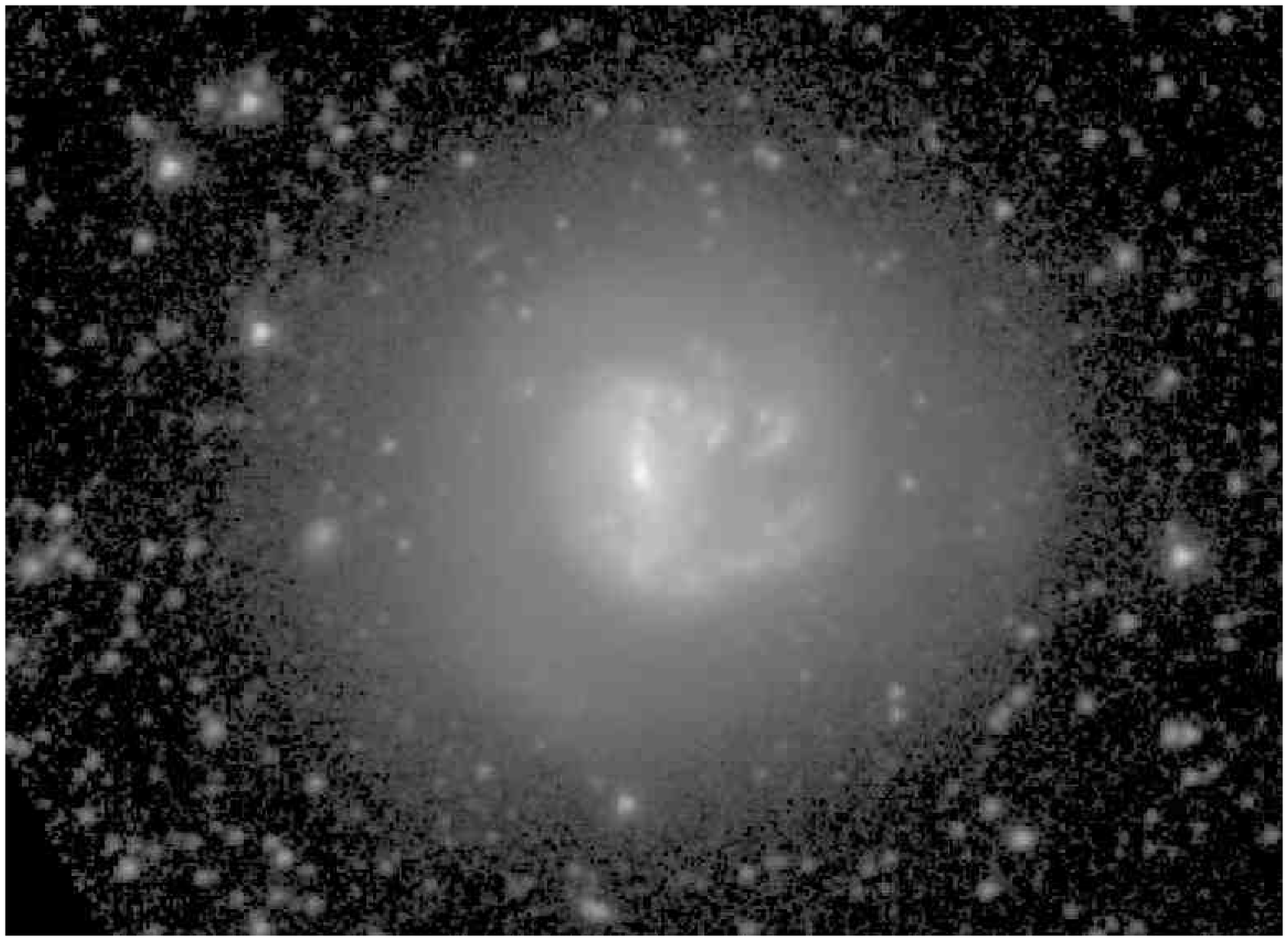}
 \vspace{2.0truecm}
 \caption{
{\bf NGC  5713   }              - S$^4$G mid-IR classification:    (R$^{\prime}$)SB(rs)ab: pec                                     ; Filter: IRAC 3.6$\mu$m; North: left, East: down; Field dimensions:   4.9$\times$  3.5 arcmin; Surface brightness range displayed: 13.0$-$28.0 mag arcsec$^{-2}$}                 
\label{NGC5713}     
 \end{figure}
 
\clearpage
\begin{figure}
\figurenum{1.149}
\plotone{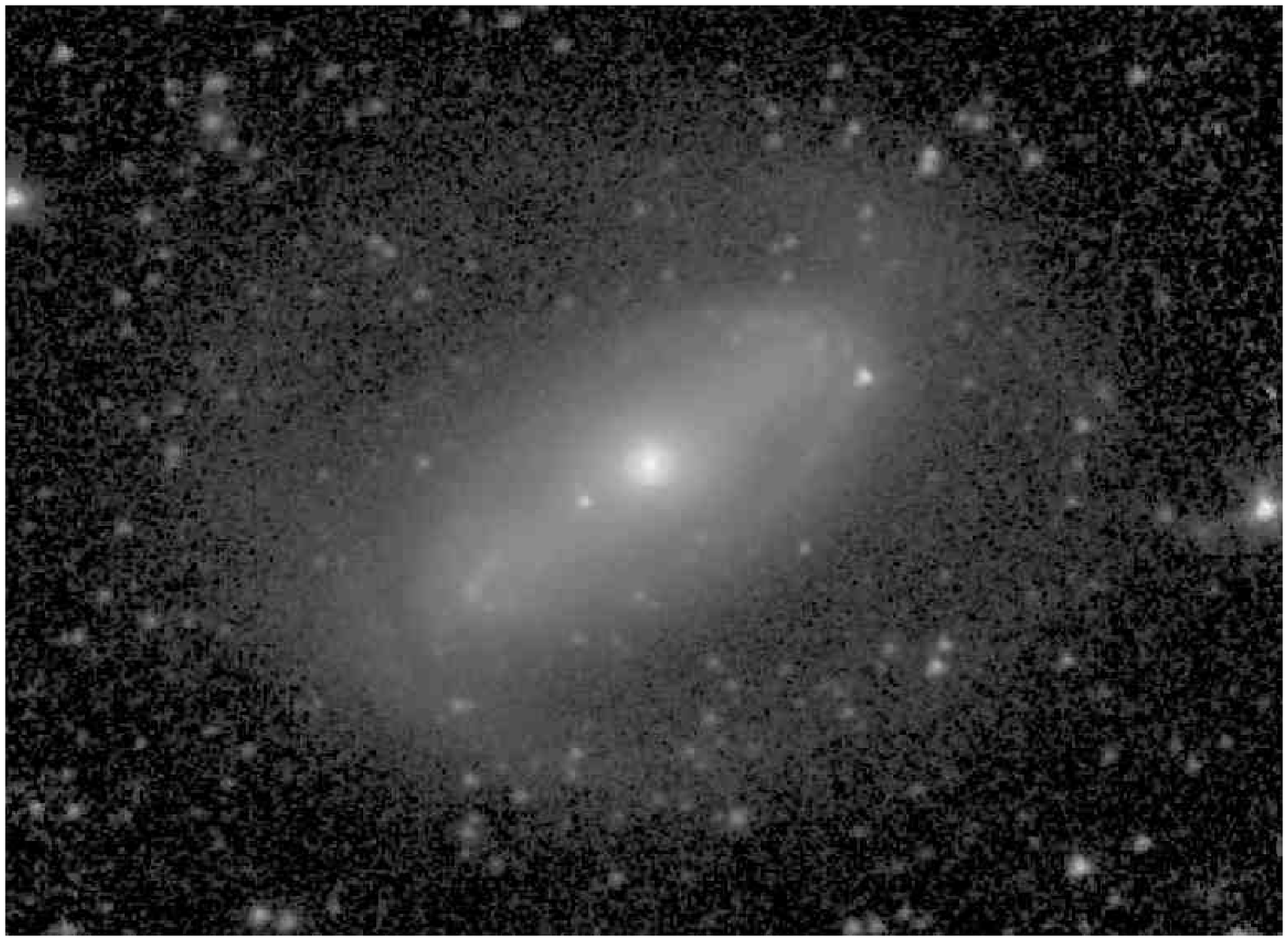}
 \vspace{2.0truecm}
 \caption{
{\bf NGC  5728   }              - S$^4$G mid-IR classification:    (R$_1$)SB($\underline{\rm r}$s,nr,nb)0/a              ; Filter: IRAC 3.6$\mu$m; North: left, East: down; Field dimensions:   5.3$\times$  3.8 arcmin; Surface brightness range displayed: 13.0$-$28.0 mag arcsec$^{-2}$}                 
\label{NGC5728}     
 \end{figure}
 
\clearpage
\begin{figure}
\figurenum{1.150}
\plotone{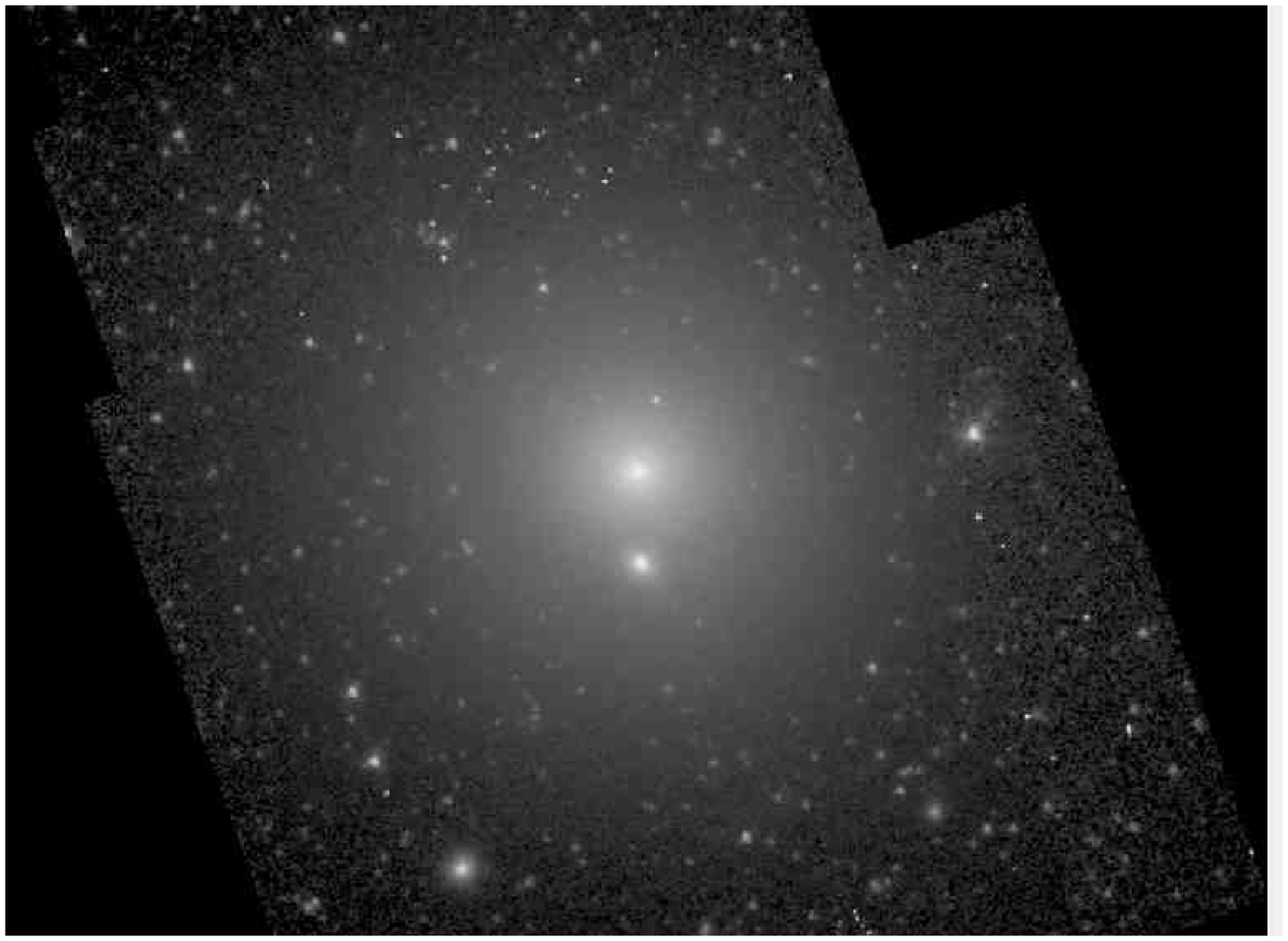}
 \vspace{2.0truecm}
 \caption{
{\bf NGC  5846   }              - S$^4$G mid-IR classification:    E$^+$0/SA0$^-$                                        ; Filter: IRAC 3.6$\mu$m; North:   up, East: left; Field dimensions:   8.9$\times$  6.5 arcmin; Surface brightness range displayed: 13.5$-$28.0 mag arcsec$^{-2}$}                 
\label{NGC5846}     
 \end{figure}
 
\clearpage
\begin{figure}
\figurenum{1.151}
\plotone{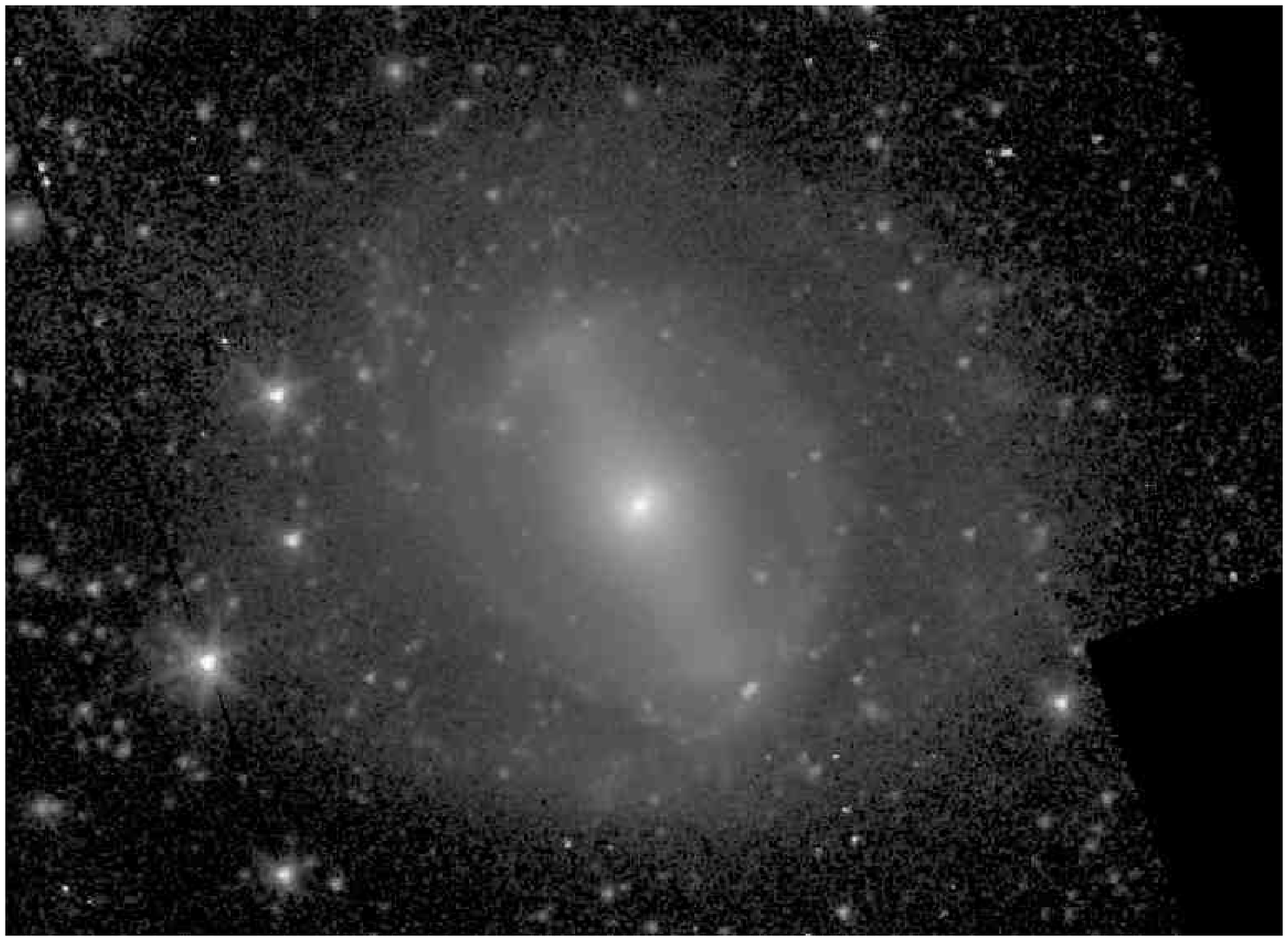}
 \vspace{2.0truecm}
 \caption{
{\bf NGC  5850   }              - S$^4$G mid-IR classification:    (R$^{\prime}$)SB(r,nr,nb)ab                                     ; Filter: IRAC 3.6$\mu$m; North: left, East: down; Field dimensions:   7.0$\times$  5.1 arcmin; Surface brightness range displayed: 13.5$-$28.0 mag arcsec$^{-2}$}                 
\label{NGC5850}     
 \end{figure}
 
\clearpage
\begin{figure}
\figurenum{1.152}
\plotone{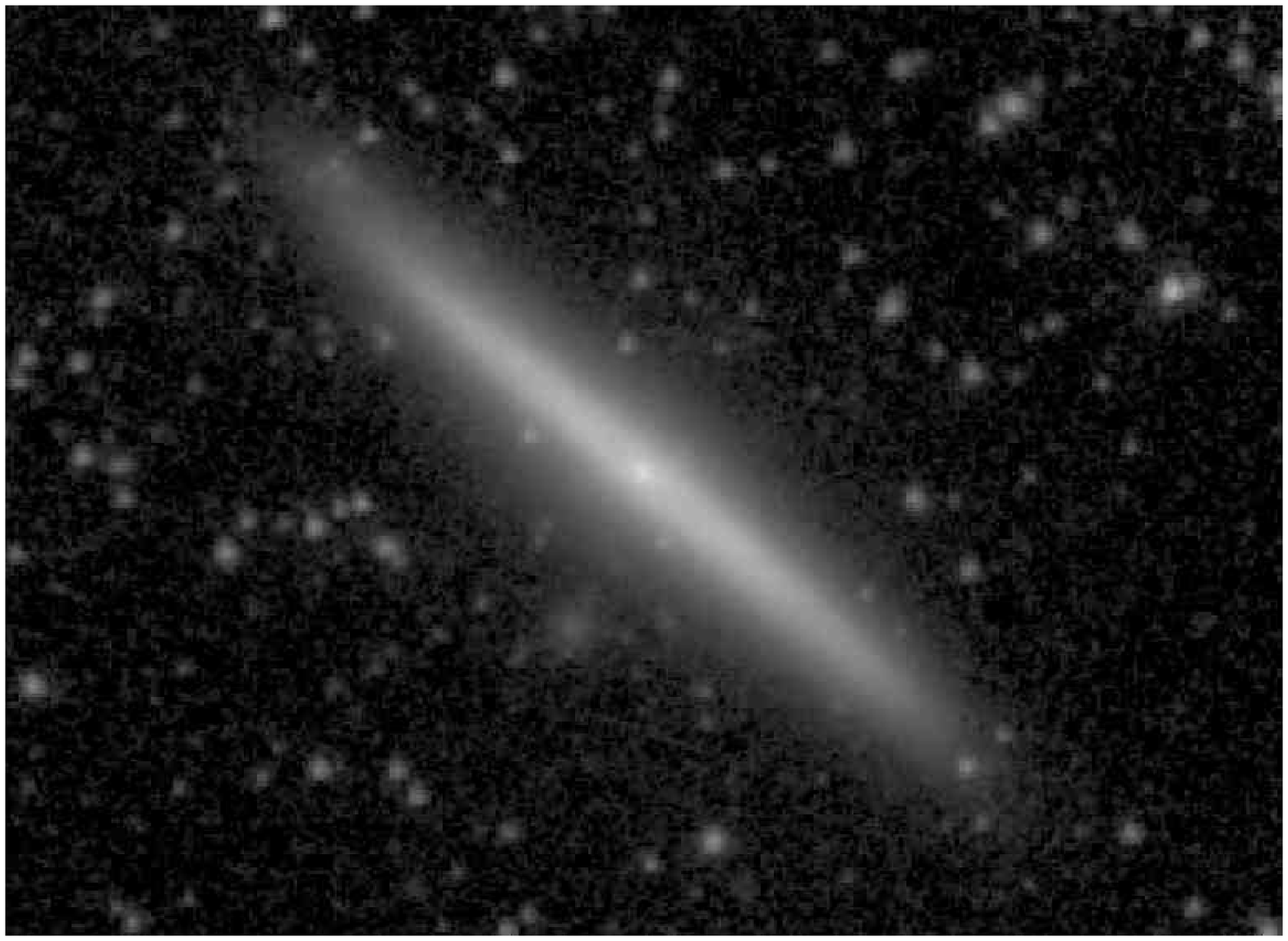}
 \vspace{2.0truecm}
 \caption{
{\bf NGC  5981   }              - S$^4$G mid-IR classification:    S0$^+$ sp                                             ; Filter: IRAC 3.6$\mu$m; North: left, East: down; Field dimensions:   4.0$\times$  2.9 arcmin; Surface brightness range displayed: 14.5$-$28.0 mag arcsec$^{-2}$}                 
\label{NGC5981}     
 \end{figure}
 
\clearpage
\begin{figure}
\figurenum{1.153}
\plotone{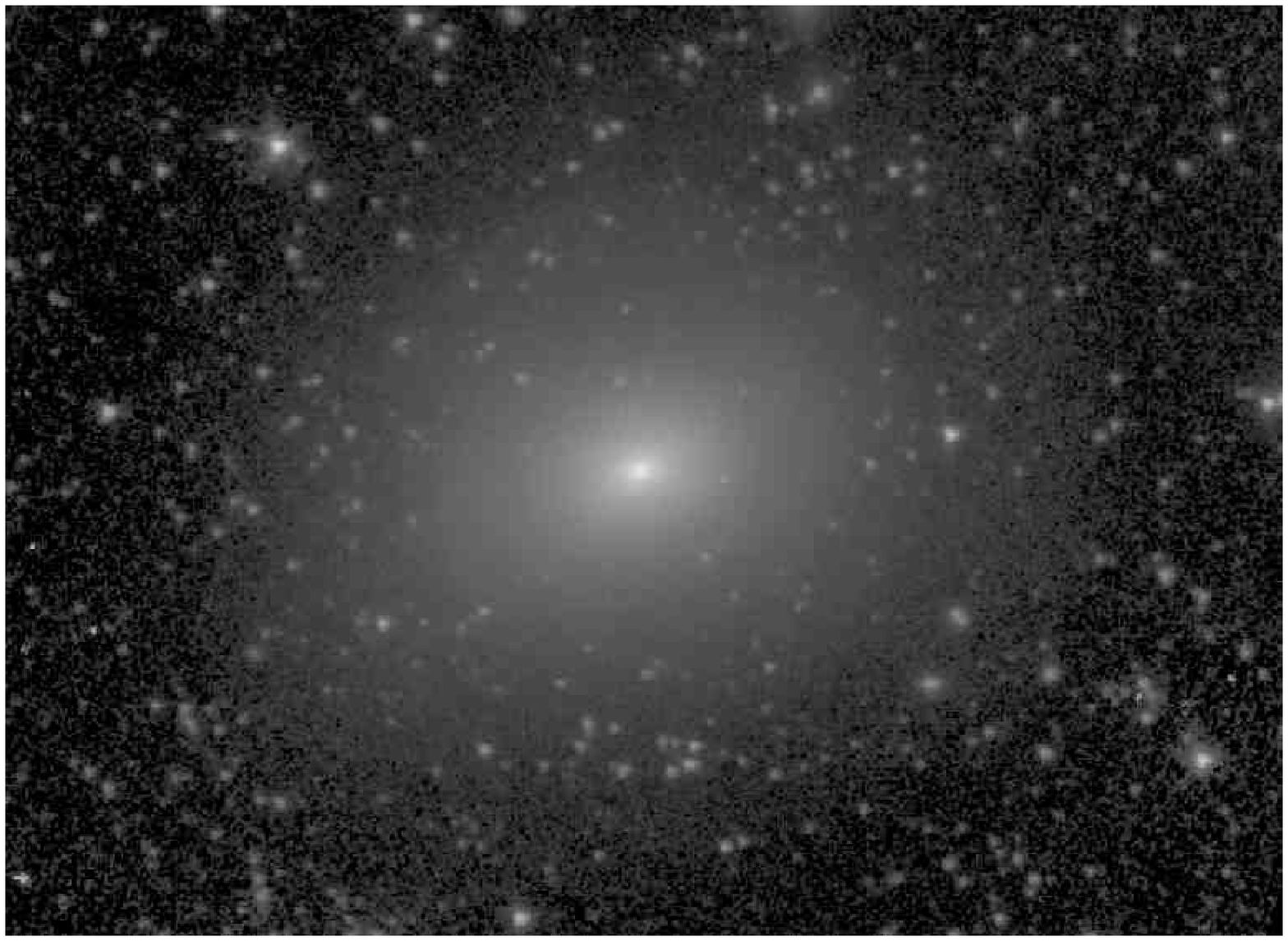}
 \vspace{2.0truecm}
 \caption{
{\bf NGC  5982   }              - S$^4$G mid-IR classification:    E2                                                    ; Filter: IRAC 3.6$\mu$m; North:   up, East: left; Field dimensions:   7.0$\times$  5.1 arcmin; Surface brightness range displayed: 13.0$-$28.0 mag arcsec$^{-2}$}                 
\label{NGC5982}     
 \end{figure}
 
\clearpage
\begin{figure}
\figurenum{1.154}
\plotone{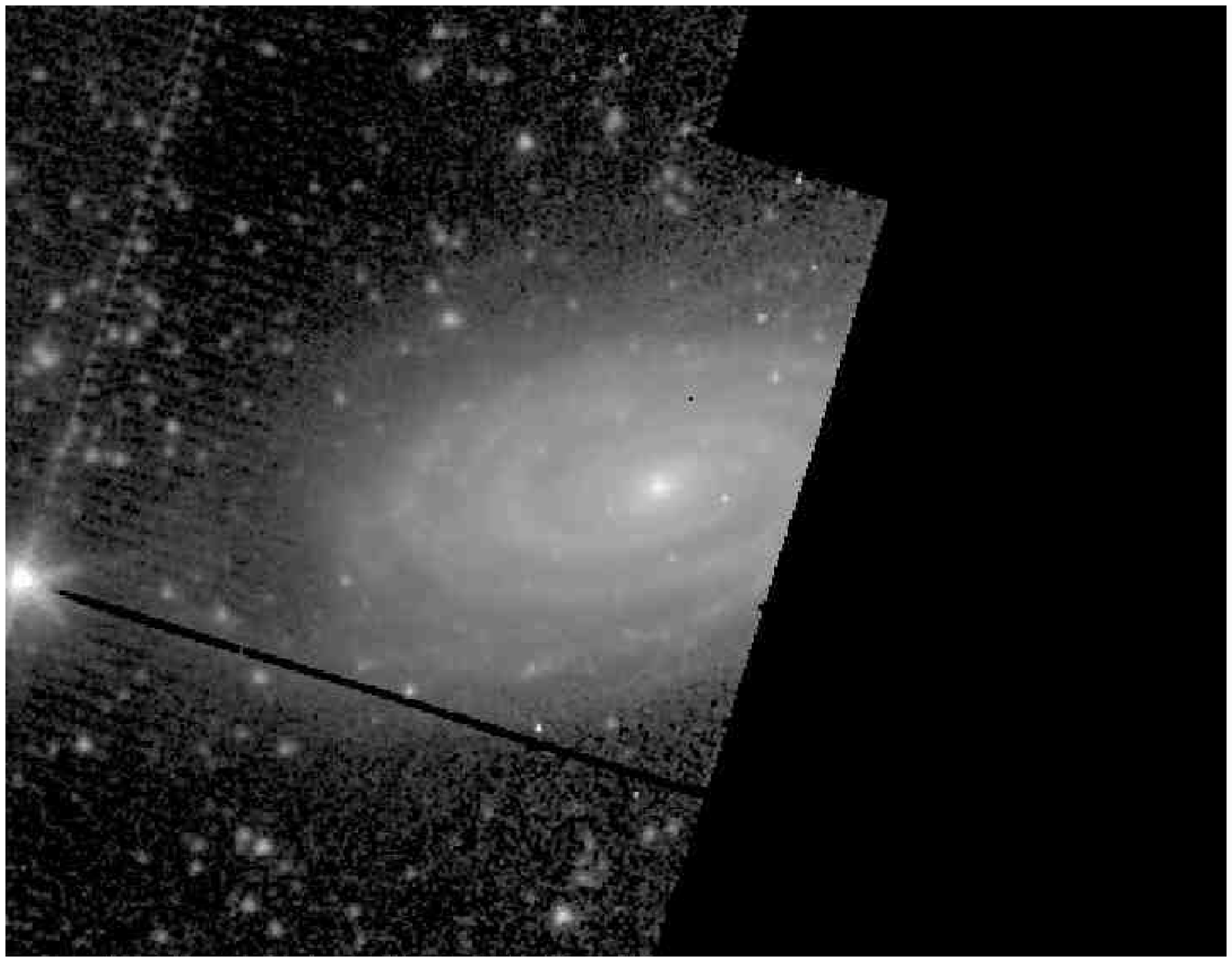}
 \vspace{2.0truecm}
 \caption{
{\bf NGC  5985   }              - S$^4$G mid-IR classification:    SAB(s)ab                                              ; Filter: IRAC 3.6$\mu$m; North: left, East: down; Field dimensions:   5.9$\times$  4.3 arcmin; Surface brightness range displayed: 14.5$-$28.0 mag arcsec$^{-2}$}                 
\label{NGC5985}     
 \end{figure}
 
\clearpage
\begin{figure}
\figurenum{1.155}
\plotone{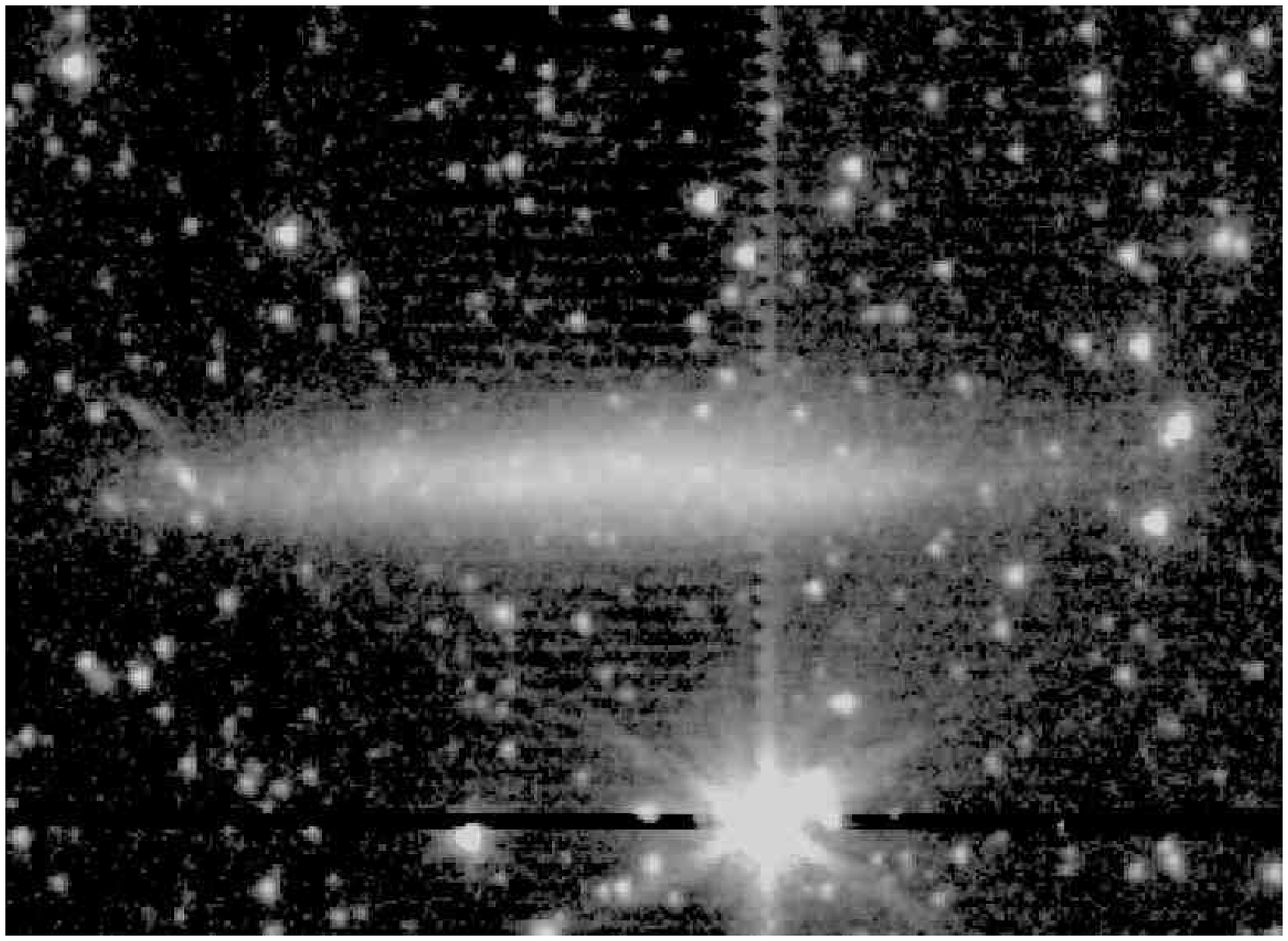}
 \vspace{2.0truecm}
 \caption{
{\bf NGC  7064   }              - S$^4$G mid-IR classification:    Sd sp                                                 ; Filter: IRAC 3.6$\mu$m; North:   up, East: left; Field dimensions:   4.5$\times$  3.3 arcmin; Surface brightness range displayed: 18.0$-$28.0 mag arcsec$^{-2}$}                 
\label{NGC7064}     
 \end{figure}
 
\clearpage
\begin{figure}
\figurenum{1.156}
\plotone{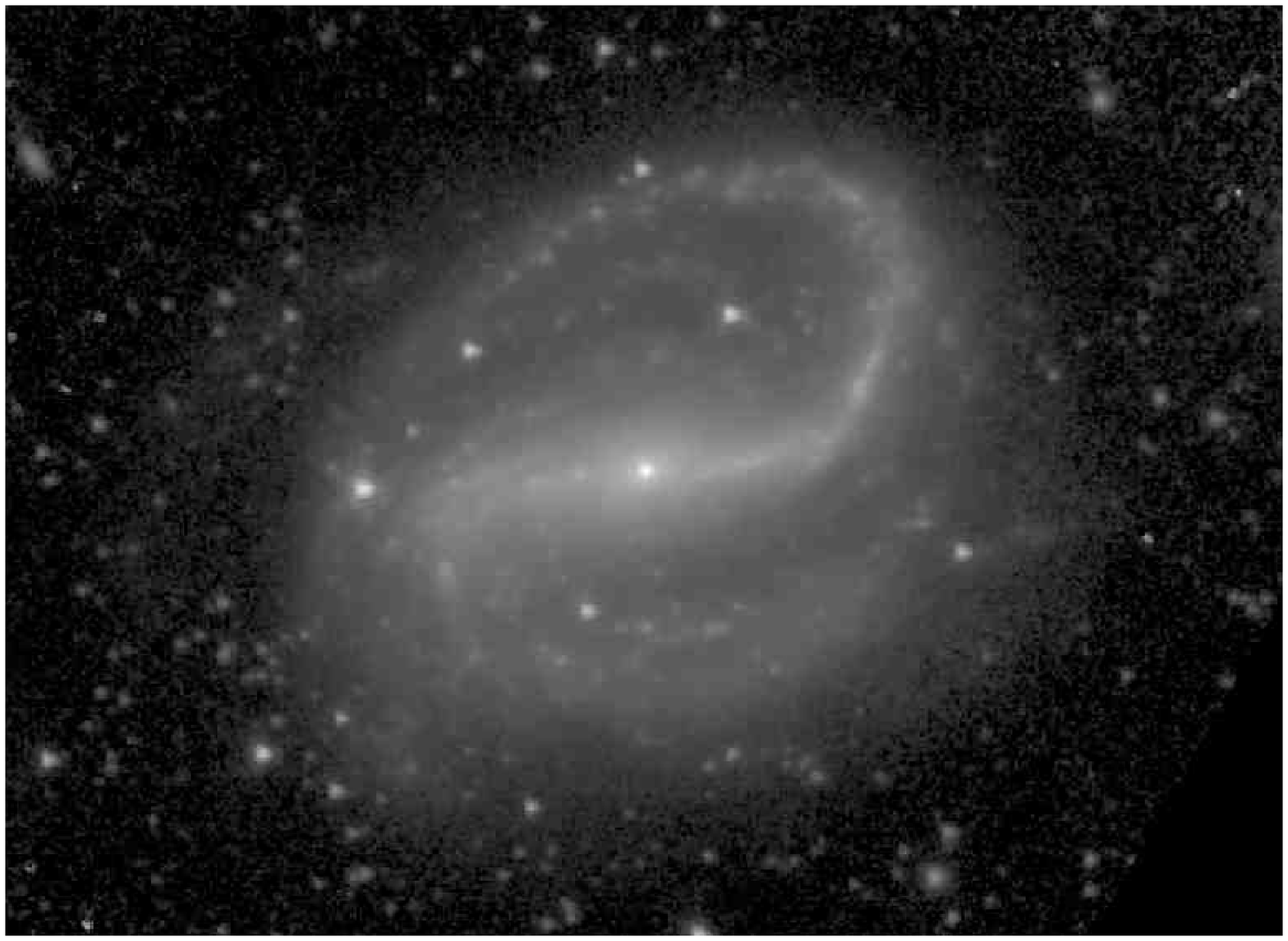}
 \vspace{2.0truecm}
 \caption{
{\bf NGC  7479   }              - S$^4$G mid-IR classification:    (R$^{\prime}$)SB(s)b                                            ; Filter: IRAC 3.6$\mu$m; North: left, East: down; Field dimensions:   5.8$\times$  4.2 arcmin; Surface brightness range displayed: 12.0$-$28.0 mag arcsec$^{-2}$}                 
\label{NGC7479}     
 \end{figure}
 
\clearpage
\begin{figure}
\figurenum{1.157}
\plotone{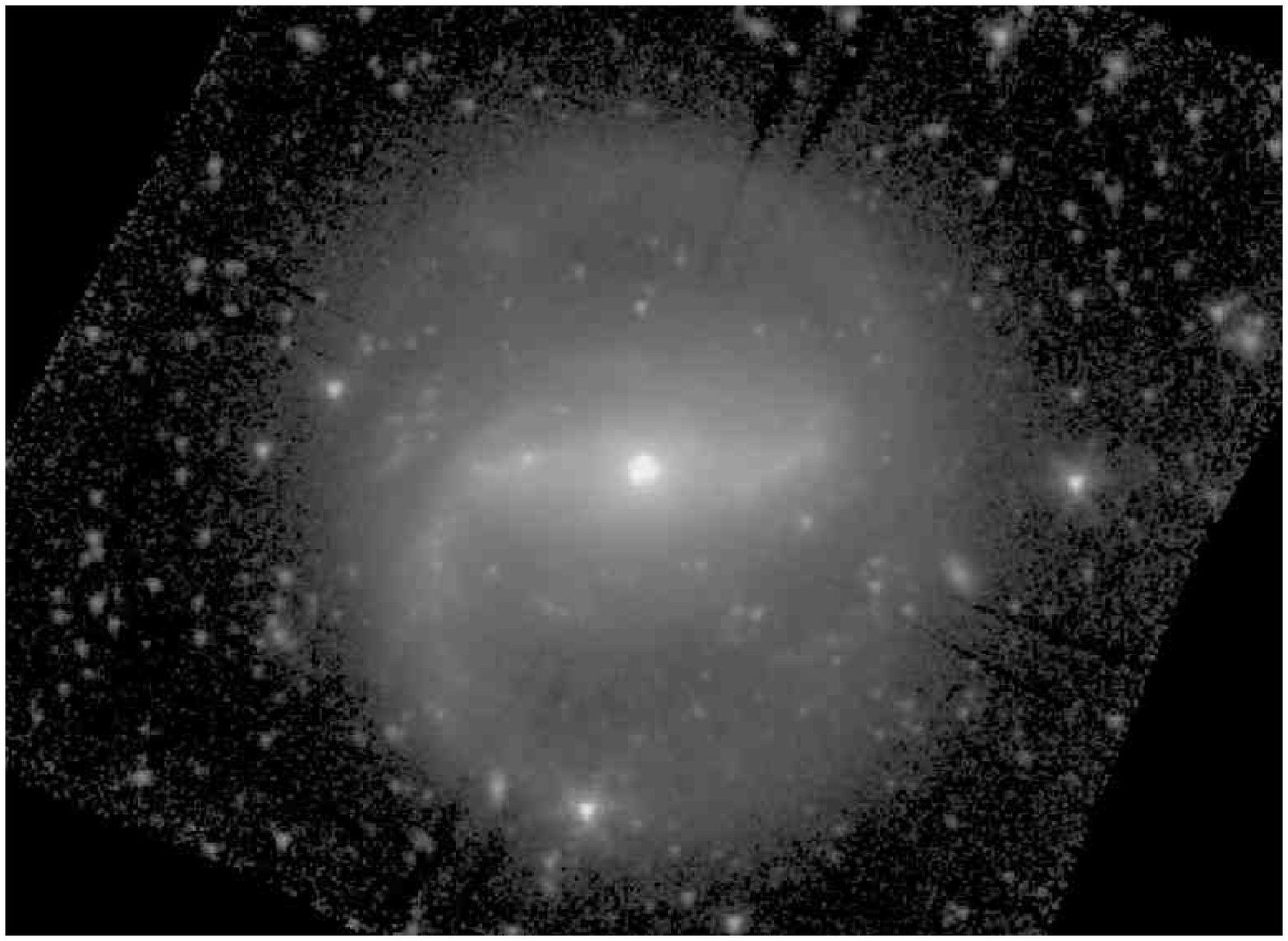}
 \vspace{2.0truecm}
 \caption{
{\bf NGC  7552   }              - S$^4$G mid-IR classification:    (R$_1^{\prime}$)SB(r$\underline{\rm s}$,nr)a                  ; Filter: IRAC 3.6$\mu$m; North:   up, East: left; Field dimensions:   6.3$\times$  4.6 arcmin; Surface brightness range displayed: 12.0$-$28.0 mag arcsec$^{-2}$}                 
\label{NGC7552}     
 \end{figure}
 
\clearpage
\begin{figure}
\figurenum{1.158}
\plotone{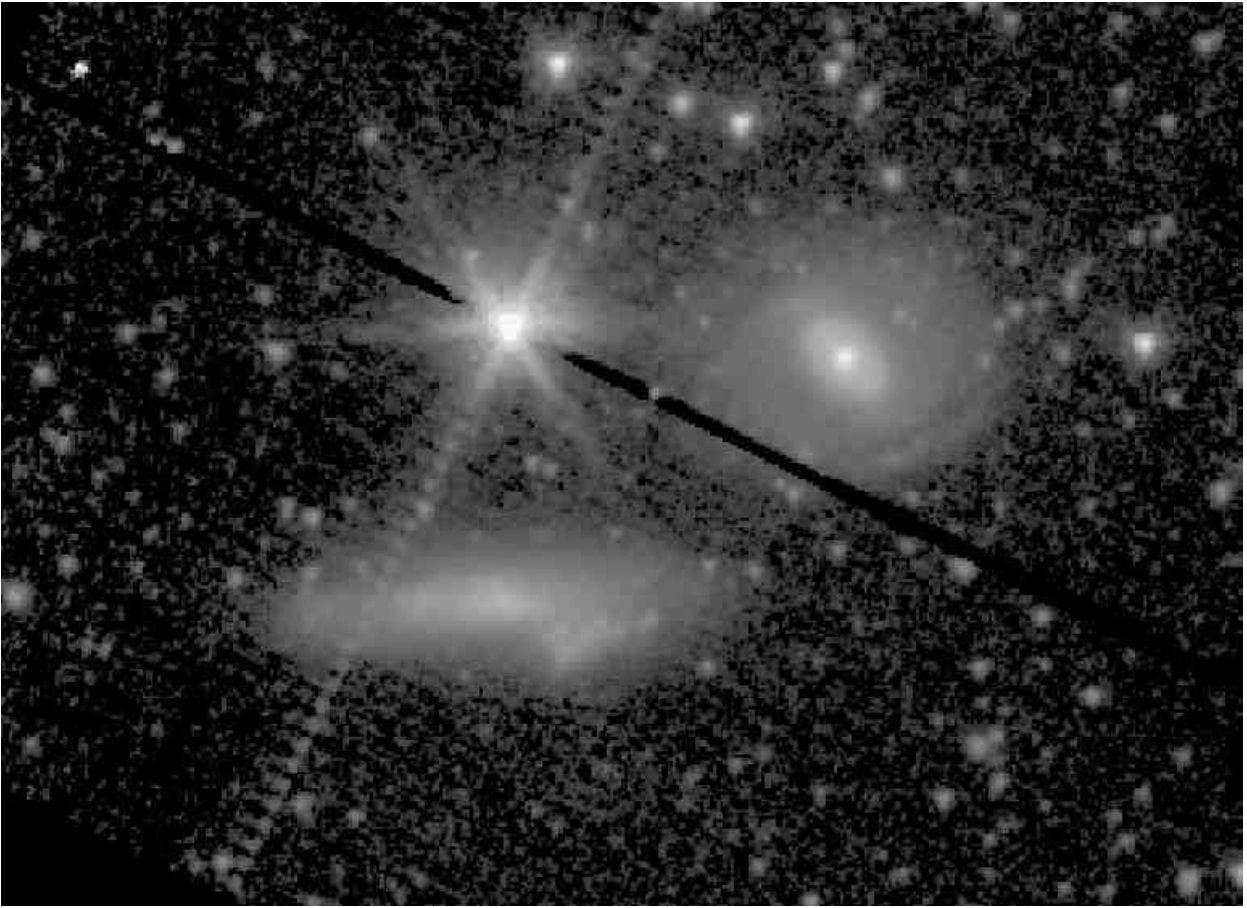}
 \vspace{2.0truecm}
 \caption{
{\bf NGC  7731} (upper right) and {\bf NGC  7732} (lower left) - S$^4$G mid-IR classifications: (R$_1$R$_2^{\prime}$)S$\underline{\rm A}$B(r)a, SBd: sp,
respectively; Filter: IRAC 3.6$\mu$m; North:   up, East: left; Field dimensions:   4.5$\times$  3.3 arcmin; Surface brightness range displayed: 14.0$-$28.0 mag arcsec$^{-2}$}               
\label{NGC7731}     
 \end{figure}
 
\clearpage
\begin{figure}
\figurenum{1.159}
\plotone{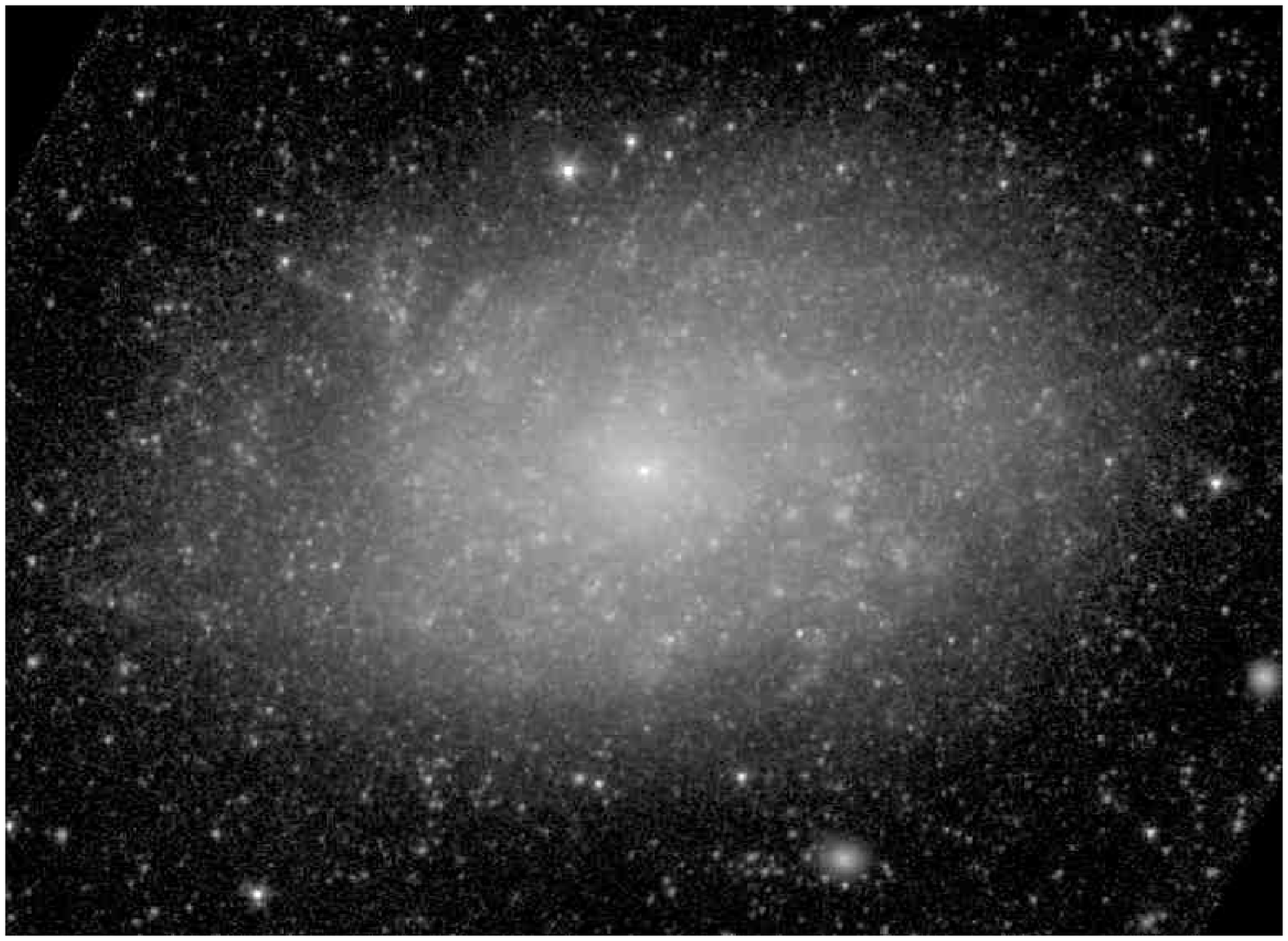}
 \vspace{2.0truecm}
 \caption{
{\bf NGC  7793   }              - S$^4$G mid-IR classification:    SA(s)c                                                ; Filter: IRAC 3.6$\mu$m; North:   up, East: left; Field dimensions:  11.7$\times$  8.5 arcmin; Surface brightness range displayed: 14.0$-$28.0 mag arcsec$^{-2}$}                 
\label{NGC7793}     
 \end{figure}
 
\clearpage
\begin{figure}
\figurenum{1.160}
\plotone{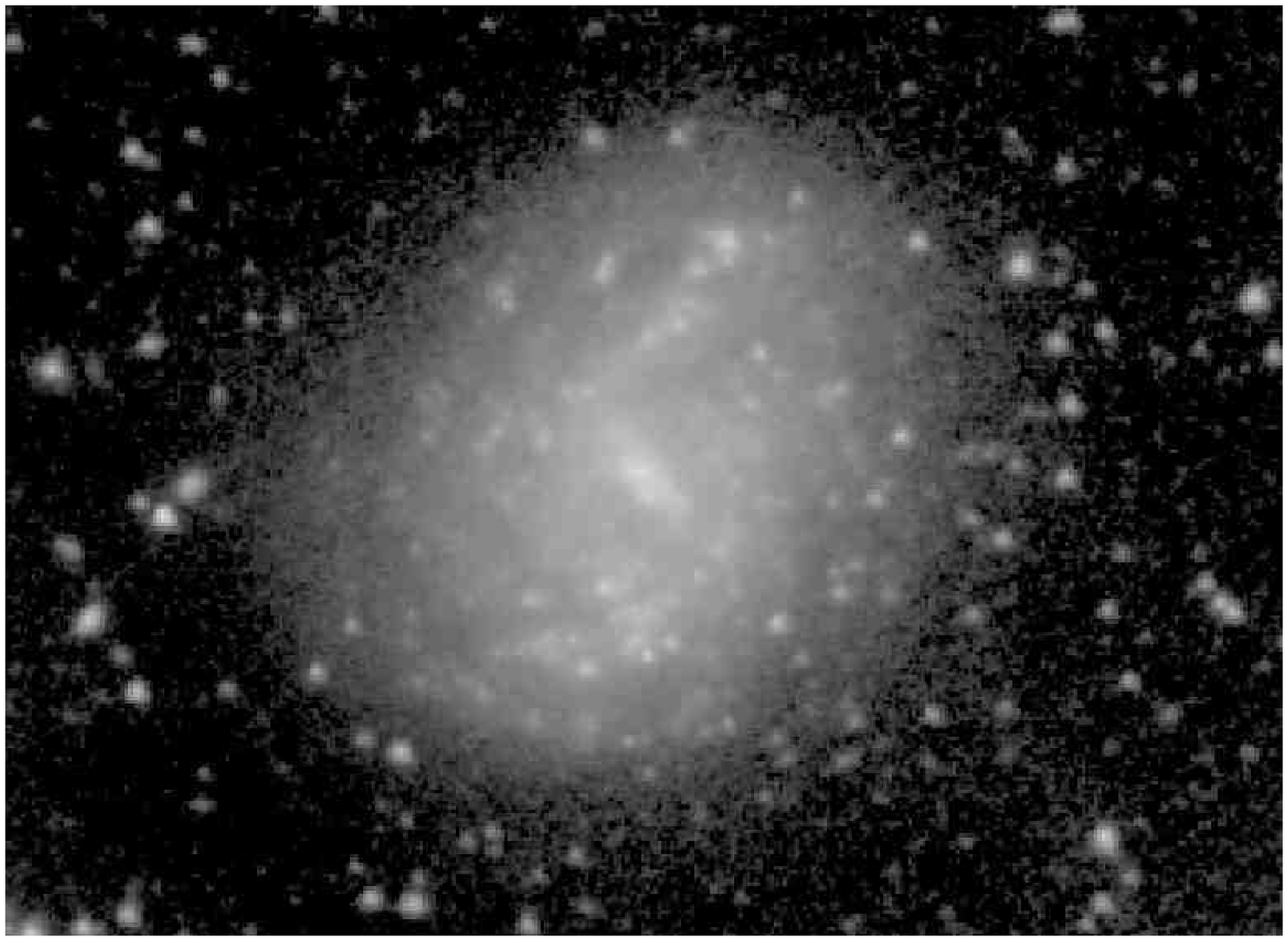}
 \vspace{2.0truecm}
 \caption{
{\bf IC    749   }              - S$^4$G mid-IR classification:    SB(rs)cd                                              ; Filter: IRAC 3.6$\mu$m; North:   up, East: left; Field dimensions:   4.0$\times$  2.9 arcmin; Surface brightness range displayed: 16.5$-$28.0 mag arcsec$^{-2}$}                 
\label{IC0749}      
 \end{figure}
 
\clearpage
\begin{figure}
\figurenum{1.161}
\plotone{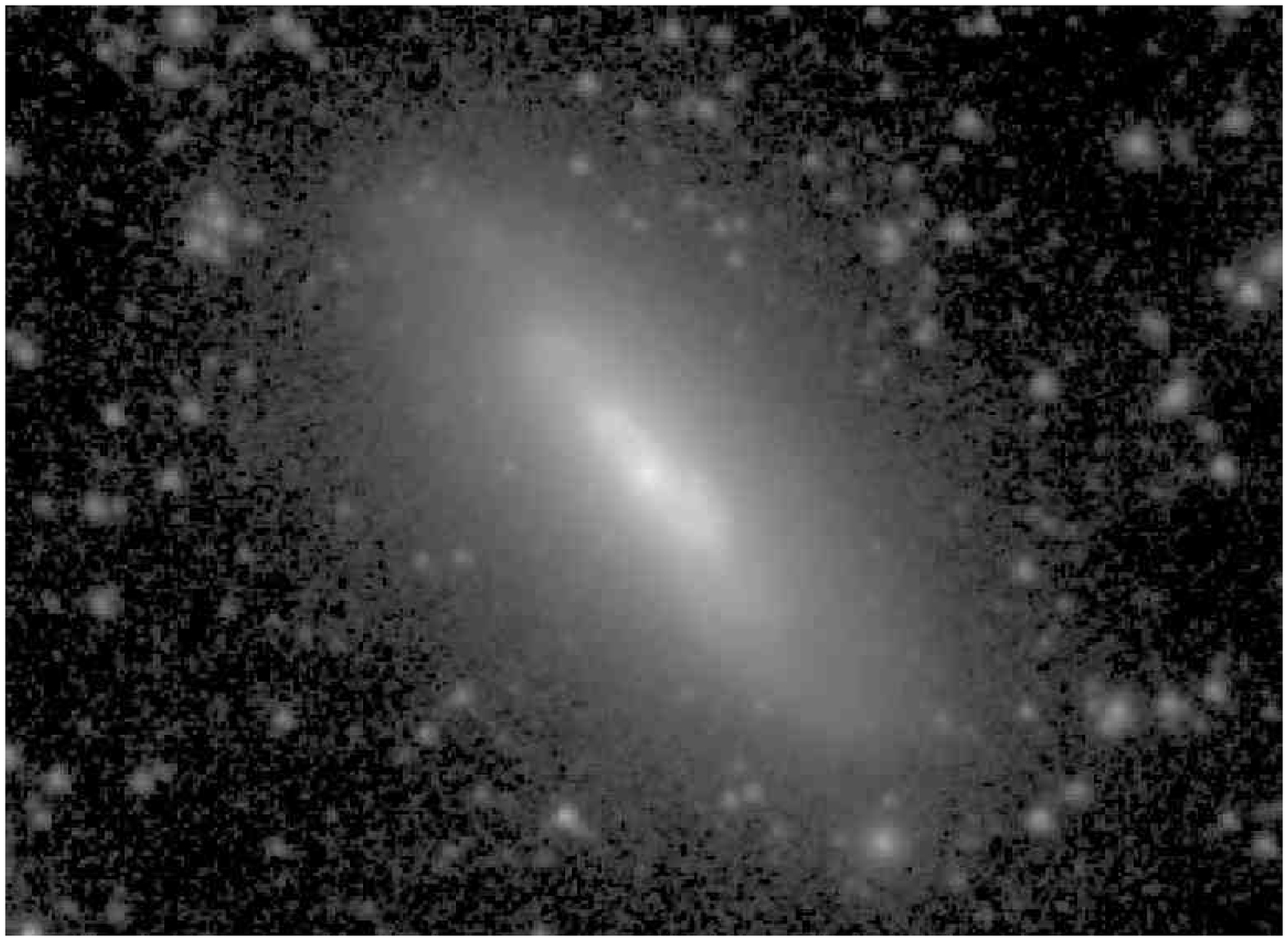}
 \vspace{2.0truecm}
 \caption{
{\bf IC    750   }              - S$^4$G mid-IR classification:    SA(s)a                                                ; Filter: IRAC 3.6$\mu$m; North:   up, East: left; Field dimensions:   4.0$\times$  2.9 arcmin; Surface brightness range displayed: 13.0$-$28.0 mag arcsec$^{-2}$}                 
\label{IC0750}      
 \end{figure}
 
\clearpage
\begin{figure}
\figurenum{1.162}
\plotone{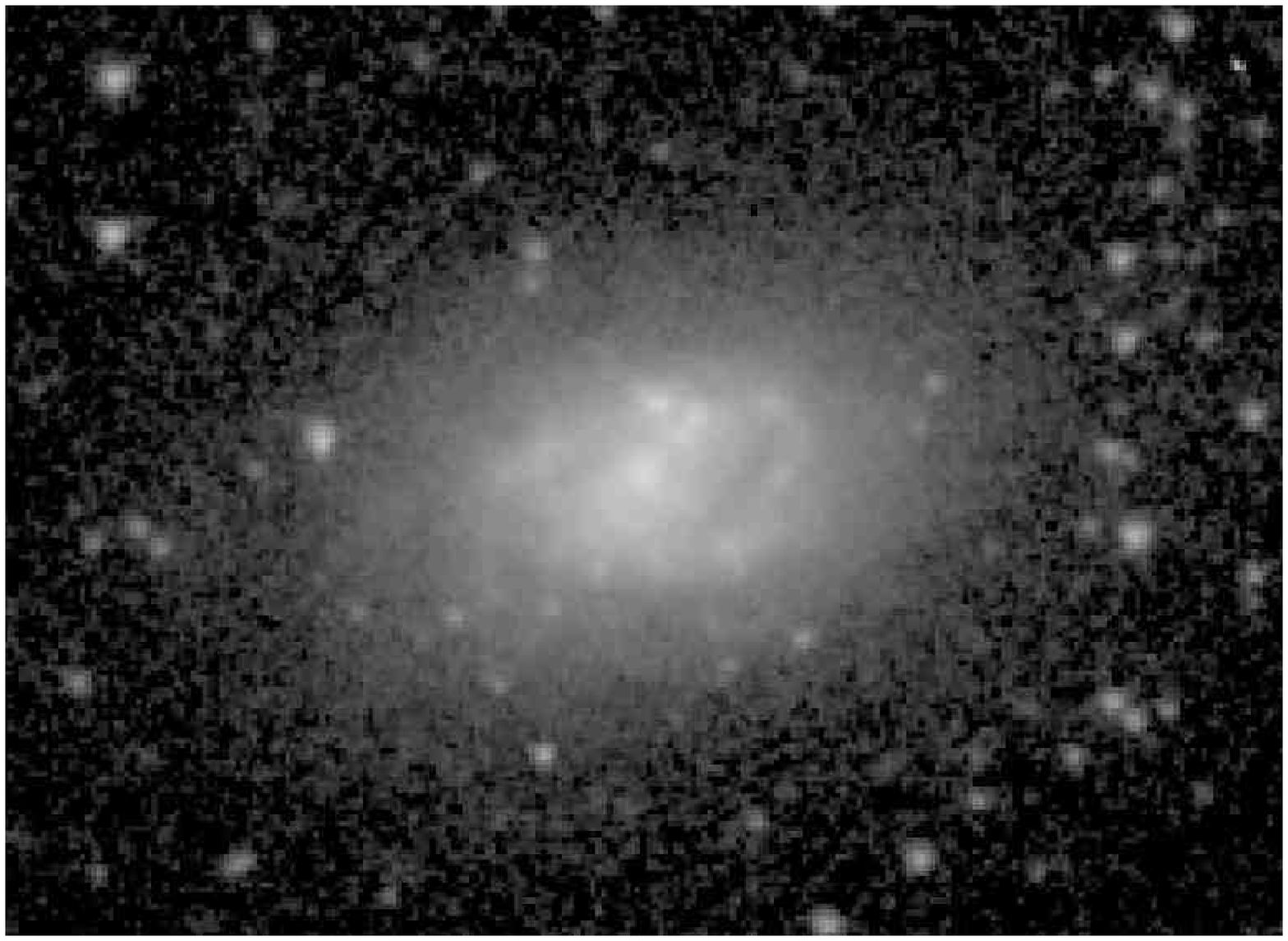}
 \vspace{2.0truecm}
 \caption{
{\bf IC    797   }              - S$^4$G mid-IR classification:    SAB(s)dm:                                             ; Filter: IRAC 3.6$\mu$m; North:   up, East: left; Field dimensions:   2.9$\times$  2.1 arcmin; Surface brightness range displayed: 17.0$-$28.0 mag arcsec$^{-2}$}                 
\label{IC0797}      
 \end{figure}
 
\clearpage
\begin{figure}
\figurenum{1.163}
\plotone{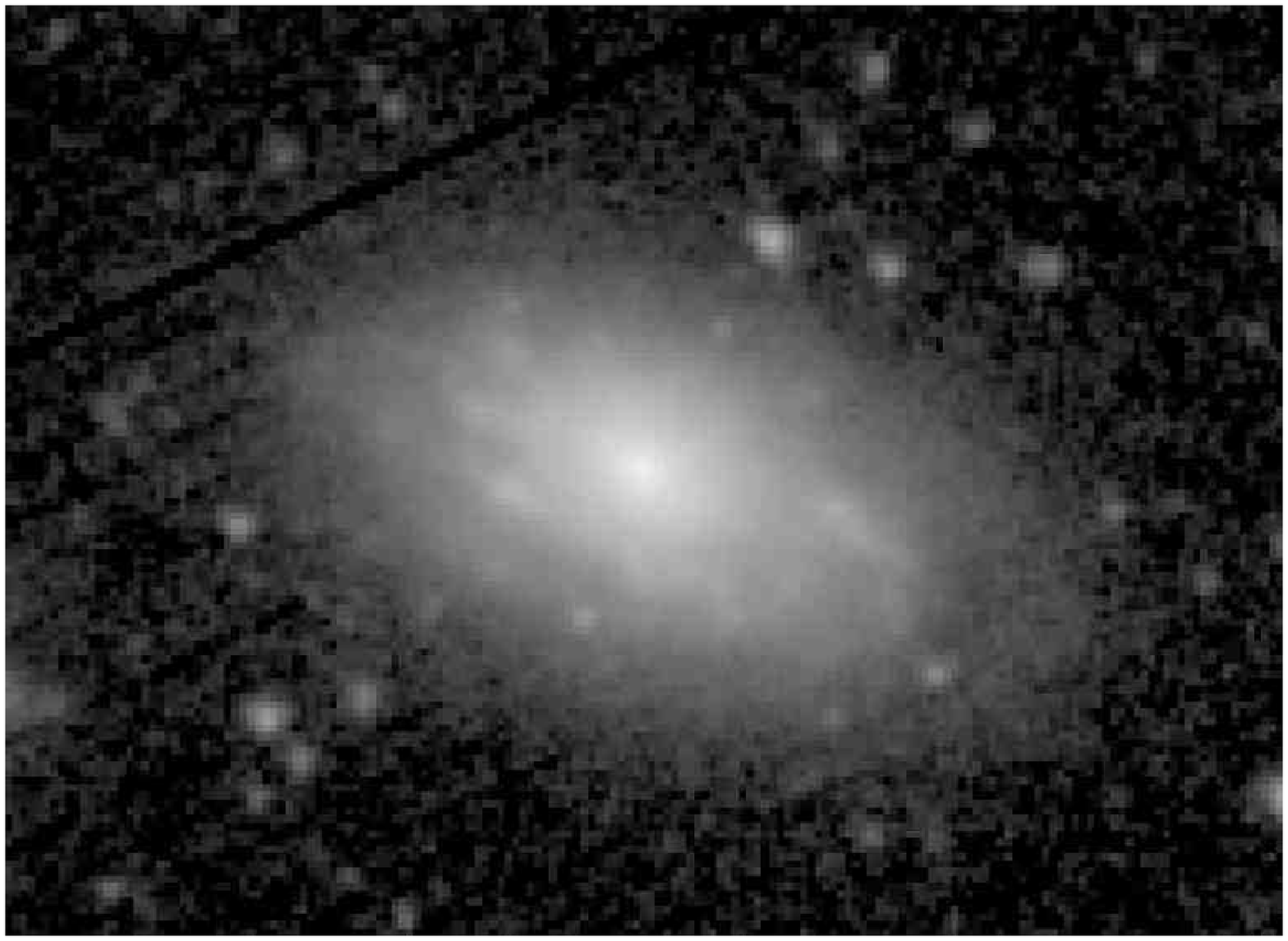}
 \vspace{2.0truecm}
 \caption{
{\bf IC   1066   }              - S$^4$G mid-IR classification:    SAbc:                                                 ; Filter: IRAC 3.6$\mu$m; North:   up, East: left; Field dimensions:   2.1$\times$  1.5 arcmin; Surface brightness range displayed: 16.0$-$28.0 mag arcsec$^{-2}$}                 
\label{IC1066}      
 \end{figure}
 
\clearpage
\begin{figure}
\figurenum{1.164}
\plotone{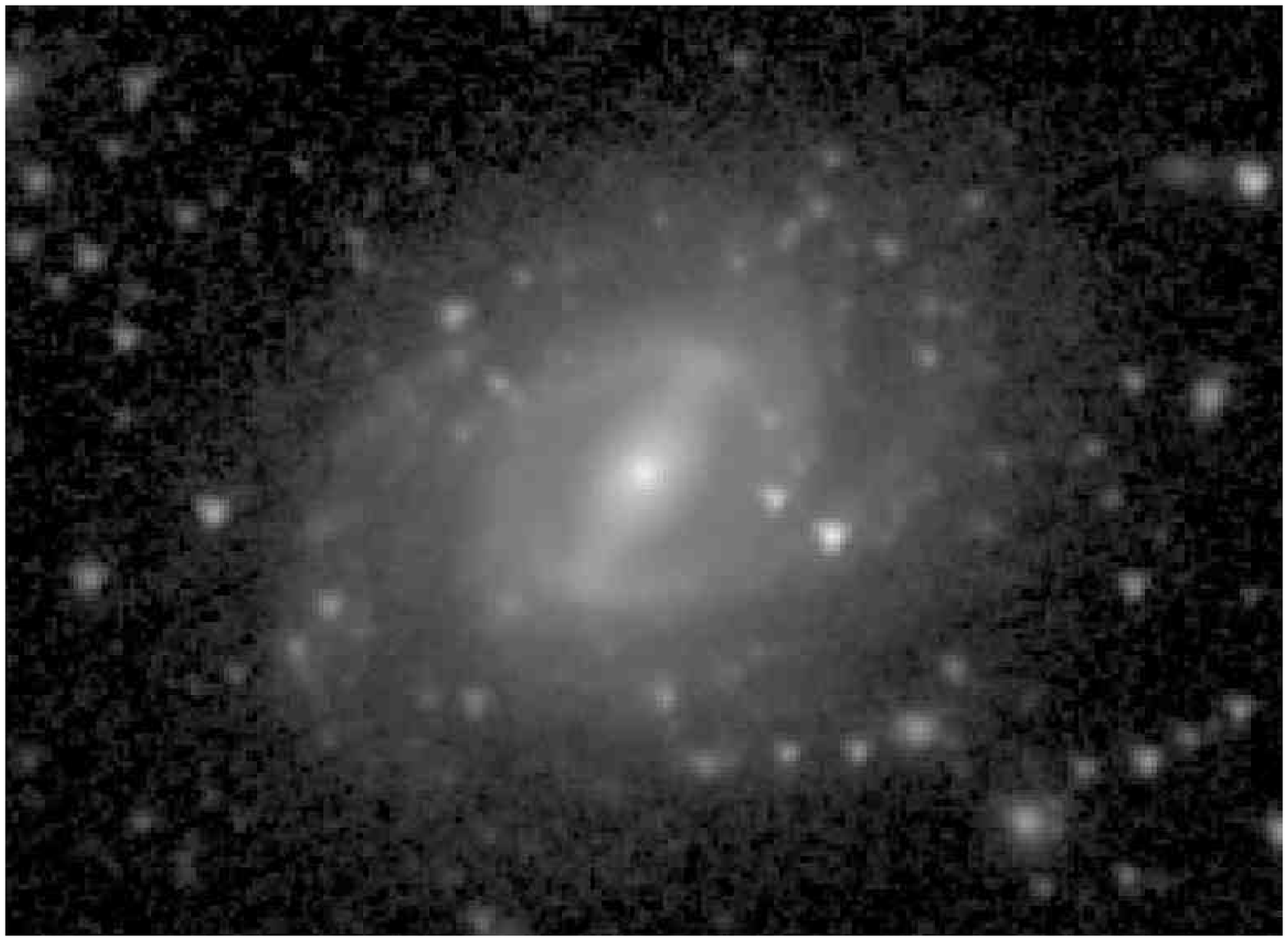}
 \vspace{2.0truecm}
 \caption{
{\bf IC   1067   }              - S$^4$G mid-IR classification:    SB(r)b                                                ; Filter: IRAC 3.6$\mu$m; North:   up, East: left; Field dimensions:   2.9$\times$  2.1 arcmin; Surface brightness range displayed: 14.5$-$28.0 mag arcsec$^{-2}$}                 
\label{IC1067}      
 \end{figure}
 
\clearpage
\begin{figure}
\figurenum{1.165}
\plotone{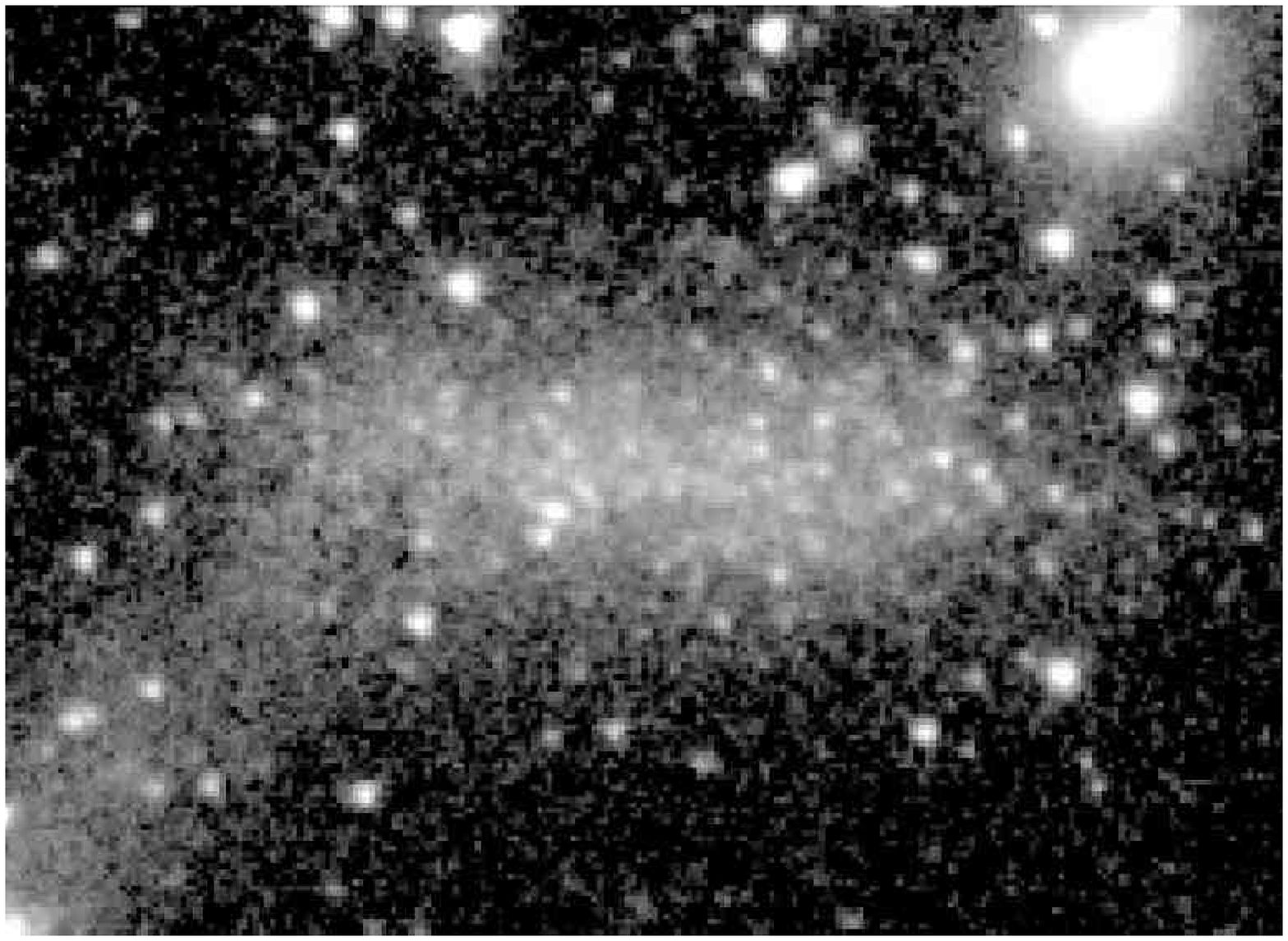}
 \vspace{2.0truecm}
 \caption{
{\bf IC   1574   }              - S$^4$G mid-IR classification:    IBm sp?                                               ; Filter: IRAC 3.6$\mu$m; North: left, East: down; Field dimensions:   2.9$\times$  2.1 arcmin; Surface brightness range displayed: 18.5$-$28.0 mag arcsec$^{-2}$}                 
\label{IC1574}      
 \end{figure}
 
\clearpage
\begin{figure}
\figurenum{1.166}
\plotone{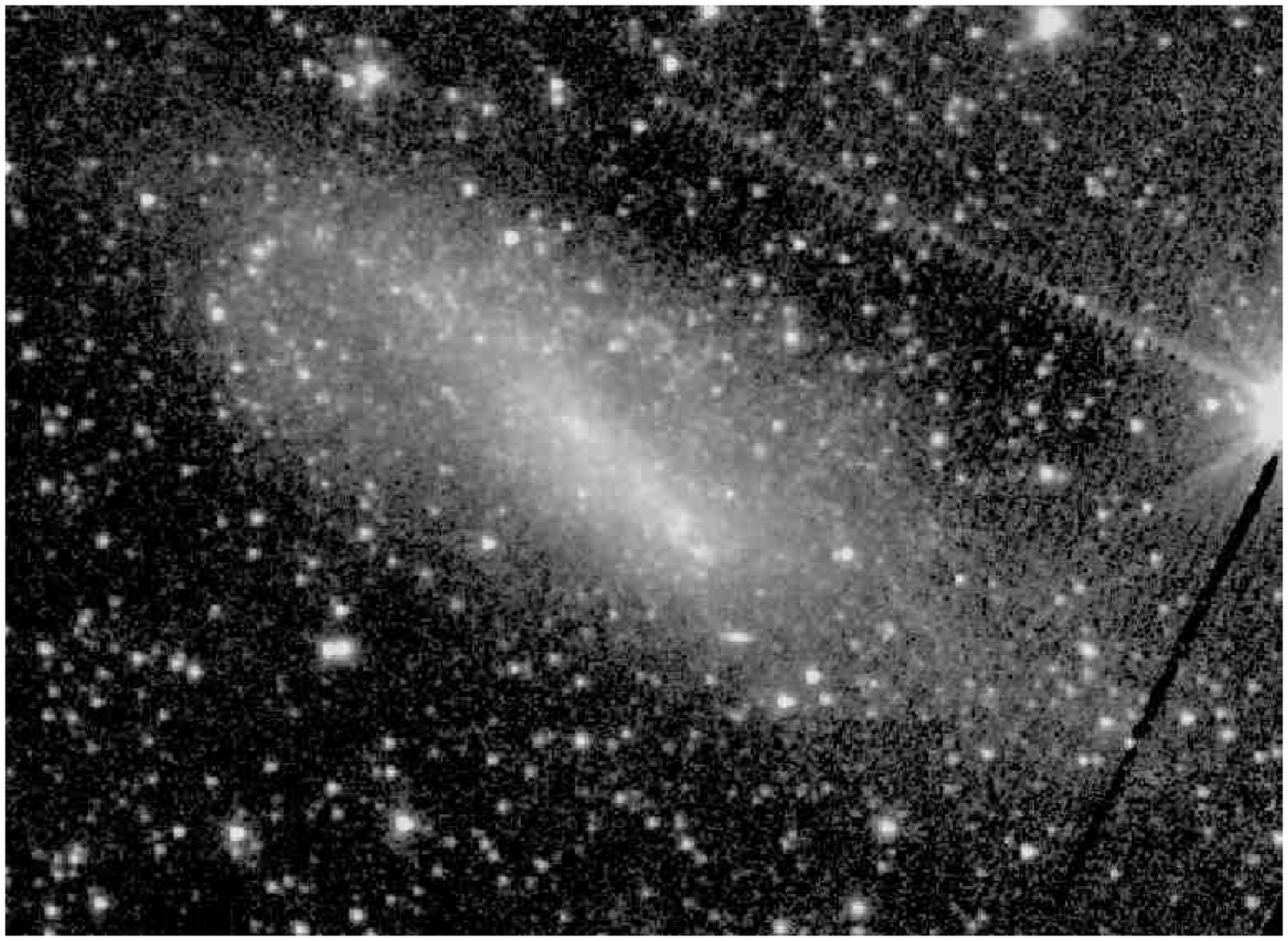}
 \vspace{2.0truecm}
 \caption{
{\bf IC   1727   }              - S$^4$G mid-IR classification:    SB(s)m                                                ; Filter: IRAC 3.6$\mu$m; North: left, East: down; Field dimensions:   7.2$\times$  5.2 arcmin; Surface brightness range displayed: 18.5$-$28.0 mag arcsec$^{-2}$}                 
\label{IC1727}      
 \end{figure}
 
\clearpage
\begin{figure}
\figurenum{1.167}
\plotone{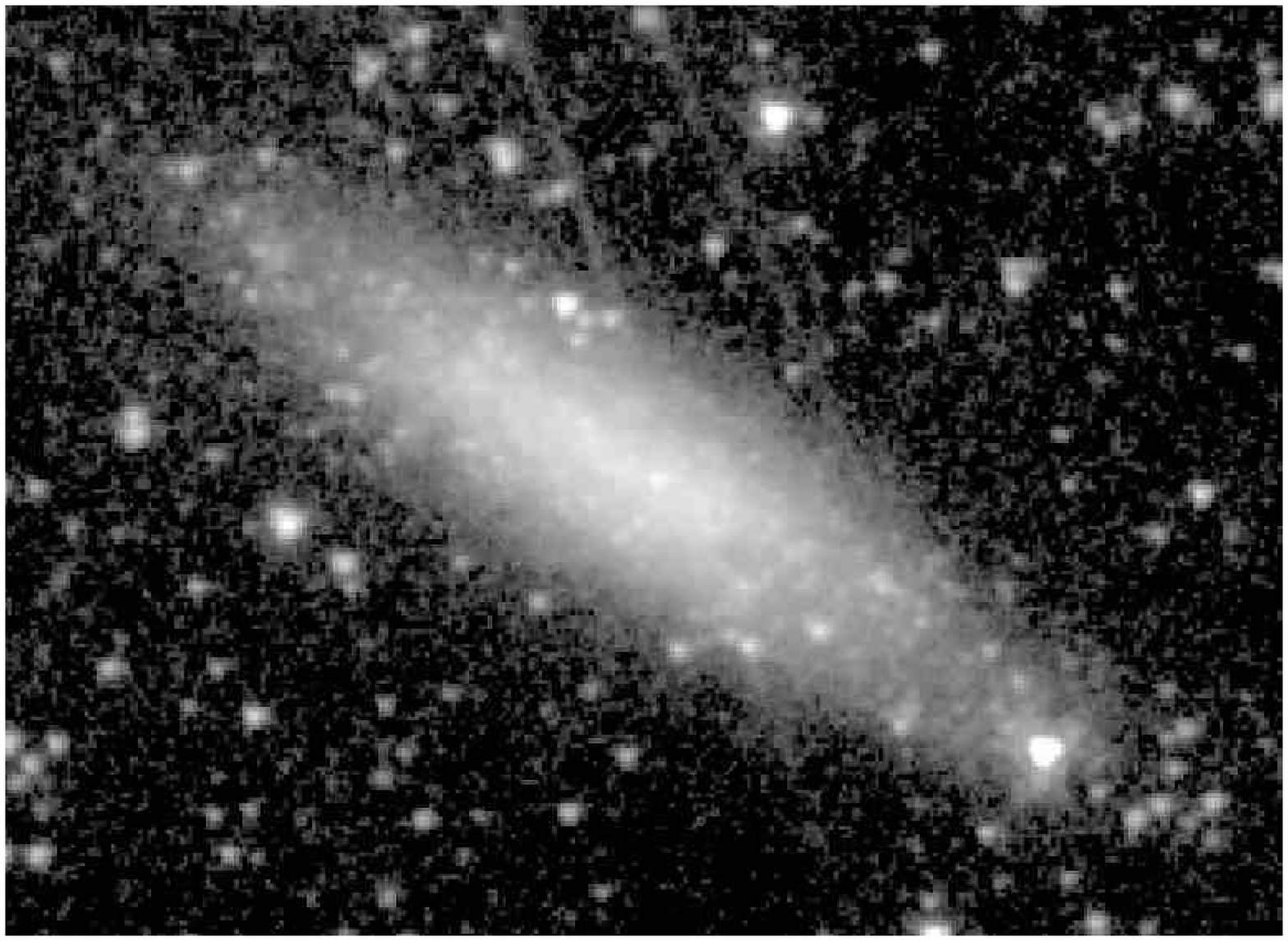}
 \vspace{2.0truecm}
 \caption{
{\bf IC   1959   }              - S$^4$G mid-IR classification:    SBd sp                                                ; Filter: IRAC 3.6$\mu$m; North: left, East: down; Field dimensions:   3.5$\times$  2.6 arcmin; Surface brightness range displayed: 17.5$-$27.0 mag arcsec$^{-2}$}                 
\label{IC1959}      
 \end{figure}
 
\clearpage
\begin{figure}
\figurenum{1.168}
\plotone{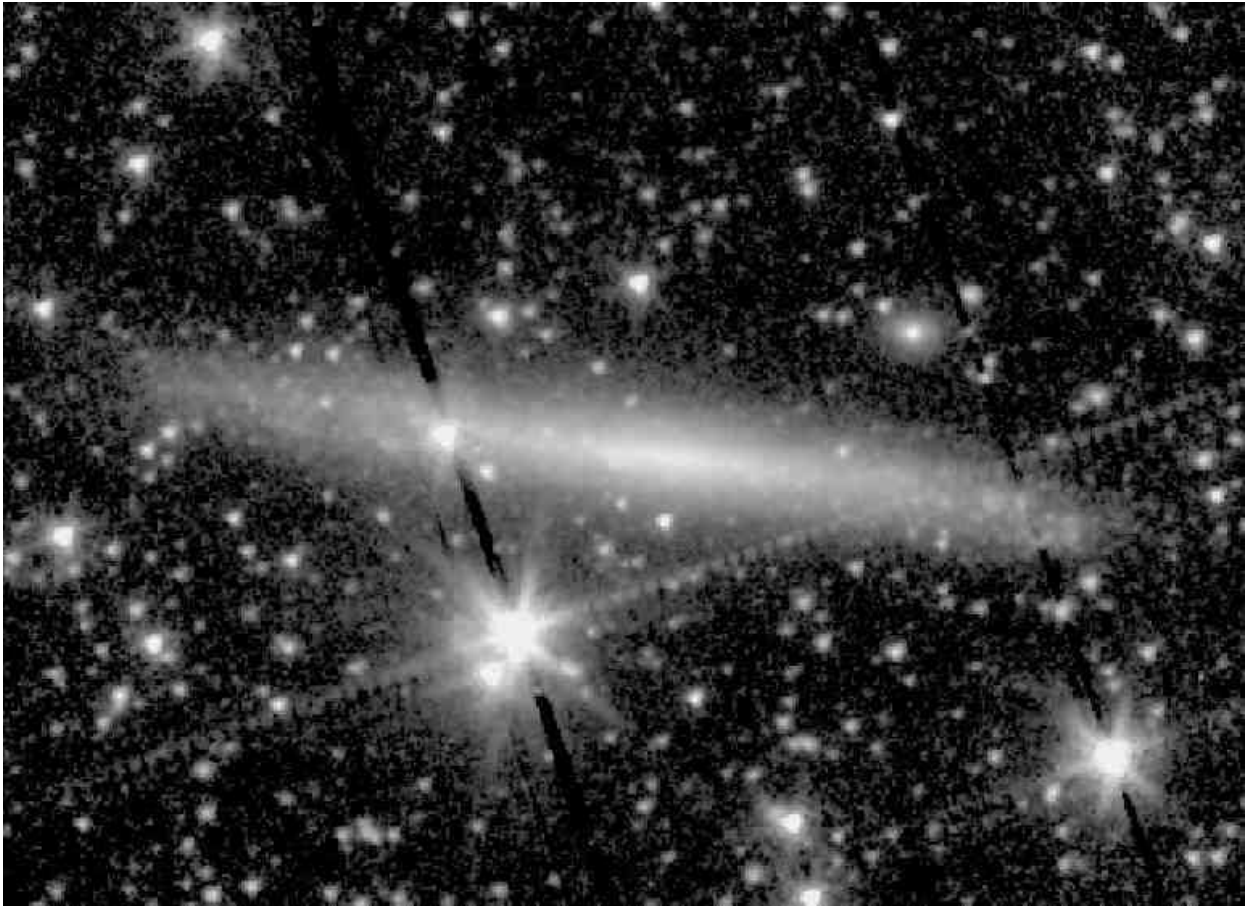}
 \vspace{2.0truecm}
 \caption{
{\bf IC   2233   }              - S$^4$G mid-IR classification:    SBdm sp                                               ; Filter: IRAC 3.6$\mu$m; North: left, East: down; Field dimensions:   5.7$\times$  4.2 arcmin; Surface brightness range displayed: 17.5$-$28.0 mag arcsec$^{-2}$}                 
\label{IC2233}      
 \end{figure}
 
\clearpage
\begin{figure}
\figurenum{1.169}
\plotone{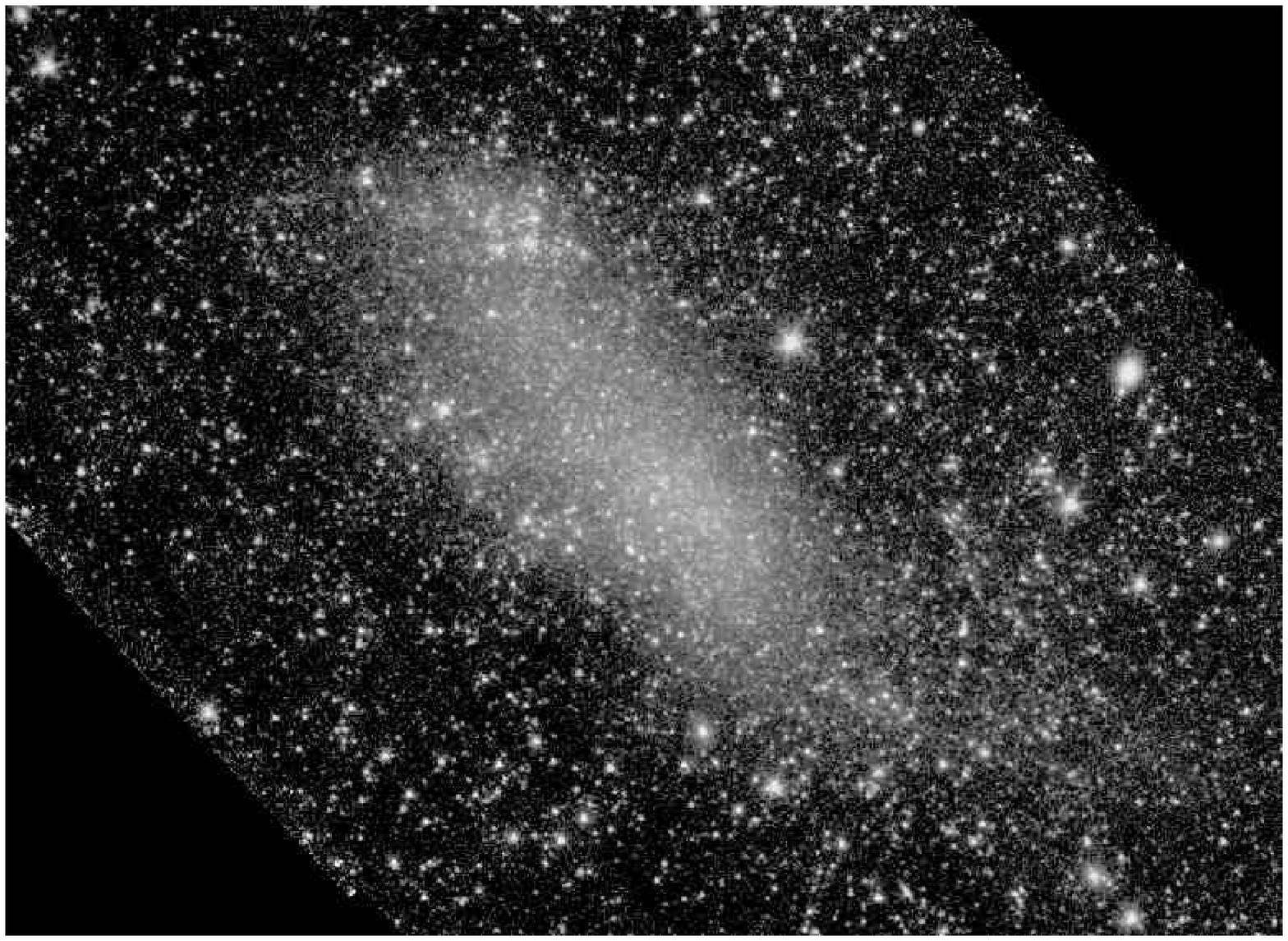}
 \vspace{2.0truecm}
 \caption{
{\bf IC   2574   }              - S$^4$G mid-IR classification:    IB(s)m                                                ; Filter: IRAC 3.6$\mu$m; North:   up, East: left; Field dimensions:  15.8$\times$ 11.5 arcmin; Surface brightness range displayed: 18.5$-$28.0 mag arcsec$^{-2}$}                 
\label{IC2574}      
 \end{figure}
 
\clearpage
\begin{figure}
\figurenum{1.170}
\plotone{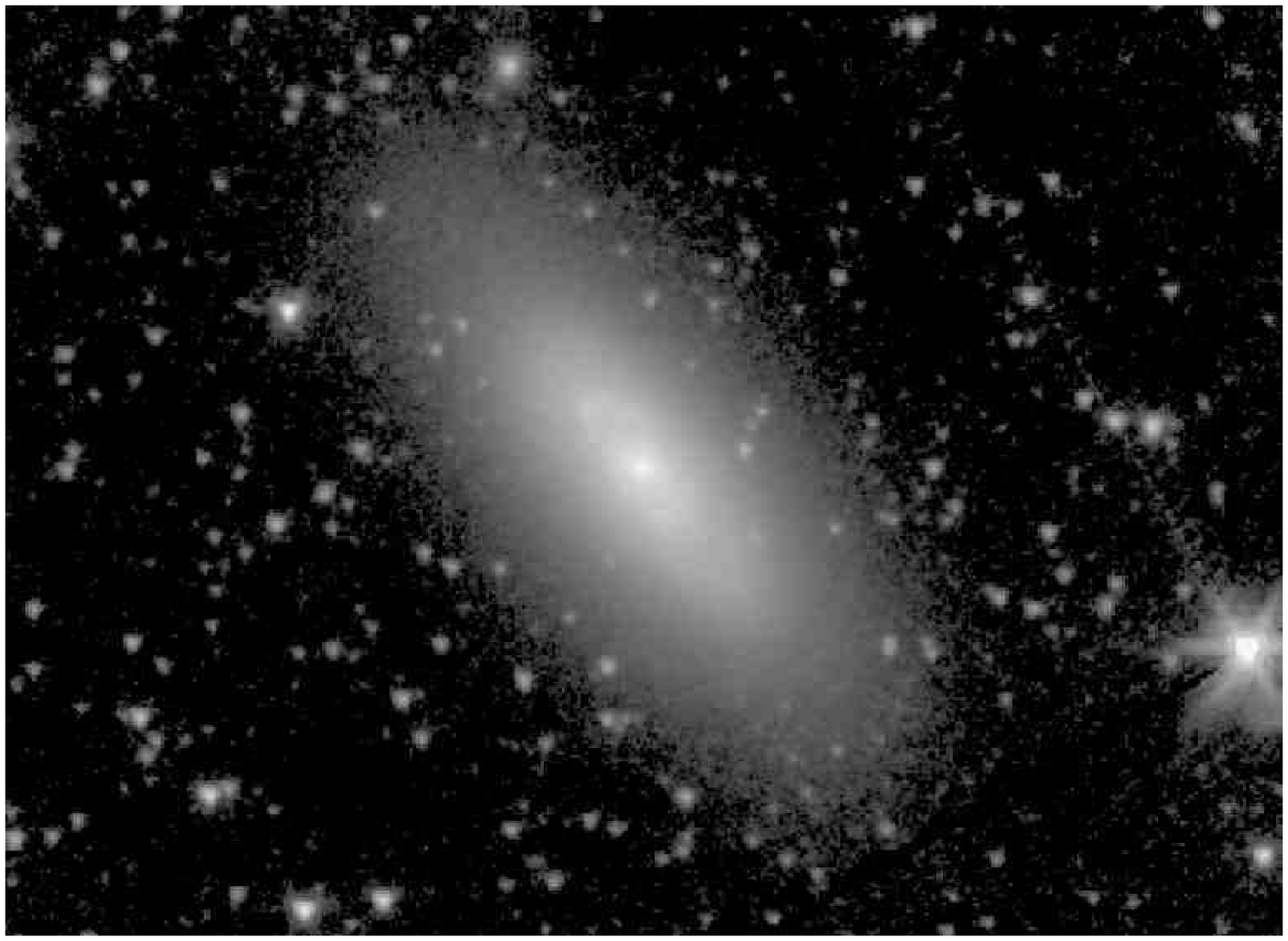}
 \vspace{2.0truecm}
 \caption{
{\bf IC   3392   }              - S$^4$G mid-IR classification:    SA(rs)ab                                              ; Filter: IRAC 3.6$\mu$m; North:   up, East: left; Field dimensions:   5.3$\times$  3.8 arcmin; Surface brightness range displayed: 15.0$-$28.0 mag arcsec$^{-2}$}                 
\label{IC3392}      
 \end{figure}
 
\clearpage
\begin{figure}
\figurenum{1.171}
\plotone{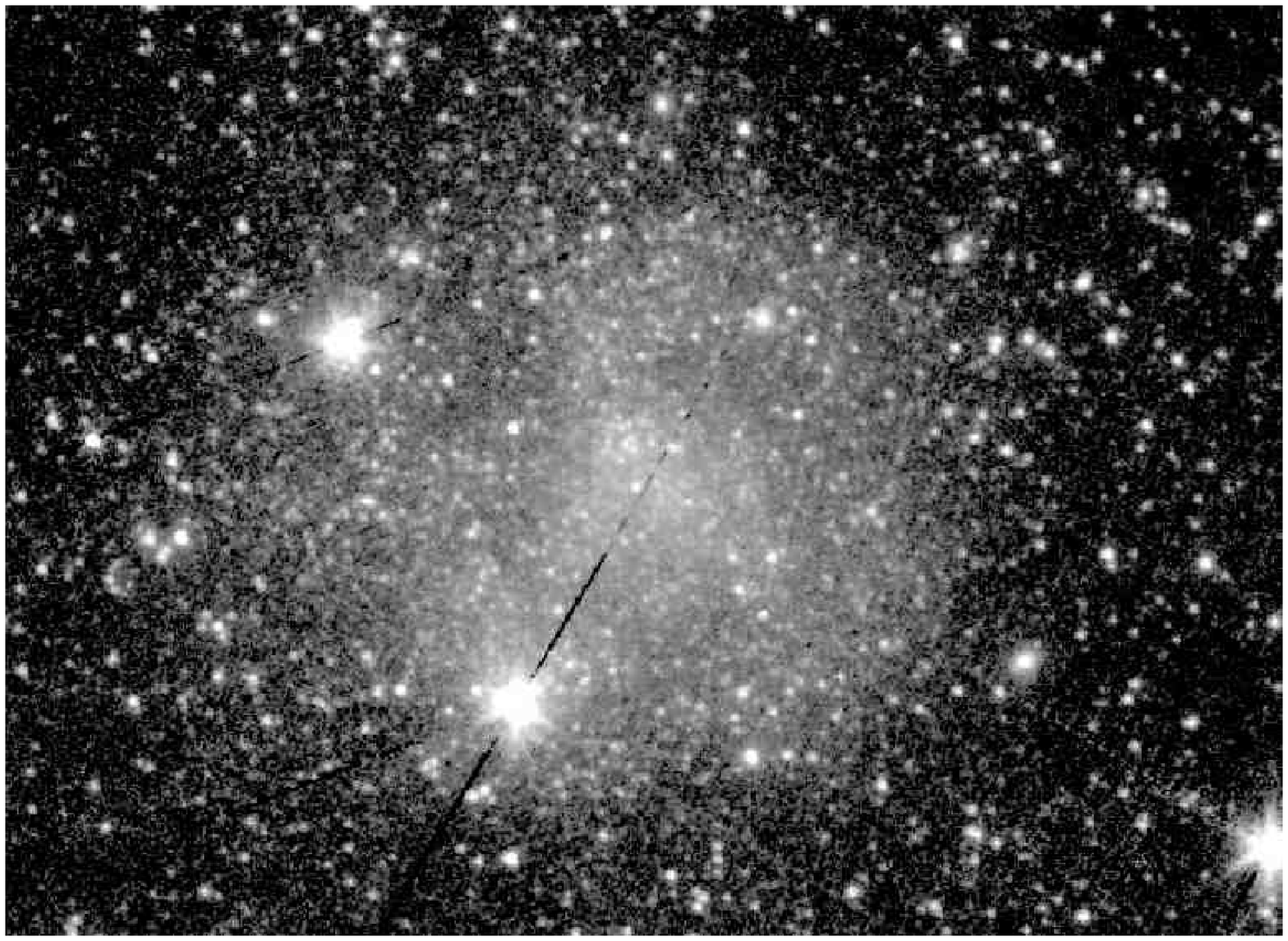}
 \vspace{2.0truecm}
 \caption{
{\bf IC   4182   }              - S$^4$G mid-IR classification:    SA(s)m                                                ; Filter: IRAC 3.6$\mu$m; North:   up, East: left; Field dimensions:   7.9$\times$  5.8 arcmin; Surface brightness range displayed: 18.5$-$28.0 mag arcsec$^{-2}$}                 
\label{IC4182}      
 \end{figure}
 
\clearpage
\begin{figure}
\figurenum{1.172}
\plotone{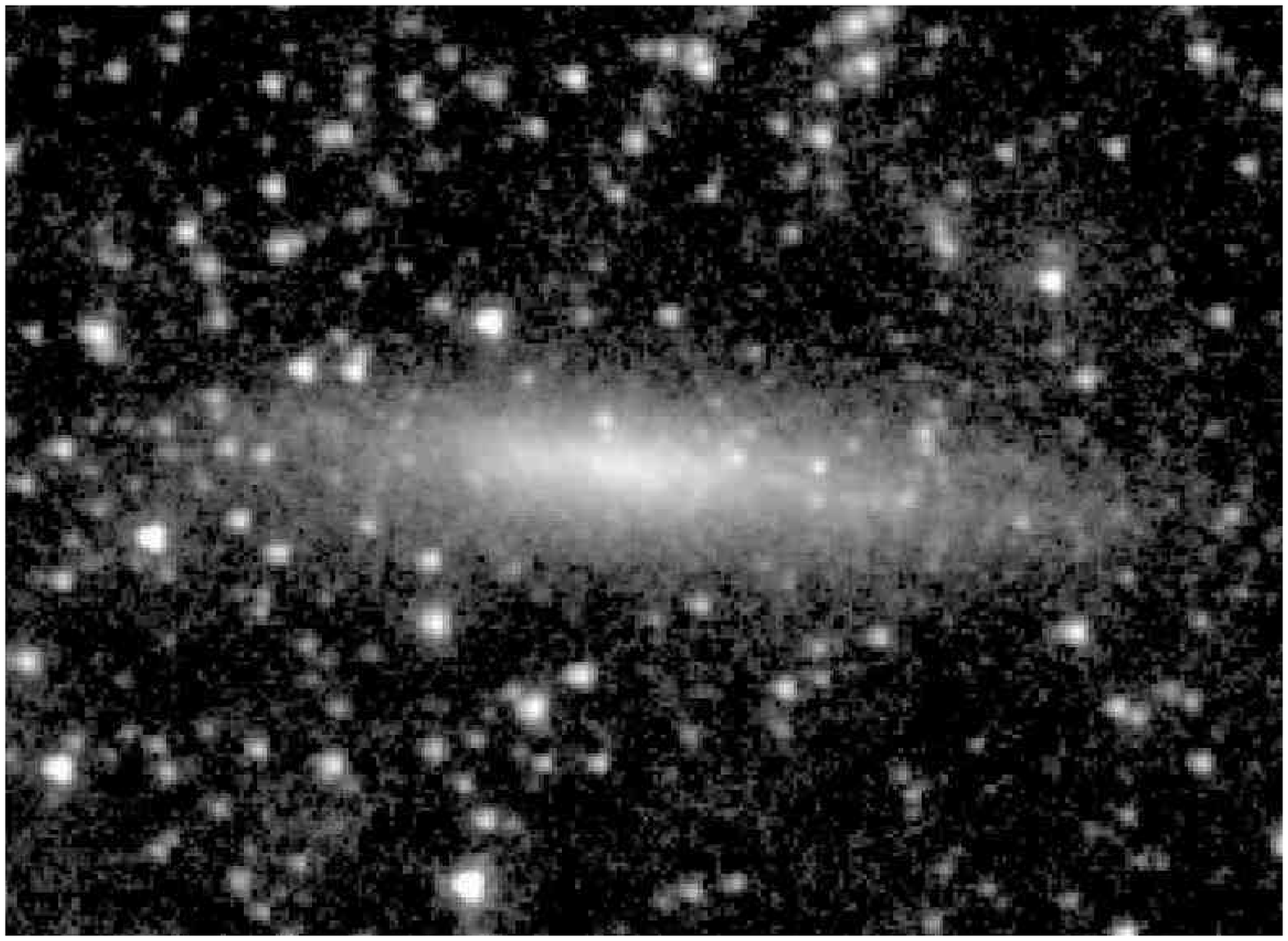}
 \vspace{2.0truecm}
 \caption{
{\bf IC   4951   }              - S$^4$G mid-IR classification:    SB(s)dm: sp                                           ; Filter: IRAC 3.6$\mu$m; North: left, East: down; Field dimensions:   3.5$\times$  2.6 arcmin; Surface brightness range displayed: 18.0$-$28.0 mag arcsec$^{-2}$}                 
\label{IC4951}      
 \end{figure}
 
\clearpage
\begin{figure}
\figurenum{1.173}
\plotone{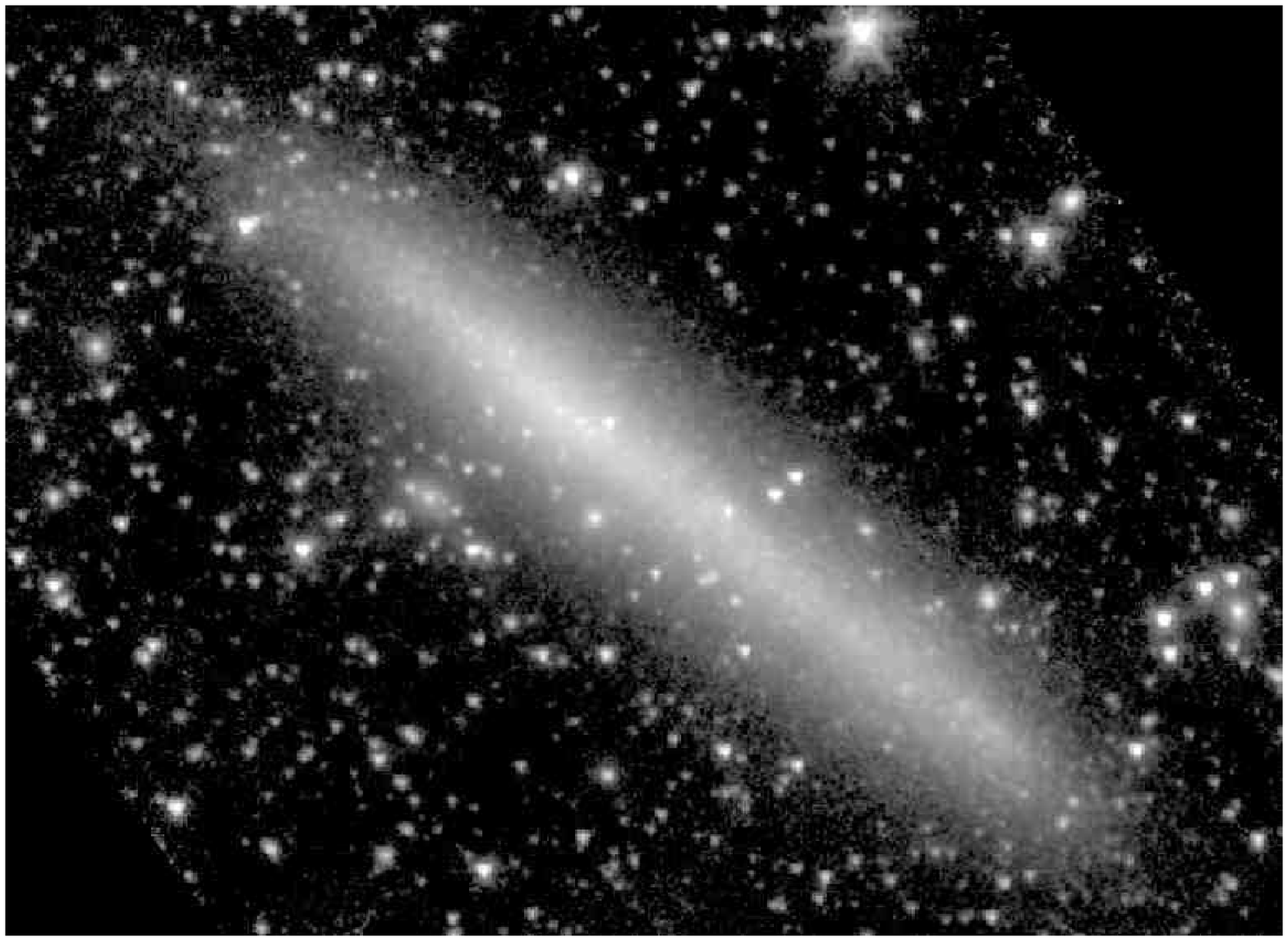}
 \vspace{2.0truecm}
 \caption{
{\bf IC   5052   }              - S$^4$G mid-IR classification:    Sd sp                                                 ; Filter: IRAC 3.6$\mu$m; North: left, East: down; Field dimensions:   7.0$\times$  5.1 arcmin; Surface brightness range displayed: 16.5$-$28.0 mag arcsec$^{-2}$}                 
\label{IC5052}      
 \end{figure}
 
\clearpage
\begin{figure}
\figurenum{1.174}
\plotone{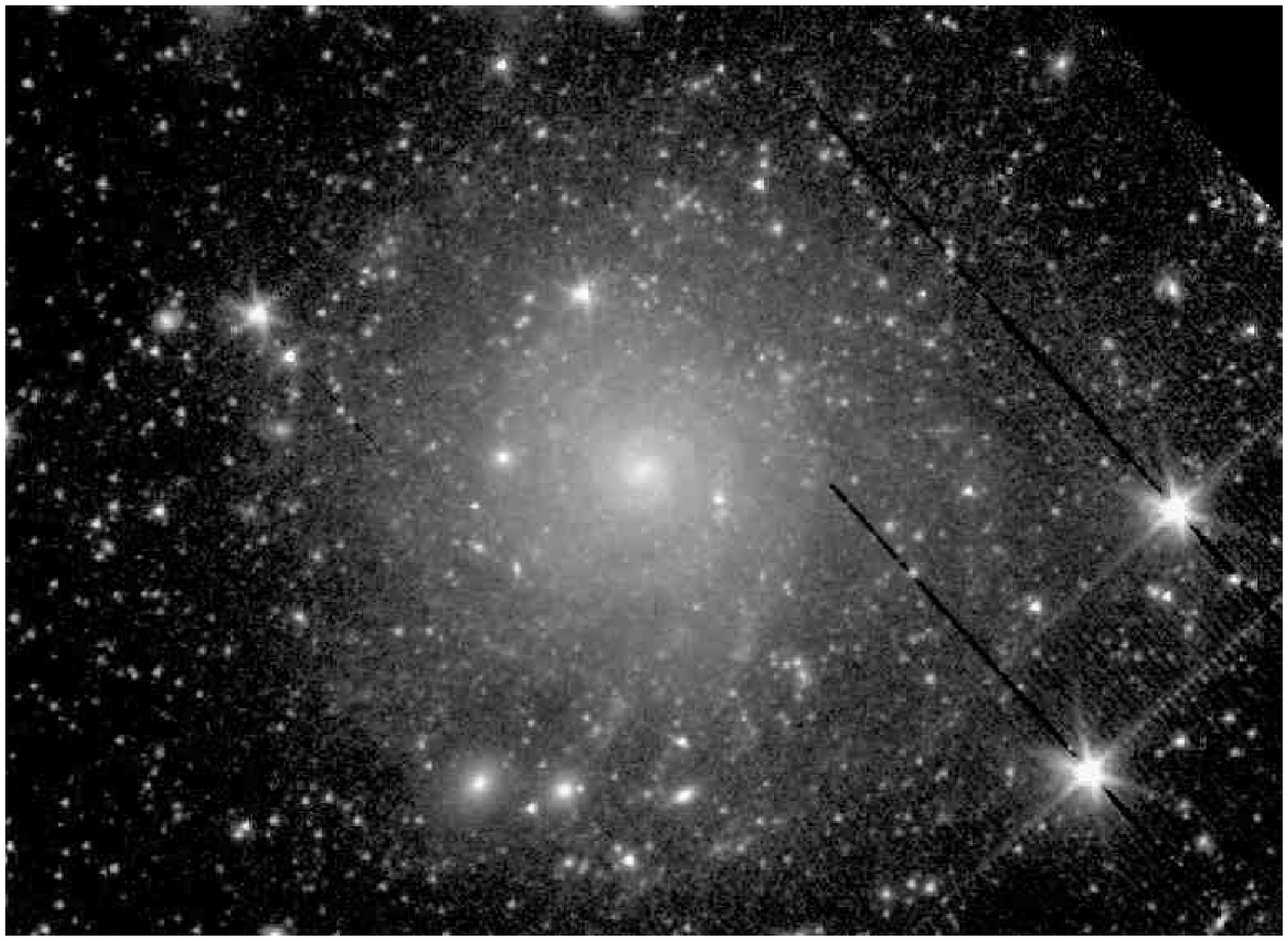}
 \vspace{2.0truecm}
 \caption{
{\bf IC   5332   }              - S$^4$G mid-IR classification:    S$\underline{\rm A}$B(s)cd                            ; Filter: IRAC 3.6$\mu$m; North:   up, East: left; Field dimensions:  10.5$\times$  7.7 arcmin; Surface brightness range displayed: 17.0$-$28.0 mag arcsec$^{-2}$}                 
\label{IC5332}      
 \end{figure}
 
\clearpage
\begin{figure}
\figurenum{1.175}
\plotone{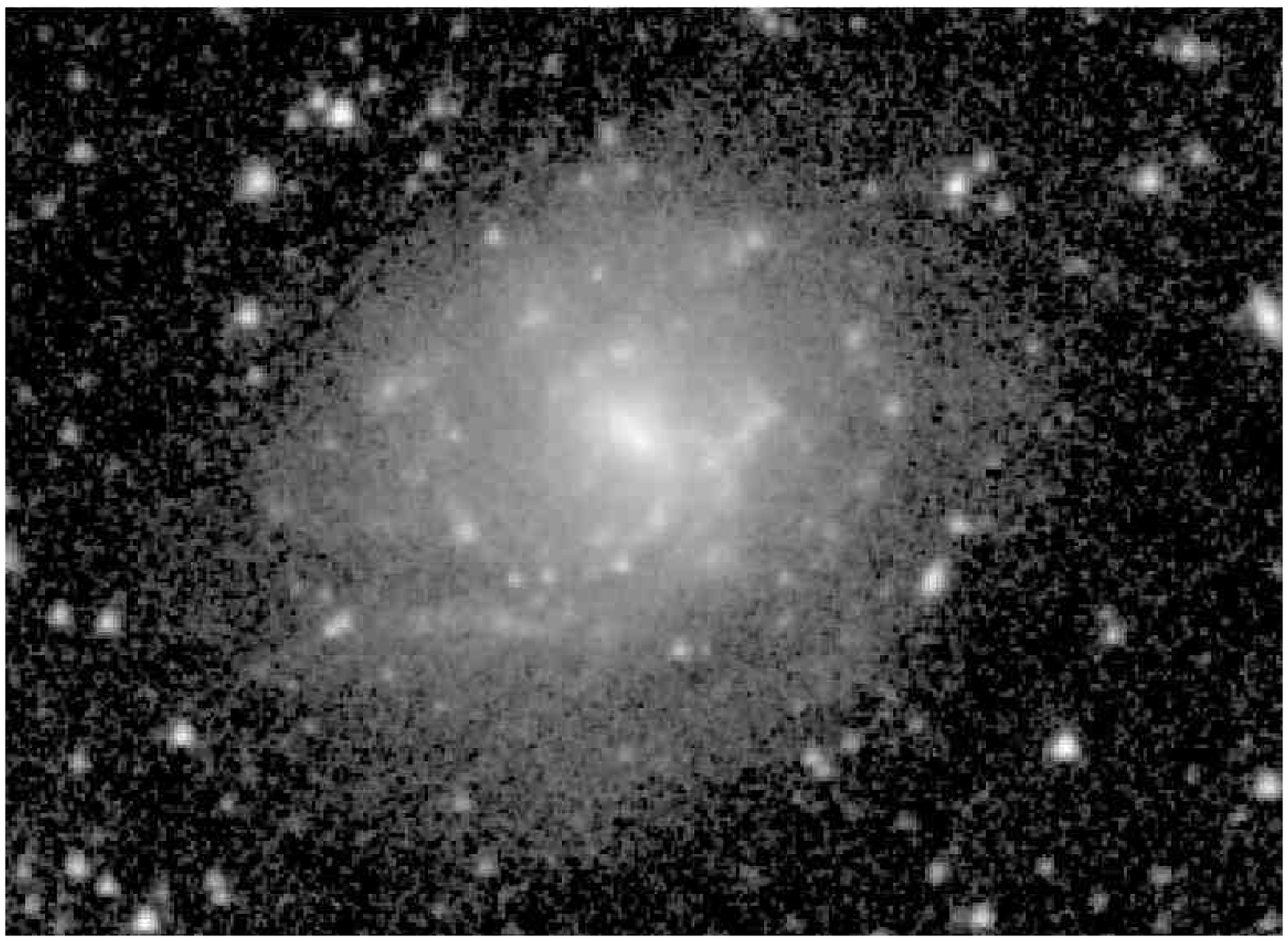}
 \vspace{2.0truecm}
 \caption{
{\bf PGC  6667   }              - S$^4$G mid-IR classification:    SB(s)dm                                               ; Filter: IRAC 3.6$\mu$m; North:   up, East: left; Field dimensions:   4.0$\times$  2.9 arcmin; Surface brightness range displayed: 17.5$-$28.0 mag arcsec$^{-2}$}                 
\label{PGC006667}   
 \end{figure}
 
\clearpage
\begin{figure}
\figurenum{1.176}
\plotone{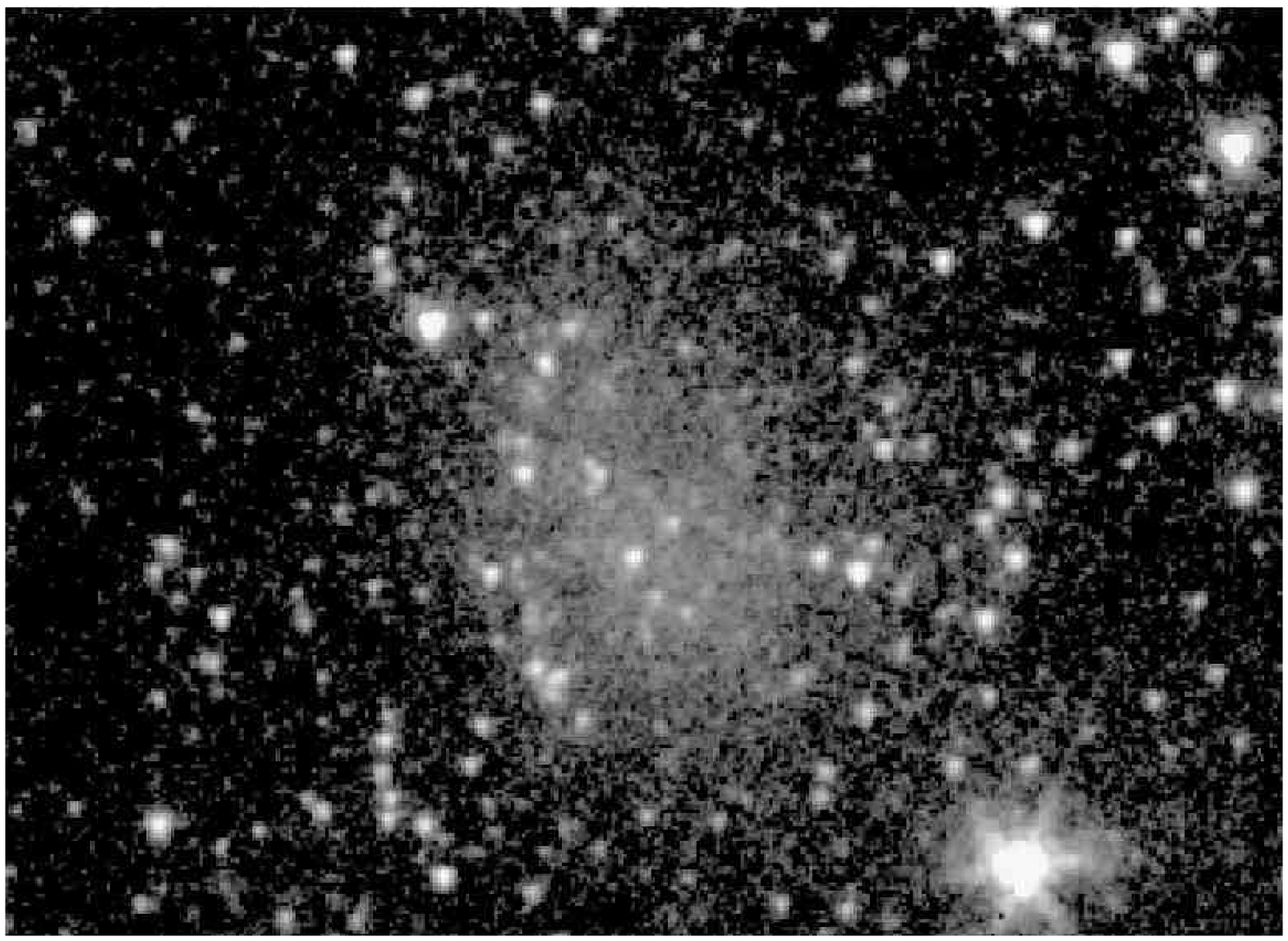}
 \vspace{2.0truecm}
 \caption{
{\bf UGC  1176   }              - S$^4$G mid-IR classification:    Im                                                    ; Filter: IRAC 3.6$\mu$m; North:   up, East: left; Field dimensions:   4.0$\times$  2.9 arcmin; Surface brightness range displayed: 18.5$-$28.0 mag arcsec$^{-2}$}                 
\label{UGC01176}    
 \end{figure}
 
\clearpage
\begin{figure}
\figurenum{1.177}
\plotone{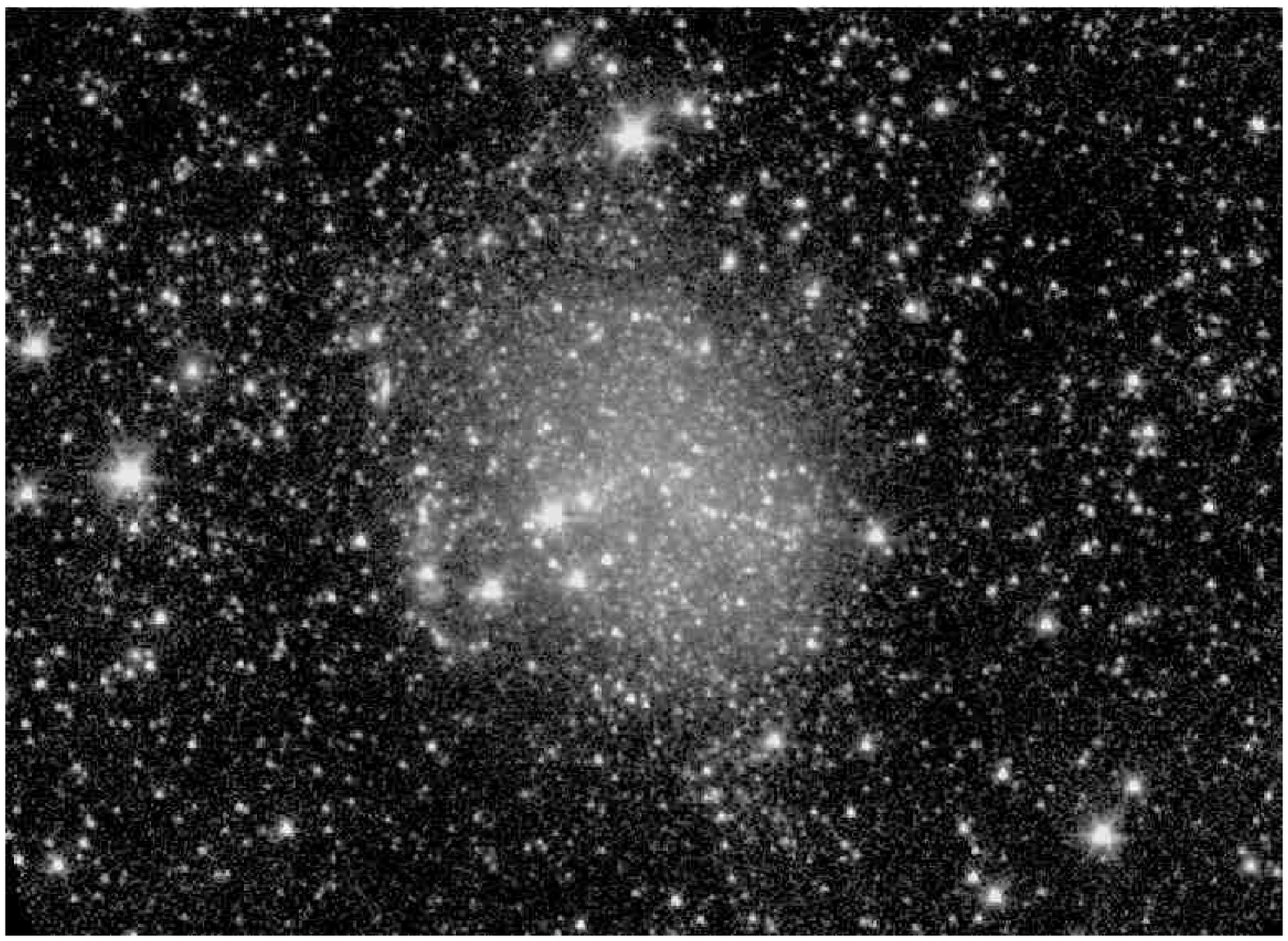}
 \vspace{2.0truecm}
 \caption{
{\bf UGC  4305   }              - S$^4$G mid-IR classification:    Im                                                    ; Filter: IRAC 3.6$\mu$m; North:   up, East: left; Field dimensions:  10.5$\times$  7.7 arcmin; Surface brightness range displayed: 18.5$-$28.0 mag arcsec$^{-2}$}                 
\label{UGC04305}    
 \end{figure}
 
\clearpage
\begin{figure}
\figurenum{1.178}
\plotone{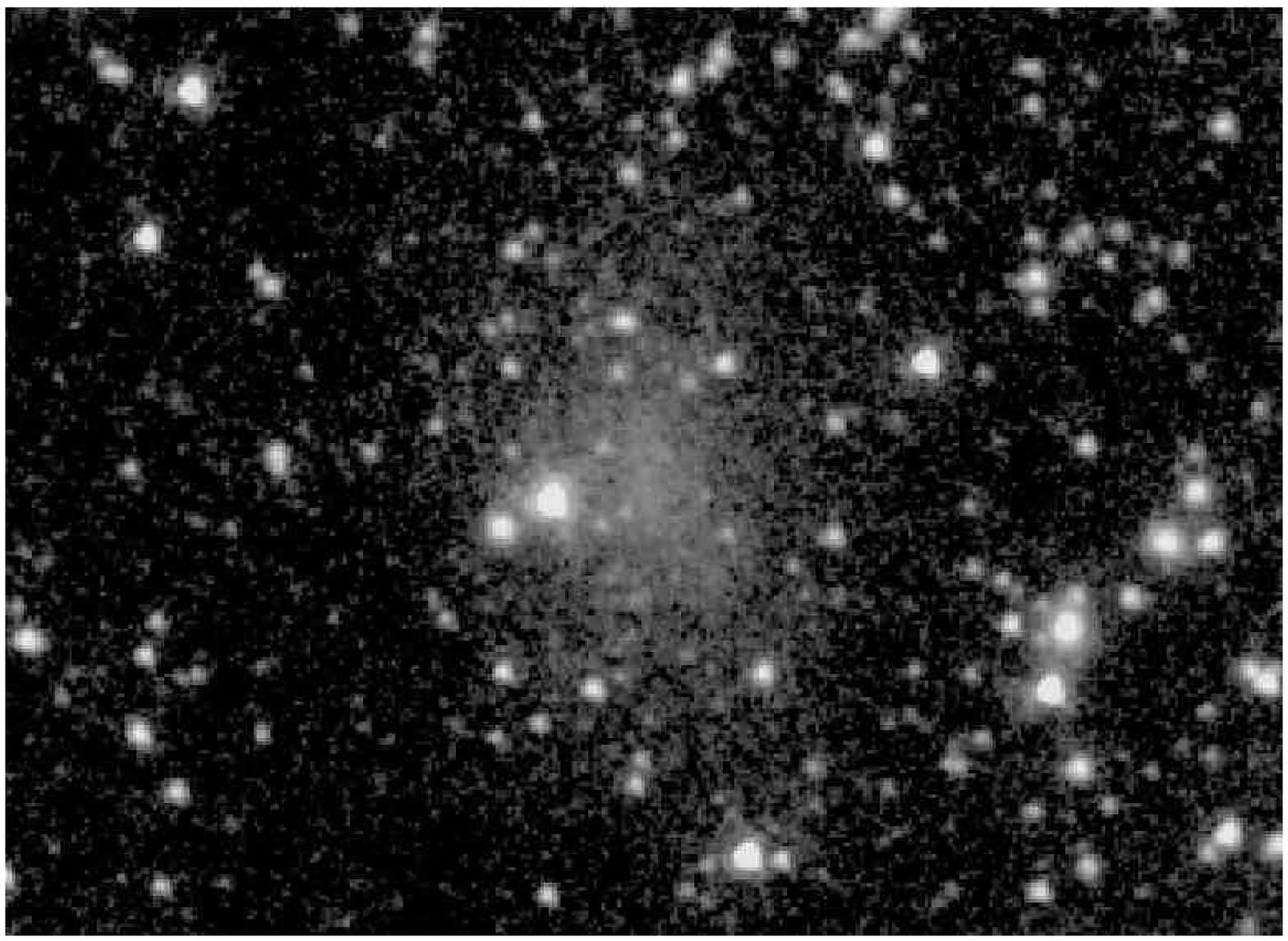}
 \vspace{2.0truecm}
 \caption{
{\bf UGC  4426   }              - S$^4$G mid-IR classification:    Im                                                    ; Filter: IRAC 3.6$\mu$m; North:   up, East: left; Field dimensions:   3.9$\times$  2.9 arcmin; Surface brightness range displayed: 18.5$-$28.0 mag arcsec$^{-2}$}                 
\label{UGC04426}    
 \end{figure}
 
\clearpage
\begin{figure}
\figurenum{1.179}
\plotone{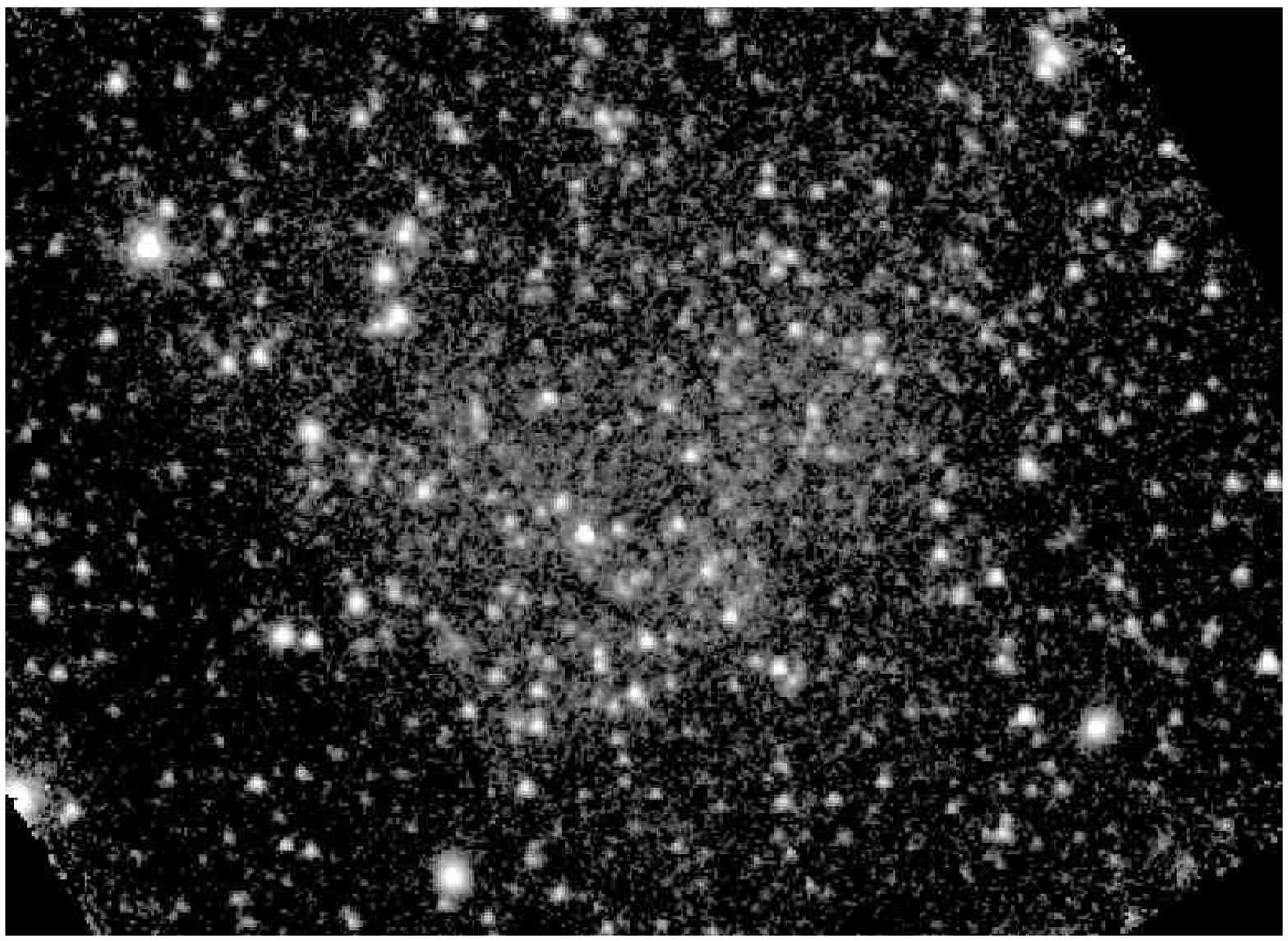}
 \vspace{2.0truecm}
 \caption{
{\bf UGC  5139   }              - S$^4$G mid-IR classification:    Im                                                    ; Filter: IRAC 3.6$\mu$m; North:   up, East: left; Field dimensions:   5.3$\times$  3.8 arcmin; Surface brightness range displayed: 18.0$-$28.0 mag arcsec$^{-2}$}                 
\label{UGC05139}    
 \end{figure}
 
\clearpage
\begin{figure}
\figurenum{1.180}
\plotone{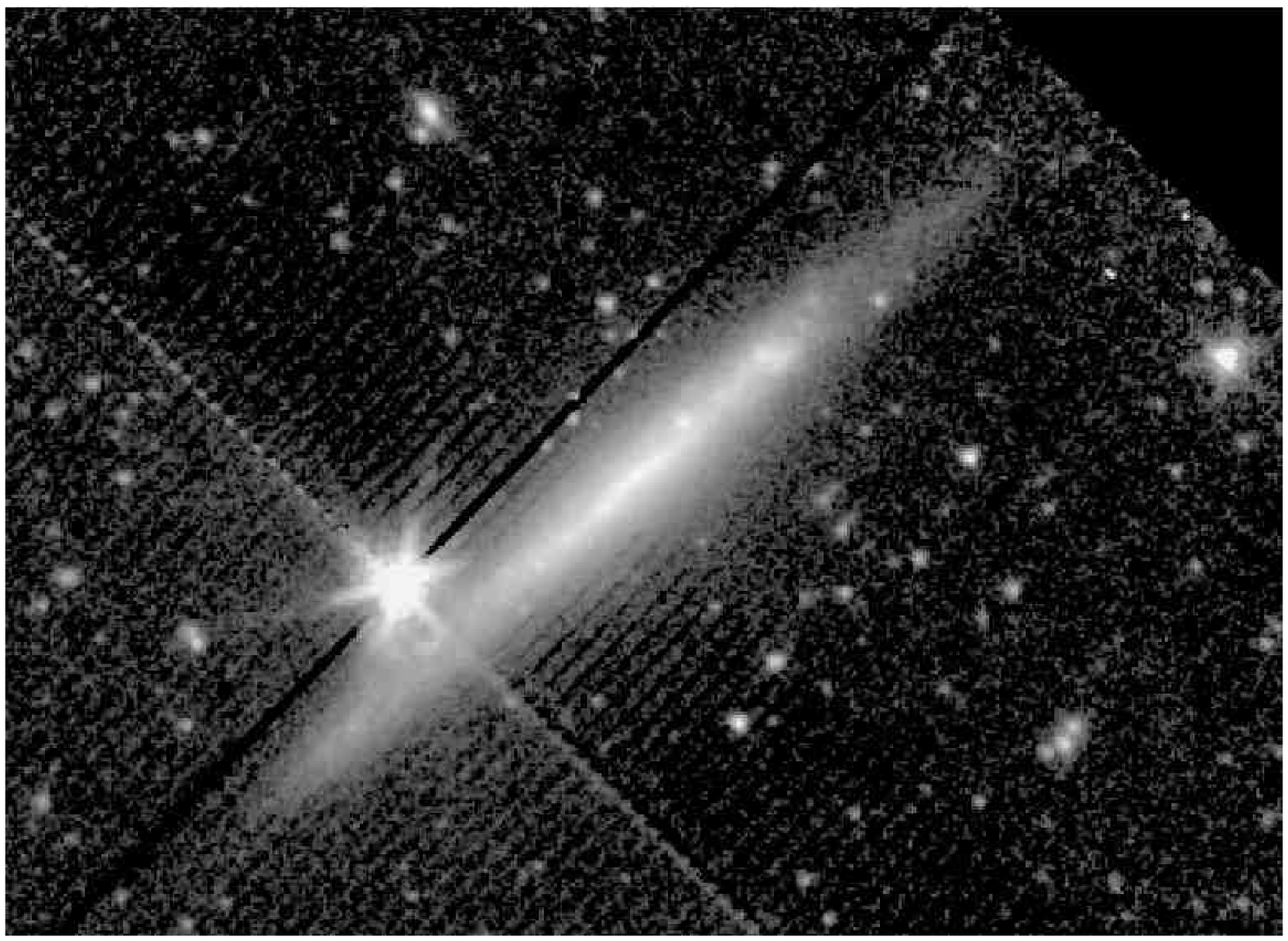}
 \vspace{2.0truecm}
 \caption{
{\bf UGC  5459   }              - S$^4$G mid-IR classification:    SBdm sp                                               ; Filter: IRAC 3.6$\mu$m; North:   up, East: left; Field dimensions:   5.3$\times$  3.8 arcmin; Surface brightness range displayed: 16.5$-$28.0 mag arcsec$^{-2}$}                 
\label{UGC05459}    
 \end{figure}
 
\clearpage
\begin{figure}
\figurenum{1.181}
\plotone{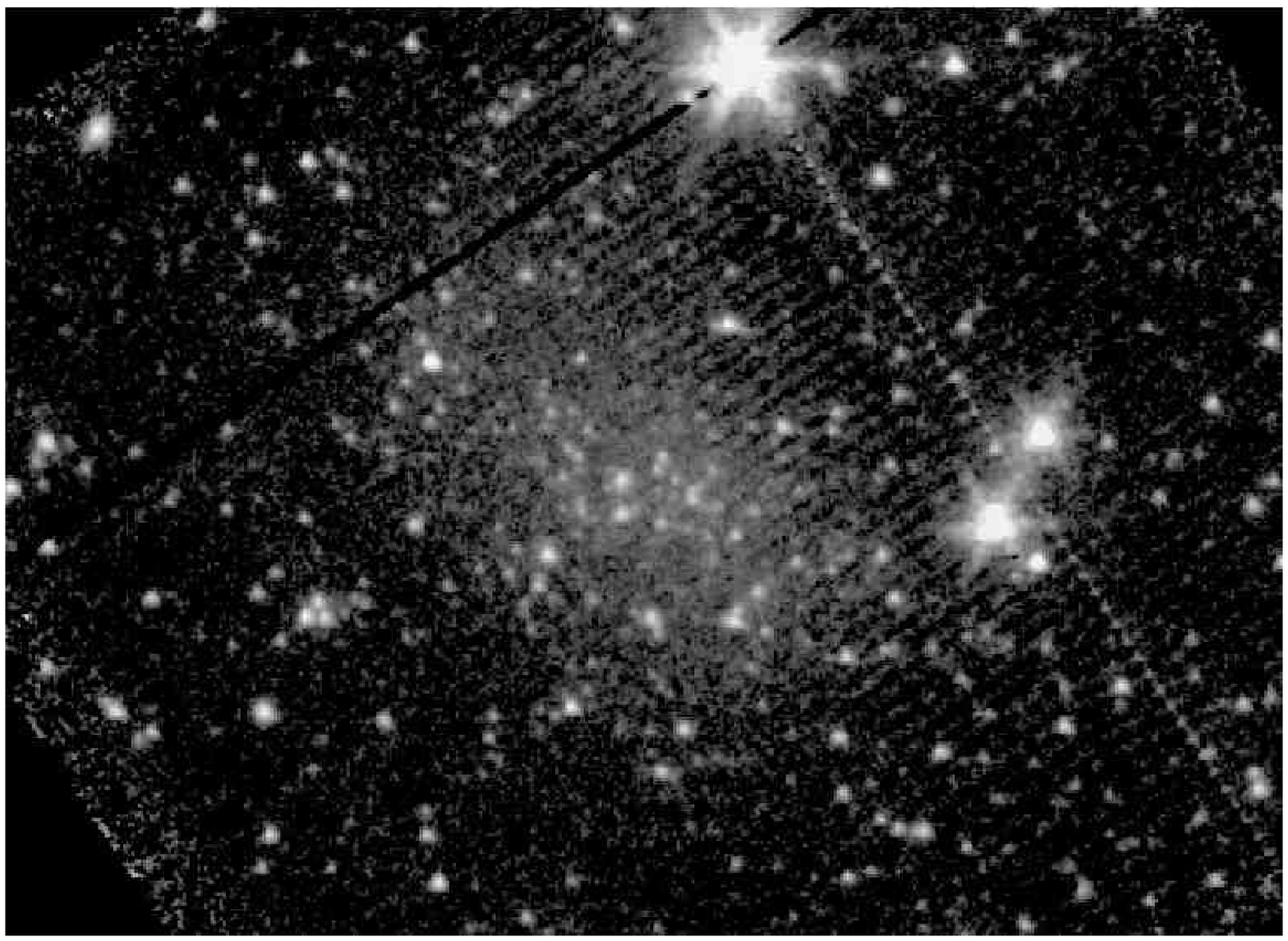}
 \vspace{2.0truecm}
 \caption{
{\bf UGC  6817   }              - S$^4$G mid-IR classification:    Im                                                    ; Filter: IRAC 3.6$\mu$m; North:   up, East: left; Field dimensions:   5.3$\times$  3.8 arcmin; Surface brightness range displayed: 18.0$-$28.0 mag arcsec$^{-2}$}                 
\label{UGC06817}    
 \end{figure}
 
\clearpage
\begin{figure}
\figurenum{1.182}
\plotone{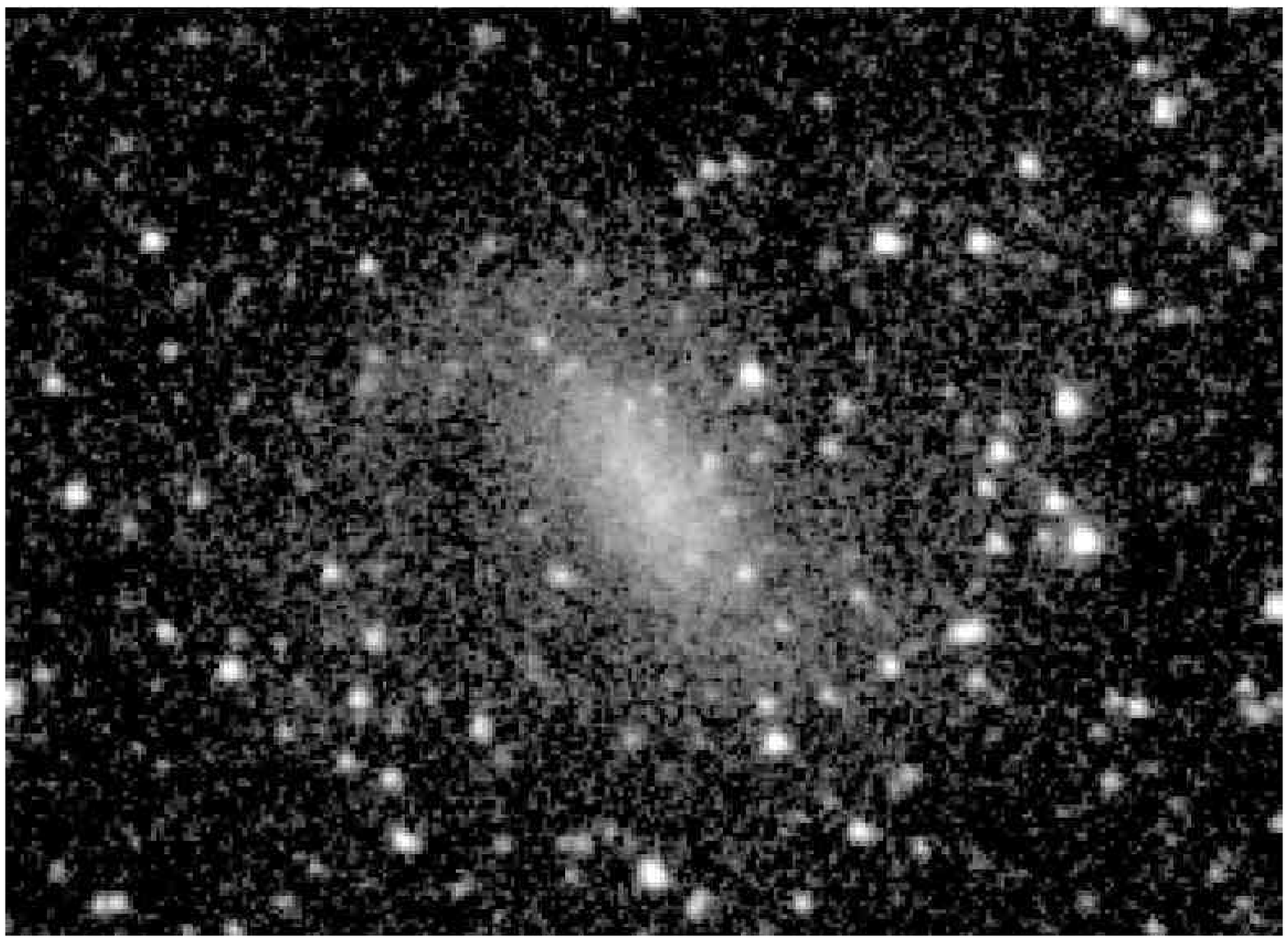}
 \vspace{2.0truecm}
 \caption{
{\bf UGC  6956   }              - S$^4$G mid-IR classification:    SB(s)dm:                                              ; Filter: IRAC 3.6$\mu$m; North:   up, East: left; Field dimensions:   3.5$\times$  2.6 arcmin; Surface brightness range displayed: 18.5$-$28.0 mag arcsec$^{-2}$}                 
\label{UGC06956}    
 \end{figure}
 
\clearpage
\begin{figure}
\figurenum{1.183}
\plotone{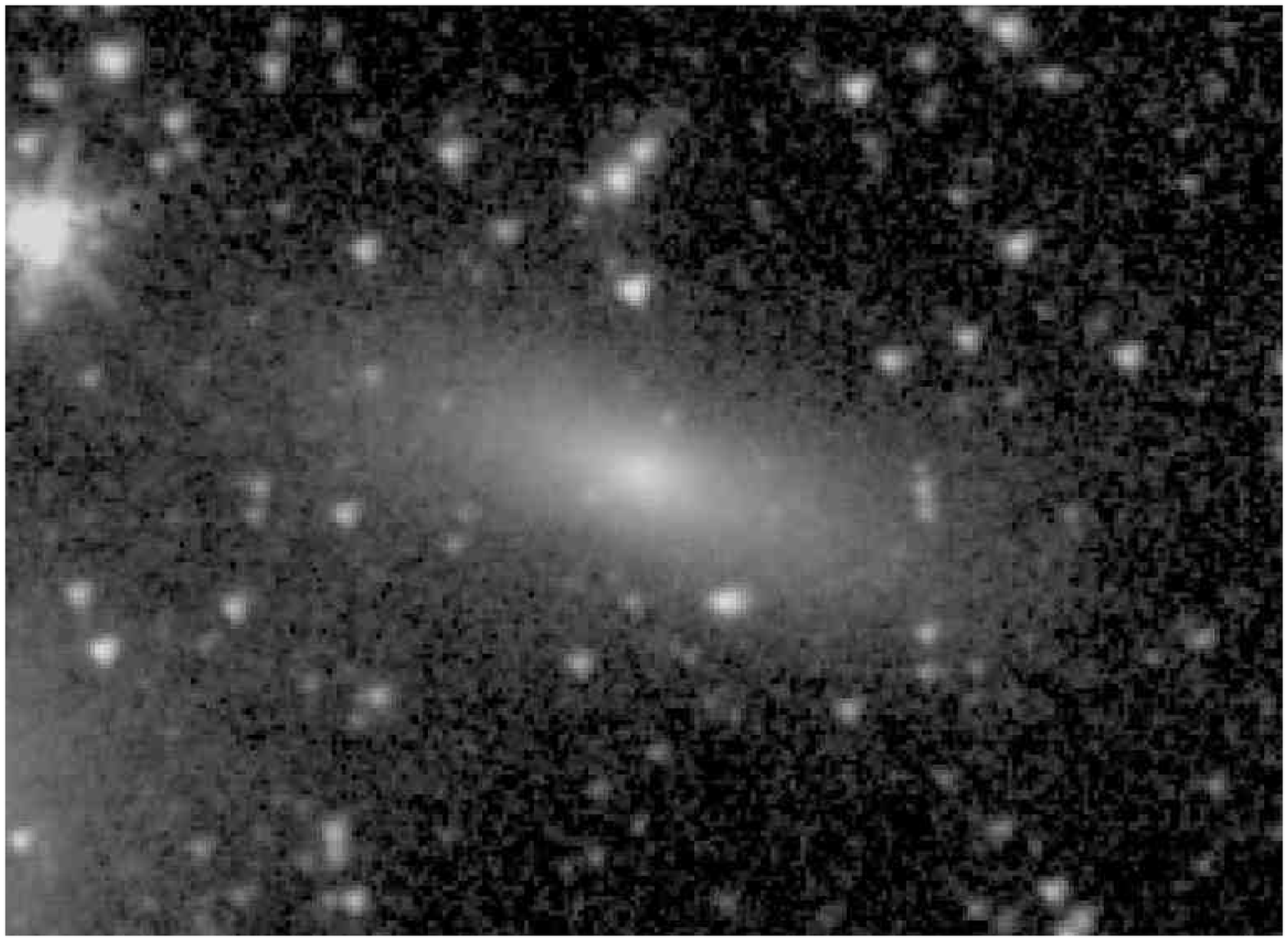}
 \vspace{2.0truecm}
 \caption{
{\bf UGC  7504   }              - S$^4$G mid-IR classification:    S0$^-$                                                ; Filter: IRAC 3.6$\mu$m; North: left, East: down; Field dimensions:   3.2$\times$  2.3 arcmin; Surface brightness range displayed: 18.0$-$28.0 mag arcsec$^{-2}$}                 
\label{UGC07504}    
 \end{figure}
 
\clearpage
\begin{figure}
\figurenum{1.184}
\plotone{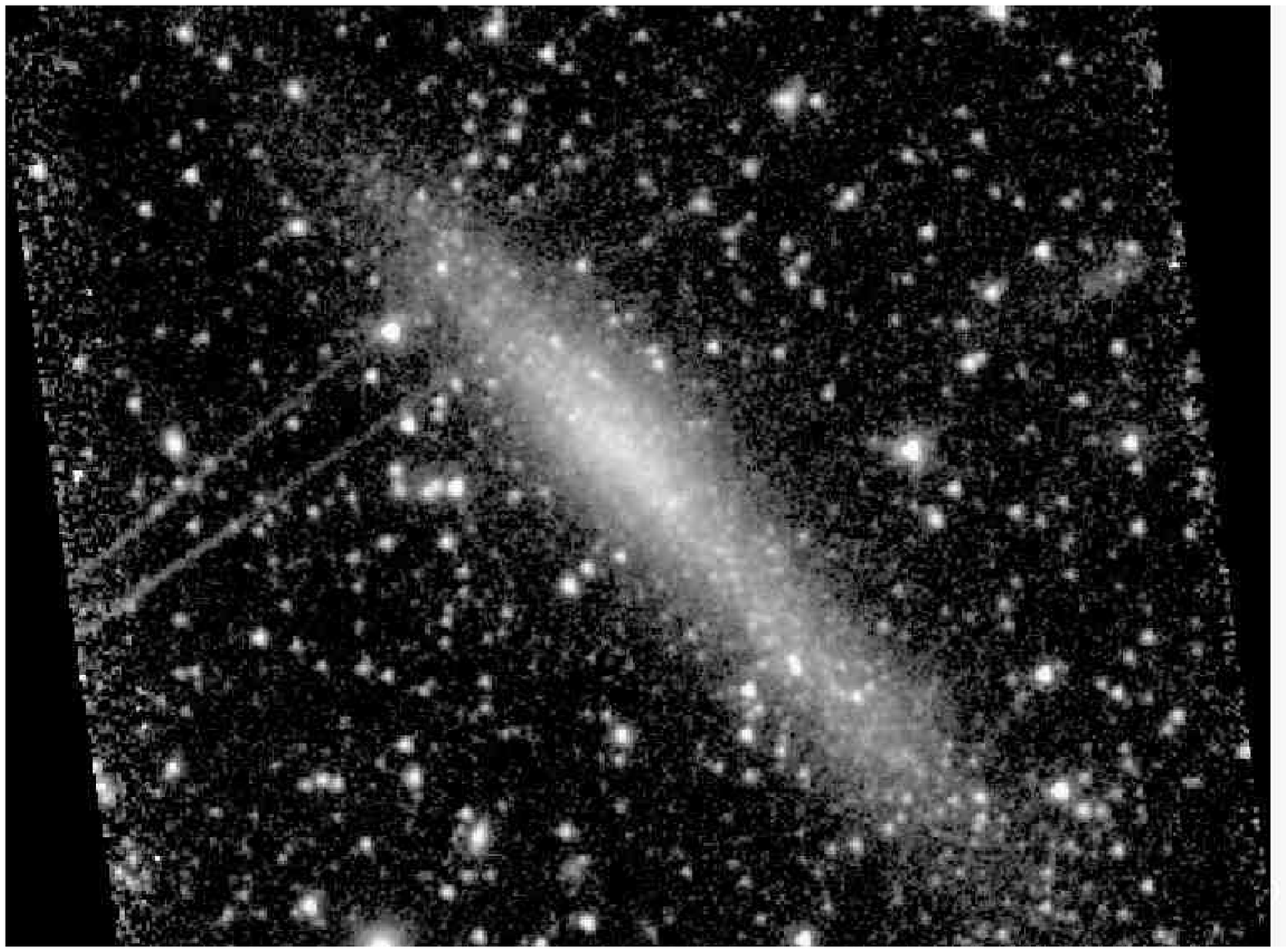}
 \vspace{2.0truecm}
 \caption{
{\bf ESO  115- 21}              - S$^4$G mid-IR classification:    SB(s)m sp                                             ; Filter: IRAC 3.6$\mu$m; North:   up, East: left; Field dimensions:   6.2$\times$  4.5 arcmin; Surface brightness range displayed: 18.5$-$28.0 mag arcsec$^{-2}$}                 
\label{ESO115-021}  
 \end{figure}
 
\clearpage
\begin{figure}
\figurenum{1.185}
\plotone{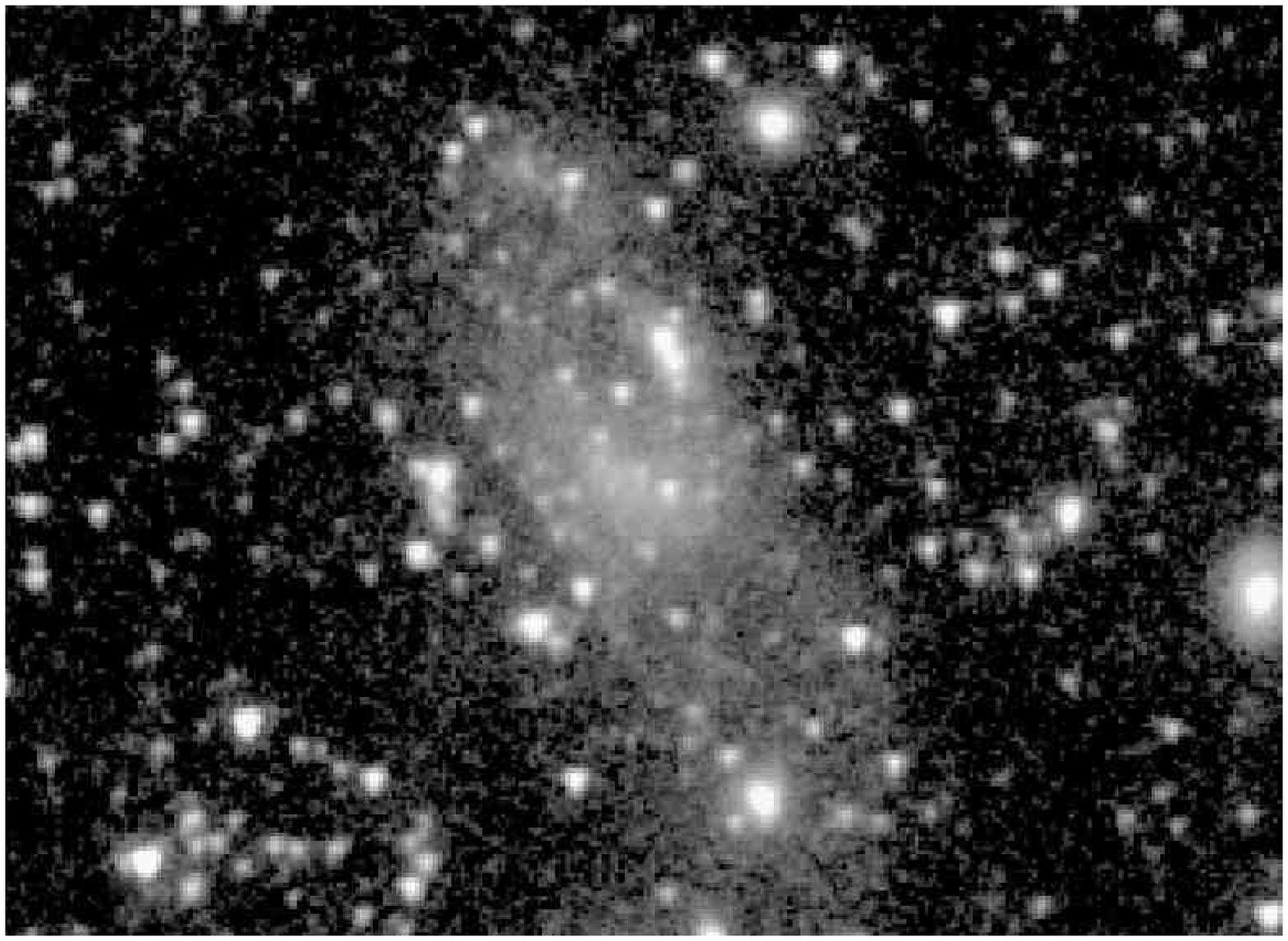}
 \vspace{2.0truecm}
 \caption{
{\bf ESO  119- 16}              - S$^4$G mid-IR classification:    (R$^{\prime}$)SAB(s)dm:                                         ; Filter: IRAC 3.6$\mu$m; North:   up, East: left; Field dimensions:   3.5$\times$  2.6 arcmin; Surface brightness range displayed: 18.5$-$28.0 mag arcsec$^{-2}$}                 
\label{ESO119-016}  
 \end{figure}
 
\clearpage
\begin{figure}
\figurenum{1.186}
\plotone{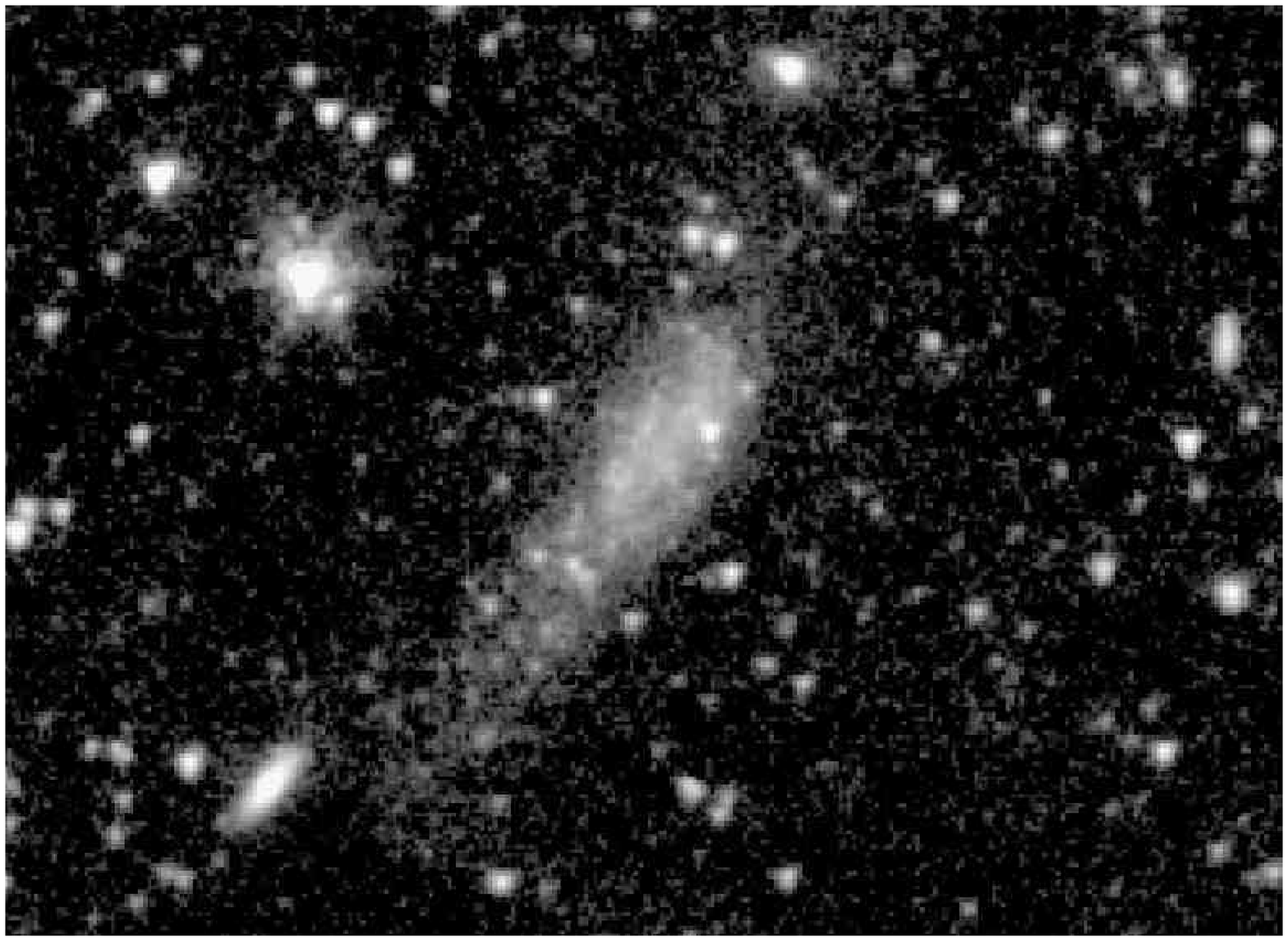}
 \vspace{2.0truecm}
 \caption{
{\bf ESO  149-  3}              - S$^4$G mid-IR classification:    IBm: sp                                               ; Filter: IRAC 3.6$\mu$m; North:   up, East: left; Field dimensions:   3.5$\times$  2.6 arcmin; Surface brightness range displayed: 18.5$-$28.0 mag arcsec$^{-2}$}                 
\label{ESO149-003}  
 \end{figure}
 
\clearpage
\begin{figure}
\figurenum{1.187}
\plotone{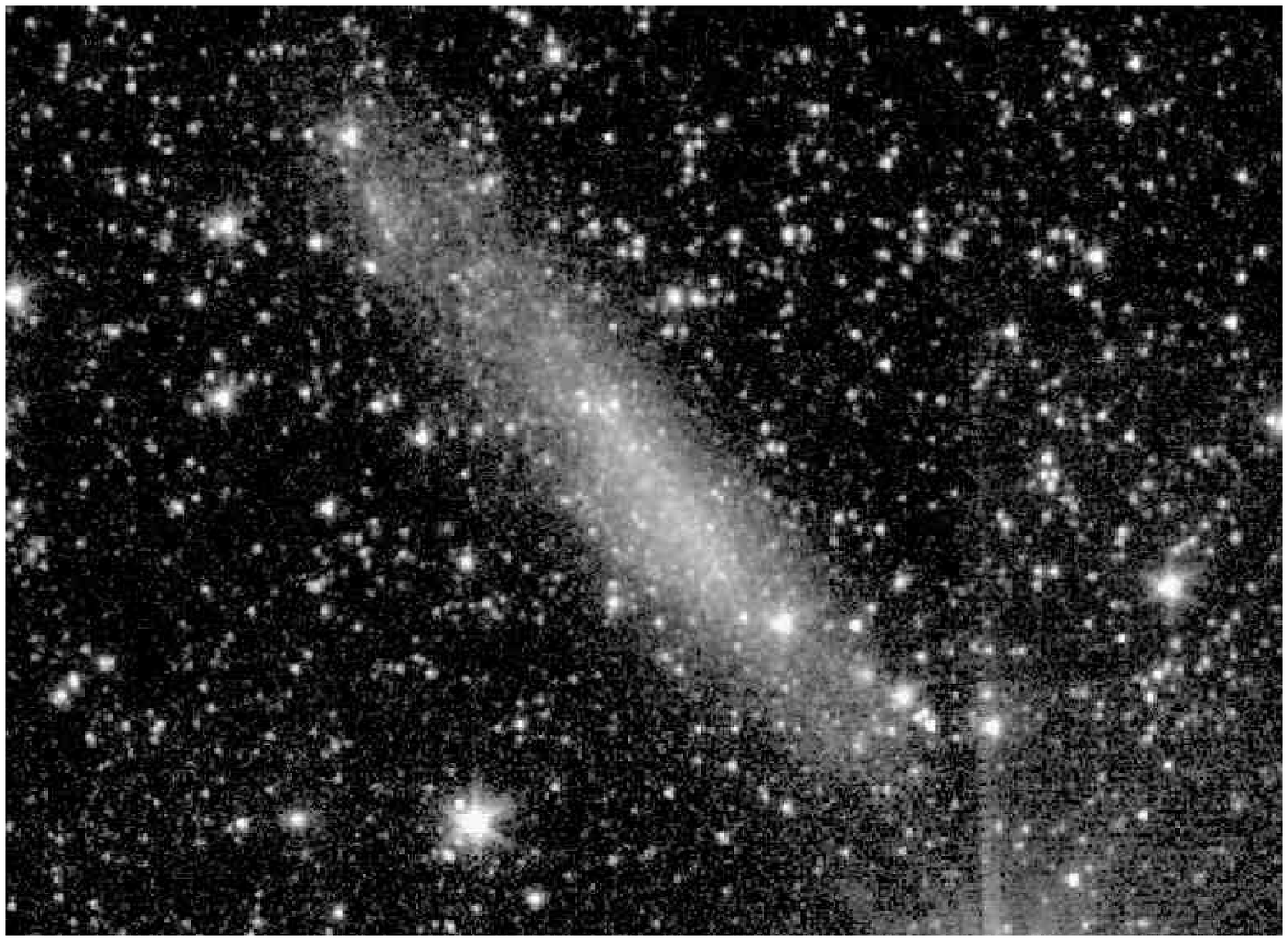}
 \vspace{2.0truecm}
 \caption{
{\bf ESO  154- 23}              - S$^4$G mid-IR classification:    SB(s)d sp                                             ; Filter: IRAC 3.6$\mu$m; North:   up, East: left; Field dimensions:   9.0$\times$  6.6 arcmin; Surface brightness range displayed: 18.5$-$28.0 mag arcsec$^{-2}$}                 
\label{ESO154-023}  
 \end{figure}
 
\clearpage
\begin{figure}
\figurenum{1.188}
\plotone{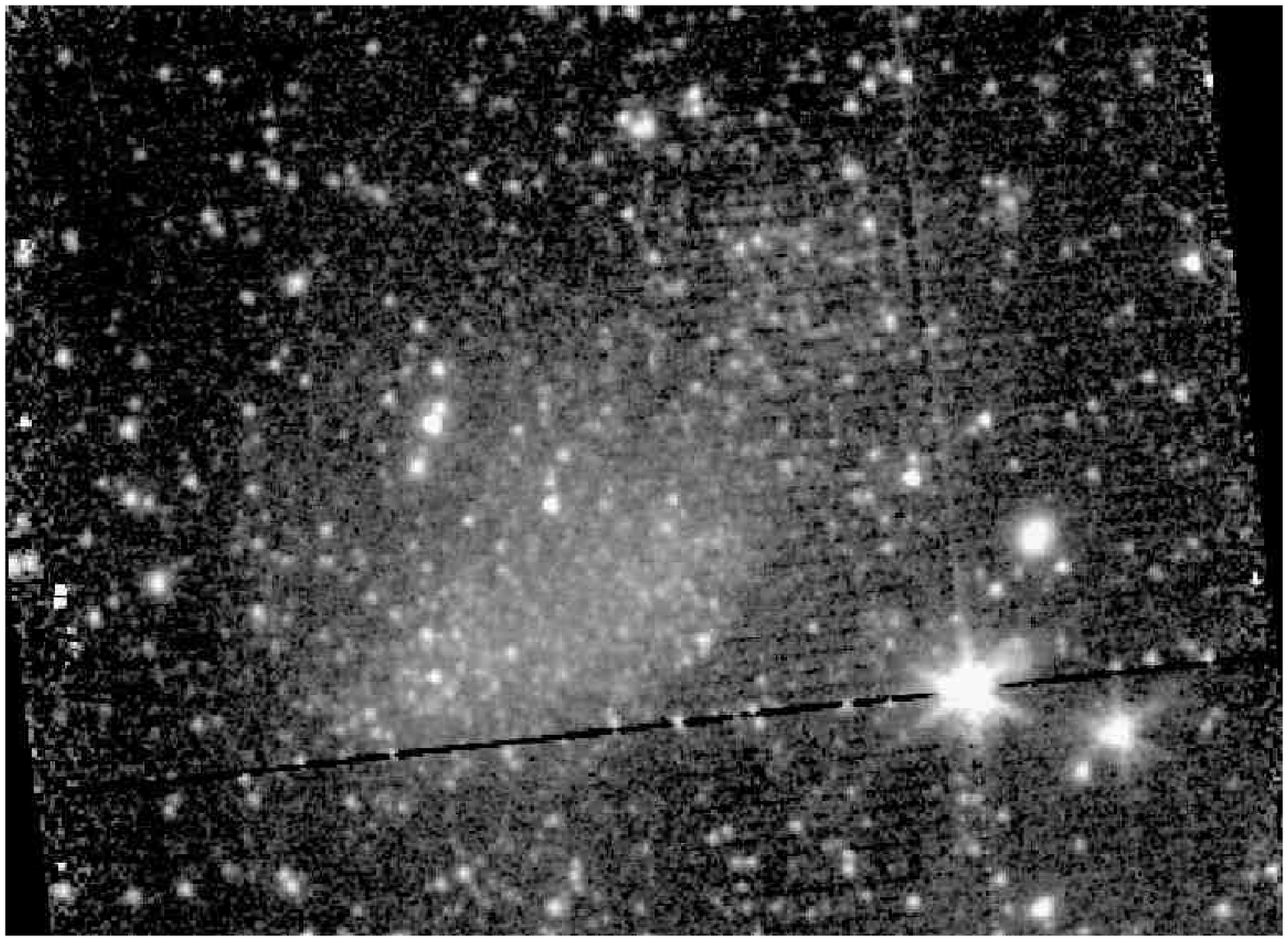}
 \vspace{2.0truecm}
 \caption{
{\bf ESO  245-  5}              - S$^4$G mid-IR classification:    IAB(s)m                                               ; Filter: IRAC 3.6$\mu$m; North:   up, East: left; Field dimensions:   5.7$\times$  4.2 arcmin; Surface brightness range displayed: 18.5$-$28.0 mag arcsec$^{-2}$}                 
\label{ESO245-005}  
 \end{figure}
 
\clearpage
\begin{figure}
\figurenum{1.189}
\plotone{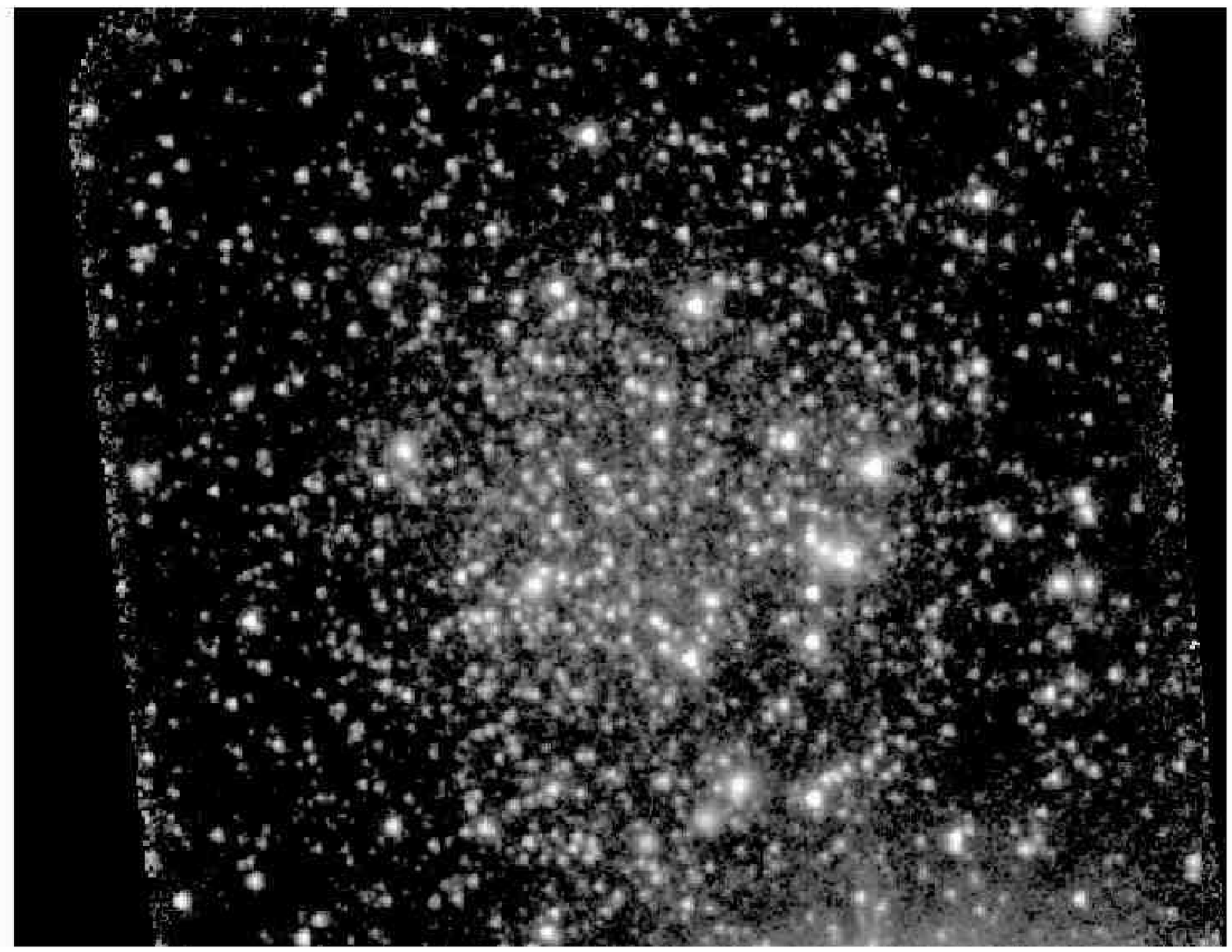}
 \vspace{2.0truecm}
 \caption{
{\bf ESO  245-  7}              - S$^4$G mid-IR classification:    Im                                                    ; Filter: IRAC 3.6$\mu$m; North:   up, East: left; Field dimensions:   6.6$\times$  4.8 arcmin; Surface brightness range displayed: 18.5$-$28.0 mag arcsec$^{-2}$}                 
\label{ESO245-007}  
 \end{figure}
 
\clearpage
\begin{figure}
\figurenum{1.190}
\plotone{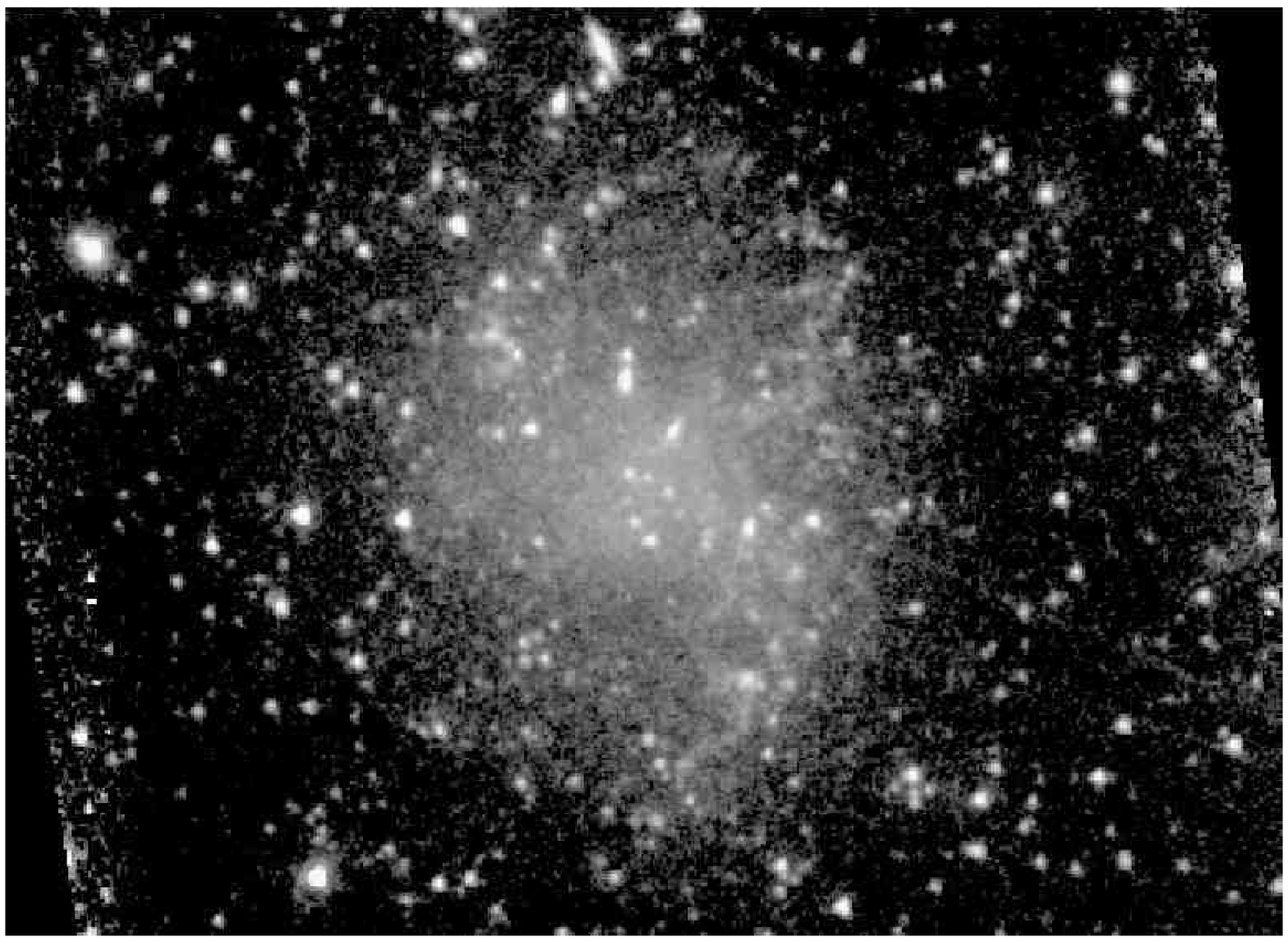}
 \vspace{2.0truecm}
 \caption{
{\bf ESO  362-  9}              - S$^4$G mid-IR classification:    S$\underline{\rm A}$B(s)dm                            ; Filter: IRAC 3.6$\mu$m; North:   up, East: left; Field dimensions:   5.8$\times$  4.2 arcmin; Surface brightness range displayed: 18.5$-$28.0 mag arcsec$^{-2}$}                 
\label{ESO362-009}  
 \end{figure}
 
\clearpage
\begin{figure}
\figurenum{1.191}
\plotone{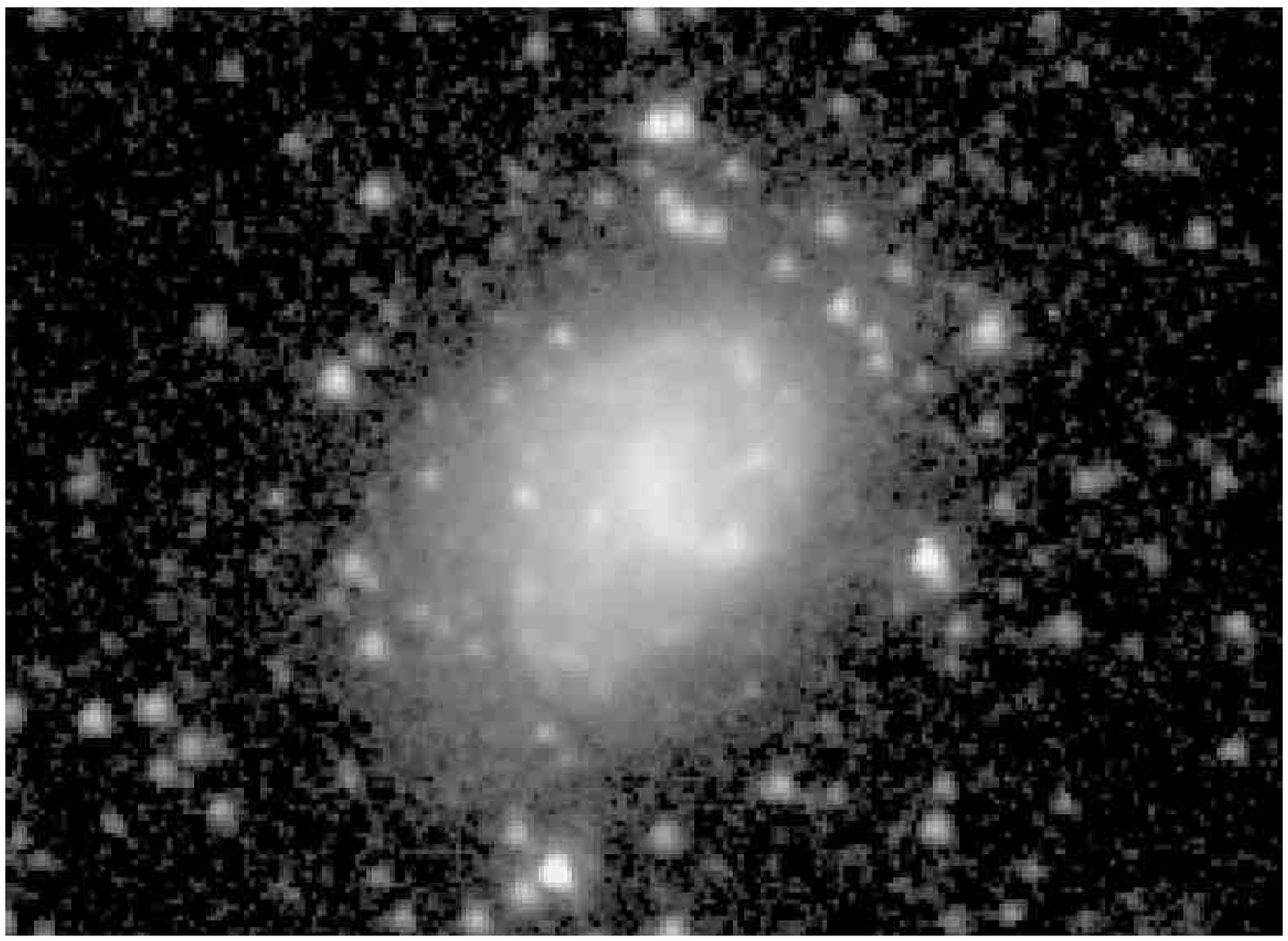}
 \vspace{2.0truecm}
 \caption{
{\bf ESO  418-  8}              - S$^4$G mid-IR classification:    (R$^{\prime}$)SB(s)dm:                                          ; Filter: IRAC 3.6$\mu$m; North:   up, East: left; Field dimensions:   2.9$\times$  2.1 arcmin; Surface brightness range displayed: 17.5$-$28.0 mag arcsec$^{-2}$}                 
\label{ESO418-008}  
 \end{figure}
 
\clearpage
\begin{figure}
\figurenum{1.192}
\plotone{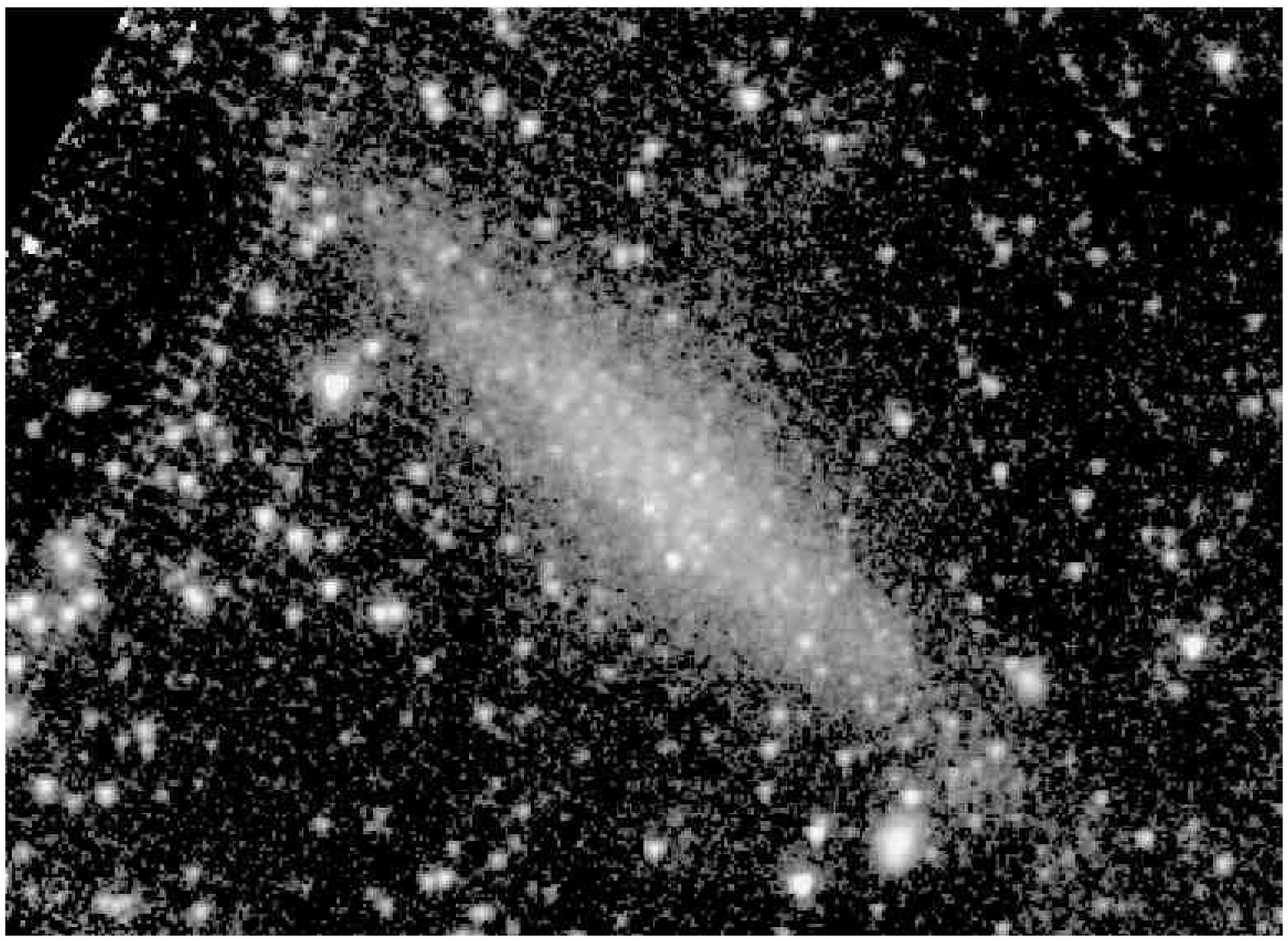}
 \vspace{2.0truecm}
 \caption{
{\bf ESO  471-  6}              - S$^4$G mid-IR classification:    SBd: sp                                               ; Filter: IRAC 3.6$\mu$m; North:   up, East: left; Field dimensions:   4.9$\times$  3.6 arcmin; Surface brightness range displayed: 18.5$-$28.0 mag arcsec$^{-2}$}                 
\label{ESO471-006}  
 \end{figure}
 
\clearpage
\begin{figure}
\figurenum{1.193}
\plotone{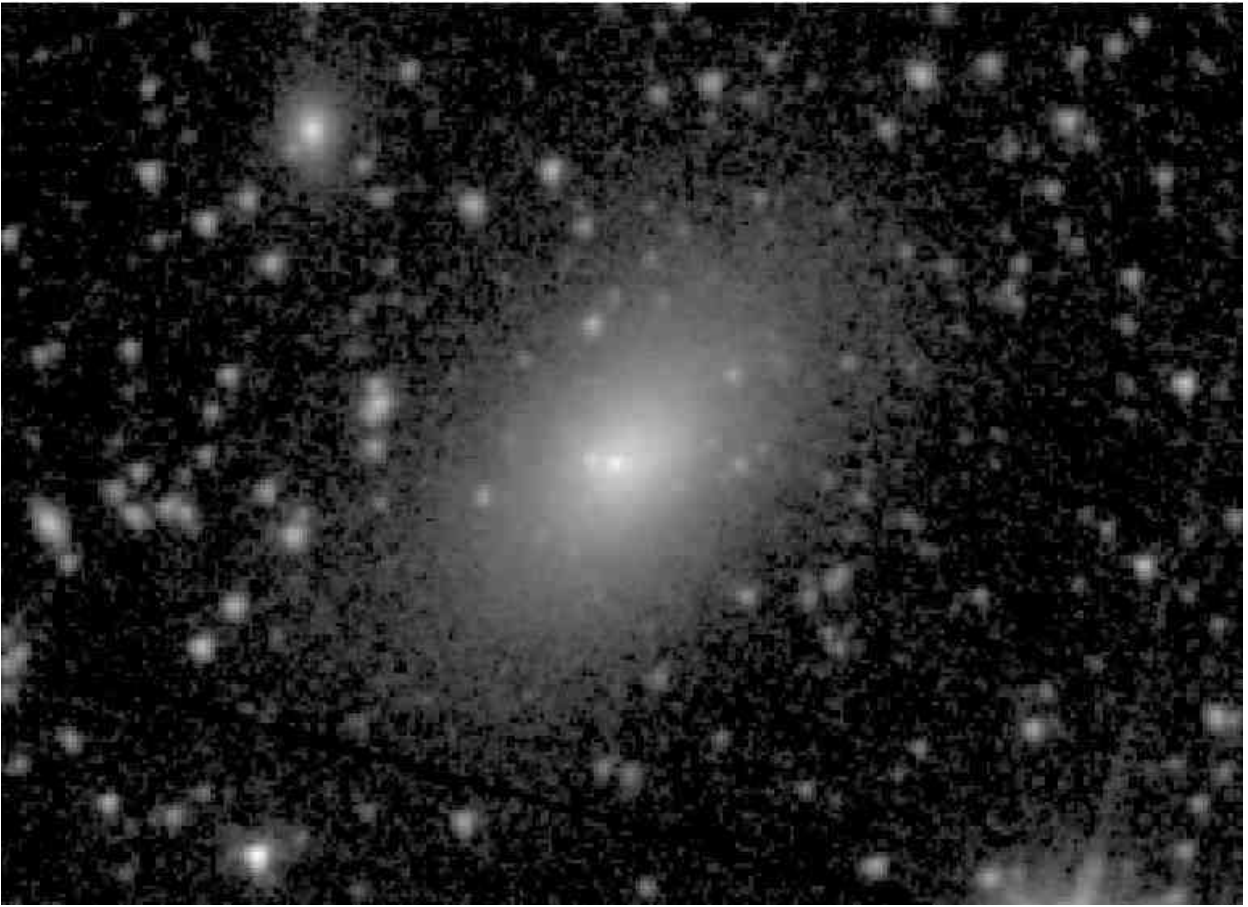}
 \vspace{2.0truecm}
 \caption{
{\bf ESO  483- 13}              - S$^4$G mid-IR classification:    dE4,N/SA0$^-$                                         ; Filter: IRAC 3.6$\mu$m; North:   up, East: left; Field dimensions:   3.7$\times$  2.7 arcmin; Surface brightness range displayed: 15.5$-$28.0 mag arcsec$^{-2}$}                 
\label{ESO483-013}  
 \end{figure}
 
\clearpage
\begin{figure}
\figurenum{1.194}
\plotone{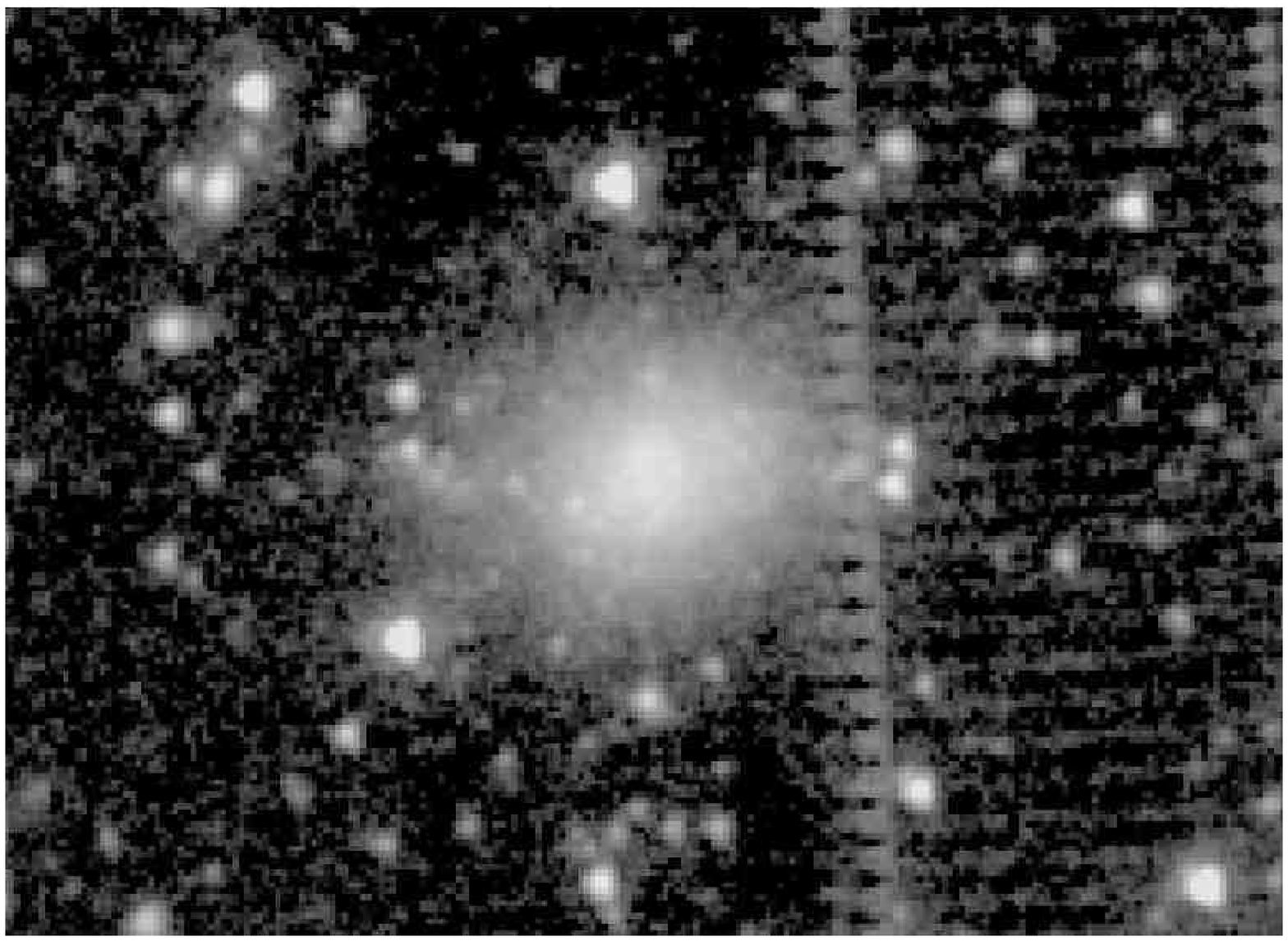}
 \vspace{2.0truecm}
 \caption{
{\bf ESO  486- 21}              - S$^4$G mid-IR classification:    dE (Im)                                               ; Filter: IRAC 3.6$\mu$m; North:   up, East: left; Field dimensions:   2.6$\times$  1.9 arcmin; Surface brightness range displayed: 18.0$-$26.0 mag arcsec$^{-2}$}                 
\label{ESO486-021}  
 \end{figure}
 
\clearpage
\begin{figure}
\figurenum{1.195}
\plotone{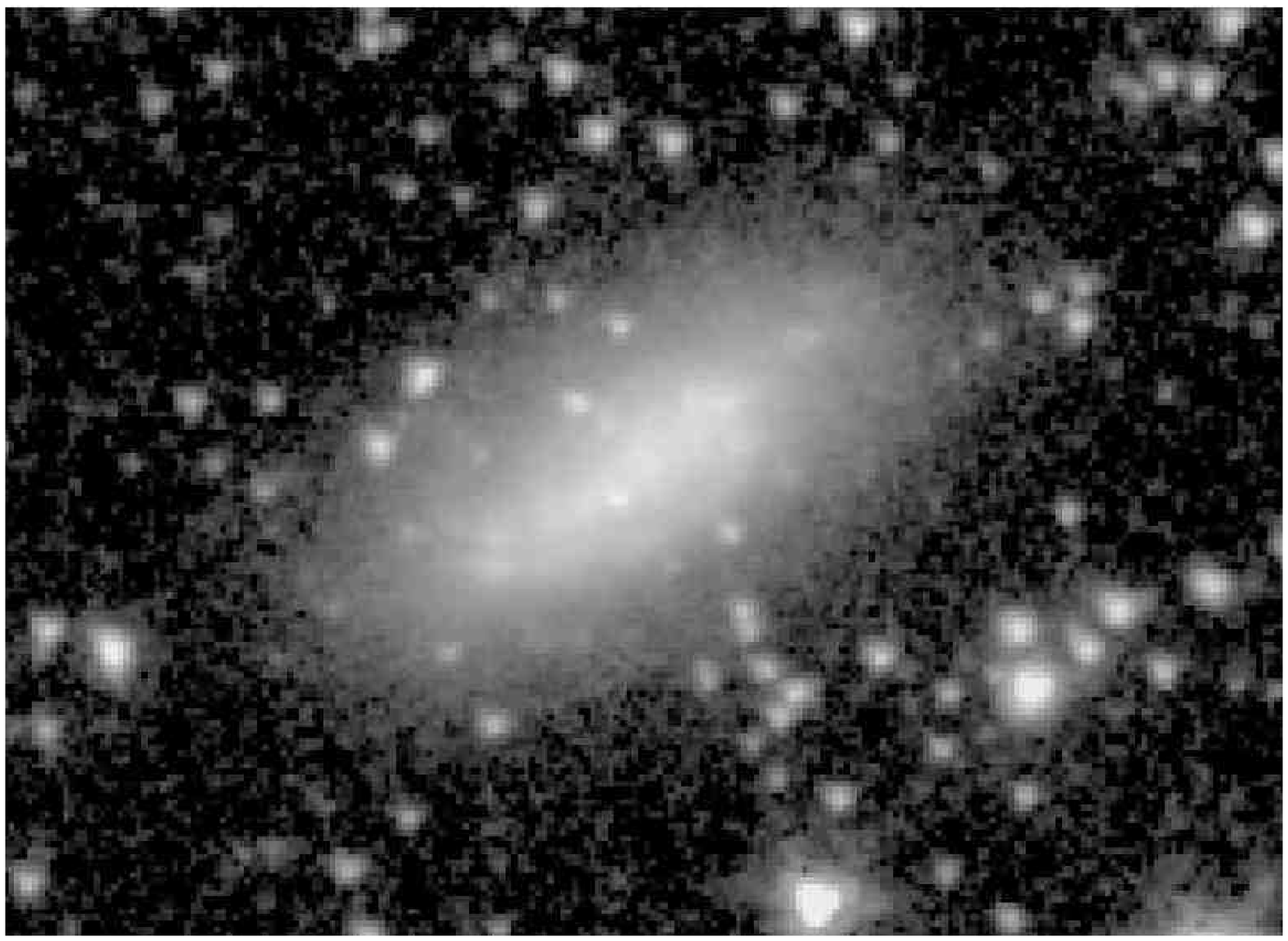}
 \vspace{2.0truecm}
 \caption{
{\bf ESO  503- 22}              - S$^4$G mid-IR classification:    SB(s)m:                                               ; Filter: IRAC 3.6$\mu$m; North:   up, East: left; Field dimensions:   2.6$\times$  1.9 arcmin; Surface brightness range displayed: 18.0$-$28.0 mag arcsec$^{-2}$}                 
\label{ESO503-022}  
 \end{figure}
 
\clearpage
\begin{figure}
\figurenum{1.196}
\plotone{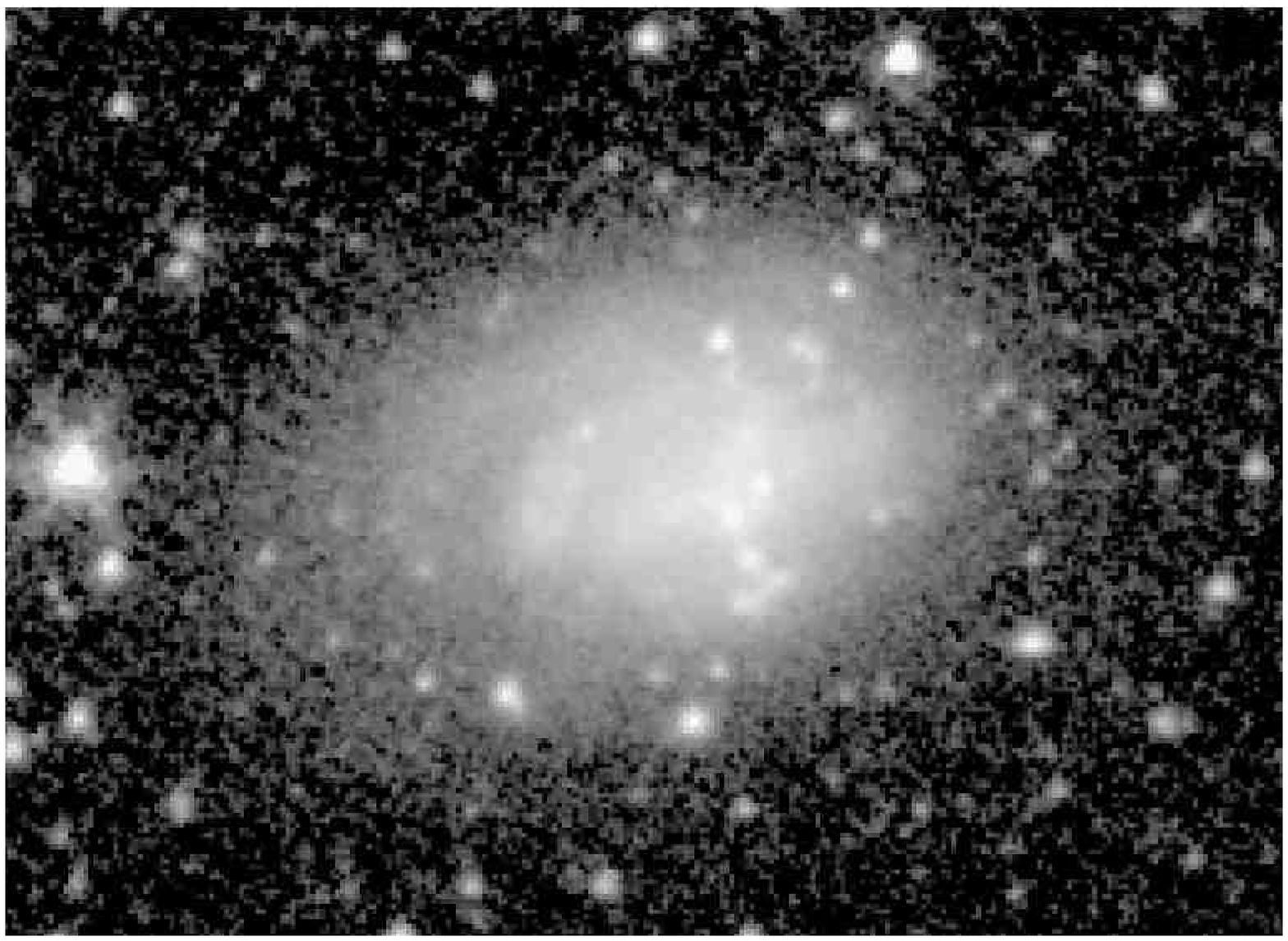}
 \vspace{2.0truecm}
 \caption{
{\bf ESO  544- 30}              - S$^4$G mid-IR classification:    SAB(s)m                                               ; Filter: IRAC 3.6$\mu$m; North:   up, East: left; Field dimensions:   3.2$\times$  2.3 arcmin; Surface brightness range displayed: 18.0$-$28.0 mag arcsec$^{-2}$}                 
\label{ESO544-030}  
 \end{figure}
 
\clearpage
\begin{figure}
\figurenum{1.197}
\plotone{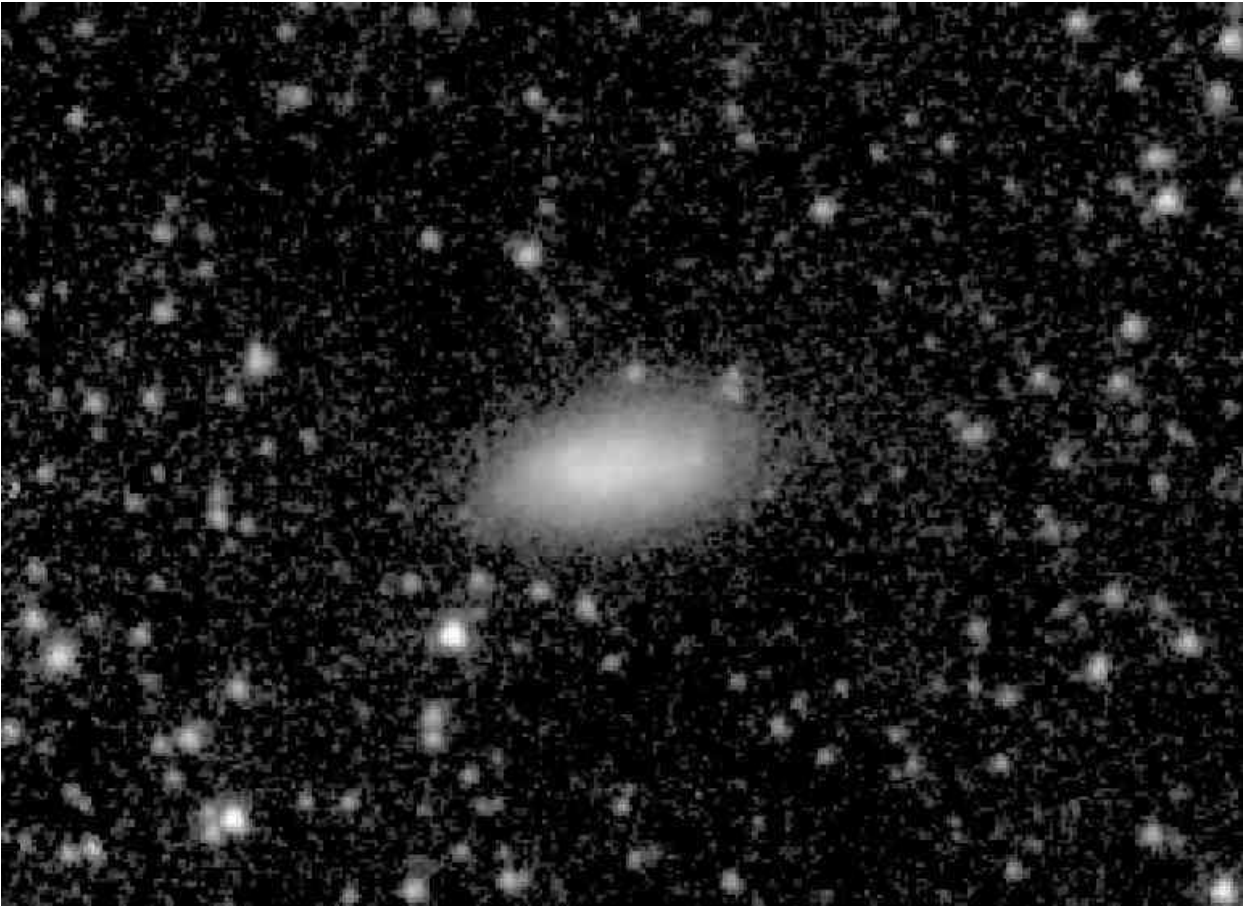}
 \vspace{2.0truecm}
 \caption{
{\bf CGCG265-55  }              - S$^4$G mid-IR classification:    dE (Im)                                               ; Filter: IRAC 3.6$\mu$m; North:   up, East: left; Field dimensions:   4.0$\times$  2.9 arcmin; Surface brightness range displayed: 16.5$-$28.0 mag arcsec$^{-2}$}                 
\label{CGCG265-55}  
 \end{figure}
 
\clearpage
\begin{figure}
\figurenum{2}
\vspace{-0.25truein}
\plotone{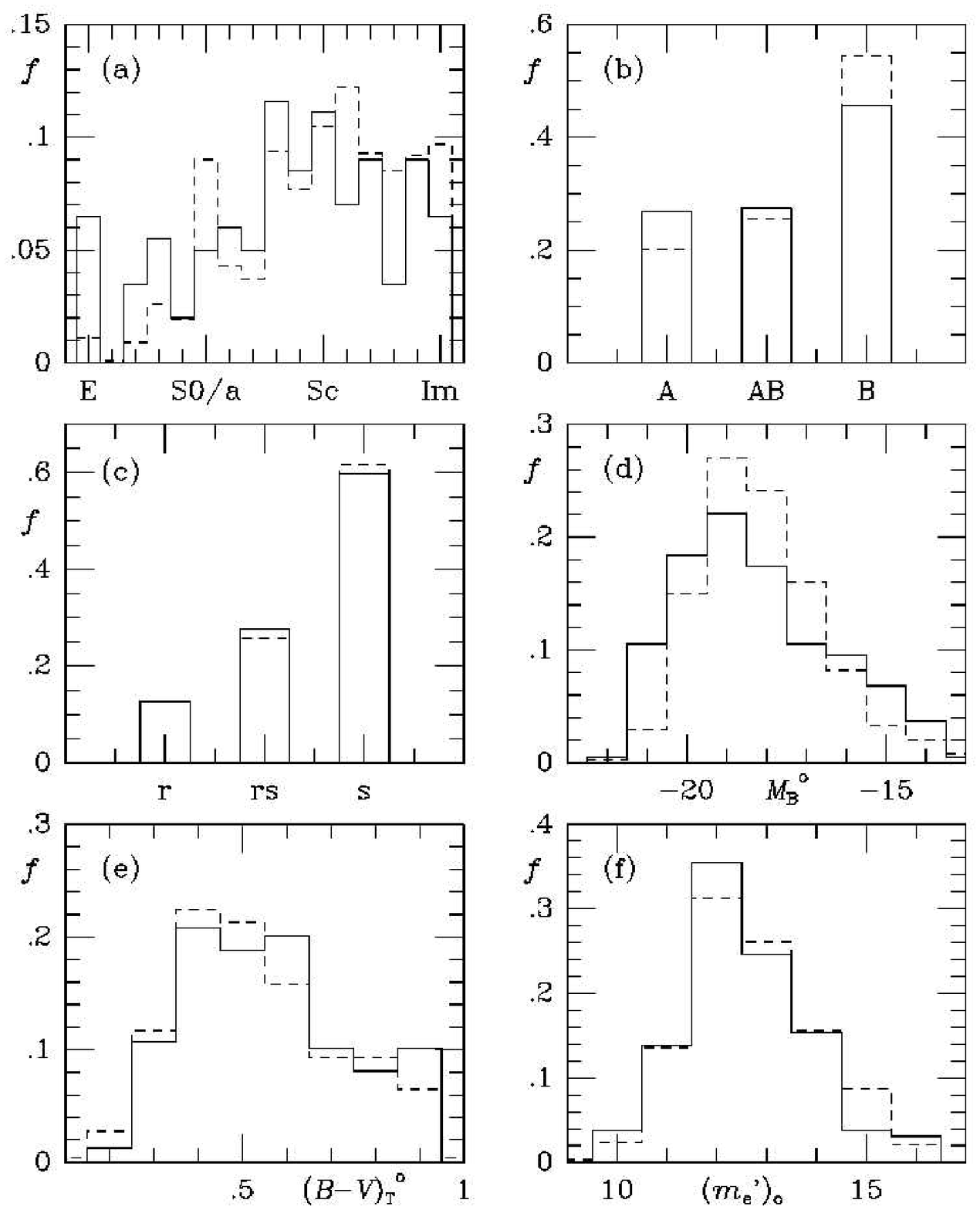}
\caption{Histograms of the relative numbers of galaxies in our subset
(solid lines) and the full S$^4$G sample (dashed lines) versus RC3 (a) 
stage; (b) family classification; (c) variety classification;
(d) absolute blue magnitude (based on the corrected RC3 total
magnitude, a distance derived from the radial velocity relative to the
Galactic Standard of rest, and a Hubble constant of 73 km s$^{-1}$
Mpc$^{-1}$); (e) total extinction-corrected $B-V$ color index;
and (f) extinction-corrected mean effective blue light surface
brightness (mag arcmin$^{-2}$).}
\label{histos}
\end{figure}

\clearpage
\begin{figure}
\figurenum{3}
\vspace{-1.5truein}
\plotone{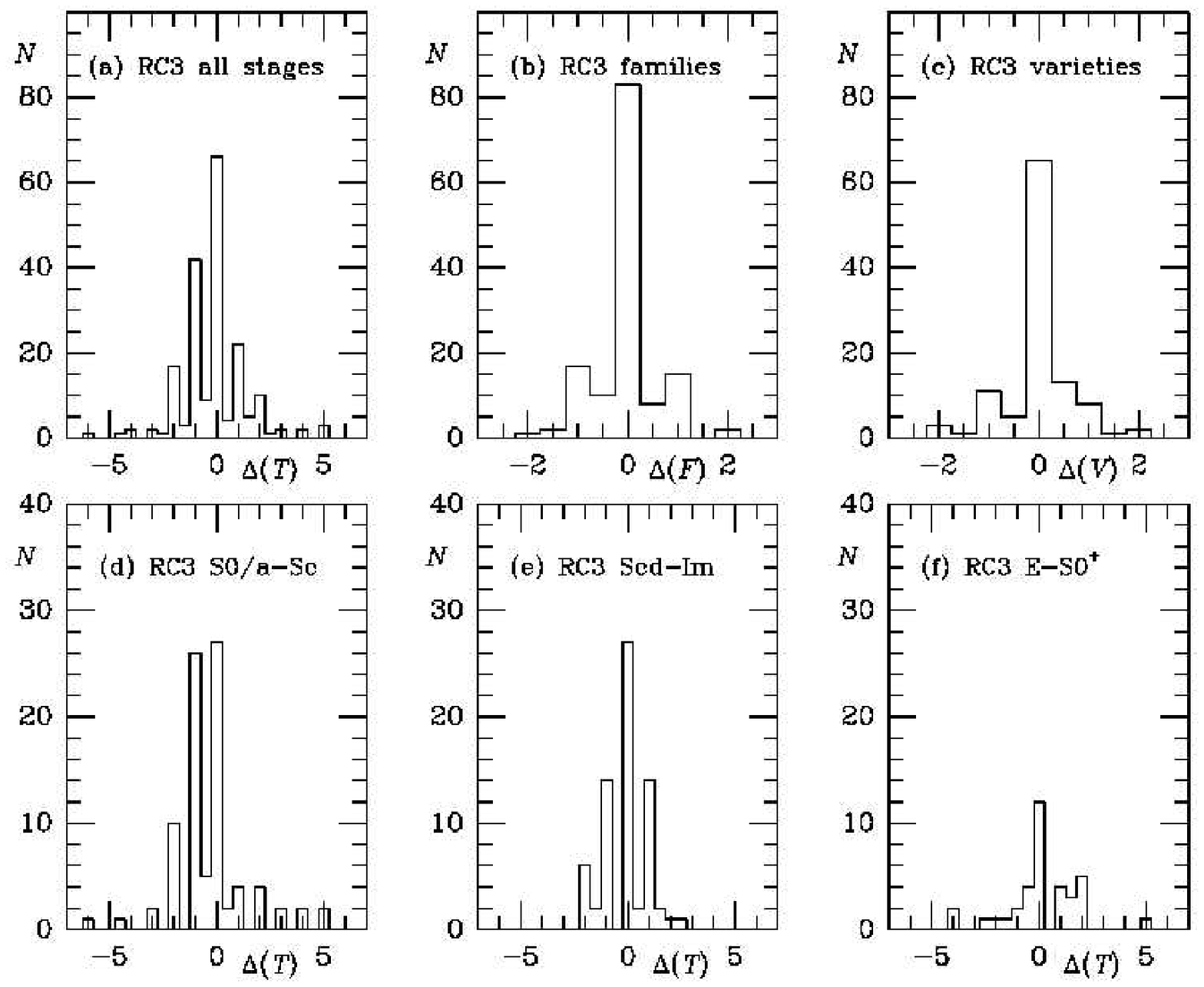}
\caption{Histograms of morphological index differences $\Delta (T,F,V)
= T,F,V(S^4G)-T,F,V(RC3)$, where $T$ is the numerical stage index, $F$
is the numerical family index, and $V$ is the numerical variety index.
The stage index ranges from $T$=$-$5 for E galaxies to +10 for
magellanic irregulars, Im. The family index is $F$=$-$1 for SA
galaxies, 0 for SAB galaxies, and +1 for SB galaxies.  The variety
index is $V$=$-$1 for (r), 0 for (rs), and +1 for (s) galaxies.
Types with underlines (e.g., Sb$\underline{\rm c}$, SA$\underline{\rm B}$,
($\underline{\rm r }$s), etc.) are assigned half steps. Varieties
like (rl) or (l) are not assigned numerical indices.}
\label{rc3comp}
\end{figure}

\clearpage
\begin{figure}
\figurenum{4}
\vspace{2.0truein}
\plotone{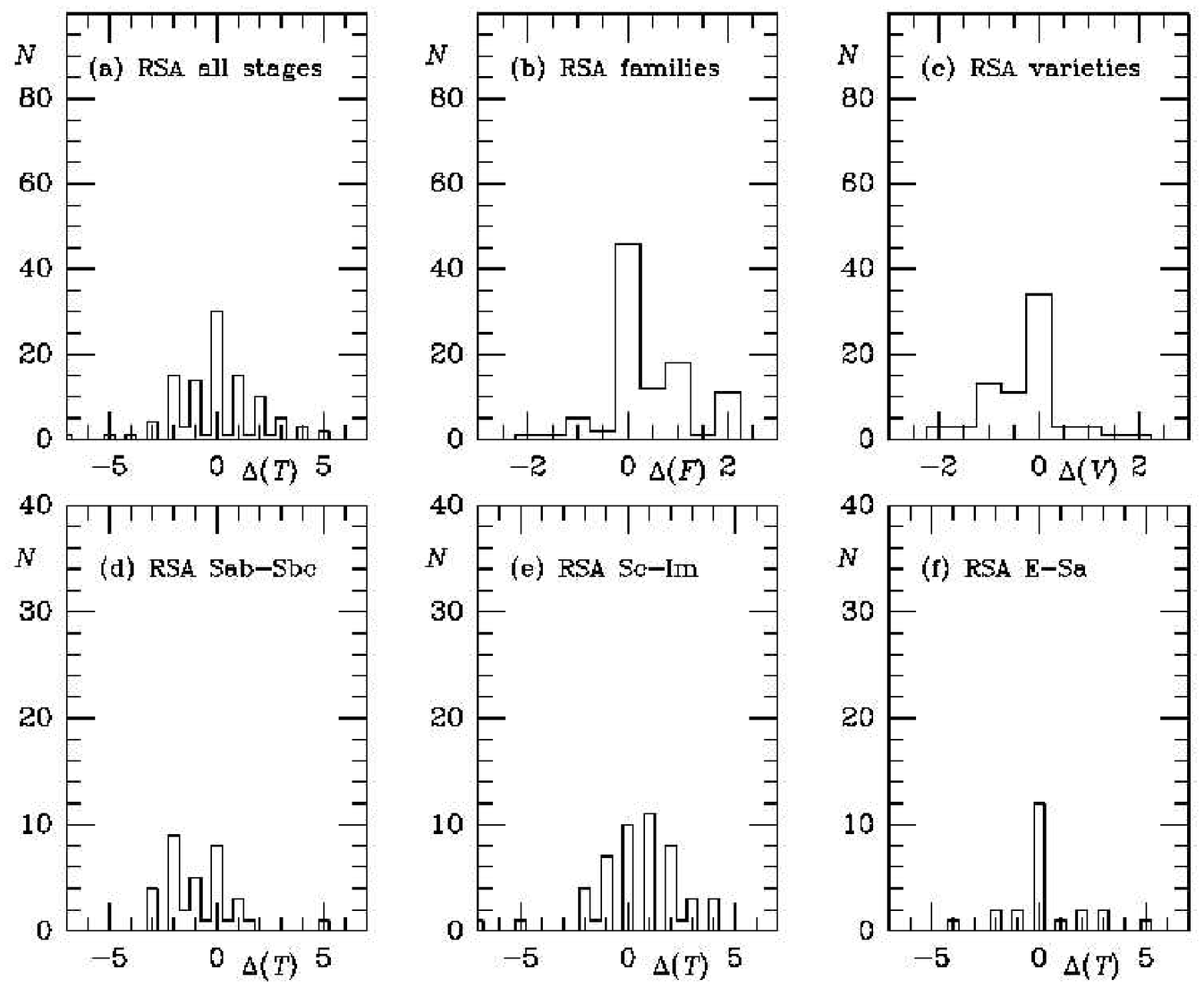}
\caption{Same as Figure 3, but for RSA types, families, and varieties.}
\label{rsacomp}
\end{figure}

\clearpage
\begin{figure}
\figurenum{5}
\vspace{2.0truein}
\plotone{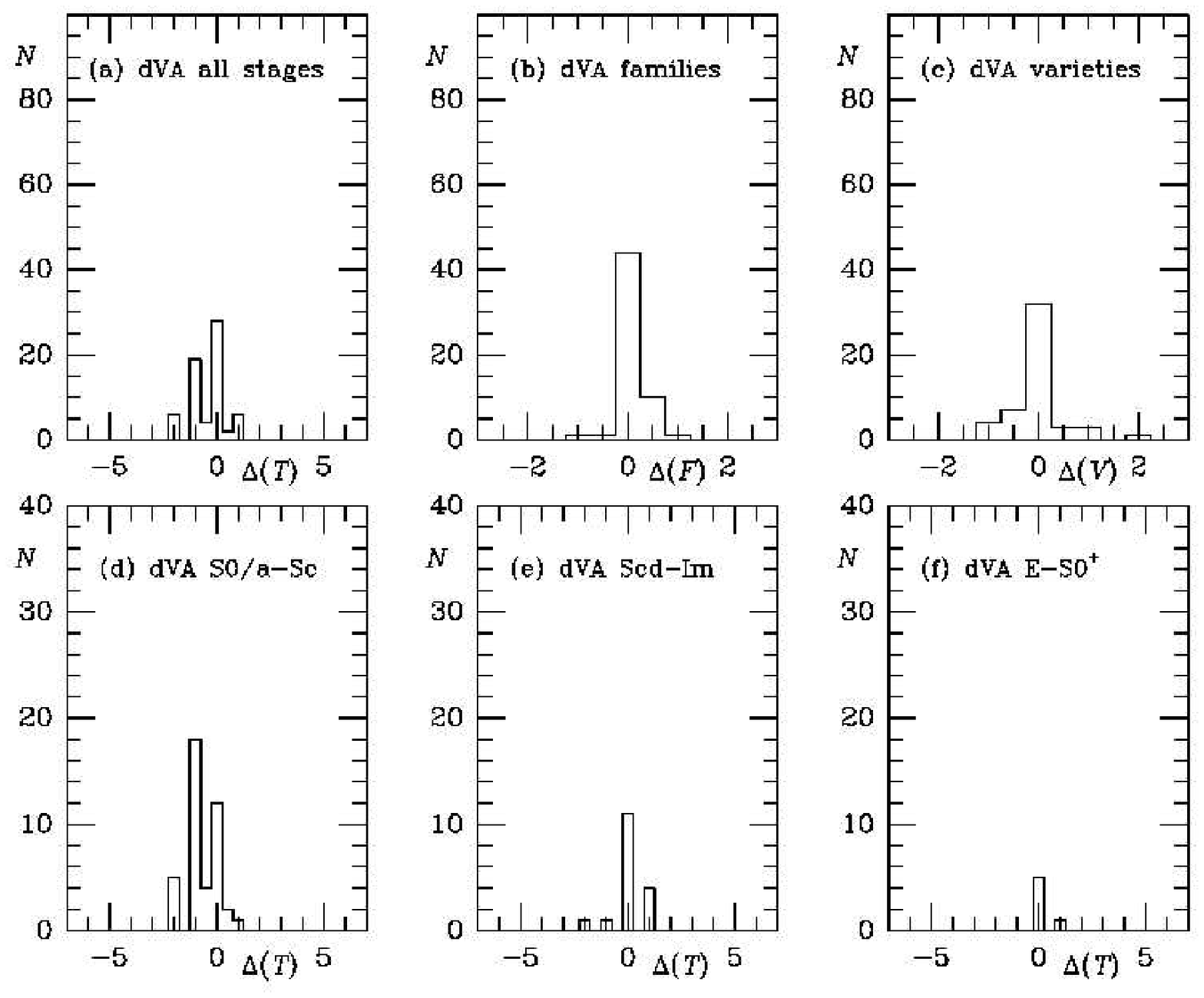}
\caption{Same as Figure 3, but for dVA types, families, and varieties.}
\label{dvacomp}
\end{figure}

\clearpage
\begin{figure}
\figurenum{6}
\vspace{0.0truein}
\plotone{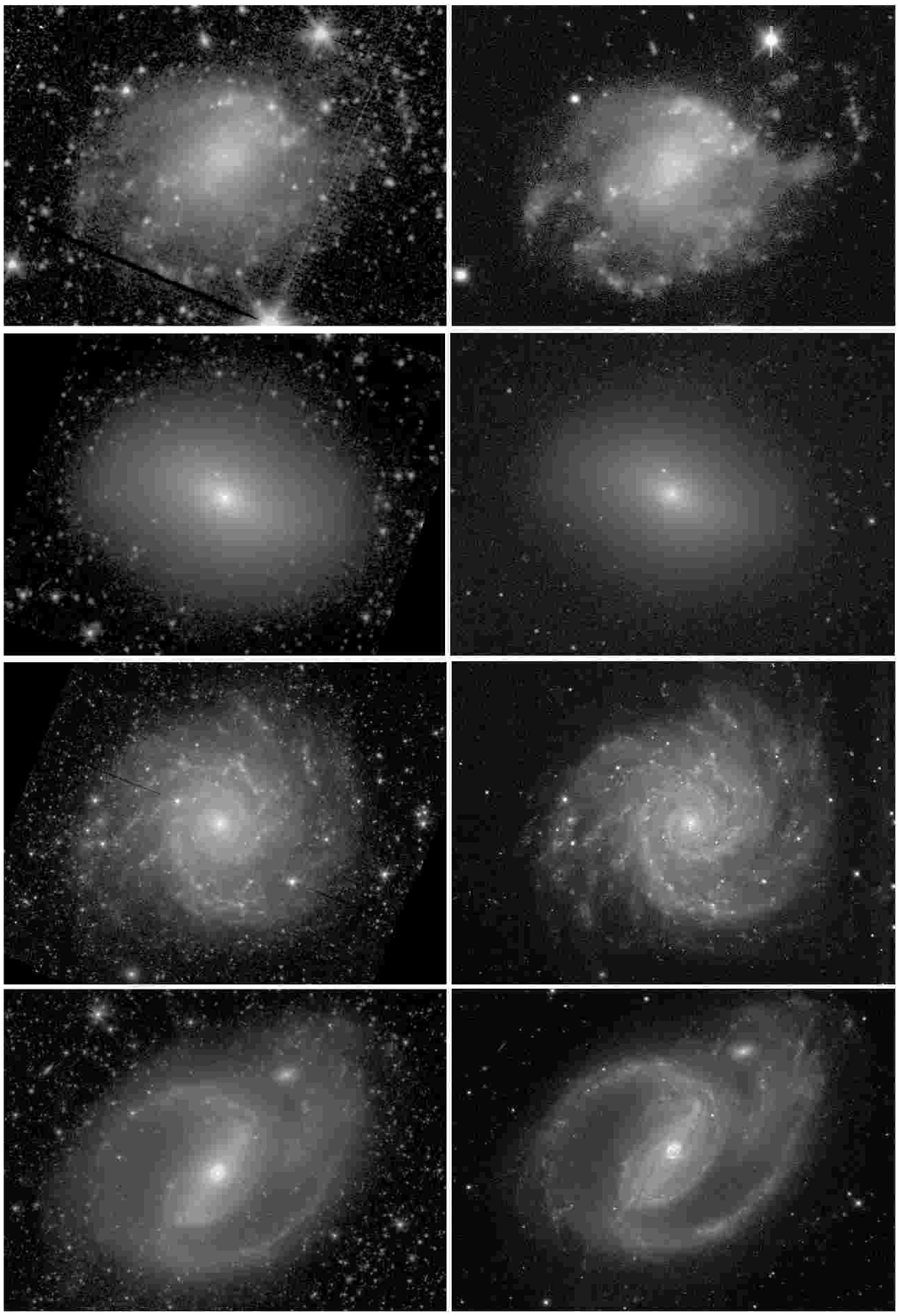}
\caption{}
\label{montage1}
\end{figure}
\begin{figure}
\figurenum{6 (cont.)}
\caption{Comparison between 3.6$\mu$m images (left) and $B$-band images (right) for
(top-to-bottom) NGC 428, NGC 584, NGC 628, and NGC 1097. All of the images are
in units of mag arcsec$^{-2}$, and the $B$-band images are from the dVA.}
\end{figure}

\clearpage
\begin{figure}
\figurenum{7}
\vspace{0.0truein}
\plotone{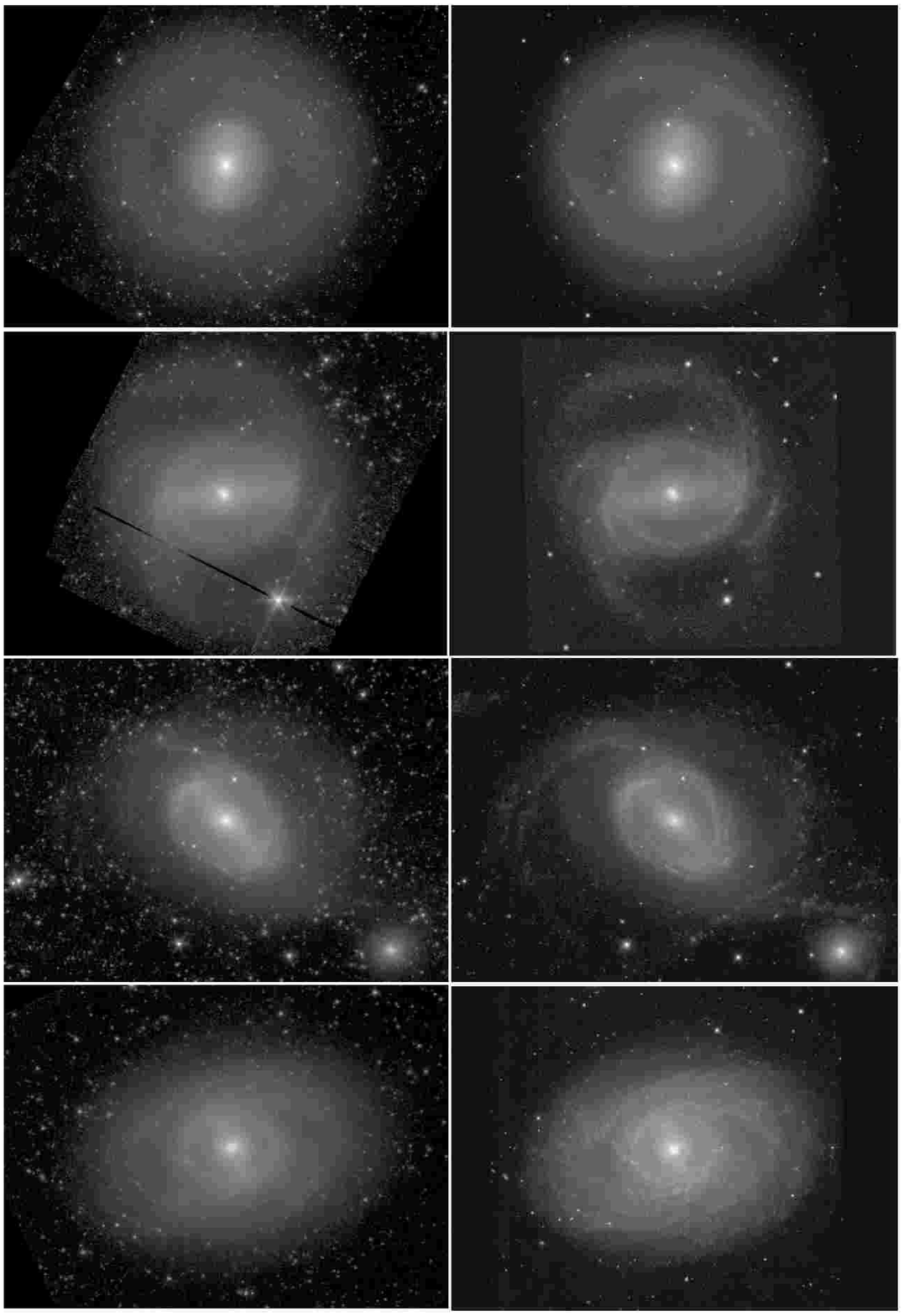}
\caption{}
\label{montage2}
\end{figure}
\begin{figure}
\figurenum{7 (cont.)}
\caption{Comparison between 3.6$\mu$m images (left) and $B$-band images
(right) for four ringed galaxies (top-to-bottom):  NGC 1291, NGC 1433,
NGC 1512, and NGC 3351. All of the images are in units of mag
arcsec$^{-2}$, and the $B$-band images are from the dVA.}
\end{figure}

\clearpage
\begin{figure}
\figurenum{8}
\vspace{0.0truein}
\plotone{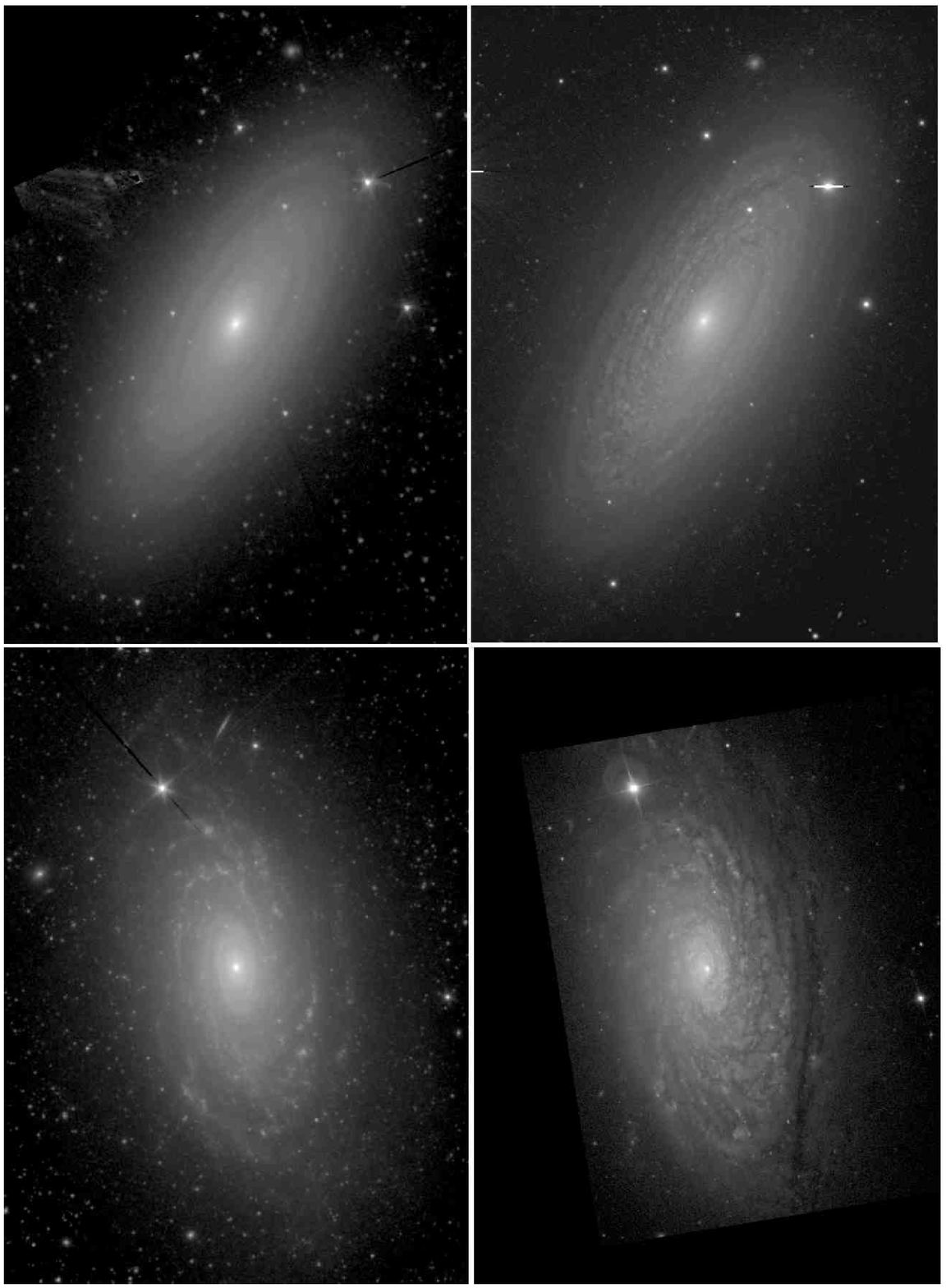}
\caption{}
\label{montage3}
\end{figure}
\begin{figure}
\figurenum{8 (cont.)}
\caption{Comparison between 3.6$\mu$m images (left) and optical images
(right) for two flocculent spirals (top-to-bottom):  NGC 2841, NGC
5055. All of the images are in units of mag arcsec$^{-2}$.
The optical image of NGC 2841 is $B$-band and is from the dVA. The optical
image of NGC 5055 is $g$-band and is from the SDSS.}
\end{figure}

\clearpage
\begin{figure}
\figurenum{9}
\vspace{0.0truein}
\plotone{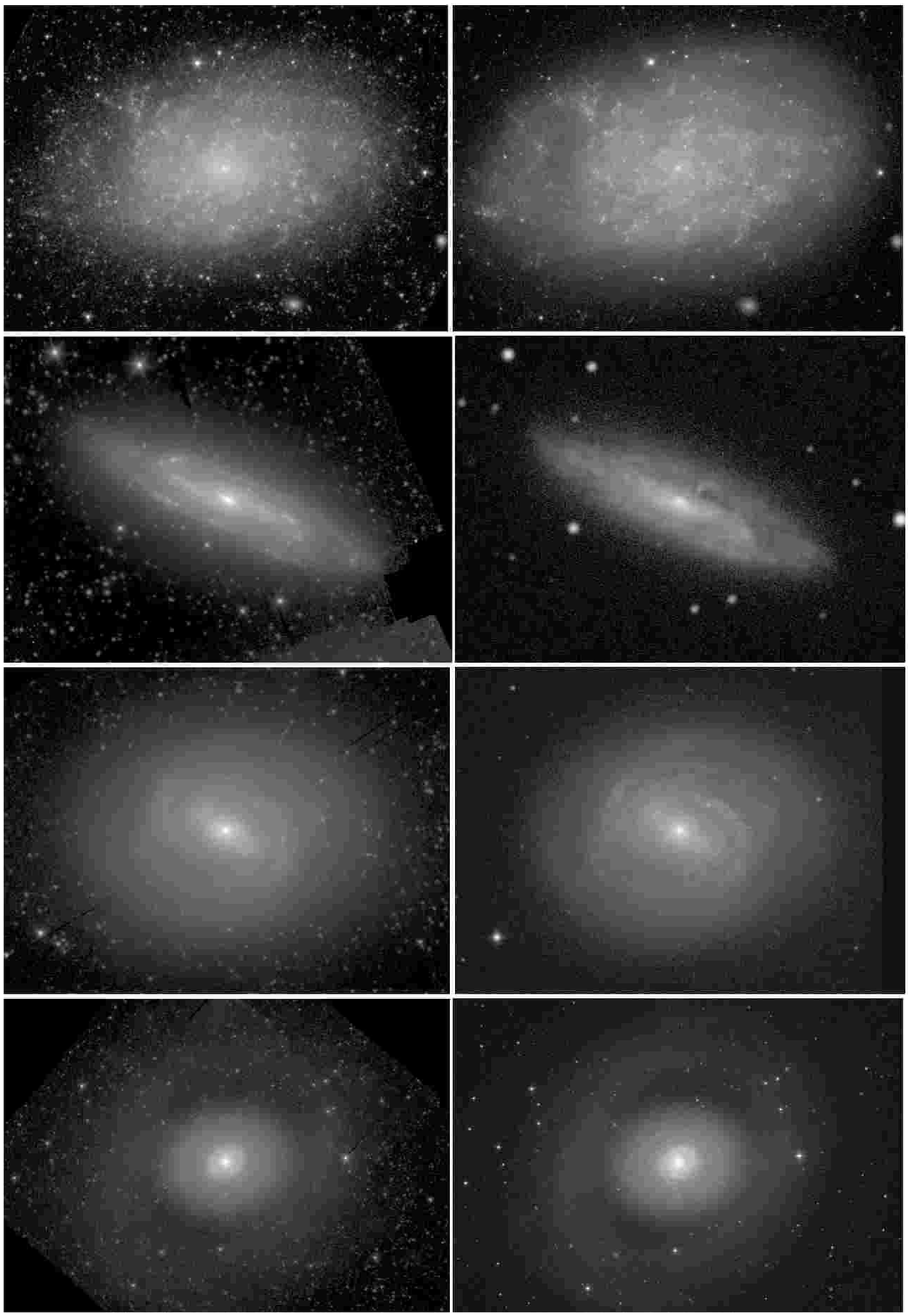}
\caption{}
\label{montage4}
\end{figure}
\begin{figure}
\figurenum{9 (cont.)}
\caption{Comparison between 3.6$\mu$m images (left) and $B$-band images
(right) for four late to early-type galaxies (top-to-bottom): NGC 7793, NGC
4527, NGC 4579, and NGC 4736. All of the images are in units of mag
arcsec$^{-2}$. The $B$-band images are from the dVA.}
\end{figure}

\clearpage
\begin{figure}
\figurenum{10}
\vspace{0.0truein}
\plotone{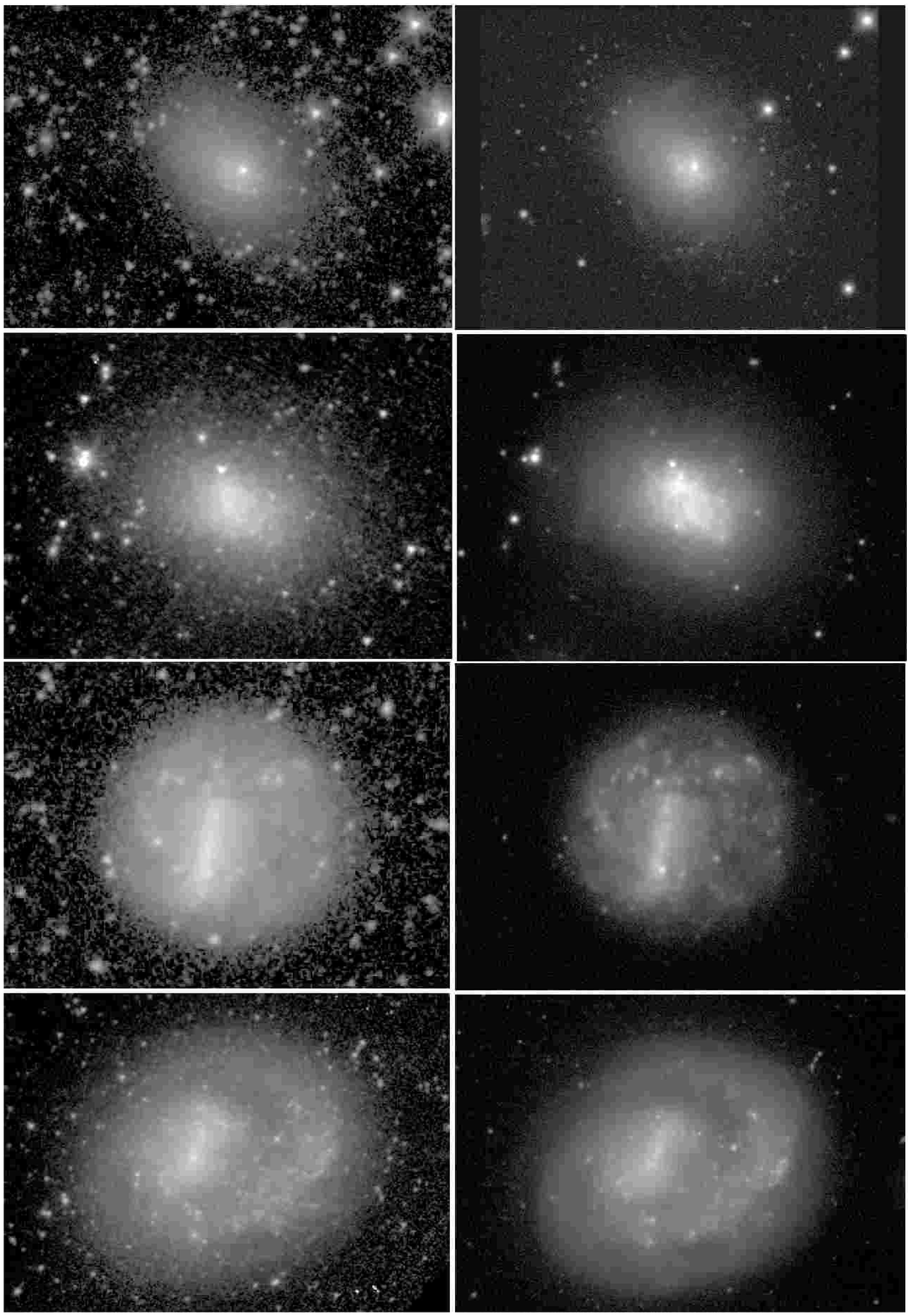}
\caption{}
\label{montage5}
\end{figure}
\begin{figure}
\figurenum{10 (cont.)}
\caption{Comparison between 3.6$\mu$m images (left) and $B$-band images
(right) for four late-type galaxies (top-to-bottom): NGC 1705, NGC
3738, NGC 3906, and NGC 4618. All of the images are in units of mag
arcsec$^{-2}$. The $B$-band images of NGC 1705, 3906, and 4618 are
from the dVA. The $B$-band image of NGC 3738 is from Taylor et al. (2005).}
\end{figure}

\clearpage
\begin{figure}
\figurenum{11}
\vspace{0.0truein}
\plotone{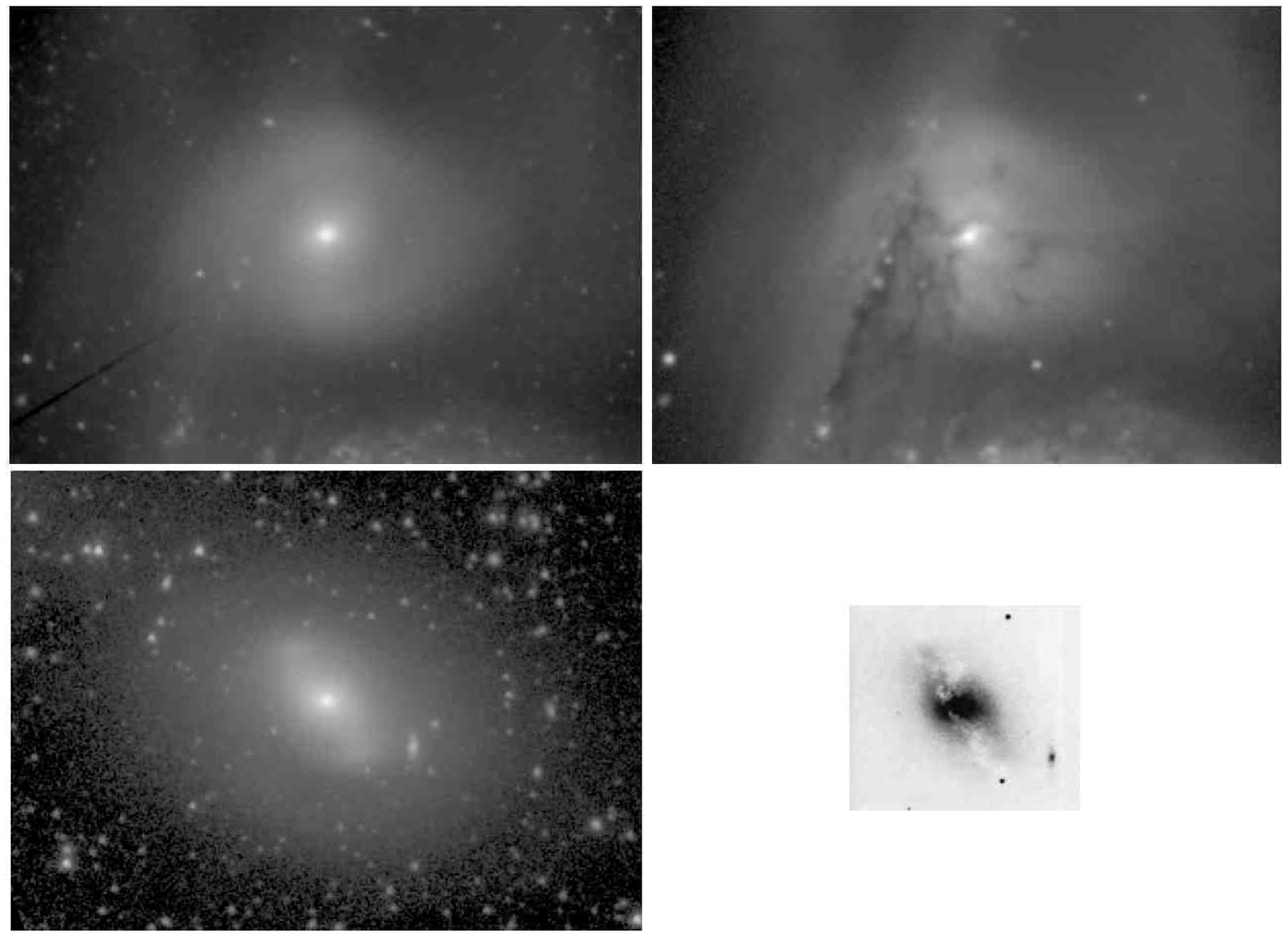}
\caption{}
\label{montage6}
\end{figure}
\begin{figure}
\figurenum{11 (cont.)}
\caption{Comparison between 3.6$\mu$m images (left) and $B$-band
images (right) of two I0 galaxies: top panels: NGC 5195; bottom
panels: NGC 2968. The $B$-band image of NGC 2968 is from the Carnegie
Atlas of Galaxies (Sandage \& Bedke 1994).} 
\end{figure}

\clearpage
\begin{figure}
\figurenum{12}
\vspace{0.0truein}
\plotone{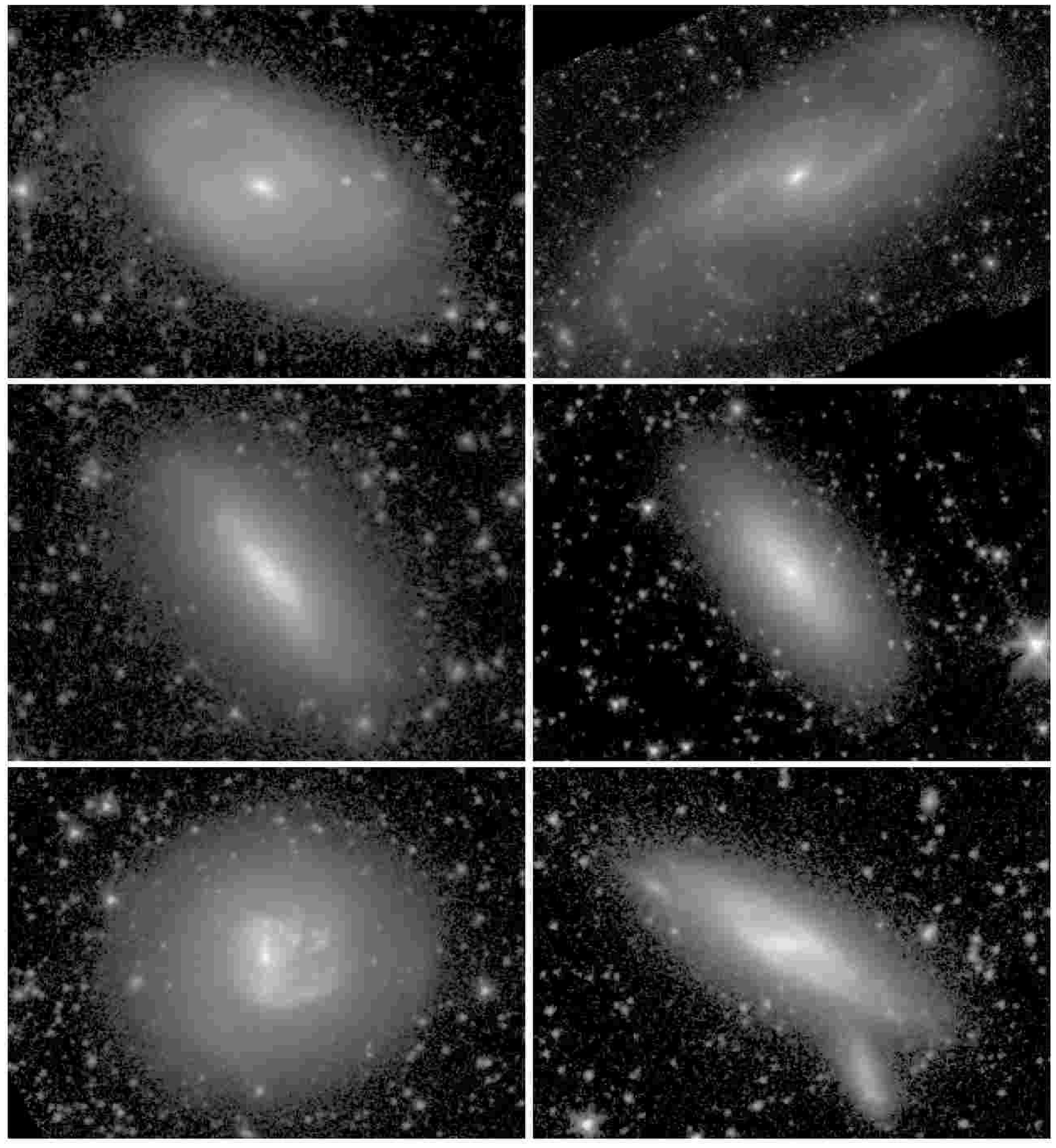}
\caption{}
\label{montage7}
\end{figure}
\begin{figure}
\figurenum{12 (cont.)}
\caption{3.6$\mu$m images of six special cases: top panels: NGC 470 (left)
and 4536 (right), showing prominent, highly-flattened pseudobulges. Middle
panels: IC 750 (left) and IC 3392 (right), two ``double-stage" galaxies
having bright inner spirals imbedded in an S0 or S0/a-like background disk.
Bottom panels: NGC 5713 (left) and NGC 3769 (right), two more double-stage
galaxies.}
\end{figure}

\clearpage
\begin{figure}
\figurenum{13}
\vspace{2.0truein}
\plotone{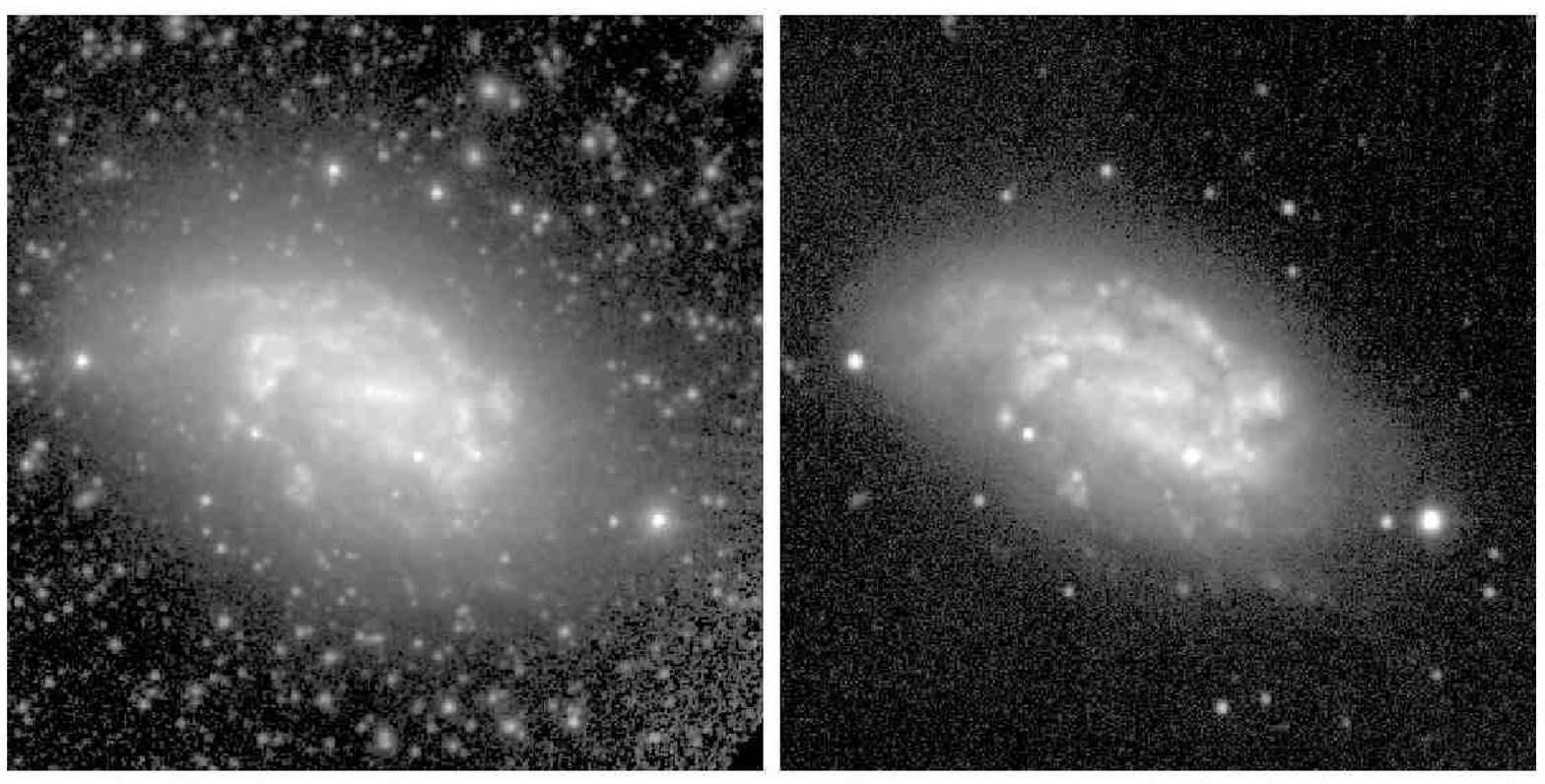}
\caption{Comparison between 3.6$\mu$m (left) and $B$-band images
of the late-type spiral galaxy NGC 1559, showing the strong
correspondence
between 3.6$\mu$m resolved sources and $B$-band star-forming regions.
The $B$-band image is from the OSUBSGS (Eskridge et al. 2002).}
\label{ngc1559}
\end{figure}

\clearpage
\begin{figure}
\figurenum{14}
\vspace{2.0truein}
\plotone{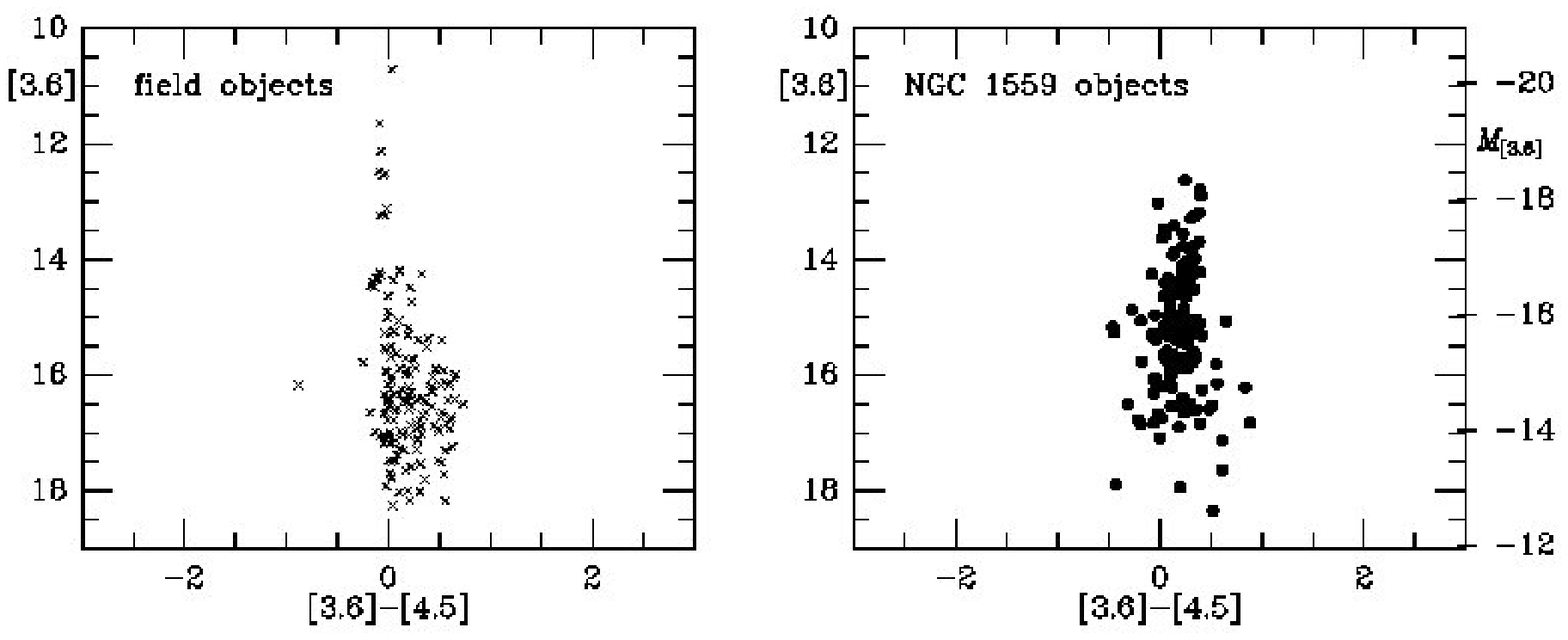}
\caption{Color-magnitude diagrams for foreground stars and background galaxies
(left panel) and resolved objects in the spiral arms of NGC 1559 (right panel).}
\label{ngc1559objects}
\end{figure}

\clearpage
\begin{figure}
\figurenum{15}
\plotone{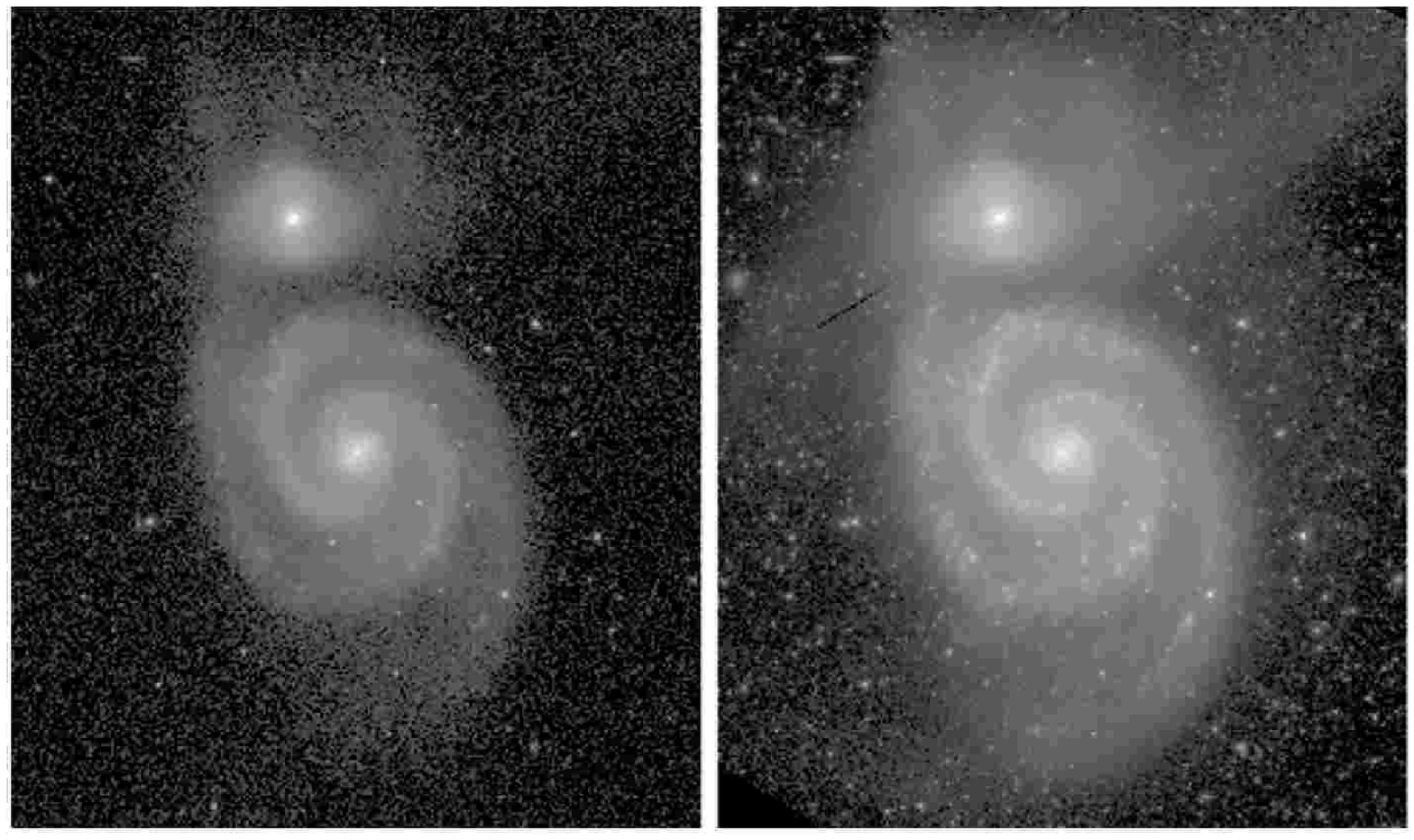}
\caption{A comparison of the morphology of the M51 system
in the 2.2$\mu$m $K_s$ band (left frame) and the IRAC 3.6$\mu$m
band.}
\label{m51}
\end{figure}

\end{document}